\documentclass[12pt,reqno]{article}

\usepackage[numbers,sort&compress]{natbib}
\usepackage{amsthm, amsmath, amsfonts, amssymb, amscd, mathtools, youngtab, euscript, mathrsfs, verbatim, enumerate, multicol, multirow, bbding, color, babel, esint, geometry, tikz, tikz-cd, tikz-3dplot, array, enumitem, hyperref, thm-restate, thmtools, datetime, graphicx, tensor, braket, slashed, standalone, pgfplots, ytableau, subfigure, wrapfig, dsfont, setspace, wasysym, pifont, float, rotating, adjustbox, pict2e,array}
\usepackage[para]{threeparttable}
\usepackage{longtable}
\usepackage{tabularx}
\usepackage{caption}
\usepackage{booktabs}
\usepackage{amsmath}
\usepackage{etoolbox}
\usepackage{makecell}
\usepackage{blindtext}
\usepackage[all]{xy}
\usepackage[utf8]{inputenc}
\usepackage[normalem]{ulem}
\makeatletter
\numberwithin{equation}{section}
\numberwithin{figure}{section}
\usetikzlibrary{arrows, positioning, decorations.pathmorphing, decorations.pathreplacing, decorations.markings, matrix, patterns}
\hypersetup{colorlinks=true}
\hypersetup{linkcolor=black}
\hypersetup{citecolor=black}
\hypersetup{urlcolor=black}
\theoremstyle{plain}
\newtheorem*{thm*}{\protect\theoremname}
\theoremstyle{definition}
\newtheorem*{defn*}{\protect\definitionname}
\providecommand{\definitionname}{Definition}
\providecommand{\theoremname}{Theorem}
\newcommand{\farsquare}[2]{#1\,{\mathpalette\far@square{#2}}}
\newcommand{\far@square}[2]{%
  \mathop{\vcenter{\hbox{%
    \sbox\z@{$\m@th#1\sum$}%
    \setlength{\unitlength}{2\dimexpr\ht\z@+\dp\z@}%
    \begin{picture}(1,1)
    \roundjoin
    \polyline(0,0)(0,1)(1,1)(1,0)(0,0)(0,0.5)
    \end{picture}%
  }}}\limits_{#1#2}%
}
\makeatother
\tikzset{big arrow/.style={
    decoration={markings,mark=at position 1 with {\arrow[scale=1.5,#1]{>}}},
    postaction={decorate},
    shorten >=0.4pt},
  big arrow/.default=black}
\allowdisplaybreaks
\numberwithin{equation}{part}

\numberwithin{figure}{part}

\numberwithin{table}{part}

\numberwithin{section}{part}
\renewcommand{\thesection}{\thepart.\arabic{section}}

\newcommand{\p}{\partial}
\newcommand{\pb}{\bar{\partial}}

\newcommand{\thetabar}{{\bar{\theta}}}
\newcommand{\jbar}{{\bar{j}}}
\newcommand{\ibar}{{\bar{i}}}
\newcommand{\partialbar}{{\bar{\partial}}}
\newcommand{\zbar}{{\bar{z}}}

\newcommand{\ZZ}{\mathbb{Z}}
\newcommand{\HH}{\mathbb{H}}
\newcommand{\RR}{\mathbb{R}}

\DeclareMathOperator{\tr}{tr}

\DeclareMathOperator{\vol}{vol}

\newcommand{\PP}{\mathds{P}}

\newcommand{\String}{\text{String}}
\newcommand{\Pin}{\text{Pin}}
\newcommand{\pt}{\text{pt}}
\newcounter{tempsection}
\newcommand{\sectionbypart}{%
  \setcounter{tempsection}{\value{section}}
  \setcounter{section}{0}
  \renewcommand{\thesection}{\arabic{section}}
  }
    {\endtcolorbox}
\usepackage{mdframed}
\mdfdefinestyle{statementstyle}{
   frametitlerule = true,
   frametitlefont = {\normalfont\bfseries},
   frametitlebackgroundcolor = blue!10,
}
\mdtheorem[style=statementstyle]{statement}{The Swampland Program}
\mdtheorem[style=statementstyle]{statement1}{Higher-form gauge symmetries}
\mdtheorem[style=statementstyle]{statement2}{Boundary symmetries in non-compact spaces}
\mdtheorem[style=statementstyle]{statement3}{Net gauge charge in compact spaces}
\mdtheorem[style=statementstyle]{statement4}{Baby universe hypothesis}
\mdtheorem[style=statementstyle]{statement6}{Distance conjecture}
\mdtheorem[style=statementstyle]{statement5}{Weak Gravity Conjecture (basic version)}
\mdtheorem[style=statementstyle]{statement7}{Particles vs instantons}
\mdtheorem[style=statementstyle]{statement8}{de Sitter conjecture}
\mdtheorem[style=statementstyle]{statement9}{Trans-Planckian censorship conjecture (TCC)}

\begin{document}\begin{titlepage}
\vspace*{-3cm} 
\begin{flushright}
{\tt CALT-TH-2022-042}\\
\end{flushright}
\begin{center}
\vspace{2.5cm}
{\LARGE\bfseries Lectures on the string landscape\\ and the Swampland \\  }
\vspace{2cm}
{\large
Nathan Benjamin Agmon{$^1$}, Alek Bedroya{$^1$}, Monica Jinwoo Kang{$^2$}, and Cumrun Vafa{$^1$} \\}
\vspace{.6cm}
{ $^1$ Physics Department, Harvard University, Cambridge, MA 02138, U.S.A.}\par
\vspace{.2cm}
{ $^2$ Walter Burke Institute for Theoretical Physics, California Institute of Technology}\par\vspace{-.3cm}
{  Pasadena, CA 91125, U.S.A.}\par
\vspace{.6cm}

\scalebox{.9}{\tt nagmon@g.harvard.edu, abedroya@g.harvard.edu, monica@caltech.edu, vafa@g.harvard.edu}\par
\vspace{1cm}
{\bf{Abstract}}\\
\end{center}
We provide an overview of the string landscape and the Swampland program. Our review of the string landscape covers the worldsheet and spacetime perspectives, including vacua and string dualities. We then review and motivate the Swampland program from the lessons learned from the string landscape. These lecture notes are aimed to be self-contained and thus can serve as a starting point for researchers interested in exploring these ideas.

These notes are an expanded version of two courses \emph{The String Landscape and the Swampland} taught by C.~Vafa at Harvard University in 2018 with a focus on the landscape, written by M.~J.~Kang with additional material from N.~B.~Agmon, and in 2022 with a focus on the Swampland, written by A.~Bedroya.

\vfill 
\date{}
\end{titlepage}
\begin{spacing}{0.8}
\tableofcontents
\end{spacing}

\pagebreak

\section*{Introduction}
\label{sec:intro}
\addcontentsline{toc}{section}{\nameref{sec:intro}}

After more than five decades of research, string theory has emerged as the most promising candidate for describing the connection between the observed universe and a theory of quantum gravity. It offers our deepest understanding of how quantum theory of gravity works. We have learned how to construct large classes of vacua in various string theories, including solutions which contain standard model matter content, such as non-abelian gauge fields, chiral fermions, and multiple generations of matter fields.  We have also discovered that string dualities lead not only to connections between various string theories, but also to the discovery of new quantum systems decoupled from gravity, in up to six dimensions.

The remarkable success of string theory may have led to the misunderstanding that any quantum field theory that appears consistent without gravity can be coupled to quantum gravity with string theory serving as the bridge. This expectation cannot be further from the truth. As we currently understand, only a few special quantum field theories emerge as the low energy limits of string theory. Essentially no generic quantum field theories can emerge, and only very special ones do! That may sound unnatural from the viewpoint of effective field theory but that is the lesson string theory is teaching us.

On the other hand, particle phenomenology and cosmology face a crisis of naturalness.  Parameters and choices of theories needed to explain our universe seem highly fine-tuned and unnatural.  It is not difficult to imagine that the reason they look fine-tuned is because the consistency of quantum gravity, as string theory solutions offer, is not incorporated into the notion of naturalness. With a correct prior, namely being able to couple the QFT to gravity, the naturalness criteria changes dramatically enough to make QFTs that describe our universe not as fine-tuned as they appear. 

The Swampland program aims to delineate conditions on effective field theories which distinguish the ``good ones'' (those that can couple to gravity consistently) from the ``bad ones''.  The aim of the courses serving as the basis for these lecture notes was to introduce this topic to students interested in doing research in this direction.

These lecture notes contain two basic parts. The first part includes several chapters dealing with an overview of string theory, with a focus on the landscape from both the worldsheet and spacetime perspectives. It is rather brief, but we try to be self-contained. The topics we cover serve as useful background for ideas in the Swampland program. The second part of these notes provides motivation for the program and explains its guiding principles.

\pagebreak 

\sectionbypart
\part{The string landscape}

\section{Bosonic string theory}

We begin with a lightning review of perturbative bosonic string theory. String theory is a theory of $1+1$ dimensional relativistic fundamental objects that propagate in some target space with an action defined on the worldsheet traced out by the strings. 

\subsection{Conventions}

Before diving into the details, it is useful to set some of the conventions first. Our discussion closely follows the modern textbook route \cite{Polchinski:1998rq, Polchinski:1998rr, Kiritsis:2007zza, Becker:2007zj} with many of the conventions of \cite{Polchinski:1998rq, Polchinski:1998rr}. We use $(\sigma^1, \sigma^2)$ to parameterize the the string worldsheet $\Sigma$ in Euclidean signature, where $\sigma^2$ plays the conventional role of Euclidean time. We often extend the domain of physical operators on the worldsheet by analytically extending them to complex values of $\sigma^1$ and $\sigma^2$. A particularly useful coordinate system for the resulting $\mathds{C}\times\mathds{C}$ space is 
\begin{align}
    z = \sigma^1 + i \sigma^2,\quad \bar{z} = \sigma^1 - i \sigma^2. 
\end{align}
Note that $z$ and $\bar z$ are not necessarily complex conjugates\footnote{We denote the complex conjugate by $z^*$.}. We can set the signature of the theory by restricting to the appropriate 2d subspace of $\mathds{C}\times\mathds{C}$. For example, the $z^*=\bar z$ subspace corresponds to the Euclidean parametrization of the worldsheet, while the $z,\bar z\in\mathds{R}$ subspace corresponds to Lorentzian signature. 

Since the functions are meromorphic in $z$ and $\bar z$, complex derivatives $\partial_z$ and $\partial_{\bar z}$ are well defined. We refer to them respectively as holomorphic and anti-holomorphic derivatives. When there is no possibility for confusion, we drop the subscripts and denote the associated derivatives by 
\begin{align}
    \p \equiv \frac{\p}{\p z}\quad\text{and}\quad {\bar \p} \equiv \frac{\p}{\p {\bar{z}}}.
\end{align}
They satisfy $\p z = \bar{\p} \bar{z} = 1$ and $\p \bar{z} = \bar{\p}z = 0$. Holomorphic functions (${\bar \p} f(z,\bar z) = 0$) are written as $f(z)$ whereas anti-holomorphic functions  (${ \p} f(z,\bar z) = 0$) are denotes by $f(\bar z)$.

The measure $d^2z = dz \wedge d\zbar$ satisfies $d^2z = 2 d\sigma^1 d\sigma^2$, with the Jacobian factor included. The integral over a closed contour in the complex plane is taken to satisfy 
\begin{align}
    \oint \frac{1}{z} = 2\pi i. 
\end{align}
For convenience, we work in units where the string length $\ell_s = 1$, \emph{i.e.} where the string tension is 
\begin{align}
    T = \frac{1}{2\pi}.
\end{align}

\subsection{Freely propagating strings}

At its core, perturbative string theory is a theory of fundamental $1+1$-dimensional objects moving in a target space. If we view the amplitude as a function of the worldsheet of the string, string theory is a 2d field theory where the amplitude of a given worldsheet configuration is given by $e^{iS}$ where $S$ is the action of the corresponding two dimensional field theory.

\begin{figure}[H]
    \centering
\tikzset{every picture/.style={line width=0.75pt}} 
\begin{tikzpicture}[x=0.75pt,y=0.75pt,yscale=-1,xscale=1]
\draw    (99.5,37) .. controls (176,62) and (309.5,148) .. (513.5,34) ;
\draw    (92.5,41) .. controls (309.5,129) and (295.5,140) .. (109.5,234) ;
\draw    (108.5,239) .. controls (265.5,160) and (364.5,153) .. (512.5,227) ;
\draw    (516.5,223) .. controls (285.5,127) and (408.5,96) .. (519.5,41) ;
\draw (243,41.4) node [anchor=north west][inner sep=0.75pt]    {$\mathcal{M} \sim \exp( iS_{worldsheet})$};
\end{tikzpicture}
    \caption{A 2 $\rightarrow$ 2 scattering event of strings represented by a single string worldsheet. The action on the worldsheet determines the scattering amplitude $\mathcal{M}$. }
\end{figure}
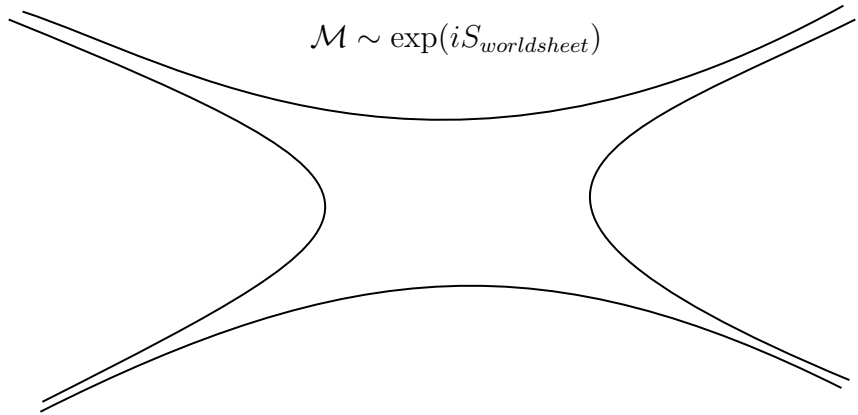

Recall that the action for a relativistic particle of mass $m$ is defined to be proportional to the proper length of its worldline $\gamma$, \emph{i.e.}
\begin{align}
    S=-m\int_\gamma d \tau \sqrt{\partial_\tau X^\mu \cdot \partial_\tau X_\mu} .
\end{align}
The same line of reasoning suggests a natural candidate for strings, namely introducing an action  proportional to the area of string worldsheet. This action is known as the \textit{Nambu--Goto action}. For a string that traces $X^\mu$ in a Minkowski background, the Nambu--Goto action reads \cite{Polchinski:1998rq}
\begin{align}
S_{NG} = -T \int_\Sigma d^2 \sigma \sqrt{\det \left(\partial_a X^\mu\cdot \partial_b X_\mu \right)}
\end{align}
where $T = 1/(2\pi)$ is the string tension, \emph{i.e.} mass per unit length, and $\Sigma$ is the worldsheet of the string. A priori, this action is difficult to quantize due to the presence of the square root. By introducing a dynamical worldsheet metric $g_{ab}$, we can instead consider the simpler \textit{Polyakov action} \cite{Brink:1976sc,Deser:1976eh,Polyakov:1981rd},
\begin{align}\label{eq:polyakovAction}
S_P = -\frac{T}{2} \int_\Sigma d^2 \sigma \sqrt{g} g_{ab} \p^a X^\mu \cdot\p^b  X_\mu .
\end{align}
Note that the metric $g_{ab}$ is an auxiliary variable and we can solve for it up to an overall scaling from the equation of motion. It is straightforward to verify that the two theories lead to the same equations of motion for $X^\mu$ and are therefore classically equivalent. The classical theory of \eqref{eq:polyakovAction} enjoys several internal symmetries. The worldsheet theory is invariant under global target space translations and Lorentz transformations,
\begin{align}
X'^\mu(\sigma) =  \Lambda^\mu_{\:\:\: \nu} X^\nu(\sigma) + v^\mu, \quad \Lambda \in SO(d-1,1).
\end{align}
Locally, this two-dimensional field theory is also invariant under several gauge symmetries, including \emph{diffeomorphisms} (\emph{i.e.} reparametrizations),
\begin{align}
X'^\mu(\sigma') = X^\mu(\sigma) , \quad
g'_{ab}(\sigma') = \frac{\partial \sigma^c}{\partial \sigma'^a} \frac{\partial \sigma^d}{\partial \sigma'^b}   g_{cd}(\sigma),
\end{align}
with $\sigma'^a = f^a(\sigma^b)$, and \textit{Weyl transformations} (\emph{i.e.} local rescalings),
\begin{align}
g_{ab}'(\sigma) = e^{2\phi(\sigma)} g_{ab}(\sigma).
\end{align}
The reason the above symmetries must be gauge symmetries and not ordinary symmetries is to ensure that the metric $g$ is a fictitious field and the Nambu--Gotu and Polyakov theories are equivalent. 

We perform a Wick rotation of the worldsheet theory to work in the Euclidean signature. The Wick rotation will remove the minus sign in front of the Lorentzian action. The quantized 2d Polyakov theory is then defined in terms of a path integral over all field configurations as
\begin{align}\label{eq:polyakovPathIntegral}
Z= \int \frac{DX Dg}{V_{\text{Diff} \times \text{Weyl}}} e^{-S_P[X,g]},
\end{align}
where we have divided by the volume of the gauge group to render the expression finite. The expression in \eqref{eq:polyakovPathIntegral} includes an implicit sum over inequivalent topologies (\emph{i.e.} manifolds which are not connected through the action of the gauge group). To make progress in the quantum theory, we utilize the standard Fadeev-Popov gauge-fixing procedure \cite{Faddeev:1967fc}, which localizes the path integral to a single gauge slice at the cost of introducing a set of anti-commuting $b$,$c$ ghost fields. There is a single $c$ ghost for every gauge parameter, and one $b$ ghost for every gauge-fixing condition. A particularly convenient choice of gauge-fixing condition is the \textit{conformal gauge}\footnote{Although it is always possible to choose $g_{ab} = \delta_{ab}$ locally in each coordinate patch of $\Sigma$, there can be global obstructions to this gauge choice. We postpone the study of these global issues to the discussion of moduli spaces of Riemann surfaces.}
\begin{align}
g_{ab} = \delta_{ab}.
\end{align}
An infinitesimal gauge transformations is specified by three worldsheet functions: infinitesimal coordinate transformations $\delta\sigma^1(\sigma)$, $\delta\sigma^2(\sigma)$ as well as the Weyl rescaling function $\phi(\sigma)$. Therefore, we expect to have three ghost $c$ fields in total. Moreover, every pair of worldsheet indices ${a,b}$ corresponds to a gauge-fixing condition $g_{ab}=\delta_{ab}$ and should lead to a ghost field $b_{ab}$. As the gauge-fixing conditions are symmetric in $a$ and $b$, $b_{ab}$ must also be symmetric.

It turns out the ghost field associated with Weyl transformations is easy to integrate out since it only appears as a quadratic term in the action. Doing so imposes a tracelessness condition on $b_{ab}$. Thus, we end up having two ghost fields $c^a$ corresponding to the coordinates reparameterizations $\delta\sigma^a$, a traceless symmetric tensor ghost field $b_{ab}$, and $d$ free massless scalars. The action for the massless scalars is
\begin{align}\label{free}
S_X = \frac{T}{2} \int_\Sigma d^2 \sigma \partial_a X^\mu \partial_a X_\mu,	
\end{align}
while the ghosts $c^a$, $b_{ab}$ are governed by the action
\begin{align}\label{ghost}
S_{gh} = \frac{1}{2\pi} \int_\Sigma d^2 \sigma b_{ab} \partial^a c^b . 
\end{align}
The path integral of the gauge-fixed theory thus reduces to
\begin{align}
Z=\int DX DbDc \: e^{-S_X-S_{gh}} .
\end{align}

\subsection{Basics of conformal field theory}\label{ssCFT}

It turns out that our choice of conformal gauge does not completely eliminate all of the gauge redundancy. There are still residual gauge transformations corresponding to combinations of diffeomorphisms and Weyl transformations that leave the metric invariant, referred to collectively as the group of \textit{conformal transformations}. In the conventional definition of conformal theories in flat space, a diffeomorphism is accompanied by a rescaling of other fields rather than the metric. It is straightforward to see these two definitions are equivalent and the rescaling of metric could be absorbed in the rescaling of the fields. For example, in the bosonic string action \eqref{free}$+$\eqref{ghost}, a rescaling of the metric under the residual conformal transformation 
\begin{align}
g_{ab}\rightarrow e^{2\omega(\sigma)}g_{ab}
\end{align}
could be replaced with a transformation of the fields given by
\begin{align}
    (X,b_{ab},c^{a})\rightarrow (X,b_{ab}e^{2\omega(\sigma)},c^ae^{-\omega(\sigma)})
\end{align}

As a result, we can view the worldsheet theory as a theory in 2d flat space which has conformal symmetry \cite{Friedan:1985ge,DiFrancesco:1997nk}. However, it is important to remember that the conformal symmetry is not a global symmetry, but a gauge symmetry of the theory! This is an important distinction. There is no a priori reason why the conformal symmetry of a classical theory must be preserved at the quantum level. However, in the case of string worldsheet, this is a requirement for the mathematical consistency of the theory. 

Since the dynamics of the strings are captured by a conformal theory, it is useful to briefly review the basic properties of CFTs in two dimensions.\footnote{Most of the essential CFT features in string theory are covered or at least mentioned in \cite{Polchinski:1998rq,Polchinski:1998rr}. For a more complete reference, see the yellow book \cite{DiFrancesco:1997nk}.}. Every local QFT by definition has a conserved operator known as the stress-energy tensor $T_{ab}$. For theories which admit a Lagrangian description, the stress tensor can be defined unambiguously by minimally coupling the theory to gravity\footnote{Minimal coupling refers to the gauging of the local Lorentz symmetry $SO(1,d-1)$.}
\begin{align}
T_{ab} = -4\pi g^{-1/2} \frac{\delta S}{\delta g^{ab}},
\end{align}
where the overall coefficient is a matter of convention. In scale invariant theories the trace of the energy momentum tensor vanishes on shell \cite{Pons:2009nb}. In conformal field theories, one can show that the energy momentum tensor can be improved by adding a total derivative $\partial_c\partial_d N^{abcd}$, such that the trace vanishes off shell as well. To ensure that $T^{ab}$ remains symmetric and conserved, $N$ needs to be symmetric under $a\leftrightarrow b$ and $c\leftrightarrow d$ and antisymmetric under $\{a~\text{or}~b\} \leftrightarrow \{c ~\text{or}~d\}$. In fact, such an improvement of the energy momentum tensor can be achieved by adding a boundary term to the action \cite{Pons:2009nb}. The boundary term ensures the action is Weyl invariant. If we work with this action, then $T^a_a$ must vanish identically, which in complex coordinates reads 
\begin{align}
T_{z \bar{z}} = 0 .
\end{align}
Similarly, the conservation equation can be written as
\begin{align}
 \bar{\p} T_{zz}=\p T_{\bar{z}\bar{z}}=0,
 \end{align}
which implies that the diagonal components $T \equiv T_{zz}(z)$ and $\bar{T} = T_{\bar{z}\bar{z}}(\bar{z})$ are holomorphic and anti-holomorphic, respectively. It follows that both admit Laurent expansions around the origin
\begin{align}\label{eq:stressTensorLaurent}
&T(z) =\sum_n L_n z^{-n-2}, \quad  \bar{T}(\bar{z}) =\sum_n \bar{L}_n \bar{z}^{-n-2},
\end{align}
where the overall factor $z^{-2}$ ($\bar{z}^{-2}$) is chosen to agree with its scaling dimension $\Delta = 2$. We know that the energy momentum tensor is the generator of coordinate transformations, and conformal transformations are a special type of coordinate transformation in flat space. Thus, we should expect the generators of conformal transformations can be expressed in terms of $T(z)$ and $\bar T(\bar z)$. In fact, $L_n$ is the generator of $(z,\bar z)\rightarrow (z+\epsilon z^n,\bar z)$ and $\bar L_n$ is the generator of  $(z,\bar z)\rightarrow (z,\bar z+\epsilon \bar z^n)$.\footnote{Conformal transformations in two-dimensions generically take the form of holomorphic functions $f(z)$. } These generators satisfy the following classical commutation relations,
\begin{align}
[L_m, L_n] &= (m-n)L_{m+n},
\end{align}
and are known as the Witt algebra. After quantizing the theory, the algebra acquires a central extension $c$. The centrally extended algebra, known as the Virasoro algebra $\text{Vir}_{c}$, takes the following form \cite{Virasoro:1969zu}
\begin{align}\label{eq:VirasoroAlgebra}
[L_m, L_n] &= (m-n)L_{m+n} + \frac{c}{12} (m^3-m)\delta_{m+n,0} .
\end{align}
Similarly, the anti-holomorphic modes $\bar{L}_n$ generate an independent copy $\text{Vir}_{\bar{c}}$ with central charge $\bar{c}$ not necessarily equal to $c$. The central charges $c$ and $\bar c$ represent the breaking of the conformal symmetry at the quantum level. 

All CFT states on the cylinder, which are dual to local operators at the origin of $\mathds{C}$ via the \emph{state-operator correspondence}, can written in terms of states with definite weights $(h,\bar{h})$ under $L_0, \bar{L}_0$. For unitary theories, these can be further decomposed into linear combinations of primary and descendant states. A primary $\ket{\psi}$ satisfies
\begin{align}\label{eq:primaryCond}
L_n \ket{\psi} =  \bar{L}_n \ket{\psi} = 0, \quad n > 0,	
\end{align}
while a descendant takes the general form
\begin{align}
\ket{\psi}' = L_{-n_1} \cdots L_{-n_k} \ket{\psi} , \quad n_1, \dots n_k > 0
\end{align}
Analogously, a primary operator ${\cal O}$ of weight $(h,\bar{h})$ obeys the operator product expansion (OPE)
\begin{align}
	T(z){\cal O}(0) = \frac{h}{z^2} {\cal O}(0) + \frac{1}{z}\partial {\cal O}(0) + \text{non-singular},
\end{align}
and similarly for $\bar{T} {\cal O}$. In particular, higher-order singularities are absent. This implies that under a conformal transformation, $\cal{O}$ transforms as
\begin{align}
    \mathcal{O}(z',\bar z')=(\partial z')^{-h}(\bar\partial \bar{z'})^{-\bar h}\mathcal{O}(z,\bar z) .
\end{align} 
where $z' =f(z)$ and $\bar{z}' = \bar{f}(\bar{z})$ are the associated conformal maps.

It is easy to verify this definition is equivalent to \eqref{eq:primaryCond} via \eqref{eq:VirasoroAlgebra}. The stress tensor is not a primary, but nonetheless satisfies the OPE:
\begin{align}
T(z) T(0) = \frac{c}{2z^4} + \frac{2}{z^2} T(0) + \frac{1}{z} \partial T(0) + \text{non-singular},
\end{align}
which through \eqref{eq:primaryCond} leads to the aforementioned Virasoro algebra \eqref{eq:VirasoroAlgebra}.

\subsection{Worldsheet gauge anomalies and the critical string}

We now return to the gauge-fixed worldsheet theory, whose action in complex coordinates takes the form
\begin{align}
S = \frac{1}{2\pi} \int d^2 z \left( \p X^\mu \bar{\p} X_\mu + b {\bar \p} c + {\bar b} \p {\bar c} \right) ,
\end{align}
where $c = c^z(z)$, $b = b_{zz}(z)$, $\bar{c} = c^{\bar{z}}(\bar{z})$, $\bar{b} = b_{\bar{z}\bar{z}}(\bar{z})$. The energy-momentum tensor for the matter fields $X^\mu$ is\footnote{All operator products are assumed to be normal ordered with respect to some procedure, such as creation-annihilation ordering (with creation operators on the left and annihilation operators on the right) or conformal normal ordering.}
\begin{align}
T_m = - \partial X^\mu \partial X_\mu, \quad \bar{T}_m = -\bar{\p}X^\mu \bar{\p}X_\mu,
\end{align}
with central charge $c = \bar{c} = d$, where $\mu=1,\hdots,d$. This agrees with the notion that the central charge is a rough measure of the number of degrees of freedom in a CFT. The ghost theory admits an independent stress tensor of the form
\begin{align}
T_{gh}=2(\p c)b+c\p b,\quad \bar{T}_{gh}=2(\pb\bar{c})\bar{b}+\bar{c}\pb\bar{b}
\end{align}
with central charge $c = \bar{c} = -26$. Therefore, the central charge of the combined matter+ghost CFT is
\begin{align}
    c_{\text{tot}} = d-26.
\end{align}

Recall that before gauge-fixing, the classical worldsheet theory possessed a $[\text{Diff}\times\text{Weyl}]$ gauge symmetry. The quantum theory, as defined by the Polyakov path integral, can suffer from both global and gauge anomalies. Indeed, on a general curved worldsheet, bosonic string theory suffers from a Weyl anomaly, which manifests itself through the non-vanishing of the trace of the stress tensor\footnote{Note that any CFT with $c \neq \bar{c}$ necessarily has a gravitational (diffeomorphism) anomaly, and so we will usually assume $c = \bar{c}$ unless explicitly stated.}
\begin{align}
T_{z\bar{z}}= -\frac{c}{24} \mathcal{R} .
\end{align}
Here, $\mathcal{R}$ is the 2d Ricci scalar on the worldsheet, which completely captures the geometry of a 2d manifold. To preserve the full $[\text{Diff}\times\text{Weyl}]$ gauge symmetry and consistently quantize the theory, we must therefore take the number of scalar fields to be equal to 26. This in turn implies that the dimensionality of spacetime is \cite{Lovelace:1971fa}
\begin{align}
d = 26.
\end{align}
The resulting Weyl-invariant theory is known as \textit{critical bosonic string theory}.

To summarize, the only dimension $d$ where the $d$-dimensional Euclidean space could be the target space of a non anomalous 2d CFT is $d=26$.

\subsection{BRST formalism}

As we discussed in subsection~ \ref{ssCFT}, the conformal gauge choice $g_{ab}=\delta_{ab}$ does not completely eliminate the gauge redundancy. As a consequence, the naive Hilbert space of the theory over-counts the physical states. The BRST formalism is a systematic method that allows us to restrict  to a faithful subspace of the Hilbert space where every physical state is represented exactly once. It is based on introducing a new global fermionic symmetry known as the \textit{BRST symmetry} \cite{Becchi:1975nq,Tyutin:1975qk}.\footnote{BRST refers to Becchi, Rouet, Stora, and Tyutin.} For the combined matter+ghost CFT, the current $j_B^a$ associated with this symmetry is \cite{Kato:1982im}
\begin{align}
j_B =c\cdot T_m+\frac{1}{2}c\cdot T_{gh} , \quad \bar{j}_B = \bar{c}\cdot\bar{T}_m + \frac12 \bar{c}\cdot\bar{T}_{gh},
\end{align}
where $j_B \equiv j_B^z$ and $\bar{j}_B \equiv j_B^{\bar{z}}$. Its components are are (anti-)holomorphic like the stress tensor, which implies $j_B^a$ is conserved
\begin{align}
    \partial_a j_B^a=\bar\partial j_B+\partial \bar j_B=0.
\end{align}
The associated topological charge operator acting on a local operator ${\cal O}(0)$ is given by
\begin{align}
Q_B = \oint_0 dz j_B(z) + \oint_0 \bar{dz} \bar{j}_B(\bar{z}),
\end{align}
where the contour necessarily surrounds the origin. The action of the BRST symmetry can be thought of as a gauge transformation with the gauge parameters replaced with $c$ ghosts. Note that the BRST symmetry acts on the $b$ ghost as
\begin{align}\label{eq:bToStressTensor}
Q_B \cdot b(z) = T(z) .
\end{align}

A key property of the BRST formalism is that $Q_B$ is nilpotent, \emph{i.e.}
\begin{align}
Q_B^2 = 0,
\end{align}
and so admits a cohomology of states. By gauge-fixing the worldsheet theory, we have enlarged the original space of states for the matter CFT to include additional states generated by free field oscillators built from the ghost fields. Naively, one might think that all such states contribute to on-shell scattering processes in spacetime; as it turns out, many of these states lead to equivalent spacetime physics. To account for this redundancy, we must restrict the set of physical states to be cohomology classes of $Q_B$. That is, a physical state is the coset of $Q_B$-closed states modulo $Q_B$-exact states:\footnote{To obtain a sensible unitary S-matrix, we must also supply an extra constraint on physical states: $b_0 \ket{\Psi} = \bar{b}_0 \ket{\Psi} = 0$, where $b_0 = \oint dz z b(z)$ and $\bar{b}_0 = \oint d\bar{z} \bar{z} \bar{b}(\bar{z})$. In the string field theory terminology, this constraint is known as Siegel gauge. }
\begin{align}
Q_B \ket{\Psi} = 0, \quad \ket{\Psi} \simeq \ket{\Psi} + Q_B \ket{\Lambda},
\end{align}
where $\ket{\Lambda}$ is an arbitrary state. The statement that $Q_B  = 0$ on physical states can be seen as reformulation of the fact that all physical observables must be gauge-invariant.

\subsection{Quantum string}

We now turn to analyzing the spectrum of physical states of the (free) closed bosonic string. In conformal gauge, we found that such states are in one-to-one correspondence with BRST cohomology classes. Using \eqref{eq:bToStressTensor}, it is clear that all physical states satisfy $L_0 = \bar{L}_0 = 0$. In principle, we could use this to derive the masses of closed string excitations by choosing a particular BRST representative for each cohomology class (for a detailed analysis, see \cite{Polchinski:1998rq}) and extracting the oscillators and their algebras from the matter and ghost fields.  A faster approach to obtain the spectrum, which easily generalizes to more complicated string theories, is to abandon conformal gauge and completely fix the gauge. A convenient gauge choice that has this property is the so-called light-cone gauge.

\subsubsection*{Light-cone quantization}

Light-cone (LC) gauge quantization is a particular approach to the quantization of sigma models whose target space admits a light-like Killing vector. For string theory, the key is to gauge-fix Diff$\times$Weyl such that there is a map from a worldsheet light-like Killing vector to one in the target spacetime. This removes the longitudinal degrees of freedom. As a bonus, the associated $bc$ ghosts completely decouple and can be neglected in the quantum theory. The price we must pay for these simplifications is the loss of manifest Lorentz invariance at the level of the classical worldsheet theory. Furthermore, we must also verify that it remains a symmetry in the quantum theory.

Working with a worldsheet in Lorentzian signature, we define null spacetime and worldsheet coordinates as, 
\begin{equation}
X^\pm:= \frac{X^0 \pm X^1}{\sqrt{2}}, \quad \sigma^\pm := \frac{\sigma^0 \pm \sigma^1}{\sqrt{2}}.
\end{equation}
Recall that taking the worldsheet metric flat, $g_{ab} = \eta_{ab}$, does not completely fix the gauge redundancy. The remaining gauge degrees of freedom can be eliminated by choosing the \textit{light-cone gauge},
\begin{equation}\label{eq:LCgauge}
X^+ = x^+ + \alpha' p^+ \sigma^+ , \quad \sigma^\pm = \sigma^0 \pm \sigma^1 ,
\end{equation}
for some constants $x^+$ and $p^+$ which ties  $\sigma^+$ to $X^+$. We can set $x^+$ to zero by reparametrizing $X^+\rightarrow X^+-x^+$. Note that $X^+$ is no longer a dynamical field. Naively this gauge-fixing condition would appear to leave a free field theory with $d-1 = 25$ free bosons. However, this is not the case since the equation for the metric forces us to set
\begin{equation}\label{eq:stressTensorConstraint}
	T_{ab} = 0
\end{equation}
 as a constraint which must be imposed by hand. This set of constraints are collectively referred to as \textit{Virasoro constraints}. In light-cone gauge, they take the form
\begin{align}
T_{++}&=  \p_+ X^+\cdot\p_+ X^--\p_+ X^i\cdot\p_+ X^i=0 , \quad i=1,\cdots,24,
\end{align}
with a similar expression for $T_{--}$. Together with \eqref{eq:LCgauge} we then have 
\begin{align}\label{eq:oscillatorConstraint}
\partial X^- = \frac{1}{p^+} \partial X^i \cdot\partial X^i ,
\end{align}
and so the nonzero oscillatory modes of $X^-$ also decouple. Thus, by choosing light-cone gauge we have explicitly removed the longitudinal degrees of freedom; only the 24 transverse fields $X^i$ remain. The quantum theory of the free string in lightcone gauge can be defined by canonical quantization. We can expand the transverse fields in terms of their Fourier modes on the cylinder, \emph{i.e.}
\begin{align}
X^i(\sigma) = x^i + \frac{p^i}{p^+} \sigma^0 + \frac{i}{\sqrt{2}}\sum_{n \neq 0} \frac{1}{n} \left( \alpha_n^i e^{in \sigma^-} + \bar{\alpha}_n^i e^{-in \sigma^+} \right).
\end{align}
The Hamiltonian $H$, which can now be identified with $p^-$, is given by
\begin{equation}
H = \frac{p^i p^i}{2p^+} + \frac{1}{2p^+} \left( \sum_{n =1}^\infty \alpha_{-n}^i \alpha_n^i + \bar{\alpha}^i_{-n} \bar{\alpha}^i_n + A + \bar{A} \right), 
\end{equation}
where $A$ and $\bar{A}$ are constants to be computed from the usual ordering ambiguity once we quantize the theory. We can proceed by imposing equal-time commutation relations on the $X^i$ and their conjugate momenta, or equivalently the following commutation relations on the oscillators:
\begin{align}
[\alpha^i_n, \alpha^j_m] = [\bar{\alpha}^i_n, \bar{\alpha}^j_m] = n \delta^{ij} \delta_{n,-m}, \quad [x^i, p^j] = i \delta^{ij}.
\end{align}
Up to an overall normalization, the $\alpha_n^i$ and $\bar{\alpha}_n^i$ satisfy the bosonic creation-annihilation commutation relations, with $n < 0$ corresponding to creation operators and $n > 0$ to annihilation operators.  There are still two unconstrained modes, the zero mode $x^-$ of $X^-$ and $p^+$, which satisfy
\begin{align}
[x^-, p^+] = -i.	
\end{align}
The minus sign reflects the fact that the flat metric $\eta_{ab}$ in the conformal coordinates $\sigma^{\pm}$ takes the form  $\eta^{-+} = -1$. Thus, the $x^\mu$ describe the center-of-mass coordinates of the string, while the oscillators describe its vibrational modes. We can decompose the Hilbert space into representations of the Lorentz group $SO(1,d-1)$. In particular, the vacua of the harmonic oscillators furnish an irreducible representation with states $\ket{p}$ labeled by a center-of-mass momentum vector $p \equiv (p^+, p^i)$. They satisfy 
\begin{equation}
	\alpha^j_{n} \ket{p} = \bar{\alpha}^j_n \ket{p} = 0, \quad n > 0. 
\end{equation}
The Fock space ${\cal H}$ of free string states can be constructed by repeated applications of the creation operators to the boosted ground states above,
\begin{equation}
	\ket{\mathbf{ N}, \mathbf{\bar{N}}; p} = \left(\prod_{j=1}^{24} \prod_{n_j = 1}^\infty (\bar{\alpha}_{-\bar{n}}^j)^{N_{j\bar{n}}} \right)\left( \prod_{i=1}^{24} \prod_{n = 1}^\infty  (\alpha_{-n}^i)^{N_{in}} \right) \ket{p} 
\end{equation}
Here, $\textbf{N} = ( N_{in} )$ and $\mathbf{\bar N}~ = ( N_{i\bar{n}} )$ are ordered collections of non-negative integer values that determine the excitation number of each mode. Occasionally, we will also take $N$ and $\bar {N}$  to be the sums $\sum_{i=1}^{24} \sum_n n N_{in}$ and $\sum_{i=1}^{24}  \sum_{\bar{n}} \bar n N_{i\bar{n}}$ respectively. They are referred to as the \textit{levels} of the string state $\ket{\mathbf{ N}, \mathbf{\bar{N}}; p}$. 

Note that we have allowed ourselves to be slightly imprecise in describing the gauge-fixing procedure. For the closed string, the light-cone gauge-fixing conditions described above leave some residual gauge freedom -- namely, translations in the $\sigma$ direction. The actual space of physical states ${\cal H}_{\text{phys}}$ consists of  gauge-invariant states in ${\cal H}$. These are the states uncharged under translations in $\sigma$, \emph{i.e.} those that obey the level-matching condition $N = \bar{N}$ \cite{Goddard:1973qh}. The closed string Hilbert space is therefore
\begin{equation}
{\cal H}_{\text{phys}} = \text{Span}\{\ket{ \mathbf{N}, \mathbf{\bar N}; p}\} .
\end{equation}
where $N=\bar N$. From now on we will simply refer to $N$ as the level of the state.

To determine the masses of the closed string states in ${\cal H}_{\text{phys}}$, notice that there is a spacetime Lorentz invariant
\begin{equation}
m^2 \equiv 2p^- p^+ - p^i p^i,
\end{equation}
which is just the usual mass squared invariant of relativistic particles. Using the fact that $p^- = H$, we can work systematically work out the closed string masses. Recall that the Hamiltonian $H$ includes two unfixed constants, $A$ and $\bar{A}$. It can be shown that these quantities necessarily take the values
\begin{equation}
A = \bar{A} = -1 
\end{equation}
for spacetime Lorentz invariance to be preserved. It follows that the closed string masses are
\begin{equation}
	m^2 \ell_s^2  = 4(N-1) , \quad N = 0, 1, 2, \ldots
\end{equation}
 where we have temporarily restored the string length $\ell_s = 1$.

\subsubsection*{Low-lying spectrum and NLSMs}

Let us now analyze the states at each level (\emph{i.e.} for each mass squared value) in detail and discuss their spacetime interpretation. At level $N=0$ we have a single state
\begin{align}
\ket{T} = \ket{p}, \quad m^2 \ell_s^2 = -4,
\end{align}
which behaves as a spacetime scalar with negative mass squared, \emph{i.e.} a tachyon. Soon, we will introduce string interactions through a perturbative expansion. Although the free theory is fully consistent, we will see that the presence of the tachyon introduces an incurable IR divergence in loop diagrams, thus spoiling the theory's consistency at the perturbative level. This is not unlike what happens in QFT whenever the potential $V(T)$ of a scalar field $T$ develops a local maximum at $T=0$ instead of a local minimum. The quantized theory yields a scalar particle with negative mass squared that inevitably introduces divergences in Feynman diagrams with loops. The solution in this case is to expand around a stable vacuum (global minimum)  where $T = T_0$ and instead consider the quanta of its fluctuations around this point.  The phenomenon of tachyon condensation and whether there exists such a vacuum remains an open problem for bosonic closed string theory.

The states at level $N=1$ are given by
\begin{equation}
  \zeta_{ij} \alpha^i_{-1} \bar{\alpha}^{j}_{-1} \ket{p} , \quad m^2 = 0, 
\end{equation}  
where $\zeta_{ij}$ is a rank-2 tensor with no constraints. According to Wigner's classification, massless particles in spacetime are associated with finite-dimensional irreducible representations (irreps) of the little group $SO(d-2)$. We therefore decompose $\zeta_{ij}$ into $SO(24)$ irreps,
\begin{align}
\zeta_{ij} = (\zeta_{(ij)} - \zeta \delta_{ij}) + \zeta_{[ij]} + \zeta \delta_{ij},
\end{align}
which correspond to a symmetric-traceless irrep, an anti-symmetric irrep, and a (trace) singlet, respectively. The symmetric-traceless states are simply the propagating modes of a massless spin 2 particle, \emph{i.e.} the graviton. The others corresponding to an antisymmetric rank 2 tensor and a scalar are less familiar in Einstein gravity, but ubiquitous in theories of gravity arising from stings. The tensor $\zeta_{ij}$ represents the collective polarization of these three states.

To each massless particle of the string we can associate a spacetime field in 26 dimensions,
\begin{align}\label{eq:masslessFields}
G_{\mu\nu}, \quad B_{\mu\nu}, \quad \phi,
\end{align}
which are, respectively, the spacetime metric, a 2-form gauge potential (known as the \textit{Kalb-Ramond field} or B-field, for short), and a scalar field (the \textit{dilaton}). As we will explain below, one can think of these fields as coherent states of the massless string states. Since coherent states satisfy the classical equations of motion, we expect these fields to be governed by an effective action that captures the low energy physics of these particles \cite{Callan:1985ia,Callan:1986jb}. The states beyond level $N=1$ have masses on the order of the string scale $\ell_p^{-1}$, and so their dynamics are irrelevant for low energies $E \ll \ell_p^{-1}$, and can be safely neglected in the low energy effective action. 

Recall that the photon is the quantum of the electromagnetic field, and that a nonzero classical background can be generated via coherent states of photons. Charged particles couple minimally to this field via a worldline action $i\int_\gamma A_\mu(X) dX^\mu$. Analogously, coherent states of the massless particles in string theory generate nonzero backgrounds for \eqref{eq:masslessFields}. These fields couple minimally to the string worldsheet, and so it is permissible to generalize the Polyakov formalism to include the effects of these backgrounds. The usual Polyakov action in flat Minkowski space generalizes to
\begin{equation}
	S_G = \frac{1}{4\pi} \int_\Sigma G_{\mu \nu}(X) dX^\mu \otimes dX^\nu .
\end{equation}
The other massless fields introduce new terms that take the form
\begin{align}
	S_B = \frac{i}{4\pi} \int_\Sigma B_{\mu\nu}(X) dX^\mu \wedge dX^\nu , \quad
	S_\phi = \frac{1}{4\pi} \int_\Sigma R(g) \Phi(X) ,
\end{align}
where $R(g)$ is the scalar curvature of the worldsheet.  The first term is similar to the worldline action of a charged particle. Indeed, we say that the string is (electrically) charged under a one-form gauge symmetry associated with the B-field. For $\Phi$ constant, the second term is a topological invariant of the worldsheet, and is proportional to the Euler characteristic $\chi$. For a compact, oriented worldsheet, it is given by
\begin{align}
	\chi = 2-2g,
\end{align}
where $g$ is its genus.

By pulling back the spacetime fields to the worldsheet, we find that the generalized Polyakov action takes the form of a nonlinear sigma model (NLSM) on the worldsheet \cite{Friedan:1980jf,Callan:1985ia},
\begin{equation}\label{eq:NLSM}
	S_{NLSM} = \frac{1}{4\pi} \int_\Sigma d^2 \sigma \sqrt{g} \: \left[ \left( G_{\mu\nu}(X) g^{ab} + i B_{\mu\nu} \epsilon^{ab} \right) \p_a X^\mu \p_b X^\nu + R(g) \Phi(X) \right] .
\end{equation}
Some brief comments are in order. The background fields are functions of the target space coordinates $X^\mu$, and so \eqref{eq:NLSM} is generically a strongly coupled field theory, and with infinitely many terms in the polynomial expansion! This makes analyzing anything but the simplest of closed string backgrounds a daunting task. A more fundamental problem is the lack of a guarantee that this theory is Weyl invariant in the quantum level.  Weyl invariance plays a crucial role in the consistency of BRST quantization and therefore in the very fabric of the critical bosonic string. Using standard field theory techniques, it is possible to calculate the beta functions of the nonlinear sigma model order by order in $\ell_s^2$. Weyl invariance requires that the beta functions vanish, and so give constraints on the allowed string backgrounds. Miraculously, the lowest order constraints reproduce Einstein's field equations for $G_{\mu\nu}$ (as well as equations of motion for $B$ and $\phi$). Of course, there are also higher ordering corrections in $\ell_s^2$ that affect the high energy physics (after all, Einstein gravity is not a UV complete theory, so this better not be the end of the story). Remarkably, the effective action obtained by imposing Weyl symmetry can also be directly calculated from string theory amplitudes -- and the two expressions perfectly match \cite{Tong:2009np}.

\subsection{String perturbation theory}

Given the quantum closed string, we now want to introduce string interactions via splitting and joining of strings. For strings in a general background (with an asymptotic dilaton value $\Phi_0$) with action given by \eqref{eq:NLSM}, this is described by the Polyakov path integral
\begin{equation}\label{eq:pathIntegralSeries}
  Z = \sum_{g=0}^\infty \lambda_0^{2g-2} \int DXDg  \exp \left(- S_G[X;g] - S_B[X;g] \right) \,,
\end{equation}
where we have now explicitly included the sum over worldsheets of different topology, which for 2d geometries reduces to a sum over discrete topologies labeled by the genus $g \in \mathbb{N}$ (\emph{i.e.} the number of holes). We are only considering interactions among closed strings, for which the number of worldsheet boundaries $b = 0$. Note the distinction between the genus $g$ in the sum and the worldsheet metric $g_{ab}$. From the Gauss-Bonnet theorem we see that the background dilaton $\exp(\Phi_0)$ and the bare coupling constant $\lambda_0$ both contribute in the same fashion, \emph{i.e.} as $\exp(\Phi_0)^{2g-2}$ and $\lambda_0^{2g-2}$ respectively. Thus, we can combine them into a single renormalized coupling constant $\lambda$ and redefine the dilaton field such that its VEV reads
\begin{equation}
	\lambda = \langle e^\Phi \rangle.
\end{equation}

Notice that \eqref{eq:pathIntegralSeries} takes the form of a perturbative series expansion, where $\lambda$ plays the role of the coupling ``constant'' of the theory. This is a slight misnomer, since one key property of $\lambda$ is that it is \textit{not} a free parameter, bur rather is fixed dynamically. As we will see, this is a universal feature of string theory backgrounds, whose parameters are generated dynamically via the VEVs of scalar fields.  

This is a remarkable feature of string theory! Even when we start with free string theory ($\lambda_0=0$), the spectrum includes coherent backgrounds where the perturbative description is given by strings with non zero coupling $\lambda=\exp{\Phi_0}$. In other words, the free theory prescribes the structure of the interacting theory!

\subsubsection*{Scattering states in conformal gauge}

For asymptotically flat backgrounds, e.g. $G_{\mu\nu} = \eta_{\mu\nu}$, we can define an S-matrix describing the scattering of closed string states. First, let us revisit the BRST formalism for the closed string in conformal gauge. Recall that an on-shell physical string state in this background is a weight (0,0) $Q_B$ cohomology class that obeys the Siegel constraints. All states in the class satisfy the level-matching conditions, and so the complete set of constraints on any state $\ket{\psi}$ in the class is given by
\begin{align}
&b_0 \ket{\psi} = \overline{b}_0\ket{\psi} = 0, \\
&p^2 = -m^2 = 4(1-N), \quad N \in \mathbb{N}\cup\{0\},
\end{align}
where the mass-shell constraint now arises from BRST invariance. We can always choose their representatives to be states of the form $c_{1} \bar{c}_1 \ket{V}$, where $c_1$ and $\bar{c}_1$ are free field oscillators of the ghost fields $c(z)$ and $\bar{c}(\bar{z})$, and $\ket{V}$ is a weight (1,1) state built purely from the fields in the matter CFT (\emph{i.e.} the oscillators of $X^\mu$). Using the state-operator correspondence, these are dual to vertex operators of the form
\begin{equation}\label{eq:phsicalVertexOp}
c(z) \bar{c}(\bar{z}) V(z,\bar{z}), \quad V(z,\bar{z}) = \partial^{n_1} X(z) \cdots \bar{\partial}^{m_1} X(\bar{z}) \cdots e^{ip \cdot X(z,\bar{z})},
\end{equation}
where all operator products are assumed to be normal ordered. Note that BRST invariance implies additional constraints on the tensor coefficients of these operators.

It is straightforward to determine the vertex operators (states) at each level in conformal gauge. For instance, the tachyon is represented by the vertex operator
\begin{equation}
	V^{(0)} = e^{ip \cdot X}, \quad p^2 = 4,
\end{equation}
Similarly, the massless states at level 1 are associated with the vertex operator
\begin{align}
V^{(1)} = \epsilon_{\mu \nu} \p X^\mu {\bar \p}X^\nu e^{ip \cdot X}, \quad p^2 = 0, \quad \epsilon_{\mu\nu}p^\mu = \epsilon_{\mu\nu}p^\nu = 0,
\end{align}
where $\epsilon_{\mu\nu}$ is a polarization tensor. Its transversality property arises from BRST-invariance, and ensures that the longitudinal degrees of freedom decouple -- this is the expected result for massless quanta of gauge fields in spacetime. Note that $V^{(1)}$ packages together the graviton, dilaton, and B-field.

\subsubsection*{The string S-matrix}

For a scattering process of $n$ strings, the worldsheet takes the form of some compact manifold $\Sigma_g$ glued to a set of $n$ infinitely long cylinders carrying the asymptotic closed string states. The Diff$\times$Weyl symmetry can be used to map this surface to the same $\Sigma_g$, but with $n$ punctures (\emph{i.e.} marked points). This transformation effectively carries out the state-operator mapping \eqref{eq:phsicalVertexOp}, with the ghosts removed. Each such operator has a natural pairing with $dz \wedge d\bar{z}$ which should be inserted in the gauge-fixed version of \eqref{eq:pathIntegralSeries}. These objects have conformal weight $(0,0)$, and so preserve our choice of conformal gauge. It turns out that while $V$ itself is not BRST-invariant, its transformation is a total derivative on the moduli space of inequivalent Riemann surfaces, and so its integrated version is BRST-invariant. We therefore have that S-matrix elements take the schematic form \cite{Scherk:1974ca}
\begin{equation}\label{eq:sloppyAmplitude}
  {\cal A}_{n} \sim \sum_{g=0}^\infty \lambda^{2g-2} \left\langle \prod_{i=1}^n \int_{\Sigma_{g}} V_i(z_i,\bar{z}_i) dz_i \wedge d\bar{z}_i \right\rangle .
\end{equation}
There are two primary issues with formula \eqref{eq:sloppyAmplitude}.

First, the summation over Riemann surfaces is incomplete. For example, for $n=0$ and $g \neq 1$, it is not possible to take the metric to be flat globally. However, we can make it flat in local patches with coordinates $(\sigma^1_i,\sigma^2_i)$ that cover the whole worldsheet.  Then the transition functions between two overlapping coordinates $(\sigma^1_i,\sigma^2_i)$ are given by conformal transformations. If we consider the complex coordinates $(z_i,\bar z_i)$ associated with each coordinate patch $(\sigma^1_i,\sigma^2_i)$, the conformal transition maps between $\sigma$s take the form of holomorphic maps between the complex coordinates of each patch. Therefore, we can think of the worldsheet as a 2 dimensional surface with a global complex structure, \emph{i.e.} a Riemann surface. However, not all Riemann surfaces of the same genus are equivalent. We can label different Riemann surfaces by continuous parameters, called moduli, whose values do not change under infinitesimal Diff$\times$Weyl transformations. To account for this in \eqref{eq:sloppyAmplitude}, we must integrate over the moduli space ${\cal M}_{g}$ of genus $g$ surfaces in addition to summing over $g$. The real dimension of ${\cal M}_{g}$  is
\begin{equation}
\text{dim}_{\mathbb{R}} \: {\cal M}_{g}	= \begin{cases}
\ 0 \quad & \text{if }g = 0, \\
\ 2 \quad & \text{if }g = 1, \\
\ 6g-6 \quad & \text{if }g > 1 . 
 \end{cases}
\end{equation}
For nonzero $n$, the moduli space includes the locations of the integrated vertex operators giving $2n$ additional parameters -- this is the moduli space of genus $g$ Riemann surfaces with $n$ punctures.

Second, the Polyakov path integral in conformal gauge possesses some residual gauge symmetry associated with global conformal transformations of $\Sigma_g$. These transformations are generated by \textit{conformal Killing vectors} (CKVs), which are global vector fields $\xi^a$ that satisfy the conformal Killing equation
\begin{equation}
\nabla_a \xi_b + \nabla_b \xi_a = g_{ab} \nabla_c \xi^c .
\end{equation}
CKVs correspond to conformal transformations that do not change the geometry of the surface. We can use (and eliminate) these extra gauge redundancies to fix the position of several vertex operators.

The way in which the moduli and CKVs enter into the path integral relates directly to the ghost path integral. The ghost action admits a $bc$ ghost number symmetry which is anomalous on a general Riemann surface. The anomaly requires that all non-vanishing correlation functions have ghost charge $6g-6$, where $b$ has charge $-1$ and $c$ has charge $+1$. The result of gauge-fixing then implies that every modulus comes with a $b$ insertion, while every CKV comes with a $c$ insertion. Thus, for each complex CKV we insert $c \bar{c} V(z,\bar{z})$ instead of $\int V dz \wedge d\bar{z}$. Overall, this result of gauge-fixing the string theory path integral can be viewed as the string theory derivation of the Riemann-Roch theorem, which states that the number of real moduli minus the number of CKVs of any Riemann surface is equal to $6g-6$. 

The precise nature of the $b$ ghost insertions traces back to the gauge-fixing procedure. Recall that the moduli of $\Sigma_g$ are encoded in the gauge-fixed metric $g_{ab}(t)$. For each modulus $t^i$, it can be shown that this gauge-fixing procedure results in the insertion
\begin{align}\label{eq:bInsertion}
{\cal B}(t) = \frac{1}{2\pi} \int d^2 z \left( b_{zz} (\mu_t)_{\bar{z}}^{\,\, z} + {b}_{\bar z\bar z} (\mu_t)_z^{\,\, \bar{z}} \right)\,,
\end{align}
where $\mu$ is the Beltrami differential; its components can be computed directly in terms of the metric and its derivatives,
\begin{align}
(\mu_t)_a^{\,\,\, b} = \frac12 g^{bc}(t) \frac{\p}{\p t} g_{ac}(t),
\end{align}
and can be viewed as deforming the complex structure of the worldsheet ($\bar\partial\rightarrow \bar\partial+\mu_{\bar z}^z\partial$) leading to an insertion of $\int b_{zz}\mu_{\bar z}^z+c.c.$. 

In summary, the $n$-point scattering amplitude of bosonic strings is given by\footnote{Moduli space integrals for multi-loop amplitudes are rigorously derived in \cite{DHoker:1985een}.}
\begin{align}
\begin{aligned}
  {\cal A}_n \propto & \sum_{g=0}^\infty \lambda^{2g-2} \left\langle \prod_{a=1}^{\text{dim}({\cal M}_g)=6g-6} \int dt^a {\cal B}(t^a) \prod_{b=1}^{\text{dim}(\text{CKG})} c \bar{c} V_b(z_b, \bar{z}_b)\right.\\ 
  & \qquad\qquad\qquad\qquad\qquad\left.
  \prod_{i=\text{dim}(\text{CKG})+1}^n \left( \int_{\Sigma_{g}} V_j(z_j,\bar{z}_j) dz_i \wedge d\bar{z}_i \right) \right\rangle ,
\end{aligned}
\end{align}
where $t_i$ are the moduli of genus $g$ surfaces and the overall coefficient can be fixed by demanding unitarity. For genus $g$, there are $3g-3$ complex moduli, or $6g-6$ real moduli.

\subsection{Tree-level scattering}

\begin{figure}[H]
\begin{center}
\scalebox{1.15}{
\begin{tikzpicture}
  \shade[ball color = gray!40, opacity = 0.4] (0,0) circle (2cm);
  \draw (0,0) circle (2cm);
  \draw (-2,0) arc (180:360:2 and 0.6);
  \draw[dashed] (2,0) arc (0:180:2 and 0.6);
  \fill[fill=black] (0,0) circle (1pt);
          \node[font=\scriptsize] (c) at (0,-0.3) {$\int  V$};
    \fill[fill=black] (-1,-1) circle (1pt);
        \node[font=\scriptsize] (c) at (-1,-1.3) {$\int  V$};
            \fill[fill=black] (1,-1) circle (1pt);
        \node[font=\scriptsize] (c) at (1,-1.3) {$c \bar{c} V$};
    \fill[fill=black] (-1,1) circle (1pt);
        \node[font=\scriptsize] (c) at (-1,0.7) {$c\bar{c} V$};
            \fill[fill=black] (1,1) circle (1pt);
        \node[font=\scriptsize] (c) at (1,0.7) {$c \bar{c} V$};
\end{tikzpicture}
}
\caption{A five-point tree level amplitude requires three fixed vertex operators $c \bar{c} V$ and two integrated vertex operators $\int d^2z V$ on the Riemann sphere. The value of the amplitude is unaffected by the choice of location of the fixed vertex operators.}
\label{fig:ccV}
\end{center}
\end{figure}
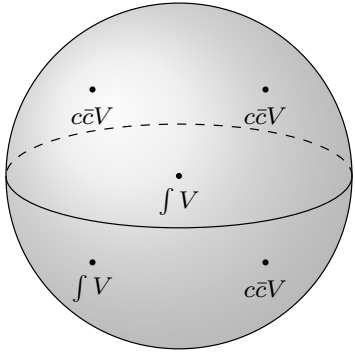

The discussion in the previous discussion is somewhat formal, so let us consider some concrete examples of string scattering amplitudes. The simplest string diagrams arise at tree level ($g = 0$). The unique genus $0$ Riemann surface is the Riemann sphere $\mathds{C}^* = \mathds{C} \cup \{ \infty \}$ which has no moduli. The CKVs can easily be found by searching for global vector fields $\xi(z) \partial_z$ and $\bar{\xi}(\bar{z}) \partial_{\bar z}$ that are sufficiently regular at $z = \infty$. The solution $\xi(z) = a + b z + cz^2$ corresponds to 3 independent complex CKVs, which together generate the Mobius group PSL(2,$\mathds{C}$) of fractional linear transformations
\begin{equation}
z \to \frac{az + b}{cz+d}, \quad ad - bc = 1 , \quad a,b,c,d \in \mathds{C}.
\end{equation}
We must therefore fix the locations of three vertex operators. The $n$-point scattering amplitude thus takes the form \cite{Friedan:1985ge}\footnote{By treating the vertex operator positions as moduli, the $n$-point amplitude can be written in a more symmetric form solely in terms of integrated vertex operators. See \cite{Polchinski:1998rq} for details.}
\begin{align}
{\cal A}^{(0)}_n = \left\langle \prod_{i=1}^3 c(z_i)\bar{c}(\bar{z}_i) V_i(z_i,\bar{z}_i) \prod_{j=4}^n \int d^2 z_j V_j(z_j, \bar{z}_j)  	\right\rangle_{S^2},
\end{align}
where $\langle \cdots \rangle_{S^2}$ indicates the correlation function on the Riemann sphere in the full matter and ghost CFTs. For instance, the $n=5$-point scattering amplitude is depicted in Figure \ref{fig:ccV}.

As a warm up exercise, we consider the scattering of three tachyons. The amplitude can be computed using any free field theory techniques of choice. Up to an overall constant fixed by unitarity, the result is 
\begin{align}
{\cal A}^{(0)}_{T^3} = g_s \left\langle \prod_{i=1}^3 c(z_i)\bar{c}(\bar{z}_i) e^{ip_i \cdot X(z_i,\bar{z}_i)} \right\rangle_{S^2} \simeq g_s\delta^{26}(p_1+p_2+p_3),
\end{align}
Here, $g_s$ is the string coupling constant, which is proportional to the dilaton VEV $\lambda$. More interesting is the four-tachyon amplitude,
\begin{align}
{\cal A}^{(0)}_{T^4} =  \left\langle \prod_{i=1}^3 c(z_i)\bar{c}(\bar{z}_i) e^{ip_i \cdot X(z_i,\bar{z}_i)} \int d^2z e^{ip_4 \cdot X(z,\bar{z})} \right\rangle_{S^2} \simeq g_s^2 \delta^{26}(p_1+p_2+p_3+p_4) A(s,t,u).
\end{align}
where $A$ is the famous  Virasoro-Shapiro amplitude \cite{Virasoro:1969me,Shapiro:1970gy}
\begin{align}\label{eq:virasoroShaprio}
A(s,t,u) = \int d^2 z |z|^{-2u-4}|1-z|^{-2t-4} = 2\pi \frac{\Gamma\left(-1-\frac{s}{4}\right) \Gamma\left(-1-\frac{t}{4}\right)  \Gamma\left(-1-\frac{u}{4}\right) }{\Gamma\left(2+\frac{s}{4}\right) \Gamma\left(2+\frac{t}{4}\right)  \Gamma\left(2+\frac{u}{4}\right)} .
\end{align}
Here, $s$, $t$, and $u$ are the usual Mandelstam invariants
\begin{align}
s = -(p_1+p_2)^2, \quad t = -(p_1+p_3)^2, \quad u = -(p_1+p_4)^2,
\end{align}
which satisfy $s+t+u = 4m_T^2$ for the tachyon mass squared $m_T^2 = -4$. The amplitude in \eqref{eq:virasoroShaprio} has many interesting properties, such as duality, which we do not have time to properly cover. Most important is that the 4-point amplitude obeys the unitarity properties expected of an S-matrix element \cite{Virasoro:1969me}. In particular, we expect the 4-point diagram to factorize into two 3-point diagrams along with a simple pole via a unitarity cut in the $s$-channel (\emph{i.e.} separating $p_1,p_2$ from $p_3,p_4$). Indeed, we see that as an intermediate state goes on-shell with $s = m_N^2 = 4(N-1)$, the 4-point amplitude reduces to
\begin{align}
A(s,t,u) \xrightarrow[]{s=4(N-1)} \frac{P_{2N}(t)}{s-4(N-1)},
\end{align}
where $P_{2N}(t)$ is a degree $2N$ polynomial in $t$ that accounts for which spins up to $2N$ are exchanged, consistent with maximum spin at level $N$.

\subsection{One-loop scattering}

Now let us consider 1-loop amplitudes, which correspond to genus $g=1$. Such worldsheets have the topology of a torus $T^2$. It is worthwhile to spend some time discussing the geometry of the torus and the structure of its moduli space. We can construct a torus as a quotient $\mathds{C}/\Lambda$, where $\Lambda$ is a lattice generated from two basis vectors $e_1$ and $e_2$. Its fundamental domain consists of the parallelogram with two of its sides given by $e_i$, with opposite sides identified. As Riemann surfaces, two tori are equivalent if they are related by a conformal transformation. For us, this means a particular torus can be described by a single complex modulus $\tau$, known as a \textit{Techm\"uller parameter}, which corresponds to $e_1/e_2$. Without loss of generality, such a torus can be constructed from the associated parallelogram with sides given by $e_1 = 1$ and $e_2 = \tau$, as shown in Figure  \ref{fig:torus}. 
\begin{figure}[H]
\begin{center}
\scalebox{.85}{
\begin{tikzpicture}
\node[draw=none,thick,scale=0.2,fill=black,label={[label distance=1mm]south:$0$}] (A2) at (0,0) {};
\node[draw=none,thick,scale=0.2,fill=black,label={[label distance=1mm]south:$1$}] (A4) at (3,0) {};
\node[draw=none,thick,scale=0.2,fill=black,label={[label distance=1mm]west:$\tau$}] (A3) at (1.2,3) {};
\node[draw=none,thick,scale=0.2,label={[label distance=1mm]east:$\mathbb{R}$}] (A3) at (5,0) {};
\node[draw=none,thick,scale=0.2,label={[label distance=1mm]north:$i\mathbb{R}$}] (A3) at (0,4.5) {};
\draw[->] (0,0)--(5,0);
\draw[->] (0,0)--(0,4.5);
 \draw[fill=green!10] (1.2,3) -- (4.2,3) -- (3,0) -- (0,0) -- cycle;
\end{tikzpicture}}
\caption{The torus $T^2$ with complex structure $\tau$ as described by the quotient $\mathds{C}/\Lambda$, where $\Lambda$ is the lattice generated by $\{1, \tau\}$. The fundamental domain is shaded in green, with opposite sides identified.}
\label{fig:torus}
\end{center}
\end{figure}
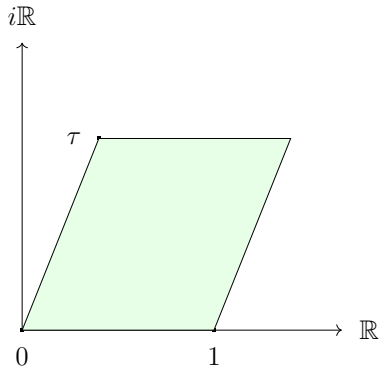
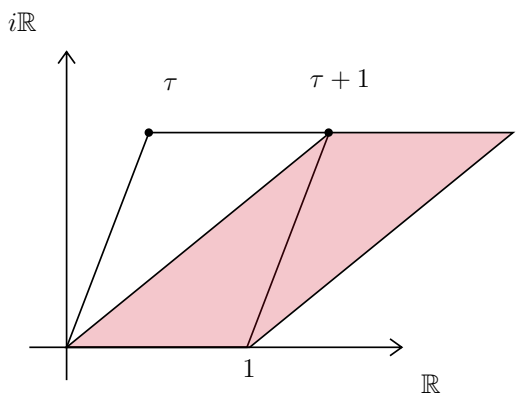
\begin{figure}[H]
\centering
\tikzset{every picture/.style={line width=0.75pt}} 
\scalebox{.85}{
\begin{tikzpicture}[x=0.75pt,y=0.75pt,yscale=-1,xscale=1]
\draw   (259,86) -- (366,86) -- (317.1,213.5) -- (210.1,213.5) -- cycle ;
\draw  (188,213.5) -- (409,213.5)(210.1,38) -- (210.1,233) (402,208.5) -- (409,213.5) -- (402,218.5) (205.1,45) -- (210.1,38) -- (215.1,45)  ;
\draw  [fill={rgb, 255:red, 0; green, 0; blue, 0 }  ,fill opacity=1 ] (256.87,86) .. controls (256.87,84.9) and (257.77,84) .. (258.87,84) .. controls (259.97,84) and (260.87,84.9) .. (260.87,86) .. controls (260.87,87.1) and (259.97,88) .. (258.87,88) .. controls (257.77,88) and (256.87,87.1) .. (256.87,86) -- cycle ;
\draw  [fill={rgb, 255:red, 208; green, 2; blue, 27 }  ,fill opacity=0.22 ] (366.36,86) -- (475,86) -- (318.74,213.5) -- (210.1,213.5) -- cycle ;
\draw  [fill={rgb, 255:red, 0; green, 0; blue, 0 }  ,fill opacity=1 ] (363.59,86) .. controls (363.59,84.9) and (364.49,84) .. (365.59,84) .. controls (366.69,84) and (367.59,84.9) .. (367.59,86) .. controls (367.59,87.1) and (366.69,88) .. (365.59,88) .. controls (364.49,88) and (363.59,87.1) .. (363.59,86) -- cycle ;
\draw (174,13.4) node [anchor=north west][inner sep=0.75pt]    {$i\mathbb{R}$};
\draw (266,52.4) node [anchor=north west][inner sep=0.75pt]    {$\tau $};
\draw (419,228.4) node [anchor=north west][inner sep=0.75pt]    {$\mathbb{R}$};
\draw (313,219.4) node [anchor=north west][inner sep=0.75pt]    {$1$};
\draw (353,47.4) node [anchor=north west][inner sep=0.75pt]    {$\tau +1$};
\end{tikzpicture}}
\caption{Different parameterizations of the same torus are related by the action of the modular group $PSL(2.\mathbb{Z})$. A torus with modulus $\tau$ can also be described with modulus $\tau+1$.}
\label{fig:torus2}
\end{figure}
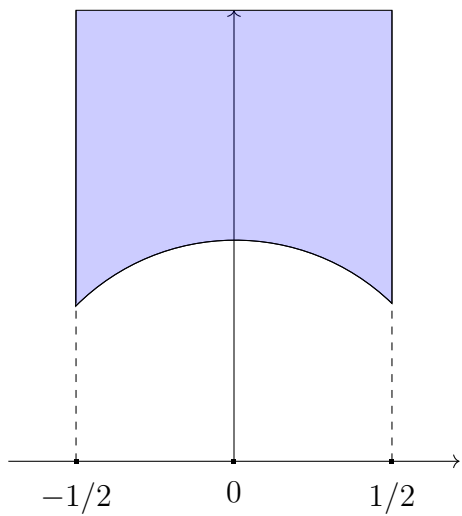
\begin{figure}[H]
\begin{center}
\begin{tikzpicture}
\node[draw=none,thick,scale=0.2,fill=black,label={[label distance=1mm]south:$0$}] (A2) at (0,0) {};
\node[draw=none,thick,scale=0.2,fill=black,label={[label distance=1mm]south:$-1/2$}] (A3) at (-2.1,0) {};
\node[draw=none,thick,scale=0.2,fill=black,label={[label distance=1mm]south:$1/2$}] (A4) at (2.1,0) {};
\draw[->] (-3,0)--(3,0);
\draw[->] (0,0)--(0,6);
\draw (-2.1,2.1)--(-2.1,6);
\draw (2.1,2.1)--(2.1,6);
\draw (2.1,2.1) arc (46:135:3);
\draw[dashed] (-2.1,0)--(-2.1,2.1);
\draw[dashed] (2.1,0)--(2.1,2.1);
\draw[fill=blue, fill opacity=0.2] (2.1,6) -- (2.1,2.1) arc (46:135:3) -- (-2.1,6)--cycle;
\end{tikzpicture}
\caption{A fundamental domain ${\cal F}$ of the modulus space of $T^2$. The two vertical sides are identified, as are the two circular arcs. The image of ${\cal F}$ under $SL(2,\mathbb{Z})$ is the entirety of $\mathbb{H}$.}
\label{fig:FDtorus}
\end{center}
\end{figure}
Naively, each torus can be parametrized by a modulus $\tau$ taking value in the upper-half plane $\mathbb{H}$. However, there is in fact an equivalence class of $\tau$ values which correspond to tori that can be mapped to one another by different choice of basis for $\Lambda$. The group which relates these different parameterizations is known as the \textit{modular group}, which for $T^2$ is isomorphic to $PSL(2,\mathbb{Z})$, and is generated by $\tau\rightarrow \tau+1$ and $\tau\rightarrow -1/\tau$. The modular group acts on a given modulus as\footnote{The correct modular group is $PSL(2,\mathbb{Z})$ since a given element $A \in SL(2,\mathbb{Z})$ and its additive inverse $-A$ both yield the same transformed value $\tau'$.}
\begin{align}
\tau \longrightarrow \tau' = \frac{a\tau + b}{c\tau+d} , \quad \begin{pmatrix}
	a & b \\
	c & d
\end{pmatrix} \in SL(2,\mathbb{Z}) .
\end{align}
To avoid over-counting in the path integral, we should really be integrating over values of $\tau$ in a single fundamental domain of $\mathbb{H}/PSL(2,\mathbb{Z})$ such as
\begin{align}
{\cal F} = \left\{\tau \in \mathbb{H} \, \bigg|  -\frac12 \leq \text{Re}(\tau) \leq \frac12,  |\tau| \ge 1 \right\},
\end{align}
shaded in blue in Figure \ref{fig:FDtorus}.

We are now ready to talk about the genus 1 contribution to worldsheet amplitudes. Recall that the nature of the ghost insertion is determined by the number of conformal Killing group (CKV) and moduli.  First, we find the number of $c$-ghosts required. The conformal Killing group (CKG) of the torus, $T^2$, is easy to describe: it consists of translations. There is thus a single (complex) CKV, which leads to the insertion of a single pair $c \bar{c}$ in all scattering amplitudes. Intuitively, their location can be fixed using the translation symmetry. Second, we work out the number of $b$-ghost insertions and their form. Earlier, we found that the torus had a single complex modulus $\tau = \tau_1 + i \tau_2$. For convenience, we encode the moduli dependence of $T^2$ in the metric,
\begin{align}
ds^2 = |d\sigma^1 + \tau d\sigma^2 |^2 
\end{align}
such that the coordinates maintain the usual periodicity properties
\begin{equation}
(\sigma^1, \sigma^2) \sim (\sigma^1,\sigma^2) + (2\pi, 2\pi).
\end{equation}
The associated $b$-ghost insertions are then given by
\begin{equation}
\frac12 {\cal B}_\tau {\cal B}_{\bar \tau} = \frac{i}{4\pi \tau_2} \int d^2z  b(z) \frac{i}{4\pi \tau_2} \int d^2 \bar{w} \bar{b}(\bar{w}),
\end{equation}
where the factor of 1/2 accounts for the fact that $d\tau d\bar{\tau} = 2d\tau_1 d\tau_2$. Note that $\text{Vol}(T^2) = (2\pi)^2 \tau_2$. From our previous discussion, the integration range of $\tau$ should be restricted to the chosen fundamental domain ${\cal F}$; we must also include an extra factor of 1/2 in the measure to account for the $\mathbb{Z}_2$ in the CKG. Putting everything together, the 1-loop scattering amplitude takes the form
\begin{align}\label{eq:torusAmplitude}
{\cal A}_n^{(1)} = \int_{\cal F} \frac{d^2 \tau}{2(4\pi \tau_2)^2} \int d^2 z d^2 w \left\langle b(z) \bar{b}(\bar{w}) c\bar{c} V_1(z_1, \bar{z}_1) \prod_{j=2}^n \int_{T^2} d^2 z_j V_j(z_j,\bar{z}_j) \right\rangle_{T^2} ,
\end{align}
where $\langle \cdots \rangle_{T^2}$ indicates the path integral over the fields on $T^2$ with implicit modulus $\tau$. 

The 1-point amplitude can be recast in a more symmetric form by recognizing that the path integral associated to the correlation function in \eqref{eq:torusAmplitude} is independent of the ghost positions: this permits the substitution $\int d^2z b(z) = 2\text{Vol}(T^2) b(0)$ for both $b$-ghost insertions. Since the full CKG $T^2 \rtimes \mathbb{Z}_2$ is finite (with volume $2 \text{Vol}(T^2)$), it is equivalent to inserting $c$-ghosts at arbitrary positions, using only integrated vertex operators, and then dividing by the volume of the CKG to account for the gauge redundancy. The result is
\begin{align}\label{eq:vacuumAmplitude}
{\cal A}_n^{(1)} = \int_{\cal F} \frac{d^2 \tau}{4\tau_2} \left\langle b(0) \bar{b}(0) c(0) \bar{c}(0) \prod_{j=1}^n \int_{T^2} d^2 z_j V_j(z_j,\bar{z}_j) \right\rangle_{T^2} .
\end{align}

\subsubsection*{The torus partition function} 

Consider the vacuum amplitude ${\cal A}^{(1)}_0$, which for the torus is well-defined and physically meaningful. We will temporarily ignore the moduli space integration; the integrand 
\begin{equation}
Z_{T^2}(\tau) = \langle b(0) \bar{b}(0) c(0) \bar{c}(0) \rangle_{T^2}
\end{equation}
is simply the path integral over the matter and ghost fields with some extra ghost insertions to saturate the fermionic zero modes. In the QFT language, this is often referred to as the torus partition function. It is most readily calculated using the path integral formalism for states in the Hilbert space. In this formalism, the path integral on a (Euclidean) cylinder of length $T$ simply computes the overlap $\langle \psi_f | e^{-H T} | \psi_i \rangle$,  where $H$ is the Hamiltonian and $\ket{\psi_{i,f}}$ are the final and initial states, respectively, which appear through the boundary conditions implemented on each end of the cylinder. This line of reasoning extends to the torus, which can be thought of as a cylinder of length $2 \pi \tau_2$ whose boundaries are glued after a $2\pi \tau_1$ twist. This twist is accounted for by inserting $e^{2\pi i \tau_1 P}$ in the overlap, where $P$ is the generator of translations along $\sigma^1$, \emph{i.e.} the momentum. The two operators $H$ and $P$ generate the isometries of the cylinder, and can be expressed in terms of the Virasoro generators as
\begin{equation}
  H = L_0 + \overline{L}_0 \,, \qquad \text{and} \qquad P = L_0 -
  \overline{L}_0 \,.
\end{equation}
Finally, the operation of gluing becomes the trace over all states in the Hilbert space $\mathcal{H}$ with periodic boundary condition for ghosts.  The path integral can thus be rewritten as
\begin{equation}
  Z_{T^2}(\tau) = \text{Tr}_{\mathcal{H}} \left( (-1)^F b_0 \bar{b}_0 c_0 \bar{c}_0 q^{L_0} \overline{q}^{\overline{L}_0} \right) \,, \quad q = e^{2 \pi i \tau} \,, \quad \overline{q} = e^{- 2 \pi i \overline{\tau}}\,,
\end{equation}
where $(-1)^F$, by definition, anti-commutes with the ghost fields and commutes with the matter fields (this determines its action on all of the states via the state-operator mapping). The full vacuum amplitude is given by reintroducing the moduli space integration:
\begin{align}
{\cal A}_0^{(1)} = \int_{\cal F} \frac{d^2\tau}{2\tau_2}  \text{Tr}_{\mathcal{H}} \left( (-1)^F b_0 \bar{b}_0 c_0 \bar{c}_0 q^{L_0} \overline{q}^{\overline{L}_0} \right) .
\end{align}

While we now have all the ingredients to compute the vacuum amplitude for the bosonic string, we pause briefly to flesh out some of details of the CFT partition function; this will come in handy later when more general matter CFTs arise, such as in string compactifications. The full state space of the bosonic string splits into the direct sum $\mathcal{H}_m \oplus \mathcal{H}_{gh}$, where the subscripts denote the matter and ghost spaces, respectively. Consequently, the partition function factorizes into separate ghost and matter contributions. The matter CFT contributes
\begin{equation}
Z_m(\tau) = \text{Tr}_{\mathcal{H}_m} \left(q^{L_0-c/24} \overline{q}^{\overline{L}_0-\bar{c}/24} \right) \,, 
\end{equation}
where the central charges $c = \bar{c} = 26$ now contribute due to the conformal anomaly on the cylinder. This equation is often the more familiar one in the CFT context, and is equal to $\langle 1 \rangle_{T^2}$ when there is no possibility for fermionic zero modes. Due to the trace, the partition function above clearly gives a sum over the matter states, weighted by the conformal weights assigned to each state. One might initially expect that such a sum is unconstrained, though thanks to string theory we know this to be untrue. Given a parametrization $\tau$ of the torus, there is a whole family of equivalent moduli generated by the action of $PSL(2,\mathbb{Z})$. Since the measure $\frac{d^2 \tau}{\tau_2}$ is invariant under this action, we therefore expect that $Z(\tau)$ should generically be modular invariant as well in any consistent string theory.\footnote{Modular invariance at the level of the CFT translates to invariance of the worldsheet string theory under large Diff$\times$Weyl gauge transformations.}

Modular invariance has dramatic consequences at the level of the states in the CFT (and by extension the string spectrum). Using the trace form of $Z(\tau)$ reveals constraints on the operator spectrum. For instance, invariance under $\tau \to \tau +1$ implies that all local operators must have integer spins, \emph{i.e.} $h - \bar{h} \in \mathbb{Z}$. Invariance under $\tau \to -1/\tau$ constrains the high energy spectrum in terms of the low energy data, which leads to various universal features such as Cardy's formula for the density of states. 

We now return to the calculation of the closed string vacuum amplitude in \eqref{eq:vacuumAmplitude}, focusing first on the CFT partition function. The matter oscillators contribute 
\begin{equation}
	\text{Tr}'_{\mathcal{H}_m} q^{L_0 -26/24} \bar{q}^{\bar{L}_0 - 26/24} = \left( |q|^{-2/24} \prod_{n=1}^\infty |1-q^n|^{-2} \right)^{26} = \eta^{-26} \bar{\eta}^{-26},
\end{equation}
where $\eta = q^{1/24} \prod_n (1-q^n)$ is the Dedekind eta function. The trace over zero modes (momentum eigenstates) gives
\begin{equation}
	i V_{26} \int \frac{d^{26} p}{(2\pi)^{26}} (q \bar{q})^{p^2/4} = i V_{26} \int \frac{d^{26} p}{(2\pi)^{26}} e^{-\pi \tau_2 p^2} = iV_d (4\pi^2 \tau_2)^{-13} ,
\end{equation}
where $V_{26}$ is the 26-dimensional spacetime volume. The factor of $i$ comes from Wick rotating the $X^0$ field. The trace over ghost oscillators gives the additional factor
\begin{equation}
	\text{Tr}_{\mathcal{H}_{gh}} (-1)^F b_0 \bar{b}_0 c_0 \bar{c}_0 q^{L_0 + 26/24} \bar{q}^{\bar{L}_0 + 26/24} = \eta^2 \bar{\eta}^2,
\end{equation}
which serves to cancel the two longitudinal degrees of freedom. This is consistent with light-cone quantization, where only the transverse degrees of freedom are dynamical. In addition, it is important to note that the cancellation between longitudinal and ghost degrees of freedom is only possible for a single timelike direction in the target space \cite{Frenkel:1986aa}. In a unitary quantum theory, the Hilbert space of states by definition should have a positive-definite inner product. Consider the usual worldsheet string theory with 26 bosons and the $bc$ ghost system, but now suppose $r$ of the bosons are timelike. The space of states for this theory includes negative norm states due to the timelike and ghost oscillators that should drop out when we project to the BRST cohomology. Therefore, in a unitary string theory the partition function should be unaffected by the insertion of $(-1)^\sigma$ into the trace, where $(-1)^\sigma = +1$ for spacelike oscillators and $(-1)^\sigma = -1$ for timelike. The ghost oscilators have signature $(1,1)$. Thus, nonzero modes with this factor contribute
\begin{equation}
	\text{Tr}'_{\mathcal{H}} \left( (-1)^\sigma (-1)^F b_0 \bar{b}_0 c_0 \bar{c}_0 q^{L_0} \bar{q}^{\bar{L}_0} \right) = \left| \frac{\prod_n (1-q^n)(1+q^n)}{\prod_n (1-q^n)^{26-r}(1+q^n)^r}\right|^2 = |(1-q^n)^{-24}|^2,
\end{equation}
where the last equality follows assuming unitarity. Therefore, we must take $r =1$ and so there is only a single timelike direction compatible with positive definite inner product on BRST cohomology states.

Combining everything yields the vacuum amplitude \cite{Polchinski:1985zf}
\begin{equation}
{\cal A}_0^{(1)} = i V_{26}  \int_{\cal F} \frac{d^2\tau}{2\tau_2} (4\pi^2 \tau_2)^{-13} |\eta|^{-48}  .
\end{equation}
Given that ${d^2 \tau}/{\tau_2}$ and $\tau_2 |\eta|^4$ are themselves modular invariant, clearly the full amplitude is as well. Note that the region $\tau_2 \to 0$ is absent from our domain of integration. This regime describes ultraviolet (UV), or high energy, processes. Without restricting the $\tau$ integration to ${\cal F}$, the integration would produce a UV divergence due to the $\tau_2 \to 0$ region. Unlike in QFT, this type of divergence is not present for the string vacuum amplitude.  The modular invariance of the torus effectively acts as a UV cutoff that renders the theory UV finite at one loop. This behavior turns out to be true for general string amplitudes, and thus perturbation of the closed bosonic string is expected to be UV finite. That being said, if we expand the integrand as a power series in $q$, we find a term that behaves like  $q^{-1}$. This is due to the tachyon, and ultimately gives rise to an IR divergence. This is the failure of perturbation theory and so bosonic string amplitudes are only formally defined. Our efforts have not truly ended in failure, however, since these IR divergences will turn out to be absent for certain superstring theories.

\section{Bosonic string compactifications}
\label{sec:stringCompactifications}

Up to now we have considered closed strings propagating in flat Minkowski spacetime. However, with the NLSM worldsheet theory introduced in Section 2.2, it is natural to consider more general target spaces for the worldsheet fields. For example, it could be that the target space of the worldsheet theory has a component of $\mathds{R}^{k-1,1}$ where the $SO(1,k-1)$ isometry of $\mathds{R}^{k-1,1}$ extends to an exact global symmetry of the worldsheet theory. Such a theory could potentially describe a string theory in a $k$ dimensional Minkowski spacetime. We can go one step further and replace $\mathds{R}^{k-1,1}$ with a space with non-zero curvatures associated with $G_{ij}$ and $B_{ij}$. Such a theory can potentially describe a theory in a curved spacetime with torsion! In general we can think of any modular invariant conformal field theory with zero total central charge as a string theory. Depending on the structure of the target space of the theory and its global symmetries, the theory may or may not have a spacetime interpretation. In the following we review some important examples of 2d CFTs and methods to construct them.

\subsection{Geometric CFTs} 

Circling back to the Minkowski spacetime, an important class of string theory backgrounds arise from worldsheet theories with the form of a product CFT
\begin{equation}
\mathbb{R}^{k-1,1}\times\mathscr{C}_{c=26-k} ,
\end{equation}
where $\mathbb{R}^{k-1,1}$ refers to the $X$ CFT with $k$ free noncompact scalars, and $\mathscr{C}_{c=26-k}$ is some compact CFT with central charge $c =26-k$. The condition $c=26-k$ ensures that after taking the ghost theory into account, the total central charge vanishes. Often times this auxiliary CFT can be expressed as a NLSM with a compact target space (though in general the compact CFT need not have a geometric interpretation).  Sometimes these backgrounds can be thought of as $k$ large spacetime dimensions with $26-k$ curled up dimensions where the spectrum of $\mathscr{C}$ is discrete. In this subsection we study some examples where this geometric interpretation is explicit in the sense that the $\mathscr{C}_{c=26-k}$ theory is a $26-k$ dimensional NLSM. These theories are often called compactifications of the $26$ dimensional theory and the target space of the NLSM is called the internal geometry. Typically, the geometric parametries such as the size of the cycles become dynamical fields. Thus, the case where the $26$ dimensions are flat corresponds to a particular limit of the field space where the cycles have infinite radii.

The simplest string compactification that we can consider is one with a one-dimensional internal geometry, a circle. In what follows, we will explain some aspects regarding such theories. 

\subsubsection*{Kaluza-Klein mechanism}

First recall field theories coupled to classical gravity compactified on a circle of radius $R$. Suppose the spacetime has $d+1$ dimensions where $x^\mu$ for $\mu\in\{0,...,d-1\}$ are non-compact while $x^d$ is the compact coordinate with the periodicity condition $x^d\sim x^d+2\pi R$. Since the spacetime momentum $p_d$ is generator of translation and a translation of $2\pi R$ along the circle is identity.
\begin{align}
    e^{i2\pi R~\hat{\bf p}_d}\equiv\mathds{1}.
\end{align}
Therefore, the eigenvalues of momentum along the circle $p_d = n/R$ are quantized with $n \in \mathbb{Z}$. Moreover, we can decompose a massless $d+1$ dimensional field $\phi$ into an infinite set of Fourier modes along the compact dimension. 
\begin{align}
    \phi(x^0,...,x^d)=\sum_{n\in\mathbb{Z}} \phi_n(x^0,...,x^{d-1})e^{in(x^d/R)}.
\end{align}
The coefficients $\phi_n$ only depend on the non-compact coordinates and hence are $d$ dimensional fields. The spacetime excitation of $\phi_n$ has $p_d=n/R$. And the $(d+1)$-dimensional mass formula
\begin{align}
    E^2=\sum_{\mu\leq d}p_\mu p^\mu
\end{align}
becomes 
\begin{align}
    E^2=n^2/R^2+\sum_{\mu\leq d-1}p_\mu p^\mu.
\end{align}
In other words, the fields develop nonzero masses via the compact momentum, \emph{i.e.}  
\begin{align}
    m^2 = n^2/R^2.
\end{align}
Such modes are ubiquitous in general compactifications, where they are referred to as Kaluza-Klein (KK) modes. In string theory, circle compactifications are more interesting because the string can wind multiple times around the circle. 

\subsubsection*{The compactified worldsheet theory}

The worldsheet theory consists of 25 noncompact scalars $X^\mu$ as well as a periodic scalar $X^{25}: \Sigma \to S^1_R$.  For a string which winds $w$ times around the circle we have
\begin{align}
    X^{25}(\sigma+2\pi,\tau) = X(\sigma,\tau) + 2\pi R w.
\end{align}
These winding modes are topologically distinct and can be thought of as solitons of the worldsheet theory. The spacetime momentum is quantized according to
\begin{align}
	p^{25} = \frac{1}{2\pi}\int_{0}^{2\pi} d\sigma \partial_\tau X^{25} = \frac{n}{R} .
\end{align}
We must include both the KK and winding modes to ensure modular invariance on the torus. The momentum and winding constraints imply that the mode expansion of $X^{25}$ on the cylinder takes the form \cite{Green:1982sw,Narain:1986am}
\begin{align}
	X^{25}(\sigma,\tau) =  wR\sigma+ \frac{n}{R}\tau + \frac{i}{\sqrt{2}}\sum_{n \neq 0} \left( \frac{\alpha_n^{25}}{n} e^{-in(\sigma+\tau)}+\frac{\overline{\alpha}^{25}_n}{n} e^{in(\sigma-\tau)} \right),
\end{align}
where $\alpha^{25}_n, \overline{\alpha}^{25}_n$ are oscillator modes that obey the usual free oscillator algebra. The quantized theory is very similar to that of the noncompact free boson, except for one crucial difference. Previously, the ground state of the worldsheet theory had a continious degenerecy labeled by $26$ dimensional momentum of the Tachyon. One can think of the degenrate vacua as a family of states that transform to each other under the spacetime Lorentz transformations which are global symmetries of the worldsheet. Now that one of the components of the momentum is quantized, the worldsheet ground states are labeled by $\ket{p^\mu; n,w}$. The integer $n$ is associated with the quantization of momentum which reflects the symmetry of the worldsheet theory generated by discrete translation along $X^{25}$. But how about $w$? Could it be that the presence of another integer signals additional features? 

The answer is yes, and global symmetries of the worldsheet lead to gauge symmetry in the spacetime! The pair $(n,w)$ are in fact charges of a $U(1)_n\times U(1)_w$ global symmetry on the worldsheet. In addition to the massless string states that give the 25-dimensional graviton, B-field, and dilaton, we now have additional massless vectors generated by
\begin{align}
	\bar{\alpha}_{-1}^{[25} \alpha_{-1}^{\mu]}\ket{p^\mu; 0,0} \,, \quad 
	\bar{\alpha}_{-1}^{(25} \alpha_{-1}^{\mu)}\ket{p^\mu; 0,0} \,.
\end{align}
These states are the ``photons'' of the two $U(1)$ factors. From the 26d point of view, the gauge field for $U(1)_n$ arises from the metric $G_{25\mu}$, while that of $U(1)_w$ arise from the B-field $B_{25\mu}$. Clearly the $U(1)_w$ symmetry is a purely stringy effect, since strings not particles are charged under the B-field. Note that there  is another massless state $\bar{\alpha}_{-1}^{25}\alpha_{-1}^{25}\ket{p}$ that leads to a massless scalar in spacetime known as a moduli field. It arises from $G_{2525}$, and so changes the size of the circle, which is dynamical.

Another difference in the worldsheet theory of the compact boson is that the oscillators $\alpha_0^{25}$ and $\overline{\alpha}_0^{25}$ are distinct. We can see this by writing the zero modes as
\begin{equation}
	X^{25}(\sigma,\tau) = \frac{1}{2}[p_L(\tau+\sigma)+p_R(\tau-\sigma)] ,
\end{equation}
where we have introduced the left/right-moving momenta
\begin{equation}
	\alpha_0^{25}p_L=\frac{n}{R}+wR, \quad \bar\alpha_0^{25}=p_R=\frac{n}{R}-wR.
\end{equation}
These quantities naturally form a vector 
\begin{align}
    l = \frac{1}{\sqrt{2}}(p_L,p_R) 
\end{align}
which generates a 2d lattice $p (1,0) + q (0,1)$ for integers $p,q$. The lattice has some special properties in terms of the following inner product.
\begin{align}\label{IP}
	l^1 \cdot l^2 = l_L^1 l_L^2 - l_R^1 l_R^2 . 
\end{align}
With this norm we can check that $l\cdot l'\in\mathbb{Z}$. Such lattices are called integral. Moreover, the norm
\begin{align}\label{EL}
	l \cdot l = 2nw \in 2 \mathbb{Z}.
\end{align}
is an even integer. Lattices with this property are referred to as \emph{even}. Moreover, if  $l'\cdot l\in \mathbb{Z}$ for all $l$ in the lattice, then $l'$ is also in the lattice. This implies that the lattice is \emph{self-dual}. Such lattices are called Narain lattices \cite{Narain:1985jj}. 

The inner product \eqref{IP} is preserved under $SO(1,1)$. Therefore, we can find a Narain lattice by applying elements of $SO(1,1)$ on another one. We can generalize this to higher dimensional lattices. Suppose we have a Narain lattice where $p_L$ takes values in a $p$ dimensional lattice and $p_R$ in a $q$ dimensional lattice such that the overall lattice is even self-dual. We denote such a lattice by $\Gamma^{p,q}$. Applying every element of $SO(p,q)$ to a Narain lattice gives another Narain lattice. However, not all of these transformations lead to distinct lattices. For example, if we act on $p_L$ and $p_R$ separately with elements of $SO(p)$ and $SO(q)$ respectively, the lattices will not change, and they simply reflect a change of coordinates. In fact, up to a discrete quotient, the moduli space of $(p,q)$ Narain lattices is $\frac{SO(p,q)}{SO(p)\times SO(q)}$. The dimension of this space is $pq$. In the special case of $p=q=1$, this space is can be parametrized by one variable which is the radius $R$. The radius $R$ serves as the $SO(1,1)$ boost parameter. 
	
\subsubsection*{T-duality}

The self-dual property of the lattice has important physical implications. Consider the $\mathbb{Z}_2$ self-dual transformation
\begin{equation}\label{Tdual}
	R\rightarrow \frac{1}{R}, \quad n\leftrightarrow w,
\end{equation}
which can be recast as the map
\begin{equation}
	p_L\rightarrow p_L, \quad p_R\rightarrow -p_R.
\end{equation}
The invariance of string theory under the duality transformation \eqref{Tdual} is referred to as T-duality \cite{Sathiapalan:1986zb}.  

Instead of $R$, we can consider a parameter $\lambda$ more closely related to the moduli field,
\begin{equation}
	R= e^{\lambda} .
\end{equation}
T-duality now takes the form $\lambda \to -\lambda$. Under T-duality, the radius of the circle is no longer an invariant notion. More surprisingly, very small radii are mapped to large radii, which implies there is no absolute notion of distance. To resolve these issues, we must consider what it means to have a circle of small radius. From the quantum viewpoint, it is unclear how to resolve distances which are smaller than the size of a typical wavepacket. It is natural then to define the physics of small radii in terms of the dual large radii theory. 

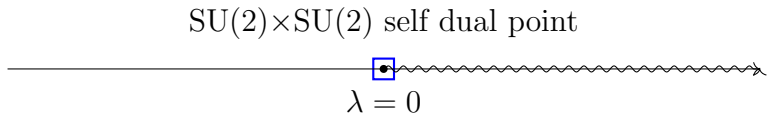
\begin{figure}[H]
\begin{center}
\vspace{2mm}
\begin{tikzpicture}
\node[draw,rectangle,thick,scale=1,blue,label={[label distance=1mm]north:SU(2)$\times$SU(2) self dual point}] (A1) at (0,0) {};
\node[draw,circle,thick,scale=0.2,fill=black,label={[label distance=1mm]south:$\lambda=0$}] (A2) at (0,0) {};
\draw[->] (-5,0)--(5,0);
\tikzset{decoration={snake,amplitude=.4mm,segment length=2mm,post length=0mm,pre length=0mm}}	
\draw[decorate,->] (0,0)--(5,0);
\end{tikzpicture}
\vspace{2mm}
\caption{The space of compactified bosonic string theories on $S^1$. Here, $\lambda = 0$ is the SU(2)$\times$SU(2) self-dual point, so only theories with $\lambda \ge 0$ are considered inequivalent.}
\label{fig:lambda}
\end{center}
\end{figure}

In other words, as seen in Figure \ref{fig:lambda}, the space of inequivalent theories is the half-line $R\geq 1$.  Equivalently, we could take the set of inequivalent theories to be $0\leq R\leq 1$, but it is more natural to think in terms of the larger of the two equivalent radii: momenta continua are more familiar than winding number continua. In particular, questions of locality are clearer in the larger radius picture. For this parametrization, there is no radius smaller than the self-dual radius:
\begin{equation}
	R_{self-dual}=R_{\text{SU}(2)\times\text{SU}(2)}=\sqrt{\alpha'}.
\end{equation}

Clearly $\lambda = 0$ is invariant under this transformation and hence labels the self-dual point. This allows us to simply the diagram of inequivalent theories in terms of $\lambda$ as in Figure \ref{fig:lambda}.\\

\begin{figure}[H]
\begin{center}
\vspace{2mm}
\begin{tikzpicture}
\draw[->] (4,0) arc (0:180:4);
\draw[->] (2,0) arc (0:180:2);
\draw[->] (0.5,0) arc (0:180:0.5);
\draw[->,blue] (0,0) arc (180:260:1);
\node[draw,rectangle,thick,scale=1,blue,label={[label distance=9mm]-40:SU(2)$\times$SU(2) gauge symmetry}] (A1) at (0,0) {};
\node[draw,circle,thick,scale=0.2,fill=black,label={[label distance=1mm]south:$R^2=1$}] (A2) at (0,0) {};
\draw[->] (5,0)--(-5,0);
\end{tikzpicture}
\caption{At the self-dual point $R=\sqrt{\alpha'}$, there is an SU(2)$\times$SU(2) gauge symmetry.}
\label{fig:su2}
\end{center}
\end{figure}
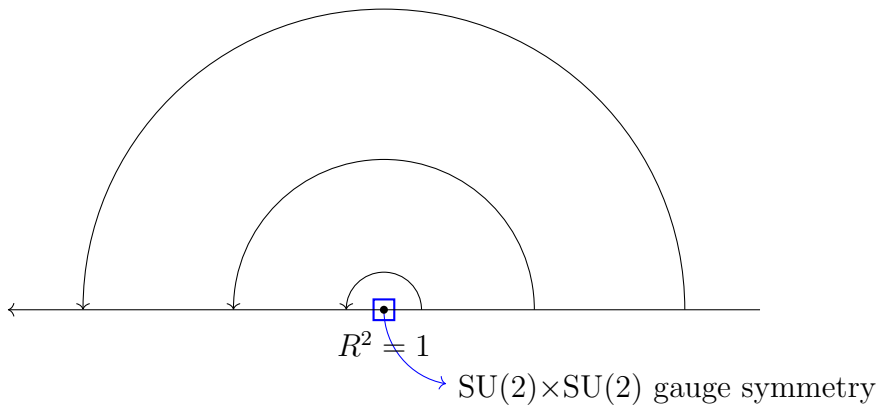

The self-dual point is related to the emergence of new symmetries. As it turns out, at the self-dual point $R = \alpha'/R$ the duality transformation becomes a bonafide symmetry! In particular, there is an emergent SU(2)$\times$SU(2) gauge symmetry in spacetime, as shown in Figure  \ref{fig:su2}.

It is fairly straightforward to derive the emergent symmetry from the worldsheet theory, where we expect the appearance of new conserved currents. The mass-shell conditions for the compactified theory at arbitrary radius are \cite{Polchinski:1998rq}
\begin{align}
    \frac{1}{2}p_L^2&-\frac{1}{2}p_R^2+\frac{4}{\alpha'}(N_L-N_R)=0 \\
    \frac{1}{2}m^2&=\frac{1}{2}p_L^2+N_L-1=\frac{1}{2}p_R^2+N_R-1.
\end{align}
For the self-dual radius $R=\sqrt{\alpha'}$ and $m^2 =0$, these take the simplified form
\begin{align}
	0 = n^2+w^2	+ 2(N_L+N_R - 2)
\end{align}
Let us focus on a single sector, say the left-moving one with $N_L > 0$ and $N_R = 0$. Then the massless states are given by
\begin{align}
	(n,w) = (1,1), \quad N_L = 1, N_R = 0 \,,	 \\
	(n,w) = (-1,-1), \quad N_L = 1, N_R = 0 .
\end{align}
These two states correspond to the left-moving $W^{\pm}$ bosons! The right-moving states are recovered from $N_R = 1$ and $N_L = 0$. The $Z$ bosons arise from states corresponding to the aforementioned $U(1)$ fields that emerge due to the KK compactification at any radius. Altogether, these particles combine into two independent SU(2) triplets. 

There is a nice connection between the duality group $\mathbb{Z}_2$ and the gauge group $SU(2)$. At the self-dual point, the duality group becomes a symmetry. Therefore, the $\mathbb{Z}_2$ must map a representation of $SU(2)$ to another representation of $SU(2)$ with same dimensions. Since every representation is specified by its weights, the symmetry group $\mathbb{Z}_2$ must map a weight lattice of $SU(2)$ to another weight lattice. In fact, the $\mathbb{Z}_2$ is the Weyl group of reflection symmetries of the root lattice of $SU(2)$. This connection extends to many other examples. At the self-dual point, a subgroup of the duality group is the symmetry group of the root lattice of the emergent gauge group.

\paragraph{Optional exercise}\hspace{-8pt}: Consider a compactification on the torus $T^n(G,B)$. Show that the partition function takes the form
\begin{equation}
	Z=\frac{\sum q^{\frac{1}{2}p_L^2}\bar{q}^{\frac{1}{2}p_R^2}}{\eta^n\bar{\eta}^n},\quad\text{for}\quad \eta=q^{1/24}\prod_n (1-q^n).
\end{equation}
Check that it is invariant under modular transformations, \emph{i.e.} $\tau \to \tau + 1$ and $\tau \to -1/\tau$. Show that invariance under T-transformations implies that the lattice of momenta is even, and invariance under S-transformations implies that the lattice is self-dual.

\noindent\textbf{Exercise 1: } Suppose we want to compactify $d$ dimensions where $d$ is the rank of an $ADE$ group $G$. Show that you can choose the background fields $B_{ij}$ and $G_{ij}$ such that the resulting lattice of $(P_L,P_R)$ of the momenta in the compact directions is a weight lattice $(w,w')$ of $G$ where the difference $w-w'$ is in the root lattice of $G$. Using this, show that this lattice is even and self dual. Now show that we can choose the size of the compact dimensions such that the compactified theory has a $G\times G$ gauge symmetry.

\subsection{WZW models and current algebras}

More generally, we can talk about geometric string compactifications via NLSMs with an arbitrary spacetime metric and H-flux where $H=dB$. For instance, we can take the target space to be some group manifold $G$. The NLSM action can be rewritten in Wess-Zumino-Witten (WZW) form \cite{Witten:1983ar}
\begin{equation}
S = -\frac{k}{8\pi}\int_{\Sigma = \p M^3}\text{Tr}\left(g^{-1}d g\right)^2 +ik S_{WZ},
\end{equation}
where the target space coordinates are embedded in the group element $g = \exp(i T^a X^a)$ with $T^a$ the generators of the Lie algebra $Lie(G)$. The Wess-Zumino term \cite{Wess:1971yu}
\begin{equation} 
S_{WZ} = \frac{1}{24\pi}\int_{M^3} \text{Tr}(g^{-1}dg)^3
\end{equation}
arises from the B-field contribution, rewritten in terms of the H-flux via Stoke's theorem. Both terms in the WZW action are necessary to preserve conformal invariance to $O(\alpha'^2)$ \cite{Witten:1983tw}.

\paragraph{Optional exercise}\hspace{-8pt}: Show that the Wess-Zumino term arises from the coupling of the B-field to the worldsheet.

\subsubsection*{Current algebras}

The WZW action is invariant under the the transformation 
\begin{equation}
	g \to L g R^{-1} ,
\end{equation}
where $L$ and $R$ lie in separate copies $G_L$ and $G_R$ of the group manifold. The theory thus admits a $G_L \times G_R$ global symmetry with conserved currents
\begin{equation}
J^i(z)=\text{Tr}\left(\p g g^{-1} T^a\right), \quad \bar{J}^j(\bar{z}) = \text{Tr}\left(\bar{\p} g g^{-1} T^a\right) .
\end{equation}
In the quantum theory, the WZ term leads to an infinite-dimensional enhancement of this global symmetry. The currents satisfy a particular type of OPE \cite{Knizhnik:1984nr}
\begin{equation}
J^a(z) J^b(w)\sim \frac{k\delta^{ab}}{(z-w)^2}+\frac{if^{abc}J^c}{z-w},
\end{equation}
which leads to an extension of the Virasoro algebra known as a current algebra. 
Here, $f^{abc}$ are the structure constants for $Lie(G)$ and the value $k$ is quantized due to the Dirac quantization condition for $B$. As it is an extension of Virasoro, the currents completely specify the stress tensor and its associated central charge
\begin{equation}
c=\frac{k\ \text{dim }G}{k+\hat{h}_G} ,
\end{equation}
where $\hat{h}_G$ is the dual Coxeter number for $Lie(G)$. The classical limit $k \to \infty$ corresponds to the case where the group manifold becomes flat. In this limit, $c \sim \text{dim}\: G$, coinciding with the fact that the theory describes $\text{dim} \: G$ free bosons. 
\vspace{10pt}

\noindent\textbf{Exercise 2: }Show that $rank(G)\leq c\leq dim(G)$. For simply laced the left hand side saturates. 
\vspace{10pt}

As a simple example, we consider a $\hat{\mathfrak{u}}(N)$ current algebra at level $k=1$ constructed from $N$ complex Weyl fermions $\psi^a$. The worldsheet action on the complex plane takes the form
\begin{align}
	S = \int d^2 z 	\bar\psi^{\bar a} \bar{\partial} \psi^a ,
\end{align}
which have a singular OPE
\begin{align}
	\Psi^a(z) \Psi^{\bar{b}}(0) \sim \frac{\delta^{a \bar{b}}}{z}.
\end{align}
Despite notation, both $\Psi$ and $\bar{\Psi}$ are holomorphic fields.
Their stress tensor is given by
\begin{align}
	T = - \frac12 \delta^{a \bar b} \Psi^a \partial \bar{\Psi}^{\bar b}
\end{align}
which fixes the central charge to be $c=N$, \emph{i.e.} each complex fermion contributes $c=1$. Note that $\bar{T} = 0$ and so $\bar{c} = 0$ as well. The theory is naturally invariant under $U(N)$ rotations, under which $\Psi$ ($\bar{\Psi}$) transforms as the fundamental representation (anti-fundamental). The associated conserved currents take the form
\begin{align}
	J^{a \bar{b}} = \Psi^a \Psi^{\bar{b}}.
\end{align}
It is a straightforward exercise to verify they satisfy the $\hat{\mathfrak{u}}(N)$ current algebra with $k=1$.

\subsubsection*{Coset models}

Another collection of models we can consider are the so-called \emph{coset models} (gauged WZW models) \cite{Gawedzki:1991yu}. Consider a WZW model with group $G$, stress tensor $T_G$, and central charge $c_G$. Supposing that $G$ contains some subgroup $H \subset G$, there is an additional set $(T_H, c_H)$ for the WZW model for $H$. We can then ``gauge out'' $H$ by constructing a new theory with a stress tensor and central charge given by
\begin{align}
	T_{G/H} = T_G - T_H \,, \quad c_{G/H} = c_G - c_H .	
\end{align}
This corresponds to taking the original WZW action and gauging the subgroup $H$.

\subsection{Orbifolds}
\label{sec:orbifolds}

Thus far we have studied toroidal compactifications of string theory, with the key stringy ingredient being the emergence of winding modes, \emph{i.e.} the string wrapping various one-cycles. We also saw that in general it may be necessary to turn on H-flux (in addition to the nontrivial metric) to preserve conformal invariance on the worldsheet. In this section, we consider another example of worldsheet CFT, orbifolds, which are manifolds with singular curvature \cite{Dixon:1985jw,Dixon:1986jc,Dijkgraaf:1989hb}.

\subsubsection*{Orbifold CFTs}

Our first experience with orbifolds will be through \emph{orbifold CFTs}, which arises from gauging discrete worldsheet symmetries. As we will see, the singular geometry of the target space will be emergent from this point of view. 

Consider a unitary CFT invariant under some discrete symmetry group $G$, which we label by  CFT$_G$. By definition, the group is represented unitarily, \emph{i.e.} the theory admits a $G$-action on states of the form
\begin{align}
g: \ket{\psi} \to U(g) \ket{\psi},	\quad g \in G,
\end{align}
where $U(g)$ is a unitary operator. When there is no room for ambiguity, we write $g$ as a shorthand for $U(g)$. A natural attempt at gauging the symmetry is to project the Hilbert space to a $G$-invariant subspace:
\begin{equation}
\mathcal{H}\longrightarrow \mathcal{H}^G:=P_G\mathcal{H}, \quad
P_G=\frac{1}{|G|}\sum_{g\in G}g .
\end{equation}
Indeed, the map $P_G$ is a Hermitian projection operator
\begin{equation}
\ln P_G=P_G, \quad P_G^2=P_G.
\end{equation}
The evaulation of partition function of this theory simply amounts to inserting $P_G$ in the trace
\begin{equation}
Z_G =\frac{1}{|G|}\sum_{g \in G}\mathrm{Tr}_{\mathcal{H}} g\ q^{L_0-c/24}\bar{q}^{\bar{L}_0-\bar{c}/24}.
\end{equation}

An immediate problem is that $Z_G$ is not modular invariant, and so our naive gauging procedure is incompatible with string perturbation theory. Consider an $S$-transformation that interchanges the temporal and spatial circles. Before, we could think of $g$ as acting on states every time we wind around the temporal circle. Now, it must act when we wind around the spatial circle. We can no longer think about $g$ as acting on states but rather as implementing different boundary conditions for local operators
\begin{align}
{\cal O}(\sigma+2\pi,\tau) = U_g {\cal O}(\sigma,\tau) U_g^\dagger := g \cdot {\cal O}(\sigma,\tau).
\end{align}
Under the state-operator mapping, these operators correspond correspond to states in some new space ${\mathscr H}_g$ known as a \emph{twisted sector}. Note that the twisted sector coincides with neither ${\mathscr H}$ nor $P_G{\mathscr H}$, where all of the (bosonic) operators are periodic.
	 
We should therefore modify the gauged theory CFT$_G$/$G$ to include all $G$-invariant states, including the twisted sectors
\begin{equation}
{\mathscr H}_{\text{CFT}_G/G} =  \bigoplus_{g \in G} P_G {\mathscr H}_g .
\end{equation}
We are then left with a well-defined, modular invariant partition function
\begin{align}
Z_{\text{CFT}_G/G} = \frac{1}{|G|} \sum_{g,h \in G| [g,h]=1} \mathrm{Tr}_{\mathcal{H}_h} g\ q^{L_0-c/24}\bar{q}^{\bar{L}_0-\bar{c}/24},
\end{align}
	where in the h-twisted sector we have restricted to symmetries $g$ that do not change the sector ($gh=hg$).
\begin{figure}[H]
    \centering

\tikzset{every picture/.style={line width=0.75pt}} 

\begin{tikzpicture}[x=0.75pt,y=0.75pt,yscale=-1,xscale=1]

\draw   (339.93,21) -- (430,21) -- (430,100.95) -- (339.93,100.95) -- cycle ;

\draw (185,43.4) node [anchor=north west][inner sep=0.75pt]  [font=\large]  {$Z=\frac{1}{|G|}\sum _{gh=hg}$};
\draw (380.92,107.37) node [anchor=north west][inner sep=0.75pt]    {$h$};
\draw (322.97,51.27) node [anchor=north west][inner sep=0.75pt]    {$g$};

\end{tikzpicture}
    \caption{Partition function of an orbifold theory including the twisted sectors.}
\end{figure}

We can assign different phases to each sector which corresponds to turning on the B field. However, theses phases, which are also called discrete torsion, cannot be assigned arbitrarily. The phase assignment must not break modular invariance \cite{Vafa:1986wx}. Of course just as in any gauging, there cannot be anomalies which renders it inconsistent. In string theory context, this amounts to the left-right level matching condition.

\subsubsection*{Geometric orbifolds}

With the understanding of how to construct orbifold theories on the worldsheet, we now investigate the consequences on the spacetime geometry. 

As a simple example, we begin with toroidal orbifolds. Take a torus $T^d$ equipped with a group action for some discrete group $G$. We can consider the quotient space $T^d/G$, which generically has a non-trivial topology. Any fixed points of $G$ are necessarily singular. For instance, consider the unit circle $S^1$ whose center coincides with the origin of the $(x,y)$ plane. It admits a $\mathbb{Z}_2$ symmetry that sends $y \to -y$.  The space $S^1/\mathbb{Z}_2$ is clearly an orbifold, since the symmetry leaves the two points at $y=0$ and $y=\frac{1}{2}$ fixed. We can visualize $S^1/\mathbb{Z}_2$ as a line segment stretching from $y-0$ to $y=\frac{1}{2}$, with two singularities at its endpoints. 

We now seek the CFT construction of geometric orbifolds. A NLSM with some target space manifold $M$ will generically be invariant under a discrete subgroup $G$ of the full isometry group of $M$. As we saw in the previous section, from the CFT perspective, discrete gauging requires that orbifold states are singlets under $G$. Moreover, to ensure modular invariance we must include all of the twisted sectors. For the closed string, it is easy to see that the twisted sectors correspond to identifying spacetime points $X^i$ with the same $G$-orbit. Now, an open string stretched between these two points is, in fact, a closed string. This leads to new string states which can be precisely identified with the twisted sector states. 

A particularly illuminating example is the bosonic string on the torus $T^2$. The theory has a $\mathbb{Z}_2$ symmetry that maps $x \mapsto -x$ for $x \in T^2$. On the worldsheet, the $\mathbb{Z}_2$-action is given by
\begin{align}
	\partial X^{1,2} \mapsto - \partial X^{1,2}, \quad \partial X^\mu \to \partial X^i, \quad \mu \neq 1,2 \,,
\end{align}
where the $X^\mu$ correspond to the transverse spacetime directions (which we henceforth drop). Note that $(p_L,p_R) \mapsto (-p_L,-p_R)$ and $\alpha_n \mapsto -\alpha_n$ under this map. In the untwisted sector, the $\mathbb{Z}_2$-invariant states are thus given by acting with an even/odd number of oscillators, respectively, on
\begin{align}
	\ket{\pm} = \frac{1}{\sqrt{2}} \left( \ket{p_L,p_R} \pm \ket{-p_L,-p_R} \right)	.
\end{align}
Recall that the partition function of the ungauged theory is
\begin{align}
	Z_{T^2} = \sum_{(p_L,p_R) \in \Lambda}\frac{q^{\frac{1}{2}p_L^2}\bar{q}^{\frac{1}{2}p_R^2}}{\left|
  q^{2/24}\prod_n (1-q^n)^2\right|^2} \,.
\label{eq:Trqqbar}
\end{align}
Once we insert the action of $g$ into the trace, the ground state now contributes
\begin{equation}
	\bra{p_L,p_R} \mathrm{Tr} \ q^{L_0}\bar{q}^{\bar{L}_0} \ket{-p_L,-p_R} ,
\end{equation}
which is non-zero only for $p_L=p_R=0$. We thus only keep the states with $p_L = p_R = 0$. Given that $g$ acts as $q \to -q$, we easily find the partition function for the untwisted sector
\begin{equation}
 Z = \frac12 Z_{T^2} + \frac12 \frac{1}{\left| q^{2/24}\prod_n (1+q^n)^2\right|^2},
\label{eq:Trgqqbar}
\end{equation}
where the factor of $1/2$ arises from the $1/2$ in $P_g = (1+g)/2$. 

The twisted sectors correspond to the choice of boundary conditions
\begin{equation}
	X^{1,2}(\sigma+2\pi, \tau) = \pm X^{1,2}(\sigma,\tau) .
\end{equation}
These periodicity conditions lead to a new ground state as well as a different set of oscillators with half-integer grading. There are four fixed points of the $\mathbb{Z}_2$ action on $T^2$, which simply produces a multiplicative factor in the twisted sector partition function. The ground state energy in this sector can be calculated from the regularized sum
\begin{equation}
  \frac{1}{2}\sum_{n=0}^{\infty}
  (n+\eta)=-\frac{1}{24}+\frac{1}{4}\left(\eta(1-\eta)\right) \,.
\end{equation}
The modes here are half-integers, so $\eta = 1/2$ and $E = +1/48$. This gives the trace in the twisted sector,
\begin{equation}\label{T2Z2twistedSector}
\text{tr}_{\mathcal{H}_{\text{twisted}}} q^{L_0-c/24} \bar{q}^{\bar{L}_0-\bar{c}/24} =  \frac{4}{\left|q^{-2/48}\prod_n (1-q^{n+1/2})^2\right|^2}
\end{equation}
Of course, this trace only gives the partition function \emph{before} gauging, \emph{i.e.} we're missing $\text{tr}_{\mathcal{H}_{\text{twisted}}}(g (\cdots))$. This is given by 
\begin{equation}
\text{tr}_{\mathcal{H}_{\text{twisted}}} g q^{L_0-c/24} \bar{q}^{\bar{L}_0-\bar{c}/24} =  \frac{4}{\left|q^{-2/48}\prod_n (1+q^{n+1/2})^2\right|^2}.
\end{equation}
The complete $T^2/\mathbb{Z}_2$ partition function is thus
\begin{equation}\label{T2Z2orbifold}
	Z_{T^2/\mathbb{Z}_2} = \frac12 Z_{T^2} + \frac12 \frac{1}{\left| q^{2/24}\prod_n (1+q^n)^2\right|^2} + \frac{2}{\left|q^{-2/48}\prod_n (1-q^{n+1/2})^2\right|^2} + \frac{2}{\left|q^{-2/48}\prod_n (1+q^{n+1/2})^2\right|^2}.
\end{equation}
\vspace{10pt}

\noindent{\bf Exercise 3:} Verify that the last term in \eqref{T2Z2orbifold} can be obtained by an appropriate modular transformation on \eqref{T2Z2twistedSector}. Also verify \eqref{T2Z2orbifold} is invariant under $\tau\rightarrow\tau+1/2$.
\vspace{10pt}

\noindent{\bf Optional exercise:} Consider the orbifold $T^2/\mathbb{Z}_3$. The $\mathbb{Z}_3$ action on $T^2$ can be written as
\begin{equation}
  z=x^1+ix^2,\quad z\rightarrow \omega z,\quad \omega^3=1 \,.
\end{equation}
Construct the partition function of this theory; it is helpful to note that the twisted sectors are complex conjugates of one another ($\omega$ and $\omega^{2} = \omega^{-1}$).
\vspace{10pt}

There are also perfectly sensible string theory vacua which are not bonafide geometries, known as \emph{non-geometric compactifications} \cite{Narain:1986qm,Harvey:1987da,Narain:1990mw}. These can arise when considering orbifold theories which cannot be interpreted as target space geometries of the form $M/G$. As an example, we can compactify on a torus $T^d$ to get states whose momenta $(p_L,p_R)$ lie in a $2d$-dimensional lattice $\Lambda$. For certain choices of metric and B-flux, it is possible to arrange things such that the resulting theory has a larger symmetry group than that of the original torus. This is due to the fact that the dimension of the lattice is what constrains the possible global symmetries of the CFT (\emph{i.e.} group actions). Continuing with the two-dimensional example, such a lattice can have $\mathbb{Z}_2$, $\mathbb{Z}_3$, $\mathbb{Z}_4$, or $\mathbb{Z}_6$ symmetry. However, for $T^2$, the Narain lattice is four-dimensional; it thus could potentially have a $\mathbb{Z}_{12}$ symmetry, which is \emph{not} a symmetry of $T^2$. If we quotient by $\mathbb{Z}_{12}$, the resulting CFT evidently does not possess the interpretation of a NL$\sigma$M with target space $T^2/G$, so it is non-geometric. These constructions have been studied in \cite{Harvey:1987da}. 

\subsubsection*{Monstrous moonshine}

Even before phycisists had encountered orbifolds, mathematicians had been independently studying asymmetric orbifold constructions with the aim of understanding conjectures about the so-called  monster group - the largest sporadic finite simple group (e.g. \cite{Conway:1979qga}). Recall that all simple Lie groups belong to either an infinite series, $SU(n)$, $SO(n)$, $Sp(n)$, or one of the exceptional groups $E_6$, $E_7$, $E_8$, $F_4$, or $G_2$. This is similar to the classification of finite simple groups, which can be split into a set of infinite families together with a finite number of additional groups (known as the sporadic groups). The monster group happens to be the sporadic group with the maximum order, namely 
\begin{equation}
  |G| \sim 8\times 10^{53} \,.
\end{equation}
For comparison, there are around $10^{80}$ atoms in the universe. 

The monster group makes an unexpected appearance in the study of modular functions which are complex functions with specific transformation properties under $SL(2,\mathbb{Z})$ transformations. An important modular function is called the j-function. The j-function is a meromorphic invariant under $SL(2,\mathbb{Z})$ with a single simple pole in the fundamental region of $SL(2,\mathbb{Z})$ at $\text{Im}(\tau)\rightarrow \infty$. 

The expansion of $j$ reads
\begin{align}
    j(q)=q^{-1}+744+196884q+21493760q^2+\hdots,
\end{align}

where the first term represents the simple pole at $\text{Im}(\tau)\rightarrow\infty$. It turns out that the expansion coefficients have a mysterious connection with the Monster group: Every coefficient except the constant term $744$, could be written as elementary sums of the dimensions of some representations of the Monster group! For example, the smallest two irreducible representations of the Monster group have dimensions 1 and 196883 which add up to 196884. 
\begin{equation}
  {\bf 196884} = {\bf 196883} \oplus {\bf 1} \,,
\end{equation}

In the 1970s, mathematicians noticed this connection \cite{10.2307/23296} which motivated them to construct the Monstrous Moonshine representation of the Monster group.

Now, let us go back to physics and see how string theory helps us to understand this unexpected connection. Modular functions have a very natural place in string theory. Because of modular invariance, they show up as partition functions of the worldsheet theory. If j-invariant were to be the partition function of a worldsheet theory, the Monstrous Moonshine suggests that states of the theory fall into representations of the Monster group. In other words, the theory probably has a Monster symmetry. Moreover, the $q^{-1}$ term signals the existance of a Tachyon, just like the Bosonic string theory. In fact, it turns out there is an orbifold construction that reproduces the j-function as the worldsheet partition function!

This construction involves compactifying the 26d target space of the bosonic string to 2d \cite{Schellekens:1991ct}. The resulting 24d internal space is chosen such that the left- and right-moving momenta are elements of a particular 24d lattice, \emph{i.e.}
\begin{equation}
  \ket{p_L,p_R}\in\Gamma_L^{24} \otimes \Gamma_R^{24} \,.
\end{equation}
The relevant conditions are that the lattice is even and self-dual, as required by modular invariance, as well as that the shortest non-trivial vector has length $p_L^2 = 4$. Recall that the root lattices associated with the simple Lie algebras have vectors with a minimum length of $2$. There is exactly one even self-dual lattice in 24 dimensions whose vectors all have length $p_L^2 > 2$, known as the \emph{Leech lattice} $\Gamma_L^{24}$ \cite{leech1967notes}. Given the string compactification, we are free to gauge the $\mathbb{Z}_2$ symmetry (the same symmetry present for $T^2$) and study the orbifold theory:
\begin{equation}
  \Gamma_L^{24}/\mathbb{Z}_2,\quad p_L\longrightarrow -p_L \,.
\end{equation}
This is an asymmetric action which satisfies the level matching conditions in the twisted sector. As before, the orbifold theory has a partition function of the form
\begin{align}
Z=\frac{1}{2}\ \left(
\raisebox{-1cm}{\scalebox{1}{
\begin{tikzpicture}
\node[draw=none,scale=0.2] (A1) at (0,0) {};
\node[draw=none,scale=0.2,label={[label distance=2mm]230:$1$}] (A2) at (1,0) {};
\node[draw=none,scale=0.2,label={[label distance=2mm]230:$1$}] (A3) at (0,1) {};
\node[draw=none,scale=0.2] (A4) at (1,1) {};
\draw (A1)--(A2)--(A4)--(A3)--(A1);
\end{tikzpicture}}}
\quad +\quad
\raisebox{-1cm}{\scalebox{1}{
\begin{tikzpicture}
\node[draw=none,scale=0.2] (A1) at (0,0) {};
\node[draw=none,scale=0.2,label={[label distance=2mm]230:$1$}] (A2) at (1,0) {};
\node[draw=none,scale=0.2,label={[label distance=2mm]230:$g$}] (A3) at (0,1) {};
\node[draw=none,scale=0.2] (A4) at (1,1) {};
\draw (A1)--(A2)--(A4)--(A3)--(A1);
\end{tikzpicture}}}\quad \right)
+\frac{1}{2}\ \left(
\raisebox{-1cm}{\scalebox{1}{
\begin{tikzpicture}
\node[draw=none,scale=0.2] (A1) at (0,0) {};
\node[draw=none,scale=0.2,label={[label distance=2mm]230:$g$}] (A2) at (1,0) {};
\node[draw=none,scale=0.2,label={[label distance=2mm]230:$1$}] (A3) at (0,1) {};
\node[draw=none,scale=0.2] (A4) at (1,1) {};
\draw (A1)--(A2)--(A4)--(A3)--(A1);
\end{tikzpicture}}}
\quad +\quad
\raisebox{-1cm}{\scalebox{1}{
\begin{tikzpicture}
\node[draw=none,scale=0.2] (A1) at (0,0) {};
\node[draw=none,scale=0.2,label={[label distance=2mm]230:$g$}] (A2) at (1,0) {};
\node[draw=none,scale=0.2,label={[label distance=2mm]230:$g$}] (A3) at (0,1) {};
\node[draw=none,scale=0.2] (A4) at (1,1) {};
\draw (A1)--(A2)--(A4)--(A3)--(A1);
\end{tikzpicture}}}\quad \right) .
\end{align}
In fact, it can be computed explicitly: 
\begin{equation}
  Z = q^{-1}+ A_1 q^1+\cdots\,, \quad A_1=196884 \,.
\label{eq:qexpansion}
\end{equation}
There is a single tachyon in the spectrum, hence the $q^{-1}$ term with unit coefficient. The fact that there is no $O(q^0)$ term indicates that there are no massless particles in the spectrum. 

The uniqueness properties of the j-function makes it easy to relate the partition function to the  $j$-function,
\begin{equation}
  Z(q)= j(q)-744 \,.
\end{equation}

At the level of the string theory, this strongly suggest that the particle content of the theory organizes into irreducible representations of the monster group -- that is, the monster group is the symmetry group of this system. It was later proven by Richard Borcherds that the full vertex operator algebra (VOA) of the CFT is a generalized Kac–Moody algebra with the Monster group acting on it \cite{borcherds1992monstrous}. 

\subsection{Noncritical string theory}

A key problem of the critical bosonic string is the existence of the tachyon, which spoils the validity of perturbation theory. We seek a new background which admits a stable string vacuum. Thus far, we have considered backgrounds with nontrivial curvature and B-flux, but with a constant dilaton profile. Let's now consider a $d$-dimensional flat Minkowski background ($G_{\mu\nu} = \eta_{\mu\nu}$ and $B_{\mu\nu} = 0$) but with a linear dilaton profile along a particular spatial direction $X^i$,
\begin{equation}
	\phi(X) = Q X^i .
\end{equation}
The nonlinear sigma model that corresponds to this choice of background is given by the action
\begin{equation}
	S = \frac{1}{4\pi} \int_{\Sigma} d^2 \sigma \sqrt{g} \left(g^{ab} \eta_{\mu\nu} \p_a X^\mu \p_b X_\nu + Q R(g) X^i \right) .
\end{equation}
In conformal gauge, the action reduces to that of $d$ free bosons with the caveat that the stress tensor is modified to
\begin{equation}
	T_m = - (\partial X^\mu \partial X_\mu) + Q \partial^2 X^i .
\end{equation}
The central charge changes accordingly:
\begin{equation}
	c_m = d + 6Q^2 .
\end{equation}
To cancel the ghost contribution, we set $c_m = 26$ as usual. This fixes the value of $Q$ to
\begin{equation}
	Q = \sqrt{\frac{26-d}{6}} .
\end{equation}
We can now use our favorite quantization method (e.g. BRST or light-cone) to determine the string spectrum in this background. Crucially, we now find that the mass squared of the tachyon (\emph{i.e.} lowest mode) is
\begin{equation}
	m_T^2 = -\frac{d-2}{24},
\end{equation}
which vanishes for $d=2$. We have thus found a string background where the tachyon is stable!\footnote{More precisely, the lowest mode of the bosonic string transforms as a massless scalar, which is erroneously but typically referred to as the tachyon.} Such string theories where the initial background is not of critical dimension $d=26$ are known as the \textit{noncritical string theory} \cite{Polyakov:1981rd}.\footnote{An equivalent approach to the noncritical string is to sacrifice Weyl invariance and promote the Weyl mode on the worldsheet to a dynamical field.}  

In two dimensions, the moduli cannot be frozen because the boundary conditions of scalar fields can be dynamically changed with finite energy. Thus, even though the theory has a stable vacuum, we must now contend with a variable coupling without a convergent boundary condition. This will create a problem since in regions where the effective coupling $\lambda(x) = e^{\phi(X)}$ grows large, we lose perturbative control of the theory. By imposing Weyl invariance on the worldsheet of the theory we can find the tachyon profile of the background. A profile that preserves Weyl invariance is
\begin{equation}
	\delta S = \frac{1}{4\pi} \int d^2 z \sqrt{g} \mu e^{\alpha X^i} \,,
\end{equation}
where $\alpha = Q^{1/2}$ is fixed by Weyl invariance (but $\mu$ is not). In this background, the region of large coupling is now cutoff due to the tachyon background. The upshot is we can now do perturbation theory again, although the theory now depends on an additional parameter $\mu$ set by the VEV of the tachyon field. Moreover, the worldsheet theory is now described by an interacting CFT known as \textit{Liouville field theory} \cite{Dorn:1994xn,Zamolodchikov:1995aa}.

\section{Superstring theory} \label{sec:superstrings}

The quantum bosonic string exhibits many interesting physical phenomena, but cannot be probed beyond tree level due the presence of the tachyon. In this section, we will consider a set of alternative models, collectively known as the superstring, with a supersymmetric worldsheet theory. The superstring admits the requisite features of any quantum string theory: a matter CFT, a ghost system, and BRST symmetry. Crucially, certain superstring theories admit consistent truncations which render them free of anomalies and tachyons. There are also certain models which exhibit spacetime supersymmetry. 

\subsection{Basics of the NSR formalism}

The action of the superstring, in Polyakov form, describes ${\cal N}=(1,1)$ supergravity on the worldsheet \cite{Neveu:1971rx,Ramond:1971gb}. It is quite complicated, so we only focus on the action after gauge-fixing. Similar to the bosonic string, we have $d$ free scalars $X^\mu$ and a pair of $bc$ ghosts with spins $(2,-1)$ that arise from worldsheet diffeomorphisms and Weyl rescalings. We additionally have  $d$ pairs of Majorana--Weyl fermions $\psi^\mu$ and $\bar{\psi}^\mu$ which transform as spacetime vectors. There are also new \textit{commuting} ghosts, namely the $\beta\gamma$ ghosts with spins $(\frac{3}{2},-\frac{1}{2})$, which arise from gauge-fixing  super-diffeomorphisms and super-Weyl transformations. The fields and their data are described in Table~\ref{table:superstring}. 

Altogether, they sit inside a free worldsheet action given by 
\begin{equation}\label{eq:NSRsuperstring}
	S_\text{NSR} = \frac{1}{2\pi} \int d^2z \left(\p X^\mu {\bar \p} X_\mu + \frac12 \psi^\mu {\bar \p} \psi_\mu + \frac12 \bar{\psi}^\mu \p \bar{\psi}_\mu + b {\bar \p} c + \bar{b} \p c + \beta {\bar \p} \gamma + \bar{\beta} \p \bar{\gamma} \right).
\end{equation}
In total, the matter fields contribute $c_m = {3d}/{2}$ and the ghosts $c_{gh} = -15$. To avoid the appearance of a Weyl anomaly, we must take $c_m = 15$ and so the superstring naturally lives in $d=10$ spacetime dimensions. 

\begin{table}[H]
\centering
\renewcommand{\arraystretch}{1.5}
\begin{threeparttable}
\begin{tabular}{ccc}
\toprule
Field content & Conformal weight $h$ & Central charge $c$ \\\midrule
$X^\mu$ & $0$ & $d$ \\
$\psi^\mu$ & $\dfrac12$ & $\dfrac{d}{2}$ \\
$(b,c)$ & $(2,-1)$ & $-26$ \\
$(\beta,\gamma)$ & $(\frac32,-\frac12)$ & $11$ \\\bottomrule
\end{tabular}
\end{threeparttable}
\caption{Field content in the gauge-fixed NSR worldsheet theory. The right-moving counterparts have the analogous weights $\bar{h}$ and central charges $\bar{c}$.}
\label{table:superstring}
\end{table} 

The gauge-fixed worldsheet theory admits a residual symmetry known as ${\cal N}=(1,1)$ \textit{superconformal symmetry}, which pairs every operator of weight $h$ with another operator of weight $h+1/2$ with opposite statistics. Both $\psi^\mu$ and $\bar{\psi}^\mu$ are superpartners of $X^\mu$, and similarly for $\beta\gamma$ and $bc$. The stress tensor $T$ is no exception, and has a fermionic superpartner $G$ of weight $3/2$ known as the super stress tensor or \textit{supercurrent}. For the matter CFT, we have 
\begin{equation}
	T_m = - \p X^\mu \p X_\mu - \frac12 \psi^\mu \p \psi_\mu , \quad
	G_m = i \sqrt{2} \psi^\mu \partial X_\mu .
\end{equation}
From this we see that $G_m$ admits the OPE
\begin{align}
	&T_m(z) G_m(0) \sim \frac{3}{2z^2} T_m(0) + \frac{1}{z} \p G_m(0) , \\
	&G_m(z) G_m(0) \sim \frac{d}{z^3} + \frac{2}{z} T_m(0) .
\end{align}
The modes of $G$ together with those of $T$, and their anti-holomorphic counterparts, then obey the ${\cal N}=(1,1)$ \textit{super-Virasoro algebra} with central charge $c=15$ \cite{Friedan:1985ge}. Similarly, the ghost stress tensor and supercurrent obey the same algebra, but central charge $c=-15$.\footnote{The superconformal algebra naturally leads to the notion of a \textit{superconformal primary} $\Phi$, which is annihilated by all of the raising modes of $T$ and $G$. A superconformal primary sits in a multiplet with operators of the form $G_{-n} \Phi$ (with $n >0$) known as superconformal descendants. A number of these operators, including the superconformal primary, are themselves conformal primaries. For instance, $\p X^\mu$ is a descendant of $\psi^\mu$, and so is $b$ with $\beta$.}

Similar to the bosonic string, the gauge-fixed theory of the superstring possesses a BRST symmetry with nilpotent charge
\begin{equation}
Q_B=cT+\gamma G,\quad Q_B^2=0,
\end{equation}
where $T,G$ are the matter+ghost (super) stress tensor.

Similar to the free boson, we can expand $\psi^\mu$ in modes $\psi^\mu_r$ on the cylinder, which obey the fermionic oscillator algebra:
\begin{align}
    \{\psi^\mu_r, \psi^\nu_s\} = \eta^{\mu\nu}\delta_{r,s} .
\end{align}
However, unlike the free boson, the free fermion admits two types of boundary conditions consistent with BRST symmetry: namely periodic or Ramond (R) boundary conditions and anti-periodic or Neveu-Schwarz (NS) boundary conditions \cite{Ramond:1971gb}. The associated modes are different for the two choices, with $r \in \mathbb{Z}$ for R boundary conditions and $r \in \frac12\mathbb{Z}$ for NS boundary conditions, and so lead to separate Hilbert spaces. Furthermore, the choice of boundary condition is independent for $\psi^\mu$ and $\bar{\psi}^\mu$, and so the Hilbert space of the free fermions decomposes into a total of four sectors, which we label by the choice of boundary conditions: (NS,NS), (NS,R), (R,NS), and (R,R).\footnote{The boundary conditions for $\psi^\mu$ must be the same for all values of $\mu$ to preserve spacetime Lorentz invariance as well as BRST symmetry on the worldsheet.}

As was the case for the bosonic string, it is simplest to extract the mass spectrum of the superstring via lightcone quantization. In short, the ghosts and longitudinal modes of the matter fields decouple, and we are left with the transverse fields $(X^i, \psi^i, \bar{\psi}^i)$ for $i = 1,\ldots, 8$. The 8 scalars contribute $-8/24$ to the ground state energy, as before. The fermions contribute different energies depending on the choice of boundary conditions. The NS sector ground state, dual to the identity operator, contributes $-8/48$, which together with the bosonic contribution adds up to $-1/2$. The mass of the NS sector ground state is thus
\begin{align}
\frac{\alpha'}{4} m^2 = - \frac12 .	
\end{align}
We seem to have encountered the same tachyon problem as the bosonic string. Additionally, the oscillators $\psi^i_{-r}$ increase the weight by a half-integer $r$, and so  the theory cannot be modular invariant. As it turns out, both issues can be solved via a method known as the GSO projection \cite{Gliozzi:1976qd} which is forced on us by modular invariance. 

\subsection{Modular invariance and the GSO projection}

While the different sectors are naively independent at the level of states, modular invariance on the torus places additional constraints on which states are allowed. On the torus, the fermions can be periodic (P) or antiperiodic (AP) along the temporal circle. This leads to four choices of boundary condition, two for each circle. The different boundary conditions are related by modular transformations. For instance, it is easy to see that a $T$ transformation maps AA boundary conditions to PA boundary conditions:
\begin{equation}
\raisebox{-1cm}{\scalebox{1}{
\begin{tikzpicture}
\node[draw=none,scale=0.2] (A1) at (0,0) {};
\node[draw=none,scale=0.2,label={[label distance=2mm]230:$A$}] (A2) at (1,0) {};
\node[draw=none,scale=0.2,label={[label distance=2mm]230:$A$}] (A3) at (0,1) {};
\node[draw=none,scale=0.2] (A4) at (1,1) {};
\draw (A1)--(A2)--(A4)--(A3)--(A1);
\end{tikzpicture}}} \quad
\xrightarrow[]{\tau \to \tau +1}
\raisebox{-1cm}{\scalebox{1}{
\begin{tikzpicture}
\node[draw=none,scale=0.2] (A1) at (0,0) {};
\node[draw=none,scale=0.2,label={[label distance=2mm]230:$A$}] (A2) at (1,0) {};
\node[draw=none,scale=0.2,label={[label distance=2mm]230:$P$}] (A3) at (0,1) {};
\node[draw=none,scale=0.2] (A4) at (1,1) {};
\draw (A1)--(A2)--(A4)--(A3)--(A1);
\end{tikzpicture}}} \quad .
\end{equation}
An S-transformation always swaps the two circles, for instance:
\begin{equation}
\raisebox{-1cm}{\scalebox{1}{
\begin{tikzpicture}
\node[draw=none,scale=0.2] (A1) at (0,0) {};
\node[draw=none,scale=0.2,label={[label distance=2mm]230:$A$}] (A2) at (1,0) {};
\node[draw=none,scale=0.2,label={[label distance=2mm]230:$P$}] (A3) at (0,1) {};
\node[draw=none,scale=0.2] (A4) at (1,1) {};
\draw (A1)--(A2)--(A4)--(A3)--(A1);
\end{tikzpicture}}} \quad
\xrightarrow[]{\tau \to -1/\tau}
\raisebox{-1cm}{\scalebox{1}{
\begin{tikzpicture}
\node[draw=none,scale=0.2] (A1) at (0,0) {};
\node[draw=none,scale=0.2,label={[label distance=2mm]230:$P$}] (A2) at (1,0) {};
\node[draw=none,scale=0.2,label={[label distance=2mm]230:$A$}] (A3) at (0,1) {};
\node[draw=none,scale=0.2] (A4) at (1,1) {};
\draw (A1)--(A2)--(A4)--(A3)--(A1);
\end{tikzpicture}}} \quad .
\end{equation}
Altogether, we see that AA, PA, and AP boundary conditions are interchanged under the modular group of the torus, while PP boundary conditions are invariant. 

Clearly modular invariance forces us to include all four sectors and as we will see including everything fixes the problem. As we will now show the above sum leads to a consistent truncation of states known as the Gliozzi-Scherk-Olive (GSO) projection \cite{Gliozzi:1976qd}. We introduce an operator $G = \pm 1$ known as the $G$-parity which implements changing boundary conditions for fermions. The GSO projection amounts to keeping states with $G = 1$ and discarding those with $G = -1$. 

We define $G$-parity in the NS sector via
\begin{align}
G=(-1)^{F+1},
\end{align}
where $F$ is the fermion number (modulo 2). It can be expressed in terms of the fermionic oscillators as
\begin{equation}
F=\sum_{r >0} \psi^i_{-r} \psi^i_r,
\end{equation}
The NS sector ground state $\ket{p}$ is defined to have $F = 0$, and so is thrown out. Excited states receive $F = +1$ from each fermionic oscillator $\psi^i_{-r}$ and $F=0$ from each bosonic oscillator $\alpha^i_{-n}$. Note that the first excited state, $\psi^i_{-1/2} \ket{p}$, is preserved by the GSO projection. It has $m^2=0$ and transforms as an $SO(8)$ vector $\mathbf{8_v}$.

Next we analyze the R sector. Unlike the NS sector, the R sector possesses zero modes which satisfy the Clifford algebra
\begin{equation}
\{\psi^i_0, \psi^j_0 \} = \delta^{ij},	
\end{equation}
and so the ground state transforms as a Dirac spinor $\mathbf{8_s}\oplus \mathbf{8_c}$ under $SO(8)$. In the R-sector, the fermions contribute $+8/24$ to the ground state energy, and so the ground state is massless. We define the $G$-parity of the R-sector through
\begin{align}
G=
\Gamma^{11} (-1)^F,
\end{align}
where the chirality matrix $\Gamma^{11}$ is given by
\begin{equation}
\Gamma^{11}= 2^4 \ \psi_0^2\ \psi_0^3\ \psi_0^4\ \psi_0^5\ \psi_0^6\ \psi_0^7\ \psi_0^8\ \psi_0^9.
\end{equation}
By definition, it acts as $\Gamma^{11} = \pm 1$ on states of definite chirality, and so the GSO projection removes either $\mathbf{8_s}$ or $\mathbf{8_c}$. Without loss of generality, we can choose $G$ to act as $+1$ on the $\mathbf{8_s}$ in the holomorphic sector, removing the $\mathbf{8_c}$. For the antiholomorphic sector, we then have a choice as to which chirality is removed from the R sector. This leads to two (seemingly) inequivalent string theories, a nonchiral theory (type IIA) with $\mathscr{N}=(1,1)$ spacetime supersymmetry and a chiral theory (type IIB) with $\mathscr{N}=(2,0)$.  Note that for both choices of GSO projection, \textit{all} states of half-integer spin are removed, as required of a modular invariant theory. Furthermore, spacetime supersymmetry is \textit{emergent} (and is notably before the GSO projection is enforced). 

In total, there are four sectors of states in each theory depending on the choice of boundary conditions, \emph{i.e.} (NS,NS), (R,R), (NS,R), and (R,NS). The (NS,NS) and (R,R) sectors have states with integer spins (spacetime bosons), whereas the (NS,R) and (R,NS) sectors have states with half-integer spins (spacetime fermions). The massless states of the type IIA theory consist of states in $(\mathbf{8_v}\oplus \mathbf{8_s}) \otimes (\mathbf{8_v}\oplus \mathbf{8_c})$, whereas the type IIB theory has states in $(\mathbf{8_v}\oplus \mathbf{8_s})^2$, which can further be decomposed into the various irreducible representations of Spin(8). 

Let's first focus on the spacetime bosons. The (NS,NS) sector leads to states of the form
\begin{equation}\label{decom}
\mathbf{8_v}\otimes\mathbf{8_v}=\mathbf{1}+\frac{8\cdot 7}{2}+\left(\frac{8\cdot 9}{2}-1\right)=\mathbf{1}+\mathbf{28}+\mathbf{35},
\end{equation}
which are just the usual massless particles of the bosonic string, namely the dilaton, Kalb-Ramond field, and dilaton, respectively. The (R,R) sector consists of states in the tensor product of two 8-dimensional spinor representations. The bosonic represntations that appear can be understood from sandwiching the SO(8) gamma matrices $\gamma^i$ between two spinors $\psi, \chi$, forming the invariants $\overline{\chi} \gamma^{i_1} \cdots \gamma^{i_p} \psi$. An odd number of $\gamma$ matrices leads to a type IIA irrep, whereas an even number leads to a type IIB irrep. All of them are massless particles corresponding to some $p$-form gauge field. For type IIA, we find
\begin{align}
\mathbf{8_s}\otimes\mathbf{8_c}= \mathbf{8} \oplus \mathbf{56}
\end{align}
corresponding to a 1-form $C_\mu$ and a 3-form $C_{\mu\nu\sigma}$. Recall that the Dynkin diagram of SO(8) has a $\mathbb{Z}_3$ symmetry which translates into a $\mathbb{Z}_3$ symmetry of the root system. The action of the $\mathbb{Z}_3$ could be extended to the weights of any representation, mapping any representation to another one with the same dimension. This transformation is called triality and it permutes the three 8 dimensional representations $\mathbf{8_v}$, $\mathbf{8_s}$, and $\mathbf{8_c}$. By applying the triality transformation on \eqref{decom}, for type IIB, we find
\begin{align}
	\mathbf{8_s}\otimes\mathbf{8_s}= \textbf{1} \oplus \textbf{28}' \oplus \textbf{35}',
\end{align}
corresponding to a 0-form $\lambda$, an R–R 2-form $C_{\mu\nu}$, and a 4-form $C_{\mu\nu\sigma\rho}$. Notice that a 4-form naively has $70$ degrees of freedom. The 4-form corresponding to the $\textbf{35}'$ in type IIB is in fact self-dual, which removes the other half.

The torus partition function of the type IIA/B string theories can be determined after imposing the relevant GSO projections. Note that for all cases, the NS and R sectors contribute 
\begin{align}
Z^{\text{NS}}&=\frac{1}{2}\left[ \frac{q^{-1/2}\prod_n (1+q^{n+1/2})^8}{\prod_n (1-q^n)} - \frac{q^{-1/2}\prod_n (1-q^{n+1/2})^8}{\prod_n (1-q^n)}\right]=\frac{1}{2}(\hspace{-7pt}\raisebox{-1cm}{\scalebox{1}{
\begin{tikzpicture}
\node[draw=none,scale=0.2] (A1) at (0,0) {};
\node[draw=none,scale=0.2,label={[label distance=2mm]230:$A$}] (A2) at (1,0) {};
\node[draw=none,scale=0.2,label={[label distance=2mm]230:$A$}] (A3) at (0,1) {};
\node[draw=none,scale=0.2] (A4) at (1,1) {};
\draw (A1)--(A2)--(A4)--(A3)--(A1);
\end{tikzpicture}}}+\hspace{-7pt}\raisebox{-1cm}{\scalebox{1}{
\begin{tikzpicture}
\node[draw=none,scale=0.2] (A1) at (0,0) {};
\node[draw=none,scale=0.2,label={[label distance=2mm]230:$A$}] (A2) at (1,0) {};
\node[draw=none,scale=0.2,label={[label distance=2mm]230:$P$}] (A3) at (0,1) {};
\node[draw=none,scale=0.2] (A4) at (1,1) {};
\draw (A1)--(A2)--(A4)--(A3)--(A1);
\end{tikzpicture}}}) ,\\
Z^{\text{R}}&=\frac{1}{2}\left[ 16 \frac{\prod_n (1+q^{n})^8}{\prod_n (1-q^n)}+0\right] =\frac{1}{2}(\hspace{-7pt}\raisebox{-1cm}{\scalebox{1}{
\begin{tikzpicture}
\node[draw=none,scale=0.2] (A1) at (0,0) {};
\node[draw=none,scale=0.2,label={[label distance=2mm]230:$P$}] (A2) at (1,0) {};
\node[draw=none,scale=0.2,label={[label distance=2mm]230:$A$}] (A3) at (0,1) {};
\node[draw=none,scale=0.2] (A4) at (1,1) {};
\draw (A1)--(A2)--(A4)--(A3)--(A1);
\end{tikzpicture}}}+\hspace{-7pt}\raisebox{-1cm}{\scalebox{1}{
\begin{tikzpicture}
\node[draw=none,scale=0.2] (A1) at (0,0) {};
\node[draw=none,scale=0.2,label={[label distance=2mm]230:$P$}] (A2) at (1,0) {};
\node[draw=none,scale=0.2,label={[label distance=2mm]230:$P$}] (A3) at (0,1) {};
\node[draw=none,scale=0.2] (A4) at (1,1) {};
\draw (A1)--(A2)--(A4)--(A3)--(A1);
\end{tikzpicture}}}).
\end{align}
We can see that the inclusion of sectors with every boundary condition makes the overall partition function modular invariant. The factor of 16 in $Z^{\text{R}}$ arises from the ground state degeneracy, while the factors of $\frac12$ in both come from the GSO projection operator $P = \frac12(1+G)$, which is inserted in the trace. Ultimately, we find that the partition function is not just modular invariant, but it identically vanishes:
\begin{equation}
Z=Z^{\text{NS}}-Z^{\text{R}}=0.
\end{equation}
This is expected of the vacuum amplitude of a supersymmetric theory. Using similar logic, it is possible to conclude that it vanishes for \textit{all} genera.

\subsection{Green-Schwarz superstring}

Although the superstring enjoys spacetime supersymmetry, this is by no means obvious from the point of view of the NSR formalism. Indeed, spacetime supersymmetry is visible in the spectrum only after imposing the chiral GSO projection together on the matter and ghost fields. The Green-Schwarz model of the superstring is an approach to formulate an equivalent string theory on the worldsheet where spacetime supersymmetry is manifest \cite{Green:1983wt}.

In this formalism, the Nambu--Goto action for the bosonic string is replaced with a suitable supersymmetric extension, where the matter field content consists of the usual bosonic fields $X^\mu$ as well as 32 anticommuting scalars $S^\alpha_A$ for $\alpha = 1,\ldots, 16$ and $A=1,2$. Taken together, these fields are interpreted as superspace coordinates such that the $X^\mu$ transform as an SO(1,9) vector $\mathbf{8_v}$ and the $S^\alpha_A$ as two Majorana--Weyl spinors, either $\mathbf{16}$ or $\mathbf{16'}$. Here, the choice of string theory (type IIA or IIB) is chosen at the classical level by fixing the relative chiralities of the $S_A$. 

The action is readily quantized in lightcone gauge, where the theory reduces to that of 8 free scalars $X^i(z,\bar{z})$ and a set of free \textit{worldsheet} fermions, $S^a_1(z)$ and $S^{\dot{a}}_2(\bar{z})$, where $a,\dot{a} = 1,\ldots, 8$. Unlike for the NSR string, the GS string does not require a GSO-like projection. In fact, spacetime supersymmetry requires that the fermions be periodic on the cylinder. Let's first study the ground state energy. Each boson contributes $-1/24$ and each fermion contributes $+1/24$, which cancel to give zero Casimir energy, as expected of a supersymmetric theory. The zero modes of $S^a_1$ and $S^{\dot{a}}_2$, which satisfy the Clifford algebra, 
\begin{align}
    \{ S^a_{1,0}, S^b_{1,0} \} = \delta^{ab}, 
\end{align}
commute with $m^2$, and so lead to a ground state degeneracy. For the holomorphic sector, the ground state state consists of the vector $\ket{i}$ for $i =1, \ldots, 8$ and a spinor $\gamma^i_{a\dot{b}}S^a \ket{i}$. In total, this corresponds to $\mathbf{8_v} \oplus \mathbf{8_s}$. For the antichiral sector, we get $\mathbf{8_v} \oplus \mathbf{8_s}$ (type IIB) or $\mathbf{8_v} \oplus \mathbf{8_c}$ (type IIA). Thus, we see that we have reproduced the massless string spectrum as the ground states of the worldsheet theory.

\subsection{Examples of superstring compactifications}
\label{sec:superCompactifications}

Recall that we can consider string compactifications on the circle, \emph{i.e.} $\mathbb{R}^{1,8} \times S^1$. Under T-duality, the type IIA theory at radius $R$ is mapped to IIB at radius $1/R$. However, as we just learned type IIB is a chiral theory whereas IIA is not. \cite{Dai:1989ua} At first glance, it seems strange that the two theories could be dual under a change of the spacetime geometry. The resolution is that, from the nine-dimensional perspective, there is no longer a well-defined notion of chirality. 
\vspace{10pt}

\noindent\textbf{Exercise 4: } Does IIA or IIB have parity symmetry, and if so which one? 
\vspace{10pt}

\noindent{\bf Optional exercise}: Try to make a simple argument why IIA on a circle of radius $R$ is equivalent to IIB on one of radius $1/R$. 
\vspace{10pt}

Up to this point we have not been so careful about the circle boundary conditions for the spacetime fermions, but in fact we have been secretly choosing periodic conditions to preserve supersymmetry. If we choose anti-periodic boundary conditions for fermions two things happen; we break the supersymmetry and we get a tachyon at some radius \cite{Scherk:1978ta}. The tachyon would be a winding mode. Note that there would be no problem with this in field theory because there would be no winding modes that could become tachyonic. This is an example of how SUSY breaking has a lot more consequences in string theory than in field theory.

\vspace{10pt}

\noindent{\bf Optional exercise}:  Consider string theory on $\mathbb{R}^{1,8} \times S^1$. Show that if the fermions are antiperiodic on $S^1$, then for sufficiently small radius there is a tachyon in the spectrum. 
\vspace{10pt}

In fact, is is more common than not for tachyons to arise whenever supersymmetry is broken, although the two are not a priori related. For instance, string theory on $\mathbb{R}^{1,8} \times S^1$ with antiperiodic boundary conditions for the fermions leads to either tachyonic modes at some unstable radius or an unstable dilaton suffering from tadpoles.

As a preview of what is to come later, we can also consider string theory compactified on $M^{3,1}\times T^6$. This theory has ${\cal N} = 8$ supersymmetry, which in the low energy limit becomes ${\cal N} = 8$ supergravity. 
\vspace{10pt}

\noindent{\bf Optional exercise}:  Consider a 4d string compactification with $\mathcal{N}=8$ supersymmetry. How many scalar are there? Verify the counting from type II theory.

\subsection{The type I string} 

In our discussion of relativistic strings, we have thus far only mentioned closed strings whose spacetime coordinates satisfy periodic boundary conditions
\begin{equation}
	X^\mu(\tau,\sigma+2\pi) = X^\mu(\tau,\sigma) .
\end{equation}
We can also talk about open strings, which topologically are equivalent to a line segment with two endpoints. By convention, we take the open string worldsheet to be parameterized by coordinates $(\sigma, \tau)$ in $[0,\pi] \times \mathbb{R}$. More generally, the open string worldsheet is given by a Riemann surface with boundary. For sake of clarity we will focus only on the bosonic fields, though the same logic can be applied to the fermions and ghosts as well. Under a general variation, the gauge-fixed Polyakov action now admits a boundary term
\begin{equation}
	\delta S \supset \frac{1}{2\pi} \int d\tau \: \delta X^\mu \partial^\sigma X_\mu \bigg|_{\sigma=0}^{\sigma=\pi} .
\end{equation}
Requiring $\delta S = 0$ as usual leads to two types of boundary conditions:
\begin{equation}
	\partial_\sigma X^\mu(\sigma=0,\pi, \tau) = 0 \quad \text{(Neumann)}, \quad X^\mu(\sigma=0,\pi,\tau) = \text{constant} \quad \text{(Dirichlet)}.
\end{equation}
Neumann (N) boundary conditions imply that there is no momentum flow across the endpoints of the string, whereas Dirichlet (D) boundary conditions mean the string endpoints are fixed in space and/or time. We are free to take a mixture of both types of boundary conditions at each endpoint and for each $X^\mu$. Notice that (D) boundary conditions break Poincare invariance by selecting a preferred point in spacetime. Although this may be somewhat awkward, we will later discover that they arise naturally in the context of opens strings ending on extended objects known as D-branes. For sake of brevity, we will only focus on Neumann boundary conditions in this section.

Quantization of the open string is straightforward and closely parallels that of the closed string. One primary difference is that now the $X^\mu$ and $(\psi^\mu, \bar{\psi}^\mu)$ theories each admit only a single set of oscillators. In light-cone gauge, we can label them as $\alpha^i_n$ and $\psi^\mu_r$. There are still two sectors for the fermions, an NS sector ($r \in \mathbb{Z}+1/2$) and an R sector ($r \in \mathbb{Z}$). Determining the physical states of the theory is straightforward and follows the standard procedures (BRST, lightcone, etc).  We are primarily interested in a closed+open string theory, which is only consistent if we include a GSO projection on the open string sector \cite{Sagnotti:1987tw}. In short, this follows from the closure of the the OPE of open+closed string vertex operators (otherwise two open string vertex operators could lead to a closed string vertex operator that gets projected out). The resulting massless spectrum is
\begin{equation}
	\mathbf{8_v \oplus 8_c},
\end{equation}
which parallels the left- and right-moving sectors of the type IIB string (except now there is no tensor product). This is an ${\cal N} = 1$ gauge multiplet, which includes a gauge boson in the $\mathbf{8}_v$ and a gaugino in the $\mathbf{8}_s$. One can check that this leads to a $U(1)$ gauge theory in spacetime. 

The endpoints of the open string can be supplied with additional pointlike \textit{Chan--Paton} degrees of freedom \cite{Paton:1969je}. A generic open string state $\ket{\psi; i,j}$ is now labeled by two indices $i,j = 1, \ldots, N$, representing the degrees of freedom at each endpoint. A conjugate state should describe the same open string, and so these states are naturally captured by $N \times N$ Hermitian matrices $H_{ij}$. Of course, invariance of the inner product leads to a $U(N)$ symmetry that acts as $H \mapsto U H U^\dagger$. This is just the adjoint representation of $U(N)$. Although the worldsheet CFT is unaffected by this method (the stress tensor, for instance, is unchanged), the spacetime physics is completely different. For instance, there are now $N^2$ copies of each gauge boson and gaugino, which together transform as an ${\cal N}=1$ vector multiplet in the adjoint of $U(N)$: that is, there is a $U(N)$ gauge symmetry in spacetime! 

The appearance of gauge theory for open strings with Chan--Paton factors is more than a happy coincidence and turns out to be at the heart of non-perturbative effects in string theory. Let us explore the physical meaning of this gauge theory. The Neumann/Dirichlet boundary conditions that we impose on the end points of open strings in different directions confine the endpoints to a submanifold in spacetime. This submanifold is called the D-brane \cite{Dai:1989ua,Horava:1989ga,Polchinski:1995mt}. In perturbative string theory we view D-branes as part of the background rather than dynamical objects. However, the ingredients of the string string theory background are tightly constrained. For example, as we saw in bosonic string theory, the Weyl invariance of the worldsheet CFT imposes certain equations of motion between the background fields. Similarly, adding the D-brane imposes specific equations on the background fields and in some way sources them. 

To see why D-branes are fixed ingredients of the background, we can do a Heuristic calculation to estimate the tension of D-branes. If the tension is very large in string units, it is reasonable to approximate them as non-dynamical ingredients of the background. 

The amplitudes in backgrounds with a D-branes involve vacuum diagrams with discs ending on D-branes. 
\begin{figure}[H]
    \centering
\tikzset{every picture/.style={line width=0.75pt}} 
\begin{tikzpicture}[x=0.75pt,y=0.75pt,yscale=-1,xscale=1]
\draw   (182,256.56) -- (182,85.9) -- (254.5,9.34) -- (254.5,180) -- cycle ;
\draw [line width=1.5]    (221.5,95) .. controls (205.5,110) and (204.5,156) .. (220.5,148) ;
\draw    (221.5,95) .. controls (257.5,123) and (242.5,138) .. (220.5,148) ;
\draw [line width=1.5]  [dash pattern={on 5.63pt off 4.5pt}]  (220.5,148) .. controls (224.5,141) and (233.5,114) .. (221.5,95) ;
\draw (100,127.4) node [anchor=north west][inner sep=0.75pt]    {$\mathcal{A}_{Connected}$};
\draw (334,131.4) node [anchor=north west][inner sep=0.75pt]  [font=\Large]  {$\sim $};
\draw (437,109.4) node [anchor=north west][inner sep=0.75pt]  [font=\Large]  {$\dfrac{1}{g_{s}}$};
\end{tikzpicture}
    \caption{Vacuum amplitude in a background with a D-brane.}
\end{figure}
The one disc amplitude has a connected worldsheet with genus zero and one hole $(g,b)=(0,1)$. Therefore, the amplitude goes like 
\begin{align}
    \mathcal{A}\sim g_s^{2g-2+b}\sim1/g_s. 
\end{align}
Summing over all disconnected worldsheets to find the D-brane effective action is equivalent to exponentiating the connected part \cite{Polchinski:1994fq}. We find
\begin{align}
    \exp(-S_\text{D-brane})\sim\exp(-g_s^{-1}\cdot\text{Vol}_{\text{D-brane}}).
\end{align} 
Therefore, the effective action has a prefactor that is proportional to $1/g_s$. In other words, the tension of the D-brane is $T\sim1/g_s$ in string units. Note that for small $g_s\ll 1$ where the string perturbation is expected to work, D-branes are very massive and can be approximated to be non-dynamical. 

It is also worth noting that Dirichlet boundary conditions and D-branes are required to extend T-duality to open strings. This is due to the fact that T-duality acts on the left and right moving components of the compact coordinate as 
\begin{align}
    (X_L,X_R)\rightarrow (X_L,-X_R). 
\end{align}
Therefore, T-duality swaps $\partial_\sigma X$ and $\partial_\tau X$ up to some factors. Consequently, T-duality swaps a Dirichlet boundary condition along the compact direction with a Neumann boundary condition and vice versa. In terms of D-branes, this means that a wrapped D-brane goes to an unwrapped D-brane and vice versa \cite{Polchinski:1996fm}. 

Now let us imagine N coincident D-branes. Then each end of the open string can end on any of the copies. In that case, we need a label in $\set{1,...,n}$ to specify the boundary condition of each open string state. These labels are the Chan--Paton factors! This observation teaches us a very important lesson; Given that fields associated with open strings are confined to the D-branes, we learn that there is a gauge theory living on the D-brane. 

We now return to the type II closed string theories, which will soon be connected to the open string story. The type II theories \eqref{eq:NSRsuperstring} respect a $\mathbb{Z}_2$ symmetry $\Omega$ that interchanges left- and right-moving fields, known as \textit{worldsheet parity}. It acts on the closed string oscillators as
\begin{equation}\label{eq:worldsheetParity}
	\Omega: \alpha_n^\mu \leftrightarrow \bar{\alpha}_n^\mu, \quad \psi_n^\mu \to \bar{\psi}_n^\mu , \quad \bar{\psi}_n^\mu \to - \psi_n^\mu .
\end{equation}
Taking into account the GSO projection, only the type IIB theory continues to respect this symmetry. We can then try to construct an unoriented string theory by trying to gauge this symmetry. It can be shown that $\Omega$ preserves the NS–NS and R–R sector ground states, whereas it swaps the NSR and RNS ground states. After projecting onto $\Omega$-invariant states, we are left with a new massless spectrum, which we can individually analyze in each of the four spacetime sectors. In the NS–NS sector, the graviton and the dilaton survive whereas the B-field is projected out. In the R–R sector, only the two-form gauge field remains, with the zero-form and self-dual four-form gone. Since the NSR and RNS sectors are exchanged under the parity transformation, only a linear combination of the two survives the orbifold procedure. The associated massless fermions that remain are a single Majorana--Weyl graviton and a Majorana--Weyl fermion. Altogether, the theory therefore has ${\cal N} = (1,0)$ supersymmetry.

This is not, however, the end of the story. This unoriented theory of closed strings is inconsistent due to a one loop divergence. Moreover, a spacetime analysis of this chiral theory reveals the presence of a gravitational anomaly. To cure both of these problems, we can introduce unoriented, open strings. The open string theory of the previous section also respects worldsheet parity, which nows acts on the oscillators as a phase rotation. The projection to $\Omega$-invariant states adds an extra constraint for the Chan--Paton degrees of freedom. In particular, there are two allowed symmetry groups: $SO(N)$ or $Sp(N)$. As it turns out, this theory of unoriented open strings is also inconsistent due to a gauge anomaly. Miraculously, there is a method to cancel \textit{both} the gravitional and gauge anomalies, known as the Green-Schwarz anomaly-cancellation mechanism \cite{Green:1984sg}. The gauge and gravitional anomalies cancel if the gauge group is $SO(32)$ or $E_8 \times E_8$. Thus, the open+closed unoriented string theory with gauge group $SO(32)$ is tachyon-free, anomaly-free, and supersymmetric! It is commonly referred to as the \textit{Type I} string. From the above construction we can see that type I string theory is the orientifold of type IIB theory with 32 space-filling half D9-branes. For convenience, we summarize the massless contents of its spectrum in Table \ref{table:typeI}.

\begin{table}[h!]
\centering	
\renewcommand{\arraystretch}{1.2}
\begin{threeparttable}
\begin{tabular}{p{3cm} p{5cm} p{4cm}}
 \toprule
 \multicolumn{3}{c}{Massless spectrum (Type I)} \\\midrule
 sector  &  $SO(8) \times SO(32)$ irrep & particle content \\\midrule
closed (NS, NS)  & $\mathbf{(1,1) \oplus (35,1)}$ & dilaton + graviton  \\ 
closed (R,R)  & $\mathbf{(35',1)}$ & 2-form R–R field  \\ 
closed mixed  & $\mathbf{(8_c,1) \oplus (56_c,1)}$ & dilatino + gravitino  \\\midrule
open NS & $\mathbf{(8_v,496)}$ & gauge bosons  \\
open R & $\mathbf{(8_c,496)}$ & gauginos  \\\bottomrule
\end{tabular}
\end{threeparttable}
\caption{Massless spectrum of the $SO(32)$ Type I string.}
\label{table:typeI}
\end{table} 

\subsection{The heterotic string}

In the previous section, we discovered that attaching Chan--Paton degrees of freedom to the endpoints of open strings gave rise to gauge theories in spacetime. A natural question is whether we can get gauge theories in ten dimensions from closed superstrings alone. In our discussion on toroidal compactifications, we learned that global symmetries on the worldsheet are associated with gauge symmetries in spacetime. Unfortunately, for the superstring, there is no room for additional unitary degrees of freedom since the matter CFT already saturates the central charge with $c_m = 15$. What about the bosonic string? We can always take the matter CFT to consist of 10 noncompact bosons and a unitary CFT with $c = 16$. However, as was the case for toroidal compactifications, these extra degrees of freedom still usually have the interpretation of (compact) spatial directions. To free ourselves of these constraints, we consider a \textit{heterosis} of the bosonic and superstring theories \cite{Gross:1984dd}. The worldsheet theory consists of a left-moving bosonic CFT $(c_m = 26)$ and a right-moving SCFT $(\bar{c}_m = 15)$. The matter content consists of 10 noncompact bosons $X^\mu$, 10 right-moving fermions $\bar{\psi}^\mu$, and 16 \textit{chiral} (left-moving) bosons. This theory is properly interpreted as living in ten spacetime dimensions, with additional internal degrees of freedom.

\subsubsection*{Heterotic $SO(32)$} 

Although this definition of the heterotic string is perfectly fine, in these notes we will use a more convenient formulation that replaces the chiral bosons with 32 left-moving free fermions $\lambda^A$ with $A = 1,\ldots,32$. At first glance, this appears to be an absurd thing to do: we're exchanging fields with different worldsheet statistics! However, this is feasible in 2d due to a boson-fermion duality known as \textit{bosonization}. The duality at least seems plausible, since a free boson and Dirac fermion both have $c=\bar{c}=1$.  This correspondence continues to hold at level of the chiral boson and two Weyl fermions, though the actual details of the duality are somewhat involved. In any case, the worldsheet action is given by \cite{Gross:1984dd}
\begin{equation}
	S_m = \frac{1}{4\pi} \int d^2 z \left(2 \partial X^\mu \bar{\partial} X_\mu + \bar{\psi}^\mu \p \bar{\psi}_\mu + \lambda^A \bar{\p} \lambda^A \right) .
\end{equation}
As was the case for the $\psi$ CFT, this is not quite enough data to fix the theory: we still need to specify the boundary conditions of the $\lambda^A$. The action is invariant under a global $O(32)$ symmetry that acts on the fields as
\begin{equation}
	U(O) \lambda^A U^\dagger(O) = O^{AB}\lambda^B, \quad O \in O(32).
\end{equation}
Requiring $O(32)$ invariance on the cylinder thus permits the boundary conditions
\begin{equation}
	\lambda^A(\sigma+2\pi,\tau) = O^{AB} \lambda^{B}(\sigma,\tau)
\end{equation}
for any orthogonal matrix. Unlike the $\psi$ CFT, we cannot use Lorentz invariance to restrict the choice of this matrix. However, there are still other consistency conditions such as the requirement of a modular invariant, tachyon-free theory. To achieve this, we need to implement an additional chiral GSO projection on the states created by the $\lambda^A$. For the heterotic string, it turns out that there are a total of 9 consistent choices of boundary conditions and GSO projections. Three of these are tachyon free, and only two of the three have ${\cal N} = 1$ spacetime supersymmetry.\footnote{The third choice of GSO projection yields a non-supersymmetric string theory with gauge group $SO(16)\times SO(16)$. It is chiral, anomaly-free, and tachyon-free \cite{Dixon:1986iz, AlvarezGaume:1986jb}.} These two theories, distinguished by their symmetry groups, are called heterotic $SO(32)$ (HO) and heterotic $E_8 \times E_8$ (HE), respectively.\footnote{The emergence of both gauge groups is manifest in the bosonic formulation. The chiral boson CFT is associated with a 16-dimensional \textit{Euclidean} lattice, which is even and self-dual by modular invariance. Miraculously, there only two such lattices: the root lattices of $\text{Spin}(32)/\mathbb{Z}_2$ and $E_8 \times E_8$! The non-supersymmetric theory can then be constructed as a $\mathbb{Z}_2$ orbifold of the $E_8 \times E_8$ theory.} 

The HO theory is singled out by choosing identical boundary conditions on all of the fermions,
\begin{equation}
	\lambda^A(\sigma+2\pi,\tau) = \epsilon \lambda^A(\sigma,\tau), \quad \epsilon = \pm .
\end{equation}
We refer to the two sectors $\epsilon = \pm$ as periodic (P) and antiperiodic (A), to distinguish them from the R and NS sectors of the right-moving sector. This choice of BCs ensures that the entire $SO(32)$ symmetry is preserved.\footnote{A more careful analysis of the representation theory shows that the global symmetry group is actually $\text{Spin}(32)/\mathbb{Z}_2$, where the $\mathbb{Z}_2$ differs from that of $SO(32) \simeq O(32)/\mathbb{Z}_2$. However,  the difference $SO(32)$ and $\text{Spin}(32)/\mathbb{Z}_2$ only appears through the allowed representations.}. There is a $(-1)^F$ fermion number symmetry,
\begin{equation}
	\{ (-1)^F , \lambda^A \} = 0, 
\end{equation}
that commutes with all of the right-moving fields as well as $\partial X^i$. For now, let's focus on the left-moving sector. The GSO projection we want to take is
\begin{equation}
	(-1)^F  = +1,
\end{equation}
where the A vacuum $\ket{0}$ is conventionally defined to have $(-1)^F = +1$. We now analyze the spectrum of the theory. In light-cone gauge, the left-moving part of the space of states is constructed from the $\p X^i$ oscillators $\alpha^i_n$ as well as the $\lambda^A$ oscillators $\lambda^A_r$, where by definition the subscript indicates the oscillator weight. The ground state energy receives $-1/24$ for each boson and $-1/48$ (AP) or $+1/24$ (P) for each fermion. The ground states energies are thus
\begin{equation}
	E^{HO}_{AP} = -\frac{8}{24} - \frac{32}{48} = -1, \quad E^{HO}_P = -\frac{8}{24} + \frac{32}{24} = +1.
\end{equation}
Naively, we then seem to conclude that there is a tachyon $\ket{0}$ in the theory. However, the level-matching condition forbids this state since there is no tachyon in the GSO-projected right-moving sector. We also see that the P sector does not contribute any massless degrees of freedom. The relevant (massless) states coming from the AP sector are
\begin{equation}
	\alpha_{-1}^i \ket{0}, \quad \lambda_{-1/2}^A \lambda^{B}_{-1/2} \ket{0} ,
\end{equation}
Under the $SO(8) \times SO(32)$ symmetry, these transform as $\mathbf{(8_v,1)}$ and $\mathbf{(1,496)}$, respectively. The analysis of the right-moving sector follows that of the NSR construction. The NS sector has a single ground state $\ket{0}$, and the R sector ground state transforms as an $SO(8)$ spinor $\ket{\bar{s}}$ that transforms as 
\begin{align}
    \mathbf{16} = \mathbf{8_s \oplus 8_c}. 
\end{align}
The GSO projection removes the would-be tachyon $\ket{0}$ as well as the antichiral spinor $\mathbf{8}_c$. The right-moving contribution to the massless spectrum is thus 
\begin{align}
    \mathbf{8_v}\oplus\mathbf{8_s}. 
\end{align}
The massless particle content is given by a tensor product of the two sets, which we list in Table \ref{table:HO}. From
 \begin{equation}
 	\mathbf{(8_v,1)} \otimes \mathbf{(\mathbf{8_v} \oplus \mathbf{8_s},1)},
\end{equation}
 we recover the ${\cal N}=1$ SUGRA multiplet, which contains the usual massless bosonic fields (graviton, dilaton, B-field) as well as a single gravitino. Additionally, the massless particles include contributions from
 \begin{equation}
 	\mathbf{(1,496)} \otimes \mathbf{(\mathbf{8_v} \oplus \mathbf{8_s},1)} = \mathbf{(8_v,496)} \oplus \mathbf{(8_s,496)} .
\end{equation}
Here, we have a vector and a spinor transforming in the adjoint representation of the gauge group, which form an ${\cal N}=1$ vector multiplet. We have therefore found a tachyon-free theory of closed strings with ${\cal N}=1$ spacetime supersymmetry and an $SO(32)$ gauge symmetry!\\

\begin{table}[H]
\centering	
\renewcommand{\arraystretch}{1.4}
\begin{threeparttable}
\begin{tabular}{p{1.5cm} p{4cm} p{4.3cm} p{5cm}}
\toprule
\multicolumn{4}{c}{Massless spectrum (HO)} \\\midrule
Sector & States  &  $SO(8) \times SO(32)$ irreps & Particle content \\\midrule
($-$, NS) & $\alpha_{-1}^i \bar{\psi}^j_{-1/2} \ket{0;0} $ & $\mathbf{(1,1) \oplus (28,1) \oplus (1,1)}$ & dilaton + B-field + graviton  \\
\: & $\lambda^A_{-1/2}\lambda^{B}_{-1/2} \bar{\psi}^j_{-1/2} \ket{0;0} $ & $\mathbf{(8_v,496)}$ & gluons \\\midrule
($-$, R) & $\alpha_{-1}^i \ket{0;\bar{s}} $ & $\mathbf{(8_c,1)} \oplus \mathbf{(56_c,1)}$  & dilatino+gravitino \\
\: & $\lambda^A_{-1/2}\lambda^{B}_{-1/2} \ket{0;\bar{s}} $ & $\mathbf{(8_s,496)}$ & gluinos  \\\bottomrule
\end{tabular}
\end{threeparttable}
\caption{Massless spectrum of the $SO(32)$ heterotic string. }
\label{table:HO}
\end{table} 

\subsubsection*{Heterotic $E_8 \times E_8$} 

Constructing the heterotic $E_8 \times E_8$ theory is only slightly more involved. We now allow for different boundary conditions between two equally partitioned sets of the fermions:
\begin{align}
	\lambda^{A_1}(\sigma+2\pi,\tau) = \epsilon_1 \lambda^{A_1}(\sigma,\tau), \quad A_1 = 1, \ldots, 16 , \\
	\lambda^{A_2}(\sigma+2\pi,\tau) = \epsilon_2 \lambda^{A_2}(\sigma,\tau), \quad A_2 = 1, \ldots, 16 ,
\end{align}
where $\epsilon_{1,2} = \pm 1$. Consequently there are a total of four left-moving sectors, labeled by $(\epsilon_1, \epsilon_2)$. This generically breaks the $SO(32)$ symmetry to an $SO(16) \times SO(16)$ subgroup. There are two independent fermion number symmetries on the left, $(-1)^{F_1}$ and $(-1)^{F_2}$, which commute with the opposite set of fermions, e.g. 
\begin{align}
    [(-1)^{F_1}, \lambda^{A_2}] = 0. 
\end{align}
On the left, we take the GSO projection
\begin{equation}
	(-1)^{F_1} = (-1)^{F_2} = +1.
\end{equation}
We now analyze the massless content of the theory. The tachyon of the $(-,-)$ sector is still projected out due to the level-matching condition and the $(+,+)$ sector is purely massive. However, now the $(\pm, \mp)$ sectors contribute to the massless sector; indeed, their ground state energies vanish by
\begin{equation}
	E^{HE}_{(\pm, \mp)} = -\frac{8}{24} \pm \frac{16}{48} \mp \frac{16}{24} = 0 .
\end{equation}
In the $(-,-)$ sector, the relevant states are
\begin{equation}
	\alpha_{-1}^i \ket{0}, \quad \lambda^{A_i}_{-1/2} \lambda^{B_i}_{-1/2}\ket{0},
\end{equation}
where the fermionic oscillators must be taken from the same set due to the GSO projection. In terms of $SO(8) \times SO(16)_1 \times SO(16)_2$ representations, the $(-,-)$ sector therefore contributes $\mathbf{(8_v,1,1)}$ and $\mathbf{(1,120,1)}$, and $\mathbf{(1,1,120)}$, where $\mathbf{120}$ is the adjoint of $SO(16)$. The $(+,-)$ ground state $\ket{\sigma}$ transforms as an $SO(16)_1$ Dirac spinor $\mathbf{256}$, which is projected to one of the chiral spinors $\mathbf{128}$ by the GSO projection (this mirrors the R sector ground state of the right-moving sector). Under the full symmetry group, it transforms as the $\mathbf{(1,128,1)}$. Similarly, the $(-,+)$ ground state transforms as the $\mathbf{(1,1,128)}$. Tensoring with the right-moving sector gives the overall massless particle content,
\begin{equation}
	\mathbf{(8_v \oplus 8_s,120,1)} \oplus 	\mathbf{(8_v \oplus 8_s,1,120)} \oplus \mathbf{(8_v \oplus 8_s,128,1)} \oplus 	\mathbf{(8_v \oplus 8_s,1,128)},
\end{equation}
where we have left out the usual ${\cal N}=1$ gravity multiplet. The key step is now identifying the $\mathbf{120 \oplus 128}$ irrep  of $SO(16)$ as the adjoint $\mathbf{248}$ of $E_8$. Therefore, these particles form an ${\cal N}=1$ vector multiplet in the adjoint of $E_8 \times E_8$! We list the full massless content of the theory in Table \ref{table:HE}.

\begin{table}[H]
\centering	
\renewcommand{\arraystretch}{1.2}
\begin{threeparttable}
\begin{tabular}{p{2cm}p{3.5cm}p{4.5cm}p{5cm}}
 \toprule
 \multicolumn{4}{c}{Massless spectrum (HE)} \\\midrule
 Sector & States  &  $SO(8) \times (E_8 \times E_8)$ irreps & Particle content \\\midrule
($\pm, \pm$, NS) & $\alpha_{-1}^i \bar{\psi}^j_{-1/2} \ket{0;0} $ & $\mathbf{(1,1) \oplus (28,1) \oplus (35,1)}$ & dilaton + B-field + graviton  \\
\: & $\lambda^{A_i}_{-1/2}\lambda^{B_i}_{-1/2} \bar{\psi}^j_{-1/2} \ket{0;0} $ $\bar{\psi}^j_{-1/2} \ket{\sigma;0} $ & $\mathbf{(8_v,248,1) \oplus (8_v,1,248)}$ & gluons \\\midrule
($\pm, \pm$, R) & $\alpha_{-1}^i \ket{0;\bar{s}} $ & $\mathbf{(8_c,1)} \oplus \mathbf{(56_c,1)}$  & dilatino + gravitino \\
\: & $\lambda^{A_i}_{-1/2}\lambda^{B_i}_{-1/2} \ket{0; \bar{s}} $ \quad\quad $ \ket{\sigma; \bar{s}} $ & $\mathbf{(8_s,248,1)\oplus(8_s,1,248)}$ & gluinos  \\\bottomrule
\end{tabular}
\end{threeparttable}
\caption{Massless spectrum of the $E_8 \times E_8$ heterotic string.}
\label{table:HE}
\end{table}

\subsection{Superstring compactifications}
\label{sec:superstringCompactifications}

Now that we've constructed all of the known consistent 10d superstring theories (type I and II, heterotic), we can consider compactifications of spacetime of the form
\begin{equation}
  \mathbb{R}^{1,9-d} \times M_d \,,
\end{equation}
where $M_d$ is a compact d-dimensional oriented Riemannian manifold, known as the internal space. It can be shown that the compactified background preserves some of the supersymmetry if there exists a covariantly constant spinor on $M_d$; that is, there exists some spinor $\xi$ such that
\begin{equation}
  \nabla_I \xi = 0 \,,
\end{equation}
where $\nabla_I$ is the covariant derivative and $I$ is an internal space vector index \cite{yau1978ricci}. A convenient way to determine if such a spinor exists is to analyze the holonomy group of $M_d$. Recall that the generic holonomy group for a Riemannian $d$-dimensional manifold is SO($d$). Special choices of the internal space will lead to ``reduced holonomy,'' where the holonomy group is is some subgroup of SO($d$). A covariantly constant spinor on $M_d$ exists if the minimal spinor representation $\mathbf{8_s}$ of 10d contains a singlet in the decomposition under the holonomy group. We now list some of the relevant manifolds (and their subgroups), enumerated by their dimensionality, which preserve SUSY. 

\begin{itemize}
\item $d = 1$: All 1d manifolds have trivial holonomy and so preserve supersymmetry. This case has been covered previously.

\item $d = 2$: The holonomy group is $SO(2) \simeq U(1)$, and only the trivial subgroup gives rise to a covariantly constant spinor. The only $M_2$ with trivial holonomy is $T^2$, and in general the only $M_d$ with trivial holonomy is the $d$-torus, $T^d$, which will always admit covariantly constant spinors.

\item $d = 3$: The holonomy group is $SO(3) \simeq SU(2)$ which has no relevant non-trivial subgroup. The only SUSY-preserving 3-fold is $T^3$. 

\item $d = 4$: The holonomy group is $SO(4) \simeq  SU(2) \times SU(2)$. The minimal spinor representations are the chiral $\mathbf{(2,1)}$ and antichiral $\mathbf{(1,2)}$. Each transforms as a singlet under the other's $SU(2)$. A 4-fold with SU(2)$\simeq$ Sp(1) holonomy is hyperK\"ahler, and is known as a K3 surface. One such example is the orbifold $T^4/\mathbb{Z}_2$. Thus, the two SUSY-preserving options are $T^4$ and K3. 

\item $d = 5$: The holonomy group is $SO(5)$. The analysis is identical to $d=4$, where we simply tensor with a circle (the holonomy subgroup is unchanged). The two options are $K3 \times S^1$ (or some twisted product) and $T^5$. 

\item $d = 6$: The holonomy group is $SO(6) \simeq SU(4)$. The obvious examples are $T^6$ and $K3 \times T^2$. Another interesting holonomy subgroup which admits a singlet in the branching is $SU(3) \subset SO(6)$. Manifolds of dimension $2N$ with $SU(N)$ holonomy are known as Calabi--Yau $N$-folds $CY_N$ and admit a Ricci-flat K\"ahler metric (as proved by Yau). The options are thus $CY_3$, $T^6$ and $K3 \times T^2$ (or some twisted product).\footnote{For instance, we can consider the heterotic string theory compactified on $CY_3$, see e.g. \cite{Candelas:1985en}.}

\item $d = 7$: The holonomy group is $SO(7)$. The obvious examples here are $T^7$, $K3 \times T^3$, and $CY_3 \times S^1$. There is an additional manifold, the so-called $G_2$-manifold, whose holonomy group is $G_2 \subset SO(7)$. For such groups, the spinor irrep decomposes as ${\bf 8_s} \rightarrow {\bf 7} \oplus {\bf 1}$. Manifolds $M_7$ with such reduced holonomy are called $G_2$-manifolds.

\item $d = 8$: The holonomy group is $SO(8)$. In this case, there are two reduced holonomy subgroups: $SU(4)$, which corresponds to $CY_4$, and $\text{Spin}(7)$, which leads to $\text{Spin}(7)$-manifolds. Additionally, we have the obvious product manifolds such as $G_2 \times S^1$ and $CY_3 \times T^2$.
\end{itemize}

\noindent\textbf{Exercise 5:} Show that a holonomy group of $\text{Spin}(7)$ of $SO(8)$ for the compact manifold preserves $1/8$ of the supersymmetry. 

\noindent\textbf{Exercise 6:} Show that the holonomy group of $T^4/\mathbb{Z}_2$ where the $\mathbb{Z}_2$ acts as $x_i\sim -x_i$ is $SU(2)$.

\noindent {\bf Optional exercise}: Determine the massless spectrum of type IIA on $T^4/\mathbb{Z}_2$. Note that there are 16 fixed points of the $\mathbb{Z}_2$ action.

\noindent {\bf Optional exercise}: Show that $M_8 = G_2 \times S^1$ with $G_2$ holonomy and $M_8 = CY_4$ with $SU(4)$ holonomy preserve the same amount of supersymmetry.

\begin{table}[H]
\begin{center}
\renewcommand{\arraystretch}{1.4}
\begin{threeparttable}
\begin{tabular}{cccccc}
\toprule
$10-d$ & $M_d$ & \begin{tabular}{@{}c@{}}
Fraction of\\[-8pt]
supercharges \end{tabular} & Type IIA & Type IIB & Heterotic / Type I \\\midrule
10 & & & ${\cal N}_{10} = (1,1)$ &  ${\cal N}_{10} = (2,0)$ & ${\cal N}_{10} = (1,0)$\\ 
6 & $K3$ & 1/2 & ${\cal N}_{6} = (1,1)$ &  ${\cal N}_{6} = (2,0)$ & ${\cal N}_{6} = (1,0)$\\
5 & $K3 \times S^1$ & 1/2 & ${\cal N}_{5} = 2$ &  ${\cal N}_{5} = 2$ & ${\cal N}_{5} = 1$\\ 
4 & $K3 \times T^2$ & 1/2 & ${\cal N}_{4} = 4$ &  ${\cal N}_{4} = 4$ & ${\cal N}_{4} = 2$\\
 & $CY_3$ & 1/4 & ${\cal N}_{4} = 2$ &  ${\cal N}_{4} =2$ & ${\cal N}_{4} = 1$\\ 
 3 & $CY_3 \times S^1$ & 1/4 & ${\cal N}_{3} = 4$ &  ${\cal N}_{3} = 4$ & ${\cal N}_{3} = 2$\\
 & $G_2$ & 1/8 & ${\cal N}_{3} = 2$ &  ${\cal N}_{3} = 2$ & ${\cal N}_{3} = 1$\\ \bottomrule
\end{tabular}
\end{threeparttable}
\caption{Reduced SUSY in various superstring compactifications}
\label{tb:SCFTs}
\end{center}
\end{table}

By using the supersymmetric compactification in Table \ref{tb:SCFTs} we can construct theories with $4,8,16,32$ super charges in four dimensions. These are $\mathcal{N}=\{1,2,4,8\}$. What about the other degrees of supersymmetries,  $\mathcal{N}=\{3,5,6,7\}$? It turns out the algebra of an interacting $\mathcal{N}=7$ supergravity automatically enhances to $\mathcal{N}=8$. The other supersymetries can be constructed by asymmetric compactification on the left and right moving sector of the worldsheet theory each giving rise to $\mathcal{N}=\{1,2,4\}$ supersymmetry.
\begin{align}
    &\mathcal{N}=3:~~(\mathcal{N}_L,\mathcal{N}_R)=(1,2)\nonumber\\
    &\mathcal{N}=5:~~(\mathcal{N}_L,\mathcal{N}_R)=(1,4)\nonumber\\
    &\mathcal{N}=6:~~(\mathcal{N}_L,\mathcal{N}_R)=(3,3)\text{ or }(2,4).
\end{align}

Note that the L-R subscript refers to the left-moving and right-moving sectors of the worldsheet theory, and is not related to spacetime parity.

\subsection{Model building}
\label{sec:modelBuilding}

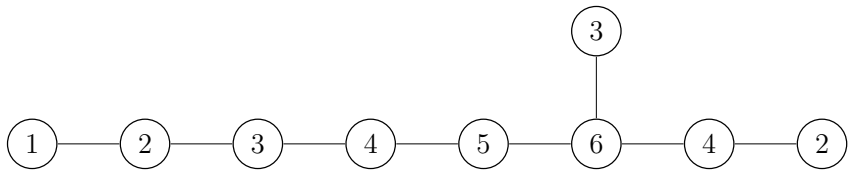
\begin{figure}[H]
\begin{center}
\scalebox{0.6}{
\begin{tikzpicture}
\node[draw,circle,thick,scale=1.5] (A1) at (0,0) {$1$};
\node[draw,circle,thick,scale=1.5] (A2)at (2.5,0) {$2$};
\node[draw,circle,thick,scale=1.5] (A3) at (5,0) {$3$};
\node[draw,circle,thick,scale=1.5] (A4) at (7.5,0) {$4$};
\node[draw,circle,thick,scale=1.5] (A5) at (10,0) {$5$};
\node[draw,circle,thick,scale=1.5] (A6) at (12.5,0) {$6$};
\node[draw,circle,thick,scale=1.5] (A7) at (15,0) {$4$};
\node[draw,circle,thick,scale=1.5] (A8) at (17.5,0) {$2$};
\node[draw,circle,thick,scale=1.5] (A9) at (12.5,2.5) {$3$};
\draw (A1)--(A2)--(A3)--(A4)--(A5)--(A6)--(A7)--(A8);
\draw (A6)--(A9);
\end{tikzpicture}}
\caption{A Dynkin diagram of an $E_8$.}
\label{fig:E8Dynkin}
\end{center}
\end{figure}

Treating the left- and right-moving modes separately, we get get chiral matter and gauge theories in spacetime. However, we cannot construct a perturbative type II theory specifically with chiral matter transforming in the $SU(3) \times SU(2) \times U(1)$ gauge group, and so we will not discuss such theories here.\footnote{In this context, we are considering only perturbative vacua here. The introduction of non-perturbative methods like S-duality enables study of such theories.} Instead, we can construct the relevant theories using the $E_8 \times E_8$ heterotic string. Focusing only on a single $E_8$, we have the decomposition
\begin{equation}
SU(3)\times SU(2)\times U(1)\subset SU(5)\subset SO(10)\subset E_6.
\label{eq:SMtoE6}
\end{equation}
Note that the fundamental representation of $E_6$ is a 27-dimensional representation whose branching rule to an $SO(10)$ is given by 
\begin{equation}
\mathbf{27}=\mathbf{10}+\mathbf{16}+\mathbf{1},
\end{equation}
where $\mathbf{16}$ is the representation in which the standard model with neutrinos transform.\footnote{In particular for $SO(10)$ GUT, the matter content of the standard model comes from the $\mathbf{16}$, which is a spinor representation of $SO(10)$ \cite{slansky1981group}.} Recall that the heterotic theory has an $E_8\times E_8$ gauge symmetry (see the Dynkin diagram for $E_8$ in Figure \ref{fig:E8Dynkin}). Using the Dynkin diagram, we can get  the $E_7$  and $E_6$ subgroups by contracting nodes. Casually speaking, the $E_8$ by itself is not of phenomenological interest but only via additional operations or such as breaking to smaller subgroups.

By decomposing the adjoint representation $\mathbf{27}$ of $E_6$ under the standard model subgroup, one can get the corresponding representation as in \eqref{eq:SMtoE6}. It is therefore rather elegant to build a standard model from $E_6$, where the fundamental representation is $\mathbf{16}$. We can match this $\mathbf{16}$ as the standard model matter with neutrinos.

Let us briefly investigate the equations of motion for the $H$ flux, which in the low energy effective theory satisfy 
\begin{align}
\begin{aligned}
    H &= dB+\frac{1}{2}\omega_{cs}(\omega_{spin})-\frac{1}{2}\omega_{cs}(A), \\ 
    dH &=\frac{1}{16\pi^2}\left(R\wedge R-F\wedge F\right),
\end{aligned}
\end{align}
where $\omega$ is the Chern-Simons terms, $\omega_{spin}$ is the spin connection, $R$ is the Ricci 2-form, and $F$ is the field strength for the Heterotic gauge group. 
Note that there are two terms that compose $dH$: the first Pontryagin class 
\begin{align}
    p_1(M) \sim R \wedge R
\end{align}
and the second Chern class 
\begin{align}
    c_2(V) \sim F \wedge F .
\end{align} 
If $\int dH = 0$, then 
\begin{align}
    p_1(M)=c_2(V). 
\end{align}
In other words, string perturbation theory breaks down due to the presence of tadpole if $R\wedge R = 0$ and $F\wedge F \neq 0$, or vice versa.

To preserve some supersymmetry, we take the internal space to have an $SU(3)$ holonomy. This implies that the spin connection $\omega^{ij}_\mu$ transforms in the adjoint of $SU(3)$, which enters into the $R \wedge R$ term through 
\begin{align}
    R = d\omega + \omega \wedge \omega . 
\end{align}
We can make the judicious choice 
\begin{align}
    \omega = A^{SU(3)}
\end{align}
for the gauge connection of $SU(3) \subset E_8$, referred to as the identification of the gauge and spin connection. With this construction, $dH = 0$ is automatically enforced. Note that this is identical to the level-matching we covered in type II theories in Section \ref{sec:superstrings}. Hence, we can conclude that we are in the same vacuum as in type II theories.

Recall that the adjoint representation of the $E_8$ is $\mathbf{248}$ and so there are massless fields transforming under $SO(8) \times E_8$ as
\begin{equation}
(\mathbf{8_v}\oplus\mathbf{8_s})\otimes\mathbf{248}.
\end{equation}
Note that $E_8$ can be broken as
\begin{equation}
E_8\longrightarrow E_6\oplus SU(3),
\end{equation}
under which the $\mathbf{248}$ decomposes as
\begin{equation}
\mathbf{248}\longrightarrow (\mathbf{78},\mathbf{1})\oplus(\mathbf{27},\mathbf{3})\oplus(\overline{\mathbf{27}},\overline{\mathbf{3}})\oplus(\mathbf{1},\mathbf{8}),
\end{equation}
where the first representation is the adjoint representations of $E_6$ and the last representations is the adjoint representation of $SU(3)$. The two representations in the middle are of the interest as they are the bifundamental representations. More precisely, the second and third representations transform as 27-dimensional irreducible representations of $E_6$, where each of them decomposes to $SO(10)$ as 
\begin{equation}
\mathbf{27}\longrightarrow \mathbf{10}+\mathbf{16}+\mathbf{1}.
\end{equation}
Note that from considering the vectors on the right, we had
\begin{equation}
(\mathbf{8_v}\oplus\mathbf{8_s})\otimes\mathbf{8_v},
\end{equation}
where the counting of the zero modes are given by the cohomologies of Calabi--Yau such that $\#\mathbf{3}-\#\overline{\mathbf{3}}$ from the contribution from 
\begin{align}
(\mathbf{27},\mathbf{3})\oplus(\overline{\mathbf{27}},\overline{\mathbf{3}}).
\end{align}
One remark to make is that the internal Calabi--Yau only sees SU(3) part and this SU(3) does not affect $E_6$.

The chiral spinors in 10d decompose in the compactification as
\begin{equation}
  (10d) = (4d) \otimes (6d) \,,
\end{equation}
where we will write $\gamma_1, \cdots, \gamma_{10}$ for the 10d
$\gamma$-matrices. The product of the chirality of 4d and 6d spinors is $+1$.
${\bf 3}$ and ${\bf \overline{3}}$ have opposite chirality spinors.

Here we have $SO(4) \oplus SO(6)$ and we count the net zero-modes of ${\bf 3}$ and ${\bf \overline{3}}$ from this opposite chirality. 

Recall that
\begin{equation}
\slashed{D} \psi = (\slashed{D}_4 + \slashed{D}_6)\psi \Rightarrow (\slashed{D}_4 + \lambda) \psi_\lambda = 0 \,. 
\end{equation}
The eigenvalues $\lambda$ act like effective mass parameters; the number of zero-modes is given by the counting of the solutions when $\lambda = 0$. 

The Calabi--Yau manifold is by definition K\"ahler. The K\"ahler function $K(z^i, \overline{z}^j)$ leads to a metric 
\begin{align}
    g_{i\overline{j}} = \partial_i \overline{\partial}_j K. 
\end{align}
K\"ahler manifolds have $U(d)$ holonomy, where $d$ is the complex dimension of the manifold. Calabi--Yau manifolds require $SU(n)$ holonomy, which further requires the manifold to have trivial first Chern class, 
\begin{align}
    c_1(M) = 0. 
\end{align}
We have $dz_1 \wedge \cdots \wedge dz_n$ well-defined as this means $U(1) \subset U(n)$ is trivial and hence is fixed, \emph{i.e.} $dz_1 \wedge \cdots \wedge dz_n$ is locally a constant, which implies the existence of a holomorphic $n$-form $\Omega^{(n)}$. For a Calabi--Yau threefold we have a holomorphic three-form $\Omega^{(3)}$.

We can define complexified combinations of the $\gamma$-matrices, applicable in the case of a complex manifold like a Calabi--Yau threefold:
\begin{equation}
    \gamma_i + i \gamma_{i + 3} = \gamma_{\widetilde{i}} \,, \quad
    \gamma_i - i \gamma_{i+3} = \gamma_{\widetilde{i}}^\dagger \,.
\end{equation}
The Clifford algebra is
\begin{equation}
  \{\gamma_{\widetilde{i}}, \gamma_{\widetilde{j}}\} = 0 \,, \quad
  \{\gamma_{\widetilde{i}}^\dagger, \gamma_{\widetilde{j}}^\dagger\} = 0 \,, \quad
  \{\gamma_{\widetilde{i}}, \gamma_{\widetilde{j}}^\dagger\} = \delta_{\widetilde{i}\widetilde{j}} \,.
\end{equation}
We can construct the following using the gamma matrices:
\begin{align}
\ket{0},\quad n \ \gamma_{\widetilde{i}}^\dagger\ket{0},\quad nCi\ \gamma_{\widetilde{i}}^\dagger\gamma_{\widetilde{j}}^\dagger\ket{0}, \cdots.
\end{align}
Also we can build
\begin{align}
1,\quad dz^{\bar{i}},\quad dz^{\bar{i}}\wedge dz^{\bar{j}},\quad dz^{\bar{i}}\wedge dz^{\bar{j}}\wedge dz^{\bar{k}} .
\end{align}
We have $(\mathbf{27},\mathbf{3})$ where $\mathbf{3}$ gives triplets. Then
\begin{equation}
\psi_{\overline{i} j}(z)\ dz^{\overline{i}}\wedge dz^j
\end{equation}
gives the number of zero modes. This allows us to have the hodge number $h^{1,1}$ to be given as the number of zero modes in $\mathbf{3}$, which is identical to the number of zero modes in $\mathbf{27}$.

Recall that Calabi--Yau threefold has a well-defined three-form $\Omega_{ijk}$. Using $(\mathbf{3}\times\mathbf{3})_{Antisym}=\overline{\mathbf{3}}$, we can get 
\begin{align}
    \psi_{\overline{i} j}(z) dz^{\overline{i}}\wedge dz^j\wedge dz^k
\end{align}
along with the aforementioned threeform. Then we can conclude that $h^{1,2}$ is given by the number of zero modes in $\overline{\mathbf{27}}$. The net number of zero modes is given by 
\begin{align}
    h^{1,1}-h^{2,1}=\frac{\chi (Y)}{2}, 
\end{align}
where $\chi (Y)$ is the Euler characteristic of the Calabi--Yau threefold. The ambiguity of the choice for $\mathbf{3}$ and $\overline{\mathbf{3}}$ interchangeably results in a mirror symmetry.

A strict Calabi--Yau threefold has two independent Hodge numbers. We have
\begin{equation}
  \begin{aligned}
    h^{0,0} = h^{3,0} = h^{0,3} = h^{3,3} = 1 \cr
    h^{1,0} = h^{2,0} = h^{0,2} = h^{0,1} = 0 \cr
    h^{1,2} = h^{2,1} \cr
    h^{2,2} = h^{1,1}\,,
  \end{aligned}
\end{equation}
and the Euler characteristic is given by
\begin{equation}
  \chi = 2(h^{1,1} - h^{2,1}) \,.
\end{equation}

\noindent {\bf Optional exercise}: Consider the type II string on $T^6/\mathbb{Z}_2$. Show that the theory is non-supersymmetric and compute the massless particle content of the twisted sector. [Note: the theory will be tachyonic. It is not known how to obtain a non-supersymmetric theory from perturbative string theory that does not have either tachyons or rolling problems.]
\vspace{10 pt}

\noindent {\bf Optional exercise}: Consider heterotic string theory compactified on $T^6/\mathbb{Z}_3$, where the $\mathbb{Z}_3$ acts as $z_i \rightarrow \omega z_i$ with $\omega^3 = 1$. The Ramond
sector gives the cohomology of the internal manifold $T^6/\mathbb{Z}_3$. Show that the Hodge numbers are
\begin{equation}
  h^{1,1} = 0 \,, \quad \text{and} \quad h^{2,1} = 36 \,,
\end{equation}
where the twisted sector contributes $27$ to $h^{2,1}$ and untwisted sector $9$. The number of massless modes of the resulting theory is determined by the
cohomology of the internal Calabi--Yau manifold.
\vspace{10 pt}

\noindent {\bf Optional exercise:} Recall an exercise that if is antiperiodic around circle, for a sufficiently small radius, we get tachyons. Now add onto this exercise. Show that a winding mode tachyon emerges at Hagedorn transition. Hagedorn transition is defined to be at $T_H=\frac{1}{\beta_H}$ and is related to the asymptotic degeneracy
\begin{align}
&n(N)=e^{\alpha\sqrt{CN}},\quad m^2\simeq CN,\nonumber\\
&n(m)\sim e^{\beta_H m}.
\end{align}
The partition function 
\begin{align}
    Z\sim\sum_m n(m)e^{-\beta m}
\end{align}
diverges at $\beta=\beta_H$.

\section{String dualities}

As we discussed before, lowering the cutoff of a theory can lead to a set of different low-energy theories with disconnected moduli. The reverse of this process (\emph{i.e.} increasing the cut-off in the UV completion) is expected to unify different low-energy theories. In other words, two seemingly different low energy theories are in fact perturbative expansion of a single theory at different corners of its moduli space. This idea is known as duality.  By their nature, dualities allow us to push past the perturbation theory and learn more about the underlying non-perturbative theory, even if we do not know its exact formulation. String theory is filled with dualities to the extent that whenever two lower dimensional constructions have the same asymptotic geometry (e.g. Minkowski or AdS), gauge groups of the same rank, and the same number of supercharges, they often turn out to be dual to each other. In the following section, we will do a quick review of dualities in string theory and the lessons learned from the underlying non-perturbative theory.

\subsection{Supergravities in \texorpdfstring{$d\geq10$}{Lg}}

Let us start with a quick summary of supergravities in dimensions $d\geq 10$. We have found five 10d supergravities in previous sections by looking at the low-energy effective actions of different string theories. These theories have different gauge groups, chiralities, and levels of supersymmetries. In addition to these five 10d supergravities, there is a unique 11d supergravity which we have not provided any string theory description of so far. These theories and their field content are summarized in Table \ref{T1}.\footnote{As we go to higher dimensions, the representations of supersymmetry get bigger. Unbroken supersymmetry in dimensions higher than eleven would require massless particles with spins greater than two which would violate the Weinberg-Witten theorem \cite{Nahm:1977tg,Weinberg:1980kq}.  In 11d, there is only a single supermultiplet with particles with spins $\leq 2$. Thus, the field content is unique in 11d supergravity. Cremmer, Julia, and Scherk were able to write down the action for this supermultiplet; \cite{Cremmer:1978km} and many subsequent works have shown that the theory is almost unique. See \cite{Freedman:2012zz} for a review of this theory. }

\begin{center}
\small
\addtolength{\tabcolsep}{1pt}  \captionsetup{width=.9\linewidth}
\renewcommand{\arraystretch}{1.5}
\begin{threeparttable}
\begin{longtable}{c|cccc}\toprule
\diaghead{xxxxxxxxxxxxxxxxx}{Theories}{Properties} & \makecell{SUSY} & \makecell{Dimension} & \makecell{Bosonic content} & \makecell{Gauge group}\\\midrule
 IIB &\makecell{$\mathcal{N}=(2,0)$\\$N_{SUSY}=32$} &\makecell{$d=10$}& \makecell{NS–NS: $\phi,B_{\mu\nu},g_{\mu\nu}$\\RR: $\tilde\phi, \tilde B_{\mu\nu},$\\$ \tilde D_{\mu\nu\rho\sigma}(F=F^*)$}&\textendash\\\midrule
IIA &\makecell{$\mathcal{N}=(1,1)$\\$N_{SUSY}=32$} &\makecell{$d=10$} & \makecell{NS–NS: $\phi,B_{\mu\nu},g_{\mu\nu}$\\RR: $C_{1~\mu}, C_{3~\mu\nu\rho} $} &\textendash\\\midrule
\makecell{Heterotic\\$E_8\times E_8$} &\makecell{$\mathcal{N}=(1,0)$\\$N_{SUSY}=16$} &\makecell{$d=10$} & \makecell{NS–NS: $\phi,B_{\mu\nu},g_{\mu\nu}$\\RR: $A_{\mu}\in\mathfrak{e}_8\oplus \mathfrak{e}_8$} &\makecell{$E_8\times E_8$}\\\midrule
\makecell{Heterotic\\$SO(32)$} & \makecell{$\mathcal{N}=(1,0)$\\$N_{SUSY}=16$} & \makecell{$d=10$}& \makecell{NS–NS: $\phi,B_{\mu\nu},g_{\mu\nu}$\\RR: $A_{\mu}\in\mathfrak{so}(32)$} & \makecell{$\text{Spin}(32)/\mathbb{Z}_2$}\\\midrule
Type I &  \makecell{$\mathcal{N}=(1,0)$\\$N_{SUSY}=16$} & \makecell{$d=10$}& \makecell{NS–NS: $\phi, g_{\mu\nu}$\\RR: $B_{\mu\nu}$\\NS+: $A_{\mu}\in\mathfrak{so}(32)$} & \makecell{$\text{Spin}(32)/\mathbb{Z}_2$}\\\midrule
\makecell{11d\\supergravity}& \makecell{$\mathcal{N}=1$\\$N_{SUSY}=32$} &  \makecell{$d=11$}& \makecell{$g_{\mu\nu},C_{\mu\nu\rho}$}& \textendash\\\bottomrule
\caption{All tensors except $g_{\mu\nu}$ are antisymmetric. $\tilde\phi$ is a pseudo scalar.}
\label{T1}
\label{table2}
\end{longtable}
\end{threeparttable}
\end{center}
We will use Table \ref{T1} throughout this section to check dualities between the low energy effective field theories of different string theories. 

\subsection{T-duality for superstring theories}

Let us start with T-duality \cite{Sathiapalan:1986zb}. In previous sections, we learned that compactifying the bosonic string theory on a circle has a duality that switches momentum and winding modes with each other. Let us try to extend the T-duality to supersymmetric string theories. From the bosonic string theory we know how T-duality acts on the left and right moving part of the bosonic worldsheet fields. Suppose $X^{9}(\sigma,\tau)$ is the compact coordinate and 
\begin{align}
    X^{9}=X^{9}_L+X^{9}_R ,
\end{align}
where $X_L$ and $X_R$ are respectively the left-moving and right-moving parts of $X^{9}$. By looking at the action of T-duality on the momentum and winding numbers $n, w$, and the radius $R$ we can see that T-duality acts on 
\begin{align}
    X_R^{9}=n\cdot\frac{\tau}{2R}-w\cdot2R\sigma+\cdots
\end{align}
as
\begin{align}
    X_R^{9}\leftrightarrow -X_R^{9}.
\end{align}
while keeping $X_L^9$ fixed. Given that worldsheet supersymmetry maps $X_R^{9}$ to $\tilde \psi^{10}$, T-duality must act on the fermionic fields as
\begin{align}
    (\psi^\mu,\tilde\psi^\mu)\rightarrow    (\psi^\mu,-\tilde\psi^\mu).
\end{align}

\subsubsection*{Type II theories}
Remember that the right-moving Ramond sector had degenerate ground states that decomposed into two representation of the spacetime Lorentz group; $\textbf{8}_s$ and $\textbf{8}_c$. Each of the two ground state are annihilated by different operators. One by $\tilde\psi^8+i\tilde\psi^9$ and the other by $\tilde\psi^8-i\tilde\psi^9$. Thus, T-duality effectively switched the spacetime chirality of the ground state of the right-moving Ramond sector. By definition, this maps IIA theory to IIB and vise versa. 

\subsubsection*{Heterotic theories}

Since in heterotic theories, the worldsheet does not have any right-moving fermions, T-duality only acts on the bosonic worldsheet fields. If we look at the conjugate momenta associated to the bosonic fields, $P_L$ is a 17 dimensional vector (16 from the 10d theory and 1 from the extra circle) and $P_R$ is a one-dimensional vector corresponding to the compact circle. The two vectors together must form an even self dual lattice $\Gamma^{17,1}$. However, since the space of such lattices is connected, we find that the two theories share the same moduli space and are the same. The two heterotic compactifications are dual theories expanding around different corners of the moduli space of $\Gamma^{17,1}$ Narain lattices. 

\subsubsection*{Questions}

Among the five ten dimensional supergravities in \ref{T1}, we were able to connect two pairs of them by T-duality. The theories in each pair lie at different corners of the moduli space of a unifying lower dimensional theory. This unification raises several natural questions. Can we extend the web of dualities to the point that all perturbative constructions are connected? 

Our study of T-duality was based on splitting the worldsheet fields into left and right moving components. However, this cannot be done for open strings of the type I theory. In fact the worldsheet parity acts on the boundary conditions of the open strings. In that case, is there a way to T-dualize type I theory? 

The dualities we studied so far are perturbative dualities, in the sense that the dual theories become weakly/strongly interacting together. In other words, the moduli space of the theories are overlapping and the duality can be proven to any order of perturbation. The price of this control is that the dualities we discussed so far teach us little about non-pertubative ophysics. Can we potentially connect different corners with non-pertubative (strong/weak) dualities?

We saw how moving towards the asymptotic of a lower dimensional theory can lead to higher dimensional theory and unify them. Can we somehow apply the same idea to 10d theories and find the 11d supergravity in the moduli space?  

To answer these question we need to go back and do a closer examination of R–R (gauge) fields in different string theories. 

\subsection{Branes}

Let us begin with a general gauge theory in a spacetime of dimension $d$. From previous experience we know that a $U(1)$ gauge symmetry has an associated one-form gauge potential 
\begin{align}
    A = A_\mu dx^\mu .
\end{align}
It transforms under the gauge transformation as 
\begin{align}
    A \to A + d\alpha
\end{align}
where $d\alpha(x)$ is the exterior derivative of an arbitrary function $\alpha(x)$ with appropriate boundary conditions. From this we can construct the gauge-invariant, two-form field strength $F = dA$, which consists of the electric and magnetic fields. It satisfies the homogeneous Maxwell equations 
\begin{align}
    dF = 0 
\end{align}
by construction since $d^2 = 0$.

The gauge field $A_\mu$ naturally couples to the worldline $\gamma$ of a point-like (one-dimensional) electric charge $q$ through
\begin{equation}\label{eq:electriccoupling}
\exp \left( i q \int_\gamma A \right).
\end{equation}
It is straightforward to check that this quantity is both Lorentz invariant and gauge invariant. 

Given a gauge potential $A$, we can construct its magnetic dual by taking the Hodge dual $\tilde{F} = \star F$ and solving for $\tilde{F} = d\tilde{A}$. From the properties of the Hodge star, we observe that $\tilde{F}$ is a $(d-2)$-form and so $\tilde{A}$ is a $(d-3)$-form. In analogy with the point particle case, $\tilde{A}$ should couple to magnetic objects of charge $p$ whose volume form matches the form of $\tilde{A}$, \emph{i.e.} objects with a $(d-3)$-dimensional worldvolume $\Sigma$. It is easy to guess that such a coupling should take the form
\begin{equation}
\exp \left( i p \int_\Sigma \tilde{A} \right).
\end{equation}
It appears that magnetically charged objects dual to electric point particles should extend over $d-4$ spatial directions. This is consistent with the well-studied case of $d=4$, where electric and magnetic charges are both point-like with zero spatial extent. For this case, we also have that the sum of spatial dimensions of electric objects and their magnetic dual objects is $d-4$, which happens to hold for higher dimensions as well.

How does one measure the charges of objects in $d$ dimensions? In the case of a point-like particle, to find its total charge we surround it with a sphere $S^{d-2}$ and integrate the inhomogeneous part of Maxwell's equations $\star J = d\star F$ over the ball comprising its interior. Since $F$ is a two-form, it is easy to see that $J$ is necessarily a $(d-2)$-form. We thus find that the total charge of a point-like particle in $d$ dimensions is
\begin{equation}
Q_e = \int_{S^{d-2}} \star F .
\end{equation}
This is just Gauss's law for electric charge in $d$ dimensions. By a similar argument, it is straightforward to see that the magnetic dual of an electric particle is surrounded by a two-sphere $S^2$, and so its magnetic charge is measured by
\begin{equation}
Q_m = \int_{S^{2}} F .
\end{equation}

Next let us turn our attention to extended objects with electric charge. Recall that a $p$-brane is a $(p+1)$-dimensional extended objects with worldvolume denoted by $\Sigma_{p+1}$. It naturally couples to a $(p+1)$-form gauge field $A_{p+1}$ as
\begin{equation}
\exp \left( i Q_e \int_{\Sigma_{p+1}} A_{p+1} \right).
\end{equation}
Note that the gauge field transforms under $p$-form gauge transformations
\begin{align}
    A_{p+1}(x) \to A_{p+1}(x) + d\Lambda_{p}(x) .
\end{align}
That is, the gauge transformations are now generated by a $p$-form $\Lambda_{p}(x)$, again with suitable boundary conditions. In this case, we say that the $p$-brane is electrically charged under a $(p+1)$-form gauge symmetry. We can measure the charge of the brane by a generalized version of Gauss's law, namely
\begin{equation}
Q_e = \int_{S^{d-p-2}} \star F_{p+2}
\end{equation}
where $F_{p+2} = dA_{p+1}$ is the $(p+2)$-dimensional field strength. From the point of view of the directions orthogonal to the brane, the sphere $S^{d-p-2}$ surrounds a point charge $Q_e$.  

The dual magnetic objects are defined in the same manner as before, coupling to a $(d-p-1)$-form magnetic potential $\tilde{A}_{d-p-1}$, with $\star F_{p+2} = d \tilde{A}_{d-p-3}$. It follows that the objects should have dimension $d-p-3$, and in particular they are $(d-p-4)$-branes. (As promised, the electric spatial dimension $p$ and magnetic spatial dimension $d-p-4$ add up to $d-4$.) To measure the magnetic charge, we surround the brane within a $(p+2)$-sphere $S^{p+2}$ and integrate over the flux, \emph{i.e.}
\begin{equation}
Q_m = \int_{S^{p+2}} F_{p+2} .
\end{equation}

\subsubsection*{Branes in string theory}

The previous discussion was true of general gauge theories with charged, extended objects. Let us now return to the appearance of branes in string theory. The massless $p$-form fields in string theory give rise to gauge symmetries in spacetime. It remains to determine what objects are charged under these fields. 

All of the perturbative string theories (except type I) admit the ``bosonic" cannon of massless fields: the dilaton field, the metric, and the Kalb-Ramond NS–NS field also known as the B-field (in type I, the B field comes from the R–R sector). The latter is a two-form gauge field $B_{\mu\nu}$ which couples directly (electrically) to the string worldsheet $\Sigma$ as
\begin{equation}
\exp \left( i \int_{\Sigma} B \right).
\end{equation}
The B-field transforms under a spacetime gauge symmetry 
\begin{align}
    B \to B + d\Lambda
\end{align}
where $\Lambda$ is a one-form gauge parameter. The gauge invariant field strength $H=dB$ is often referred to as the H-field or H-flux. From this point of view, the string is electrically charged under the B-field. We can see this in more familiar language by taking the theory compactified on a circle, say in the $x_9$ direction, in which case the winding number of the string serves as the conserved charge associated with a one-form $B_{\mu9}$. Note that the absence of a B-field in NS–NS sector in type I theory implies that the winding number of the fundamental string is not a conserved quantity in the compactified theory.  

We saw that every electric object had a magnetic partner. It is natural to ask what objects are magnetically charged under the B-field. The dual field strength is given by $\tilde{h} = \star H$, and so the dual gauge potential $\tilde{B}$ is necessarily a six-form field. Thus, the string is magnetically dual to a five-dimensional object known as an NS five-brane. Unlike D-branes, which have a tension proportional to $1/\lambda$, it turns out that such objects have a tension that goes like $1/\lambda^2$ where $\lambda$ is the string coupling. 

Now we move onto the massless gauge fields originating in the R–R sector.
On the worldsheet they correspond to the R–R vertex operators, and so they do not naturally couple to worldsheet. In other words, the string is not charged under these fields in the usual sense, and whatever is electrically charged must necessarily be comprised of nonperturbative objects.  This is also true of the dual R–R fields, and hence of the associated magnetic objects as well. 

The type I string is especially simple. It only has a single R–R two-form field $C_{\mu\nu}$ that couples electrically to 1d strings and magnetically to five-dimensional objects. The type IIA theory has a one-form field $C_\mu$ (with 0d electric objects and 6d magnetic objects) and a three-form field $C_{\mu\nu\rho}$ (with 2d electric objects and 4d magnetic objects). Similarly, the type IIB string instead has even-form gauge potentials, with a two-form field $C_{\mu\nu}$ (with 1d electric objects and 5d magnetic objects), a four-form field $C_{\mu\nu\sigma\rho}$ (with 3d electric and magnetic objects), and a scalar $C_0$ (with 7d magnetic objects and (-1)d electric objects).\footnote{These are called D(-1) branes which are not true states of the theory but rather a special class of instantons which give non-perturbative corrections, beginning at order $1/\lambda$, to the perturbative series.}.

We can also consider the 11-dimensional supergravity theory, which turns out to have deep connections with the 10d string theories. The 11d theory admits a single gauge potential, namely the three-form $C^{(11)}_{\mu\nu\rho}$. The objects charged under the three-form potential are necessarily two-dimensional, and are generically referred to as membranes or M2 branes, for short. From the usual dual argument, we also find that there should be five-dimensional branes, referred to as M5 branes, which are the magnetic dual of M2 branes.

The appearance of these higher form gauge symmetries suggest the existence of some charged branes. But how could we see these EFT branes in our microscopic perturbative string description of supergravities? 

To answer this question let us go back to T-duality and the only 10d supergravity that was left un-T-dualized: the type I theory. This also happens to be the only 10d theory with open strings. Open strings requrie boundary conditions and the reason T-duality is more challenging for open strings is that it affects that boundary condition. T-dualizing in a given compact direction switches the Dirichlet and Neumann boundary conditions with each other. Therefore, suppose the T-dual of type I theory exist, it must include D(richlet)-branes of lower dimension to account for the boundary conditions. This teaches us that D-branes are essential to completing the picture of dualities. Luckily, these "fundamental" objects turn out to be the same as the macroscopic charged branes in the effective field theories. For example, we can calculate the $p$-form charge of a D(p-1)-brane to see if it matches with the fundamental EFT charge. It turns out it does \cite{Johnson:2003glb}. Note that D-branes are part of the background, therefore they are not perturbative objects. However, this does not mean they are not dynamical objects either! The vanishing of the Weyl anomaly for the worldsheet theory imposes certain equations on a background to be consistent. We can think of these equations as equations of motions for background fields and D-branes. Equivalently, we can study the interactions between D-branes by looking at the tree-level amplitudes involving D-branes exchanging strings. For example, for two D-branes the leading contribution would involve a cylinder connecting the two branes. Since the amplitude of a given worldsheet scales like $\lambda^{2g-2+b}$, at $g=0, b=1$ we get $\sim\lambda^{-1}$. This signals that the tension of the fundamental brane which, plays the role of the gravitational mass, in the Einstein frame is $\sim\lambda^{-1}$ (see \cite{Johnson:2003glb} for detailed calculation). 

The fact that the tension of the D-brane is inversely proportional to the coupling constant is another evidence that why they are non-perturbative objects. This is similar to the case of gauge theory instantons which have actions proportional to $1/g^2$ and therefore, are non-perturbative objects. One can estimate the action of non-perturbative objects from the divergences of the perturbation series. Typically, in the presence of non-perturbative objects, the perturbative expansion diverges after some point and the smallest term determines the maximum resolution of the perturbation theory due to non-pertuabtive effects. For example, going back to gauge theories, there are $\sim \exp(\mathcal{O}(l\ln(l)))$ graphs with $l$ loops and each of them carry a factor of $g^{2l}$. Therefore, the $l$-loop amplitude goes like $\sim g^{2l}\sim \exp(\mathcal{O}(l\ln(l)))$ which minimizes at a value of $\exp(\mathcal{O}(-1/g^2))$. This is the amplitude of the gauge theory instantons! Before the discovery of D-branes as non-perturbative ingredients of string thery, Shenker did a similar calculation to show that their tension must go like $1/\lambda$ \cite{Shenker:1990uf}.

In the following, we use our knowledge of D-branes as non-perturbative objects of the underlying theory to complete the web of dualities.

\subsection{M-theory}

Let us start with the 11d supergravity. Could it be that the 11d supergravity is just the low-energy EFT of some corner of string moduli space? If so, it must be connected to theories with the same number of supercharges like type II theories. It seems the answer to this question is yes and the conjectural theory that has the 11d supergravity as its low energy limit is called M-theory. The 11d supergravity theory consists of an 11d metric $G^{(11)}_{\mu\nu}$, a three-form gauge potential $C^{(11)}_{\mu\nu\rho}$, and a gravitino $\psi_{\mu\alpha}$. It turns out that the EFT of the 11d supergravity compactified on a circle yields the type IIA supergravity, \emph{i.e.} the low energy limit of the type IIA string theory \cite{Horava:1995qa}.

\subsubsection*{M-theory and type IIA}

The first item on our checklist is to match the (bosonic) field content of the two theories. Recall that the massless content of the type IIA theory consists of a metric tensor $G_{\mu \nu}$, the B-field $B_{\mu \nu}$, the dilaton $\phi$, and the odd-valued gauge R–R gauge forms $C_{\mu}$ and $C_{\mu \nu \rho}$. In the compactified theory, the type IIA fields $B_{\mu\nu}$ and $C_{\mu\nu\rho}$ originate from the dimensional reduction of the 11d gauge potential. On the other hand, the dimensional reduction of the 11d metric $G_{\mu\nu}^{(11)}$ should yield the 10d metric, a scalar, and a vector field. These correspond precisely to $G_{\mu\nu}$, $\phi$, and $C_\mu$, respectively.

Next we can try and match the objects (perturbative and non-perturbative) in the two theories. A string in 10d ought to be an M2 in 11d wrapped on a circle. Its radius $R_{11}$ turns out to be directly related to the expectation value of $\phi$, \emph{i.e.} the type IIA coupling. Reducing the 11d gravity action on the circle (and only keeping the Einstein-Hilbert term) yields
\begin{equation}
R_{11} \int d^{10} x \, e^{-2\phi_M} \sqrt{G_M} R_M,
\end{equation}
where $G_M$ and $R_M$ are the 10d metric and its associated scalar curvature, as constructed from the zero-momentum KK modes of the 11d metric. The field $\phi_M$ is a scalar arising from the metric along the compact directions. Note that we have set the 11d mass scale to $M_{pl} = 1$. The resulting theory is nonchiral, and so it is expected to be equal to the action of type IIA supergravity
\begin{equation}
\frac{1}{\lambda^2} \int d^{10} x \, e^{-2\phi} \sqrt{G_s} R_s,
\end{equation}
where $\lambda = e^{\langle \phi \rangle}$ is the type IIA string coupling. Here we have written the type IIA action in the string frame, where the dilaton multiplies the Einstein-Hilbert term. Weak coupling in particular corresponds to small $\lambda$. It is clear that $\phi_M = \phi$. The metrics $G_{\mu \nu \, M}$ and $G_{\mu \nu \, s}$ should be related by some field redefinition. Note that $\sqrt{G_M} R_M$ scales as $G_M^4$, and so if we write $G_{\mu \nu \, M} = f G_{\mu \nu \, s}$ then it follows that
\begin{equation}
R_{11} f^{4} = \frac{1}{\lambda^2}.
\end{equation}
Thus, the two metrics are related by a rescaling of the form
\begin{equation}
G_{\mu \nu \, M} = \left(\frac{1}{R_{11} \lambda^2 } \right)^{\frac{1}{4}} G_{\mu \nu \, s}.
\end{equation}
This has a consequence on energy measurements with respect to the two metrics. Let $E_M$ be an energy measured using $G_M$ and $E_s$ an energy measured using $G_s$. From the above expression we see that the two energies are related by
\begin{equation} 
E_s \lambda^{\frac{1}{4}} = (R_{11})^{- \frac{1}{8}} E_M. 
\label{eq:energyrelations} \end{equation}

Now let us return to the case of an M2 brane wrapping the circle, which results in a string in 10d with some tension $T_M$ as measured in the M-theory frame (\emph{i.e.} with respect to the metric $G_M$). This compares to the tension $T_s$ in the string frame as
\begin{equation}
T_s \lambda^{\frac{1}{2}} = T_M (R_{11})^{- \frac{1}{4}}.
\end{equation}
which follows from \eqref{eq:energyrelations} by dimensional analysis. In M-theory, there is only one scale (the Planck scale). So, an M2 brane has a tension equal to one in Planck units. The circle has radius $R_{11}$, and so the tension of the wrapped brane is given by $T_M = R_{11}$. Note that the string tension in the string frame is $T_s = 1$. We conclude that the radius of the circle and the string coupling are related as\footnote{We have not been careful in tracking powers of $2\pi$ as well as the string length $\ell_s$. However, one can check that all powers of $\ell_s$ cancel out, and furthermore that equation \eqref{eq:radiuscouplingrelation} is correct including numerical factors in Planck units.}
\begin{equation} R_{11}^3 = \lambda^2. \label{eq:radiuscouplingrelation}\end{equation}
We know that string perturbation theory in 10d breaks down as $\lambda$ becomes large. This is also the limit in which the eleventh dimension becomes relevant. So it makes sense that $R_{11}$ increases with $\lambda$.

Recall that the KK reduction of the 11d metric yields a $U(1)$ gauge field which is none other than the type IIA R–R one-form field $C_{\mu}$. The KK modes with nonzero (quantized) momenta along the circle are charged under this symmetry, with the number of quanta being the conserved charge. In the 10d nonchiral supersymmetry algebra of type IIA, this charge appears as a central extension (it can also be obtained by dimensionally reducing the 11d supersymmetry algebra), which roughly speaking takes the form
\begin{align}
    \{Q_\alpha^1, \bar{Q}_\beta^2\} = -2 P_M
\end{align}
where $Q_\alpha^{i}$ for $i=1,2$ are the two sets 10d supercharges with chiral indices and $P_M$ is the KK momentum. Following the supersymmetry algebra, one can derive a BPS bound relating the charges of states in the theory to their masses. We will focus only on BPS states for which the energy equals the KK momentum. In the M-theory frame, the energy of a KK excitation with $n$ units of momentum around the circle is thus
\begin{equation}
E_M = \frac{n}{R_{11}}.
\end{equation}
The energy in the string frame is then
\begin{equation} E_s = \lambda^{-1/4} R_{11}^{-1/8}\ \frac{n}{R_{11}} = \frac{n}{\lambda}. \label{eq:energyd0brane} \end{equation}
As $\lambda$ goes to zero, this blows up, so we do not see these states in string perturbation theory. In closed string perturbation theory, the string coupling constant only appears as $\lambda^{2g-2+b}$. If we took the worldsheet to be a disk, we get $\lambda^{-1}$. In Type IIA, there are also D-branes which can be treated as infinitely massive sources in open-string perturbation theory, where the worldsheet has boundaries. These objects have been shown to be charged under the R–R one-form as expected. The $n$ in equation \eqref{eq:energyd0brane} corresponds to number of D-branes. The states described by equation \eqref{eq:energyd0brane} are D0-branes because they couple to the R–R 1-form.

Now we consider the unwrapped M2-brane extending in the 10-dimensions of type IIA. In the M-theory frame, its tension is
\begin{equation}
T^{M2}_M = 1.
\end{equation}
Using equation \eqref{eq:energyrelations}, the tension of this object in the string frame is
\begin{equation}
T_s^{M2} = \lambda^{-1}.
\end{equation}
This object is a D2-brane.

The main lesson is that as the string coupling goes to infinity, the type IIA theory becomes eleven-dimensional. The Lorentz group SO(9,1) becomes extended to SO(10,1).

The degrees of freedom of a fluctuating brane correspond to a scalar field that lives on the brane. The number of scalar fields living on a $p$-brane corresponds to the number of transverse directions, $d - (p + 1)$.

Suppose that we have a D-brane, which by definition is where open strings can end. These endpoints of the open string can carry additional pointlike degrees of freedom known as Chan--Paton factors. These degrees of freedom are charged under a one-form gauge-field $A$ which lives on the brane. Now consider the $B$-field, which couples to the string worldsheet. We expect that the transformation 
\begin{align}
    B \rightarrow B + d \Lambda 
\end{align}
for a one-form $\Lambda$ should be a gauge symmetry that leaves the worldsheet action invariant. This is true for closed strings, but fails to hold for worldsheets with boundary. The resolution is that $A$ must live on the boundary, \emph{i.e.} on the D-branes, and transform in an opposite manner, 
\begin{align}
    A \rightarrow A - \Lambda, 
\end{align}
to compensate. For a stack of $N$ coincident D-branes, we can use a naive counting argument to show that the gauge symmetry should be at least $U(1)^N$. It turns out that the gauge group is actually $U(N)$. This follows from the fact that there are $N^2$ different ways for the open string to end on the branes at the level of perturbation theory. The massless states of the perturbative open string in the type II theories contains a single $\mathbf{8_v}$, which is governed by a 10-component field. On a $Dp$-brane, this field splits into a $p+1$-dimensional gauge field living on the brane and $9-p$ scalars which describe the fluctuations of the brane. This gauge field transforms under the nonabelian $U(N)$ gauge symmetry. 

Recall that we have constructed an eleven-dimensional M-theory which when compactified on an $S^1$, gives the type IIA theory. We have identified the radius of the circle and the IIA string coupling constant are identified as 
\begin{equation}
R^3 \sim g_s^2.
\end{equation}
The three-form field $C_{\mu\nu\rho}$ is associated to the M2-brane, which is a 3-dimensional world volume object. Then this M2-brane wraps around the $S^1$ yielding a string, which is identified as IIA string. Similarly, we can have an M2-brane not wrapping around, which as discussed above has a tension $\sim\lambda^{-1}$ and corresponds to D2 brane. 

The magnetic version of the M2-brane is the M5-brane in M-theory. Let us consider an M5-brane around this $S^1$. Then we get D4-branes. When M5-branes are not wrapping around the circle, we can consider the other 5-brane in M-theory: NS 5-brane, which is a magnetic dual object to the string. In M-theory, there are also D6-branes and D8-branes. First of all, D6-brane is the Kaluza-Klein monopole, which happens when the circle shrinks at a point. 

\begin{figure}[H]
\centering
\scalebox{1.2}{
\begin{tikzpicture}
\node[circle,thick,scale=0.5,fill=black,label={[label distance=2mm]east:$\mathbb{R}^3\times S^1$}] (A1) at (1,2) {};
\node[circle,thick,scale=0.5,fill=black,label={[label distance=1mm]south:D6-brane}] (A2) at (1,0) {};
\node[draw=none,opacity=0,thick,scale=0.1,fill=black,label={[label distance=2mm]40:$\mathbb{R}^3$}] (A3) at (1,0) {};
\node[draw=none,opacity=0,thick,scale=0.1,fill=black,label={[label distance=5mm]east:Taub-Nut}] (A5) at (3,2) {};
\draw (A1)--(A2);
\draw (3,2.5) arc (90:120:4);
\draw (1,2) arc (240:270:4);
\draw (-0.2,1)--(2.5,1)--(2.1,-1)--(-0.6,-1)--(-0.2,1);
\draw (3,2) ellipse (0.15cm and 0.5cm);
\end{tikzpicture}
}
\caption{D6-brane is the object where the circle shrinks from the Taub-Nut space.}
\label{fig:D6brane}
\end{figure}
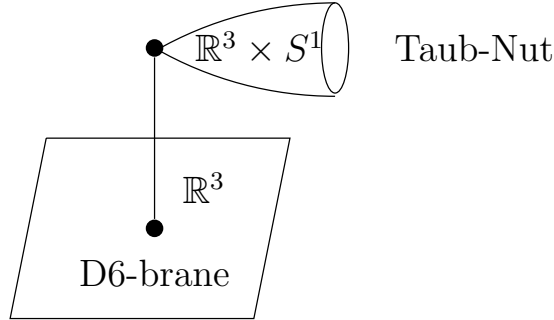

Consider the Taub-Nut geometry, which looks like $\mathbb{R}^3\times S^1$ asymptotically, where the circle shrinks in the middle. Then such a point on $\mathbb{R}^3$ is the D6-brane, as represented in Figure \ref{fig:D6brane}. The D6-brane is a codimension-three object. On the other hand, D8-brane is a codimension-one object, which dramatically impacts the global topology of the space and is more difficult to describe in M-theory.

When we consider $N$ number of D-branes, we get theory with a U(N) gauge symmetry. The fact that one of them having a gauge symmetry is related to the fact that the $B$-field that the strings couples to has a gauge symmetry. On the boundary of the worldsheet, we get a gauge field and the strings are ending with s Dirichlet boundary condition. This is then related to the twisted sectors between the branes. D-branes breaks half of the supersymmetry. Thus we get the analog of having 16 supercharges via
\begin{align}
(\mathbf{8_v}+\mathbf{8_s})\otimes(\mathbf{8_v}+\mathbf{8_s}),
\end{align}
with only half conserved. For example, D3-branes have a vector and six scalars:
\begin{align}
2_{\mathbf{V}}+6(\mathbf{0}),
\end{align}
which adds up to have eight degrees of freedom. This gives in turn an $\mathcal{N}=4$ super Yang--Mills with a U(N) gauge group.

\subsubsection*{M theory and type IIB}

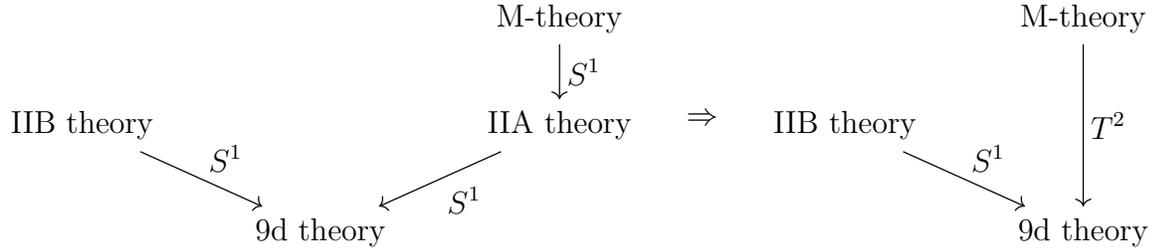
\begin{figure}
\begin{tikzcd}[column sep=normal]
 & & \text{M-theory} \arrow[rightarrow]{d}{\displaystyle{S^1}} \\
\text{IIB theory}\arrow[rightarrow]{rd}{\displaystyle{S^1}} & & \text{IIA theory}\arrow[rightarrow]{ld}{\displaystyle{S^1}} \\
 & \text{9d theory} &
\end{tikzcd}
$\quad\Rightarrow\quad$
\begin{tikzcd}[column sep=normal]
 & \text{M-theory} \arrow[rightarrow]{dd}{\displaystyle{T^2}} \\
\text{IIB theory}\arrow[rightarrow]{rd}{\displaystyle{S^1}} & \\
 & \text{9d theory}
\end{tikzcd}
\caption{M-theory and IIB theory yield the same 9d theory}
\label{fig:MandIIB}
\end{figure}

Now that we connected M-theory and IIA, we know that they live in the same moduli space. However, we know that IIB and IIA share that property too because they are T-dual to each other. Therefore, IIB and M-theory must also share the same moduli space. Let us try to connect them directly to each other instead of taking the long route through IIA  \cite{Schwarz:1995jq}. In order to find the connection we follow the duality chain between M-theory on $T^2$ to IIB on $S^1$ as represented in Figure \ref{fig:MandIIB}.

From Table \ref{T1} we know that type IIB theory has the following bosonic field content,
\begin{align}
\lambda_s,\quad B_{\mu\nu},\quad g_{\mu\nu},\quad \chi_0,\quad \widetilde{B}_{\mu\nu},\quad \widetilde{D}_{\mu\nu\rho\lambda},
\end{align}
where the first three are from the NS–NS sector and the last three aref from the R–R sector that are coupled to D-1-brane, D1-brane, and D3-brane respectively. The scalar fields can naturally be composited and to become complexified as
\begin{align}
\tau=\chi+\frac{i}{\lambda_s}.
\end{align}
Then we have three real parameters for a type IIB to be compactified on a circle -- one complex coupling parameter and one real parameter that is the radius
\begin{align}
(R^{\text{IIB}}\ ,\ \tau ).
\end{align}
On the other hand, M-theory is also parametrized by three real parameters: 
\begin{align}
(A\ ,\ \widetilde{\tau}),
\end{align}
where $A$ is the area and $\widetilde{\tau}$ is the Teichmüller parameter of torus and has a symmetry under $\widetilde{\tau}\longrightarrow\widetilde{\tau}+1$.
This $\chi$ is the gauge degrees of freedom that is periodically valued. As we go around the 7-brane (which has a codimension-two world volume), $\chi$ shifts by 1. Hence, $\tau$ shifts by 1. It follows that this $\tau$ respects the symmetry of $\tau\longrightarrow\tau +1$. 
Then it has the same symmetry with $\widetilde{\tau}$ from M-theory and hence we relate these two $\tau$s:
\begin{align}
\widetilde{\tau}=\tau.
\end{align}
Then we relate the other real variables,
\begin{align}
\frac{1}{R}=A,
\end{align}
where $R$ is the radius from IIB theory and $A$ is the area from M-theory.

In fact, the $\widetilde{\tau}$ has a full $SL(2,\mathbb{Z})$ symmetry, not just a shift symmetry. Suppose we take $\chi=0$, then 
\begin{align}
\lambda\longrightarrow\frac{1}{\lambda},
\end{align}
which gives a strong-weak duality. Indeed, type IIB theory is self-dual under the strong-weak duality.

Recall that Type IIB had two-form fields $B_{\mu\nu}$ and a $\widetilde{B}_{\mu\nu}$, which, respectively, coupled electrically to the fundamental string (also called the F1-brane) and the D1-brane. Under the $SL(2, \mathbb{Z})$ symmetry transformation
\begin{equation}
\label{eqn:almostS}
\lambda_s \rightarrow \frac{1}{\lambda_s} \,,
\end{equation}
these two fields are exchanged. This transformation swaps the tensions of the fundamental string and the D1-brane, and thus it is necessary that the potentials coupled to these extended objects are also exchanged. One the other hand, there is only a single four-form field, $\widetilde{D}_{\mu\nu\rho\lambda}$, and thus it must map to itself under the transformation (\ref{eqn:almostS}). That is, the D3-brane which couples to the four-form must be self-dual under the given transformation. As a consequence of this it is immediate that the $U(N)$ $\mathcal{N} = 4$ super Yang--Mills theory living on the worldvolume of a stack of $N$ D3-branes must be invariant under an $SL(2,\mathbb{Z})$ symmetry, as the D3-branes are simply mapped onto themselves. The worldvolume theory enjoys invariance under the transformations
\begin{equation}
\tau \rightarrow \tau + 1 \,, \quad \tau \rightarrow - 1/\tau \,,
\end{equation}
of its complexified coupling constant, $\tau$. This gives the S-duality of the $\mathcal{N}=4$ $U(N)$ SYM in $d=4$. Since the D5-brane and the NS5-brane are the magnetic duals of the fundamental string and of the D1-brane they are also necessarily exchanged by the transformation (\ref{eqn:almostS}).

This duality, which is called the S-duality of Type IIB string theory, is changing the fundamental string into what one would perturbatively think of as a composite, heavy, object, the D1-brane. We note that because the duality replaces fundamental objects with composite objects one has to be careful with the definition of a Feynman path integral -- to perform the integral one must pick a particular duality frame, and integrate over the fundamental degrees of freedom in that frame. The notion of what is light, or fundamental, that goes into the definition of the path integral is not necessarily a duality invariant notion. 

We now consider how these two different two-form fields of Type IIB arise by considering M-theory compactified in a torus. That is, we will consider a compactification of Type IIB on an $S^1$, and look at the components of the two 10d two-forms that do not extend along the $S^1$, and thus they behave like 9d two-forms. These equally can be understood from the perspective of the M-theory three-form, $C_{\mu\nu\rho}$. To get a two-form one of the directions in $C_{\mu\nu\rho}$ must lie along one of the directions inside the $T^2$ on which
we are compactifying. A $T^2$ has two distinct one-cycles, the $A$ and $B$ cycles, and the two two-forms $B_{\mu\nu}$ and $\widetilde{B}_{\mu\nu}$ arise from wrapping $C_{\mu\nu\rho}$ with one direction along each of these cycles. Realizing the $SL(2,\mathbb{Z})$ transformation (\ref{eqn:almostS}) as a modular transformation of the complex structure of the torus, one can see that the $A$ and $B$ cycles are exchanged.  Generally this process involves wrapping the M2-brane, which couples electrically to the three-form, along the cycle $pA + qB$ of the torus. For $(p,q) = (1,0)$ the resulting string is the fundamental string, and $(p,q) = (0,1)$ is the D1-brane. The $SL(2,\mathbb{Z})$ symmetry of the torus can be thought of, in this way, as generating the $SL(2,\mathbb{Z})$ self-duality symmetry of Type IIB from M-theory.  M2 branes wrapping on a general $(p,q)$-cycle (with $p$ and $q$ coprime) gives rise to a bound state of F-strings and D1-branes. 

\noindent {\bf Exercise 1}: show that upon compactification on an $S^1$ and T-duality, D-brane dimensions change up or down by one unit, depending on whether or not the brane wraps the $S^1$. That is, how do the D-branes of Type IIA transform into the D-branes of Type IIB under T-duality? [Hint: if you consider a Dp-brane in one Type II theory and compactify on an $S^1$ which is orthogonal to the Dp-brane worldvolume then the $p$ does not change. However Type IIA admits only supersymmetric branes with $p$ even, and Type IIB only with $p$ odd; thus the T-duality should change the parity of the dimension of the brane worldvolume.]

\noindent {\bf Note}: it is not true that the 11d M-theory picture is always the most useful way to study string theory. In M-theory on $T^2$ with area $A$ gets mapped to Type IIB on an $S^1$ with radius 
\begin{equation}\label{eqn:RA}
R = 1/A \,.
\end{equation}
This is consistent with the fact that the Kaluza--Klein modes of Type IIB, which are the winding modes in Type IIA language, which in the M-theory uplift is then a M2-brane wrapping also the other cycle, \emph{i.e.}~ wrapping the entire torus. If we consider M-theory on a $T^2$ and we shrink $A \rightarrow 0$ then the theory is not 9d, as one would naively think, but because of (\ref{eqn:RA}) it is, in fact, 10d Type IIB string theory.  In this way M-theory would naively miss Type IIB, and knowing about T-duality of Type IIA is an additional
ingredient which is obscured from the pure M-theory point of view.

Note that something strange happened in the type IIB/M-theory duality: they both appear in an asymptotic corner of the moduli space of a 9d theory ($\ln(A)\sim -\ln(R_{IIB})\rightarrow \pm\infty$). However, the dimensions of these two theories are different! In one corner we get a 10d theory while in the other corner we get an 11d theory! This shows that dimension is not a good guiding principle to classify disconnected theories in quantum gravity.

\subsection{Completing the web of dualities for \texorpdfstring{$N_{SUSY}=16$}{Lg}}

Thus far we have connected together all of the maximal supergravity theories (32 supercharges), being (the massless sector of) Type IIA, Type IIB, and M-theory. We now want to include the 10d supergravities with $\mathcal{N} = 1$ supersymmetry, that is, the Type I and heterotic theories. We have already established that the two heterotic theories, with gauge groups $SO(32)$ and $E_8 \times E_8$ are related via $S^1$ compactification to 9d. Both Type I and heterotic $SO(32)$ have an $SO(32)$ gauge group, which motivates us to search for a relationship amongst these theories already in 10d. 

To show this we first give a different perspective on Type I. 

\subsubsection*{Type IIB construction of the type I theory}

Type I is type IIB in 10d, orientifolded by the parity operator $\Omega$, that reverses the orientation on the worldsheet \cite{Horava:1989vt}. The invariant subspace of the spacetime is the ``orientifold plane'', which is 1+9 dimensional in this context, and thus is known as the O9-plane. The O9-plane carries $-32$ units of D9-brane charge, and thus the theory would be inconsistent via a Gauss-like law unless the charge is cancelled off by the inclusion of $32$ D9-branes.  This is the brane
interpretation for why Green--Schwarz found that the requirement for anomaly cancellation is was an $SO(32)$ gauge group -- the $32$ D9-branes naively generate a $U(32)$ gauge group, but this is quotiented to $SO(32)$ by the parity reversal of the O9-plane.

\subsubsection*{Type I and heterotic \texorpdfstring{$SO(32)$}{Lg}}

Now we are ready to discuss the relation between Type I and heterotic $SO(32)$ \cite{Horava:1995qa}. The latter has an effective action from the genus zero worldsheet like
\begin{equation}
\frac{1}{\lambda_h^2} \int \left( R_h + F_h^2 + \cdots  \right) \,,
\end{equation}
where $F_h^2 = g^{\nu\beta}g^{\mu\alpha}F_{\mu\nu}F_{\alpha\beta}$ comes from
the field strength of the $SO(32)$ gauge potential. In Type I the $SO(32)$
gauge symmetry comes from the open string sector and so the effective action is
instead of the form
\begin{equation}
  \frac{1}{\lambda_I^2} \int R_I + \frac{1}{\lambda_I} \int F_I^2 \,.
\end{equation}
The different scaling of the coefficients of the $F^2$ terms tells us
immediately that these theories cannot be directly the same, so let us compare
the scaling relations. Since 
\begin{equation}
  R \sim g^4 \,, \quad F \sim g^3 \,,
\end{equation}
it can be seen that the scaling gives
\begin{equation}
  \frac{g_h^4}{\lambda_h^2} = \frac{g_I^4}{\lambda_I^2} \quad \Rightarrow \quad \frac{\lambda_I}{\lambda_h} = \left(\frac{g_I}{g_h}\right)^2 \,,
\end{equation}
and
\begin{equation}
  \frac{g_h^3}{\lambda_h^2} = \frac{g_I^3}{\lambda_I} \quad \Rightarrow \quad \frac{\lambda_I}{\lambda_h^2} = \left(\frac{g_I}{g_h}\right)^3 \,.
\end{equation}
Putting this altogether one finds
\begin{equation}
  \lambda_I = \frac{1}{\lambda_h} \,.
\end{equation}
This tells us that if we begin with the weakly coupled heterotic $SO(32)$ theory and we move to strong coupling then there is a dual weakly coupled description in terms of Type I theory, and vice versa. This is an S-duality between the two theories, which is ideal as we have a method to understand the theory at both strong and weak coupling. 

In fact, we cannot have two good description at the same limit of the moduli of the weakly coupled theory. In other words, we always have separate understandings at the opposite limits. 

From a type I theory, consider an example where we stretch a D1-brane on the $(1+1)$d spacetime with 32 D9-branes from the theory. Then there is a string coupling the D1 brane and D9 branes. By looking into this sector, we can study the lightest mode to be a fermion for each D9-brane coupled to the D1-brane. Therefore, the D9 branes give us 32 fermions living on the D1-brane. Moreover, on the D1-brane we have $(0,8)$ supersymmetry, which we can identify to the right movers of the heterotic string theory. Hence we can get a glimpse of the duality between the type I theory and the heterotic theory from this picture as well where the type I D1 brane is dual to the heterotic string.

So far we have connected type I theory and heterotic theory with $SO(32)$ via duality, but not for the heterotic theory with $E_8\times E_8$. We can connect both heterotic theories since when compactified on a circle, they result in the same 9d theory. However, heterotic theory with $E_8\times E_8$ is not directly linked with type I, which requires a perturbative understanding of the theory. This is because its string coupling constant diverges to describe its relation to type I theory.

\subsubsection*{M-theory and heterotic}

First we showed that all the theories with 32 supercharges share the same moduli space in the sense that they correspond to different corners of a single moduli space. Next we showed that the 10d theories with 16 supercharges share the same moduli space as well. Since type I has an orientifold construction from type IIB theory, it is natural to ask if one can construct theories with 16 supercharges directly from IIA and M-theory as well. If the dualities are real, such a construction should follow from the chain of dualities. Hořava--Witten proposed an M-theory construction that is placed in the same moduli space of $E_8\times E_8$ Heterotic theory \cite{Horava:1995qa,Horava:1996ma}. The Hořava--Witten theory is thought to describe the strong coupling limit of the $E_8\times E_8$ Heterotic theory.

\begin{figure}[H]
\centering
\begin{tikzcd}[column sep=normal]
\text{Het}_{SO(32)}\arrow[rightarrow]{rd}[swap]{\displaystyle{S^1}} & & \text{Het}_{E_8\times E_8} \arrow[rightarrow]{ld}{\displaystyle{S^1}} \\
 & \text{9d theory}
\end{tikzcd}
\caption{Heterotic string theories on a circle are identical}
\label{fig:Hets}
\end{figure}
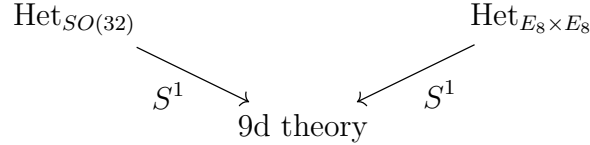

From M-theory on a circle, we get a type IIA theory by construction. Let us consider to mod out the circle by $\mathbb{Z}_2$ and orbifold the theory. With this setup one can check that half of the supersymmetry survives:
\begin{align}
X^{11}\longrightarrow -X^{11},\quad
\gamma^{11}\longrightarrow -\gamma^{11}
\end{align}
in the language of spinors. Then such a $\gamma$ will project out half of them, resulting in 16 supercharges from 32 supercharges before. We have two singularities as orbifold points. some localized degrees of freedoms can live on these points. there are two $(9+1)$-dimensional spaces. We can have each $E_8$ to live on each wall. This is called Hořava--Witten construction. The coupling of the heterotic strings is the radius. 

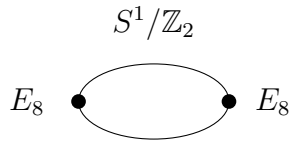
\begin{figure}[H]
\centering
\begin{tikzpicture}
\node[circle,thick,scale=0.5,fill=black,label={[label distance=2mm]west:$E_8$}] (A1) at (-1,0) {};
\node[circle,thick,scale=0.5,fill=black,label={[label distance=1mm]east:$E_8$}] (A2) at (1,0) {};
\node[draw=none,opacity=0,thick,scale=0.5,fill=black,label={[label distance=1mm]north:$S^1/\mathbb{Z}_2$}] (A2) at (0,0.5) {};
\draw (0,0) ellipse (-1cm and 0.5cm);
\end{tikzpicture}
\caption{An $S^1$ with $\mathbb{Z}_2$ orbifold points for Hořava--Witten construction}
\label{fig:Hořava--Witten}
\end{figure}

M-theory has a membrane, M2-brane. Heterotic theory on the other hand has no membranes. So one might be puzzled how they can be mapped. In 11d supergravity, we have $C_{\mu\nu\rho}$. What happens to this in 10d? In order to keep the term $C\wedge G_4\wedge G_4$ to be constant under $\mathbb{Z}_2$, we have $C\rightarrow -C$ while preserving $C_{11\nu\rho}$. This $C_{11\nu\rho}$ is $B_{\nu\rho}$ in heterotic theory. M2-branes wrap the fundamental string that sees both $E_8$ on the orbifold points, and hence has $E_8\times E_8$ current algebra. It follows that the wrapped M2-branes survive but not the unwrapped M2-branes, and for M5 branes, the opposite happens.

\subsubsection*{Extending dualities to lower dimensions}

So far we have connected M-theory, type IIA theory, type IIB theory, heterotic theories, and type I theory. We can use string duality further down to lower dimensional theories. For example, we can consider an M-theory compactified on K3 surfaces. This gives a 7d theory with half amount of supersymmetries. Similarly, heterotic theory compactified on a $T^3$ gives a 7d theory. Either heterotic theories result in the same theory in 7d theories as they were already the same in 9d. We can construct to see if these 7d theories are dual to each other. First of all, they have the same number of supersymmetry. The moduli space of the the heterotic theory on a $T^3$ is given by
\begin{align}
\frac{SO(19,3)}{SO(19)\times SO(3)}\times R^+,
\end{align}
where the first part provides Lorentzian Narain lattice and $R^+$ controls the strings coupling. On the other side, M-theory on K3 surfaces, we can see the Hodge numbers for the K3 surfaces to be
\begin{align}
h^{1.1}=20=19+1,
\end{align}
so there are 19 complex deformations as one is trivialized by hyperk\"ahler rotations mixing $h^{1,0}$, $h^{1,1}$, $h^{1,1}$. Hence, we have 19 complex and 19 real deformations with 1 hyperk"ahler rotation. Hence we have $(19\times 3)+1$. This looks like we get the same moduli space
\begin{align}
\frac{SO(19,3)}{SO(19)\times SO(3)}\times R^+,
\end{align}
where we map both $R^+$ of both moduli spaces. The bigger volume then means bigger coupling for the other theory, which is reasonable. Hence we can have a map between M-theory on K3 and heterotic theory on $T^3$.

Let's go down to six dimensions now. We can now ask about Type IIA compactified on a K3 surface. This gives a 6d theory which we have already observed has a moduli space
\begin{equation}
  \frac{SO(20,4)}{SO(20) \times SO(4)} \times R^+ \,.
\end{equation}
Similarly one observes that heterotic on $T^4$ has the same moduli space. These two six-dimensional theories are related by a strong-weak duality that maps between Type IIA on K3 and heterotic on $T^4$. This is particularly interesting as it relates a compactification on the curved internal manifold, the K3, with one on a flat manifold, the $T^4$; in this way, all of the intricate geometry of the K3 surface is captured dually in a straightforward toroidal compactification of the heterotic string. 

One immediate question is how the non-abelian gauge symmetry of the heterotic theory is replicated in the compactification of the type IIA theory, which has no perturbative non-Abelian gauge symmetry, on K3. We will explain this later in this section.

\subsection{F-theory}
We have seen that we have dualities from heterotic theory on $T^4$ and type IIA theory on K3; similarly, from heterotic theory on $T^3$ and M-theory on K3. One can speculate as to whether this pattern uplifts further, and whether there is a duality between heterotic on $T^2$ and some uplifted 12d theory on K3. Naively this will not be possible as there does not exist a 12d supergravity theory that we can compactify on the K3 for the right-hand-side of the duality, however, there is a hint that this might be possible. 
We take M-theory on $T^2$ which we said is related to Type IIB on an $S^1$. There is a limit where the raduis of the $S^1$ goes to infinity, which is where the area of the $T^2$ goes to zero. While the $T^2$ is of zero volume, the data of it is not completely absent from the Type IIB, for instance the complex structure mode of the torus is a part of the IIB theory. In this way one can think of 
Type IIB has a 12d theory, where two of the directions look like a zero-area torus. 

Now we can consider a compactification of M-theory on some manifold which has a torus fibration over some base space, $B$. When we take the limit where the area of the torus fiber shrinks to zero volume we recover Type IIB compactified on $B$. If we now take a K3 surface which admits a torus fibration over a 2d manifold then we can do this procedure to get an 8d theory which is the compactification of IIB on such a 2d manifold -- this would be the candidate theory for the dual to heterotic on $T^2$ that we speculated about above.

In fact, there exist K3 surfaces which are torus (or elliptic) fibrations over $\mathds{P}^1$. Such a K3 surface has a realization via a Weierstrass equation
\begin{equation}
y^2 = x^3 + f_8(z_1, z_2) x + g_{12}(z_1, z_2) \,,
\end{equation}
where $f_8$ and $g_{12}$ are degrees 8 and 12 homogeneous polynomials in the projective coordinates, $[z_1 : z_2]$ of the base $\mathds{P}^1$. To match with heterotic on $T^2$ we must count the moduli of this K3 surface. A polynomial of degree $d$ has $d+1$ parameters because there are $(d+1)$ coefficients in the generic polynomial, which tells us that we naively have $9 + 13 = 22$ parameters from the complex structure moduli of the K3. An overall rescaling will remove one of these parameters, and then there is an $SL(2,\mathds{C})$ action which removes three more parameters. Thus the K3 surface has 18 complex parameters. The K\"ahler parameters are just the volumes of the $T^2$ fiber and the $\mathds{P}^1$ base, however, since the theory requires the fiber to shrink into zero volume, the volume of $T^2$ is not a part of the theory. In turn, there is just one real K\"ahler parameter controlling the size of the $\mathds{P}^1$. 

The parameter space of the K3 is then given by
\begin{equation}
\frac{SO(18,2)}{SO(18) \times SO(2)} \times R^+ \,,
\end{equation}
which is exactly the Narain moduli space for the compactification of heterotic on $T^2$. Furthermore, we see that 
\begin{equation}
A \sim \lambda_h \,,
\end{equation}
so increasing the area of the $\mathds{P}^1$ makes the heterotic theory strongly coupled.

We have just described a funny compactification of Type IIB on a $\mathds{P}^1$ which preserves only half of the supersymmetry, as it is dual to a heterotic compactification. By writing K3 as a torus fibration we allowed the complex structure modulus, $\tau$, to depend on the holomorphic coordinate on the $\mathds{P}^1$, $z$. Since
\begin{equation}
  \tau = \chi + \frac{i}{\lambda_s} \,,
\end{equation}
we see that $\chi$ and $\lambda_s$ now depend also on $z$. We do not usually consider such compactifications in superstring perturbation theory as there may be a point on spacetime where the string coupling becomes large, and then perturbation theory breaks down. We can determine the points in $\mathds{P}^1$ where $\tau \rightarrow \infty$, where we have no perturbative control of the theory. It turns out that when $\tau$ becomes infinite is exactly where the torus fibers of the elliptic fibration degenerate. This occurs at the discriminant locus, which is when 
\begin{equation}
\Delta = 4 f_8^3 + 27 g_{12}^2 = 0 \,.
\end{equation}
Since $\Delta$ is a degree $24$ polynomial then there are $24$ zeros of $\Delta$ distributed over the $\mathds{P}^1$. We note 
that $\tau$ is actually not a well-defined function of $z$, as it undergoes $SL(2,\mathbb{Z})$ transformations when one 
moves on a path around one of these zeros of $\Delta$. 

Then we have constructed a theory in 12d by identifying the $\tau$ as a shrunk torus on top of IIB theory. This theory is called F-theory and can be viewed as a non-perturbative compactification of type IIB \cite{Vafa:1996xn}. This is a new type of compactification coming from string duality. For example, at points where $\tau\to\infty$, we have monodromies given by $SL(2,\mathbb{Z})$ action. We know that a D7 produces a monodromy of $\tau\to\tau +1$ as we go around it. The more complicated monodromies are sourced by bound states of $SL(2,\mathbb{Z})$ images of D7 branes. Hence we can determine that at the zeros of $\Delta$, there are non-perturbative 7-brane characterized via their $SL(2,\mathbb{Z})$ monodromy
\begin{align}
\renewcommand*{\arraystretch}{1.1}
SL(2,\mathbb{Z})\ \text{action} : \quad \begin{pmatrix} p & r\\ q & s \end{pmatrix}
\begin{pmatrix} 1\\ 0 \end{pmatrix} = \begin{pmatrix} p\\ q \end{pmatrix}, \quad qs=rq=1,
\end{align}
we call the non-perturbative brane that sources the above monodromy the $(p,q)$ 7-branes.

Having 24 zeros of $\Delta$ then yields 24 7-branes. However, we should not be allowed to have 24 7-branes as the charge does not cancel. The reason why we could have 24 D7-branes is because we are in a non-abelian theory from $SL(2,\mathbb{Z})$ action.

Note that a stack of $N$ D7-brane has a $U(N)$ gauge group. Therefore, when the pinched points are brought together, we can an enhancement of the gauge group. This can be easily demonstrated via looking into pinched points. Each pinching point has a local description
\begin{align}
xy=z.
\end{align}
Then all pinched points can be written as
\begin{align}
xy=z(z-a_1)(z-a_2)\cdots .
\end{align}
When these pinched points are brought together, the local geometry becomes
\begin{align}
xy=z^n,
\end{align}
and hence we get an SU(N) gauge group when we have $A_{N-1}$ singularity. 

In order to have all heterotic theory gauge symmetries from this construction, we need to be able to put some 7-branes together to build an $E_8$. We can expect such arrangements to be possible by using singularities of type $D$ and $E$. 

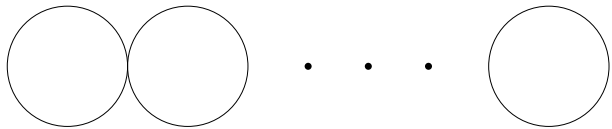
\begin{figure}[H]
\centering
\scalebox{.8}{
\begin{tikzpicture}
\draw (0,0) circle (1);
\draw (2,0) circle (1);
\node[circle,thick,scale=0.3,fill=black,label={[label distance=1mm]0:}] (A3) at (4,0) {};
\node[circle,thick,scale=0.3,fill=black,label={[label distance=1mm]0:}] (A3) at (5,0) {};
\node[circle,thick,scale=0.3,fill=black,label={[label distance=1mm]0:}] (A3) at (6,0) {};
\draw (8,0) circle (1);
\end{tikzpicture}}
\caption{$(N-1)$ spheres touching each other to form an $A_{N-1}$ type singularity}
\label{fig:Atype}
\end{figure}

Let's consider M-theory on K3 that admits singularities of ADE type of Lie algebra. This is dual to heterotic theory on $T^3$. Let us consider the case of $A_{N-1}$ as an example to demonstrate such is possible from M-theory. Geometry of type $\mathds{C}_2/\mathbb{Z}_N$ gives a type $A_{N-1}$ singularity. To consider the singularities of $A_{N-1}$ type, we can think of having $(N-1)$ number of spheres, 2d cycles, touching each other as in Figure \ref{fig:Atype}.\\

The size of these spheres are controlled by K\"ahler parameters of $\mathds{P}^1$s: 
\begin{align}
    \phi_1,\ \phi_2,\ \cdots,\ \phi_{N-1} .
\end{align}
Furthermore, M-theory has 3-form fields $C_{\mu\nu\rho}$ that give for every independent 2-cycle a 2-form. So we can write the three-form fields in the basis of 2-forms $\omega_{\mu\nu}^i$ as
\begin{align}
C_{\mu\nu\rho}=\sum_i A_{\mu}^i (x) \omega_{\mu\nu}^i ,
\end{align}
where $A_{\mu}^i$ depends on the direction that is not compactified. Hence for every three-form fields, we get a two-form and a one-form that is in the leftover space, which is the gauge field. This gauge field has a U(1)$^{N-1}$ symmetry automatically.

Now wrap M2-brane on one of the 2-cycles:
\begin{align}
e^{i\int_{\text{M2}}\ C_{\mu\nu\rho}} \sim e^{i \oint A_{\mu}}.
\end{align}
The 3-form field $C_{\mu\nu\rho}$ is the one that is coupled to M2-brane, and hence on this gauge field $A_{\mu}$, there will be a U(1) charge. Thus we get a charged object corresponding to this M2-brane wrapping around 2-cycle. Wrapping around just one 2-cycle, we are considering the geometry of $\mathds{C}_2/\mathbb{Z}_2$. Then we have two possibilites: we can have an M2-brane or an anti-M2-brane, wrapping around with the opposite orientation. Hence we get two states: $\pm 1$ charges of U(1) based on the orientation of M2-brane. Via quantization of M2-branes, these are gauge multiplets (vector multiplets). More precisely, they are charged massive vector multiplets $W^{\pm}$ where the mass is proportional to the area. More precisely, the mass of the vector multiplets is proportional to
\begin{align}
m\sim TA,
\end{align}
where $T$ is the tension of the M2-brane and $A$ is the area. However, it is impossible to have a charged massive vector multiplet  under U(1) unless it is non-abelian U(1). In fact that is possible as if the area shrinks to zero, we get a massless vector multiplet. In other words, we get an U(1) with two charged objects that are opposite in charge, and we can conclude that this is SU(2). Thus we see that when we wrap an M2-brane on one 2-cycle that has $A_1$ singularity, we get a non-abelian gauge symmetry SU(2). Giving vacuum expectation value to the scalar $\phi$ for this U(1), \emph{i.e.} Higgsing the U(1), is equivalent to blowing up.

Now we can consider a general case of $(N-1)$ 2-cycles that had U(1)$^{N-1}$ symmetry. This will yield an SU(N) gauge symmetry. However, we do not have enough vectors as we have only $2(N-1)$ charged and $(N-1)$ neutral objects. We are required to wrap two touching $\mathds{P}^1$s to bind and form a bound state to resolve such an issue.

\noindent{\bf Optional exercise}: Reduce M-theory on K3 and convince yourself that there are sixteen supersymmetries and that is enough amount of supersymmetries to have a scalar in the gravity multiplet. In fact, except in 10d, the gravity multiplet with 16 supercharges in all lower dimensions have scalars.

\noindent{\bf Exercise 2}: We learned that we can wrap M2-branes on touching $\mathds{P}^1$s to form a bound state. We can wrap many at once upto all $(N-1)$ of them. Each chain will give a charged object upto $\pm$ sign. Check that this gives exact dimensions of SU(N). Checking along all the chains upto the sign, show that we can get full rank of SU(N). By this way we can see the charge and degeneracies of SU(N). [Hint: Focus more on the degeneracies than charge for this exercise.]

\noindent{\bf Optional exercise}: We can also have singularities of type D and E as well. Find what the rules of binding the M2-branes for the $D$ and $E$ types in order to reproduce the rank of the gauge groups. [Hint: For example, $\dim(E_8)=248$.]

\subsection{More dualities in lower dimensions}

Suppose we have two theories A and B compactified on $M_1$ and $M_2$ respectively, yielding the same theory on $\mathbb{R}^d$. Then there are parameters corresponding to the moduli of $M1$ and $M2$ which give rise to scalar fields. We usually construct dualities by taking such parameters to be constant in the resulting d-dimensional compactified theory. However, if we can imagine these parameters vary in $\mathbb{R}^d$ we still expect to have the duality. In particular, if the parameters vary gradually, we can go between the dual frames point by point in $\mathbb{R}^d$. As long as we preserve the amount of supersymmetries, in all the examples it is shown that the duality persists even if the change of moduli breaks adiabatic principle. By considering non-constant backgrounds and compactifying them while preserving some supersymmetry, we can find more lower-dimensional theories that enjoy non-trivial dualities. The assumption that higher-dimensional dualities continues to be true for non-constant backgrounds  is called the \textit{adiabatic assumption}.

For example, Type IIB theory with varying the coupling constant over $\mathds{P}^1\times S^1$ is dual to M-theory on K3 when the parameters are mapped point by point \cite{Klemm:1995tj}. On the other hand, we take $S^1$ to have infinite size, which then corresponds to having elliptic fibration shrinking to zero size in F-theory. Note that the adiabatic principle is explicitly violated from shrinking the elliptic fibration.\footnote{We are changing the topology of the internal space and letting $\tau\to\infty$. In fact, this is the case for any duality using T-duality.} Moreover, M-theory on K3 is dual to heterotic on $T^3$. By taking heterotic on $T^2\times S^1$, we have the duality between heterotic on $T^2$ and IIB on $\mathds{P}^1$, which was explained earlier by building F-theory from IIB on $\mathds{P}^1$.\\
\begin{figure}[H]
\centering
\begin{tikzcd}[column sep=small]
\text{Heterotic on}\ T^3 & \longleftrightarrow & \text{M-theory on}\ K3 & \longleftrightarrow & \text{IIB on}\ \mathds{P}^1\times S^1
\end{tikzcd}\\
\begin{tikzcd}[column sep=small]
& \text{Heterotic on}\ T^2 & \longleftrightarrow & \text{IIB on}\ \mathds{P}^1\ \text{(F-theory on K3)}
\end{tikzcd}
\end{figure}

Now consider F-theory on an elliptic manifold $M_{\text{ell}}^d$, then the resulting theory is a $(12-d)$-dimensional theory. When we have an elliptic manifold, we can consider F-theory for any case that has a type IIB theory with a varying parameter via duality. Then we can construct the following duality that results in the same $(10-d)$-dimensional theory.\\
\begin{figure}[H]
\centering
\begin{tikzcd}[column sep=normal, row sep=huge]
\text{F-theory} \arrow[rightarrow]{rd}[swap]{\displaystyle{M_{\text{ell}}^d\times S^1\times S^1}\ } & \text{M-theory} \arrow[rightarrow]{d}{\displaystyle{M_{\text{ell}}^d\times S^1}} & \text{Type IIA theory} \arrow[rightarrow]{ld}{\ \displaystyle{M_{\text{ell}}^d}} \\
 & (10-d)\text{-dimensional theory} &
\end{tikzcd}
\end{figure}

We can study these dualities for various dimensions. We have studied 7d theories earlier, which was found to have a total of 16 or 32 supercharges to be preserved. In the 6d theory, we can also get a theory with 8 supercharges. This can be achieved via heterotic theory on K3 or F-theory on elliptically-fibered Calabi--Yau threefolds. In 5d, we also have the same amount of supercharges. For 4d, the minimal number of supercharges is 4, which corresponds to  $\mathcal{N}=1$. Such theories can be constructed by compactifing heterotic theories on Calabi--Yau threefolds \cite{Candelas:1985en}, F-theory on Calabi--Yau fourfolds \cite{Brunner:1996pk}, or M-theory on $G_2$-manifolds \cite{Awada:1982pk,Acharya:1998pm}. These are summarized in Table \ref{tb:compactified}.\\

\begin{table}[htb]
\begin{center}
\renewcommand{\arraystretch}{1.4}
\begin{threeparttable}
\begin{tabular}{c | c}
\toprule
$10-d$ & Number of supercharges \\
\midrule
$7$ & 32, 16 supercharges \\
$6$ & 32, 16, 8 supercharges \\
$5$ & 32, 16, 8 supercharges \\
$4$ & 32, 16, 8, 4 supercharges\\
\bottomrule
\end{tabular}
\end{threeparttable}
\caption{Possible number of supercharges in lower-dimension theories via compactifications. In the case of 4-dimensional theories there are more options not listed in the table.}
\label{tb:compactified}
\end{center}
\end{table}

As an example, we can demonstrate how we can get a duality between two resulting theories in 6d via heterotic theories on K3 and F-theory on elliptically-fibered Calabi--Yau threefolds. Let us recall that we have a duality between heterotic theories on $T^2$ and F-theory on K3 (or IIB theory on a $\mathds{P}^1$). Now fiber both cases over a $\mathds{P}^1$. Then we have heterotic theories on $T^2\ltimes \mathds{P}^1=K3$ and IIB on a $\mathds{P}^1$ fibered over a $\mathds{P}^1$ base. Not precisely, but in some sense we get a $\mathds{P}^1\times\mathds{P}^1$. Note that $\mathds{P}^1\times\mathds{P}^1$ is the same as $\mathbb{F}_0$. In general, we can have a various way to fiber a $\mathds{P}^1$ over a $\mathds{P}^1$ base, which produces Hirzebruch surfaces $\mathbb{F}_n$ with a variable $n$. When $n>12$, such an Hirzebruch surface is no longer Calabi--Yau and hence it only works for $0\leq n\leq 12$.

On the other side, we have to investigate heterotic theory on $K3$. Recall that we have the $H$-flux to vanish
\begin{align}
dH=\frac{1}{16\pi^2}(R\wedge R-F\wedge F),
\end{align}
and for a K3, we have a nonzero $R\wedge R$ term, which is 24. Then it follows that $F\wedge F$ term will have to include 24 instanton numbers. Note that considering $E_8\times E_8$, we can put for example 12 instantons numbers each, then this turned out to be corresponding to $\mathbb{F}_0$ from the F-theory compactification. We can consider in generality to have $(12-n)$ and $(12+n)$ instantons on each $E_8$ for $0\leq n\leq 12$. This corresponds to $\mathbb{F}_n$ from F-theory compactification, which is the exact match of the same range of $n$.

\noindent{\bf Optional exercise}: Show that $n=12$ is the maximal $n$ for the Hirzebruch surface $\mathbb{F}_n$ to be elliptically-fibered Calabi--Yau threefolds.

Additionally, M-theory on an interval $S^1/\mathbb{Z}_2\times K3$ is also dual to heterotic theory on K3 (see Figure \ref{fig:Hořava--Witten} for the interval). Then on each end of the interval we have an orbifold point. What does it mean to have $E_8$ instanton numbers on each orbifold point? Having nontrivial instanton numbers means we are turning on some excitation. Instantons shrinking into zero size is equivalent to M5-branes approaching the boundary.\footnote{This is because we are on 6d resulting theory from 11d theory and hence an instanton is equivalent to 11-6=5-dimensional brane.} Then in the equivalent picture in F-theory, we should match the M5-branes, \emph{i.e.} squeezing instantons on K3 to a point, to the Hirzebruch surfaces $\mathbb{F}_n$. This can be demonstrated with toric diagrams.

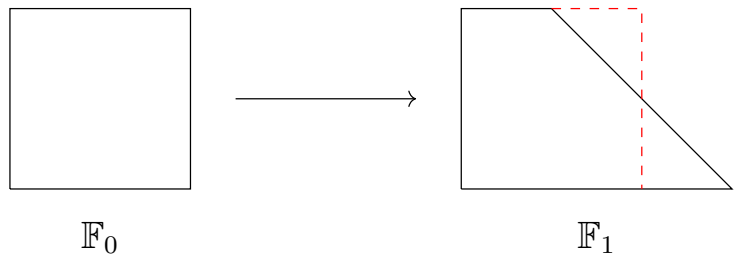
\begin{figure}[H]
\centering
\scalebox{1.2}{
\begin{tikzpicture}
\draw (0,0)--(0,2)--(2,2)--(2,0)--(0,0);
\draw[->] (2.5,1)--(4.5,1);
\draw (5,0)--(8,0)--(6,2)--(5,2)--(5,0);
\draw[dashed,color=red] (6,2)--(7,2)--(7,0);
\node[draw=none,opacity=0,thick,scale=0.1,fill=black,label={[label distance=2mm]south:$\mathbb{F}_0$}] (A1) at (1,0) {};
\node[draw=none,opacity=0,thick,scale=0.1,fill=black,label={[label distance=2mm]south:$\mathbb{F}_1$}] (A2) at (6.5,0) {};
\end{tikzpicture}
}
\caption{These are toric diagrams of $\mathbb{F}_0$ and $\mathbb{F}_1$. By blowing up a point and blowing down another $\mathds{P}^2$ on $\mathbb{F}_0$, we get $\mathbb{F}_1$.}
\label{fig:F0F1}
\end{figure}

We can have M-theory on an interval and both ends to have $(12-n)$ instantons and $(12+n)$ instantons. In order to shift from $n=0$ case to $n=1$ case, we need to squeeze one instanton out and move to the other end. Squeezing into zero size of the instanton corresponds blowing-up a point geometrically, and pushing the M5-brane to the other end of the boundary corresponds to blowing-down another point on toric diagram, as demonstrated in Figure \ref{fig:F0F1}.

In this manner we can see that all three theories, heterotic theory on a K3 surface, F-theory on an elliptically-fibered Calabi--Yau threefold, and M-theory on an $S^1/\mathbb{Z}_2$, are all dual to each other to give a six-dimensional resulting theory. Similarly, the same things happen for all the other lower dimensions listed in Table \ref{tb:compactified}, which summarizes all string dualities.

\section{Complex geometry}

\subsection{Preliminary definitions}

A {\textit{complex manifold}} is a (topological) manifold that can be covered by patches of complex coordinates $z^i = x^i + i y^i$ such that the transition functions between different patches are holomorphic. This ensures that the notion of a holomorphic function $f(z)$ is well-defined on the entire manifold regardless of the choice of coordinates\footnote{By definition, a {\textit{Riemann surface}} is a one-dimensional complex manifold.}. On such manifolds there is a natural notion of complex differential forms, with the space of $(p,q)$ forms $\Omega^{p,q}$ spanned by elements of the form $dz^{i_1} \wedge \cdots \wedge dz^{i_p} \wedge d\bar{z}^{\jbar_1} \wedge \cdots \wedge d\bar{z}^{\jbar_q}$. This leads to a refined version of the de Rham cohomology known as the Dolbeault cohomology. 

Recall that the de Rham cohomology $H^p(M)$ of a manifold $M$ is the set of closed $p$-forms modulo exact $p$-forms with respect to the exterior derivative $d$. It is isomorphic to the space of harmonic $p$-forms, \emph{i.e.} those that vanish under the Laplacian $(d + \star d \star)^2$. Their dimensions $b_p = H^p(M)$ are topological invariants known as the {\textit{Betti numbers}}, related to the Euler characteristic of $M$ by 
\begin{equation}
\chi(M) = \sum_{p=0}^{\text{dim}(M)} (-1)^p b_p .
\end{equation}
For complex manifolds, we can consider an extended set of differential operators $\p$ and $\pb$ which map $(p,q)$ forms to $(p+1,q)$ and $(p,q+1)$ forms, respectively. They satisfy the relations
\begin{equation}
d = \p + \pb , \quad \p^2 = \pb^2 = \{\p, \pb\} = 0.
\end{equation}
The {\textit{Dolbeault cohomology}} $H^{p,q}$, also a topological invariant, is defined as the space of $\pb$-closed $(p,q)$-forms modulo $\pb$-exact $(p,q)$-forms. The dimensions of these spaces, known as {\textit{Hodge numbers}}, satisfy a myriad of relations such as $h^{p,q} = h^{q,p}$. We will have more to say about this later in the context of Calabi--Yau manifolds. 

A {\textit{K\"ahler manifold}} is a Riemannian manifold with nonzero metric components $g_{i \jbar}$ which locally take the form $g_{i \jbar} = \partial_i \partial_\jbar K(z,\zbar)$ for a function $K$ called the {\textit{K\"ahler potential}}. The {\textit{K\"ahler form}} is a closed (1,1) form $k_{i \jbar}$ that is essentially the antisymmetric version of $g_{i\jbar}$. It is related to the metric as $k = \partial \partialbar K$, with components $k_{i \jbar}  = - k_{\jbar i} = g_{i \jbar}$. K\"ahler manifolds of complex dimension $n$ generically have a $U(n)$ holonomy. {\textit{Calabi--Yau}} manifolds $CY_n$ of complex dimension $n$ are special examples of K\"ahler manifolds that have a reduced SU($n$) holonomy. This implies that the curvature class of the $U(1)$ piece of the spin connection (\emph{i.e.} the first Chern class) is zero. Yau proved that a K\"ahler manifold with vanishing first Chern class admits a Ricci-flat K\"ahler metric, which has SU($n$) holonomy. As mentioned perviously, $SU(n)$ holonomy guarantees the existence of a covariantly constant spinor, and hence the preservation of supersymmetry. 

Calabi--Yau manifolds come in families depending on their metric and complex structure. The K\"ahler form is contained in the cohomology $H^{1,1}$. Its dimension $h^{1,1}$ is also the number of ways we can deform the metric. In string theory, the addition of $B_{\mu \nu}$ increases the dimensionality of the moduli space. Deformations of $G_{\mu\nu}$ and $B_{\mu\nu}$ can thus be captured by the complex combination
\begin{align}
k + i B
\end{align}
which lives in the complexified (1,1)-form cohomology with real dimension $2h^{1,1}$. In addition to deforming the metric/B-field, we can also deform the complex structure, which is equivalent to mixing $\partial_i$ and $\partialbar_\jbar$. For instance, under a generic deformation this takes the form
\begin{equation} 
\partialbar_\ibar \longrightarrow \partialbar_\ibar + \mu\indices{_\ibar^j}(z,\bar{z})\partial_j. \end{equation}
In order to maintain $\partialbar^2 = 0$, $\mu$ is required to satisfy
\begin{align}
\partialbar_{[\ibar} \mu\indices{_{\jbar]}^j} = O(z,\bar{z}). 
\end{align}
To count the number of complex structure deformations, we first note that Calabi--Yau manfiolds also have a globally defined $(n,0)$ form, represented in component form by the tensor $\epsilon_{i_1 \cdots i_n}$, which can be used to covert an upper holomorphic index to $n-1$ lower indices. In particular, it can be used to write
\begin{equation}
\mu_{\ibar j_1 \cdots j_{n-1}}=\epsilon_{j j_1 \cdots j_{n-1}} \mu\indices{_\ibar^j}.
\end{equation}
Since $\partialbar_{[\ibar} \mu\indices{_{\jbar]}^j} = 0$, the $(n-1,1)$-form $\mu_{\ibar j_1 \cdots j_{n-1}}$ is also annihilated by $\partialbar$. A $\partialbar$-exact $(n-1,1)$-form corresponds to 
\begin{equation}\mu\indices{_\ibar^j} = \partialbar_\ibar v^j(z,\bar{z})
\end{equation} 
for some vector field $v^j$, which is a trivial coordinate change (as opposed to a deformation of the complex structure). Then the number complex structure deformations is given by $h^{n-1,1} = h^{1,n-1}$.

\subsection{Examples of Calabai--Yau manifolds}

\paragraph{1-folds}

We first consider the simple example of a (non-singular) Calabi--Yau manifold, $T^2$. it is particularly nice because we known the explicit form of its metric, given by $dz d\zbar$. Its cohomology classes are spanned by the forms $1,dz,d\zbar,dz \wedge d\zbar$. It immediately follows that
\begin{align}
h^{0,0}=h^{0,1}=h^{1,0}=h^{1,1}=1.
\end{align}
From this we see that there is a single K\"ahler deformation as well as one complex structure deformation. Recall that the inequivalent tori are labeled by a single complex modulus $\tau$; complex structure deformations correspond to changing the value of $\tau$. The K\"ahler parameter (modulus) corresponds to an overall rescaling. For string theory we shall also include the $B$-field in our analysis. If we let $A$ denote the area of the torus, we can define a complex K\"ahler parameter $\rho = B + i A$. The parameter $\tau$ transforms under the usual SL(2,$\mathbb{Z}$) symmetry. Next we study the symmetries of $\rho$. There is a shift symmetry $\rho \rightarrow \rho + 1$.\footnote{Recall that in the path integral, the Kalb-Ramond field appears as $e^{2 \pi i \int B}$ with appropriately chosen units, so the theory has a shift symmetry $B \rightarrow B + 1$.} There is also another symmetry transformation on $\rho$, which is best illsutrated for the case where $B = 0$. If the torus has radii $R_1$ and $R_2$, then a T-duality transformation given by $R_1 \rightarrow \frac{1}{R_1}$ and $R_2 \rightarrow \frac{1}{R_2}$ corresponds to $\rho \rightarrow - \frac{1}{\rho}$. Combing these two symmetry transformations implies that $\rho$ also has an $SL(2,\mathbb{Z})$ symmetry, with a fundamental domain identical to that of $\tau$.

The $T^2$ manifold provides us with our first example of {\textit{mirror symmetry}}, which exchanges different Hodge numbers, and in particular the complex and K\"ahler structures. For simplicity, consider again a torus with radii $R_1$ and $R_2$ and $B = 0$. Then 
\begin{align}
|\tau| = \frac{R_2}{R_1},\quad |\rho| = R_{1} R_2.
\end{align}
If we perform a T-duality transform on just $R_1$, then $|\tau| \to \frac{1}{R_1 R_2}$ and $|\rho| \to \frac{R_2}{R_1}$; so $\tau$ and $\rho$ are exchanged under T-duality!

\paragraph{2-folds}
\label{sec:k3}
The obvious next example are Calabi--Yau 2-folds. There is of course the trivial CY given by $T^4$ with trivial holonomy, but this behaves very similarly to the previous example of $T^2$. Instead we consider the orbifold $T^4/\mathbb{Z}_2$, which has a $\mathbb{Z}_2$ holonomy and is a singular limit of K3. The $\mathbb{Z}_2$ acts on both coordinates as $z^i \to -z^i$. The metric of this space is known but singular, and so $T^4/\mathbb{Z}_2$ is not a smooth manifold. To find its cohomology, we instead consider the cohomology of $T^4$, which like $T^2$ is given by the products of $1, dz^i, d\zbar^{\jbar}$ for $i=1,2$ and $\jbar=1,2$\footnote{Technically speaking, orbifold geometries with singular points are not Calabi--Yau manifolds. Instead, one can construct a CY from an orbifold classically by smoothing any singularities. However, string theory is well-defined on orbifold geometries. After correctly accounting for the twisted sectors, the orbifold theories give results that precisely agree with the classical geometries that have been smoothed. PLEASE CHECK}. To construct the cohomology of the orbifold, we simply mod out by the $\mathbb{Z}_2$ action (which projects out the odd-dimensional forms) and include the twisted sectors. Recall the exercise of type II on $T^4/\mathbb{Z}_2$. Each twisted sector contributes a single complex scalar, where the number of twisted sectors is given by the number of fixed points of the action, which is 16 for the orbifold under consideration. The VEVs of these scalars corresponds geometrically to deformations of the metric/$B$-field, and so they contribute directly to $h^{1,1}$. The cohomology of $T^4/\mathbb{Z}_2$ is thus given by $h^{1,1} = 4+16 = 20$. We list the cohomologies of $T^4$ and $T^4/\mathbb{Z}_2$ below, represented as the matrices $h^{p,q}$:
\begin{align}
\begingroup
\renewcommand*{\arraystretch}{1.1}
T^4 : \quad 
\begin{pmatrix}
1 & 2 & 1\\
2 & 4 & 2\\
1 & 2 & 1
\end{pmatrix}, \quad\quad
T^4/\mathbb{Z}_2: \quad 
\begin{pmatrix}
1 & 0 & 1\\
0 & 20 & 0\\
1 & 0 & 1
\end{pmatrix}.
\endgroup
\label{eq:t4cohom}
\end{align}

We can now study the details of the $T^4/\mathbb{Z}_2$ geometry. Recall that $T^2$ has two complex moduli, $\tau$ and $\rho$. What are the analogous parameters for K3? Looking at the value of $h^{1,1} = 20$, it would naively seem that there are $20 + 20 = 40$ complex parameters, in analogy with $T^2$. This turns out to be correct, however we must be careful with their treatment. 

The first caveat comes from the fact that there are two complex structure deformations that preserve the metric. Let $\Omega$ denote the unique $(2,0)$-form (which exists because $h^{2,0} = 1$) and let $\bar{\Omega}$ denote the analogous $(0,2)$-form. We take $t \in \mathds{C}$ to parametrize the deformations under consideration.. Defining a complex structure on a complex manifold of dimension $n$ is equivalent to choosing an $(n,0)$ form. We can thus introduce a real two-parameter family of complex structures labeled by $\hat{\Omega}(t)$, defined as
\begin{equation} \hat{\Omega}(t) = \Omega + t k + t^2 \bar{\Omega}. \label{eq:twodeformations} \end{equation}
To show that $\hat{\Omega}$ defines a complex structure, we must verify that $\hat{\Omega}^2 = 0$. This follows from $\Omega^2 = \bar{\Omega}^2 = 0$ as well as $\Omega \wedge \bar{\Omega} = k \wedge k$. Note that $\Omega \wedge \bar{\Omega}$ is the volume form on the manifold. The equation \eqref{eq:twodeformations} then defines two deformations of the complex structure that leave the metric (e.g. $k$) fixed.

Though it would now seem that we have 40-1=39 complex moduli (after dropping those parametrized by $t$), the missing parameter is restored by correctly considering the B-field deformations. The aforementioned complex parameters only cover the subspace of the moduli space where the $B$-field is deformed by $(1,1)$-forms.\footnote{For $T^2$, $B_{\mu \nu}$ has only a single component, $B_{z \zbar}$.} We must also take into account deformations of the $B$-field by $(2,0)$ and $(0,2)$-forms. This restores the number of complex parameters to 40. 

As it turns out, the moduli space of $T^4/\mathbb{Z}_2$ is given by
\begin{align}
\frac{SO(20,4)}{SO(20) \times SO(4)}\ \Big/ \ SO(20,4,\mathbb{Z}) .
\end{align}
This is the analog of the moduli space of $\tau$ and $\rho$ for $T^2$, which live in $(\mathbb{H} \times \mathbb{H} / SL(2,\mathbb{Z}))$. It is the same moduli space as heterotic theory on $T^4$. If we turn off the $B$-field, then the moduli space reduces to
\begin{align}
\frac{SO(19,3)}{SO(19) \times SO(3)}\ \Big/ \ \left( SO(19,3,\mathbb{Z}) \times R_{\text{radius}}\right) .
\end{align}
To check that the dimensionality works out, note that there are 40 real complex structure moduli and 20 real  K\"ahler moduli (since $B = 0$). We must still subtract 2 due to the complex structure deformations which preserve the metric, given by \eqref{eq:twodeformations}. There are thus $58 = 57 + 1$ real moduli when $B = 0$. 

To summarize, K3 surfaces are the only twofolds realizing the full $SU(2)$ holonomy. In turn, we deduce that all Calabi--Yau twofolds are K3 surfaces.

\paragraph{3-folds}

Now, we move on to the 3-dimensional case, which is especially interesting. First, it is unclear whether the number of Calabi--Yau threefolds is even finite. To study these surfaces, we can work out the Hodge numbers. Note that if there is a nontrivial (\emph{i.e.} non-exact) one-form, then the fundamental group of the manifold is nontrivial. In particular, if the fundamental group of $CY_3$ is nontrivial, then the holonomy must be a proper subgroup of SU(3). It follows that $CY_3$ with SU(3) holonomy must have 
\begin{align}
h^{1,0} = h^{0,1} = 0 .
\end{align}
Generically, the cohomology of $CY_3$ with SU(3) holonomy is given by
\begin{align}
\begingroup
\renewcommand*{\arraystretch}{1.1}
\begin{pmatrix}
1 & 0 & 0 &1 \\
0 & h^{2,1} & h^{1,1}& 0\\
0 & h^{1,1} & h^{2,1}& 0\\
1 & 0 & 0 &1
\end{pmatrix} .
\endgroup
\label{eq:cy3cohom}
\end{align}
There are clearly only two numbers that characterize the basic topological structure, $h^{1,1}$ and $h^{2,1}$. However, there are in fact different $CY_3$-folds with identical Hodge numbers, so they are not sufficient to distinguish between different surfaces. Currently, the largest known Hodge numbers are on the order of 500\footnote{It is believed that the Hodge numbers of $CY_3$-folds are bounded, but this remains unproven}.

As an easy but nontrivial example, we consider the orbifold $T^6 / \mathbb{Z}_3$, where the $\mathbb{Z}_3$ acts as $(z_1,z_2,z_3) \rightarrow (\omega z_1,\omega z_2,\omega z_3)$ with $\omega^3 = 1$. Its cohomology is most easily worked out by again starting with that of $T^6$, which is simply given by combinations of $dz_i$ and $d\zbar_{\ibar}$ and their products. The $\mathbb{Z}_3$-invariant forms are $dz_1 \wedge dz_2 \wedge dz_3$, $d\zbar_1 \wedge d\zbar_2 \wedge d\zbar_3$, and all $(i,j)$-forms for $i = j$.

As in the previous example, the full cohomology receives contributions from the twisted sectors of the worldsheet orbifold CFT. For $\mathbb{Z}_3$, there is a single untwisted sector and two twisted sectors. The number of ground states in each twisted sector is given by the number of fixed points of the $\mathbb{Z}_3$ symmetry, which for a $T^6$ target space is 27. Thus, $h^{1,1}$ and $h^{2,2}$ are each increased by 27, yielding a total of 36 each. One interesting thing to note is that due to the orbifold projection, $h^{1,2} = 0$, which implies that there are no complex structure deformations. Roughly, this means that you can change the size of this manifold but not its shape. This is an example of whats known as a {\textit{rigid Calabi--Yau}}.  
For convenience, we list the Hodge numbers for both $T^6$ and $T^6/\mathbb{Z}_3$ below.

\begin{align}
\begingroup
\renewcommand*{\arraystretch}{1.1}
T^6 : \quad 
\begin{pmatrix}
1 & 3 & 3 & 1\\
3 & 9 & 9 & 3\\
3 & 9 & 9 & 3\\
1 & 3 & 3 & 1
\end{pmatrix}, \quad\quad
T^6/\mathbb{Z}_3: \quad 
\begin{pmatrix}
1 & 0 & 0 & 1\\
0 & 0 & 36 & 0\\
0 & 36 & 0 & 0\\
1 & 0 & 0 & 1
\end{pmatrix}.
\endgroup
\label{eq:t6cohom}
\end{align}

\subsection{Calabi--Yau manifolds from complex projective spaces} 
\label{sec:projectiveSpaces}

The orbifolds considered in the previous sections are but a small subset of the set of Calabi--Yau manifolds. As a non-orbifold example, consider the $n$-dimensional complex projective space, $\mathds{CP}^n$. Geometrically, the complex projective space is the set of complex lines in $\mathds{C}^{n+1}$ that pass through the origin. The most convenient description is in terms of the points $(z_1,\ldots,z_{n+1}) \in \mathds{C}^{n+1}$ identified under
\begin{equation} \label{eq:rescaling}
(z_1,\ldots,z_{n+1}) \sim \lambda (z_1,\ldots,z_{n+1}) , \quad \lambda \in \mathds{C}.
\end{equation}
There is an additional restriction that $0 \in \mathds{C}^{n+1}$ is not included. For instance, we can study $\mathds{CP}^1 \simeq S^2$. There are two coordinate patches which correspond to either $z^1 \neq 0$ or $z^2 \neq 0$ -- these are the patches on $S^2$ which cover the north and south poles, respectively. Using our knowledge of $S^2$, we observe that $\mathds{CP}^1$ does not admit a Ricci flat metric. More generally, $\mathds{CP}^n$ admits metrics with positive curvature; in particular, they do not admit Ricci flat metrics, which recall are an essential ingredient of any Calabi--Yau surface.

To construct the desired Ricci-flat manifold, we instead analyze hypersurfaces in $\mathds{CP}^n$ defined through a homogenous degree-$(n+1)$ polynomial equation $P_{n+1}(z_i) = 0$. The homogeneity property is a consistency condition which follows from \eqref{eq:rescaling}. For instance, we could consider the equation
\begin{equation} 
z_1^{n+1} + \cdots + z_{n+1}^{n+1} = 0. 
\label{eq:homogenouspolynomial} 
\end{equation}
The resulting hypersurface is a Calabi--Yau $(n-1)$-fold whose first Chern class is zero the $n=2$ case yields a $T^2$. As a special case, a $CY_2$-fold embedded in $\mathds{CP}^{3}$ can be defined by
\begin{equation} 
z_1^4 + z_2^4 + z_3^4 + z_4^4 = 0, 
\label{eq:torus} 
\end{equation} 
which is just a realization of $K3$. Similarly, the embedding of a $CY_3$-fold follows from
\begin{equation} 
z_1^5 + z_2^5 + z_3^5 + z_4^5  + z_5^5=0, 
\end{equation}
which is a 3-fold. This is known as a {\textit{quintic 3-fold}}. As an aside, quintic 3-folds have Hodge numbers 
\begin{align}
h^{2,1} = 101, \quad h^{1,1} = 1.
\end{align}
as well as an Euler characteristic given by
\begin{equation} 
\chi = 2\, (h^{1,1} - h^{1,2}) = -200. 
\end{equation}

More generally, we can obtain a $CY_{n-1}$-fold by choosing any degree-$n$ homogenous polynomial. For a 1-fold which is a $T^2$, there is only a single term $a z_1 z_2 z_3$ up to a change of coordinates. Here, $a$ is a complex modulus which parametrizes the complex structure of $T^2$. For $CY_2$, there are various terms we can add to \eqref{eq:torus}. For $CY_3$, there are 19 possible terms; this implies that such terms cannot completely capture the complex structure deformations of K3. 

\noindent {\bf Exercise 1: } By writing down allowed terms of degree five (and removing redundancies), show that for the quintic threefold, there are 101 independent complex deformations.

\subsection{Singularities of K3 surfaces}

Thus far, we have swept the issue of orbifold singularities under the rug. This was possible for the worldsheet because string theory is well defined on such geometries, even if the target space doesn't take the form of a bona fide manifold. Our first encounter with singular geometries was $T^4 / \mathbb{Z}_2$. Near each of its 16 fixed points, the space locally looks like $\mathds{C}^2 / \mathbb{Z}_2$. This follows from identifying two complex coordinates $z_1$ and $z_2$ under the $\mathbb{Z}_2$ reflection:
\begin{equation} (z_1,z_2) \sim (-z_1,-z_2). \end{equation}
It is instructive to define three new quantities
\begin{equation} 
u := z_1^2, \quad  v := z_2^2, \quad  w := z_1 z_2, 
\end{equation}
which are clearly $\mathbb{Z}_2$-invariant. There are two independent coordinates, with all three related by
\begin{equation}
u v = w^2. 
\label{eq:hypersurface} 
\end{equation}
There is thus a one-to-one correspondence between $\mathds{C}^2 / \mathbb{Z}_2$ and the hypersurface in $\mathds{C}^3$ defined by \eqref{eq:hypersurface}. This geometry becomes singular at the origin $z_1 = z_2 = 0$. We can resolve this singularity by modifying \eqref{eq:hypersurface} to
\begin{equation}\label{eq:resolvedsing} 
u v = (w - \sqrt{\mu})(w + \sqrt{\mu}) 
\end{equation}
for some parameter $\mu$. 

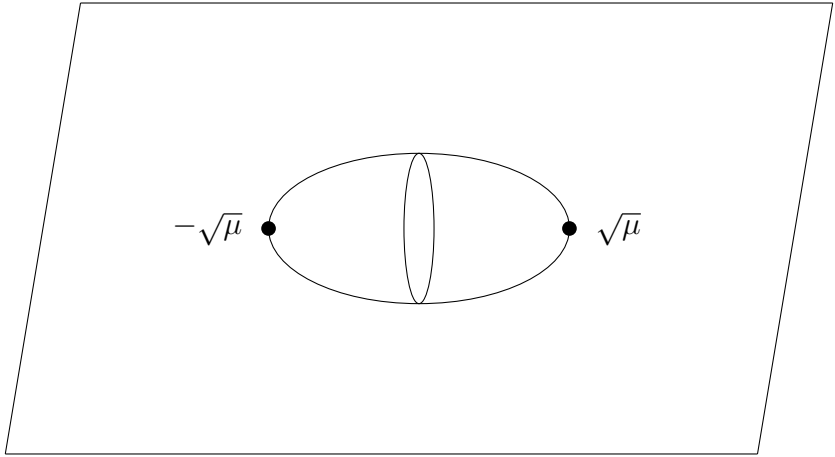
\begin{figure}[H]
\centering
\begin{tikzpicture}
\node[circle,thick,scale=0.5,fill=black,label={[label distance=1mm]west:$-\sqrt{\mu}$}] (A1) at (-2,0) {};
\node[circle,thick,scale=0.5,fill=black,label={[label distance=1mm]east:$\sqrt{\mu}$}] (A2) at (2,0) {};
\draw (-4.5,3)--(5.5,3);
\draw (-5.5,-3)--(4.5,-3);
\draw (-4.5,3)--(-5.5,-3);
\draw (5.5,3)--(4.5,-3);
\draw (0,0) ellipse (2cm and 1cm);
\draw (0,0) ellipse (0.2cm and 1cm);
\end{tikzpicture}
\caption{A plot of the $w$ plane, with an $S^2$ at the origin. The $S^1$ clearly shrinks to zero at the points $w = \pm \sqrt{\mu}$, which in effect defines the $S^2$.}
\label{fig:sphereinwplane}
\end{figure}

Resolving the singularity leads to the emergence of an $S^2$ at the origin of $\mathds{C}^2/\mathbb{Z}_2-\{0\}$. Note that the left-hand side of \eqref{eq:resolvedsing} is invariant under the phase rotation $v \rightarrow e^{-i \theta} v, u \rightarrow e^{i \theta} u$. This defines an $S^1$ which shrinks as $w$ approaches $\pm \sqrt{\mu}$. The emergent $S^2$, with its $S^1$ subspace, is depicted graphically in \ref{fig:sphereinwplane}. Let's return our attention to the $T^4/\mathbb{Z}_2$ example. Each of the 16 singularities is replaced by an $S^2$, which each contribute $+1$ to $h^{1,1}$. This is exactly what we found in Section \ref{sec:k3} when considering the twisted sectors on the string worldsheet. 

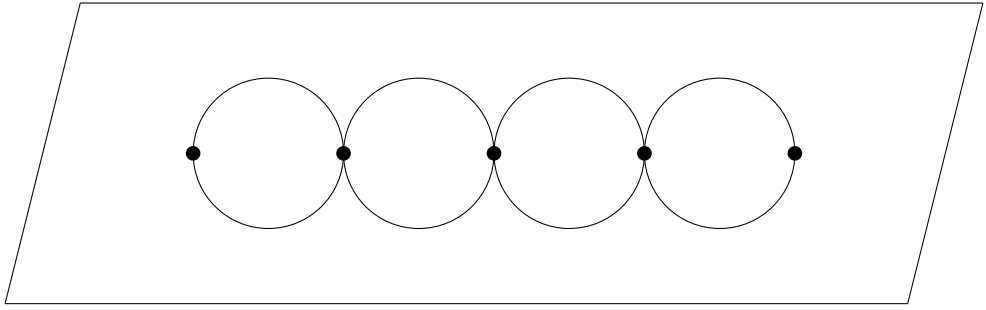
\begin{figure}[H]
\centering
\begin{tikzpicture}
\node[circle,thick,scale=0.5,fill=black,label={[label distance=1mm]west:$$}] (A1) at (-3,0) {};
\node[circle,thick,scale=0.5,fill=black,label={[label distance=1mm]east:$$}] (A2) at (-1,0) {};
\node[circle,thick,scale=0.5,fill=black,label={[label distance=1mm]west:$$}] (A3) at (1,0) {};
\node[circle,thick,scale=0.5,fill=black,label={[label distance=1mm]east:$$}] (A4) at (3,0) {};
\node[circle,thick,scale=0.5,fill=black,label={[label distance=1mm]east:$$}] (A5) at (5,0) {};
\draw (-4.5,2)--(7.5,2);
\draw (-5.5,-2)--(6.5,-2);
\draw (-4.5,2)--(-5.5,-2);
\draw (7.5,2)--(6.5,-2);
\draw (-2,0) circle (1);
\draw (0,0) circle (1);
\draw (2,0) circle (1);
\draw (4,0) circle (1);
\end{tikzpicture}
\caption{Resolution of $\mathds{C}^2 / \mathbb{Z}_n$ singularity. Here, each singular point is replaced by a copy of $\mathds{CP}^1$ of nonzero size.}
\label{fig:spheresinwplane}
\end{figure}

We can generalize the above example by instead considering a $\mathds{C}^2/\mathbb{Z}_n$ orbifold singularity. The $\mathbb{Z}_n$ symmetry acts on $z_1,z_2$ as
\begin{equation} 
(z_1,z_2) \rightarrow (\omega z_1, \omega^{-1} z_2), \quad \omega^n = 1.
\end{equation}
The $\mathbb{Z}_n$ is a discrete subgroup of $SU(2)$, which is also true for the holonomy group. In analogy with the $\mathbb{Z}_2$ case, we consider a $\mathds{C}^3$ space parametrized by
\begin{equation} 
u := z_1^n, \quad v := z_2^n, \quad w := z_1 z_2. 
\end{equation}
The embedding of $\mathds{C}^2/\mathbb{Z}_n$ in $\mathds{C}^3$ is specified by the constraint
\begin{equation} 
u v = w^n. 
\end{equation}
Under the appropriate change of variables, this equation can be rewritten as
\begin{equation}
  u^2 + v^2 = w^n \,,
\end{equation}
The singularity at $w = 0$ can be resolved in a similar manner at the previous case, replacing $w^n$ with a product:
\begin{equation} 
  u^2 + v^2 = \prod_{i = 1}^n (w - \alpha_i) \,. 
 \label{eq:c2modznresolved} 
\end{equation}
If two of the $\alpha_i$ parameters are brought close together, then equation \eqref{eq:c2modznresolved} describes the $\mathds{C}^2 / \mathbb{Z}_2$singularity in a neighborhood. So, equation \eqref{eq:c2modznresolved} smooths out the $\mathds{C}^2 / \mathbb{Z}_n$ by introducing $(n-1)$ $\mathds{P}^1$'s, as shown in Figure  \ref{fig:spheresinwplane}. 

\noindent {\bf Optional Exercise: } Show that the hypersurface defined by equation \eqref{eq:hypersurface} is topologically the same as the cotangent bundle of $\mathds{P}^1$, denoted by $T^* \mathds{P}^1$.

The $(n-1)$ $\mathds{P}^1$s, which we refer to as $C_i$ for $i = 1,\cdots,n-1$, are a basis for the homology of the deformed space. Their intersection numbers can be organized into the following matrix (where empty entries are taken to be zero):
\begin{equation}
\renewcommand*{\arraystretch}{1.1}
C_i \cdot C_j = \left(\begin{array}{cccc}
-2 & 1 &  &  \\ 
1 & -2 & 1 &  \\ 
& 1 & \ddots & 1 \\ 
&  & 1 & -2
\end{array} \right)_{ij} \,.
\end{equation}
If we represent each sphere by a node and each intersection between two $\mathds{P}^1$s by a line between such nodes, then Figure \ref{fig:spheresinwplane} looks like the $A_{n-1}$ Dynkin diagram. In fact, the matrix of intersection numbers is just the Cartan matrix (multiplied by $-1$) associated to the $A_{n-1}$ Lie algebra. 

The resolution of the $\mathds{C}^2 / \mathbb{Z}_n$ singularity thus appears to be connected to the $A_{n-1}$ Lie algebra. This is a particular example of a more general phenomenon where resolved singularities are defined by discrete subgroups of $SU(2)$. Generically, these discrete subgroups fall into several infinite families, $A_n$ and $D_n$, as well as a few exceptional subgroups $E_6, E_7, E_8$. Similar to the $A_{n-1}$ example for $\mathds{C}^2/\mathbb{Z}_n$, in the general case the associated Dynkin diagrams indicate how the deformed singularity behaves. 

We may think of the deformations of equation \eqref{eq:c2modznresolved} as complex structure deformations, although the distinction between deformations of the complex structure and the K\"ahler structure is ambiguous for K3 due to its  hyperK\"ahler structure.

Let us return to the example of $\mathds{C}^2 / \mathbb{Z}_2$ and its connection to stringy geometry \cite{Dixon:1985jw}. The orbifold point geometrically corresponds to where the area of the $\mathds{P}^1$ shrinks to zero. As mentioned previously, the CFT with an orbifold target space is well-behaved, and so too is this limit in the CFT moduli space. Note that we could have turned on a $B$ field, which in certain units is only defined modulo $2 \pi$. If the area of the $\mathds{P}^1$ is $A$, then $A + i B$ forms a complexified K\"ahler parameter. We should then upgrade $\mu$ in equation \eqref{eq:resolvedsing} to a complex parameter and identify 
\begin{equation}
  \mu = A + i B \,. 
\end{equation}
It turns out that the orbifold point corresponds to $A = 0$ and $B = \pi$. In one of the previous exercises, we learned that the $\mathbb{Z}_2$ orbifold CFT has a $\tilde{\mathbb{Z}}_2$ symmetry. This symmetry only exists on the worldsheet when $B$ takes the values of $0$ or $\pi$. The $\tilde{\mathbb{Z}}_2$ symmetry then acts as $B \rightarrow - B$, which is compatible with the definition of the $B$-field. When $B = 0$, the path integral of the worldsheet wrapping the $\mathds{P}^1$ behaves like
\begin{equation}
  \sum_{n > 0} e^{-n A} \xrightarrow{A \rightarrow 0} \infty \,.
\end{equation}
 When $B = \pi$, the path integral instead behaves like
 \begin{equation}
   \sum_{n > 0} e^{-n A + i \pi n} < \infty \,,
 \end{equation}
so the conformal theory is well-defined. The orbifold limit corresponds to $B=\pi$.
 
The resolution of orbifold singularities also has physical implications for the resulting string theory. Only certain kinds of singularities can appear in the moduli space of a compact K3 surface. This comes from the fact that the second homology of K3 is given by 
\begin{equation}
  h^{0,2} + h^{1,1} + h^{2,0} = 1 + 20 + 1 = 22 \,.
\end{equation}
It follows that it is not possible to have an arbitrarily large number of $\mathds{P}^1$s which arise from deforming singularities. Recall that we can only compactify string theory on a $T^4$ or a K3 surface if we want to preserve some supersymmetry. The types of physical theories which result are thus constrained by the aforementioned bound on the number of $\mathds{P}^1$s. 

\subsection{Singularities of Calabi--Yau threefolds}

There are two methods to deal with singularities: {\textit{deformations}} and {\textit{resolutions}}.\footnote{For rigorous definitions of deformations and desingularizations, as well as more modern ways of utilizing them for Calabi--Yau threefold compactifications, see \cite{Esole:2014dea,Esole:2017kyr,Esole:2017rgz,Esole:2017qeh,Esole:2017hlw,Esole:2018csl,Esole:2018mqb,Esole:2019asj}.} 

We first consider deformations, which amounts to giving the singular $S^3$ a nonzero size. The 3-dimensional analog of equation \eqref{eq:resolvedsing} is
\begin{equation}
\tilde{z}_1^2 + \tilde{z}_2^2 + \tilde{z}_3^2 + \tilde{z}_4^2 = \mu,
\label{eq:3folddef}
\end{equation}
which describes $T^* S^3$, where the $S^3$ has size $\mu$. The geometry is singular for the case where the size of the $S^3$ vanishes. One can explicitly write down a Ricci flat K\"ahler metric, and thus the total space of the cotangent bundle is a non-compact Calabi--Yau threefold. The singularity appears at $\mu = 0$. Note that this singularity is \emph{not} an orbifold singularity, since it does not take the local form $\mathds{C}^3/G$ for some discrete isometry group $G$. We can deform it by considering $\mu \neq 0$, which in particular corresponds to a complex structure deformation. 
 
We now consider resolutions, or giving the singular $S^2 \simeq \mathds{P}^1$ a nonzero size. By a linear change of coordinates, we can rewrite equation \eqref{eq:3folddef} (with $\mu = 0$) as
\begin{equation}\label{eq:detformula}
\renewcommand*{\arraystretch}{1.1}
\det \left(\begin{array}{cc}
 z_1 & z_3 \\ 
 z_4 & z_2
\end{array} \right) =   z_1 z_2 - z_3 z_4 = 0 \,.
\end{equation}
Let $v = (\alpha,  \beta)^T$ be a nonzero vector annihilated by this matrix. 
be a nonzero vector that is annihilated by the above matrix. Because its overall normalization is irrelevant, we can think of $v$ as an element of $\mathds{P}^1$. In a coordinate patch $z$ where $\beta \neq 0$, we can write
\begin{equation}
z := \frac{\alpha}{\beta} \,.
\end{equation}
We may parameterize the singular manifold of \eqref{eq:detformula} by the variables $z_1,z_4,z$. The singularity is located at $z_1 = z_4 = 0$, where the $\mathds{P}^1$ parameterized by $z$ shrinks to zero. Resolving (or blowing-up) the singularity  amounts to giving $\mathds{P}^1$ a nonzero area, so this is an example of a K\"ahler deformation. The geometry of the resolved singularity is a sphere $\mathds{P}^1$ parameterized by $z$; the directions $z_1,z_4$ are normal to th sphere. Formally, we say that the entire geometry is an $O(-1) \oplus O(-1)$ line bundle over $\mathds{P}^1$. The $O(-1)$ notation means that the number of zeros minus the number of poles of any holomorphic section of the line bundle is minus one.

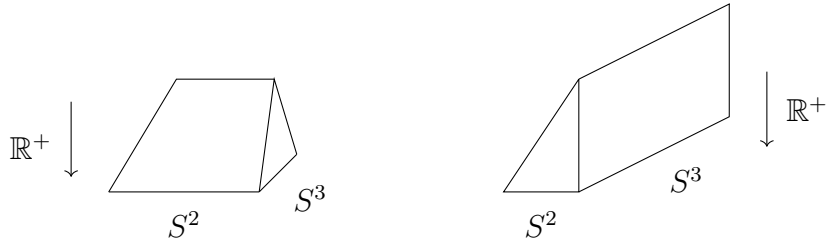
\begin{figure}[H]
\centering
\begin{tikzpicture}
\node[draw=none,opacity=0,thick,scale=0.1,fill=black,label={[label distance=1mm]south:$S^2$}] (A1) at (0,0) {};
\node[draw=none,opacity=0,thick,scale=0.1,fill=black,label={[label distance=0.5mm]-20:$S^3$}] (A2) at (1.25,0.25) {};
\node[draw=none,opacity=0,thick,scale=0.1,fill=black,label={[label distance=1mm]west:$\mathbb{R}^+$}] (A2) at (-1.5,0.6) {};
\draw (-1,0)--(1,0);
\draw (1,0)--(1.5,0.5);
\draw (1,0)--(1.2,1.5)--(1.5,0.5);
\draw (1.2,1.5)--(-0.1,1.5)--(-1,0);
\draw[->] (-1.5,1.2)--(-1.5,0.2);
\end{tikzpicture}
\quad\quad\quad\quad\quad
\begin{tikzpicture}
\node[draw=none,opacity=0,thick,scale=0.1,fill=black,label={[label distance=1mm]south:$S^2$}] (A1) at (0,0) {};
\node[draw=none,opacity=0,thick,scale=0.1,fill=black,label={[label distance=0.5mm]-20:$S^3$}] (A2) at (1.5,0.5) {};
\node[draw=none,opacity=0,thick,scale=0.1,fill=black,label={[label distance=1mm]east:$\mathbb{R}^+$}] (A2) at (3,1.1) {};
\draw (0.5,1.5)--(-0.5,0)--(0.5,0)--(0.5,1.5);
\draw (0.5,0)--(2.5,1)--(2.5,2.5)--(0.5,1.5);
\draw[->] (3,1.6)--(3,0.6);
\end{tikzpicture}
\caption{The left diagram corresponds to the (resolved) blown-up singularity, while the right diagram corresponds to the deformed singularity.}
\label{fig:conifold}
\end{figure}

In summary, we have shown two different ways of smoothing out the same singularity. The deformation method gives an $S^3$ nonzero size, while the blow-up/resolution method gives an $S^2$ nonzero size. As mentioned previously, these procedures
correspond to complex structure and K\"ahler deformations, respectively. To see how they are related, note that $T^* S^3$ has topology $S^3 \times \mathbb{R}^3$, or $S^3 \times S^2 \times \mathbb{R}^+$. The $S^2$ is
contractible, while the $S^3$ has nonzero size. The singularity is restored by letting the $S^3$ shrink as $\mathbb{R}^+$ goes to zero. Likewise, the line bundle considered above locally looks like $S^2 \times \mathbb{R}^4$, or $S^2 \times S^3 \times \mathbb{R}^+$. In this case, the singularity is restored by letting $S^2$ shrink as $\mathbb{R}^+$ goes to zero. The relation between these two methods of smoothing a singular geometry is known as the {\bf conifold transition}, as illustrated in Figure \ref{fig:conifold}.

String perturbation theory breaks down near the singular point, since the size of the sphere becomes small relative to the string scale.

Calabi--Yau manifolds with different Hodge numbers can be related through transitions like the conifold transition. It is conjectured that the number of Calabi--Yau threefolds is finite. For known examples of Calabi--Yau manifolds, the largest Hodge numbers are on the order of  $h^{1,1} + h^{2,1} \sim 500$. Thus, conjecturally there is a bound on the number of massless fields that can arise from supersymmetric string compactifications on complex threefolds. 

\subsection{Toric geometry}

Toric geometry is the study of algebraic varieties which are equipped with an embedded algebraic torus $(\mathds{C}^*)^p$, such that the group action of the torus extends to the entire variety. Toric spaces are a particularly tractable example of more general spaces as their topological and geometric data can be understood through combinatorics. Many familiar spaces in physics are in fact toric spaces, as we shall see.

\begin{figure}[H]
\centering
\scalebox{1.2}{
\begin{tikzpicture}
\node[circle,thick,scale=0.5,fill=black,label={[label distance=1mm]south:$0$}] (A1) at (0,0) {};
\node[draw=none,opacity=0,thick,scale=0.1,fill=black,label={[label distance=1mm]south:$|z|^2$}] (A2) at (4,0) {};
\draw[->] (A1)--(A2);
\draw (A1)--(4,1);
\draw (0.8,0.1) ellipse (0.05cm and 0.1cm);
\draw (1.6,0.2) ellipse (0.1cm and 0.2cm);
\draw (2.4,0.3) ellipse (0.15cm and 0.3cm);
\draw (3.2,0.4) ellipse (0.18cm and 0.4cm);
\end{tikzpicture}
}
\caption{Toric diagram of $\mathds{C}$. The circle, parametrized by the angle $\theta$, shrinks to zero when $|z|^2\rightarrow 0$.}
\label{fig:toricc}
\end{figure}
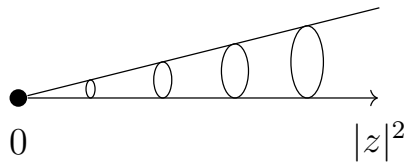

A toric diagram is a representation of a toric space, in particular a Calabi--Yau space, as a $T^p$ fibration over some base, $B$. The cycles of the torus fiber degenerate over the boundaries of $B$. One of the simplest examples of a toric space is the complex plane, $\mathds{C}$. By parametrizing $\mathds{C}$ by polar coordinates, $|z|^2$ and $\theta$, we see that $\mathds{C}$ can be viewed as an $S^1$ fibered over the positive real line, where the $S^1$ collapses at the origin, as depicted in Figure \ref{fig:toricc}. The toric diagram describing this 1 complex dimensional toric space is then just the semi-infinite line.

There is a natural symplectic form on this space. In terms of polar coordinates, it takes the form
\begin{equation}
	dz \wedge d\overline{z} = d(|z|^2) \wedge d \theta \,.
\end{equation}
Symplectic manifolds (\emph{i.e.} smooth manifolds equipped with a symplectic form) form the phase space of physical systems. Indeed, the 6d phase space of a particle moving in 3d is a Calabi--Yau threefold: locally the momentum and position are given by 
the vertical and horizontal coordinates on the cotangent bundle $T^*M$ with $M = S^3$, for example. In this way 
a Calabi--Yau space can be viewed as a compact version of the phase space where the symplectic form is the K\"ahler form.

Next, we consider $\mathds{C}^2$. Similarly to the case of $\mathds{C}$, switching to polar coordinates ($|z_i|^2, \theta_i$ for $i =1,2$) makes the fiber structure manifest. Collecting the angles and distances separately, we observe that $\mathds{C}^2$ can be realized as a $T^2$ fibration over the closed positive quadrant of $\mathbb{R}^2$. The $A$ and $B$ cycles of the torus vanish, respectively, on the $x = |z_1|^2$ and $y = |z_2|^2$ axes; both degenerate to a point at the origin. Note that the space is smooth, as expected. The toric diagram for $\mathds{C}^2$, with special degenerating points highlighted, is depicted in Figure \ref{fig:toricc2}. The generalization of this procedure to $\mathds{C}^n$ is straightforward.

\begin{figure}[H]
\centering
\begin{tikzpicture}
\node[draw=none,opacity=0,thick,scale=0.1,fill=black,label={[label distance=0.5mm]225:$0$}] (A1) at (0,0) {};
\node[draw=none,opacity=0,thick,scale=0.1,fill=black,label={[label distance=1mm]south:$|z_2|^2$}] (A2) at (4.5,0) {};
\node[draw=none,opacity=0,thick,scale=0.1,fill=black,label={[label distance=1mm]west:$|z_1|^2$}] (A3) at (0,4.5) {};
\node[circle,thick,scale=0.4,fill=black,label={[label distance=9mm]18:$(\theta_1,\theta_2)$}] (A4) at (3,3) {};
\draw[->] (A1)--(A2);
\draw[->] (A1)--(A3);
\draw (0.6,3) ellipse (0.6cm and 0.1cm);
\draw (3,0.6) ellipse (0.1cm and 0.6cm);
\draw (3.35,3.6) ellipse (0.5cm and 0.8cm);
\draw (3.35,3.9) arc (150:210:0.7);
\draw (3.3,3.25) arc (-40:40:0.4);
\end{tikzpicture}
\caption{Toric geometry of $\mathds{C}^2$.}
\label{fig:toricc2}
\end{figure}

By just drawing the base of the torus fibration, which is exactly the toric diagram, we see that it is possible to visualize an $n$-dimensional complex toric space in terms of its lower-dimensional base.

\begin{figure}[H]
\centering
\scalebox{0.8}{
\begin{tikzpicture}
\node[circle,thick,scale=0.4,fill=black,label={[label distance=4.2cm]north:$S^2$}] (A1) at (4,0) {};
\draw (4,0) circle (3.7);
\draw[->] (0.8,0) ellipse (0.05cm and 1.8cm);
\draw[->]  (1.6,0) ellipse (0.1cm and 2.8cm);
\draw[->]  (2.4,0) ellipse (0.13cm and 3.3cm);
\draw[->]  (3.2,0) ellipse (0.15cm and 3.6cm);
\draw[->]  (4,0) ellipse (0.17cm and 3.7cm);
\draw[->]  (4.8,0) ellipse (0.15cm and 3.6cm);
\draw[->]  (5.6,0) ellipse (0.13cm and 3.3cm);
\draw[->]  (6.4,0) ellipse (0.1cm and 2.8cm);
\draw[->]  (7.2,0) ellipse (0.05cm and 1.8cm);
\draw (4,0) ellipse (3.7cm and 0.5cm);
\end{tikzpicture}
}\\ \scalebox{0.8}{
\begin{tikzpicture}
\node[draw=none,opacity=0,thick,scale=0.1,fill=black,label={[label distance=1mm]south:$0$}] (A1) at (0.2,0) {};
\node[draw=none,opacity=0,thick,scale=0.1,fill=black,label={[label distance=1mm]south:$S^1$}] (A2) at (3.9,0) {};
\draw (A1)--(7.9,0);
\draw (0.8,0.2) ellipse (0.05cm and 0.21cm);
\draw (1.6,0.46) ellipse (0.1cm and 0.44cm);
\draw (2.4,0.59) ellipse (0.15cm and 0.59cm);
\draw (3.2,0.65) ellipse (0.18cm and 0.66cm);
\draw (4,0.7) ellipse (0.2cm and 0.7cm);
\draw (4.8,0.66) ellipse (0.18cm and 0.66cm);
\draw (5.6,0.61) ellipse (0.15cm and 0.61cm);
\draw (6.4,0.49) ellipse (0.1cm and 0.46cm);
\draw (7.2,0.25) ellipse (0.05cm and 0.25cm);
\draw (7.9,0) arc (50:130:6);
\end{tikzpicture}
}
\caption{An $S^2$ realized as an $S^1$ over the interval, where the circle shrinks at the ends of the interval.}
\label{fig:toricsphere}
\end{figure}
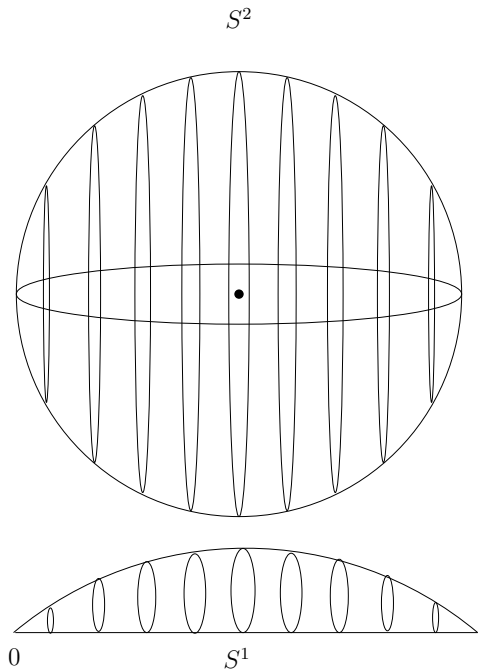

Although the above examples described toric diagrams for (non-compact) Calabi--Yau manifolds, we can also use toric diagrams to depict more generic manifolds. For example, toric geometry says that a sphere is just an interval; more specifically, an $S^2$ can be realized as an $S^1$ fibration over the interval, where the fiber collapses at the boundaries. The toric diagram for $S^2$ is depicted in Figure \ref{fig:toricsphere}.

Next we consider two-dimensional complex projective space $\mathds{CP}^2$, parametrized by coordinates $z_1$, $z_2$, and $z_3$ identified by
\begin{equation}
 (z_1, z_2, z_3) \sim \lambda (z_1, z_2, z_3) \,, \quad \lambda \in \mathds{C}^*.
\end{equation}
There are three phases, but only two are independent due to an allowed rescaling. Thus, this geometry contains two circles, which can be represented with a 2d diagram. A general point in the interior is associated with a smooth $T^2$ fiber. There is a line $\mathds{CP}^1$ where each of the three coordinates $z_1,z_2,z_3$ individually go to zero. When any two coordinates vanish, we are left with a single point. Any two such lines must meet at a corner, where both of the associated coordinates vanish. The full diagram of a $\mathds{C}\mathds{P}^2$ is given in Figure \ref{fig:toricp2}.

In order for the geometry to be smooth everywhere, it is crucial that each corner of Figure \ref{fig:toricp2} looks like $\mathds{C}^2$. That is, the two shrinking circles must form a basis for the homology of the two-dimensional torus. Thus, they must intersect once. Suppose we had a geometry where the two shrinking cycles are given by (1,0) and (1,2)\footnote{A cycle $(n_A,n_B)$ of the torus $T^2$ is one which wraps the A-cycle $n_A$ times and the B-cycle $n_B$ times}. In this case, there is a corner singularity that we can blow-up by inserting a $\mathds{P}^1$; this yields a geometry which locally looks like a toric diagram. This blow-up process is depicted in Figure \ref{fig:toricblowingup}. For a $\mathbb{Z}_n$ orbifold singularity, one applies this blow-up procedure $(n-1)$ times to obtain $(n-1)$ $\mathds{P}^1$s.

\begin{figure}[H]
\centering
\begin{tikzpicture}
\node[circle,thick,scale=0.5,fill=black,label={[label distance=4mm]south:locally $\mathds{C}^2$}] (A1) at (0,0) {};
\node[draw=none,opacity=0,thick,scale=0.1,fill=black,label={[label distance=1mm]west:$z_1=0$}] (A2) at (0,2) {};
\node[draw=none,opacity=0,thick,scale=0.1,fill=black,label={[label distance=1mm]south:$z_2=0$}] (A3) at (2,0) {};
\node[draw=none,opacity=0,thick,scale=0.1,fill=black,label={[label distance=1mm]east:$z_3=0$}] (A4) at (2,2) {};
\node[circle,thick,scale=0.5,fill=black,label={[label distance=1mm]south:$T^2$}] (A5) at (1.2,1.2) {};
\draw (4,0)--(A1)--(0,4)--(4,0);
\draw (0,0) circle (0.5);
\end{tikzpicture}
\caption{Toric diagram of $\mathds{C}\mathds{P}^2$. Each edge corresponds to a $\mathds{C}\mathds{P}^1$.}
\label{fig:toricp2}
\end{figure}
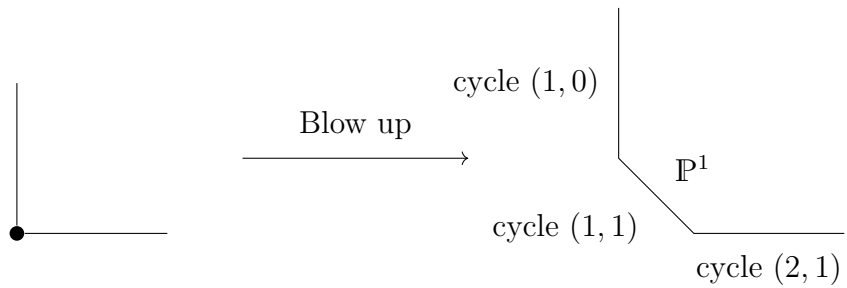
\begin{figure}[H]
\centering
\begin{tikzpicture}
\node[circle,thick,scale=0.5,fill=black] (A1) at (-3,0) {};
\node[draw=none,opacity=0,thick,scale=0.1,fill=black,label={[label distance=1mm]north:Blow up}] (A2) at (1.5,1) {};
\draw (-1,0)--(A1)--(-3,2);
\draw[->] (0,1)--(3,1);
\node[draw=none,opacity=0,thick,scale=0.1,fill=black,label={[label distance=1mm]west:cycle $(1,0)$}] (A3) at (5,2) {};
\node[draw=none,opacity=0,thick,scale=0.1,fill=black,label={[label distance=1mm]220:cycle $(1,1)$}] (A4) at (5.5,0.5) {};
\node[draw=none,opacity=0,thick,scale=0.1,fill=black,label={[label distance=1mm]30:$\mathds{P}^1$}] (A5) at (5.5,0.5) {};
\node[draw=none,opacity=0,thick,scale=0.1,fill=black,label={[label distance=1mm]south:cycle $(2,1)$}] (A6) at (7,0) {};
\draw (5,3)--(5,1)--(6,0)--(8,0);
\end{tikzpicture}
\caption{Toric representation of blowing up a $\mathds{C}^2 / \mathbb{Z}_2$ singularity to obtain a $\mathds{P}^1$. After the blowup, the intersection number of the cycles at each corner is 1, so there are no singularities. This blowup resolves $A_1$ singularity.}
\label{fig:toricblowingup}
\end{figure}

We can also represent $S^3$ via a toric action, despite the fact that $S^3$ is not a toric, or even a complex, space. Given two complex coordinates $z_1,z_2$, $S^3$ is defined to be the locus
\begin{equation} 
  |z_1|^2 + |z_2|^2 = 1 \,.
\end{equation}
The coordinates $|z_1|$ and $|z_2|$ take values on the interval $[0,1]$. At each point on the interval for $|z_1|$, there is an associated $T^2$. At $|z_1| = 0$ one circle vanishes, whereas at $|z_1| = 1$ the other circle vanishes. So we can think of a shrinking circle as filling in one of the cycles of the torus. Combining the two coordinates together, we land at the fact that $S^3$ is essentially two solid tori glued together. 

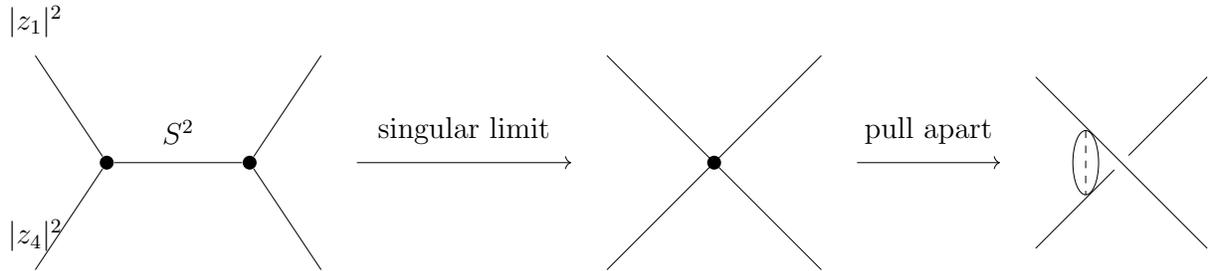
\begin{figure}[H]
\centering
\scalebox{0.95}{
\begin{tikzpicture}
\node[circle,thick,scale=0.5,fill=black] (A1) at (-4,0) {};
\node[circle,thick,scale=0.5,fill=black] (A2) at (-2,0) {};
\node[draw=none,opacity=0,thick,scale=0.1,fill=black,label={[label distance=1mm]north:$|z_1|^2$}] (A3) at (-5,1.5) {};
\node[draw=none,opacity=0,thick,scale=0.1,fill=black,label={[label distance=1mm]north:$|z_4|^2$}] (A4) at (-5,-1.5) {};
\node[draw=none,opacity=0,thick,scale=0.1,fill=black,label={[label distance=1mm]north:$S^2$}] (A5) at (-3,0) {};
\draw (-1,1.5)--(A2)--(-1,-1.5);
\draw (A1)--(A2);
\draw (-5,1.5)--(A1)--(-5,-1.5);
\draw[->] (-0.5,0)--(2.5,0);
\node[draw=none,opacity=0,thick,scale=0.1,fill=black,label={[label distance=1mm]north:singular limit}] (A6) at (1,0) {};
\node[circle,thick,scale=0.5,fill=black] (A7) at (4.5,0) {};
\draw (3,1.5)--(6,-1.5);
\draw (3,-1.5)--(6,1.5);
\draw[->] (6.5,0)--(8.5,0);
\node[draw=none,opacity=0,thick,scale=0.1,fill=black,label={[label distance=1mm]north:pull apart}] (A8) at (7.5,0) {};
\draw (9,1.2)--(11.4,-1.2);
\draw (9,-1.2)--(10.1,-0.1);
\draw (10.3,0.1)--(11.4,1.2);
\draw[dashed] (9.7,0.45)--(9.7,-0.45);
\draw (9.7,0) ellipse (1.8mm and 4.5mm);
\end{tikzpicture}
}
\caption{Resolving and deforming a singular conifold.}
\label{fig:singconifold}
\end{figure}

We next turn to studying the conifold using the toric language. We can depict the resolution and deformation of a singular conifold in Figure \ref{fig:singconifold}. The intersection points in the toric diagram of the singular conifold signal the location of singularities, as can be determined from the degenerating cycles of the torus fibers. The resolution of a singular point, as explained above, involves the creation of a $\mathds{P}^1$, which effectively replaces the singularity. This is represented by the central interval on the left-hand-side of Figure \ref{fig:singconifold}. The right-hand side of the figure represents the deformed conifold, which recall involves replacing the singularity with an $S^3$ of nonzero size. The deformation involves pulling the lines of the singular diagram apart, which then has the structure of a $T^2$ fibered over an interval; the degenerations at the endpoints are just an $S^3$. In this way the two different methods for smoothing the conifold singularity can be understood from the point of view of toric diagrams.

\begin{figure}[H]
\centering
\scalebox{1}{
\begin{tikzpicture}
\node[circle,thick,scale=0.5,fill=black] (A1) at (-0.5,0) {};
\draw (2.5,0)--(A1)--(-0.5,3);
\draw (A1)--(-2.3,-1.8);
\draw[->] (3,0)--(5,0);
\node[draw=none,opacity=0,thick,scale=0.1,fill=black,label={[label distance=1mm]north:Blow up}] (A2) at (4,0) {};
\draw (7.5,3)--(7.5,0.5)--(8,0)--(10.5,0);
\draw (7.5,0.5)--(7.2,-0.3)--(5.3,-1.8);
\draw (8,0)--(7.2,-0.3);
\node[draw=none,opacity=0,thick,scale=0.1,fill=black,label={[label distance=3mm]45:$\mathds{P}^2$}] (A3) at (7.5,0) {};
\end{tikzpicture}
}
\caption{Toric diagram of resolving the $\mathds{C}^3 / \mathbb{Z}_3$ singularity to obtain $\mathds{CP}^2$.}
\label{fig:chopcorner}
\end{figure}
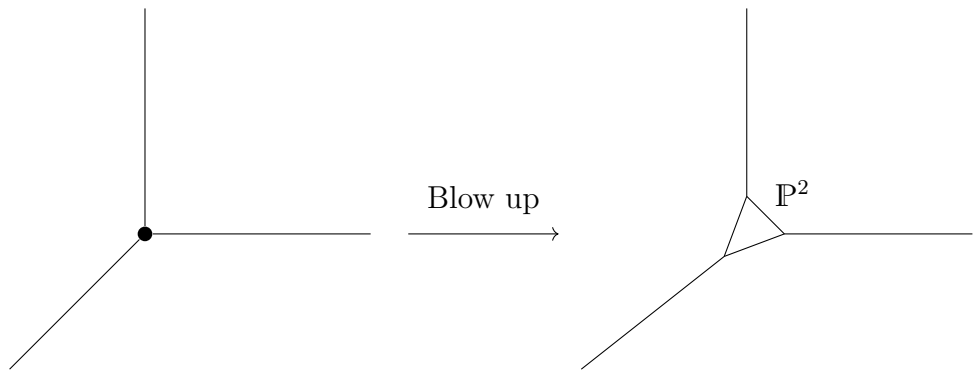

Lastly, we discuss the $\mathds{C}^3 / \mathbb{Z}^3$ singularity, where $(z_1,z_2,z_3)$ is identified with  $\omega (z_1,z_2,z_3)$ for $\omega^3 = 1$. This singularity can only be smoothed with K\"ahler deformations, not complex deformations. This is because $h^{2,1} = 0$ for $\mathds{C}^3 / \mathbb{Z}_3$ as discussed before. The toric diagram is three-dimensional and the singularity is located in the corner. The blown-up singularity looks like chopping off a corner and replacing it with a triangle, which is $\mathds{CP}^2$. See Figure
\ref{fig:chopcorner}.
 
\section{Sigma models}

\subsection{Supersymmetric sigma models and mirror symmetry}

Thus far, we have talked about the geometric structures that appear in superstring theory from the spacetime perspective. To describe such geometries via the worldsheet, we consider an ${\cal N}=(1,1)$ $\sigma$-model with target space $M$, \emph{i.e.} the supersymmetric version of \eqref{eq:NLSM}, whose action is given by \cite{Callan:1989nz,Alvarez-Gaume:1981exv}\footnote{In \eqref{eq:SUSY-NLSM} and the equations which follow, purely anti-holomorphic operators will typically be adorned with an overline to distinguish them from their holomorphic counterparts. In particular we reserve $^*$ for complex conjugation.}
\begin{equation}\label{eq:SUSY-NLSM}
S = \frac{1}{4\pi} \int d^2 z (G_{\mu \nu} + B_{\mu\nu}) \p X^\mu {\bar \p} X^\nu + G_{\mu\nu}\psi^\mu  \left( \psi^\nu \nabla_{\zbar} \psi^\nu + \bar{\psi}^\mu \nabla_z \bar{\psi}^\nu \right) + \frac12 R_{\mu\nu\sigma\rho} \psi^\mu \psi^\nu \bar{\psi}^\rho \bar{\psi}^\sigma .
\end{equation}
Here, $R_{\mu\nu\sigma\rho}(X)$ is the Riemann curvature of $M$ and $\nabla_a$ is the spin covariant derivative, which acts on the fermions as
\begin{equation}
\nabla_z \psi^\mu = \p \psi^\mu + \left(\Gamma^\mu_{\nu\sigma} + \frac12 H^\mu_{\nu\sigma}\right) \p X^\nu \psi^\sigma .
\end{equation}

As a trivial example, one may choose the target space to be $\mathds{C}^n$ with zero $H$-flux, which yields a theory of free bosons and free fermions. This is just the standard worldsheet action of the type II superstring in flat Minkowski space. Its global symmetry group includes a U(1) factor with current
\begin{equation} 
J = g_{i \jbar} \psi^i \psi^{\jbar}.
\end{equation}
Under this U(1) symmetry, $\psi^i$ has charge $+1$, while $\psi^{\ibar}$ has charge $-1$. Even if we take the the target space be a general Calabi--Yau manifold, the worldsheet SCFT preserves this symmetry because the U(1) holonomy of the manifold is trivial. That is, the U(1) piece of the spin connection has no curvature, so the associated current $J$ still exists for a generic choice of Calabi--Yau manifold. In fact, there are independent U(1) currents associated with left-movers and right-movers:\footnote{The minus sign is a convention. See for instance the bottom of page 385 in \cite{Polchinski:1998rr}.}
\begin{equation} J = g_{i \jbar} \psi^i \psi^{\jbar}, \quad
\bar{J} = -g_{i \jbar} \bar{\psi}^i \bar{\psi}^{\jbar}. \end{equation}
Recall that for a single chiral supersymmetry (which we take to be left-moving for concreteness), there is a supercurrent $G(z)$ with weight $(\frac{3}{2},0)$ given by
\begin{equation} G_L = g_{i \jbar} ( \psi^i \p X^{\jbar} + \psi^{\jbar} \p X^{i}). \end{equation}
The two terms on the right-hand-side above have respective charges $\pm 1$ under $J$. Thus, for the Calabi--Yau SCFT, the supersymmetry is enhanced to (2,2) because there are two supercurrents $G^{\pm}$ on each side, where $\pm$ denotes the charge under $J$. To be precise, we write
\begin{align}
 G^+ &= g_{i \jbar} \psi^i_L \p X^{\jbar}, \quad
 G^- = g_{i \jbar} \psi^\jbar_L \p X^{i}, \\
 \bar{G}^- &= g_{i \jbar} \bar{\psi}^i \pb X^{\jbar}, \quad
 \bar{G}^+ = g_{i \jbar} \bar{\psi}^\jbar \pb X^{i}.
\label{eq:spin32currents}
\end{align}
The full ${\cal N} =  (2,0)$ superconformal algebra is generated by $T(z), J(z), G^\pm(z)$ \cite{Ademollo:1975an}. In addition to the internal U(1) generated by $J$, there is a $\mathbb{Z}_2$ auto-automorphism which acts on the supercurrents $G^1 = G^+ + G^-$ and $G^2 = i(G^+ - G^-)$ as
\begin{equation}
\mathbb{Z}_2: G^1 \to G^1, \quad G^2 \to - G^2\Leftrightarrow G^+\leftrightarrow G^-.
\end{equation}

Having established the existence of the supercurrents $G^{\pm}$ for the left- and right-movers, we now discuss chiral and antichiral operators.  On the left side, a {\textit{chiral operator}} $\phi_c(z)$ satisfies
\begin{equation} 
G^+(z) \phi_c(0) \sim 0. \label{eq:chiral} 
\end{equation} 
Note that operators $\phi_c\equiv G^+\tilde\phi$ are trivially chiral, and we define chiral operators by quotienting by such trivial operators. Likewise, an {\textit{antichiral operator}} $\phi_a(z)$ satisfies
\begin{equation} 
G^-(z) \phi_a(0) \sim 0. 
\end{equation}
One can use the right-moving supercurrents to also define chiral and antichiral operators on the right. So we have four possibilities: $(c,a)$,$(c,c)$,$(a,c)$, and $(a,a)$. Note that PCT symmetry implies that the number of $(c,a)$ operators equals the number of $(a,c)$ operators, and the number of $(c,c)$ operators equals the number of $(a,a)$ operators. So one can independently specify the number of $(c,a)$ and $(c,c)$ operators.

For simplicity, let us specialize to the trivial target space manifold $\mathds{C}^n$. An example of a $(c,c)$ operator is given by
\begin{equation} 
\phi^{++}_{\frac{1}{2},\frac{1}{2}} = k_{i \jbar} \psi^i \bar{\psi}^{\jbar}, 
\end{equation}
where $k_{i \jbar}$ is the K\"ahler 2-form. This operator is $(c,c)$ because $G^+ \psi^i \sim 0$ and $\bar{G}^+ \bar{\psi}^\jbar \sim 0$. Note that by acting with the $G^-$ supercurrents, we obtain a marginal operator that can be added to the action of the $\sigma$-model. The relevant OPE is
\begin{equation} 
G^-(z) \bar{G}^- (\zbar) \, \phi^{++}_{\frac{1}{2},\frac{1}{2}}(0,0) \sim \frac{1}{z \zbar} k_{i \jbar} \p X^i(0) \pb X^{\jbar}(0). 
\end{equation}
Note that $\phi^{++}_{\frac{1}{2},\frac{1}{2}}$ has weight $(\frac{1}{2},\frac{1}{2})$ and J-charge $(1,1)$, so that $L_0 = \frac{1}{2}J_0$ on the left and right sides.

The relation $L_0 = \frac{1}{2}J_0$ holds more generally for all chiral fields (for antichiral fields, we have $L_0 = -\frac{1}{2}J_0$). Given a general $(c,c)$ field of weight $(\frac{1}{2},\frac{1}{2})$, we may act with $G^-_{-\frac{1}{2}} \bar{G}^-_{-\frac{1}{2}}$ to obtain a weight $(1,1)$ operator that is neutral under $J, \bar{J}$. Such operators can be added to the action and correspond to deformations of the Calabi--Yau manifold. Likewise, we can also obtain Calabi--Yau deformations from $(c,a)$ operators of weight $(\frac{1}{2},\frac{1}{2})$, denoted by $\phi^{+-}_{\frac{1}{2}, \frac{1}{2}}$, by acting with $G^-_{-\frac{1}{2}} \bar{G}^+_{-\frac{1}{2}}$. The $(c,c)$ deformations are related to K\"ahler deformations, while the $(c,a)$ operators are related to complex structure deformations. The correspondence associates the charge of an operator under $J, \bar{J}$ with raised or lowered indices. That is, a $+1$ $J$ ($\bar{J}$) charge is associated with a lowered $i$ ($\ibar$) index while a $-1$ $J$ ($\bar{J}$) charge is associated with a raised $i$ ($\ibar$) index. For example, a $(1,1)$-form $dz_i \wedge d\zbar_\ibar$ corresponds to an operator of charge $(1,1)$, while the tensor $\mu\indices{_\ibar^j}$ is associated with an operator of charge $(1,-1)$.

Note that the overall sign of $\bar{J}$ is a convention. If we flip its sign, then chiral operators become antichiral operators and vice versa (note that we always define $G^{\pm}$ to have the J-charge given by its superscript). This ambiguity means that given an abstract CFT, it is not possible to precisely determine the manifold. The {\textit{mirror symmetry}} conjecture asserts that, a Calabi--Yau CFT is equivalent to another Calabi--Yau CFT on a manifold with the Hodge numbers $h^{1,1}$ and $h^{n-1,1}$ swapped \cite{Candelas:1990rm}. The same CFT can lead to two different Calabi--Yaus. One may immediately identify $T^6 / \mathbb{Z}^3$ as a counterexample.  See Figure \ref{fig:t6modz3cohomcounterexample}. However, this is an exception and generally there are mirror pairs of CY. In general dimensions, mirror symmetry relates a Calabi--Yau manifold with its mirror manifold which has swapped Hodge numbers $h^{p,q}$ and $h^{n-p,q}$.

\begin{figure}[H]
\centering
\begin{tabular}{|c|c|c|c|}
	\hline
	1 & 0  & 0 &1\\
	\hline 
	0 & 0  & 36 & 0\\
	\hline 
	0 & 36 & 0  &  0  \\
	\hline 
	1 & 0 & 0 & 1  \\ 
	\hline
\end{tabular}
$\longleftrightarrow$
\begin{tabular}{|c|c|c|c|}
	\hline
	1 & 0  & 0 &1\\
	\hline 
	0 & 36  & 0 & 0\\
	\hline 
	0 & 0 & 36  &  0  \\
	\hline 
	1 & 0 & 0 & 1  \\ 
	\hline
\end{tabular}
\caption{A possible counterexample to the mirror symmetry conjecture. The left hand side is the cohomology of $T^6 / \mathbb{Z}^3$. Because $T^6 / \mathbb{Z}_3$ has $h^{2,1} = 0$, a putative mirror manifold would have $h^{1,1} = 0$, which contradicts the existence of the K\"ahler form. Yet, there exists a sense in which mirror symmetry holds.}
\label{fig:t6modz3cohomcounterexample}
\end{figure}

\subsection{Supersymmetric minimal models}

We will now learn more about Calabi--Yau CFTs at certain points on the moduli space. The $\mathcal{N} = (2,2)$ supersymmetric minimal models are an important part of the discussion \cite{Gepner:1987qi}. First, recall that the unitary non-supersymmetric minimal models admit a Lagrangian description given by the Landau-Ginzburg theory. They organize into an ADE classification.  The A-series models are labeled by an integer $m = 2,3,4,\ldots$, with central charge
\begin{equation} 
c_m = 1 - \frac{6}{m(m+1)}. 
\end{equation}
All primaries have the same $L_0$ and $\bar{L}_0$ weights, and hence are scalars. The set of weights is
\begin{equation} 
\left\{  \frac{((m+1)r - m s)^2 - 1}{4 m (m + 1)} \, : \, r,s \in \mathbb{N} \, \,  1 \leq r \leq m -1 , 1 \leq s \leq m \right\}. 
\end{equation}
 One can then check that there are $2 m - 3$ relevant operators. The Landau-Ginsparg theory has a single scalar field with Lagrangian
\begin{equation} 
\mathcal{L} = \partial \phi \partialbar \phi + g (\phi^2)^{m-1}. 
\end{equation}
The field $\phi$ has zero classical scaling dimension but it acquires a positive quantum scaling dimension. The relevant operators of this Lagrangian are $1,\phi,\phi^2,\ldots,\phi^{2m - 4}$. The equations of motion imply that $\phi^{2m - 3}$ is proportional to $\partial \partialbar \phi$, which is irrelevant. Thus, the number of relevant operators in the Landau-Ginsparg theory equals the number of relevant operators of the corresponding minimal model, which is evidence for the proposed Lagrangian description.

Note that in these minimal models, $c < 1$. This is reasonable because in the absence of any potential, the central charge would be 1. A potential effectively kills some degrees of freedom (the field cannot go all the way to infinity).

To obtain a Lagrangian description of the (2,2) supersymmetric minimal models, we use supersymmetric Landau-Ginsparg theory \cite{Witten:1993jg}. We will write down a Lagrangian and then flow to an IR fixed point, so that the manifest supersymmetry is in fact part of a larger superconformal symmetry. 

Also note that 2d\, $(2,2)$ supersymmetry has the same number of supersymmetries as in 4d\, $\mathcal{N} = 1$ theory.\footnote{See \cite{Krippendorf:aa} for 4d\, $\mathcal{N} = 1$ theory.} We promote the scalar field $\phi$ to a chiral superfield $\phi$. The chiral superfield $\phi(z,\zbar; \theta^\pm, \bar{\theta}^\pm)$ depends on $z,\zbar$ and Grassmann numbers $\theta^{\pm},\thetabar^{\pm}$ in such a way that
\begin{equation} D^+ \phi = \bar{D}^+ \phi = 0, \end{equation}
where
\begin{equation} D^+ \phi = \frac{\partial}{\partial \theta^+} +  \theta^- \frac{\partial}{\partial z}, \end{equation}
\begin{equation}\bar{D}^+ \phi = \frac{\partial}{\partial \thetabar^+} +  \bar{\theta}^- \frac{\partial}{\partial \zbar}. \end{equation}
The $\pm$ indicates the charge under a global U(1) symmetry (R-charge), which we have called $J$ and $\bar{J}$ earlier. In order for $D^+$ and $\bar{D}^+$ to have J-charges $(1,0)$ and $(0,1)$ respectively, $\theta^+$ and $\bar{\theta}^+$ must have J-charges $(-1,0)$ and $(0,-1)$ respectively.
 Note that
 \begin{equation} (D^+)^2 = (\bar{D}^+)^2 = 0. \end{equation}
 The action is comprised of a D-term (also called the K\"ahler potential) and an F-term (also called the superpotential): 
\begin{align} \label{eq:susyaction} \int d^2 z \, d^4 \theta \, K(\phi,\bar\phi) + \int d^2 z d^2 \theta^+ \, W(\phi) + \int d^2 z d^2 \theta^- \,  \overline{W(\phi)}. \end{align}
As per usual, the notation $W(\phi)$ implies that $W$ is a holomorphic function of $\phi$. The action \eqref{eq:susyaction} is manifestly supersymmetric in terms of the supercharges $Q^\pm$ and $\bar{Q}^\pm$. Integrating over the Grassmann coordinates yields a functional of finitely many fields which depends solely on the bosonic variables $z,\zbar$. In particular, the potential for the complex scalar $\phi(z,\zbar)$ embedded in the chiral superfield $\phi$ is given by the superpotential:
\begin{equation} V = \left| \frac{\partial W}{\partial \phi} \right|^2. \end{equation}
The combined central charges of the complex scalar and its superpartner is 3; it is convenient to define
\begin{equation} \hat{c} := \frac{c}{3}. \end{equation}
to match $\hat{c}=1$ of the supersymmetric free theory with $c=1$ of the standard free boson. Because of the superpotential, the supersymmetric minimal models we describe have $\hat{c} < 1$. The simplest choice of $W$ is
\begin{equation} W(\phi) = \phi^n. \label{eq:superpotential} \end{equation}
A nonrenormalization theorem says that this superpotential receives no quantum corrections \cite{Gates:1983nr}. The same cannot be said of the D-term, which generically receives quantum corrections. The flowed theory of equations \eqref{eq:susyaction} and \eqref{eq:superpotential} has central charge
\begin{equation} \hat{c} = 1 - \frac{2}{n}. \end{equation}
For the case $n = 2$, the theory is a free massive theory which flows to a trivial theory in the IR, consistent with $\hat{c} = 0$. Because $(D^+)^2 = 0$, we can use $D^+$ to define a cohomology. The {\textit{chiral ring}} is defined to be the space of $(D^+)$-closed fields modulo $(D^+)$-exact fields. Due to the equations of motion,
\begin{equation} D^+ \bar{D}^+ \bar{\phi} \propto \frac{\partial W}{\partial \phi} \propto \phi^{n-1}. \end{equation}
Thus, the chiral ring is generated by the fields
\begin{equation} 1,\phi,\phi^2,\ldots,\phi^{n-2}. \label{eq:chiralring}
\end{equation} 
These fields correspond to (c,c) operators in the SCFT\footnote{More generally, the chiral ring is generated by the (c,c) operators in any ${\cal N} = 2$ SCFT. The OPE between (c,c) operators is non-singular, which allows us to consider associative products of the form $\phi_i(0) \phi_j(0) = c_{ijk} \phi_k(0)$. The coefficients $c_{ijk}$ define the associated algebraic structure. It does not admit an inverse, leading to its identification as a ring and not a group.}. In order for the F-term to be invariant under the $U(1) \times U(1)$ symmetry, $\phi^n$ must have J-charge (1,1) which implies that $\phi$ has J-charge $(\frac{1}{n},\frac{1}{n})$. Thus, the maximum charge of a (c,c) operator in this theory is
\begin{equation} (n - 2) \cdot \frac{1}{n}. \end{equation}
It turns out that this must be equal to $\hat{c}$: this is plausible because earlier when we considered the Calabi--Yau $\sigma$-model we saw how the chiral operators mimic the cohomology of the target space manifold. The maximum charge thus corresponds to the dimension of the manifold, which is $\hat{c}$.

Lastly, we note that all $\mathcal{N} = (2,2)$ minimal models also fit into an ADE classification, which can be specified by the choice of superpotential of two chiral superfields $x$ and $y$.  The classification is spelled out in Table \ref{table:classification}.

\begin{table}[H]
	\begin{center}
	\renewcommand{\arraystretch}{1.4}
\begin{threeparttable}
\begin{tabular}{cc}
	\toprule 
	Superpotential & Classification \\ \midrule 
	$W = x^n$ & $A_{n-1} \quad n \ge 2$ \\ 
	$W = x^n + x y^2$ & $D_{n+1} \quad n \ge 0$ \\
	$W = x^3 + y^4$ & $E_6$ \\ 
	$W = x^3 + x y^3$ & $E_7$ \\ 
	$W = x^3 + y^5$ & $E_8$ \\ \bottomrule 
\end{tabular} 
\end{threeparttable}
\caption{Classification of $\mathcal{N} = (2,2)$ minimal models. $X$ and $Y$ are chiral superfields.}
\label{table:classification}
\end{center}
\end{table}

\subsection{Mirror symmetry in minimal models}

Consider the $A_{n-1}$ minimal model with superpotential $W = \phi^n$. It has a $\mathbb{Z}_n$ symmetry defined by $\phi \rightarrow \omega \phi$ where $\omega^n = 1$. This symmetry is generated by the operator
\begin{equation} e^{2 \pi i \frac{J_0+\bar{J}_0}{2}} \end{equation}
because $\phi$ has charge $(\frac{1}{n},\frac{1}{n})$. We believe that if we orbifold the minimal model by this $\mathbb{Z}_n$ symmetry, we obtain the same minimal model. This is plausible because the minimal models have been completely classified, and in many cases the central charge alone is enough to select a model. There are no chiral fields in the untwisted sector because none of the chiral ring elements of equation \eqref{eq:chiralring} are invariant under $\phi \rightarrow \omega \phi$. However, there are $n - 1$ twisted sectors, which is exactly the number of fields in equation \eqref{eq:chiralring}. All states in the orbifold theory must be invariant under $J_0+ \bar{J}_0$. Hence, for every chiral field in equation \eqref{eq:chiralring} of charge $(q,q)$, the orbifold theory has a (c,a) field of charge $(q,-q)$. Recall that the designation between (c,c) and (c,a) fields is ambiguous. We say that the $\phi^n$ theory is ``mirror'' to itself.

\subsection{Calabi--Yau SCFT from minimal models}
\label{sec:CalabiYauMinimalModels}

We would like to describe a $\sigma$-model superconformal field theory with a Calabi--Yau target space. We will start with a one-dimensional Calabi--Yau, \emph{i.e.} the torus $T^2$, which must have $\hat{c} = 1$. To get $\hat{c} = 1$, we will take three copies of the $A_2$ minimal model, with chiral superfields $x,y,z$. The superpotential is 
\begin{equation} 
W = x^3 + y^3 + z^3. 
\end{equation}
This looks just like equation \eqref{eq:torus}. However, the presence of chiral fields with fractional charges is undesirable, since these charges are supposed to also specify elements of the cohomology. We can simply eliminate fractional charges by modding out by a  $\mathbb{Z}_3$ symmetry generated by 
\begin{align}
e^{2 \pi i \frac{(J_0+\bar{J}_0)}{2}}. 
\end{align}
Now, this theory has a chance of being the conformal field theory with target space $T^2$. In particular, we can deform the superpotential by adding $a x y z$, where the coefficient $a$ is a complex number, since $xyz$ is a $\mathbb{Z}_3$ invariant. This deformation corresponds to a complex structure deformation. However, we cannot identify a K\"ahler parameter $\rho$ from such a superpotential. It follows that if this is the $\sigma$-model with a $T^2$ target space, it must be the theory at a particular fixed value of K\"ahler parameter. In summary, the theory is expected to have a superpotential
\begin{equation}\label{eq:torustheory}
W = \frac{x^3+y^3+z^3+ a xyz}{\mathbb{Z}_3} ,
\end{equation}
which is invariant under $\mathbb{Z}_3$. As you will show in exercise, an orbifold CFT of the form $CFT/\mathbb{Z}_n$ itself has a $\tilde{\mathbb{Z}}_n$ symmetry. From this we learn that \eqref{eq:torustheory} has a $\tilde{\mathbb{Z}}_3$ symmetry. This picks out a particular $T^2$, \emph{i.e.} selects where $\rho$ must lie in its moduli space ($\rho=e^{2\pi i/3}$).

\begin{figure}[H]
\centering
\begin{tikzpicture}
\node[draw=none,thick,scale=0.2,fill=black,label={[label distance=1mm]south:$0$}] (A2) at (0,0) {};
\node[draw=none,thick,scale=0.2,fill=black,label={[label distance=1mm]south:$-1/2$}] (A3) at (-1.5,0) {};
\node[draw=none,thick,scale=0.2,fill=black,label={[label distance=1mm]south:$1/2$}] (A4) at (1.5,0) {};
\node[circle,thick,scale=0.5,fill=black,label={[label distance=1mm]east:$\mathbb{Z}_3$ symmetry sits here}] (A5) at (1.5,1.5) {};
\draw[->] (-2,0)--(2,0);
\draw[->] (0,0)--(0,5);
\draw (-1.5,1.5)--(-1.5,5);
\draw (1.5,1.5)--(1.5,5);
\draw (1.5,1.5) arc (42:139:2);
\draw[dashed] (-1.5,0)--(-1.5,1.5);
\draw[dashed] (1.5,0)--(1.5,1.5);
\draw[fill=blue, fill opacity=0.2] (1.5,5) -- (1.5,1.5) arc (42:139:2) -- (-1.5,5)--cycle;
\end{tikzpicture}
\caption{The moduli space of $\rho$. The point with $\mathbb{Z}_3$ symmetry is labeled.}
\label{fig:rhomodulispace}
\end{figure}
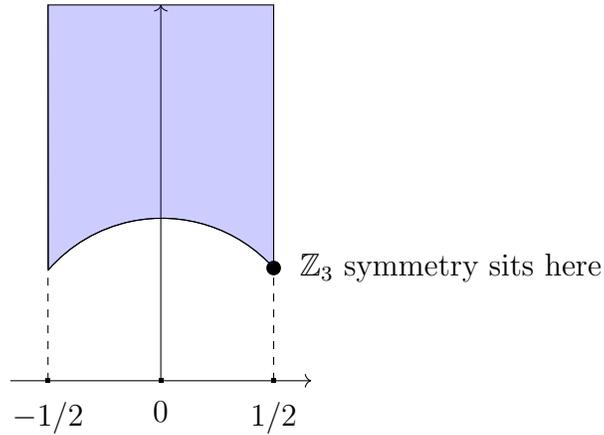

Recall that $\rho$ and $\tau$ are mirror to each other (that is, they can be swapped by T-duality). Let us swap the roles of $\rho$ and $\tau$ so that the theory we are considering is at the $\mathbb{Z}_3$-symmetric point of the $\tau$ moduli space. We may then deduce from the exercise that if we do not quotient the superpotential to orbifold the $\mathbb{Z}_3$ symmetry in equation \eqref{eq:torustheory}, we obtain the superconformal field theory with target space $T^2 / \mathbb{Z}_3$. The fractional charges of the chiral fields correspond to the twisted sectors of the $T^2 / \mathbb{Z}_3$ theory. We can see the twisted sectors in the geometry of $W$.

Since we are not modding Lagrangian by $\mathbb{Z}_3$, we can choose our superpotential to be e.g.
\begin{equation} 
W = x^3 + y^3 + z^3 + \alpha x, 
\end{equation}
wich would give a vev to fields that arise from the twisted sector. Because these fields have conformal dimensions less than 1, they lead to tachyonic modes. To avoid tachyons, it is important to have the $\mathbb{Z}_3$ quotient in equation \eqref{eq:torustheory}.

For higher dimensions, recall that in $\mathds{CP}^n$ the equation
\begin{equation} 
z_1^{n+1} + \cdots + z_{n+1}^{n+1} = 0 
\end{equation}
defines a Calabi--Yau $(n-1)$-fold. This suggests that we can build the corresponding superconformal field theory by taking $(n+1)$ minimal models, each with central charge
\begin{equation} 
\hat{c} = 1 - \frac{2}{n+1}. 
\end{equation}
The total central charge is then
\begin{align}
c=(n+1)\, \hat{c}=n - 1,
\end{align}
which matches the dimensionality of the manifold. Of course, we should also quotient by a $\mathbb{Z}_{n+1}$ symmetry. For instance, for a quintic Calabi--Yau threefold, the superpotential is given by
\begin{equation} 
W = \frac{x_1^5 + \cdots + x_5^5}{\mathbb{Z}_5},
\label{eq:superpotentialquintic} 
\end{equation}
which is invariant under $\mathbb{Z}_5$. This superpotential should correspond to a point on the moduli space of the quintic manifold. The K\"ahler parameter must be at a point with $\mathbb{Z}_5$ symmetry. Similarly as in the case of $T^2$, we expect that the manifold at this point should be at the size of the string scale. 

\noindent {\bf Exercise 1:} Consider any CFT with a $\mathbb{Z}_n$ symmetry. Show (at the level of the partition function) that the orbifold theory, denoted by $CFT / \mathbb{Z}_n$, itself has a $\mathbb{Z}_n$ symmetry, which we denote by $\tilde{\mathbb{Z}}_n$. Furthermore, show (at the level of the partition function) that if we orbifold $CFT / \mathbb{Z}_n$ by $\tilde{Z}_n$, we get the original CFT back. 

\subsection{Minimal models and Calabi--Yau \texorpdfstring{$\sigma$}{Lg}-models}
\label{sec:sigmaModels}

Until now, we have motivated the connection between minimal models and Calabi--Yau $\sigma$-models. In this section, we will actually prove this connection. Details, including the full Lagrangian of the gauged linear $\sigma$-model, are given in \cite{Witten:1993yc}. Our starting point is 2d\, $\mathcal{N} = (2,2)$ gauge theory with gauge group U(1). We have a single chiral superfield $p$ of charge $-(n+1)$ and $n+1$ chiral superfields $x_i$ with $i=1,\ldots, n+1$ of charge 1. The superpotential is chosen to be
\begin{equation} W = p \, G (x_i) ,
\label{eq:GLSMsuperpotential}
\end{equation}
where $G$ is a homogeneous polynomial of degree $(n+1)$. We also impose that $G$ satisfies
\begin{equation} 
\label{eq:condition} 
\partial_i G = 0 \quad \forall i \implies G = 0. 
\end{equation} 
That is, the gradient of $G$ only vanishes at the origin. \eqref{eq:homogenouspolynomial} is one example of such a polynomial. The potential for the complex scalar fields within the chiral superfields is a sum of nonzero terms, one of which is given by\footnote{In this section, we use $X_i$ and $P$ to denote both the chiral superfields and their complex scalar components.}
\begin{equation} D^2 \propto e^2 (\sum_i|x_i|^2 - (n+1)|p|^2 - r),
\label{eq:d2} \end{equation}
where $r$ is a real parameter coming from the Fayet-Iliopoulos (FI) term in the Lagrangian and $e$ is the charge. The superpotential makes an additional contribution to the potential:
\begin{equation} 
\sum_{i=1}^{n+1} \left|\frac{\partial W}{\partial x_i}\right|^2 + \left|\frac{\partial W}{\partial p}\right|^2. \label{eq:f2} 
\end{equation}

The corresponding CFT sits at the IR fixed point of the RG flow. At low energies, the potential for the complex scalars should vanish. This implies that equations \eqref{eq:d2} and \eqref{eq:f2} must equal zero, so \eqref{eq:f2} becomes
\begin{equation}
\sum_{i=1}^{n+1} |p|^2 |\partial_i G|^2 + |G|^2 = 0.
\end{equation}
We observe that $G = 0$ at the minimum of the potential, so \eqref{eq:condition} leads to $P = 0$. In order for \eqref{eq:d2} to vanish, we choose $r > 0$ such that
\begin{equation} 
\sum_{i=1}^{n+1} |x_i|^2 = r. 
\end{equation}
Thus, the magnitude of each of the $x_i$ is positive and bounded by $\sqrt{r}$. Because the $U(1)$ gauge group is Higgsed in this vacuum, we need to quotient out the allowed space of $X_i$ by overall phase rotations. The space of vacua is given by $\{x_i\} \in \mathds{C}^{n+1}$ subject to the identification
\begin{equation} (x_1,\ldots,x_{n+1}) \sim \lambda (x_1,\ldots,x_{n+1}), \quad \lambda \in \mathds{C}, \end{equation}
as well as the restriction
\begin{equation} G = 0, \end{equation}
which is precisely the defining equation of a Calabi--Yau in $\mathds{CP}^n$. The superpotential is invariant along the RG flow, so we conclude that the IR fixed point of this gauge theory is the supersymmetric $\sigma$-model with Calabi--Yau target space. This is the $(n-1)$-dimensional generalization of the quintic threefold.

Because $r$ fixes the overall size of the $x_i$, $r$ corresponds to a K\"ahler parameter that sets the size of the Calabi--Yau. If we want the $\sigma$-model to be weakly coupled (so that stringy corrections to the geometry are small), we must take $r$ large. Recall that for the quintic Calabi--Yau, $h^{1,1} = 1$. We need another real parameter to combine with $r$ to form the complexified K\"ahler parameter; the $\theta$ parameter of the gauge theory fills this role. The complex K\"ahler parameter $\rho$ is then given by
\begin{equation} \rho = \theta + i r. \end{equation}

We now want to study the case of $r$ being negative. If $\theta \neq 0$, we still have nonzero $\rho$ as $r$ passes through zero. If $r$ is negative, then from equation \eqref{eq:d2} we see that $p$ can no longer be zero, so $p$ acquires a vev. Because $p$ has charge $-(n+1)$, the $U(1)$ gauge symmetry is broken down to $\mathbb{Z}_{n+1}$, which is just the orbifold symmetry of the Landau-Ginsparg theories we studied earlier. Since we set the superpotential to equation \eqref{eq:GLSMsuperpotential}, we see that the theory at $r \rightarrow -\infty$ becomes the Landau-Ginsparg theory studied earlier (compare with equation \eqref{eq:superpotentialquintic}). We take the limit $r \rightarrow -\infty$ so that the contribution of the $X_i$ fields to the vev of $P$ is negligible; it is only in this limit that we obtain the Landau-Ginsparg theory with the $\tilde{\mathbb{Z}}_5$ symmetry discussed earlier. Clearly, the opposite limits $r \rightarrow \infty$ and $r \rightarrow - \infty$ probe different regions of the $\rho$ moduli space of the Calabi--Yau $\sigma$-model.

To summarize, the CFT on the quintic manifold at $r \rightarrow -\infty$ is the Landau-Ginsparg theory of  \eqref{eq:superpotentialquintic}. We know from earlier that the orbifold Landau-Ginsparg theory with superpotential
\begin{equation} 
W = \frac{x_1^5}{\mathbb{Z}_5}
\end{equation}
is the same as the Landau-Ginsparg theory with superpotential
\begin{equation} W = \tilde{x}_1^5. \end{equation}
It trivially follows that the $\mathbb{Z}_5^5$-orbifold theory with superpotential
\begin{equation} W = \frac{x_1^5}{\mathbb{Z}_5} + \frac{x_2^5}{\mathbb{Z}_5} + \cdots + \frac{x_5^5}{\mathbb{Z}_5}  \label{eq:fivecopiesorbifold} \end{equation}
is equivalent to the theory with
\begin{equation} W = \tilde{x}_1^5 + \tilde{x}_2^5 + \cdots + \tilde{x}_5^5. \label{eq:fivecopies} \end{equation} To get the quintic theory, we must mod out \eqref{eq:fivecopies} by a diagonal $\mathbb{Z}_5$. This is equivalent to ungauging one of the $\mathbb{Z}_5$ symmetries in equation \eqref{eq:fivecopiesorbifold}. If $\alpha_i$ denotes the five different phase rotations that make up the $\mathbb{Z}_5$ symmetries in equation \eqref{eq:fivecopiesorbifold} ($\alpha_i^5 = 1$), then the quintic theory corresponds to the theory of equation \eqref{eq:fivecopiesorbifold} with the additional restriction that
\begin{equation}
\alpha_1 \alpha_2 \alpha_3 \alpha_4 \alpha_5 = 1. \label{eq:z5to4}
\end{equation}
That is, to get the quintic theory we can start off with five copies of the $W = x_1^5$ theory and then orbifold by a $\mathbb{Z}_5^4$ symmetry defined by equation \eqref{eq:z5to4}. We can write this $\mathbb{Z}_5^4$ symmetry as a $\mathbb{Z}_5^3$ symmetry times one generated by $e^{2 \pi i \frac{J_0 + \bar{J}_0}{2}}$, which is the $\mathbb{Z}_5$ symmetry that corresponds to $\alpha_1 = \alpha_2 = \ldots = \alpha_5$. This means that the CFT on the quintic is isomorphic to the same theory orbifolded by the $\mathbb{Z}_5^3$ symmetry, \emph{i.e.} with superpotential 
\begin{equation} W = \frac{\left[x_1^5 + x_2^5 + \cdots + x_5^5\right] / \mathbb{Z}_5^3}{e^{2 \pi i \frac{J_0+\bar{J}_0}{2}}}. \end{equation}
The conclusion is that the quintic manifold mod $\mathbb{Z}_5^3$ is mirror to the quintic.

\noindent { \bf Exercise 2: } Show that this theory admits just one complex structure deformation, given by
\begin{equation} \alpha  x_1 x_2 x_3 x_4 x_5. \end{equation}
Thus, $h^{1,2} = 1$ as it should be, since $h^{1,1} = 1$ for the quintic threefold. The 101 K\"ahler deformations for the quintic mod $\mathbb{Z}_5^3$ manifold can be found by resolving its singularities.

\section{Black Holes and holography}

\subsection{Black holes in string theory}

We now want to try and use our microscopic understanding of branes in string theory to unravel some of the mysteries of gravity, and in particular of black holes. 

\subsubsection*{Black holes from wrapped branes}

Black holes display features of thermodynamic systems, as studied by Bekenstein and Hawking. In particular, in 4 spacetime dimensions, the thermodynamic entropy of a black hole is proportional to the area of its horizon, \cite{Bardeen:1973gs,Bekenstein:1973ur,Hawking:1975vcx}
\begin{equation}
  S = \frac{A}{4} \,,
\end{equation}
where the factor of $1/4$ is universal. Recall that we are working in Planck units.

This result, which follows from a semiclassical field theory analysis, is perhaps somewhat surprising since from a classical viewpoint black holes solutions have zero degrees of freedom. Moreover, the semiclassical entropy calculation seems to suggest that black holes in fact comprise a huge number of degrees of freedom. This leads to basic puzzle as to their origin. While in general this is not known, string theory offers a potential solution: the degrees of freedom of the black hole can be hidden in the extra compactified dimensions, for example as branes wrapping cycles in some internal geometry. 

Black holes in 3+1 dimensions in general are not well understood. However, there exist supersymmetric versions of black holes, called BPS black holes, which have certain properties protected by supersymmetry, and are therefore more straightforward to analyze \cite{Strominger:1996sh}. Due to supersymmetry, these properties can also obey rigid constraints. For instance, an extremal BPS black hole in 3+1 dimensions must have a fixed mass-to-charge ratio, \emph{i.e.} 
\begin{align}
    M = |Q|.
\end{align} 

As black holes are particle-like (\emph{i.e.} 0+1 dimensional), we can attempt to construct them by considering D$p$ branes wrapped on $p$-cycles. For example, one can consider a D3 brane wrapped on $C \times S^1$, where $C$ is a Riemann surface. In string theories compactified on some five-dimensional manifold $M_5$ with $C \times S^1 \subset M_5$, such as $M_5 = K_3 \times S^1$ with $C \subset K3$, the wrapped branes behave as particles propagating in the five noncompact dimensions. From the Kaluza-Klein reduction, the mass of the particle must be proportional to the genus of the Riemann surface, and so for surfaces with sufficiently many handles, the particle becomes very heavy and looks like a black hole. It turns out that the microscopic entropy of such particles precisely matches the macroscopic entropy of black holes as given by Bekenstein and Hawking \cite{Strominger:1996sh}.

\subsubsection*{Near-horizon limit of branes}

If we zoom in very close to the particle (black hole) in the large dimensions, we observe a horizon associated to a 5d geometry of the form
\begin{equation}
AdS_2 \times S^3 \,.
\end{equation}
In fact, the emergence of AdS geometries in the near-horizon limit of black holes was well known long before they were considered in the string theory context.  In this case, the presence of the heavy D3 branes backreacts with the surrounding geometry, curving it into a space that looks like AdS near the horizon. 

Let us instead start with a stack of $N$ D$p$ branes in type IIA or IIB string theory. In general, the presence of the branes backreacts with Minkowski space to give a new geometry with metric \cite{Duff:1993ye}
\begin{equation}\label{eqn:met}
ds^2  = H(r)^{-\frac{1}{2}}ds^2_{1,p} + H(r)^{\frac{1}{2}} ds^2_{0,9-p} \,,
\end{equation}
where $d^2_{p,q}$ is the flat Lorentzian metric in signature $(p,q)$. It naturally splits into a contribution to the metric parallel and perpendicular to the brane worldvolume. The warp factor $H(r)$ depends on the distance $r$ perpendicular to the stack of branes and is given by
\begin{equation}
H(r) = 1 + c\ \frac{g_s N}{r^{7-p}} \,.
\end{equation}
where $g_s$ is the closed string coupling and $c$ is a constant. It also contributes to the dilaton VEV $e^{-2\phi}$ as
\begin{equation}\label{eqn:coup}
e^{-2\phi} = g_s^{-2} H(r)^{\frac{p-3}{2}} \,.
\end{equation}

There are two main limits to consider. As $r \rightarrow \infty$, that is, as one is sufficiently far away from the stack of branes, the warp factor tends to $H \rightarrow 1$. Consequently the metric becomes flat and it is no longer possible to resolve the effects of the backreaction. On the other hand, as one approaches the branes, in the limit where $r \rightarrow 0$ (near-horizon limit), $H \rightarrow c g_s N r^{p-7}$, and the string coupling generically begins to run unless $p = 3$. 

However, for the stack of D3 branes, the string coupling remains constant but the metric is modified. In the near-horizon limit, the metric \eqref{eqn:met} reduces to 
\begin{equation}
ds^2 = \frac{1}{\sqrt{cg_s N}} r^2 ds^2_{1,3} + \sqrt{cg_sN} r^{-2} ds^2_{0,6} \,.
\end{equation}
We can always rewrite the perpendicular part of the metric in spherical coordinates,
\begin{equation}
ds^2_{0,6} = dr^2 + r^2 d\Omega_5^2 \,,
\end{equation}
where $d\Omega_5^2$ is the metric of the unit 5-sphere. Notice that the
the factor of $r^{-2}$ in front of the 6d metric cancels the $r^2$ in the spherical metric. In other words, in the near-horizon limit we reach a point where the size of the $S^5$ does not shrink anymore. Overall, the metric becomes
\begin{equation}
\left(\frac{1}{\sqrt{cg_s N}} r^2 ds^2_{1,3} + \sqrt{cg_sN}\frac{1}{r^2} dr^2 \right) + \sqrt{cg_sN}d\Omega_5^2 \,,
\end{equation}
which describes none other than the $AdS_5 \times S^5$ spacetime, where both the AdS and sphere radii are given by $\sqrt{c g_s N}$.

\noindent{\bf Exercise 1}: Show that we can write the AdS$_{d+1}$ metric as 
\begin{equation}
(ds^2)_{d+1}=r^2 ds^2_{1,d-1} + \frac{1}{r^2} dr^2 \,,
\end{equation}
where 
\begin{equation}
(ds^2)_{1, p+2}=\frac{-dt^2 + d\vec x^2  + dy^2}{y^2}   \,.
\end{equation}

In addition to the $AdS_5 \times S^5$ geometry, we also need to consider the effects of the R–R fluxes. The presence of the D3 branes indicates that there is an $F_5$ flux, which in this case is turned on along the $S^5$ as
\begin{equation}
\int_{S^5} F_5 = N \,.
\end{equation}
Unsurprisingly, the $N$ D3 branes lead to $N$ units of five-form flux.
This looks like type IIB theory compactified on an $S^5$, which we did not discuss previously because $S^5$ does not admit Killing spinors due to the fact that it is not a special holonomy manifold. However, in our previous analysis we restricted our search to Minkowski vacua without fluxes. It turns out that if you turn on fluxes and take positive curvature, the it sometimes happens that one can get supersymmetric AdS solutions. 

\subsection{Holography}

\subsubsection*{Holography from D3 branes}

We just discovered that a stack of branes backreacts with the geometry, leading to a spacetime that looks like AdS in the near-horizon limit. However, from the string theory perspective we also know that the fluctuations of the brane are described by open string degrees of freedom living on its worldvolume, which in the low energy limit admits a gauge theory description. This led Maldacena to conjecture that these two descriptions are in fact equivalent . In particular, the conjecture states that \cite{Maldacena:1997re}
\begin{equation}
\text{$\mathcal{N} = 4$ super Yang--Mills} \quad\longleftrightarrow\quad \text{Type IIB on $AdS_5 \times S^5$} \,,
\end{equation}
i.e. that the degrees of freedom living on the left and right are equivalent. This is known as the holographic duality, or holography for short.  The 4d spacetime where the gauge theory lives is the boundary of the $AdS_5$ and the information of the gauge theory on the boundary and the gravitational theory in the bulk are supposed to be identical.

Strictly speaking, by boundary we mean the conformal boundary of AdS$_5$. This tells us that any metric in the same conformal class should give identical results for the boundary theory. In fact, the Euclidean AdS$_5$ isometry group $SO(1,5)$ acts as the \textit{conformal} group on the boundary, and so the gauge theory is also a CFT. This had to be the case, otherwise conformal transformations would not act faithfully on observables in the theory. 

There is a rich history matching observables on both sides of the duality. The boundary theory lives on $S^4$. Every local CFT admits a local spin 2 (symmetric, traceless rank 2 tensor) conserved current $T^{\mu\nu}(x)$, \emph{i.e.} the stress tensor. Its correlation functions thus constitute a universal set of observables in any CFT. In the dual bulk (gravitational) theory, it is natural to consider gravitons scattering processes. While asymptotically AdS spacetimes do not admit an S-matrix construction, scattering amplitudes are instead captured by correlation functions of boundary operators sourcing fields in the bulk. In this case, the stress tensor correlation functions compute amplitudes of gravitons traveling to and from the boundary and scattering in the bulk. This logic extends to other observables on the two sides of the duality, and there is a precise dictionary of how to relate gravitational computations in the bulk to correlation functions in the gauge theory.

\subsubsection*{Conifold geometries}

\begin{figure}[H]
\centering
\begin{tikzpicture}
\node[draw=none,opacity=0,thick,scale=0.1,fill=black,label={[label distance=1mm]south:$S^4$}] (A1) at (0,0) {};
\node[draw=none,opacity=0,thick,scale=0.1,fill=black,label={[label distance=0.5mm]-20:$S^5$}] (A2) at (1.5,0.4) {};
\node[draw=none,opacity=0,thick,scale=0.1,fill=black,label={[label distance=0.5mm]north:brane}] (A3) at (0.5,2.1) {};
\draw (-1,0)--(1,0)--(1.8,0.8);
\draw (1,0)--(1.2,1.8)--(1.8,0.8);
\draw (1.2,1.8)--(-0.1,1.8)--(-1,0);
\draw (1.2,1.9)--(-0.1,1.9);
\draw (1.2,2)--(-0.1,2);
\draw (1.2,2.1)--(-0.1,2.1);
\draw[->] (3,0.9)--(5.5,0.9);
\end{tikzpicture}
\begin{tikzpicture}
\node[draw=none,opacity=0,thick,scale=0.1,fill=black,label={[label distance=1mm]south:$S^4$}] (A1) at (0.25,0) {};
\node[draw=none,opacity=0,thick,scale=0.1,fill=black,label={[label distance=0.5mm]-20:$S^5$}] (A2) at (1.5,0.5) {};
\node[draw=none,opacity=0,thick,scale=0.1,fill=black,label={[label distance=0.5mm]west:flux}] (A3) at (0,2) {};
\node[draw=none,opacity=0,thick,scale=0.1,fill=black,label={[label distance=1mm]east:}] (A4) at (-2,0) {};
\draw (0.3,1.5)--(-0.5,0)--(1,0)--(0.3,1.5);
\draw (1,0)--(2,1)--(1.3,2.5)--(0.3,1.5);
\draw[->] (0.2,1.7)--(0.8,1.7);
\draw[->] (0.45,1.95)--(1.05,1.95);
\draw[->] (0.7,2.2)--(1.3,2.2);
\end{tikzpicture}
\caption{Two diagrams depicting $S^4 \times S^5$ geometries, where the branes reside on the $S^4$ and the flux on $S^5$. The diagrams are related via holography due to the backreaction of the branes on the geometry. Note that $S^4$ is the boundary of the $AdS_5$.}
\label{fig:conifold2}
\end{figure}
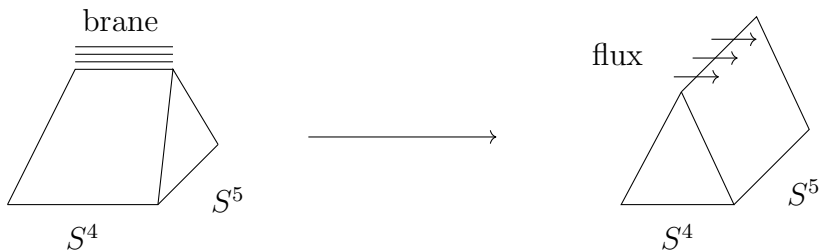

Let us now return to the brane description. In the backreacted geometry, there are two spheres of interest, namely the $S^4$ that the D3 branes wrap and the $S^5$ where the fluxes live. Far away from the branes, the $S^5$ can be shrunk to a point, whereas the $S^4$ corresponds to the worldvolume of the D3 branes and is fixed. Going to the branes, we have that the $S^5$ is stabilized by the five-form flux and so has finite size. Meanwhile the $S^4$ gets pushed to the boundary and becomes trivial. In other words, the branes get ``pushed to infinity'' and are nowhere to be found in the AdS bulk. In this sense the near-horizon limit has a geometric interpretation as exchanging the fixed sphere with the one allowed to shrink to a point. We can represent this transition as a cone with base $S^4 \times S^5$. Before the transition, the $S^4$ is smoothed out by D3 branes, while after the $S^5$ is smoothed out by the fluxes. This is depicted diagrammatically in Figure \ref{fig:conifold2} \cite{Gopakumar:1998ki}.\\

\begin{figure}[H]
\centering
\begin{tikzpicture}
\node[draw=none,opacity=0,thick,scale=0.1,fill=black,label={[label distance=1mm]south:$S^3$}] (A1) at (0,0) {};
\node[draw=none,opacity=0,thick,scale=0.1,fill=black,label={[label distance=0.5mm]-20:$S^7$}] (A2) at (1.5,0.4) {};
\node[draw=none,opacity=0,thick,scale=0.1,fill=black,label={[label distance=0.5mm]north:M2-brane}] (A3) at (0.5,2.1) {};
\draw (-1,0)--(1,0)--(1.8,0.8);
\draw (1,0)--(1.2,1.8)--(1.8,0.8);
\draw (1.2,1.8)--(-0.1,1.8)--(-1,0);
\draw (1.2,1.9)--(-0.1,1.9);
\draw (1.2,2)--(-0.1,2);
\draw (1.2,2.1)--(-0.1,2.1);
\draw[->] (3,0.9)--(5.5,0.9);
\end{tikzpicture}
\begin{tikzpicture}
\node[draw=none,opacity=0,thick,scale=0.1,fill=black,label={[label distance=1mm]south:$S^3$}] (A1) at (0.25,0) {};
\node[draw=none,opacity=0,thick,scale=0.1,fill=black,label={[label distance=0.5mm]-20:$S^7$}] (A2) at (1.5,0.5) {};
\node[draw=none,opacity=0,thick,scale=0.1,fill=black,label={[label distance=0.5mm]west:flux}] (A3) at (0,2) {};
\node[draw=none,opacity=0,thick,scale=0.1,fill=black,label={[label distance=1mm]east:}] (A4) at (-2,0) {};
\draw (0.3,1.5)--(-0.5,0)--(1,0)--(0.3,1.5);
\draw (1,0)--(2,1)--(1.3,2.5)--(0.3,1.5);
\draw[->] (0.2,1.7)--(0.8,1.7);
\draw[->] (0.45,1.95)--(1.05,1.95);
\draw[->] (0.7,2.2)--(1.3,2.2);
\end{tikzpicture}
\caption{$AdS_4\times S^7$ geometry the branes on $AdS_4$ and fluxes on $S^7$, where $\int_{S^7} \star G=N$.}
\label{fig:AdS4S7}
\end{figure}
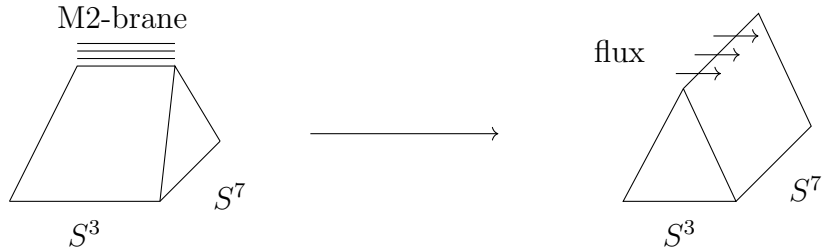

We can repeat these exercise relating spheres in different settings, such as for M2 branes and M5 branes in 11d flat space. Recall that an M2 brane has a three-dimensional worldvolume and couples electrically to the three-form gauge potential $C_3$. It can wrap an $S^3$, whie the eight transverse directions contain an $S^7$. In the ambient space, the $S^7$ is trivial whereas the $S^3$ is fixed, while in the near-horizon limit, the $S^7$ is stabilized by the seven-form flux $\star dC_3$, leading to an $AdS_4 \times S^7$ geometry. This transition is depicted in Figure \ref{fig:AdS4S7}. 

\begin{figure}[H]
\centering
\begin{tikzpicture}
\node[draw=none,opacity=0,thick,scale=0.1,fill=black,label={[label distance=1mm]south:$S^6$}] (A1) at (0,0) {};
\node[draw=none,opacity=0,thick,scale=0.1,fill=black,label={[label distance=0.5mm]-20:$S^4$}] (A2) at (1.5,0.4) {};
\node[draw=none,opacity=0,thick,scale=0.1,fill=black,label={[label distance=0.5mm]north:M5-brane}] (A3) at (0.5,2.1) {};
\draw (-1,0)--(1,0)--(1.8,0.8);
\draw (1,0)--(1.2,1.8)--(1.8,0.8);
\draw (1.2,1.8)--(-0.1,1.8)--(-1,0);
\draw (1.2,1.9)--(-0.1,1.9);
\draw (1.2,2)--(-0.1,2);
\draw (1.2,2.1)--(-0.1,2.1);
\draw[->] (3,0.9)--(5.5,0.9);
\end{tikzpicture}
\begin{tikzpicture}
\node[draw=none,opacity=0,thick,scale=0.1,fill=black,label={[label distance=1mm]south:$S^6$}] (A1) at (0.25,0) {};
\node[draw=none,opacity=0,thick,scale=0.1,fill=black,label={[label distance=0.5mm]-20:$S^4$}] (A2) at (1.5,0.5) {};
\node[draw=none,opacity=0,thick,scale=0.1,fill=black,label={[label distance=0.5mm]west:flux}] (A3) at (0,2) {};
\node[draw=none,opacity=0,thick,scale=0.1,fill=black,label={[label distance=1mm]east:}] (A4) at (-2,0) {};
\draw (0.3,1.5)--(-0.5,0)--(1,0)--(0.3,1.5);
\draw (1,0)--(2,1)--(1.3,2.5)--(0.3,1.5);
\draw[->] (0.2,1.7)--(0.8,1.7);
\draw[->] (0.45,1.95)--(1.05,1.95);
\draw[->] (0.7,2.2)--(1.3,2.2);
\end{tikzpicture}
\caption{$AdS_7\times S^4$ geometry the branes on $AdS_7$ and fluxes on $S^4$.}
\label{fig:AdS7S4}
\end{figure}
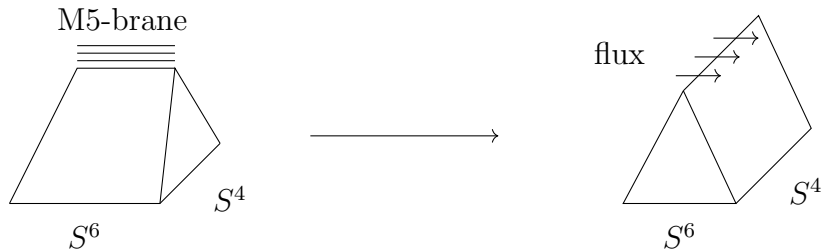

We can play the same game with M5 branes, where the 4 and 7 are essentially swapped. The conifold geometry again takes the form of a cone with base $S^6 \times S^4$, where now the branes wrap the $S^6$ and $S^4$ lives in the transverse directions. After taking the near-horizon limit, we find an $AdS_7 \times S^4$ geometry with $S^4$ stabilized by the four-form flux $dC_3$. This is represented in Figure \ref{fig:AdS7S4}.

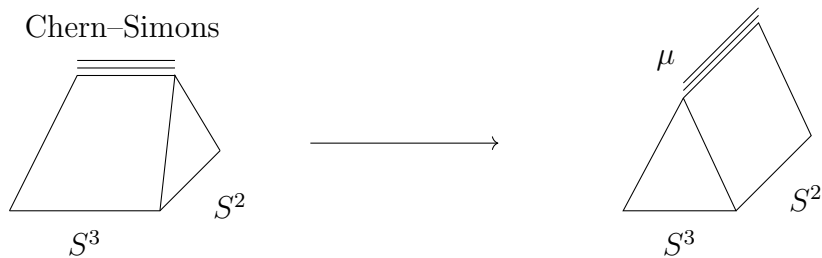
\begin{figure}[H]
\centering
\begin{tikzpicture}
\node[draw=none,opacity=0,thick,scale=0.1,fill=black,label={[label distance=1mm]south:$S^3$}] (A1) at (0,0) {};
\node[draw=none,opacity=0,thick,scale=0.1,fill=black,label={[label distance=0.5mm]-20:$S^2$}] (A2) at (1.5,0.4) {};
\node[draw=none,opacity=0,thick,scale=0.1,fill=black,label={[label distance=0.5mm]north:Chern--Simons}] (A3) at (0.5,2.1) {};
\draw (-1,0)--(1,0)--(1.8,0.8);
\draw (1,0)--(1.2,1.8)--(1.8,0.8);
\draw (1.2,1.8)--(-0.1,1.8)--(-1,0);
\draw (1.2,1.9)--(-0.1,1.9);
\draw (1.2,2)--(-0.1,2);
\draw[->] (3,0.9)--(5.5,0.9);
\end{tikzpicture}
\begin{tikzpicture}
\node[draw=none,opacity=0,thick,scale=0.1,fill=black,label={[label distance=1mm]south:$S^3$}] (A1) at (0.25,0) {};
\node[draw=none,opacity=0,thick,scale=0.1,fill=black,label={[label distance=0.5mm]-20:$S^2$}] (A2) at (1.5,0.5) {};
\node[draw=none,opacity=0,thick,scale=0.1,fill=black,label={[label distance=0.5mm]west:$\mu$}] (A3) at (0.4,2) {};
\node[draw=none,opacity=0,thick,scale=0.1,fill=black,label={[label distance=1mm]east:}] (A4) at (-2,0) {};
\draw (0.3,1.5)--(-0.5,0)--(1,0)--(0.3,1.5);
\draw (1,0)--(2,1)--(1.3,2.5)--(0.3,1.5);
\draw (0.3,1.6)--(1.3,2.6);
\draw (0.3,1.7)--(1.3,2.7);=
\end{tikzpicture}
\caption{$S^3\times S^2$ geometry with Chern--Simons theory on $S^3$ yielding a conifold with $\mu=N g_s$.}
\label{fig:S2S3}
\end{figure}
There are also similar effects that arise in the context of topological strings \cite{Gopakumar:1998ki}. Recall the conifold construction with base $S^2 \times S^3$. Adding $N \gg 1$ branes to the topology theory gives a Chern-Simons theory in target space, as represented in the left diagram of Figure \ref{fig:S2S3}. Here, the left diagram corresponds to the blown-up singularity, while the right diagram corresponds to the deformed singularity. In the right diagram, we find a large $N$ dual theory where the original branes are replaced by a size $\mu = N g_s$.

\subsubsection*{AdS and positive curvature spaces}

In all of the brane constructions considered so far we have an $AdS_{p+1} \times S^q$ spacetime in the near-horizon limit. One could imagine scenarios where the $S^q$ is replaced by another compact manifold. However, it turns out that all such replacements must have positive curvature

Negative $\Lambda$ means we need to have the curvature of the internal manifold to be positive. With negative $\Lambda$, we have an anti-de Sitter space.

\begin{figure}[H]
\centering
\scalebox{1}{
\tikzset{every picture/.style={line width=0.75pt}} 
\begin{tikzpicture}[x=0.75pt,y=0.75pt,yscale=-1,xscale=1]
\draw  (123,128) -- (524,128)(173,7) -- (173,286) (517,123) -- (524,128) -- (517,133) (168,14) -- (173,7) -- (178,14)  ;
\draw    (176,43) .. controls (178,227) and (203,239) .. (235,240) .. controls (267,241) and (360.88,142.61) .. (500,137) ;
\draw [color={rgb, 255:red, 74; green, 144; blue, 226 }  ,draw opacity=1 ] [dash pattern={on 4.5pt off 4.5pt}]  (235,128) -- (235,240) ;
\draw (529,140.4) node [anchor=north west][inner sep=0.75pt]    {$r$};
\draw (126,20.4) node [anchor=north west][inner sep=0.75pt]    {$V( r)$};
\draw (231,102.4) node [anchor=north west][inner sep=0.75pt]    {$r_{s}$};
\draw (419,150.4) node [anchor=north west][inner sep=0.75pt]    {$\sim r^{-b}$};
\draw (183,67.4) node [anchor=north west][inner sep=0.75pt]    {$\sim r^{-a}$};
\end{tikzpicture}}
\caption{The potential of the theory is given by the sum of two terms $V_G(r)\propto \int|G|^2$ and $V_\mathcal{R}\propto \int \mathcal{R}$, which has its minimum at $r_s\sim (g_s N)^{1/4}$ where the radius is stabilized.}
\label{fig:potential}
\end{figure}
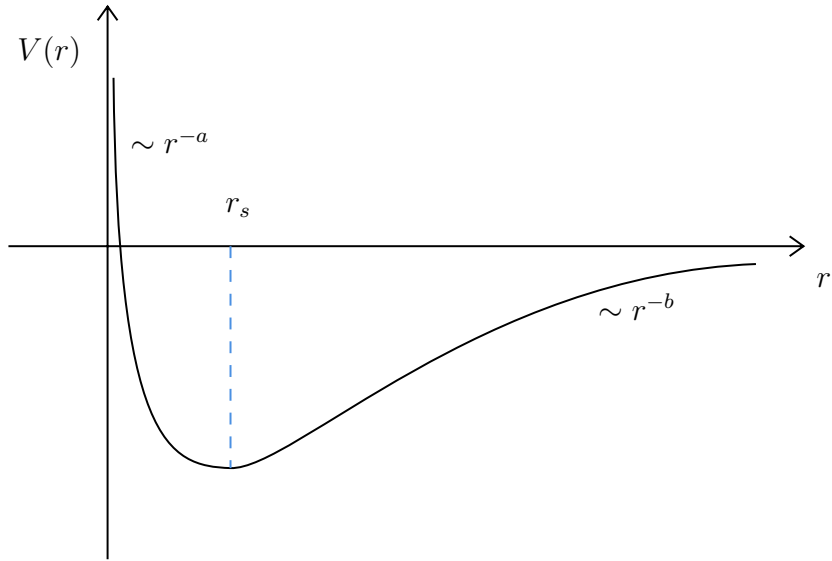

\noindent{\bf Exercise 2}: This exercise considers the case of $AdS_5\times S^5$. First we compactify type IIB theory on an $S^5$ (use D3-branes) with the 5-form R–R flux, which is self-dual, that has $N$ units around the $S^5$:
\begin{equation}
\int_S^5 G_5=N,
\end{equation}
where the internal size will be $R_{int}\sim 1/r^2$, which determines the cosmological constant to be $\Lambda\sim -1/r^2$. Now we have to go to the Einstein frame in $(4+1)$-dimensions and consider the potential $V$ that depends on $r$ such that
\begin{align}
V(r)=-\Lambda (r).
\end{align}
Then compute the potential $V(r)$ of the theory. The potential $V(r)$ will then look like in Figure \ref{fig:potential}, where its minimum is at $r=r_s$ with
\begin{align}
V(r)=\frac{A}{r^a}-Br^b,
\end{align}
where the first term comes from $\int |G|^2$ and the second term from $\int \mathcal{R}$ both in the compact manifold. Then more precisely, find $a$ and $b$ for such a potential and show that
\begin{equation}
r_s\sim (g_s N)^{1/4}.
\end{equation}
[Hint: The curvature of the radius is related to $r_s$. When $g_s\to 0$, the $AdS_5$ shrinks and only see the boundary $S^4$.]

\noindent{\bf Optional exercise}: Do the same computation for the M-theory compactification.

These are some examples of holographies we covered. In fact we can have more broad versions of holography and it requires at least a relaxed version of the Einstein's equation. The nicest thing to have is when 
\begin{align}
R_{ij}\propto g_{ij},
\end{align}
such as Sasaki--Einstein manifold. In some sense these manifolds are related to Calabi--Yau. For example in the example we covered we had $S^5$. The boundary of $S^5$ is $\mathds{C}^3$ and when orbifolded, this can be a Calabi--Yau manifold:
\begin{align}
\mathds{C}^3/\Gamma \longrightarrow S^5/\Gamma.
\end{align}
Then by acting with orbifolds on the holographic side, we get new theories. Interestingly, all the Calabi--Yau manifold we get via holography is noncompact. 

\pagebreak

\part{The Swampland program}
\sectionbypart
\section{Introduction to Swampland program}\label{secintro}

\subsection{Basic features of quantum field theories}

We begin with a quick review of what constitutes a well-defined quantum field theory (QFT). QFTs are powerful theories that allow us to compute various physical quantities with a wide range of available perturbative and non-perturbative techniques. Whenever we talk about a QFT, in addition to the computational machinery, we are also thinking of an underlying mathematical structure (e.g. the space of operators and their algebra) that satisfy some fundamental principles such as unitarity. The list of expected criteria that a good QFT must satisfy could be long, but in low dimensions such as $d\leq 4$, we have developed a good understanding of what constitutes a "healthy" QFT. In the following, we will review some of the most important features of "good" QFTs.

A quantum field theory usually comes with an action. The action $S[\Phi]$ is a classical functional of the local fields in the theory. The quantum field theory builds upon the classical theory by associating physical quantities with path integrals
\begin{align}\label{pi}
    \langle{O(\Phi)}\rangle \sim  \int \mathcal{D}\Phi e^{-S[\Phi]}O(\Phi).
\end{align}

To write down the action, we usually start with a symmetry principle and find an action that respects that symmetry (spacetime symmetries, gauge symmetries, etc.). However, the symmetries do not always hold or even make sense at the quantum level. When that happens, we say the symmetry is anomalous, and if the gauge symmetry is anomalous, the theory is inconsistent. Local anomalies in four dimensions are associated with triangle Feynman diagrams. 

\begin{figure}[H]
    \centering
\tikzset{every picture/.style={line width=0.75pt}} 
\begin{tikzpicture}[x=0.75pt,y=0.75pt,yscale=-1,xscale=1]
\draw   (319.95,53.63) -- (346.99,103.65) -- (293.66,103.65) -- cycle ;
\draw    (319.76,55.61) .. controls (318.13,53.91) and (318.16,52.24) .. (319.86,50.61) .. controls (321.56,48.98) and (321.59,47.31) .. (319.96,45.61) .. controls (318.33,43.91) and (318.36,42.24) .. (320.06,40.61) .. controls (321.76,38.98) and (321.79,37.31) .. (320.16,35.61) .. controls (318.53,33.91) and (318.56,32.24) .. (320.26,30.61) .. controls (321.96,28.98) and (321.99,27.31) .. (320.36,25.61) .. controls (318.73,23.91) and (318.76,22.24) .. (320.46,20.61) .. controls (322.16,18.98) and (322.19,17.32) .. (320.56,15.62) .. controls (318.93,13.92) and (318.96,12.25) .. (320.66,10.62) -- (320.71,8) -- (320.71,8) ;
\draw  [color={rgb, 255:red, 0; green, 0; blue, 0 }  ,draw opacity=1 ][fill={rgb, 255:red, 0; green, 0; blue, 0 }  ,fill opacity=1 ] (318.23,55.61) .. controls (318.23,54.52) and (319,53.63) .. (319.95,53.63) .. controls (320.89,53.63) and (321.66,54.52) .. (321.66,55.61) .. controls (321.66,56.7) and (320.89,57.58) .. (319.95,57.58) .. controls (319,57.58) and (318.23,56.7) .. (318.23,55.61) -- cycle ;
\draw  [color={rgb, 255:red, 0; green, 0; blue, 0 }  ,draw opacity=1 ][fill={rgb, 255:red, 0; green, 0; blue, 0 }  ,fill opacity=1 ] (343.75,102.56) .. controls (343.75,101.47) and (344.52,100.58) .. (345.47,100.58) .. controls (346.42,100.58) and (347.18,101.47) .. (347.18,102.56) .. controls (347.18,103.65) and (346.42,104.53) .. (345.47,104.53) .. controls (344.52,104.53) and (343.75,103.65) .. (343.75,102.56) -- cycle ;
\draw  [color={rgb, 255:red, 0; green, 0; blue, 0 }  ,draw opacity=1 ][fill={rgb, 255:red, 0; green, 0; blue, 0 }  ,fill opacity=1 ] (292.9,102.78) .. controls (292.9,101.69) and (293.67,100.8) .. (294.62,100.8) .. controls (295.56,100.8) and (296.33,101.69) .. (296.33,102.78) .. controls (296.33,103.87) and (295.56,104.75) .. (294.62,104.75) .. controls (293.67,104.75) and (292.9,103.87) .. (292.9,102.78) -- cycle ;
\draw    (256.5,134.42) .. controls (256.77,132.08) and (258.07,131.04) .. (260.41,131.31) .. controls (262.75,131.58) and (264.06,130.54) .. (264.33,128.2) .. controls (264.6,125.86) and (265.9,124.82) .. (268.24,125.09) .. controls (270.58,125.36) and (271.89,124.32) .. (272.16,121.98) .. controls (272.43,119.64) and (273.73,118.6) .. (276.07,118.87) .. controls (278.41,119.14) and (279.72,118.1) .. (279.99,115.76) .. controls (280.26,113.42) and (281.56,112.38) .. (283.9,112.65) .. controls (286.24,112.92) and (287.55,111.88) .. (287.82,109.54) .. controls (288.09,107.2) and (289.39,106.16) .. (291.73,106.43) .. controls (294.07,106.7) and (295.38,105.66) .. (295.65,103.32) -- (296.33,102.78) -- (296.33,102.78) ;
\draw    (343.75,102.56) .. controls (346.1,102.33) and (347.38,103.39) .. (347.61,105.74) .. controls (347.84,108.09) and (349.12,109.15) .. (351.47,108.92) .. controls (353.82,108.69) and (355.1,109.75) .. (355.32,112.1) .. controls (355.55,114.45) and (356.83,115.51) .. (359.18,115.28) .. controls (361.53,115.06) and (362.81,116.12) .. (363.04,118.47) .. controls (363.26,120.82) and (364.54,121.88) .. (366.89,121.65) .. controls (369.24,121.42) and (370.52,122.48) .. (370.75,124.83) .. controls (370.98,127.18) and (372.26,128.24) .. (374.61,128.01) .. controls (376.96,127.79) and (378.24,128.85) .. (378.46,131.2) .. controls (378.69,133.55) and (379.97,134.61) .. (382.32,134.38) -- (385.5,137) -- (385.5,137) ;
\end{tikzpicture}
\end{figure}

A simple dimensional analysis shows that anomalies in $D$ dimensions correspond to diagrams with $1+D/2$ gauge boson legs. For example, in 10 dimensions, the relevant diagrams are hexagons with six external gauge boson legs. One can use the tetrad formalism to view gravity as a gauge theory where the gauge group is the local Lorentz group and the gauge bosons are the spin connections. Therefore, the six-point gravitational amplitude is related to the breakdown of general covariance. This is called gravitational anomaly and must vanish in a consistent theory of gravity. We will talk more about such anomalies in supergravity backgrounds.
\begin{figure}[H]
    \centering
\tikzset{every picture/.style={line width=0.75pt}} 
\begin{tikzpicture}[x=0.75pt,y=0.75pt,yscale=-.6,xscale=.6]
\draw    (233.5,284) .. controls (233.25,281.65) and (234.3,280.36) .. (236.65,280.11) .. controls (238.99,279.86) and (240.04,278.57) .. (239.79,276.23) .. controls (239.54,273.88) and (240.59,272.59) .. (242.94,272.34) .. controls (245.29,272.1) and (246.34,270.81) .. (246.09,268.46) .. controls (245.84,266.11) and (246.88,264.82) .. (249.23,264.57) .. controls (251.58,264.33) and (252.63,263.04) .. (252.38,260.69) .. controls (252.13,258.34) and (253.18,257.05) .. (255.53,256.8) .. controls (257.88,256.55) and (258.92,255.26) .. (258.67,252.91) .. controls (258.42,250.56) and (259.47,249.27) .. (261.82,249.03) .. controls (264.17,248.78) and (265.22,247.49) .. (264.97,245.14) .. controls (264.72,242.8) and (265.77,241.51) .. (268.11,241.26) .. controls (270.46,241.01) and (271.51,239.72) .. (271.26,237.37) .. controls (271.01,235.02) and (272.05,233.73) .. (274.4,233.48) .. controls (276.75,233.24) and (277.8,231.95) .. (277.55,229.6) .. controls (277.3,227.25) and (278.35,225.96) .. (280.7,225.71) .. controls (283.04,225.46) and (284.09,224.17) .. (283.84,221.83) .. controls (283.59,219.48) and (284.64,218.19) .. (286.99,217.94) -- (289.47,214.88) -- (289.47,214.88) ;
\draw   (265.56,154.88) -- (289.47,94.88) -- (369.15,94.88) -- (393.06,154.88) -- (369.15,214.88) -- (289.47,214.88) -- cycle ;
\draw    (426.5,282) .. controls (424.15,281.81) and (423.07,280.55) .. (423.25,278.2) .. controls (423.43,275.85) and (422.35,274.59) .. (420,274.4) .. controls (417.65,274.21) and (416.57,272.95) .. (416.76,270.6) .. controls (416.95,268.25) and (415.86,266.98) .. (413.51,266.79) .. controls (411.16,266.6) and (410.08,265.34) .. (410.26,262.99) .. controls (410.44,260.64) and (409.36,259.38) .. (407.01,259.19) .. controls (404.66,259) and (403.58,257.74) .. (403.76,255.39) .. controls (403.95,253.04) and (402.87,251.78) .. (400.52,251.59) .. controls (398.17,251.4) and (397.09,250.14) .. (397.27,247.79) .. controls (397.45,245.44) and (396.37,244.18) .. (394.02,243.99) .. controls (391.67,243.8) and (390.58,242.53) .. (390.77,240.18) .. controls (390.95,237.83) and (389.87,236.57) .. (387.52,236.38) .. controls (385.17,236.19) and (384.09,234.93) .. (384.28,232.58) .. controls (384.46,230.23) and (383.38,228.97) .. (381.03,228.78) .. controls (378.68,228.59) and (377.6,227.33) .. (377.78,224.98) .. controls (377.96,222.63) and (376.88,221.37) .. (374.53,221.18) .. controls (372.18,220.99) and (371.1,219.73) .. (371.29,217.38) -- (369.15,214.88) -- (369.15,214.88) ;
\draw    (483.5,155) .. controls (481.83,156.66) and (480.17,156.66) .. (478.5,154.99) .. controls (476.83,153.32) and (475.17,153.32) .. (473.5,154.99) .. controls (471.83,156.65) and (470.17,156.65) .. (468.5,154.98) .. controls (466.83,153.31) and (465.17,153.31) .. (463.5,154.97) .. controls (461.83,156.64) and (460.17,156.64) .. (458.5,154.97) .. controls (456.83,153.3) and (455.17,153.3) .. (453.5,154.96) .. controls (451.83,156.62) and (450.17,156.62) .. (448.5,154.95) .. controls (446.83,153.28) and (445.17,153.28) .. (443.5,154.95) .. controls (441.83,156.61) and (440.17,156.61) .. (438.5,154.94) .. controls (436.83,153.27) and (435.17,153.27) .. (433.5,154.94) .. controls (431.83,156.6) and (430.17,156.6) .. (428.5,154.93) .. controls (426.83,153.26) and (425.17,153.26) .. (423.5,154.92) .. controls (421.83,156.59) and (420.17,156.59) .. (418.5,154.92) .. controls (416.83,153.25) and (415.17,153.25) .. (413.5,154.91) .. controls (411.83,156.57) and (410.17,156.57) .. (408.5,154.9) .. controls (406.83,153.23) and (405.17,153.23) .. (403.5,154.9) .. controls (401.83,156.56) and (400.17,156.56) .. (398.5,154.89) .. controls (396.83,153.22) and (395.17,153.22) .. (393.5,154.88) -- (393.06,154.88) -- (393.06,154.88) ;
\draw    (265.56,154.88) .. controls (263.91,156.57) and (262.24,156.58) .. (260.56,154.93) .. controls (258.88,153.28) and (257.21,153.3) .. (255.56,154.98) .. controls (253.91,156.66) and (252.24,156.67) .. (250.56,155.02) .. controls (248.88,153.37) and (247.21,153.39) .. (245.56,155.07) .. controls (243.91,156.75) and (242.24,156.77) .. (240.56,155.12) .. controls (238.88,153.47) and (237.21,153.48) .. (235.56,155.16) .. controls (233.91,156.84) and (232.24,156.86) .. (230.56,155.21) .. controls (228.88,153.56) and (227.21,153.58) .. (225.56,155.26) .. controls (223.91,156.94) and (222.24,156.95) .. (220.56,155.3) .. controls (218.88,153.65) and (217.21,153.67) .. (215.56,155.35) .. controls (213.91,157.03) and (212.24,157.05) .. (210.56,155.4) .. controls (208.88,153.75) and (207.21,153.76) .. (205.56,155.44) .. controls (203.91,157.12) and (202.24,157.14) .. (200.56,155.49) .. controls (198.88,153.84) and (197.21,153.86) .. (195.56,155.54) .. controls (193.91,157.22) and (192.24,157.23) .. (190.56,155.58) .. controls (188.88,153.93) and (187.21,153.95) .. (185.56,155.63) .. controls (183.91,157.31) and (182.24,157.33) .. (180.56,155.68) .. controls (178.88,154.03) and (177.22,154.04) .. (175.57,155.72) -- (171.12,155.77) -- (171.12,155.77) ;
\draw    (433.5,26) .. controls (433.58,28.35) and (432.44,29.57) .. (430.09,29.65) .. controls (427.73,29.73) and (426.59,30.95) .. (426.67,33.31) .. controls (426.75,35.66) and (425.61,36.88) .. (423.26,36.96) .. controls (420.91,37.05) and (419.77,38.27) .. (419.85,40.62) .. controls (419.92,42.97) and (418.78,44.19) .. (416.43,44.27) .. controls (414.08,44.35) and (412.94,45.57) .. (413.02,47.92) .. controls (413.1,50.27) and (411.96,51.49) .. (409.61,51.58) .. controls (407.26,51.66) and (406.12,52.88) .. (406.19,55.23) .. controls (406.27,57.58) and (405.13,58.8) .. (402.78,58.88) .. controls (400.43,58.97) and (399.29,60.19) .. (399.37,62.54) .. controls (399.45,64.89) and (398.31,66.11) .. (395.96,66.19) .. controls (393.6,66.27) and (392.46,67.49) .. (392.54,69.85) .. controls (392.62,72.2) and (391.48,73.42) .. (389.13,73.5) .. controls (386.78,73.58) and (385.64,74.8) .. (385.72,77.15) .. controls (385.8,79.51) and (384.66,80.73) .. (382.3,80.81) .. controls (379.95,80.89) and (378.81,82.11) .. (378.89,84.46) .. controls (378.97,86.81) and (377.83,88.03) .. (375.48,88.12) .. controls (373.13,88.2) and (371.99,89.42) .. (372.06,91.77) -- (369.15,94.88) -- (369.15,94.88) ;
\draw    (223.5,20) .. controls (225.85,20.15) and (226.96,21.4) .. (226.81,23.75) .. controls (226.66,26.1) and (227.76,27.35) .. (230.11,27.5) .. controls (232.46,27.65) and (233.57,28.91) .. (233.42,31.26) .. controls (233.27,33.61) and (234.37,34.86) .. (236.72,35.01) .. controls (239.07,35.16) and (240.18,36.41) .. (240.03,38.76) .. controls (239.88,41.11) and (240.98,42.36) .. (243.33,42.51) .. controls (245.68,42.66) and (246.79,43.91) .. (246.64,46.26) .. controls (246.49,48.61) and (247.59,49.86) .. (249.94,50.01) .. controls (252.29,50.16) and (253.4,51.42) .. (253.25,53.77) .. controls (253.1,56.12) and (254.2,57.37) .. (256.55,57.52) .. controls (258.9,57.67) and (260.01,58.92) .. (259.86,61.27) .. controls (259.71,63.62) and (260.81,64.87) .. (263.16,65.02) .. controls (265.51,65.17) and (266.62,66.42) .. (266.47,68.77) .. controls (266.32,71.12) and (267.42,72.38) .. (269.77,72.53) .. controls (272.12,72.68) and (273.23,73.93) .. (273.08,76.28) .. controls (272.93,78.63) and (274.03,79.88) .. (276.38,80.03) .. controls (278.73,80.18) and (279.84,81.43) .. (279.69,83.78) .. controls (279.54,86.13) and (280.64,87.38) .. (282.99,87.53) .. controls (285.34,87.68) and (286.45,88.93) .. (286.3,91.28) -- (289.47,94.88) -- (289.47,94.88) ;
\end{tikzpicture}
\end{figure}
Another important feature of QFTs is the dependence of physical quantities on the energy scales. This dependence is often captured by the RG flow. There is a class of renormalizable field theories that become free theories at high energies. These theories are called asymptotically free and we can describe their UV completion without appealing to a cut-off or anything beyond QFT. Another group of such promising QFTs are scale invariant theories. 

For a generic QFT we are not this lucky and we usually need to define a cut-off. Depending on whether the theory is renormalizable or not, we need finite or infinite number of parameters to define the theory below the cut-off. This approach is called effective field theory (EFT). The idea is that we can capture the physics at a certain energy scale by an effective field theory and if we are lucky enough (the theory is renormalizable) we can find the corresponding effective theory at other energy scales below the cutoff as well. We will come back to this approach later.

\subsection{Quantum gravity vs quantum field theory}

So far we talked about "\textit{good}" QFTs. Now we turn to an important question; what is the role of gravity here? If we add it to a good field theory, could it still be a good QFT? 

The idea to incorporate gravity into QFT is to view the metric as a field that interacts with other fields and proceed with the quantization procedure. For example, the action of a free massless scalar can be covariantized to the following action that includes interaction with the gravitational field (metric $g_{\mu\nu}$).

\begin{align}
    S[\phi]=\int_{M^4}d^4x\sqrt{g}[\frac{\mathcal{R}}{2\kappa}+\frac{1}{2}\partial_\mu\phi\partial_\nu\phi g^{\mu\nu}],
\end{align}

where $\kappa=8\pi G$. If we decide to follow the path integral formulation \ref{pi} for the above action, we need to mod out by the diffeomorphisms. Schematically, we can write that as

\begin{align}
   \int \frac{\mathcal{D}g}{\text{Vol}[Diff_{\mathcal{M}}]}\mathcal{D}\Phi  e^{-S[\Phi,g]}.
\end{align}
But does this formula make computational sense? Feynman tried to use the tools of perturbative QFT for gravity but he quickly ran into problems \cite{Feynman:1963ax}. Let us demonstrate those problems. Every theory has relevant and irrelevant operators depending on whether the coefficient of the corresponding operator goes to $\infty$ or $0$ as the energy scale decreases. If the theory does not have irrelevant operators, we are in a good shape. For example the coupling constants in QCD is defined by a specific relevant operator and the rest of the observables are calculable from that coupling constant, which can be traded with an energy scale by dimensional transmutation.

For gravity, the effective coupling for the graviton scattering goes like 
\begin{align}g_{eff}(E)^2\sim\frac{E^2}{M_P^2}.\end{align}

\begin{figure}[H]
    \centering

\tikzset{every picture/.style={line width=0.75pt}} 

\begin{tikzpicture}[x=0.75pt,y=0.75pt,yscale=-1,xscale=1]

\draw    (274,58) .. controls (275.67,56.33) and (277.33,56.33) .. (279,58) .. controls (280.67,59.67) and (282.33,59.67) .. (284,58) .. controls (285.67,56.33) and (287.33,56.33) .. (289,58) .. controls (290.67,59.67) and (292.33,59.67) .. (294,58) .. controls (295.67,56.33) and (297.33,56.33) .. (299,58) .. controls (300.67,59.67) and (302.33,59.67) .. (304,58) .. controls (305.67,56.33) and (307.33,56.33) .. (309,58) .. controls (310.67,59.67) and (312.33,59.67) .. (314,58) .. controls (315.67,56.33) and (317.33,56.33) .. (319,58) .. controls (320.67,59.67) and (322.33,59.67) .. (324,58) .. controls (325.67,56.33) and (327.33,56.33) .. (329,58) .. controls (330.67,59.67) and (332.33,59.67) .. (334,58) .. controls (335.67,56.33) and (337.33,56.33) .. (339,58) .. controls (340.67,59.67) and (342.33,59.67) .. (344,58) .. controls (345.67,56.33) and (347.33,56.33) .. (349,58) .. controls (350.67,59.67) and (352.33,59.67) .. (354,58) .. controls (355.67,56.33) and (357.33,56.33) .. (359,58) .. controls (360.67,59.67) and (362.33,59.67) .. (364,58) .. controls (365.67,56.33) and (367.33,56.33) .. (369,58) .. controls (370.67,59.67) and (372.33,59.67) .. (374,58) .. controls (375.67,56.33) and (377.33,56.33) .. (379,58) .. controls (380.67,59.67) and (382.33,59.67) .. (384,58) .. controls (385.67,56.33) and (387.33,56.33) .. (389,58) -- (389,58) ;
\draw    (237,19) .. controls (239.35,19.07) and (240.5,20.28) .. (240.44,22.63) .. controls (240.38,24.98) and (241.53,26.19) .. (243.88,26.25) .. controls (246.23,26.32) and (247.38,27.53) .. (247.32,29.88) .. controls (247.26,32.24) and (248.41,33.45) .. (250.77,33.51) .. controls (253.12,33.58) and (254.27,34.79) .. (254.21,37.14) .. controls (254.15,39.49) and (255.3,40.7) .. (257.65,40.76) .. controls (260,40.83) and (261.15,42.04) .. (261.09,44.39) .. controls (261.03,46.74) and (262.18,47.95) .. (264.53,48.02) .. controls (266.88,48.09) and (268.03,49.3) .. (267.97,51.65) .. controls (267.91,54) and (269.06,55.21) .. (271.41,55.27) -- (274,58) -- (274,58) ;
\draw    (274,58) .. controls (273.94,60.35) and (272.73,61.5) .. (270.38,61.44) .. controls (268.02,61.38) and (266.81,62.53) .. (266.75,64.89) .. controls (266.69,67.24) and (265.48,68.39) .. (263.13,68.33) .. controls (260.78,68.27) and (259.57,69.42) .. (259.5,71.77) .. controls (259.44,74.12) and (258.23,75.27) .. (255.88,75.22) .. controls (253.53,75.16) and (252.32,76.31) .. (252.25,78.66) .. controls (252.19,81.01) and (250.98,82.16) .. (248.63,82.11) .. controls (246.28,82.05) and (245.07,83.2) .. (245,85.55) .. controls (244.94,87.9) and (243.73,89.05) .. (241.38,88.99) .. controls (239.02,88.93) and (237.81,90.08) .. (237.75,92.44) .. controls (237.69,94.79) and (236.48,95.94) .. (234.13,95.88) -- (234,96) -- (234,96) ;
\draw    (424,98) .. controls (421.65,97.89) and (420.53,96.65) .. (420.64,94.3) .. controls (420.75,91.95) and (419.63,90.71) .. (417.28,90.6) .. controls (414.93,90.48) and (413.81,89.24) .. (413.92,86.89) .. controls (414.03,84.54) and (412.91,83.3) .. (410.56,83.19) .. controls (408.21,83.08) and (407.08,81.84) .. (407.19,79.49) .. controls (407.3,77.14) and (406.18,75.9) .. (403.83,75.79) .. controls (401.48,75.68) and (400.36,74.44) .. (400.47,72.09) .. controls (400.58,69.74) and (399.46,68.5) .. (397.11,68.39) .. controls (394.76,68.27) and (393.64,67.03) .. (393.75,64.68) .. controls (393.86,62.33) and (392.74,61.09) .. (390.39,60.98) -- (388.1,58.46) -- (388.1,58.46) ;
\draw    (388.1,58.46) .. controls (387.99,56.11) and (389.11,54.87) .. (391.46,54.76) .. controls (393.81,54.65) and (394.94,53.41) .. (394.83,51.06) .. controls (394.72,48.71) and (395.84,47.47) .. (398.19,47.36) .. controls (400.54,47.25) and (401.67,46.02) .. (401.56,43.67) .. controls (401.45,41.32) and (402.57,40.08) .. (404.92,39.97) .. controls (407.27,39.86) and (408.4,38.62) .. (408.29,36.27) .. controls (408.18,33.92) and (409.3,32.68) .. (411.65,32.57) .. controls (414,32.46) and (415.13,31.22) .. (415.02,28.87) .. controls (414.91,26.52) and (416.03,25.28) .. (418.38,25.17) .. controls (420.73,25.06) and (421.86,23.83) .. (421.75,21.48) -- (424,19) -- (424,19) ;

\draw (321,31.4) node [anchor=north west][inner sep=0.75pt]    {$E$};

\end{tikzpicture}
\end{figure}

Where $M_P$ is the reduced Planck mass\footnote{Planck units conventions: $l_P=(\hbar G/c^3)^{1/(D-2)},~t_P=l_P/c,~m_P=\hbar/(cl_P),~M_P=m_p/(8\pi)^{1/(D-2)},~T_P=m_pc^2/K_B$\label{Planck}}. This is different from the usual logarithmic energy dependence $1/g(E)^2\sim\ln(E)$ and becomes strong for $E>>M_P$. One might be tempted to take the EFT approach explained earlier and input enough coupling constants from experiment to find couplings at other energy scales. However, it turns out you have infinitely many vertex operators with UV-divergent amplitudes that need to be kept track of. In other words, our theory needs infinitely many parameters as input and is not predictive! So QFT+gravity, at least in the most naive sense, seems to be problematic.  But where is the problem really coming from? 

An important remark before we go further:

As long as we are dealing with energies below the Planck scale $E<<M_P$, the changes in the  couplings are small and it is reasonable to believe that we have a nice classical effective field theory. In other words, the Planck mass $M_P$ introduces a natural scale beyond which the EFT should start to break down.

\subsection{Why is gravity special?}

To see why the canonical method of QFT could not have worked for gravity we need to take a detour through the realm of black holes. 

Black holes are solutions to the classical equations of motion for gravity. Large 4d black holes of mass $M$ have radius of $\sim M l_P/M_P$ and the spacetime outside their horizon is weakly curved $\mathcal{R}l_P^2\lesssim M_P^2/M^2$ for $M\gg M_P$. Since the curvature is very small ($\mathcal{R}l_P^2\ll 1$) we can think of black holes as IR backgrounds where Einstein's classical equations are reliable. 

\begin{figure}[H]
    \centering

\tikzset{every picture/.style={line width=0.75pt}} 

\begin{tikzpicture}[x=0.75pt,y=0.75pt,yscale=-1,xscale=1]

\draw  [fill={rgb, 255:red, 0; green, 0; blue, 0 }  ,fill opacity=1 ] (350,124) .. controls (350,77.06) and (388.06,39) .. (435,39) .. controls (481.94,39) and (520,77.06) .. (520,124) .. controls (520,170.94) and (481.94,209) .. (435,209) .. controls (388.06,209) and (350,170.94) .. (350,124) -- cycle ;

\draw (398,231) node [anchor=north west][inner sep=0.75pt]   [align=left] {4d black hole};
\draw (21,17.4) node [anchor=north west][inner sep=0.75pt]    {$r_{H} \sim \ \frac{M}{M_{P}} l_{P},$};
\draw (18,66.4) node [anchor=north west][inner sep=0.75pt]    {$Area\ \sim r_{H}^{2} \sim \frac{M^{2}}{M_{P}^{2}} l_{P}^{2},$};
\draw (16,109.4) node [anchor=north west][inner sep=0.75pt]    {$Curvature\sim \mathcal{R} \sim \frac{1}{r_{H}^{2}} \sim \left(\frac{M_{P}}{M}\right)^{2} l_{P}^{-2},$};
\draw (16,176.4) node [anchor=north west][inner sep=0.75pt]    {$Temperature\sim \kappa \ \sim \frac{M_{P}}{M} T_{P},$};
\draw (18,235.4) node [anchor=north west][inner sep=0.75pt]    {$Entropy\sim Area\sim \frac{M^{2}}{M_{P}^{2}},$};

\end{tikzpicture}
\end{figure}

Black holes in $d=4$ are very simple in the sense that a classically stationary black hole is described by only three parameters: mass, charge, and angular momentum. Bekenstein speculated that there should be more degrees of freedom for black holes. Otherwise, by dropping a thermal system with entropy into a black hole we can decrease universe's entropy and violate the second law of thermodynamics. In 1971 Hawking had shown the overall areas of the black holes horizons always increases in collisions \cite{Hawking:1971tu}. This result sounded very similar to the second law of thermodynamics to Bekenstein and it motivated him to propose that black holes must carry an entropy proportional to the area of their horizons \cite{Bekenstein:1972tm}. If black holes really have entropy, we can also define a temperature for them according to the first law of thermodynamics $dE=TdS$. Hawking managed to find the temperature of black holes by showing that they emit an almost thermal radiation with temperature $T=\frac{\kappa}{2\pi}$ in Planck units where $\kappa$ is the surface gravity of black holes \cite{Hawking:1975vcx,Hawking:1976de}. This equation also fixed the proportionality constant between black hole's entropy and area and lead to the following equation in Planck units.

\begin{align}
    S=\frac{A}{4}.
\end{align}

This single equation is the starting point of many of the strange aspects of quantum gravity! This equation implies that despite the classical uniqueness of black holes, there must be a huge number of degree of freedom represented by a black hole of mass M. In particular, this implies the number of high energy bound states grows as $\exp(S(E))\sim \exp(c E^2)$ for some constant $c$.

But the fact that IR object such as large black hole have so much information about very UV states is very strange. It implies that extremely low energy physics (\emph{i.e.} large black holes) and extremely high energy states somehow know about each other which is completely contrary to the EFT perspective. This UV-IR dependence is a remarkable failure of UV-IR decoupling used in EFT. With this knowledge, it is much easier to see why Feynman's approach to quantize gravity could have never worked, because the premise of EFT and renormalization is to neglect the UV physics by renormalizing IR parameters, but a quantum treatment of gravity even at large distances requires incorporating the high energy degrees of freedom. Another reflection of this is that scattering of gravitons at very high energies (UV) proceeds via large intermediate black holes (IR).

\textbf{Exercise 1: }
\vspace{10pt}

\textbf{Part 1: } Find the relationship between the mass of the black hole and the radius of its horizon in any dimension $d>3$. Numerical coefficients are not necessary. Just find the correct power law dependence.

Hint: You can use find the exponent of $r$ in $g_{tt}$ of higher dimensional Schwarzschild solution using gravitational Green's function in higher dimensions. 
\vspace{10pt}

\noindent\textbf{Part 2: } Assuming that entropy $S$ of a black hole of mass $E$ grows like the volume of the horizon, find $a$ such that $S(E)\sim (E/m_P)^a$ and show that $a>1$ in $d>3$\footnote{Many of the equations and inequalities break down in $d\leq3$ because the gravitational degrees of freedom are topological.}. Argue that in quantum gravity (QG), the number of single particle states (number of states with $\langle{X^i}\rangle=\langle{P^i}\rangle=0$ for an arbitrarily small but fixed IR cut-off) of mass $M$ must grow like $\rho(M)\gtrsim e^{b(M/m_P)^a}$ for some constants $b>0$ and $a>1$. 
\vspace{10pt}

\textbf{Part 3: } Show that in any (non-gravitational) theory where the theory at any finite energy range is described by an EFT, $\rho(M)\lesssim \exp{b M}$ for some constant $b>0$. Note that we are assuming there is no energy scale where EFT approach does not work, so that the high energy limit makes sense. 
\vspace{10pt}

Hint: consider the thermodynamic partition function of this theory in a big box at small temperatures. 

\textbf{Exercise 2: } How do you think gravity could avoid the upper bound of the previous exercise?
\vspace{10pt}

From the exercises, we see that Hawking's semi-classical calculation has dramatic implications and that QFTs are very different from QGs. Moreover, it seems that black holes play a key role in highlighting that difference. As we will see in the course, black holes are the star of the show in many aspects of quantum gravity.

Luckily, we do not have to rely entirely on semi-classical celculations to study QG because we already know some examples of QG. An existing consistent theory of quantum gravity is string theory. The reason we think it is consistent is that we have consistent perturbative descriptions of it in addition to non-perturbative dualities that relate those perturbative descriptions to each other. So string theory is at least "an" example of QG. But string theory is not a theory of particles and this suggests another explanation for why Feynman's argument failed.

The fundamental classical objects in string theory are extended objects such as strings. These strings can oscillate in different ways and the amplitudes of these oscillations can be viewed as different degrees of freedom. This suggests, the degrees of freedom of a black hole cannot be entirely comprised of particles, but we also need stringy oscillations. String theory is strange in that it is unusually specific. For example super string theories must be ten dimensional. Moreover, the set of theories is very restricted as opposed to EFTs where we have so much freedom in choosing the gauge group, dimension, matter content, etc. Also, supersymmetry seems to be more than a nice accessory in string theory since the only well-understood stable examples of string theory are supersymmetric!

One might wonder how could a ten dimensional theory describe our four dimensional universe? One answer is that the extra dimensions could be compact and small and there are limited options for how they could look. We will discuss this in more detail in the course. 
\vspace{10pt}

\textbf{Question :} We do not see low-energy supersymmetry in our universe.  So if supersymmetry is a necessary feature of string theory are we not in trouble?  
\vspace{10pt}

Luckily, supersymmetry is not required and can be broken in string theory. However, usually the breaking of supersymmetry comes with losing some sort of controlablity. For example, you lose stability in the sense that the value of the coupling constants may vary over time. Also applied to our universe, which seems to be non-supersummetric, string theory suggest our universe should decay in some sense. We will talk more about this in the course.

\subsection{Problems of treating gravitational theories as EFTs}

Let us review some of the basic features of the EFT approach. In the EFT approach, we start with a symmetry and write an action that includes all the relevant terms that are consistent with that symmetry. EFTs usually come with a cutoff based on the premise that the low-energy physics can be described independently from the high energy physics. This principle is called \textbf{\textit{UV/IR decoupling}}.

We typically assume that the coupling constants are of order one in the appropriate mass scale of the theory. This principle is called \textbf{\textit{naturalness}}. Naturalness implies that if some of these parameters is unusually small or large, there has to be a good explanation for it, e.g. a missed symmetry. This is a cherished principle in particle physics.

The combination of UV/IR decoupling and naturalness is very powerful and has lead to many successful predictions in standard model which is why particle physicist have so much confidence in these principles. 

In the following, we list some of the important questions that arise due to tensions between experimental observations and the principles of the EFT approach. 
\vspace{10pt}

\textbf{Question 1:} Why is dimension of spacetime four in our universe?
\vspace{10pt}

This is a question that is not commonly asked. However, from the naturalness point of view, it is strange that we are living in such a small dimension. If every dimension from 1 to $\infty$ is allowed, $d=4$ seems a very unnatural choice.

One potential answer could be that QFTs have special properties in low dimensions, especially 4. For starters, we do not even know of any consistent UV complete QFT in more than 6 dimensions. So there could be some truth to this argument.
\vspace{10pt}

\textbf{Question 2:} Why is the rank of the standard model gauge group so small?
\vspace{10pt}

Again, if every rank is allowed, why 4? An EFTheorist might view this as a question about naturalness since we usually fix our symmetries and then proceed with finding the right action. Nonetheless, this is an interesting question with no clear answer from the field theory point of view.
\vspace{10pt}

\textbf{Question 3:} Suppose we have fixed the dimension to $4$ and the gauge group to $U(1)\times SU(2)\times SU(3)$. Why are the representations of the gauge group that appear in our universe so small (\emph{i.e.} fundamental and adjoint)? 
\vspace{10pt}

One might argue that this could be related to asymptotic freedom since including arbitrarily large representations change the sign of the beta function and destroys asymptotic freedom. However, from the EFT perspective, why should we care about asymptotic freedom? Even if our theory is not asymptotically free, we can always put a cutoff and proceed with calculation as long as the theory is renormalizable. Furthermore, from string theory, we have examples that show us asymptotic freedom for QFT is not a requirement for UV completion. So in quantum gravity, asymptotic freedom is not a good guide and it cannot explain the smallness of representations.
\vspace{10pt}

\textbf{Question 4 (the hierarchy problem):} Why is the vev of the Higgs field so much smaller than $M_P$?
\vspace{10pt}

There is no principle for the mass of the Higgs field to be so small compared to the cut off of the theory. EFTheorists have tried to find a symmetry-based explanation for this. For example, one proposal is to explain via weakly broken supersymmetry (SUSY) since SUSY implies non-renormalization theorems that prevents $m$ from running to $\mathcal{O}(M_P)$. However, none of the explanations so far have been quite successful in providing a natural explanation for this puzzle while keep being compatible with incoming experimental results. 
\vspace{10pt}

\textbf{Question 5 (CC problem):} From the EFT perspective, there is no apriori reason for the cosmological constant $\Lambda$ to not be of the order of $\Lambda_{EFT}^2$ where $\Lambda_{EFT}$ is the cutoff. However, in our universe, we have $\Lambda\sim 10^{-122}M_P^4$! $10^{-122}$ is not order one by any means.
\vspace{10pt}

A popular solution to remedy this problem in the EFT picture is the Anthropic argument. The argument roughly goes like this: If $\Lambda$ were much greater than its measured value, the expansion of the universe would be too fast for large gravitationally bound structures such as large galaxies to exits. Without large galaxies, there would be no star formations and heavy elements hence life as we know it could not exist. So if there were many many possible theories (universes) with different values of $\Lambda$, the existence of the humankind that poses this question, already puts an upper bound on $\Lambda$ which is just a few orders of magnitude higher than its measure value. Weinberg estimated this upper bound before $\Lambda$ was measured and the upper bound turned out to be just couple of orders of magnitude away from the measured value \cite{Weinberg:1987dv}. This is a powerful example of scientific methodology. However, is this argument enough or there is a more fundamental reason for the smallness of $\Lambda$? Maybe, or maybe not...
\vspace{10pt}

\textbf{Question 6 (another hierarchy problem):} Why are the masses of some particles orders of magnitudes different from each other? For example, the mass of neutrinos are very small compared to the Higgs mass. Interestingly, the mass of the neutrinos is almost $m_\nu\sim\Lambda^{1/4}$ in Planck units where $\Lambda$ is the cosmological constant. Is this a coincident?
\vspace{10pt}

\textbf{Question 7 (dark energy):}  The common interpretation of dark energy is the value of the scalar potential. If we are stuck in a local minimum of the potential, we must have $\nabla V(\Phi)=0$. The existing experiments can only tell us that $|\nabla V|/V$ is smaller than some $\mathcal{O}(1)$ constant. So, if $|\nabla V|$ is not zero, it must be extremely small and finely tuned which would pose another naturalness problem. 
\vspace{10pt}

\begin{figure}[H]
    \centering

\tikzset{every picture/.style={line width=0.75pt}} 

\begin{tikzpicture}[x=0.75pt,y=0.75pt,yscale=-1,xscale=1]

\draw  (116,225.9) -- (520,225.9)(156.4,63) -- (156.4,244) (513,220.9) -- (520,225.9) -- (513,230.9) (151.4,70) -- (156.4,63) -- (161.4,70)  ;
\draw    (214,52) .. controls (253,77) and (238,158) .. (304,167) ;
\draw  [dash pattern={on 4.5pt off 4.5pt}]  (304,167) .. controls (359,174) and (366,195) .. (431,195) ;
\draw [color={rgb, 255:red, 155; green, 155; blue, 155 }  ,draw opacity=1 ]   (307,169) -- (307,223) ;
\draw [shift={(307,225)}, rotate = 270] [color={rgb, 255:red, 155; green, 155; blue, 155 }  ,draw opacity=1 ][line width=0.75]    (10.93,-3.29) .. controls (6.95,-1.4) and (3.31,-0.3) .. (0,0) .. controls (3.31,0.3) and (6.95,1.4) .. (10.93,3.29)   ;
\draw [shift={(307,167)}, rotate = 90] [color={rgb, 255:red, 155; green, 155; blue, 155 }  ,draw opacity=1 ][line width=0.75]    (10.93,-3.29) .. controls (6.95,-1.4) and (3.31,-0.3) .. (0,0) .. controls (3.31,0.3) and (6.95,1.4) .. (10.93,3.29)   ;
\draw   (296,167) .. controls (296,160.92) and (300.92,156) .. (307,156) .. controls (313.08,156) and (318,160.92) .. (318,167) .. controls (318,173.08) and (313.08,178) .. (307,178) .. controls (300.92,178) and (296,173.08) .. (296,167) -- cycle ;
\draw    (316,160) -- (398.2,119.88) ;
\draw [shift={(400,119)}, rotate = 153.98] [color={rgb, 255:red, 0; green, 0; blue, 0 }  ][line width=0.75]    (10.93,-3.29) .. controls (6.95,-1.4) and (3.31,-0.3) .. (0,0) .. controls (3.31,0.3) and (6.95,1.4) .. (10.93,3.29)   ;

\draw (104,46.4) node [anchor=north west][inner sep=0.75pt]    {$V( \phi )$};
\draw (527,248.4) node [anchor=north west][inner sep=0.75pt]    {$\phi $};
\draw (286,188.4) node [anchor=north west][inner sep=0.75pt]    {$\Lambda $};
\draw (278,232) node [anchor=north west][inner sep=0.75pt]   [align=left] {\begin{minipage}[lt]{39.83pt}\setlength\topsep{0pt}
\begin{center}
{\fontfamily{ptm}\selectfont Our }\\{\fontfamily{ptm}\selectfont universe }\\{\fontfamily{ptm}\selectfont now}
\end{center}

\end{minipage}};
\draw (403,82.4) node [anchor=north west][inner sep=0.75pt]    {$\frac{|V'|}{V} < \mathcal{O}( 1)$};

\end{tikzpicture}
\end{figure}
\textbf{Question 8 (a version of coincidence problem):} The age of the universe is around the natural timescale associated with cosmological constant $\mathcal{O}(\Lambda^{-1/2})$ in Planck units. Is that a coincident or there is an explanation to it? 
\vspace{10pt}

\textbf{Question 9 (strong CP problem):} From the EFT perspective, it is natural to add a $\theta F\wedge F$ term to the Lagrangian. This term would violate CP and have experimental consequences. Experiment suggests $|\theta|<10^{-10}$ which is unnaturally small. Peccei and Quinn tried to explain this smallness by promoting $\theta$ to a dynamical variable (axion) that is dynamically fixed. Promoting $\theta$ to a field is well-motivated by quantum gravity, but it still does not explain why it should stabilize at such a small value. 
\vspace{10pt}

\textbf{Question 10 (homogeneity):} How did the universe become so homogeneous with an anisotropy that is so scale-invariant? 
\vspace{10pt}

Inflation seems to be a contender to explain this observation. The premise of inflation is that there was a long (compared to Hubble time) period of exponential expansion in the early universe that homogenized the observable universe. However, the conventional models to realize inflation are usually in tension with EFT. Inflation typically requires a potential that is very flat ($|V'|\ll V$) over very long field ranges $\Delta\phi\gg M_P$ which needs fine tuning and is unnatural.

\begin{figure}[H]
    \centering

\tikzset{every picture/.style={line width=0.75pt}} 

\begin{tikzpicture}[x=0.75pt,y=0.75pt,yscale=-1,xscale=1]

\draw    (52,46) .. controls (772,41) and (554,223) .. (622,217) ;
\draw  [dash pattern={on 4.5pt off 4.5pt}]  (66,141) -- (549,141) ;
\draw [shift={(551,141)}, rotate = 180] [color={rgb, 255:red, 0; green, 0; blue, 0 }  ][line width=0.75]    (10.93,-3.29) .. controls (6.95,-1.4) and (3.31,-0.3) .. (0,0) .. controls (3.31,0.3) and (6.95,1.4) .. (10.93,3.29)   ;
\draw [shift={(64,141)}, rotate = 0] [color={rgb, 255:red, 0; green, 0; blue, 0 }  ][line width=0.75]    (10.93,-3.29) .. controls (6.95,-1.4) and (3.31,-0.3) .. (0,0) .. controls (3.31,0.3) and (6.95,1.4) .. (10.93,3.29)   ;
\draw  (27,251) -- (649,251)(48,23) -- (48,282) (642,246) -- (649,251) -- (642,256) (43,30) -- (48,23) -- (53,30)  ;

\draw (251,23.4) node [anchor=north west][inner sep=0.75pt]  [color={rgb, 255:red, 208; green, 2; blue, 27 }  ,opacity=1 ]  {$| V'|\ll V$};
\draw (260,157.4) node [anchor=north west][inner sep=0.75pt]  [color={rgb, 255:red, 208; green, 2; blue, 27 }  ,opacity=1 ]  {$\Delta \phi \gg M_{P}$};
\draw (7,25.4) node [anchor=north west][inner sep=0.75pt]    {$V( \phi )$};
\draw (638,269.4) node [anchor=north west][inner sep=0.75pt]    {$\phi $};

\end{tikzpicture}
\end{figure}

The question we would like to study in this course is that: Could quantum gravity (QG) shed light on these problems by considering the impact of UV degrees of freedom?

So far, the only example of quantum gravity that we know of is string theory. It is reasonable to use it to get as much insight about QG from it as possible. We can study the above questions in some controlled examples in string theory to see if there are general patterns.

String theory setups usually consist of a highly constrained higher dimensional theory living in a 10 dimensional spacetime with some compact dimensions and some non-compact dimensions. To model our 4d universe, we would need six compact dimensions. 

\begin{figure}[H]
    \centering

\tikzset{every picture/.style={line width=0.75pt}} 

\begin{tikzpicture}[x=0.75pt,y=0.75pt,yscale=-1,xscale=1]

\draw    (251,61) -- (252.49,277) ;
\draw [shift={(252.5,279)}, rotate = 269.61] [color={rgb, 255:red, 0; green, 0; blue, 0 }  ][line width=0.75]    (10.93,-3.29) .. controls (6.95,-1.4) and (3.31,-0.3) .. (0,0) .. controls (3.31,0.3) and (6.95,1.4) .. (10.93,3.29)   ;
\draw [color={rgb, 255:red, 0; green, 0; blue, 0 }  ,draw opacity=1 ]   (276,71) .. controls (302,54) and (308,69) .. (317,85) .. controls (326,101) and (337,72) .. (349,76) .. controls (361,80) and (360.3,102.53) .. (357,120) .. controls (353.7,137.47) and (337,121) .. (316,112) .. controls (295,103) and (306,129) .. (289,124) .. controls (272,119) and (258,86) .. (276,71) -- cycle ;

\draw (254,13) node [anchor=north west][inner sep=0.75pt]   [align=left] {10d String theory};
\draw (223,166) node [anchor=north west][inner sep=0.75pt]  [rotate=-270] [align=left] {Compactification};
\draw (298,91.4) node [anchor=north west][inner sep=0.75pt]    {$M$};
\draw (303,184.4) node [anchor=north west][inner sep=0.75pt]    {$\times $};
\draw (302,238.4) node [anchor=north west][inner sep=0.75pt]    {$\mathbb{R}^{4}$};
\draw (280,133) node [anchor=north west][inner sep=0.75pt]   [align=left] {compact};
\draw (268,263) node [anchor=north west][inner sep=0.75pt]   [align=left] {non-compact};
\draw (413,253) node [anchor=north west][inner sep=0.75pt]   [align=left] {4d theory};

\end{tikzpicture}
\end{figure}

Depending on the choice of the compact manifold $M$, we find different 4d EFTs $M\rightarrow EFT(M)$. The geometric properties of $M$ are reflected in physical properties of $EFT(M)$. For example, as we will explain later, the easiest cases that we can study are supersymmetric EFTs that typically correspond to Calabi--Yau manifolds $M$. It turns out that if we fix the cutoff of the EFT, the number of known different supersymmetric theories that that arise in QG is finite! This is very different from the typical EFT perspective where we have continuous adjustable parameters that give us a continuous spectrum of theories.
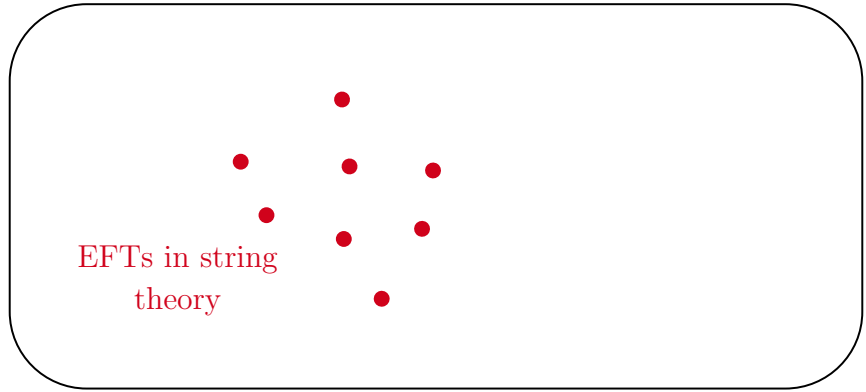
\begin{figure}[H]
    \centering

\tikzset{every picture/.style={line width=0.75pt}} 

\begin{tikzpicture}[x=0.75pt,y=0.75pt,yscale=-1,xscale=1]

\draw   (124,60.8) .. controls (124,39.37) and (141.37,22) .. (162.8,22) -- (515.2,22) .. controls (536.63,22) and (554,39.37) .. (554,60.8) -- (554,177.2) .. controls (554,198.63) and (536.63,216) .. (515.2,216) -- (162.8,216) .. controls (141.37,216) and (124,198.63) .. (124,177.2) -- cycle ;
\draw  [color={rgb, 255:red, 208; green, 2; blue, 27 }  ,draw opacity=1 ][fill={rgb, 255:red, 208; green, 2; blue, 27 }  ,fill opacity=1 ] (237,101.5) .. controls (237,99.57) and (238.57,98) .. (240.5,98) .. controls (242.43,98) and (244,99.57) .. (244,101.5) .. controls (244,103.43) and (242.43,105) .. (240.5,105) .. controls (238.57,105) and (237,103.43) .. (237,101.5) -- cycle ;
\draw  [color={rgb, 255:red, 208; green, 2; blue, 27 }  ,draw opacity=1 ][fill={rgb, 255:red, 208; green, 2; blue, 27 }  ,fill opacity=1 ] (250,128.5) .. controls (250,126.57) and (251.57,125) .. (253.5,125) .. controls (255.43,125) and (257,126.57) .. (257,128.5) .. controls (257,130.43) and (255.43,132) .. (253.5,132) .. controls (251.57,132) and (250,130.43) .. (250,128.5) -- cycle ;
\draw  [color={rgb, 255:red, 208; green, 2; blue, 27 }  ,draw opacity=1 ][fill={rgb, 255:red, 208; green, 2; blue, 27 }  ,fill opacity=1 ] (289.25,67.52) .. controls (290.68,66.22) and (292.9,66.32) .. (294.2,67.75) .. controls (295.5,69.18) and (295.39,71.39) .. (293.96,72.69) .. controls (292.53,74) and (290.32,73.89) .. (289.02,72.46) .. controls (287.72,71.03) and (287.82,68.82) .. (289.25,67.52) -- cycle ;
\draw  [color={rgb, 255:red, 208; green, 2; blue, 27 }  ,draw opacity=1 ][fill={rgb, 255:red, 208; green, 2; blue, 27 }  ,fill opacity=1 ] (293.04,101.31) .. controls (294.47,100) and (296.68,100.11) .. (297.98,101.54) .. controls (299.28,102.97) and (299.18,105.18) .. (297.75,106.48) .. controls (296.32,107.78) and (294.1,107.68) .. (292.8,106.25) .. controls (291.5,104.82) and (291.61,102.61) .. (293.04,101.31) -- cycle ;
\draw  [color={rgb, 255:red, 208; green, 2; blue, 27 }  ,draw opacity=1 ][fill={rgb, 255:red, 208; green, 2; blue, 27 }  ,fill opacity=1 ] (289,140.5) .. controls (289,138.57) and (290.57,137) .. (292.5,137) .. controls (294.43,137) and (296,138.57) .. (296,140.5) .. controls (296,142.43) and (294.43,144) .. (292.5,144) .. controls (290.57,144) and (289,142.43) .. (289,140.5) -- cycle ;
\draw  [color={rgb, 255:red, 208; green, 2; blue, 27 }  ,draw opacity=1 ][fill={rgb, 255:red, 208; green, 2; blue, 27 }  ,fill opacity=1 ] (308.18,171.36) .. controls (307.8,169.47) and (309.02,167.62) .. (310.92,167.24) .. controls (312.82,166.86) and (314.66,168.09) .. (315.04,169.99) .. controls (315.42,171.88) and (314.19,173.73) .. (312.29,174.11) .. controls (310.4,174.49) and (308.56,173.26) .. (308.18,171.36) -- cycle ;
\draw  [color={rgb, 255:red, 208; green, 2; blue, 27 }  ,draw opacity=1 ][fill={rgb, 255:red, 208; green, 2; blue, 27 }  ,fill opacity=1 ] (334.68,103.86) .. controls (335.83,102.3) and (338.02,101.97) .. (339.58,103.11) .. controls (341.13,104.26) and (341.47,106.45) .. (340.32,108.01) .. controls (339.17,109.56) and (336.98,109.9) .. (335.43,108.75) .. controls (333.87,107.6) and (333.54,105.41) .. (334.68,103.86) -- cycle ;
\draw  [color={rgb, 255:red, 208; green, 2; blue, 27 }  ,draw opacity=1 ][fill={rgb, 255:red, 208; green, 2; blue, 27 }  ,fill opacity=1 ] (329.14,133.31) .. controls (330.29,131.75) and (332.48,131.42) .. (334.04,132.56) .. controls (335.59,133.71) and (335.93,135.9) .. (334.78,137.46) .. controls (333.64,139.01) and (331.44,139.35) .. (329.89,138.2) .. controls (328.33,137.05) and (328,134.86) .. (329.14,133.31) -- cycle ;

\draw (266,233) node [anchor=north west][inner sep=0.75pt]   [align=left] {};
\draw (137,142) node [anchor=north west][inner sep=0.75pt]   [align=left] {\begin{minipage}[lt]{105.3pt}\setlength\topsep{0pt}
\begin{center}
\textcolor[rgb]{0.82,0.01,0.11}{EFTs in string theory}
\end{center}

\end{minipage}};

\end{tikzpicture}
\caption{Space of "good" EFTs.}
\end{figure}
As we will see, these examples are usually jammed in some specific corners of the theory space. This observation motivates us to think there is an underlying fundamental reason behind the patterns that we see. That maybe a theory of quantum gravity must always follow specific criteria that are not obvious from EFT perspective. However, we should be cautious that we might be misguided by the limited size of our supersymmetric sample set. This is why it is important to support any observed pattern by some more reasoning that bears on more basic physics (e.g. unitrarity or black hole physics). It is important to point out that typically both the observation of patterns and the supporting reasoning should be viewed as motivations and not a proof. In order to prove something robust about quantum gravity, we first need to have a much clearer understanding of what quantum gravity is. Even though we are not there yet, that is certainly the final goal. 

\subsection{The Swampland program}

The discreteness of the set of consistent theories makes it difficult to discern whether a given EFT has gravitational UV completion or not. In fact, you would need to measure the physical parameters with infinite precision to do that. This makes it so much more difficult to say if a theory is consistent with QG than whether it is not. The idea of the Swampland program is to rule out the inconsistent theories rathen that pin point the consistent ones.

\textbf{Swampland and Landscape:} the EFTs that are consistent based on EFT reasoning (no anomalies, etc...) but do not have a QG UV-completion are said to be in the "\textbf{\textit{Swampland}}" while the ones that do are said to belong to the "\textbf{\textit{Landscape}}". 
\vspace{10pt}

\textbf{Example: } As we will see in the class, the N=4 super Yang--Mills in $d=4$ with a gauge group of rank greater than 22 is in the Swampland \cite{Kim:2019ths}.

\begin{statement*} Finding criteria that ensure a theory belongs to the Swampland using universal observations in string theory as well as arguments based on more basic physics (unitarity, black hole physics, etc.).
\end{statement*}
\vspace{10pt}

Note that, by definition our universe is in the Landscape. So finding criteria that cut away corners of the theory space from the Landscape could lead to direct predictions about our universe. 

There are two different definitions for the Landscape. Suppose we call the previous definition the QG-Landscape, there is also a string-landscape which corresponds to the set of EFTs that are realized in string theory. Right now, since the only known well-defined quantum theory of gravity is string theory, there seems to be no distinction between them. However, generally $\text{String Landscape}\subset \text{QG Landscape}$. Recently, there has been more and more evidence emerging in support of the equality of the two sets. The conjectural equality of the two sets is often called the \textbf"string lamppost principle (SLP)" (also called string universality). If string theory turns out to be a tiny subset of the landscape, we are in a bad shape because it would be very difficult to find correct Swampland conditions. However, as we will see in the course, at least with enough supersymmetry, this does not seem to be the case.

Some of the Swampland conditions that we will discuss are also motivated by black hole physics but some others just have string theory backing. The former group thus have more evidence to be true. In addition to these two sources, the overlaps and consistencies between the different conditions is another source of assurance of their mutual validity.

If these principles are somehow all different sides of the same underlying principle, we would like to unify them and find that core principle. Unfortunately, we are not there yet, but we have a good web of statements that seem to be circling around a few more fundamental statements. 
Even if one does not think of Swampland conditions as principles, they are still very useful as organizing principles for the many examples we see in string theory. 

\begin{figure}[H]
    \centering
\tikzset{
pattern size/.store in=\mcSize, 
pattern size = 5pt,
pattern thickness/.store in=\mcThickness, 
pattern thickness = 0.3pt,
pattern radius/.store in=\mcRadius, 
pattern radius = 1pt}
\makeatletter
\pgfutil@ifundefined{pgf@pattern@name@_leab4g1ds}{
\pgfdeclarepatternformonly[\mcThickness,\mcSize]{_leab4g1ds}
{\pgfqpoint{0pt}{-\mcThickness}}
{\pgfpoint{\mcSize}{\mcSize}}
{\pgfpoint{\mcSize}{\mcSize}}
{
\pgfsetcolor{\tikz@pattern@color}
\pgfsetlinewidth{\mcThickness}
\pgfpathmoveto{\pgfqpoint{0pt}{\mcSize}}
\pgfpathlineto{\pgfpoint{\mcSize+\mcThickness}{-\mcThickness}}
\pgfusepath{stroke}
}}
\makeatother
\tikzset{
pattern size/.store in=\mcSize, 
pattern size = 5pt,
pattern thickness/.store in=\mcThickness, 
pattern thickness = 0.3pt,
pattern radius/.store in=\mcRadius, 
pattern radius = 1pt}
\makeatletter
\pgfutil@ifundefined{pgf@pattern@name@_fhsiuippu}{
\pgfdeclarepatternformonly[\mcThickness,\mcSize]{_fhsiuippu}
{\pgfqpoint{0pt}{0pt}}
{\pgfpoint{\mcSize+\mcThickness}{\mcSize+\mcThickness}}
{\pgfpoint{\mcSize}{\mcSize}}
{
\pgfsetcolor{\tikz@pattern@color}
\pgfsetlinewidth{\mcThickness}
\pgfpathmoveto{\pgfqpoint{0pt}{0pt}}
\pgfpathlineto{\pgfpoint{\mcSize+\mcThickness}{\mcSize+\mcThickness}}
\pgfusepath{stroke}
}}
\makeatother
\tikzset{every picture/.style={line width=0.75pt}} 
\begin{tikzpicture}[x=0.75pt,y=0.75pt,yscale=-1,xscale=1]
\draw  [color={rgb, 255:red, 208; green, 2; blue, 27 }  ,draw opacity=1 ][fill={rgb, 255:red, 208; green, 2; blue, 27 }  ,fill opacity=1 ] (222,139.5) .. controls (222,137.57) and (223.57,136) .. (225.5,136) .. controls (227.43,136) and (229,137.57) .. (229,139.5) .. controls (229,141.43) and (227.43,143) .. (225.5,143) .. controls (223.57,143) and (222,141.43) .. (222,139.5) -- cycle ;
\draw  [color={rgb, 255:red, 208; green, 2; blue, 27 }  ,draw opacity=1 ][fill={rgb, 255:red, 208; green, 2; blue, 27 }  ,fill opacity=1 ] (235,166.5) .. controls (235,164.57) and (236.57,163) .. (238.5,163) .. controls (240.43,163) and (242,164.57) .. (242,166.5) .. controls (242,168.43) and (240.43,170) .. (238.5,170) .. controls (236.57,170) and (235,168.43) .. (235,166.5) -- cycle ;
\draw  [color={rgb, 255:red, 208; green, 2; blue, 27 }  ,draw opacity=1 ][fill={rgb, 255:red, 208; green, 2; blue, 27 }  ,fill opacity=1 ] (274.25,105.52) .. controls (275.68,104.22) and (277.9,104.32) .. (279.2,105.75) .. controls (280.5,107.18) and (280.39,109.39) .. (278.96,110.69) .. controls (277.53,112) and (275.32,111.89) .. (274.02,110.46) .. controls (272.72,109.03) and (272.82,106.82) .. (274.25,105.52) -- cycle ;
\draw  [color={rgb, 255:red, 208; green, 2; blue, 27 }  ,draw opacity=1 ][fill={rgb, 255:red, 208; green, 2; blue, 27 }  ,fill opacity=1 ] (278.04,139.31) .. controls (279.47,138) and (281.68,138.11) .. (282.98,139.54) .. controls (284.28,140.97) and (284.18,143.18) .. (282.75,144.48) .. controls (281.32,145.78) and (279.1,145.68) .. (277.8,144.25) .. controls (276.5,142.82) and (276.61,140.61) .. (278.04,139.31) -- cycle ;
\draw  [color={rgb, 255:red, 208; green, 2; blue, 27 }  ,draw opacity=1 ][fill={rgb, 255:red, 208; green, 2; blue, 27 }  ,fill opacity=1 ] (274,178.5) .. controls (274,176.57) and (275.57,175) .. (277.5,175) .. controls (279.43,175) and (281,176.57) .. (281,178.5) .. controls (281,180.43) and (279.43,182) .. (277.5,182) .. controls (275.57,182) and (274,180.43) .. (274,178.5) -- cycle ;
\draw  [color={rgb, 255:red, 208; green, 2; blue, 27 }  ,draw opacity=1 ][fill={rgb, 255:red, 208; green, 2; blue, 27 }  ,fill opacity=1 ] (293.18,209.36) .. controls (292.8,207.47) and (294.02,205.62) .. (295.92,205.24) .. controls (297.82,204.86) and (299.66,206.09) .. (300.04,207.99) .. controls (300.42,209.88) and (299.19,211.73) .. (297.29,212.11) .. controls (295.4,212.49) and (293.56,211.26) .. (293.18,209.36) -- cycle ;
\draw  [color={rgb, 255:red, 208; green, 2; blue, 27 }  ,draw opacity=1 ][fill={rgb, 255:red, 208; green, 2; blue, 27 }  ,fill opacity=1 ] (319.68,141.86) .. controls (320.83,140.3) and (323.02,139.97) .. (324.58,141.11) .. controls (326.13,142.26) and (326.47,144.45) .. (325.32,146.01) .. controls (324.17,147.56) and (321.98,147.9) .. (320.43,146.75) .. controls (318.87,145.6) and (318.54,143.41) .. (319.68,141.86) -- cycle ;
\draw  [color={rgb, 255:red, 208; green, 2; blue, 27 }  ,draw opacity=1 ][fill={rgb, 255:red, 208; green, 2; blue, 27 }  ,fill opacity=1 ] (314.14,171.31) .. controls (315.29,169.75) and (317.48,169.42) .. (319.04,170.56) .. controls (320.59,171.71) and (320.93,173.9) .. (319.78,175.46) .. controls (318.64,177.01) and (316.44,177.35) .. (314.89,176.2) .. controls (313.33,175.05) and (313,172.86) .. (314.14,171.31) -- cycle ;
\draw  [pattern=_leab4g1ds,pattern size=6pt,pattern thickness=0.75pt,pattern radius=0pt, pattern color={rgb, 255:red, 208; green, 2; blue, 27}] (602,126) -- (165.02,17) -- (602,17) -- cycle ;
\draw   (101,18) -- (601,18) -- (601,256) -- (101,256) -- cycle ;
\draw  [pattern=_fhsiuippu,pattern size=6pt,pattern thickness=0.75pt,pattern radius=0pt, pattern color={rgb, 255:red, 74; green, 144; blue, 226}] (601,248) -- (358,18) -- (601,18) -- cycle ;
\draw (287,263) node [anchor=north west][inner sep=0.75pt]   [align=left] {Space of "healthy" EFTs};
\draw (152,179) node [anchor=north west][inner sep=0.75pt]   [align=left] {\begin{minipage}[lt]{52.64pt}\setlength\topsep{0pt}
\begin{center}
\textcolor[rgb]{0.82,0.01,0.11}{Landscape}
\end{center}
\end{minipage}};
\draw (335.81,67.74) node [anchor=north west][inner sep=0.75pt]  [rotate=-23.55] [align=left] {Swampland criteria};
\end{tikzpicture}
\end{figure}

The core of the Swampland program is the uniqueness of string theory. In string theory, as we increase the cutoff, the landscape of theories that seemed to be disconnected, become connected. It is believed that increasing cut-off high enough would lead to one single theory with a single connected moduli space. Given that different Calabi--Yau manifolds lead to different EFTs, this implies that all different Calabi--Yau manifolds must be transformable to each other using specific geometric transitions. In fact, this statement is a well-motivated math conjecture often known as Reid's fantasy. We will talk more about this in the future when we talk about dualities.

\begin{figure}[H]
    \centering
\tikzset{every picture/.style={line width=0.75pt}} 
\begin{tikzpicture}[x=0.75pt,y=0.75pt,yscale=-1,xscale=1]
\draw  (41,259.33) -- (618.5,259.33)(320.5,30) -- (320.5,288) (611.5,254.33) -- (618.5,259.33) -- (611.5,264.33) (315.5,37) -- (320.5,30) -- (325.5,37)  ;
\draw    (97.5,84) .. controls (116.33,83.23) and (167.5,182) .. (235.5,188) .. controls (303.5,194) and (311.5,82) .. (358.5,86) .. controls (405.5,90) and (481.5,232) .. (527.5,236) .. controls (573.5,240) and (608.4,123.58) .. (618.5,116) ;
\draw [color={rgb, 255:red, 208; green, 2; blue, 27 }  ,draw opacity=1 ]   (143.5,150) -- (628.5,152) ;
\draw [color={rgb, 255:red, 74; green, 144; blue, 226 }  ,draw opacity=1 ]   (95,71) -- (639.5,70) ;

\draw (581,275) node [anchor=north west][inner sep=0.75pt]   [align=left] {Moduli};
\draw (266,8) node [anchor=north west][inner sep=0.75pt]   [align=left] {Scalar potential};
\draw (227,197.4) node [anchor=north west][inner sep=0.75pt]    {$EFT_{1}$};
\draw (549.5,231.4) node [anchor=north west][inner sep=0.75pt]    {$EFT_{2}$};
\draw (109,142.4) node [anchor=north west][inner sep=0.75pt]  [color={rgb, 255:red, 208; green, 2; blue, 27 }  ,opacity=1 ]  {$\Lambda _{1}$};
\draw (60,55.4) node [anchor=north west][inner sep=0.75pt]  [color={rgb, 255:red, 74; green, 144; blue, 226 }  ,opacity=1 ]  {$\Lambda _{2}$};
\end{tikzpicture}
    \caption{In the above example, decreasing the cut off of the theory from $\Lambda_2$ to $\Lambda_1$, breaks up the moduli space into two disconnected peices. Therefore, each local minimum is described by a different EFT with cutoff $\Lambda_1$. However, by increasing the cutoff back to $\Lambda_2$, the moduli spaces of the two EFTs connect and we can describe both of them by a single EFT with cutoff $\Lambda_1$. If we keep increasing the cutoff to infinity, we expect for all the low-energy EFTs to become connected. Thus, all seemingly different EFTs are just different low-energy corners of the moduli space of a more fundamental theory.}
\end{figure}
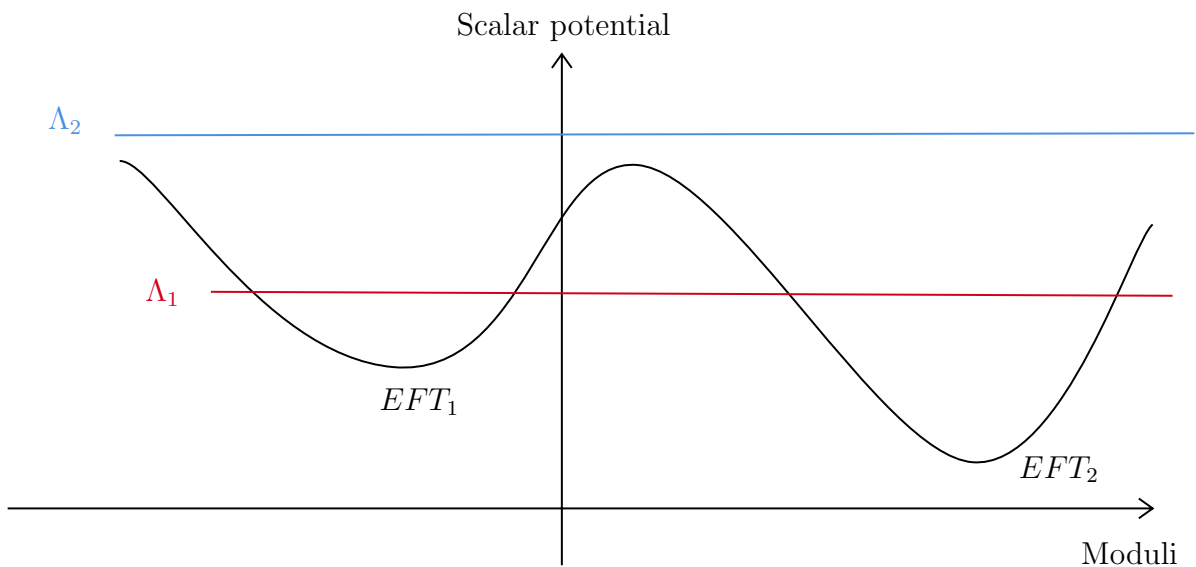

Let us summarize our introduction by going back to the differences between quantum gravity and quantum field theory and list a few major differences between quantum gravity from quantum field theory. We will sharpen these differences when we get to a detailed discussion of the Swampland criteria. 

\begin{itemize}
    \item \textbf{Locality: }Quantum field theory is predicated on an algebra of local operators. A local structure of observables is a starting point in any quantum field theory. However, dualities strongly suggest that any local structure is emergent rather than fundamental in quantum gravity. For example, in the AdS/CFT duality, the local classical theory on the gravity side could be only trusted in the strong coupling limit on the CFT side. In other words, a set of observables emerge and a local description in terms of these observables becomes a good approximate description in a particular limit. Another example is T-duality where the notion of \textit{"local"} in the compact dimension is very different depending on the frame. A local excitation in one frame is a topological winding state in the other frame and vice-versa. 
    
    \item \textbf{UV/IR decoupling: }In Quantum field theory, the IR dynamics is thought to be impacted by the UV theory in a very limited way via corrections to some terms in the effective action. However, in quantum gravity, an IR calculation can be very sensitive to UV details. The black hole entropy formula is a perfect example of this UV-IR connection.
    
    \item \textbf{Symmetries: }In effective field theory, symmetries are guiding princinples, however, this is not the case in quantum gravity. In fact, as we will see, quantum gravity avoids global symmetries and is pretty picky about its gauge symmetries. 
    
    \item \textbf{Naturalness: }If we consider a UV theory with $\mathcal{O}(1)$ couplings in the appropriate mass unit of the theory, we can estimate quantitative and qualitative properties of the low-energy field theory. In that sense, a UV theory sets a \textit{natural} expectation for the IR theory. However, what is natural in EFT, can be very unnatural in quantum gravity and vice-versa. For example, exponentially light states are unnatural in field theory while natural in quantum gravity and arbitrarily large gauge symmetries are natural in field theory while unnatural in quantum gravity.
\end{itemize}

\section{Swampland I: No global symmetry conjecture}

\subsection{No global symmetry: black hole argument}

As we discussed in the previous section, black holes are low-energy windows into UV gravitational physics. Many of the insights that we learn from black hole physics hinge on the fact that the properties of black holes are universal. This includes their entropy formula or thermal features of the Hawking radiation which does not seem to depend on the details of the effective field theory. In the early days of black holes, these universalities raised many questions including the information paradox. It was also pointed out that since the Hawking radiation seems to only depend on the near horizon geometry which only depends on gauge charges, angular momentum and mass, any other label gets lost in the black hole. This implies that if we throw a conserved charge under a global symmetry that is not protected by a gauge symmetry, it seems to get lost in the radiation and the conservation gets violated \cite{Banks:2010zn}. Hence, there can be no global symmetries in quantum gravity. 

Note that this is a separate issue from the information paradox. One might think that the resolution of information paradox restores information of the what we throw in the black hole, including the conserved charge. However, since the spectrum of the outgoing Hawking radiation is blind to the global symmetry charge, this conservation cannot hold for global symmetries. Therefore, the global symmetry is violated. 

As it is clear from the statement of the no global symmetry conjecture, there must be a difference between gauge and global symemtries. This raises an important question: What is a physical definition that can separate the two symmetries from one another? Are gauge symmetries just an artifact of our mathematical redundancy of the theory or do they have a physical meaning to them? 

One can ask similar questions about discrete symmetries. For example, what separates discrete gauge symmetries from discrete global symmetries? 

One way to think about discrete gauge symmetries is in terms of Higgsing a continuous gauge symmetries. For example if one considers a complex scalar field with a unit $U(1)$ charge, the gauge symmetry can be Higgsed to $\mathbb{Z}_n$. However, can we always think about discrete gauge symmetries in this way? The answer turns out to be no! 

Another way to think about discrete gauge symmetries is in terms of lattice gauge theories and taking the limit where the lattice spacing goes to zero. Let us see how this works in a nutshell. For a continuous gauge group, we can define the theory on a principal bundle over spacetime. The gauge field represents an infinitesimal change along the fibre as we parallel transport along an infinitesimal line in spacetime. For discrete gauge groups, we quickly run into a problem because the fibres can be discrete. Suppose each fibre is a a set of points. Then, the only non-trivial information in the bundle are the holonomies. However, in a simply connected spacetime like $\mathbb{R}^n$, all loops are contractible and all holonomies must be trivial. So it becomes unclear what the addition of gauge symmetry has to offer. However, lattice gauge theory naturally resolves this via summing over discretized spacetimes. A lattice is made up of holes and therefore the parallel transport can admit non-trivial holonomies as we move along the cells. By taking the limit where the lattice spacing goes to zero and properly regulating the physical observables, we can define a discrete gauge theory via lattice.

Although this description is very helpful, it seems more practical than fundamental and it raises the question that whether there is a more abstract and fundamental definition for symmetries? 

Another natural question is that when we have a gauge symmetry many times it is accompanied with a conservation from a global symmetry. For example, if one considers a pure $SU(3)$ gauge theory, the symmetry that acts like a gauge transformation but with constant $g(x)$ is a symmetry of the theory which does not fo to $g=1$ as $|x|\rightarrow\infty$. This symmetry maps different gluons to each other and since different particles are mapped to each other we should not think of such large gauge transformation as a gauge symmetry, but a global symmetry. In fact this is the global symmetry responsible for conservation of charges in gauge theories. But how is this different from a normal global symmetry? To give a preview of the answer, it turns out such a global symmetry can only exist in non-compact spaces. Moreover, when viewed as a symmetry on the operators rather than Hilbert space, it cannot be defined locally. It only acts on the boundary of the non-compact space. We will proceed with trying to come up with a definition of symmetries (gauge and global) that pushes aside all the non-physical formulation-dependent aspects of symmetries and focuses on the physical properties of the symmetries.

\subsection{What is a global symmetry?}

Let us start with continuous global symmetries as a case study. In that case, for every generator $\tau^a$ in the Lie algebra $\mathfrak{g}$ of the global symmetry group, there is a conserved Noether's current $j^a_{\mu}$. To be more precise, $\sum_a j^ac_a$ is the Noether's conserved current corresponding to the symmetry generated by exponentiating $\sum_a c_a\tau^a\in\mathfrak{g}$. We can think of $J=j^a_{\mu}dx^\mu$ as a $\mathfrak{g}^*$-valued one-form. This means, corresponding to every element $g=\sum_a c^a\tau_a$ of $\mathfrak{g}$ we assign a one-form $J(g)=j^a_{\mu}c_a dx^\mu$ which is conserved ($\partial_\mu j^\mu=0$)
\begin{align}\label{cons}
    d\star{J}(g)=0,
\end{align}
at points with no charge present. Now using the Stokes' theorem, we can rewrite the above constraint as 
\begin{align}
    \int_{\Sigma^{d-1}} \star J(g)=\int_\mathcal{M} d\star J(g)=0,
\end{align}
where $\Sigma^{d-1}$ is a compact $d-1$ dimensional orientable manifold and $\mathcal{M}$ is a d-dimensional region such that its boundary is $\Sigma^{d-1}$. Note that this is only true if the equation \eqref{cons} holds everywhere in $\mathcal{M}$ which means there is no charge inside $\Sigma^{d-1}$. If we take $\Sigma^{d-1}$ to be the non-compact hypersurface of constant time $t=t_0$, the integral $\int_{\Sigma^{d-1}} \star J$ becomes
\begin{align}
    Q(g)=\int dx^{d-1} j^a_0c_a,
\end{align}
which is the Noether's conserved charge. We can generalize the above definition to any boundary-less hypersurface $\Sigma$ (asymptotic or compact).
\begin{align}\label{NC}
    Q_g(\Sigma)=\int_\Sigma \star J(g),
\end{align}
Suppose $\exp(g)$ is an element of the symmetry group $G$, the conservation of the charge operator \eqref{NC} implies that the following operator is topological 
\begin{align}
    U_{\exp(g)}(\Sigma)=\exp(Q_g(\Sigma)).
\end{align}
What we mean by topological is that if we insert $U_g(\Sigma)$ in the path integral, and variate the hypersurface $\Sigma$, the result will not change until $\Sigma$ hits a charged operator. We will see an example of this in a moment.

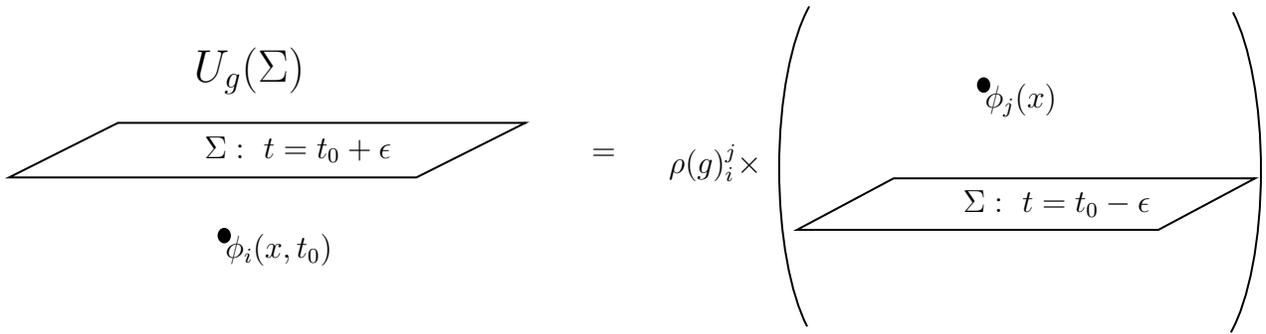
\begin{figure}[H]

    \centering

\tikzset{every picture/.style={line width=0.75pt}} 

\begin{tikzpicture}[x=0.75pt,y=0.75pt,yscale=-1,xscale=1]

\draw   (56.53,109) -- (262,109) -- (207.07,136.54) -- (1.6,136.54) -- cycle ;
\draw  [fill={rgb, 255:red, 0; green, 0; blue, 0 }  ,fill opacity=1 ] (107.28,165.89) .. controls (107.28,164.01) and (108.56,162.49) .. (110.14,162.49) .. controls (111.71,162.49) and (112.99,164.01) .. (112.99,165.89) .. controls (112.99,167.77) and (111.71,169.29) .. (110.14,169.29) .. controls (108.56,169.29) and (107.28,167.77) .. (107.28,165.89) -- cycle ;
\draw  [fill={rgb, 255:red, 0; green, 0; blue, 0 }  ,fill opacity=1 ] (490.28,89.89) .. controls (490.28,88.01) and (491.56,86.49) .. (493.14,86.49) .. controls (494.71,86.49) and (495.99,88.01) .. (495.99,89.89) .. controls (495.99,91.77) and (494.71,93.29) .. (493.14,93.29) .. controls (491.56,93.29) and (490.28,91.77) .. (490.28,89.89) -- cycle ;
\draw  [draw opacity=0] (618.87,53.22) .. controls (627.36,69.82) and (633,99.34) .. (633,133) .. controls (633,168.02) and (626.89,198.57) .. (617.83,214.74) -- (603,133) -- cycle ; \draw   (618.87,53.22) .. controls (627.36,69.82) and (633,99.34) .. (633,133) .. controls (633,168.02) and (626.89,198.57) .. (617.83,214.74) ;  
\draw  [draw opacity=0] (404.13,211.78) .. controls (395.64,195.18) and (390,165.66) .. (390,132) .. controls (390,96.98) and (396.11,66.43) .. (405.17,50.26) -- (420,132) -- cycle ; \draw   (404.13,211.78) .. controls (395.64,195.18) and (390,165.66) .. (390,132) .. controls (390,96.98) and (396.11,66.43) .. (405.17,50.26) ;  
\draw   (447.73,137) -- (630.01,137) -- (581.27,163) -- (398.99,163) -- cycle ;

\draw (294,120.4) node [anchor=north west][inner sep=0.75pt]    {$=$};
\draw (333,118.4) node [anchor=north west][inner sep=0.75pt]    {$\rho ( g)_{i}^{j} \times $};
\draw (98.71,114.2) node [anchor=north west][inner sep=0.75pt]    {$\Sigma :\ t=t_{0} +\epsilon $};
\draw (93.42,69.25) node [anchor=north west][inner sep=0.75pt]  [font=\Large]  {$U_{g}( \Sigma )$};
\draw (109.31,163.77) node [anchor=north west][inner sep=0.75pt]    {$\phi _{i}( x,t_{0})$};
\draw (492.31,87.77) node [anchor=north west][inner sep=0.75pt]    {$\phi _{j}( x)$};
\draw (481.25,141.4) node [anchor=north west][inner sep=0.75pt]    {$\Sigma :\ t=t_{0} -\epsilon $};

\end{tikzpicture}
\caption{moving a local charged operator past $U_g(\Sigma)$ will replace that operator with the action of the group on that operator. If there are no more operator insertions at earlier times, we can remove $U_g(\Sigma)$ from the right hand side.}
    \label{GTSO1}
\end{figure}

The Noether's conserved charge $Q(g)$ is said to generate the action of the symmetry group $G$. What that means is that it determines how the group $G$ acts on the local operators and Hilbert space. For example, consider a charged state in the Hilbert space which is prepared by the insertion a charged local operator $\phi_i(x)$ in the path integral at some time before $t<t_0$. By $\phi_i$ being charged we mean it transforms under some representation $\rho$ under $G$. Now we can think of insertion of $U_{g\in G}(\Sigma)$ as an operator that acts on $\phi_i$ as 
\begin{align}\label{AST}
    \int \mathcal{D}\Phi e^{iS}U_g(\Sigma)\phi_i(x)=\int \mathcal{D}\Phi e^{iS}\rho(g)^j_i\phi_j(x).
\end{align}
This is often illustrated as in Figure \ref{GTSO1}.

The equation \eqref{AST} is true for any boundary-less hypersurface $\Sigma$ enclosing a local charged operator $\phi_i(x)$. 
\begin{figure}[H]
    \centering

\tikzset{every picture/.style={line width=0.75pt}} 

\begin{tikzpicture}[x=0.75pt,y=0.75pt,yscale=-1,xscale=1]

\draw  [fill={rgb, 255:red, 0; green, 0; blue, 0 }  ,fill opacity=1 ] (149.4,144.85) .. controls (149.4,142.97) and (150.67,141.45) .. (152.25,141.45) .. controls (153.83,141.45) and (155.1,142.97) .. (155.1,144.85) .. controls (155.1,146.73) and (153.83,148.25) .. (152.25,148.25) .. controls (150.67,148.25) and (149.4,146.73) .. (149.4,144.85) -- cycle ;
\draw  [fill={rgb, 255:red, 0; green, 0; blue, 0 }  ,fill opacity=1 ] (425.28,142.89) .. controls (425.28,141.01) and (426.56,139.49) .. (428.14,139.49) .. controls (429.71,139.49) and (430.99,141.01) .. (430.99,142.89) .. controls (430.99,144.77) and (429.71,146.29) .. (428.14,146.29) .. controls (426.56,146.29) and (425.28,144.77) .. (425.28,142.89) -- cycle ;
\draw   (70,148.25) .. controls (70,102.82) and (106.82,66) .. (152.25,66) .. controls (197.68,66) and (234.5,102.82) .. (234.5,148.25) .. controls (234.5,193.68) and (197.68,230.5) .. (152.25,230.5) .. controls (106.82,230.5) and (70,193.68) .. (70,148.25) -- cycle ;
\draw    (70,148.25) .. controls (86.5,187) and (224.5,189) .. (234.5,148.25) ;
\draw  [dash pattern={on 4.5pt off 4.5pt}]  (70,148.25) .. controls (83.5,112) and (224.5,112) .. (234.5,148.25) ;

\draw (319,121.4) node [anchor=north west][inner sep=0.75pt]    {$=$};
\draw (67.42,22.25) node [anchor=north west][inner sep=0.75pt]  [font=\Large]  {$U_{g}( \Sigma )$};
\draw (157.1,148.25) node [anchor=north west][inner sep=0.75pt]    {$\phi _{i}( x)$};
\draw (430.14,146.29) node [anchor=north west][inner sep=0.75pt]    {$\rho ( g)_{i}^{j} \phi _{j}( x)$};

\end{tikzpicture}
    \caption{}
\end{figure}

This formulation of global symmetry encodes the information of a global symmetry in topological operators $U_g(\Sigma)$ associated with elements $g$ in the symmetry group $G$ that defined on $d-1$ dimensional boundary-less hypersurfaces $\Sigma$ that act on $0$-dimensional (local) operators (e.g. $\phi_i(x)$) that are enclosed by the hypersurface $\Sigma$. 

We can brush aside all the unnecessary details of the above formulation and define a global symmetry based on the action of the topological operators on local charged operators. This allows us to even generalize the notion of global symmetry to higher dimensional topological operators as follows \cite{Gaiotto:2014kfa}. 

\begin{statement*}
A $p$-form global symmetry constitutes of a set of topological operators $\{U_g(\Sigma)|g\in G\}$ defined on boundary-less orientable submanifold $\Sigma$ of dimension $d-p-1$. The operator $\{U_g(\Sigma)|g\in G\}$ is topological in the sense that it corresponds to an insertion in the path integral such that the result is independent from variation of $\Sigma$ unless it hits a defect operator $\Phi$ defined on a $p$-dimensional submanifold $\Sigma'$ that links with $\Sigma$. Passing $U_g(\Sigma)$ through $\Phi$ replaces it with another $p$-dimensional defect $\Phi_g$. In this sense, the topological operator $U_g$ acts on $p$-dimensional charged defects.

Moreover, a global symmetry $G$ must be equipped with a fusion algebra for the topological operators $U_g$ such that for every two elements $g$ and $g'$ in $G$, and homotopic surfaces $\Sigma_1,\Sigma_2$, and $\Sigma_3$, there exist an element $g"\in G$ such that $U_g(\Sigma_1) U_{g'}(\Sigma_2)=U_{g"}(\Sigma_3)$ in the absence of any charged operators between two of the $\Sigma_i\in\{1,2,3\}$ (linking with only one or two of them). See Figure \ref{Fusion}. 

For the above data to be a global symmetry, we impose that there needs to be at least one non-trivially charged defect. 
\end{statement*}

\begin{figure}[H]
    \centering

\tikzset{every picture/.style={line width=0.75pt}} 

\begin{tikzpicture}[x=0.75pt,y=0.75pt,yscale=-1,xscale=1]

\draw  [dash pattern={on 4.5pt off 4.5pt}]  (84.5,122.5) .. controls (102.88,71.16) and (294.88,71.16) .. (308.5,122.5) ;
\draw  [color={rgb, 255:red, 0; green, 0; blue, 0 }  ,draw opacity=1 ][fill={rgb, 255:red, 74; green, 74; blue, 74 }  ,fill opacity=1 ] (115,120.25) .. controls (115,74.82) and (151.82,38) .. (197.25,38) .. controls (242.68,38) and (279.5,74.82) .. (279.5,120.25) .. controls (279.5,165.68) and (242.68,202.5) .. (197.25,202.5) .. controls (151.82,202.5) and (115,165.68) .. (115,120.25) -- cycle ;
\draw [color={rgb, 255:red, 0; green, 0; blue, 0 }  ,draw opacity=1 ]   (115,120.25) .. controls (131.5,159) and (269.5,161) .. (279.5,120.25) ;
\draw  [fill={rgb, 255:red, 74; green, 74; blue, 74 }  ,fill opacity=0.53 ] (84.5,122.5) .. controls (84.5,58.16) and (134.64,6) .. (196.5,6) .. controls (258.36,6) and (308.5,58.16) .. (308.5,122.5) .. controls (308.5,186.84) and (258.36,239) .. (196.5,239) .. controls (134.64,239) and (84.5,186.84) .. (84.5,122.5) -- cycle ;
\draw [color={rgb, 255:red, 0; green, 0; blue, 0 }  ,draw opacity=1 ] [dash pattern={on 4.5pt off 4.5pt}]  (115,120.25) .. controls (128.5,84) and (269.5,84) .. (279.5,120.25) ;
\draw    (84.5,122.5) .. controls (119.5,188) and (269.5,198) .. (308.5,122.5) ;
\draw  [color={rgb, 255:red, 0; green, 0; blue, 0 }  ,draw opacity=1 ][fill={rgb, 255:red, 74; green, 74; blue, 74 }  ,fill opacity=1 ] (439.28,119.5) .. controls (439.28,68.97) and (479.62,28) .. (529.39,28) .. controls (579.16,28) and (619.5,68.97) .. (619.5,119.5) .. controls (619.5,170.03) and (579.16,211) .. (529.39,211) .. controls (479.62,211) and (439.28,170.03) .. (439.28,119.5) -- cycle ;
\draw [color={rgb, 255:red, 0; green, 0; blue, 0 }  ,draw opacity=1 ]   (439.28,119.5) .. controls (457.36,162.61) and (608.54,164.83) .. (619.5,119.5) ;
\draw [color={rgb, 255:red, 0; green, 0; blue, 0 }  ,draw opacity=1 ] [dash pattern={on 4.5pt off 4.5pt}]  (439.28,119.5) .. controls (454.07,79.17) and (608.54,79.17) .. (619.5,119.5) ;

\draw (163,54.4) node [anchor=north west][inner sep=0.75pt]  [color={rgb, 255:red, 74; green, 100; blue, 226 }  ,opacity=1 ]  {$U_{g}( \Sigma _{1})$};
\draw (49,22.4) node [anchor=north west][inner sep=0.75pt]  [color={rgb, 255:red, 74; green, 100; blue, 226 }  ,opacity=1 ]  {$U_{g'}( \Sigma _{2})$};
\draw (388.37,47.27) node [anchor=north west][inner sep=0.75pt]  [color={rgb, 255:red, 74; green, 100; blue, 226 }  ,opacity=1 ]  {$U_{g"}( \Sigma _{3})$};
\draw (354,114.4) node [anchor=north west][inner sep=0.75pt]    {$=$};

\end{tikzpicture}
    \caption{}
    \label{Fusion}
\end{figure}

Moreover, the fusion algebra is not-necessarily Abelian. However, for $p>0$ it is necessarily Abelian. This is because any two boundary-less submanifolds with co-dimension $p>1$ can be continuously permuted without intersecting each other. See Figure \ref{Linking} for an example of this. 
\begin{figure}[H]
    \centering

\tikzset{every picture/.style={line width=0.75pt}} 

\begin{tikzpicture}[x=0.75pt,y=0.75pt,yscale=-1,xscale=1]

\draw  [color={rgb, 255:red, 74; green, 101; blue, 226 }  ,draw opacity=1 ][line width=3]  (23.34,165.25) .. controls (29.27,157.35) and (65.01,150.94) .. (103.17,150.94) .. controls (141.33,150.94) and (167.46,157.35) .. (161.53,165.25) .. controls (155.61,173.15) and (119.87,179.56) .. (81.71,179.56) .. controls (43.55,179.56) and (17.42,173.15) .. (23.34,165.25) -- cycle ;
\draw  [color={rgb, 255:red, 208; green, 2; blue, 27 }  ,draw opacity=1 ][line width=3]  (7.37,165.25) .. controls (19.9,148.54) and (68.13,135) .. (115.12,135) .. controls (162.1,135) and (190.03,148.54) .. (177.5,165.25) .. controls (164.98,181.96) and (116.74,195.5) .. (69.76,195.5) .. controls (22.78,195.5) and (-5.16,181.96) .. (7.37,165.25) -- cycle ;
\draw  [color={rgb, 255:red, 208; green, 2; blue, 27 }  ,draw opacity=1 ][line width=3]  (500.34,157.25) .. controls (506.27,149.35) and (542.01,142.94) .. (580.17,142.94) .. controls (618.33,142.94) and (644.46,149.35) .. (638.53,157.25) .. controls (632.61,165.15) and (596.87,171.56) .. (558.71,171.56) .. controls (520.55,171.56) and (494.42,165.15) .. (500.34,157.25) -- cycle ;
\draw  [color={rgb, 255:red, 74; green, 101; blue, 226 }  ,draw opacity=1 ][line width=3]  (484.37,157.25) .. controls (496.9,140.54) and (545.13,127) .. (592.12,127) .. controls (639.1,127) and (667.03,140.54) .. (654.5,157.25) .. controls (641.98,173.96) and (593.74,187.5) .. (546.76,187.5) .. controls (499.78,187.5) and (471.84,173.96) .. (484.37,157.25) -- cycle ;
\draw  [color={rgb, 255:red, 208; green, 2; blue, 27 }  ,draw opacity=1 ][line width=3]  (257.56,163.59) .. controls (268.88,151.5) and (312.46,141.69) .. (354.91,141.69) .. controls (397.36,141.69) and (422.6,151.5) .. (411.28,163.59) .. controls (399.96,175.69) and (356.38,185.5) .. (313.93,185.5) .. controls (271.48,185.5) and (246.24,175.69) .. (257.56,163.59) -- cycle ;
\draw  [color={rgb, 255:red, 74; green, 101; blue, 226 }  ,draw opacity=1 ][line width=3]  (252.93,135.9) .. controls (265.27,122.7) and (312.8,112) .. (359.09,112) .. controls (405.38,112) and (432.89,122.7) .. (420.55,135.9) .. controls (408.2,149.09) and (360.67,159.79) .. (314.38,159.79) .. controls (268.1,159.79) and (240.58,149.09) .. (252.93,135.9) -- cycle ;
\draw  [fill={rgb, 255:red, 0; green, 0; blue, 0 }  ,fill opacity=1 ] (188,164.75) -- (222.5,164.75) -- (222.5,161) -- (245.5,168.5) -- (222.5,176) -- (222.5,172.25) -- (188,172.25) -- cycle ;
\draw  [fill={rgb, 255:red, 0; green, 0; blue, 0 }  ,fill opacity=1 ] (419,158.75) -- (453.5,158.75) -- (453.5,155) -- (476.5,162.5) -- (453.5,170) -- (453.5,166.25) -- (419,166.25) -- cycle ;

\end{tikzpicture}
    \caption{}
    \label{Linking}
\end{figure}

This explain the fact that all higher-form ($p>0$) symmetries (global or gauged) are always Abelian in String theory.

\subsection{Non-invertible symmetries}

Up to now we assumed that a group is behind the symmetry. However, this is not necessary and we can have a more general fusion algebra for the topological operators. For example, the topological operators can fuse as
\begin{align}
    U_\alpha U_\beta= \sum_\gamma C_{\alpha\beta\gamma}U_\gamma,
\end{align}
with some coefficients $C_{\alpha\beta\gamma}$ that do not satisfy all the properties of group multiplication such as the existence of inverse elements. Such symmetries are called non-invertible symmetries \cite{Chang:2018iay,Thorngren:2019iar,Lin:2019hks,Rudelius:2020orz,Heidenreich:2021xpr}.

Let us investigate an example of non-invertible symmetry which naturally arises in string theory. When we orbifold (mod out) a worldsheet theory by a discrete group $G$, we get a twisted sector for every conjugacy class of $G$. The first important observation is that the twisted sectors are labeled by conjugacy classes and not by the group elements. 

Consider a twisted string that its initial and final endpoints are related by the action of $g\in G$. Suppose the endpoints of the string are $x$ and $g(x)$. Take element $h\in G$ and act on the twisted string with $h$. The endpoints of the new string are $h(x)$ and $h\cdot g(x)$ and are related to each other by the action of $hgh^{-1}$. Since we have to identify the two strings under orbifolding, we find that the twisted sectors of $g$ and $h^{-1}gh$ are indistinguishable. This is why the twisted sectors are labeled by conjugacy classes rather than group elements. 

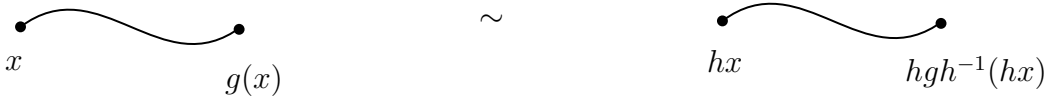
\begin{figure}[H]
    \centering

\tikzset{every picture/.style={line width=0.75pt}} 

\begin{tikzpicture}[x=0.75pt,y=0.75pt,yscale=-1,xscale=1]

\draw  [fill={rgb, 255:red, 0; green, 0; blue, 0 }  ,fill opacity=1 ] (85,67) .. controls (85,65.76) and (86.01,64.75) .. (87.25,64.75) .. controls (88.49,64.75) and (89.5,65.76) .. (89.5,67) .. controls (89.5,68.24) and (88.49,69.25) .. (87.25,69.25) .. controls (86.01,69.25) and (85,68.24) .. (85,67) -- cycle ;
\draw    (87.25,67) .. controls (127.25,37) and (158.5,97) .. (198.5,67) ;
\draw  [fill={rgb, 255:red, 0; green, 0; blue, 0 }  ,fill opacity=1 ] (195.25,68.25) .. controls (195.25,67.01) and (196.26,66) .. (197.5,66) .. controls (198.74,66) and (199.75,67.01) .. (199.75,68.25) .. controls (199.75,69.49) and (198.74,70.5) .. (197.5,70.5) .. controls (196.26,70.5) and (195.25,69.49) .. (195.25,68.25) -- cycle ;
\draw  [fill={rgb, 255:red, 0; green, 0; blue, 0 }  ,fill opacity=1 ] (439,64) .. controls (439,62.76) and (440.01,61.75) .. (441.25,61.75) .. controls (442.49,61.75) and (443.5,62.76) .. (443.5,64) .. controls (443.5,65.24) and (442.49,66.25) .. (441.25,66.25) .. controls (440.01,66.25) and (439,65.24) .. (439,64) -- cycle ;
\draw    (441.25,64) .. controls (481.25,34) and (512.5,94) .. (552.5,64) ;
\draw  [fill={rgb, 255:red, 0; green, 0; blue, 0 }  ,fill opacity=1 ] (549.25,65.25) .. controls (549.25,64.01) and (550.26,63) .. (551.5,63) .. controls (552.74,63) and (553.75,64.01) .. (553.75,65.25) .. controls (553.75,66.49) and (552.74,67.5) .. (551.5,67.5) .. controls (550.26,67.5) and (549.25,66.49) .. (549.25,65.25) -- cycle ;

\draw (78,80.4) node [anchor=north west][inner sep=0.75pt]    {$x$};
\draw (189,84.4) node [anchor=north west][inner sep=0.75pt]    {$g( x)$};
\draw (432,77.4) node [anchor=north west][inner sep=0.75pt]    {$hx$};
\draw (532,80.4) node [anchor=north west][inner sep=0.75pt]    {$hgh^{-1}( hx)$};
\draw (317,59.4) node [anchor=north west][inner sep=0.75pt]    {$\sim $};

\end{tikzpicture}
    \caption{Two twisted strings in the same conjugacy class belong to the same twisted sector.}
    \label{Orbifold conjugacy}
\end{figure}

Now let us see what happens if we fuse two particles (twisted strings) with conjugacy classes $C_i$ and $C_j$. The fusion will give us a state in tensor product of the Hilbert spaces $\mathcal{H}_{C_i}$ and $\mathcal{H}_{C_j}$ associated with each conjugacy class. The resulting Hilbert space can be decompoed into a linear combination of one-particle state as 
\begin{align}
    \mathcal{H}_{C_i}\otimes\mathcal{H}_{C_j}= \oplus_k N_{ij}^k\mathcal{H}'_{C_k}, 
\end{align}
where $\mathcal{H}'_{C_k}$ is a non-zero subspace of $H_{C_K}$ and $N_{ij}^k$ counts the multiplicity of the conjugacy class $C_k$ if we multiply the elements of $C_i$ and $C_j$ by the group product of $G$. One can think of each Hilbert space on the right hand side as a scattering channel with two incoming particles in $\mathcal{H}_{C_i}$ and $\mathcal{H}_{C_j}$.

\begin{figure}[H]
    \centering

\tikzset{every picture/.style={line width=0.75pt}} 

\begin{tikzpicture}[x=0.75pt,y=0.75pt,yscale=-1,xscale=1]

\draw    (129,76) -- (286,78) ;
\draw    (129,84) -- (286,86) ;
\draw  [fill={rgb, 255:red, 0; green, 0; blue, 0 }  ,fill opacity=1 ] (279,81.5) .. controls (279,75.7) and (283.7,71) .. (289.5,71) .. controls (295.3,71) and (300,75.7) .. (300,81.5) .. controls (300,87.3) and (295.3,92) .. (289.5,92) .. controls (283.7,92) and (279,87.3) .. (279,81.5) -- cycle ;
\draw    (289.5,81.5) -- (481,82) ;

\draw (135,53.4) node [anchor=north west][inner sep=0.75pt]    {$C_{i}$};
\draw (135,87.4) node [anchor=north west][inner sep=0.75pt]    {$C_{j}$};
\draw (492,75.4) node [anchor=north west][inner sep=0.75pt]    {$\oplus _{k} N_{i\ j}^{k} \ C_{k}$};

\end{tikzpicture}
    \caption{A scattering vertex with two incoming particles in twisted sectors $C_i$ and $C_j$ and an outgoing particle in a linear combination of 1-particle states in different conjugacy classes.}
\end{figure}
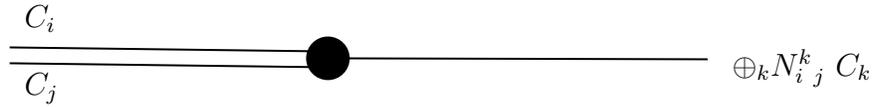

Note that the above product is typically not invertible for non-Abelian groups. For example, if you multiply the conjugacy class of $g$ by the conjugacy class of $g^{-1}$, you always get some conjugacy classes in addition to that of identity, unless $g$ is in the center of the group $Z(G)$. 
\begin{align}
    \forall g\not\in G:~~[g]\cdot [g^{-1}]\neq [\mathds{1}].
\end{align}

So far, we talked about the fusion rules for the charged operators. Now let us see if we can define any topological operator that can define a symmetry. To define the symmetry operators from fusion rules of charged operators, we review a general argument for diagonal rational CFTs that applies to orbifolds. Suppose the fusion algebra takes the following form.
\begin{align}\label{FAV}
    [i]\times[j]=\sum N_{ij}^k[k].
\end{align}
Since there is no fundamental ordering for the operators, the matrices $(N^k)_{ij}$ must be symmetric. Moreover, due to the associativity of the fusion algebra, these matrices must commute. Therefore, we can mutually diagonalize them by considering a new basis $[i]'$ of charged operators.
\begin{align}
    (N^k)'_{i,j}=\delta_{ij}\lambda_i^k,
\end{align}
For some real numbers $\lambda^k_i$. Then we can define a collection of commuting line operators $\mathcal{L}_k$ such that if $\mathcal{L}_k$ encircles $[i]'$ it multiplies it by $\lambda^k_i$. For example, if we consider a $\mathbb{Z}_n$ orbifold, the corresponding line operators will form a $\mathbb{Z}_n$ group. However, despite the commutativity of these operators, they do not necessarily form a group. It is easy to see that the fusion rules of $\mathcal{L}_k$ is the same as the fusion rules of \eqref{FAV}. These line operators are called Verlinde operators (see \cite{Moore:1988qv,Chang:2018iay}). A simple example of a non-invertible symmetry arises in the $(4,3)$ minimal model which describes the critical 2d Ising model. This theory has three primary operators $\{1,\sigma,\psi\}$ that satisfy the following OPE. 
\begin{align}
    \sigma\sigma&\sim1+\psi\nonumber\\
    \sigma\psi&\sim\sigma\nonumber\\
    \psi\psi&\sim1.
\end{align}
Since the Verlinde operators will satisfy the same fusion ring, $\mathcal{L}_\sigma$ will have no inverse, just like $\sigma$ has no inverse.

Any global symmetry on the string worldsheet can be thought of as a symemtry in the spacetime. We can define the spacetime symmetry operator to act on the string states exactly as the worldsheet symmetry operator acts on the operators that create those states on the worldsheet. However, as we will later see, these spacetime symmetries always turn out to be gauge symmetries. 

\subsection{What is a gauge symmetry?}

In the previous subsection we gave a general definition of global symmetries. Now let us revisit that discussion for gauge symmetries. We defined global symmetries based on their action on local physical operators. So to define gauge symmetries, we need to know how the gauge symmetries acts on local physical operators? But this is almost a trivial question for gauge symmetries! They must not act on physical operators at all. In other words, local physical operators must be gauge invariant. So if a symmetry $G$ is gauged, a non-trivial topological operator like $U_g(\Sigma)$ does not exist for $G$. In this language, gauging a global symmetry $G$ means to start with a set of topological operators $U_g$ and use them to restrict the spectrum of physical operators by throwing out those that are not invariant under the topological operators.

What we described above is not a satisfactory definition of gauge symmetry since it tells us what gauge symmetry is not rather than what it is. Now we know that a gauge symmetry is not a global symmetry, but can we define a gauge symmetry by its physical implications or is it just a purely mathematical feature of the formulation of a theory? 

To answer this question let us take a moment to think about a simple-looking but deep question: what is the difference between a $U(1)$ pure gauge theory and an $\mathbb{R}$ pure gauge theory? The Lie-algebras of the two groups are the same. So any difference must depend on the global structure of the group. But what kind of information is sensitive to the global structure of the groups? The answer is the representations! The representations of $U(1)$ are quantized while the representation of $\mathbb{R}$ are labeled by a continuous parameter $q$
\begin{align}
    \alpha\rightarrow e^{iq} \alpha.
\end{align}
Representations of a gauge theory naturally appear with charged operators. But if we consider a pure gauge theory (no charged operator) how is the information of the allowed representations is encoded in the theory? In other words, what are some operators that carry representations of the gauge group and come with a pure gauge theory? The answer is Wilson loops! For every closed loop $\gamma$ and a representation $\rho$ of the gauge group there is a gauge invariant operator $W_{\rho}(\gamma)$ that can be inserted in the path integral and captures the information about the curvature of the associate gauge group.  Therefore, gauge symmetries have something more than not being a global symmetry, and that is having Wilson line operators $W_{\rho}(\gamma)$ or their higher dimensional generalizations. 

\begin{statement1*}
A $p$-form gauge symmetry comes with defect operators $W_{\rho}(\Sigma)$ where $\rho$ is some representation of symmetry $G$ and $\Sigma$ is a $p+1$-dimensional compact orientable manifold. A Wilson operator can end on a p-dimensional charged operators defined on the boundaries of $\Sigma$. Such operators must carry matching representations of symmetry $G$ with that of the Wilson operator. 

Moreover, there are $d-p-1$ dimensional topological operators (e.g. $\int \star F$) that only act on Wilson lines with boundary-less domains $\Sigma$ by conjugation of $G$. 

If Wilson operators can be defined on non-compact boundary-less domains (e.g. $\Sigma$ extends to asymptotic boundary) we say the gauge symmetry is long-range \cite{Harlow:2018tng}.
\end{statement1*}

It was noted in \cite{Heidenreich:2021xpr} that in the absence of any charged operators, $p$-form gauge symmetry $G$ always comes with a $p+1$ form global symmetry $Z(G)$ (center of $G$) generated by topological operators $\int \star F$. 

\subsection{Non-compact spaces and boundary symmetries}\label{NCVC}

In pure gauge theories, there are no local physical charged operators. Therefore, generally there are no charged states either since we can think of charged states as a charged operator acting on the vacuum. However, this argument has a loop hole for non-compact spaces. In non-compact spaces it is possible to have a gauge-invariant charged operator, but the catch is the charged operator does not have a compact support. For example, consider the $SU(3)$ pure gauge theory and the total number of a particular type of gluon of a given momentum. This operator is definitely a charged operator because it changes under the $SU(3)$ (gluons get mapped to each other). However, it clearly does not have a compact support. We can understand this in the language of Wilson loops too. In non-compact spaces, you can have Wilson lines that extend all the way to infinity.

\begin{figure}[H]
    \centering

\tikzset{every picture/.style={line width=0.75pt}} 

\begin{tikzpicture}[x=0.75pt,y=0.75pt,yscale=-1,xscale=1]

\draw   (213,153.75) .. controls (213,95.9) and (259.9,49) .. (317.75,49) .. controls (375.6,49) and (422.5,95.9) .. (422.5,153.75) .. controls (422.5,211.6) and (375.6,258.5) .. (317.75,258.5) .. controls (259.9,258.5) and (213,211.6) .. (213,153.75) -- cycle ;
\draw    (364.5,248) .. controls (272.5,188) and (305.5,87) .. (359.5,57) ;
\draw  [fill={rgb, 255:red, 0; green, 0; blue, 0 }  ,fill opacity=1 ] (356.25,57) .. controls (356.25,55.21) and (357.71,53.75) .. (359.5,53.75) .. controls (361.29,53.75) and (362.75,55.21) .. (362.75,57) .. controls (362.75,58.79) and (361.29,60.25) .. (359.5,60.25) .. controls (357.71,60.25) and (356.25,58.79) .. (356.25,57) -- cycle ;
\draw  [fill={rgb, 255:red, 0; green, 0; blue, 0 }  ,fill opacity=1 ] (361.25,248) .. controls (361.25,246.21) and (362.71,244.75) .. (364.5,244.75) .. controls (366.29,244.75) and (367.75,246.21) .. (367.75,248) .. controls (367.75,249.79) and (366.29,251.25) .. (364.5,251.25) .. controls (362.71,251.25) and (361.25,249.79) .. (361.25,248) -- cycle ;

\draw (97,53) node [anchor=north west][inner sep=0.75pt]   [align=left] {Asymptotic boundary};
\draw (234,146.4) node [anchor=north west][inner sep=0.75pt]    {$W_{[ \rho ]}( \gamma )^{i} \ _{j}$};
\draw (370,35.4) node [anchor=north west][inner sep=0.75pt]    {$i$};
\draw (382,257.4) node [anchor=north west][inner sep=0.75pt]    {$j$};

\end{tikzpicture}
    \caption{Wilson lines can end on asymptotic boundary even in pure gauge theories. Insertion of such Wilson operators creates net gauge charge in non-compact space.}
    \label{AsympW}
\end{figure}
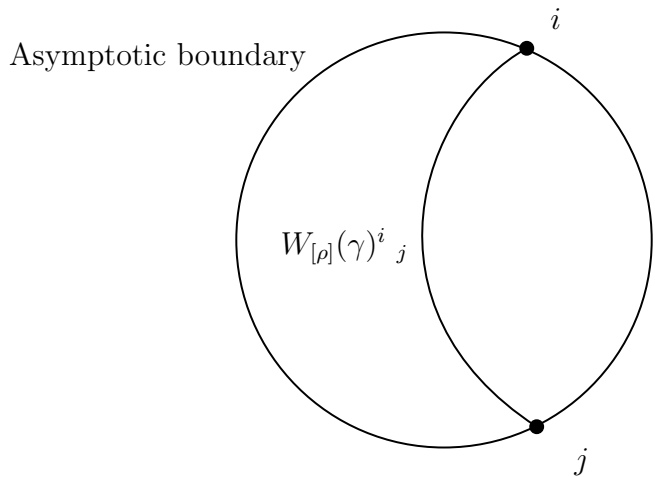

These are gauge-invariant operators in the sense that any operator $U_g(\Sigma)$ with a compact support does not change them. However, they are charged in the sense that when acted on the Hilbert space, they change the total gauge charge. Such Wilson operator creates a charged particle in space and cancelling charged particle at infinity, effectively, creating charge in the universe.

If we take $\Sigma$ to a hypersurface that extends to infinity and cuts through the Wilson line, $U_g(\Sigma)$ acts on the Wilson line by changing the labels of the end points according to the corresponding representation $\rho$.

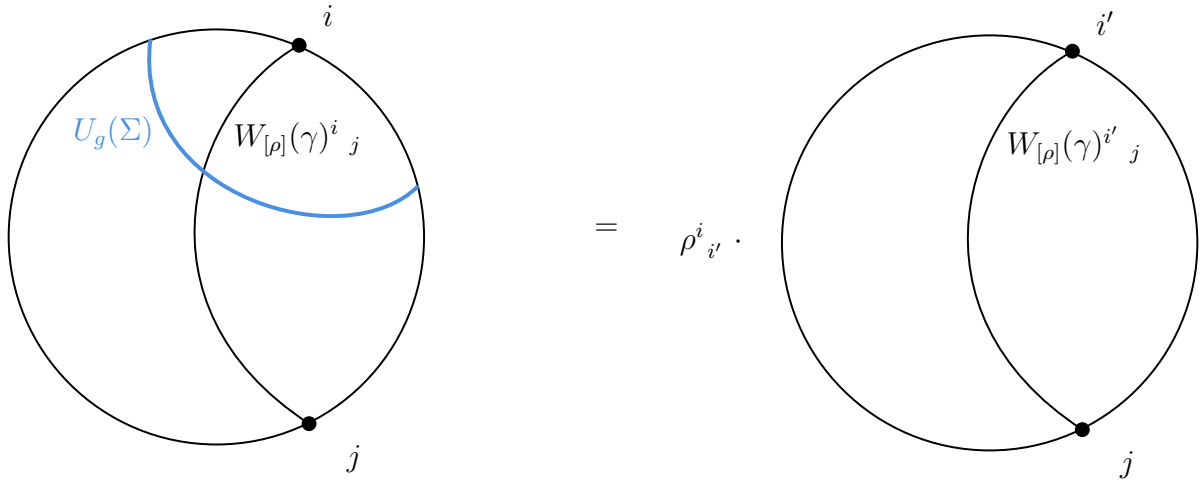
\begin{figure}[H]
    \centering

\tikzset{every picture/.style={line width=0.75pt}} 

\begin{tikzpicture}[x=0.75pt,y=0.75pt,yscale=-1,xscale=1]

\draw   (31,149.75) .. controls (31,91.9) and (77.9,45) .. (135.75,45) .. controls (193.6,45) and (240.5,91.9) .. (240.5,149.75) .. controls (240.5,207.6) and (193.6,254.5) .. (135.75,254.5) .. controls (77.9,254.5) and (31,207.6) .. (31,149.75) -- cycle ;
\draw    (182.5,244) .. controls (90.5,184) and (123.5,83) .. (177.5,53) ;
\draw  [fill={rgb, 255:red, 0; green, 0; blue, 0 }  ,fill opacity=1 ] (174.25,53) .. controls (174.25,51.21) and (175.71,49.75) .. (177.5,49.75) .. controls (179.29,49.75) and (180.75,51.21) .. (180.75,53) .. controls (180.75,54.79) and (179.29,56.25) .. (177.5,56.25) .. controls (175.71,56.25) and (174.25,54.79) .. (174.25,53) -- cycle ;
\draw  [fill={rgb, 255:red, 0; green, 0; blue, 0 }  ,fill opacity=1 ] (179.25,244) .. controls (179.25,242.21) and (180.71,240.75) .. (182.5,240.75) .. controls (184.29,240.75) and (185.75,242.21) .. (185.75,244) .. controls (185.75,245.79) and (184.29,247.25) .. (182.5,247.25) .. controls (180.71,247.25) and (179.25,245.79) .. (179.25,244) -- cycle ;
\draw [color={rgb, 255:red, 74; green, 144; blue, 226 }  ,draw opacity=1 ][line width=1.5]    (102.5,50) .. controls (92.5,131) and (202.5,160) .. (237.5,124) ;
\draw   (421,152.75) .. controls (421,94.9) and (467.9,48) .. (525.75,48) .. controls (583.6,48) and (630.5,94.9) .. (630.5,152.75) .. controls (630.5,210.6) and (583.6,257.5) .. (525.75,257.5) .. controls (467.9,257.5) and (421,210.6) .. (421,152.75) -- cycle ;
\draw    (572.5,247) .. controls (480.5,187) and (513.5,86) .. (567.5,56) ;
\draw  [fill={rgb, 255:red, 0; green, 0; blue, 0 }  ,fill opacity=1 ] (564.25,56) .. controls (564.25,54.21) and (565.71,52.75) .. (567.5,52.75) .. controls (569.29,52.75) and (570.75,54.21) .. (570.75,56) .. controls (570.75,57.79) and (569.29,59.25) .. (567.5,59.25) .. controls (565.71,59.25) and (564.25,57.79) .. (564.25,56) -- cycle ;
\draw  [fill={rgb, 255:red, 0; green, 0; blue, 0 }  ,fill opacity=1 ] (569.25,247) .. controls (569.25,245.21) and (570.71,243.75) .. (572.5,243.75) .. controls (574.29,243.75) and (575.75,245.21) .. (575.75,247) .. controls (575.75,248.79) and (574.29,250.25) .. (572.5,250.25) .. controls (570.71,250.25) and (569.25,248.79) .. (569.25,247) -- cycle ;

\draw (143,89.4) node [anchor=north west][inner sep=0.75pt]    {$W_{[ \rho ]}( \gamma )^{i} \ _{j}$};
\draw (188,31.4) node [anchor=north west][inner sep=0.75pt]    {$i$};
\draw (200,253.4) node [anchor=north west][inner sep=0.75pt]    {$j$};
\draw (62,87.4) node [anchor=north west][inner sep=0.75pt]    {$\textcolor[rgb]{0.29,0.56,0.89}{U_{g}( \Sigma )}$};
\draw (325,141.4) node [anchor=north west][inner sep=0.75pt]    {$=$};
\draw (368,141.4) node [anchor=north west][inner sep=0.75pt]    {$\rho {_{\ }^{i}}_{i'} \ \cdot $};
\draw (533,92.4) node [anchor=north west][inner sep=0.75pt]    {$W_{[ \rho ]}( \gamma )^{i'} \ _{j}$};
\draw (578,34.4) node [anchor=north west][inner sep=0.75pt]    {$i'$};
\draw (590,256.4) node [anchor=north west][inner sep=0.75pt]    {$j$};

\end{tikzpicture}
    \caption{Gauge symmetry can induce a boundary global symmetry by its action on the asymptotic endpoints of the Wilson lines that end on the asymptotic boundary.}
    \label{BoundG}
\end{figure}

However, it is important to note that not always such operators exist. Sometimes creating a charge is infinitely costly. This happens when the gauge symmetry is confined. This is why we made a distinction in for gauge symmetries with asymptotic Wilson lines by calling them long-range gauge symmetries \cite{Harlow:2018tng}. We summarize this section by highlighting the following two important results for compact and non-compact spaces.

\begin{statement3*}
There is no net gauge charge in compact spaces.
\end{statement3*}

The last statement is often stated as in compact spaces, field lines have no where to escape, therefore the charges must cancel out. 

\begin{statement2*}
Long-range gauge symmetries induce global symmetries on the asymptotic boundary.
\end{statement2*}

\subsection{No global symmetry: holographic argument}

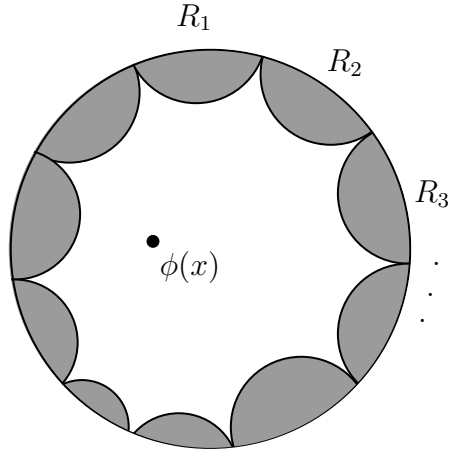
\begin{figure}[H]
    \centering\vspace{-.5cm}
\tikzset{every picture/.style={line width=0.75pt}} 
\begin{tikzpicture}[x=0.75pt,y=0.75pt,yscale=-1,xscale=1]
\draw  [fill={rgb, 255:red, 155; green, 155; blue, 155 }  ,fill opacity=1 ] (228,60.5) .. controls (228,40.89) and (243.89,25) .. (263.5,25) .. controls (283.11,25) and (299,40.89) .. (299,60.5) .. controls (299,80.11) and (283.11,96) .. (263.5,96) .. controls (243.89,96) and (228,80.11) .. (228,60.5) -- cycle ;
\draw    (536.5,146) .. controls (557.5,129) and (557.5,84) .. (523.5,73) ;
\draw  [fill={rgb, 255:red, 155; green, 155; blue, 155 }  ,fill opacity=1 ] (203,122) .. controls (203,103.77) and (217.77,89) .. (236,89) .. controls (254.23,89) and (269,103.77) .. (269,122) .. controls (269,140.23) and (254.23,155) .. (236,155) .. controls (217.77,155) and (203,140.23) .. (203,122) -- cycle ;
\draw  [fill={rgb, 255:red, 155; green, 155; blue, 155 }  ,fill opacity=1 ] (292,30.5) .. controls (292,10.89) and (307.89,-5) .. (327.5,-5) .. controls (347.11,-5) and (363,10.89) .. (363,30.5) .. controls (363,50.11) and (347.11,66) .. (327.5,66) .. controls (307.89,66) and (292,50.11) .. (292,30.5) -- cycle ;
\draw  [fill={rgb, 255:red, 155; green, 155; blue, 155 }  ,fill opacity=1 ] (360,51.5) .. controls (360,31.89) and (375.89,16) .. (395.5,16) .. controls (415.11,16) and (431,31.89) .. (431,51.5) .. controls (431,71.11) and (415.11,87) .. (395.5,87) .. controls (375.89,87) and (360,71.11) .. (360,51.5) -- cycle ;
\draw  [fill={rgb, 255:red, 155; green, 155; blue, 155 }  ,fill opacity=1 ] (399,111.5) .. controls (399,91.89) and (414.89,76) .. (434.5,76) .. controls (454.11,76) and (470,91.89) .. (470,111.5) .. controls (470,131.11) and (454.11,147) .. (434.5,147) .. controls (414.89,147) and (399,131.11) .. (399,111.5) -- cycle ;
\draw  [fill={rgb, 255:red, 155; green, 155; blue, 155 }  ,fill opacity=1 ] (399,182.5) .. controls (399,162.89) and (414.89,147) .. (434.5,147) .. controls (454.11,147) and (470,162.89) .. (470,182.5) .. controls (470,202.11) and (454.11,218) .. (434.5,218) .. controls (414.89,218) and (399,202.11) .. (399,182.5) -- cycle ;
\draw  [fill={rgb, 255:red, 155; green, 155; blue, 155 }  ,fill opacity=1 ] (345,229.5) .. controls (345,209.89) and (360.89,194) .. (380.5,194) .. controls (400.11,194) and (416,209.89) .. (416,229.5) .. controls (416,249.11) and (400.11,265) .. (380.5,265) .. controls (360.89,265) and (345,249.11) .. (345,229.5) -- cycle ;
\draw  [fill={rgb, 255:red, 155; green, 155; blue, 155 }  ,fill opacity=1 ] (288.5,252.75) .. controls (288.5,236.04) and (302.04,222.5) .. (318.75,222.5) .. controls (335.46,222.5) and (349,236.04) .. (349,252.75) .. controls (349,269.46) and (335.46,283) .. (318.75,283) .. controls (302.04,283) and (288.5,269.46) .. (288.5,252.75) -- cycle ;
\draw  [fill={rgb, 255:red, 155; green, 155; blue, 155 }  ,fill opacity=1 ] (204.25,186.75) .. controls (204.25,169.21) and (218.46,155) .. (236,155) .. controls (253.54,155) and (267.75,169.21) .. (267.75,186.75) .. controls (267.75,204.29) and (253.54,218.5) .. (236,218.5) .. controls (218.46,218.5) and (204.25,204.29) .. (204.25,186.75) -- cycle ;
\draw  [fill={rgb, 255:red, 155; green, 155; blue, 155 }  ,fill opacity=1 ] (246.38,229.31) .. controls (246.38,216.3) and (256.92,205.75) .. (269.94,205.75) .. controls (282.95,205.75) and (293.5,216.3) .. (293.5,229.31) .. controls (293.5,242.33) and (282.95,252.88) .. (269.94,252.88) .. controls (256.92,252.88) and (246.38,242.33) .. (246.38,229.31) -- cycle ;
\draw   (234,139.75) .. controls (234,84.11) and (279.11,39) .. (334.75,39) .. controls (390.39,39) and (435.5,84.11) .. (435.5,139.75) .. controls (435.5,195.39) and (390.39,240.5) .. (334.75,240.5) .. controls (279.11,240.5) and (234,195.39) .. (234,139.75) -- cycle ;
\draw  [draw opacity=0][fill={rgb, 255:red, 255; green, 255; blue, 255 }  ,fill opacity=1 ,even odd rule] (233,139.5) .. controls (233,83.72) and (278.44,38.5) .. (334.5,38.5) .. controls (390.56,38.5) and (436,83.72) .. (436,139.5) .. controls (436,195.28) and (390.56,240.5) .. (334.5,240.5) .. controls (278.44,240.5) and (233,195.28) .. (233,139.5)(81.5,139.5) .. controls (81.5,0.05) and (194.77,-113) .. (334.5,-113) .. controls (474.23,-113) and (587.5,0.05) .. (587.5,139.5) .. controls (587.5,278.95) and (474.23,392) .. (334.5,392) .. controls (194.77,392) and (81.5,278.95) .. (81.5,139.5) ;
\draw  [fill={rgb, 255:red, 0; green, 0; blue, 0 }  ,fill opacity=1 ] (303,135.75) .. controls (303,134.23) and (304.23,133) .. (305.75,133) .. controls (307.27,133) and (308.5,134.23) .. (308.5,135.75) .. controls (308.5,137.27) and (307.27,138.5) .. (305.75,138.5) .. controls (304.23,138.5) and (303,137.27) .. (303,135.75) -- cycle ;
\draw (316,14.4) node [anchor=north west][inner sep=0.75pt]    {$R_{1}$};
\draw (392,37.4) node [anchor=north west][inner sep=0.75pt]    {$R_{2}$};
\draw (436,103.4) node [anchor=north west][inner sep=0.75pt]    {$R_{3}$};
\draw (445,144.4) node [anchor=north west][inner sep=0.75pt]    {$.$};
\draw (442,160.4) node [anchor=north west][inner sep=0.75pt]    {$.$};
\draw (438,173.4) node [anchor=north west][inner sep=0.75pt]    {$.$};
\draw (307.75,139.15) node [anchor=north west][inner sep=0.75pt]    {$\phi ( x)$};
\end{tikzpicture}\vspace{-4cm}
    \caption{The action of the boundary global symmetry is restricted to the union of the entanglement wedges shown by the grey area. Therefore, the supposedly charged operator $\phi(x)$ is not acted upon by the global symmetry.}
    \label{HolographyNGS}
\end{figure}

Now that we have clear definitions for gauge and global symmetries, we can review a holographic argument by Harlow and Ooguri that shows why there can be no global symmetries in quantum gravities that admit holographic description \cite{Harlow:2018tng}. Suppose we have a global symmetry in the bulk. Consider a local operator $\phi(x)$ that is charged under the symmetry. Since the global symmetry acts on any local operator, it acts on the boundary operators as well. Therefore, the boundary CFT also has the same global symmetry. Suppose the action of the global symmetry is given by $U_g$. Take a fine partition of the boundary into small regions $\{R_i\}$. Then we can write
\begin{align}
    U_g=(\Pi_iU_i(R_i))\circ U_{edge}.
\end{align}
The effect of $U_g$ in the bulk is limited to the union of the entanglement wedges of regions $R_i$s (see Figure \ref{HolographyNGS}). For fine enough partition $\{R_i\}$, the entanglement does not contain $x$ and therefore the global symmetry does not act on $\phi(x)$ which contradicts out assumption. 
\vspace{-11.8cm}

Note that gauge symmetries avoid this argument because a charge operator cannot exist on its own. It has to be connected to a Wilson line that extends to the boundary. Therefore, the charge operator always has a point on the boundary which ensures that it is always acted on in the entanglement wedge of $\{R_i\}$.\vspace{8.5cm}

\subsection{Symmetries in string theory}

It is typically easier to verify that continuous symmetries in string theory are gauged. The more non-trivial examples usually involve discrete symmetries. In the following, we will review a few important examples of potential candidates for global symmetries in string theory.

Let us start with the 11d supergravity (M theory). If there is a 0-forms symmetry, it is likely that the 0-form symmetry will act non-trivially on a scalar in the theory. However, M-theory has no scalar fields to be acted upon by 0-form global symmetries. There are no apparent 0-form global symmetries in IIA either. The only scalar field is dilaton which does not have any symmetry. 

Now let us look at the IIB theory. IIB theory has a a complex scalar $\tau=\tau_1+i\tau_2$ which is a combination of the R–R and NS–NS scalars $\tau_1=C_0$ and $\tilde \phi\sim -\ln(\lambda)=\ln(\tau_2)$ respectively. The kinetic term of this scalar is proportional to $\partial_\mu\tau\partial_\mu \bar\tau/\tau_2^2$. You might think: Aha! There is a global symmetry $\tau_1\rightarrow \tau_1+\epsilon$. However, type IIB theory has BPS bound states of $p$ fundamental strings and $q$ D1-strings that are charged under both $B$ and $\tilde B$ and have tensions \cite{Schwarz:1995dk,Cederwall:1997ts}
\begin{align}
    T=\sqrt{(q\tau_2)^2+(p+q\tau_1)^2},
\end{align}
in string frame. The above tension formula follows from the BPS formula. Since the discrete spectrum of allowed tensions is not invariant under a continuous shift of $\tau_1$, shift symmetry is not an actual symmetry. But what about shifting $\tau_1$ by an integer and simultaneously permuting the BPS strings by a representation of $\mathbb{Z}$? Is this a global symmetry? 

It turns out the answer is still no, but for a more non-trivial reason. Consider a background with a $D7$-brane. The $D7$ brane is magnetically charged under $C_0=\tau_1$, therefore, $\oint d\tau_1$ arouund the D7-brane must be one. That means, $\tau$ picks up a monodromy and goes to $\tau+1$. Thus, the existence of $D7$ brane forces us to identify $\tau$ and $\tau+1$ rendering $\tau\rightarrow \tau+1$ to be a gauge symmetry. All in all, thanks to of D1-branes and D7-branes, there is no global symmetry in IIB theory.

Heterotic $SO(32)$ and the type I theories also do not have any apparent 0-form global symmetries. However, the situation is slightly different for the $E_8\times E_8$ Heterotic theory. This theory has a discrete $\mathbb{Z}_2$ symmetry that swaps the two $\mathfrak{e}_8$ Lie algebras and their corresponding representations. This $\mathbb{Z}_2$ seems to be a global symmetry, however, upon a closer look we can see that it is a gauge symmetry. To see why, remember the Hořava–Witten construction for the strong coupling limit of the Heterotic theory \cite{Horava:1996ma,Horava:1995qa}. In that construction, the $\mathbb{Z}_2$ symmetry at question is simply the symmetry that swaps the two endpoints of the interval and all diffeomorphisms including this $\mathbb{Z}_2$ are gauged.

Generally, whenever we compactify a theory on a symmetric manifold, the isometries of the manifold become gauge symmetries in the lower dimensional theory.

Let us finish this section with a more non-trivial example of a potential candidate for global symmetry in string theory. 

Consider compactifying IIB on K3. If there is no way of continuously deforming this background to IIB on $T^4$, we have a global symmetry! To see why, consider the following two backgorunds: IIB on $K3\times \mathbb{R}^6$ and IIB on $\mathbb{R}^4\times \mathbb{R}^6$. Now suppose we remove a small disk $D^4$ from the $K3$ and $\mathbb{R}^4$ and connect the two backgrounds on the $S^3$ boundary of the removed $D^4$. This operation is called a surgery. You can see the resulting background in the following picture. 

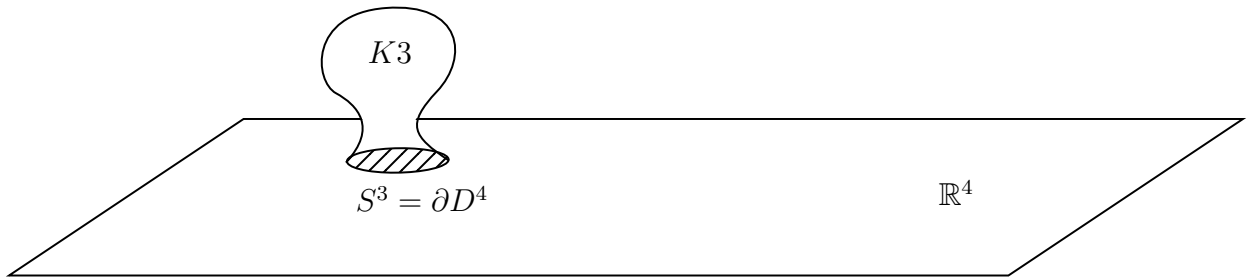
\begin{figure}[H]
    \centering

 
\tikzset{
pattern size/.store in=\mcSize, 
pattern size = 5pt,
pattern thickness/.store in=\mcThickness, 
pattern thickness = 0.3pt,
pattern radius/.store in=\mcRadius, 
pattern radius = 1pt}
\makeatletter
\pgfutil@ifundefined{pgf@pattern@name@_jrixoo7ah}{
\pgfdeclarepatternformonly[\mcThickness,\mcSize]{_jrixoo7ah}
{\pgfqpoint{0pt}{0pt}}
{\pgfpoint{\mcSize+\mcThickness}{\mcSize+\mcThickness}}
{\pgfpoint{\mcSize}{\mcSize}}
{
\pgfsetcolor{\tikz@pattern@color}
\pgfsetlinewidth{\mcThickness}
\pgfpathmoveto{\pgfqpoint{0pt}{0pt}}
\pgfpathlineto{\pgfpoint{\mcSize+\mcThickness}{\mcSize+\mcThickness}}
\pgfusepath{stroke}
}}
\makeatother
\tikzset{every picture/.style={line width=0.75pt}} 

\begin{tikzpicture}[x=0.75pt,y=0.75pt,yscale=-1,xscale=1]

\draw   (139.88,106) -- (643.92,106) -- (525.64,185) -- (21.61,185) -- cycle ;
\draw  [pattern=_jrixoo7ah,pattern size=6pt,pattern thickness=0.75pt,pattern radius=0pt, pattern color={rgb, 255:red, 0; green, 0; blue, 0}] (197.03,123.62) .. controls (205.42,120.74) and (221.46,119.87) .. (232.86,121.7) .. controls (244.26,123.52) and (246.7,127.34) .. (238.3,130.23) .. controls (229.91,133.11) and (213.87,133.97) .. (202.47,132.15) .. controls (191.07,130.33) and (188.63,126.51) .. (197.03,123.62) -- cycle ;
\draw  [draw opacity=0][fill={rgb, 255:red, 255; green, 255; blue, 255 }  ,fill opacity=1 ] (197,97.25) .. controls (197,88.28) and (204.28,81) .. (213.25,81) .. controls (222.22,81) and (229.5,88.28) .. (229.5,97.25) .. controls (229.5,106.22) and (222.22,113.5) .. (213.25,113.5) .. controls (204.28,113.5) and (197,106.22) .. (197,97.25) -- cycle ;
\draw    (243.5,127) .. controls (225.11,114.74) and (221.5,109) .. (237.5,92) .. controls (253.5,75) and (250.5,47) .. (211.5,50) .. controls (172.5,53) and (175.5,87) .. (186.5,93) .. controls (197.5,99) and (207.7,110.41) .. (191.5,128) ;

\draw (195,138.4) node [anchor=north west][inner sep=0.75pt]    {$S^{3} =\partial D^{4}$};
\draw (201,65.4) node [anchor=north west][inner sep=0.75pt]    {$K3$};
\draw (489,135.4) node [anchor=north west][inner sep=0.75pt]    {$\mathbb{R}^{4}$};

\end{tikzpicture}
    \caption{A certain compactification can be viewed as a defect in a higher dimensional theory after connecting the compact geometry to a non-compact space via surgery.}
    \label{K3 charge}
\end{figure}
This can be viewed as a defect with 6d worldvolume. And if the number of such defects is conserved, there is a global symmetry with these objects as its charged objects \cite{McNamara:2019rup}. To go back to the black hole argument, we can compactify 10d IIB on $T^5$ such that these defects become $0+1$ dimensional and now the black hole argument tells us that their number should not be conserved. So either such compactification of IIB on K3 is inconsistent (which they are) or there is a way to continuously transform such compactifications to IIB on other manifolds such as $T^4$. We will come back to this example later. 

\subsection{Cobordism conjecture}

As we discussed, the no global symmetry condition means there are no topological operators that can label different backgrounds according to their global charges. Put more generally, in quantum gravity we should not be able to tag different backgrounds. When we think of global symmetries as lables that distinguish different backgrounds, dualities become natural transitions between different backgrounds that ensure there are no such lables. In a sense, dualities are realizing the no global symmetry condition by telling you that there are no exact superselection sectors in quantum gravity. There is a formulation of this statement which allows us to sharpen this intuition.

Suppose you compactify a theory on two different compact manifolds $M$ and $N$. Then we want the two compactifications to be transportable with a finite action process (dynamically allowed). In other words, the two backgrounds $M\times\mathbb{R}^{d-k}$ and  $N\times\mathbb{R}^{d-k}$ do not tag different theories of quantum gravity. The transition between these two theories maifests itself as a finite tension domain wall in the $d-k$ dimensional theory.

\begin{figure}[H]
    \centering

 
\tikzset{
pattern size/.store in=\mcSize, 
pattern size = 5pt,
pattern thickness/.store in=\mcThickness, 
pattern thickness = 0.3pt,
pattern radius/.store in=\mcRadius, 
pattern radius = 1pt}
\makeatletter
\pgfutil@ifundefined{pgf@pattern@name@_m29dy6ye7}{
\pgfdeclarepatternformonly[\mcThickness,\mcSize]{_m29dy6ye7}
{\pgfqpoint{0pt}{-\mcThickness}}
{\pgfpoint{\mcSize}{\mcSize}}
{\pgfpoint{\mcSize}{\mcSize}}
{
\pgfsetcolor{\tikz@pattern@color}
\pgfsetlinewidth{\mcThickness}
\pgfpathmoveto{\pgfqpoint{0pt}{\mcSize}}
\pgfpathlineto{\pgfpoint{\mcSize+\mcThickness}{-\mcThickness}}
\pgfusepath{stroke}
}}
\makeatother
\tikzset{every picture/.style={line width=0.75pt}} 

\begin{tikzpicture}[x=0.75pt,y=0.75pt,yscale=-1,xscale=1]

\draw   (238.43,247) -- (656,247) -- (455.57,162) -- (38,162) -- cycle ;
\draw  [fill={rgb, 255:red, 0; green, 0; blue, 0 }  ,fill opacity=1 ] (200,196) .. controls (200,194.34) and (201.34,193) .. (203,193) .. controls (204.66,193) and (206,194.34) .. (206,196) .. controls (206,197.66) and (204.66,199) .. (203,199) .. controls (201.34,199) and (200,197.66) .. (200,196) -- cycle ;
\draw  [fill={rgb, 255:red, 0; green, 0; blue, 0 }  ,fill opacity=1 ] (452,196) .. controls (452,194.34) and (453.34,193) .. (455,193) .. controls (456.66,193) and (458,194.34) .. (458,196) .. controls (458,197.66) and (456.66,199) .. (455,199) .. controls (453.34,199) and (452,197.66) .. (452,196) -- cycle ;
\draw  [dash pattern={on 4.5pt off 4.5pt}]  (202,107) -- (203,193) ;
\draw   (170,42) .. controls (190,32) and (202,49) .. (220,70) .. controls (238,91) and (286,81) .. (246,109) .. controls (206,137) and (221,95) .. (191,98) .. controls (161,101) and (150,52) .. (170,42) -- cycle ;
\draw   (447,46) .. controls (470,43) and (505,48) .. (487,65) .. controls (469,82) and (515,90) .. (484,104) .. controls (453,118) and (422,101) .. (417,77) .. controls (412,53) and (424,49) .. (447,46) -- cycle ;
\draw  [dash pattern={on 4.5pt off 4.5pt}]  (454,110) -- (455,196) ;
\draw  [pattern=_m29dy6ye7,pattern size=3.75pt,pattern thickness=0.75pt,pattern radius=0pt, pattern color={rgb, 255:red, 128; green, 128; blue, 128}] (416,246.5) -- (436,246.5) -- (278,162.5) -- (258,162.5) -- cycle ;
\draw    (183,39) .. controls (240,49) and (374,62) .. (429,50) ;
\draw    (232,117) .. controls (279,93) and (378,96) .. (454,109) ;

\draw (45,200.4) node [anchor=north west][inner sep=0.75pt]    {$\mathbb{R}^{d-k}$};
\draw (192,66.4) node [anchor=north west][inner sep=0.75pt]    {$M$};
\draw (451,66.4) node [anchor=north west][inner sep=0.75pt]    {$N$};
\draw (385,258) node [anchor=north west][inner sep=0.75pt]   [align=left] {Domain wall};
\draw (313,65.4) node [anchor=north west][inner sep=0.75pt]    {$X_{M,N}$};

\end{tikzpicture}
    \caption{If the compact manifolds $M$ and $N$ are cobordant, they can be realized as boundaries of a higher dimensional manifold $X_{M,N}$. We can use $X_{M,N}$ to continiously transition between backgrounds with compact manifolds $M$ and $N$ as we move along a certain non-compact dimension. As shown in the figure above, in the non-compact dimensions, this configuration looks like a domain wall between two backgrounds with different comapact manifolds $M$ and $N$.}
    \label{fig:my_label1}
\end{figure}
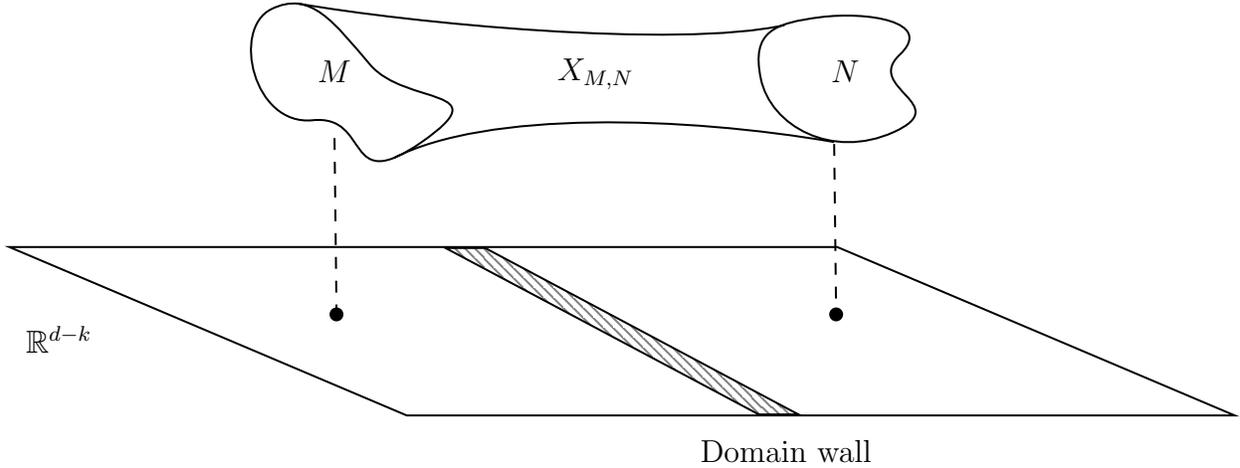

This statement has a close and natural connection to a mathematical concept called cobordism. Two $k$-dimensional orientable manifolds $M_1$ and $M_2$  are called cobordant if we can connect them via a $k+1$ dimensional manifold that has $M_1$ and $M_2$ as its boundaries. We can enrich this definition by imposing extra structure on $M_1$, $M_2$, and $\Sigma$ (e.g. spin structure). A class of manifolds that are cobordant are said to belong to the same cobordism class. Cobordism classes come with a natural Abelian group structure $+$ ,
\begin{align}
    [M_1]+[M_2]=[M_1\sqcup M_2],
\end{align}
where $\sqcup$ is the disjoint union. The identity element is the cobordism class of the empty manifold. This abelian group that captures non-equivalent calsses of $k$ dimensional manifolds carrying a structure $G$ is shown by $\Omega_k^G$. 

According to our definition above, there is a natural appearance of the notion of cobordism in quantum gravity, with the exception that rules of transition are determined by physics (existence of a finite tension domain wall). We denote this cobordism group by $\Omega_k^{QG}$. The cobordism conjecture states that $\Omega^{QG}_k$ is trivial. 

Therefore, whenever we have a non-trivial mathematical cobordism group $\Omega_k^G$, it means that the mathematical constrains do not accurately capture the physical rules. Any non-trivial class is either:
\begin{itemize}
    \item not allowed in quantum gravity (symmetry is gauged)
    \item can be transformed to the trivial class with a process that was not included in the mathematical evaluation of the cobordism group (symmetry is broken)
\end{itemize}

Note that every non-trivial cobordism class $\Omega_k$ defines a topological  $d-k-1$ defect obtained from gluing $M^k$ to $\mathbb{R}^k$ on the boundary of a small disk $D^k$. To see the global symmetry consider the group of homomorphisms from $\Omega_k$ to $U(1)$. Every element of this group acts on the $d-k-1$ defects. Therefore, non-trivial cobordism groups leads to non-trivial higher-form global symmetries. 

The cobordism conjecture (which is in essence the no global symmetry conjecture) is the ultimate duality. For example, it implies that there is a domain wall even between 10d string theories without any dimensional reduction. In particular, any QG should be realizable in a bubble inside any other QG.

Trivializing a cobordism class is equivalent to an end of the universe wall in the lower dimensional theory. In the language of condensed matter physics, this means you can always gap out any QG system by appropriate boundary conditions without any symmetry structure preventing it. Every QG admits a domain wall! 

There is a familiar challenge to have an end of the universe wall in chiral theories where we have no parity symmetry to gap out. For example take the IIB theory. What boundary condition do we put on gravitino which has a single chirality? Since the anomaly cancellation relates gravitino to the four form gauge field, their boundary conditions are likely mixed. But we do not know what that would exactly look like. There is a similar story with other string theories. But if such walls exist, how come we have not discovered them yet? 
\vspace{10pt}

\noindent\textbf{Exercise 1: }Prove that the boundary of the IIB theory in 10d (if there is one) breaks supersymmetry completely. This explains why it is difficult to construct.

\vspace{10pt}

Now let us look at an example where such end of the universe walls can preserve supersymmetry. For example in 11d where there is no chirality there is a well-known construction: the Hořava–Witten wall. 

Let us consider the cobordism classes of IIB. We do not know all the rules of quantum gravitiy for IIB in terms of what backgrounds are allowed. But IIB does require a spin structure and we can look at a subclass of rules. The cobordism classes of compact manifolds with a spin-structure is as follows.

\vspace{6pt}
 \begin{figure}[H]
\[ \arraycolsep=5pt\def\arraystretch{1.5} \begin{array}{c|cccccccc}
k & 0 & 1 & 2 & 3 & 4 & 5 & 6 & 7 \\
\hline
\Omega_k^{Spin} & \ZZ & \ZZ_2 & \ZZ_2 & 0 & \ZZ & 0 & 0 & 0 \\
\text{Gens} & \pt^+ & S^1_p & S^1_p \times S^1_p & - & K3 & - & - & - \\
 \end{array} \]
 \end{figure}

The $k=0$ case is trivialized by an end of the universe wall which in non-supersymmetric and unknown. But what about $k=1$ generated by circle? We already know how to trivialize it! We mod out by $\mathbb{Z}_2$ both the comapct circle and the base $\mathbb{R}$. This is just the orientifold plane. This construction is often called the pillowcase construction.

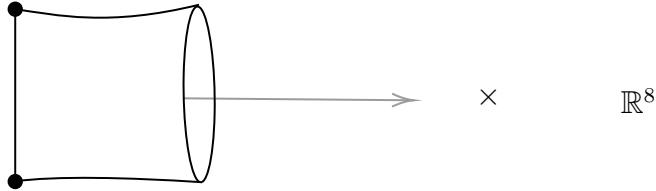
\begin{figure}[H]
    \centering

\tikzset{every picture/.style={line width=0.75pt}} 

\begin{tikzpicture}[x=0.75pt,y=0.75pt,yscale=-1,xscale=1]

\draw [color={rgb, 255:red, 155; green, 155; blue, 155 }  ,draw opacity=1 ]   (245,105) -- (359,105.98) ;
\draw [shift={(361,106)}, rotate = 180.49] [color={rgb, 255:red, 155; green, 155; blue, 155 }  ,draw opacity=1 ][line width=0.75]    (10.93,-3.29) .. controls (6.95,-1.4) and (3.31,-0.3) .. (0,0) .. controls (3.31,0.3) and (6.95,1.4) .. (10.93,3.29)   ;
\draw    (160,60) -- (160,147) ;
\draw [shift={(160,147)}, rotate = 90] [color={rgb, 255:red, 0; green, 0; blue, 0 }  ][fill={rgb, 255:red, 0; green, 0; blue, 0 }  ][line width=0.75]      (0, 0) circle [x radius= 3.35, y radius= 3.35]   ;
\draw [shift={(160,60)}, rotate = 90] [color={rgb, 255:red, 0; green, 0; blue, 0 }  ][fill={rgb, 255:red, 0; green, 0; blue, 0 }  ][line width=0.75]      (0, 0) circle [x radius= 3.35, y radius= 3.35]   ;
\draw   (253.68,59.76) .. controls (257.99,65.16) and (261.07,88.9) .. (260.56,112.79) .. controls (260.04,136.68) and (256.13,151.66) .. (251.83,146.26) .. controls (247.52,140.86) and (244.44,117.12) .. (244.95,93.23) .. controls (245.47,69.34) and (249.37,54.36) .. (253.68,59.76) -- cycle ;
\draw    (160,60) .. controls (181,63) and (207,69) .. (252.68,57.76) ;
\draw    (159,147) .. controls (180,144) and (218,145) .. (252.83,147.26) ;

\draw (391,97.4) node [anchor=north west][inner sep=0.75pt]    {$\times $};
\draw (464,98.4) node [anchor=north west][inner sep=0.75pt]    {$\mathbb{R}^{8}$};

\end{tikzpicture}
    \caption{The 'pillowcase' geometry as a IIB orientifold. The $\mathbb{Z}_2$ acts on $\mathbb{R}^9\times S^1$. It acts as a reflection on the $S^2$ and one of the $\mathbb{R}$s. It also changes the orientation of the worldsheet. Each one of the corners is an O7 plane. If we glue two of these geometries and add the correct number of D7 branes, we get type IIB on $T^2/\mathbb{Z}_2$ which looks like a full pillowcase. }
\end{figure}

The $k=2$ cobordism classes generated by torus can be trivialized in a similar way. But what about the $k=4$ classes generated by K3? This is related to our discussion in the previous section about gluing a K3 to $\mathbb{R}^4$ to get a conserved charge.
\vspace{10pt}

\noindent\textbf{Exercise 2: }Show that any end of the universe wall that could trivialize the cobordism class of $\Omega^{Spin}_4$ (generated by K3) in IIB must be non-supersymmetric. (Hint: Show that the end of the universe wall in IIB on K3 is not supersymmetric.)
\vspace{10pt}

Now let us think about the Heterotic case. There is more structure needed for it than just spin. Consider the class where $F\wedge F =0$. This equation requires the manifold to have $\mathcal{R}\wedge\mathcal{R}=0$ due to the equation of motion $dH=\frac{1}{16\pi^2}[\tr(\mathcal{R}\wedge\mathcal{R})-\tr(F\wedge F)]$. In other words, the first Pontryagin class $\frac{1}{2}P_1(\mathcal{R})$ must be trivial. Such manifolds are called string manifolds. String cobordism classes are shown in the following table.

\vspace{6pt}
\begin{figure}[H]
\[ \arraycolsep=5pt\def\arraystretch{1.5} \begin{array}{c|cccccccc}
k & 0 & 1 & 2 & 3 & 4 & 5 & 6 & 7 \\
\hline
\Omega_k^\String & \ZZ & \ZZ_2 & \ZZ_2 & \ZZ_{24} & 0 & 0 & \ZZ_2 & 0 \\
\text{Gens} & \pt^+ & S^1_p & S^1_p \times S^1_p & S^3_H & - & - & S^3_H \times S^3_H & - \\
 \end{array} \]
 \end{figure}

We can trivialize the $k=3$ case with NS5 branes. The generator is $S^3$ with unit $H$ flux. One can think of this $S^3$ as the 3-sphere that surrounds the NS5 brane. If we look at the transverse dimensions to the NS5 brane, we can write it as $\mathbb{R}_+\times S^3$, where the half-line is parametrized by the distance from the NS5 brane. As the distance decreases towards 0, the size of the 3-sphere carrying the $H$ flux shrinks until it reaches a singular point, where the NS5 brane is sitting. 

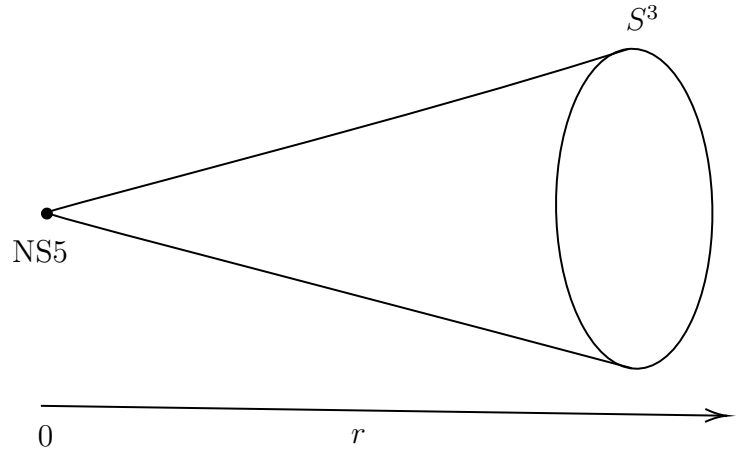
\begin{figure}[H]
    \centering

\tikzset{every picture/.style={line width=0.75pt}} 

\begin{tikzpicture}[x=0.75pt,y=0.75pt,yscale=-1,xscale=1]

\draw   (475.42,60.83) .. controls (492.19,90.23) and (494.53,141.24) .. (480.63,174.77) .. controls (466.74,208.29) and (441.88,211.64) .. (425.11,182.24) .. controls (408.34,152.84) and (406.01,101.83) .. (419.91,68.3) .. controls (433.8,34.78) and (458.66,31.43) .. (475.42,60.83) -- cycle ;
\draw    (449,202) .. controls (246,148) and (154,125) .. (154.07,123.46) .. controls (154.13,121.93) and (346.78,73.47) .. (447.25,40.88) ;
\draw  [fill={rgb, 255:red, 0; green, 0; blue, 0 }  ,fill opacity=1 ] (151.57,123.96) .. controls (151.57,122.58) and (152.69,121.46) .. (154.07,121.46) .. controls (155.45,121.46) and (156.57,122.58) .. (156.57,123.96) .. controls (156.57,125.34) and (155.45,126.46) .. (154.07,126.46) .. controls (152.69,126.46) and (151.57,125.34) .. (151.57,123.96) -- cycle ;
\draw    (151,221) -- (495,225.97) ;
\draw [shift={(497,226)}, rotate = 180.83] [color={rgb, 255:red, 0; green, 0; blue, 0 }  ][line width=0.75]    (10.93,-3.29) .. controls (6.95,-1.4) and (3.31,-0.3) .. (0,0) .. controls (3.31,0.3) and (6.95,1.4) .. (10.93,3.29)   ;

\draw (444.54,17.31) node [anchor=north west][inner sep=0.75pt]    {$S^{3}$};
\draw (135,136) node [anchor=north west][inner sep=0.75pt]   [align=left] {NS5};
\draw (306,231.4) node [anchor=north west][inner sep=0.75pt]    {$r$};
\draw (148,229.4) node [anchor=north west][inner sep=0.75pt]    {$0$};

\end{tikzpicture}
    \caption{A schematic representation of the four transverse dimensions of NS5, as a shrinking $S^3$ that carries the $H$-flux.}
    \label{fig:my_label2}
\end{figure}

We can use a similar argument to see that two intersecting NS5 branes trivialize the 6d cobordism group which is generated by $S_H^3\times S_H^3$. Consider two NS5 branes that have a 2d intersection. We can parameterize the $\mathbb{R}^8$ transverse to that intersection as $\mathbb{R}^2\times\mathbb{R}_+^2\times (S^3\times S^3)$ where each $\mathbb{R}_+$ denotes the distance from one of the NS5 branes. These distances also control the sizes of the 3-spheres that surround the fivebranes. The boundary of $\mathbb{R}^2\times\mathbb{R}_+^2$ which is made up of the fivebranes, is a domain wall for the theory on $S_H^3\times S_H^3$.

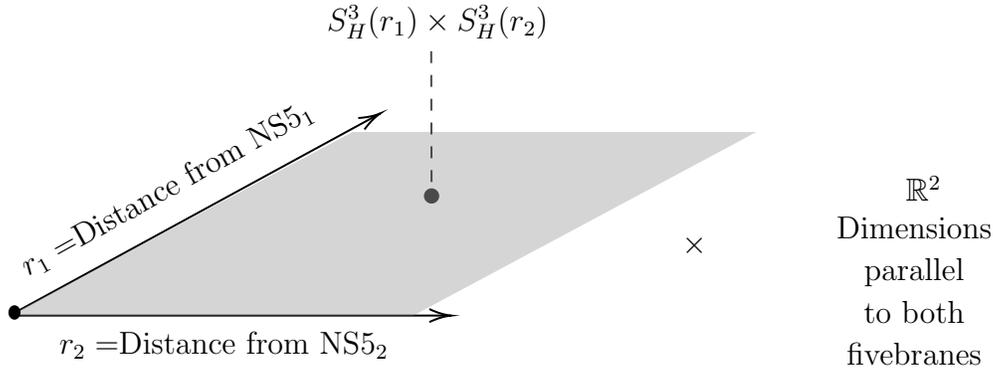
\begin{figure}[H]
    \centering

\tikzset{every picture/.style={line width=0.75pt}} 

\begin{tikzpicture}[x=0.75pt,y=0.75pt,yscale=-1,xscale=1]

\draw    (42.29,182.43) -- (263.01,182.43) ;
\draw [shift={(265.01,182.43)}, rotate = 180] [color={rgb, 255:red, 0; green, 0; blue, 0 }  ][line width=0.75]    (10.93,-3.29) .. controls (6.95,-1.4) and (3.31,-0.3) .. (0,0) .. controls (3.31,0.3) and (6.95,1.4) .. (10.93,3.29)   ;
\draw    (42.29,182.43) -- (227.62,81.33) ;
\draw [shift={(229.37,80.37)}, rotate = 151.39] [color={rgb, 255:red, 0; green, 0; blue, 0 }  ][line width=0.75]    (10.93,-3.29) .. controls (6.95,-1.4) and (3.31,-0.3) .. (0,0) .. controls (3.31,0.3) and (6.95,1.4) .. (10.93,3.29)   ;
\draw  [draw opacity=0][fill={rgb, 255:red, 155; green, 155; blue, 155 }  ,fill opacity=0.42 ] (215.31,89.79) -- (419,89.79) -- (245.98,182.43) -- (42.29,182.43) -- cycle ;
\draw  [fill={rgb, 255:red, 0; green, 0; blue, 0 }  ,fill opacity=1 ] (42.29,180.86) .. controls (42.29,179.13) and (43.43,177.72) .. (44.83,177.72) .. controls (46.24,177.72) and (47.38,179.13) .. (47.38,180.86) .. controls (47.38,182.59) and (46.24,184) .. (44.83,184) .. controls (43.43,184) and (42.29,182.59) .. (42.29,180.86) -- cycle ;
\draw [color={rgb, 255:red, 74; green, 74; blue, 74 }  ,draw opacity=1 ] [dash pattern={on 4.5pt off 4.5pt}]  (255,49) -- (255.19,122.06) ;
\draw [shift={(255.19,122.06)}, rotate = 89.85] [color={rgb, 255:red, 74; green, 74; blue, 74 }  ,draw opacity=1 ][fill={rgb, 255:red, 74; green, 74; blue, 74 }  ,fill opacity=1 ][line width=0.75]      (0, 0) circle [x radius= 3.35, y radius= 3.35]   ;

\draw (44.2,152.38) node [anchor=north west][inner sep=0.75pt]  [rotate=-330.9] [align=left] {$\displaystyle r_{1} =$Distance from $\displaystyle \text{NS5}_{1}$};
\draw (66,189) node [anchor=north west][inner sep=0.75pt]   [align=left] {$\displaystyle r_{2} =$Distance from $\displaystyle \text{NS5}_{2}$};
\draw (380,140.4) node [anchor=north west][inner sep=0.75pt]    {$\times $};
\draw (493,110.4) node [anchor=north west][inner sep=0.75pt]    {$\mathbb{R}^{2}$};
\draw (436,131) node [anchor=north west][inner sep=0.75pt]   [align=left] {\begin{minipage}[lt]{91.73pt}\setlength\topsep{0pt}
\begin{center}
Dimensions parallel\\ to both fivebranes
\end{center}

\end{minipage}};
\draw (201,24.4) node [anchor=north west][inner sep=0.75pt]    {$S_{H}^{3}( r_{1}) \times S_{H}^{3}( r_{2})$};

\end{tikzpicture}
    \caption{We view the 10d background with two intersecting fivebranes, as $\mathbb{R}^2\times\mathbb{R}_+^2\times (S^3\times S^3)$. The $\mathbb{R}^2$ represents the two dimensions parallel to the intersection of the fivebranes, while the rest is two copies of $\mathbb{R}_+\times S_H^3=\mathbb{R}^4_*$, each of which is the four transverse dimensions to a fivebrane. Therefore, we can think of this background as the 10 theory on $S_H^3\times S_H^3$ where the radii of the 3-spheres depend on the remaining four coordinates. The remaining four non-compact dimensions form a $\mathbb{R}^2\times\mathbb{R}_+^2$ which ends on a co-dimension one brane made up of fivebranes.}
\end{figure}

We showed how to trivialize the cobodism classes for $k=3$ and $6$. But what about $k=0,1, \text{and},2$? It is not difficult to see that the remaining cobordism require non-BPS defects to be trivialized \cite{McNamara:2019rup}.

Now let us study a different type of cobordism classes. M-theory has parity so we can compactify it on non-orientible manifolds which should carry a Pin structure. There are two possible Pin structures but the formulation of M-theory is only consistent on $\Pin^+$ manifolds. 

\vspace{6pt}
\begin{figure}[H]
\[ \arraycolsep=5pt\def\arraystretch{1.5} \begin{array}{c|ccccccccc}
k & 0 & 1 & 2 & 3 & 4 & 5 & 6 & 7 & 8 \\
\hline
\Omega_k^{\Pin^+} & \ZZ_2 & 0 & \ZZ_2 & \ZZ_2 & \ZZ_{16} & 0 & 0 & 0 & \ZZ_2 \times \ZZ_{32} \\
\text{Gens} & \pt & - & KB & KB \times S^1_p & \RR \PP^4 & - & - & - & \HH \PP^2, \RR \PP^8 \\
 \end{array} \]
 \end{figure}

As we mentioned above, the first class can be killed by the Hořava–Witten wall. Generators for the rest non-zero ones are respectively, $K_B$ , $K_B\times S^1$, and $\mathbb{R}\mathds{P}^4$. The first two cobordism classes cannot be killed supersymmetrically, however, the last one is possible to trivialze in a supersymmetric fashion and arises in the familiar compactification of M-theory on $\mathbb{T}^5/\mathbb{Z}_2$ \cite{Witten:1995em}. The $\mathbb{Z}_2$ acts on all coordinates of the torus by reflection, and it leaves 32 fixed points. The geometry around each fixed point is $\mathbb{R}^5/\mathbb{Z}_2$ which can be viewed as $\mathbb{R}\mathds{P}^4\times\mathbb{R}_+$. The defect sitting at the fixed point is the MO5 M-orientifold which is an end of the universe wall to M-theory on $\mathbb{RP}^4$ and therefore, trivializes its cobordism class. Similarly, the MO1 \cite{Becker:1996gj} trivializes the cobordism class of $\mathbb{R}^8$ given that its transverse geometry is given by $\mathbb{R}_*^9/\mathbb{Z}_2=\mathbb{RP}_*^9\times \mathbb{R}_+$. As for the cobordism class generated by $\mathbb{HP}^2$, we can show that it is gauged. In other words, Compactification of M-theory on $\mathbb{HP}^2$ is not allowed. This follows from the tadpole cancelation condition \cite{Becker:1996gj}. If we compactify M-theory on a compact 8-dimensional manifold $X$, we have
\begin{align}
    N_{M2}+\frac{1}{2}G_4(X)^2=I_8(X),
\end{align}
where $N_{M2}$ is the number of spacefilling M2 branes and
\begin{align}
    I_8(X)=\int\frac{p_2(R)-(p_1(R)/2)^2}{48}.
\end{align}
Therefore, if we do not turn on any gauge fields ($G_4=0$), or insert spacefilling M2 branes ($N_{M2}=0$), the only allowed compactifications are those satisfying $I_8(X)=0$. However, $I_8(\mathbb{HP}^2)=1/8\neq0$, and therefore, is not an allowed compactification. Similar to the previous cases, the remaining cobordism classes can be shown to require a non-BPS defect to be trivialized.

Let us point out that since we do not have a complete formulation of quantum gravity, we cannot exactly calculate the cobordism groups. This is why we resort to the approach that we calculate the cobordism classes with approximate rules and show that in the exact theory they indeed vanish.

The cobordism conjecture could lead to very powerful statements. For example, in \cite{Montero:2020icj}, the cobordism conjecture was used to argue that comapctifications of certain higher dimensional supersymemtric theories on $T^3/\mathbb{Z}_2$ must be allowed. By checking the anomaly cancellation in the compactified theories, one can restrict the rank of the gauge group in the original theories. In some cases these restrictions are so strong that the allowed ranks match with the existing examples in string theory.

\subsection{Baby universe hypothesis}

In section, \ref{NCVC} we saw that in the presence of gauge symmetries, there is a major difference between compact and non-compact spaces. Compact spaces cannot have states with net gauge charge while non-compact Hilbert spaces can. We saw that this is because non-compact spaces come with operators that extend to the boundary and can create net charge. In fact, this enlargement of the Hilbert space is very natural from holography's point of view. If the degrees of freedom live on the boundary, then compact spaces must have none. In other words, taking holography at its face value suggests the following hypothesis \cite{McNamara:2020uza}. 

\begin{statement4*}
The Hilbert space of quantum gravity on a compact boundary-less spaces with more than three dimensions is trivial.
\end{statement4*}

Note that boundary-less is a crucial condition for the baby universe hypothesis. For example if you remove a disc from your compact manifolds, you would expect to have degrees of freedom living on the boundary of the removed disc. 

Note that there are counterexamples to this hypothesis in 2d quantum gravities. For example a 2d quantum gravity could be described by an ensemble of many degrees of freedom given by the SYK model. Two-dimensional quantum gravities are special and often avoid Swampland conditions. For example, they can have global symmetries as well. For example, the worldsheet theory of Heterotic strings admits the spacetime gauge symmetry as a global symmetry global symmetry. 

Some of the differences between two and higher dimensions might be related to the fact that many of the Swampland conjectures are motivated by black holes which typically exist in dimensions greater than two. 

\begin{figure}[H]
    \centering

\tikzset{every picture/.style={line width=0.75pt}} 

\begin{tikzpicture}[x=0.75pt,y=0.75pt,yscale=-1,xscale=1]

\draw   (455.5,58.13) -- (455.5,230.88) .. controls (455.5,244.75) and (418,256) .. (371.75,256) .. controls (325.5,256) and (288,244.75) .. (288,230.88) -- (288,58.13) .. controls (288,44.25) and (325.5,33) .. (371.75,33) .. controls (418,33) and (455.5,44.25) .. (455.5,58.13) .. controls (455.5,72) and (418,83.25) .. (371.75,83.25) .. controls (325.5,83.25) and (288,72) .. (288,58.13) ;
\draw  [draw opacity=0][fill={rgb, 255:red, 255; green, 255; blue, 255 }  ,fill opacity=1 ] (268.5,112.5) .. controls (268.5,103.39) and (275.89,96) .. (285,96) .. controls (294.11,96) and (301.5,103.39) .. (301.5,112.5) .. controls (301.5,121.61) and (294.11,129) .. (285,129) .. controls (275.89,129) and (268.5,121.61) .. (268.5,112.5) -- cycle ;
\draw  [draw opacity=0][fill={rgb, 255:red, 255; green, 255; blue, 255 }  ,fill opacity=1 ] (274.5,192.5) .. controls (274.5,186.15) and (279.65,181) .. (286,181) .. controls (292.35,181) and (297.5,186.15) .. (297.5,192.5) .. controls (297.5,198.85) and (292.35,204) .. (286,204) .. controls (279.65,204) and (274.5,198.85) .. (274.5,192.5) -- cycle ;
\draw    (299.5,181) .. controls (254.5,192) and (252.91,133.84) .. (299.91,125.84) ;
\draw    (297.5,96) .. controls (205,104) and (230.5,220) .. (302.5,200) ;
\draw    (299.91,125.84) .. controls (305.5,119) and (304.5,105) .. (297.5,96) ;
\draw    (302.5,200) .. controls (307.5,195.5) and (306.5,180.5) .. (299.5,181) ;
\draw    (239.5,164.5) .. controls (244.5,173.5) and (263.5,172.5) .. (265.5,162.5) ;
\draw [color={rgb, 255:red, 0; green, 0; blue, 0 }  ,draw opacity=1 ] [dash pattern={on 4.5pt off 4.5pt}]  (239.5,164.5) .. controls (251.5,156.5) and (258.5,159.5) .. (265.5,162.5) ;

\draw (164,145) node [anchor=north west][inner sep=0.75pt]   [align=left] {\begin{minipage}[lt]{41.28pt}\setlength\topsep{0pt}
\begin{center}
Baby \\universe
\end{center}

\end{minipage}};
\draw (317.5,182.4) node [anchor=north west][inner sep=0.75pt]    {$\hat{a}^{\dagger }$};
\draw (309.5,98.4) node [anchor=north west][inner sep=0.75pt]    {$\hat{a}$};
\end{tikzpicture}
    \caption{}
    \label{baby}
\end{figure}
The reason the hypothesis is called the baby universe hypothesis is due to the fact that emission and absorption of compact universes is called a baby universe. Coleman studied these processes \cite{Coleman:1988cy} and showed that one can associate a creation and annihilation pair of operators $\{a_i,a_i^\dagger\}$ to any degree of freedom $i$ of a baby universe. Moreover, they modify the effective action as  $\mathcal{L}=\mathcal{L}_0(\Phi,...)+\sum_i(a_i+a_i^\dagger)\mathcal{L}_1(\Phi,...)$. A coherent state $\ket{\alpha}$ defined as $a_i\ket{\alpha}=\alpha_i\ket{\alpha}$ corresponds to summing over processes involving emission/absorption of baby universes with specific weights. Every $\ket{\alpha}$ corresponds to a different vacuum of the theory. This is very similar to the notion of $\theta$-vacua in gauge theories. Note that $\alpha$ is a parameter and not a field. The vacua labelled by different $\alpha$ belong to different superselection sectors which violates the cobordism conjecture. Therefore, if one believes in no-cobordism conjecture (\emph{i.e.} no-global symmetry), the baby universe Hilbert space must be just one-dimensional.\footnote{For a different perspective on the baby universe hypothesis see \cite{Gesteau:2020wrk}.}

\section{Swampland II: Completeness of spectrum}

In the previous section, we talked about global symmetries in quantum gravity and we argued that all the global symmetries must be gauged. We also saw that pure gauge theories make sense and have physical implication in terms of the spectrum of physical operators. We defined the gauge theories by the inclusion of Wilson loops or Wilson lines that have asymptotic endpoints. However, if one adds charged particles, we can make new gauge invariant operators by including those charged operators at the endpotins of the Wilson lines. Such an operator can be thought of as an operator that creates/annihilates charged particles at the endpoints of the Wilson line.

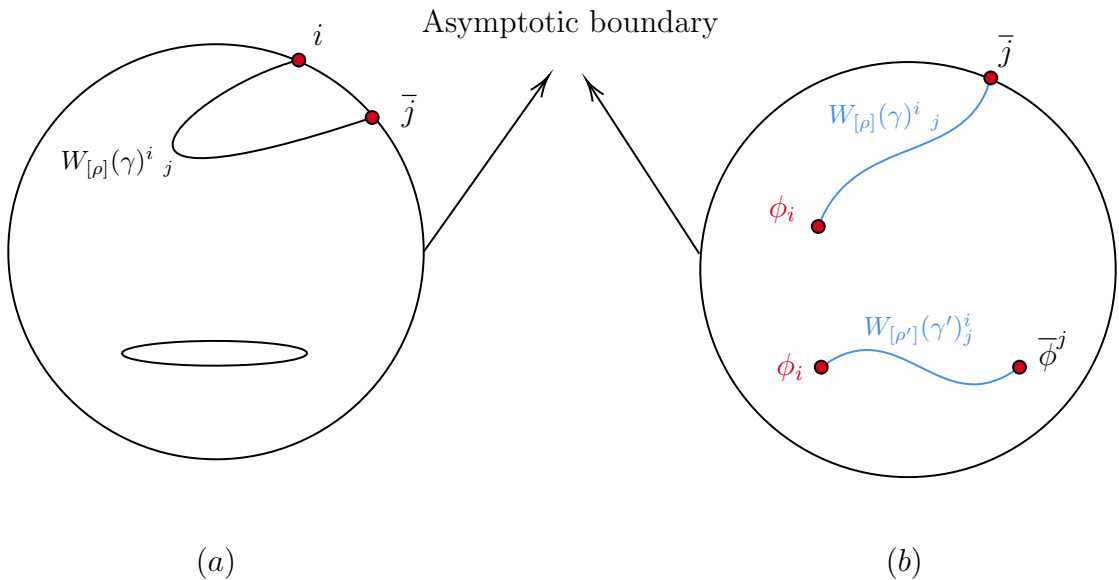
\begin{figure}[H]
    \centering

\tikzset{every picture/.style={line width=0.75pt}} 

\begin{tikzpicture}[x=0.75pt,y=0.75pt,yscale=-1,xscale=1]

\draw [color={rgb, 255:red, 74; green, 144; blue, 226 }  ,draw opacity=1 ]   (455,198) .. controls (495,168) and (515,228) .. (555,198) ;
\draw [color={rgb, 255:red, 74; green, 144; blue, 226 }  ,draw opacity=1 ]   (453.5,127) .. controls (471.5,79) and (528.5,97) .. (540.5,52) ;
\draw   (45,139.75) .. controls (45,81.9) and (91.9,35) .. (149.75,35) .. controls (207.6,35) and (254.5,81.9) .. (254.5,139.75) .. controls (254.5,197.6) and (207.6,244.5) .. (149.75,244.5) .. controls (91.9,244.5) and (45,197.6) .. (45,139.75) -- cycle ;
\draw    (228.5,72) .. controls (67.5,126) and (134.5,58) .. (191.5,43) ;
\draw  [fill={rgb, 255:red, 208; green, 2; blue, 27 }  ,fill opacity=1 ] (188.25,43) .. controls (188.25,41.21) and (189.71,39.75) .. (191.5,39.75) .. controls (193.29,39.75) and (194.75,41.21) .. (194.75,43) .. controls (194.75,44.79) and (193.29,46.25) .. (191.5,46.25) .. controls (189.71,46.25) and (188.25,44.79) .. (188.25,43) -- cycle ;
\draw  [fill={rgb, 255:red, 208; green, 2; blue, 27 }  ,fill opacity=1 ] (225.25,72) .. controls (225.25,70.21) and (226.71,68.75) .. (228.5,68.75) .. controls (230.29,68.75) and (231.75,70.21) .. (231.75,72) .. controls (231.75,73.79) and (230.29,75.25) .. (228.5,75.25) .. controls (226.71,75.25) and (225.25,73.79) .. (225.25,72) -- cycle ;
\draw   (132.9,185.15) .. controls (157.07,183.94) and (183.87,185.58) .. (192.76,188.81) .. controls (201.65,192.05) and (189.27,195.64) .. (165.1,196.85) .. controls (140.93,198.06) and (114.13,196.42) .. (105.24,193.19) .. controls (96.35,189.95) and (108.73,186.36) .. (132.9,185.15) -- cycle ;
\draw   (394,148.75) .. controls (394,90.9) and (440.9,44) .. (498.75,44) .. controls (556.6,44) and (603.5,90.9) .. (603.5,148.75) .. controls (603.5,206.6) and (556.6,253.5) .. (498.75,253.5) .. controls (440.9,253.5) and (394,206.6) .. (394,148.75) -- cycle ;
\draw  [fill={rgb, 255:red, 208; green, 2; blue, 27 }  ,fill opacity=1 ] (537.25,52) .. controls (537.25,50.21) and (538.71,48.75) .. (540.5,48.75) .. controls (542.29,48.75) and (543.75,50.21) .. (543.75,52) .. controls (543.75,53.79) and (542.29,55.25) .. (540.5,55.25) .. controls (538.71,55.25) and (537.25,53.79) .. (537.25,52) -- cycle ;
\draw  [fill={rgb, 255:red, 208; green, 2; blue, 27 }  ,fill opacity=1 ] (450.25,127) .. controls (450.25,125.21) and (451.71,123.75) .. (453.5,123.75) .. controls (455.29,123.75) and (456.75,125.21) .. (456.75,127) .. controls (456.75,128.79) and (455.29,130.25) .. (453.5,130.25) .. controls (451.71,130.25) and (450.25,128.79) .. (450.25,127) -- cycle ;
\draw  [fill={rgb, 255:red, 208; green, 2; blue, 27 }  ,fill opacity=1 ] (451.75,198) .. controls (451.75,196.21) and (453.21,194.75) .. (455,194.75) .. controls (456.79,194.75) and (458.25,196.21) .. (458.25,198) .. controls (458.25,199.79) and (456.79,201.25) .. (455,201.25) .. controls (453.21,201.25) and (451.75,199.79) .. (451.75,198) -- cycle ;
\draw  [fill={rgb, 255:red, 208; green, 2; blue, 27 }  ,fill opacity=1 ] (551.75,198) .. controls (551.75,196.21) and (553.21,194.75) .. (555,194.75) .. controls (556.79,194.75) and (558.25,196.21) .. (558.25,198) .. controls (558.25,199.79) and (556.79,201.25) .. (555,201.25) .. controls (553.21,201.25) and (551.75,199.79) .. (551.75,198) -- cycle ;
\draw    (254.5,139.75) -- (316.34,52.63) ;
\draw [shift={(317.5,51)}, rotate = 125.37] [color={rgb, 255:red, 0; green, 0; blue, 0 }  ][line width=0.75]    (10.93,-3.29) .. controls (6.95,-1.4) and (3.31,-0.3) .. (0,0) .. controls (3.31,0.3) and (6.95,1.4) .. (10.93,3.29)   ;
\draw    (393.5,140.75) -- (337.59,54.68) ;
\draw [shift={(336.5,53)}, rotate = 56.99] [color={rgb, 255:red, 0; green, 0; blue, 0 }  ][line width=0.75]    (10.93,-3.29) .. controls (6.95,-1.4) and (3.31,-0.3) .. (0,0) .. controls (3.31,0.3) and (6.95,1.4) .. (10.93,3.29)   ;

\draw (252,16) node [anchor=north west][inner sep=0.75pt]   [align=left] {Asymptotic boundary};
\draw (70,85.4) node [anchor=north west][inner sep=0.75pt]  [font=\footnotesize]  {$W_{[ \rho ]}( \gamma )^{i} \ _{j}$};
\draw (197,23.4) node [anchor=north west][inner sep=0.75pt]    {$i$};
\draw (242,58.4) node [anchor=north west][inner sep=0.75pt]    {$\overline{j}$};
\draw (475,169.4) node [anchor=north west][inner sep=0.75pt]  [font=\footnotesize,color={rgb, 255:red, 74; green, 144; blue, 226 }  ,opacity=1 ]  {$W_{[ \rho ']}( \gamma ')_{j}^{i}$};
\draw (138,287.4) node [anchor=north west][inner sep=0.75pt]    {$( a)$};
\draw (486,287.4) node [anchor=north west][inner sep=0.75pt]    {$( b)$};
\draw (431,189.4) node [anchor=north west][inner sep=0.75pt]  [color={rgb, 255:red, 208; green, 2; blue, 27 }  ,opacity=1 ]  {$\phi _{i}$};
\draw (563,178.4) node [anchor=north west][inner sep=0.75pt]    {$\overline{\phi }^{j}$};
\draw (457,62.4) node [anchor=north west][inner sep=0.75pt]  [font=\footnotesize,color={rgb, 255:red, 74; green, 144; blue, 226 }  ,opacity=1 ]  {$W_{[ \rho ]}( \gamma )^{i} \ _{j}$};
\draw (543,28.4) node [anchor=north west][inner sep=0.75pt]    {$\overline{j}$};
\draw (427,109.4) node [anchor=north west][inner sep=0.75pt]  [color={rgb, 255:red, 208; green, 2; blue, 27 }  ,opacity=1 ]  {$\phi _{i}$};

\end{tikzpicture}
    \caption{(a) Gauge invariant Wilson operators that exist in any theory with long-range gauge symmetry. (b) Additional gauge invariant Wilson operators that only exist in gauge theories with local charged operators.}
    \label{Line operators}
\end{figure}

\subsection{Completeness hypothesis}

As we will see, in quantum gravity, whenever there is a symmetry on the boundary, not only the symmetry is a gauge symmetry in the bulk, but also all the possible representations of that gauge group always appear in the spectrum of the theory (see \cite{Banks:2010zn,Polchinski:2003bq} for earlier arguments). This is called the completeness of spectrum conjecture. In principle, the question of what charged operators are allowed is in principle independent from that of whether global symmetries exist or not. However, it turns out that this conjecture is closely connected with the no-global symmetry conjecture. 

Let us emphasize two important points about what completeness of spectrum does and does not mean. 

\begin{itemize}
    \item The charged states are not necessarily low-energy states. In other words, the states that carry a certain representation of the gauge group might be very massive and not part of the low-energy EFT.
    \item The charged states are not necessarily stable one-particle states. The charged states could be multi-particle states or meta-stable bound states.
\end{itemize}

\subsection{Evidence in string theory}

Now that we know the statement of the conjecture, let us start examining this conjecture with the simplest known examples of gauge theories in string theory; the higher form gauge symmetries that typically come from the Ramond-Ramond sectors. 

In type II theories, the charged particles corresponding to such gauge symmetries are the D-branes which indeed are required to be in string theory by dualities. Moreover, for M-theory, the existence of the electric and magnetic gauge symmetries associated with the 3-form $C_{\mu\nu\rho}$ forces us to include charged objects that are 2+1 and 5+1 dimensional. There is indeed extensive evidence for such objects to exist in the non-perturbative describtion of M-theory. These extended objects are the M2 and M5 branes.

Now let us move to the Heterotic string theories. Heterotic theories have conventional 0-form gauge symmetries. Moreover, the string sees the gauge symmetry.  Let us make this terminology more precise. 

Typically, the gauge symmetry in spacetime is realized as a global symmetry on the worldsheet theory. If the action of this global symmetry on the spectrum of the worldsheet theory (string excitations) is non-trivial, we say the fundamental string sees the the gauge symmetry. This implies that there are string states that as spacetime particles are charged under the gauge symmetry. Therefore, non-trivial representations of $G$ as a global symmetry on the wolrdsheet theory usually corresponds to non-trivial representations of charged spacetime states under the spacetime gauge group $G$.

In the case of the Heterotic theory, the symmetry of the Heterotic theory is realized as an Affine Kac-Moody algebra on the worldsheet. This immediately implies that there are operators on the worldsheet that transform in the adjoint of the gauge group. These are just the spacetime gauge bosons. But how about the other representations? To check whether all the representations appear or not, first we need to know the spectrum of all representations. The set of all representations depends on the global structure of the group. In the $E_8\times E_8$ theory, the gauge group is $E_8\times E_8$ and the adjoint representation and its successive tensor products gives all. However, in the  
$SO(32)$ theory the gauge group is $\text{spin}(32)/\mathbb{Z}_2$. Therefore, the  the full set of the representations of the gauge group consists of half of representations of $\text{spin}(32)$.

In these cases, the level of the Kac-Moody algebra is 1. The fundamental representations that generate all possible representations are given by the corresponding Dynkin diagrams.
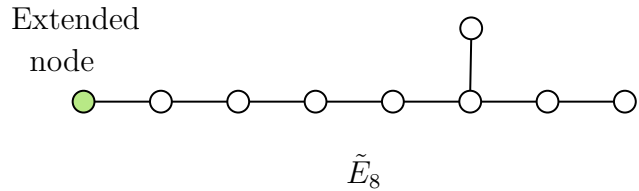
\begin{figure}[H]
    \centering

\tikzset{every picture/.style={line width=0.75pt}} 

\begin{tikzpicture}[x=0.75pt,y=0.75pt,yscale=-1,xscale=1]

\draw   (455.5,69.5) .. controls (455.5,66.46) and (457.96,64) .. (461,64) .. controls (464.04,64) and (466.5,66.46) .. (466.5,69.5) .. controls (466.5,72.54) and (464.04,75) .. (461,75) .. controls (457.96,75) and (455.5,72.54) .. (455.5,69.5) -- cycle ;
\draw   (416.5,69.5) .. controls (416.5,66.46) and (418.96,64) .. (422,64) .. controls (425.04,64) and (427.5,66.46) .. (427.5,69.5) .. controls (427.5,72.54) and (425.04,75) .. (422,75) .. controls (418.96,75) and (416.5,72.54) .. (416.5,69.5) -- cycle ;
\draw    (427.5,69.5) -- (455.5,69.5) ;
\draw   (377.5,69.5) .. controls (377.5,66.46) and (379.96,64) .. (383,64) .. controls (386.04,64) and (388.5,66.46) .. (388.5,69.5) .. controls (388.5,72.54) and (386.04,75) .. (383,75) .. controls (379.96,75) and (377.5,72.54) .. (377.5,69.5) -- cycle ;
\draw   (338.5,69.5) .. controls (338.5,66.46) and (340.96,64) .. (344,64) .. controls (347.04,64) and (349.5,66.46) .. (349.5,69.5) .. controls (349.5,72.54) and (347.04,75) .. (344,75) .. controls (340.96,75) and (338.5,72.54) .. (338.5,69.5) -- cycle ;
\draw    (349.5,69.5) -- (377.5,69.5) ;
\draw    (388.5,69.5) -- (416.5,69.5) ;
\draw   (299.5,69.5) .. controls (299.5,66.46) and (301.96,64) .. (305,64) .. controls (308.04,64) and (310.5,66.46) .. (310.5,69.5) .. controls (310.5,72.54) and (308.04,75) .. (305,75) .. controls (301.96,75) and (299.5,72.54) .. (299.5,69.5) -- cycle ;
\draw   (260.5,69.5) .. controls (260.5,66.46) and (262.96,64) .. (266,64) .. controls (269.04,64) and (271.5,66.46) .. (271.5,69.5) .. controls (271.5,72.54) and (269.04,75) .. (266,75) .. controls (262.96,75) and (260.5,72.54) .. (260.5,69.5) -- cycle ;
\draw    (271.5,69.5) -- (299.5,69.5) ;
\draw   (221.5,69.5) .. controls (221.5,66.46) and (223.96,64) .. (227,64) .. controls (230.04,64) and (232.5,66.46) .. (232.5,69.5) .. controls (232.5,72.54) and (230.04,75) .. (227,75) .. controls (223.96,75) and (221.5,72.54) .. (221.5,69.5) -- cycle ;
\draw  [fill={rgb, 255:red, 184; green, 233; blue, 134 }  ,fill opacity=1 ] (182.5,69.5) .. controls (182.5,66.46) and (184.96,64) .. (188,64) .. controls (191.04,64) and (193.5,66.46) .. (193.5,69.5) .. controls (193.5,72.54) and (191.04,75) .. (188,75) .. controls (184.96,75) and (182.5,72.54) .. (182.5,69.5) -- cycle ;
\draw    (193.5,69.5) -- (221.5,69.5) ;
\draw    (232.5,69.5) -- (260.5,69.5) ;
\draw    (310.5,69.5) -- (338.5,69.5) ;
\draw    (383.5,38) -- (383,64) ;
\draw   (378,32.5) .. controls (378,29.46) and (380.46,27) .. (383.5,27) .. controls (386.54,27) and (389,29.46) .. (389,32.5) .. controls (389,35.54) and (386.54,38) .. (383.5,38) .. controls (380.46,38) and (378,35.54) .. (378,32.5) -- cycle ;

\draw (319,94.4) node [anchor=north west][inner sep=0.75pt]    {$\tilde{E}_{8}$};
\draw (150,10) node [anchor=north west][inner sep=0.75pt]   [align=left] {\begin{minipage}[lt]{38.44pt}\setlength\topsep{0pt}
\begin{center}
Extended \\node
\end{center}

\end{minipage}};

\end{tikzpicture}
    \caption{Affine Dynkin diagram of the central extension of $\mathfrak{e}_8$.}
    \label{Affine Dynkin}
\end{figure}
The Dynkin diagrams of Kac-Moody algebras have an extra node compared to the ordinary Lie algebra due to the central extension. The fundamental representations are labled by their highest weights which satisfy
\begin{align}
    \sum n_i d_i\leq 1,
\end{align}

where $n_i$ is the coefficient of the $i$-th fundamental weight in the expansion of the highest weight and $d_i$s are the Dynkin labels. Moreover, since the contributing labels for fundamental representations are $d_i=1$, they are allowed for all levels $k$.  

In the case of the Heterotic $SO(32)$ thanks to modular invariance, all such fundamental representations are already included. This is because modular transformations map different fundamental representations to each other and upon including one of them, we are forced to include all of them. 

So far our discussion was focused on examples of the completeness of spectrum for continuous gauge groups. Suppose you have a discrete group $G$, do all representations appear?

\subsection{Completeness of spectrum for discrete symmetries}

Let us start our discussion with a special case where the discrete group is visible to the fundamental string. As we explained above, this means that there are charged strong excitations or in other words, the discrete gauge group realizes as a non-trivial global symmetry on the worldsheet theory. If so, we argue that the string spectrum must include all of the representations. Again, the key is the modular invariance!

Suppose you take a torus diagram and you put the a twisted boundary condition in the time direction. 

\begin{figure}[H]
    \centering

\tikzset{every picture/.style={line width=0.75pt}} 

\begin{tikzpicture}[x=0.75pt,y=0.75pt,yscale=-1,xscale=1]

\draw   (304.25,55) -- (456.5,55) -- (391.25,195) -- (239,195) -- cycle ;

\draw (255,28.4) node [anchor=north west][inner sep=0.75pt]    {$\text{Im}( \tau ) =\beta $};
\draw (257,98.4) node [anchor=north west][inner sep=0.75pt]    {$g$};
\draw (321,199.4) node [anchor=north west][inner sep=0.75pt]    {$e$};

\end{tikzpicture}
    \caption{The twisted sector corresponding to twisted boundary condition in the Euclidean time direction.}
    \label{Twist}
\end{figure}
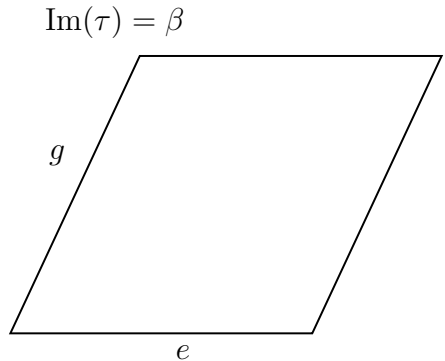

Note that we are not orbifolding the theory, but just computing a certain partition function,
\begin{align}\label{P1}
  \frac{Tr \tilde U_g e^{-\beta H}}{Tr e^{-\beta H}}.
\end{align}
where $\tilde U_g$ enforces the twisted boundary condition. We claim that the above twisted partition function, in the $  \lim_{\beta\rightarrow 0}$ vanishes unless $g$ is the identity element. Let us give two different arguments.
 
\noindent\textbf{Argument 1: }In the $\beta\rightarrow 0$ limit, the worldsheet fields must have a very strong gradient in the Euclidean time direction so to ensure the initial and final state differ by the action of $g$. Therefore, the kinetic part of the worldsheet scalars in the action contribute as
 \begin{align}
    e^{ -\int d\sigma\int_0^\beta|\nabla\phi|^2d\tau+\hdots}= e^{-\mathcal{O}(\frac{\beta}{\beta^2})}\rightarrow 0
 \end{align}
 \noindent\textbf{Argument 2: } Modular transformation $\tau\rightarrow 1/\tau$ maps the partition function \ref{P1} to the amplitude over on the $g$-twisted ground state sector.
 
 \begin{figure}[H]
     \centering

\tikzset{every picture/.style={line width=0.75pt}} 

\begin{tikzpicture}[x=0.75pt,y=0.75pt,yscale=-1,xscale=1]

\draw   (73,127) -- (210.5,127) -- (210.5,177) -- (73,177) -- cycle ;
\draw   (472,40) -- (609.5,40) -- (609.5,264) -- (472,264) -- cycle ;
\draw   (63.5,131) .. controls (58.83,131) and (56.5,133.33) .. (56.5,138) -- (56.5,141.61) .. controls (56.5,148.28) and (54.17,151.61) .. (49.5,151.61) .. controls (54.17,151.61) and (56.5,154.94) .. (56.5,161.61)(56.5,158.61) -- (56.5,167) .. controls (56.5,171.67) and (58.83,174) .. (63.5,174) ;
\draw    (244,158) -- (407.5,157.01) ;
\draw [shift={(409.5,157)}, rotate = 179.65] [color={rgb, 255:red, 0; green, 0; blue, 0 }  ][line width=0.75]    (10.93,-3.29) .. controls (6.95,-1.4) and (3.31,-0.3) .. (0,0) .. controls (3.31,0.3) and (6.95,1.4) .. (10.93,3.29)   ;
\draw   (464.5,42) .. controls (459.83,42) and (457.5,44.33) .. (457.5,49) -- (457.5,148) .. controls (457.5,154.67) and (455.17,158) .. (450.5,158) .. controls (455.17,158) and (457.5,161.33) .. (457.5,168)(457.5,165) -- (457.5,255) .. controls (457.5,259.67) and (459.83,262) .. (464.5,262) ;

\draw (31,145.4) node [anchor=north west][inner sep=0.75pt]    {$\beta $};
\draw (215,142.4) node [anchor=north west][inner sep=0.75pt]    {$g$};
\draw (134,178.4) node [anchor=north west][inner sep=0.75pt]    {$e$};
\draw (288,164.4) node [anchor=north west][inner sep=0.75pt]    {$\tau \rightarrow 1/\tau $};
\draw (423,137.4) node [anchor=north west][inner sep=0.75pt]    {$\frac{1}{\beta }$};
\draw (534,268.4) node [anchor=north west][inner sep=0.75pt]    {$g$};
\draw (625,131.4) node [anchor=north west][inner sep=0.75pt]    {$e$};

\end{tikzpicture}
     \caption{Modular transformation maps two twsited partition functions to one another.}
     \label{Modular}
 \end{figure}
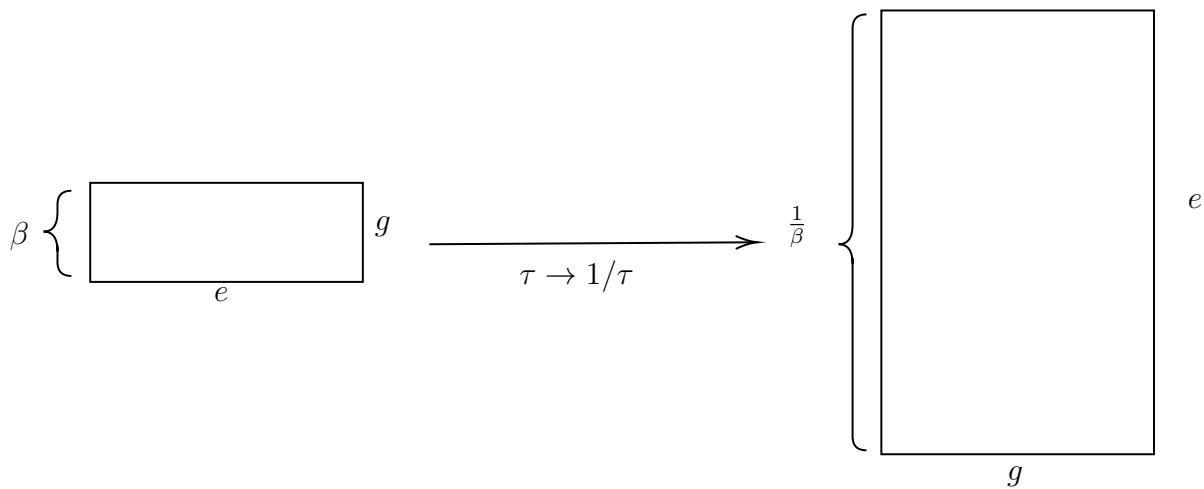
 
 Suppose the ground state of the twisted sector has energy $E_g\neq 0$, which is always the case for non-degenerate CFT's where the vacuum ($E=0$) is unique. We find
 \begin{align}
     Z_g\sim e^{-E_g\frac{1}{\beta}}
 \end{align}
 
 Both arguments show that at high energies (small $\beta$), the states furnish a very special representation such that the partition function \ref{P1} vanishes for all elements other than the identity element. For this to be true, the high energy representation of $G$ must be very special. It turns out there is exactly one such representation and that is the regular representation which permutes the group elements by left or right group action which leads to isomorphic representations. The regular representation of a finite group $G$, has a dimension $|G|$ and it acts on the basis vectors $\{e_g\}_{g\in G}$ as $\rho(s)e_g=e_{s\circ g}$. It is easy to see that the character of the regular representation $\chi_\rho(s)=\tr[\rho(s)]$ vanishes unless $s$ is the identity. This is exactly what we observed must happen to the states of theory at high energies.

 The regular representation is also special in a second way, in that it decomposes into all irreducible representations. Every irreducible representation $R\alpha$ appears with multiplicity $\dim(R)$ in the decomposition of the regular representation.
 \begin{align}
     \rho\equiv\oplus_\alpha \dim(R_\alpha)\cdot R_\alpha.
 \end{align}

So we know that all of them appear and we also now how often they appear. Note that since the argument 1 applies to all quantum field theories, above statement is more general than 2d CFTs. For any QFT, if one representation of a discrete global symmetry appears, all of them must appear. In fact, this argument works for continuous groups as well! For  example, take the spin group $\text{Spin}(3)$ which is the universal cover of $SO(3)$. Did we have to have fermionic representations as well as bosonic representations? And if they do, how frequent should they be compared to the bosonic representations? 
 
 Suppose the $\text{Spin}$ group acts faithfully on the Hilbert space (\emph{i.e.} there is at least one frmionic representation), then we can use the above argument to deduce that the following partition function must vanish at high temperatures. 
 
 \begin{align}
     \frac{Tr (-1)^F e^{-\beta H}}{Tr e^{-\beta g}},
 \end{align}
 
where $F$ is the number of spacetime fermions. If we rewrite this in terms of the ratio of bosonic and fermionic degrees of freedom we find

\begin{align}
    \sim(\frac{n_B-n_f}{n_B+n_f})\sim e^{-\frac{1}{\beta}}.
\end{align}

Thus, even though high energy supersymmetry is not universal, there is some sense of equality of fermionic and bosonic degrees of freedom at high energies that is universal.

\subsection{Completeness of spectrum: arguments}

So far we gave evidence for the completeness of spectrum in string theory, now we  will give a general explanation for it. Let us start with a continuous $U(1)$. If there are no charges, we have $dF=d\tilde F=0$. However, existence of the electric of magnetic sources will break this vanishings. Cobordism conjecture tells you that these equations give you global symmetry. So the existence of electric and magnetic states make sure that there are no global symmetries (in this case 1-form symmetries) \cite{Montero:2020icj}. This argument is in fact correct for higher form gauge symmetries as well. 

However, this argument just tells us that we must have charged state, but those charges do not have to be minimal charges. What if we only have certain multiples of the fundamental charge? Then the spectrum would be incomplete. Let us see why this cannot happen. 

Suppose the only charges that appear are multiples of $ke$ where $k>1$ is a natural number and $e$ is the fundamental charge. We show that such a theory has a $\mathbb{Z}_k$ 1-form global symmetry. The corresponding topological operator is $\int_\Sigma \tilde F\mod k$. This operator measures the charges of the Wilson lines that link with $\Sigma$ mod $k$ which is conserved. 
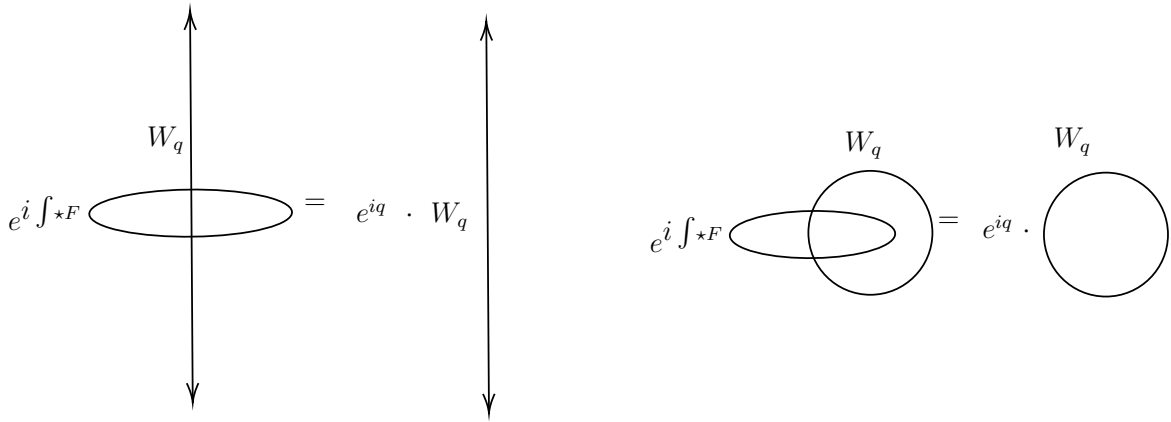
\begin{figure}[H]
    \centering
\tikzset{every picture/.style={line width=0.75pt}} 
\scalebox{.9}{
\begin{tikzpicture}[x=0.75pt,y=0.75pt,yscale=-1,xscale=1]
\draw    (103.51,36) -- (104.99,252) ;
\draw [shift={(105,254)}, rotate = 269.61] [color={rgb, 255:red, 0; green, 0; blue, 0 }  ][line width=0.75]    (10.93,-3.29) .. controls (6.95,-1.4) and (3.31,-0.3) .. (0,0) .. controls (3.31,0.3) and (6.95,1.4) .. (10.93,3.29);
\draw [shift={(103.5,34)}, rotate = 89.61] [color={rgb, 255:red, 0; green, 0; blue, 0 }  ][line width=0.75]    (10.93,-3.29) .. controls (6.95,-1.4) and (3.31,-0.3) .. (0,0) .. controls (3.31,0.3) and (6.95,1.4) .. (10.93,3.29);
\draw   (70.63,137.27) .. controls (96.09,132.79) and (131.59,133.83) .. (149.93,139.6) .. controls (168.27,145.38) and (162.49,153.69) .. (137.03,158.17) .. controls (111.57,162.65) and (76.07,161.6) .. (57.73,155.83) .. controls (39.39,150.06) and (45.17,141.75) .. (70.63,137.27) -- cycle ;
\draw    (269.51,42) -- (270.99,258) ;
\draw [shift={(271,260)}, rotate = 269.61] [color={rgb, 255:red, 0; green, 0; blue, 0 }  ][line width=0.75]    (10.93,-3.29) .. controls (6.95,-1.4) and (3.31,-0.3) .. (0,0) .. controls (3.31,0.3) and (6.95,1.4) .. (10.93,3.29);
\draw [shift={(269.5,40)}, rotate = 89.61] [color={rgb, 255:red, 0; green, 0; blue, 0 }  ][line width=0.75]    (10.93,-3.29) .. controls (6.95,-1.4) and (3.31,-0.3) .. (0,0) .. controls (3.31,0.3) and (6.95,1.4) .. (10.93,3.29);
\draw   (425.22,149.27) .. controls (445.98,144.79) and (474.95,145.83) .. (489.91,151.61) .. controls (504.87,157.38) and (500.15,165.69) .. (479.39,170.17) .. controls (458.62,174.65) and (429.65,173.6) .. (414.69,167.83) .. controls (399.74,162.05) and (404.45,153.74) .. (425.22,149.27) -- cycle ;
\draw   (582,159.75) .. controls (582,140.56) and (597.56,125) .. (616.75,125) .. controls (635.94,125) and (651.5,140.56) .. (651.5,159.75) .. controls (651.5,178.94) and (635.94,194.5) .. (616.75,194.5) .. controls (597.56,194.5) and (582,178.94) .. (582,159.75) -- cycle ;
\draw   (450,158.75) .. controls (450,139.56) and (465.56,124) .. (484.75,124) .. controls (503.94,124) and (519.5,139.56) .. (519.5,158.75) .. controls (519.5,177.94) and (503.94,193.5) .. (484.75,193.5) .. controls (465.56,193.5) and (450,177.94) .. (450,158.75) -- cycle ;
\draw (78,97.4) node [anchor=north west][inner sep=0.75pt]    {$W_{q}$};
\draw (0,136.4) node [anchor=north west][inner sep=0.75pt]    {${\textstyle e^{{\textstyle i\int } \star F}}$};
\draw (166,138.4) node [anchor=north west][inner sep=0.75pt]    {$=$};
\draw (237,140.4) node [anchor=north west][inner sep=0.75pt]    {$W_{q}$};
\draw (195,138.4) node [anchor=north west][inner sep=0.75pt]    {$e^{iq} ~~\cdot $};
\draw (469,100.4) node [anchor=north west][inner sep=0.75pt]    {$W_{q}$};
\draw (359.72,148.4) node [anchor=north west][inner sep=0.75pt]    {${\textstyle e^{{\textstyle i\int } \star F}}$};
\draw (521,148.4) node [anchor=north west][inner sep=0.75pt]    {$=$};
\draw (586,98.4) node [anchor=north west][inner sep=0.75pt]    {$W_{q}$};
\draw (546,147.4) node [anchor=north west][inner sep=0.75pt]    {$e^{iq} ~\cdot $};
\end{tikzpicture}}
    \caption{In the absence of charged operators, there is a 1-form global symemtry associated with the topological operator $\exp(i\int\star F)$.}
    \label{ofgs}
\end{figure}

Let us explain it slightly differently which is a useful logic. Consider a Wilson line. Define an operator that when linked with the Wilson line that carries charge $q$, it multiples it by $e^{2\pi i \frac{q}{k}}$. Wilson loops with any charge $q$ can appear but only then ones with $q=nk$ can end. However, changing $q$ by a multiple of $k$ does not change this $k$-th root of unity. Thus, we find a $\mathbb{Z}_k$ global 1-form symmetry. This shows us that no-global symmetry implies the completeness of spectrum at least for the case of $U(1)$.

But how about continuous non-Abelian cases? For non-Abelian symmetries there is an even easier way of doing it. Multiples of every allowed weight in the weight lattice correspond to the spectrum of representations of a $U(1)$ subgroup that is generated by an element in the Cartan subalgebra. From the Abelian argument we know that all of them must be occupied. Therefore, all the weights in the weight lattice of any non-Abelian group must be occupied.  
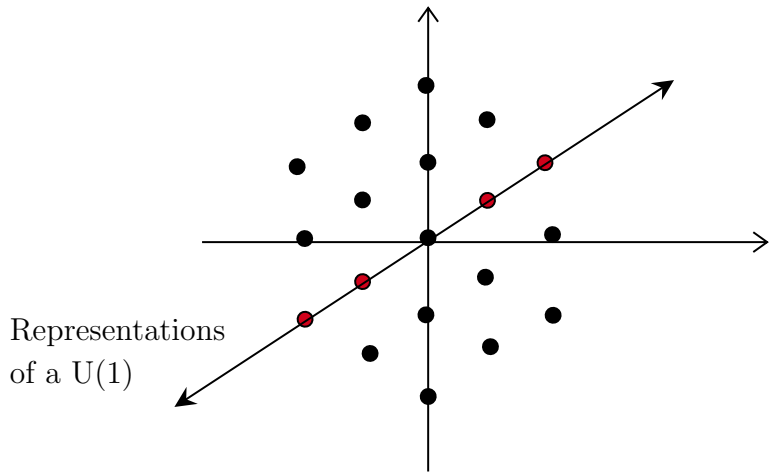
\begin{figure}[H]
    \centering

\tikzset{every picture/.style={line width=0.75pt}} 

\begin{tikzpicture}[x=0.75pt,y=0.75pt,yscale=-1,xscale=1]

\draw  (197.46,137.79) -- (482.46,137.79)(311.49,19.57) -- (311.49,253.57) (475.46,132.79) -- (482.46,137.79) -- (475.46,142.79) (306.49,26.57) -- (311.49,19.57) -- (316.49,26.57)  ;
\draw  [fill={rgb, 255:red, 0; green, 0; blue, 0 }  ,fill opacity=1 ] (372.26,137.19) .. controls (370.49,136.12) and (369.93,133.82) .. (371,132.04) .. controls (372.07,130.27) and (374.38,129.71) .. (376.15,130.78) .. controls (377.92,131.85) and (378.49,134.16) .. (377.41,135.93) .. controls (376.34,137.7) and (374.04,138.27) .. (372.26,137.19) -- cycle ;
\draw  [fill={rgb, 255:red, 0; green, 0; blue, 0 }  ,fill opacity=1 ] (340.96,193.77) .. controls (339.18,192.7) and (338.62,190.4) .. (339.69,188.62) .. controls (340.76,186.85) and (343.07,186.29) .. (344.84,187.36) .. controls (346.61,188.43) and (347.18,190.74) .. (346.11,192.51) .. controls (345.03,194.28) and (342.73,194.85) .. (340.96,193.77) -- cycle ;
\draw  [fill={rgb, 255:red, 0; green, 0; blue, 0 }  ,fill opacity=1 ] (309.49,138.79) .. controls (307.71,137.71) and (307.15,135.41) .. (308.22,133.64) .. controls (309.29,131.86) and (311.6,131.3) .. (313.37,132.37) .. controls (315.14,133.44) and (315.71,135.75) .. (314.64,137.52) .. controls (313.56,139.29) and (311.26,139.86) .. (309.49,138.79) -- cycle ;
\draw  [fill={rgb, 255:red, 208; green, 2; blue, 27 }  ,fill opacity=1 ] (343.16,120.03) .. controls (341.33,121) and (339.06,120.3) .. (338.1,118.46) .. controls (337.13,116.63) and (337.83,114.36) .. (339.67,113.4) .. controls (341.5,112.43) and (343.77,113.13) .. (344.73,114.97) .. controls (345.7,116.8) and (345,119.07) .. (343.16,120.03) -- cycle ;
\draw  [fill={rgb, 255:red, 0; green, 0; blue, 0 }  ,fill opacity=1 ] (376.26,178.03) .. controls (374.43,179) and (372.16,178.3) .. (371.2,176.47) .. controls (370.23,174.63) and (370.93,172.37) .. (372.76,171.4) .. controls (374.6,170.43) and (376.86,171.14) .. (377.83,172.97) .. controls (378.8,174.8) and (378.09,177.07) .. (376.26,178.03) -- cycle ;
\draw  [fill={rgb, 255:red, 0; green, 0; blue, 0 }  ,fill opacity=1 ] (312.12,177.81) .. controls (310.28,178.78) and (308.02,178.08) .. (307.05,176.24) .. controls (306.08,174.41) and (306.79,172.14) .. (308.62,171.18) .. controls (310.45,170.21) and (312.72,170.91) .. (313.68,172.75) .. controls (314.65,174.58) and (313.95,176.85) .. (312.12,177.81) -- cycle ;
\draw  [fill={rgb, 255:red, 208; green, 2; blue, 27 }  ,fill opacity=1 ] (372.16,101.03) .. controls (370.33,102) and (368.06,101.3) .. (367.1,99.46) .. controls (366.13,97.63) and (366.83,95.36) .. (368.67,94.4) .. controls (370.5,93.43) and (372.77,94.13) .. (373.73,95.97) .. controls (374.7,97.8) and (374,100.07) .. (372.16,101.03) -- cycle ;
\draw  [fill={rgb, 255:red, 0; green, 0; blue, 0 }  ,fill opacity=1 ] (342.12,158.81) .. controls (340.28,159.78) and (338.02,159.08) .. (337.05,157.24) .. controls (336.08,155.41) and (336.79,153.14) .. (338.62,152.18) .. controls (340.45,151.21) and (342.72,151.91) .. (343.68,153.75) .. controls (344.65,155.58) and (343.95,157.85) .. (342.12,158.81) -- cycle ;
\draw  [fill={rgb, 255:red, 0; green, 0; blue, 0 }  ,fill opacity=1 ] (339.26,79.19) .. controls (337.49,78.12) and (336.93,75.82) .. (338,74.04) .. controls (339.07,72.27) and (341.38,71.71) .. (343.15,72.78) .. controls (344.92,73.85) and (345.49,76.16) .. (344.41,77.93) .. controls (343.34,79.7) and (341.04,80.27) .. (339.26,79.19) -- cycle ;
\draw  [fill={rgb, 255:red, 0; green, 0; blue, 0 }  ,fill opacity=1 ] (276.49,80.79) .. controls (274.71,79.71) and (274.15,77.41) .. (275.22,75.64) .. controls (276.29,73.86) and (278.6,73.3) .. (280.37,74.37) .. controls (282.14,75.44) and (282.71,77.75) .. (281.64,79.52) .. controls (280.56,81.29) and (278.26,81.86) .. (276.49,80.79) -- cycle ;
\draw  [fill={rgb, 255:red, 0; green, 0; blue, 0 }  ,fill opacity=1 ] (312.16,62.03) .. controls (310.33,63) and (308.06,62.3) .. (307.1,60.46) .. controls (306.13,58.63) and (306.83,56.36) .. (308.67,55.4) .. controls (310.5,54.43) and (312.77,55.13) .. (313.73,56.97) .. controls (314.7,58.8) and (314,61.07) .. (312.16,62.03) -- cycle ;
\draw  [fill={rgb, 255:red, 0; green, 0; blue, 0 }  ,fill opacity=1 ] (280.12,119.81) .. controls (278.28,120.78) and (276.02,120.08) .. (275.05,118.24) .. controls (274.08,116.41) and (274.79,114.14) .. (276.62,113.18) .. controls (278.45,112.21) and (280.72,112.91) .. (281.68,114.75) .. controls (282.65,116.58) and (281.95,118.85) .. (280.12,119.81) -- cycle ;
\draw  [fill={rgb, 255:red, 0; green, 0; blue, 0 }  ,fill opacity=1 ] (313.12,100.81) .. controls (311.28,101.78) and (309.02,101.08) .. (308.05,99.24) .. controls (307.08,97.41) and (307.79,95.14) .. (309.62,94.18) .. controls (311.45,93.21) and (313.72,93.91) .. (314.68,95.75) .. controls (315.65,97.58) and (314.95,99.85) .. (313.12,100.81) -- cycle ;
\draw  [fill={rgb, 255:red, 0; green, 0; blue, 0 }  ,fill opacity=1 ] (280.26,197.19) .. controls (278.49,196.12) and (277.93,193.82) .. (279,192.04) .. controls (280.07,190.27) and (282.38,189.71) .. (284.15,190.78) .. controls (285.92,191.85) and (286.49,194.16) .. (285.41,195.93) .. controls (284.34,197.7) and (282.04,198.27) .. (280.26,197.19) -- cycle ;
\draw  [fill={rgb, 255:red, 208; green, 2; blue, 27 }  ,fill opacity=1 ] (251.16,180.03) .. controls (249.33,181) and (247.06,180.3) .. (246.1,178.46) .. controls (245.13,176.63) and (245.83,174.36) .. (247.67,173.4) .. controls (249.5,172.43) and (251.77,173.13) .. (252.73,174.97) .. controls (253.7,176.8) and (253,179.07) .. (251.16,180.03) -- cycle ;
\draw  [fill={rgb, 255:red, 208; green, 2; blue, 27 }  ,fill opacity=1 ] (280.16,161.03) .. controls (278.33,162) and (276.06,161.3) .. (275.1,159.46) .. controls (274.13,157.63) and (274.83,155.36) .. (276.67,154.4) .. controls (278.5,153.43) and (280.77,154.13) .. (281.73,155.97) .. controls (282.7,157.8) and (282,160.07) .. (280.16,161.03) -- cycle ;
\draw  [fill={rgb, 255:red, 0; green, 0; blue, 0 }  ,fill opacity=1 ] (313.26,219.03) .. controls (311.43,220) and (309.16,219.3) .. (308.2,217.47) .. controls (307.23,215.63) and (307.93,213.37) .. (309.76,212.4) .. controls (311.6,211.43) and (313.86,212.14) .. (314.83,213.97) .. controls (315.8,215.8) and (315.09,218.07) .. (313.26,219.03) -- cycle ;
\draw  [fill={rgb, 255:red, 0; green, 0; blue, 0 }  ,fill opacity=1 ] (247.26,139.19) .. controls (245.49,138.12) and (244.93,135.82) .. (246,134.04) .. controls (247.07,132.27) and (249.38,131.71) .. (251.15,132.78) .. controls (252.92,133.85) and (253.49,136.16) .. (252.41,137.93) .. controls (251.34,139.7) and (249.04,140.27) .. (247.26,139.19) -- cycle ;
\draw  [fill={rgb, 255:red, 0; green, 0; blue, 0 }  ,fill opacity=1 ] (247.16,103.03) .. controls (245.33,104) and (243.06,103.3) .. (242.1,101.46) .. controls (241.13,99.63) and (241.83,97.36) .. (243.67,96.4) .. controls (245.5,95.43) and (247.77,96.13) .. (248.73,97.97) .. controls (249.7,99.8) and (249,102.07) .. (247.16,103.03) -- cycle ;
\draw    (186.01,219.36) -- (432.99,57.64) ;
\draw [shift={(435.5,56)}, rotate = 146.78] [fill={rgb, 255:red, 0; green, 0; blue, 0 }  ][line width=0.08]  [draw opacity=0] (10.72,-5.15) -- (0,0) -- (10.72,5.15) -- (7.12,0) -- cycle    ;
\draw [shift={(183.5,221)}, rotate = 326.78] [fill={rgb, 255:red, 0; green, 0; blue, 0 }  ][line width=0.08]  [draw opacity=0] (10.72,-5.15) -- (0,0) -- (10.72,5.15) -- (7.12,0) -- cycle    ;

\draw (99,175) node [anchor=north west][inner sep=0.75pt]   [align=left] {Representations\\of a U(1)};

\end{tikzpicture}
    \caption{Any weight and its multiples can be viewed as the spectrum of representations of a $U(1)$ subgroup generated by an element of Cartan.}
    \label{NASW}
\end{figure}
There is a second argument for completeness of spectrum using black holes. We will first argue the completeness for Abelian groups. Then, one can use the trick in the last paragraph to generalize the result to the non-Abelian groups. Consider a Reissner-Nordstrom black hole with $M\geq Q$ in Planck units. If we take $M$ and $Q$ to be very large in Planck units, the curvature outside the black hole would be small. Thus, the Bekenstein-Hawking formula which is a semiclassical calculation is trustable. The formula tells us that there is a large number of microstates with a given charge and mass. But how do we know that small charges exist? If we consider two black holes with large charges $Q$ and $-Q+e$ the net charge of the state would be a fundamental charge of the theory.

Now let us consider discrete gauge symmetries. What goes wrong if  the spectrum of a discrete gauge group is not complete? We will use an argument similar to the continuous case to show that lack of completeness implies the existence of a 1-form global symmetry. The idea is that we can define a non-invertible topological operator for every conjugacy class $[g]$ that acts on a Wilson line carrying a representation $R$ by multiplying it by $\text{size}([g])\cdot\frac{\chi_R([g])}{\dim(R)}$, where $\chi_R$ is the character of $R$ \cite{Rudelius:2020orz,Heidenreich:2021xpr}. The number of such 1-form symmetries is equal to the number of conjugacy classes of $G$. However, for finite groups, this number is the same as the number of irreducible representations $R$. This matching is not a accidental and it can be shown that charged particles in every irreducible representation is needed to break all of these 1-form symmetries \cite{Rudelius:2020orz}.

\section{Swampland III: Weak gravity conjecture}

Now we move on to a another Swampland conjecture which is closely related to the completeness of spectrum conjecture. The completeness of spectrum conjecture deals with the existence of charged states in a theory of quantum gravity, however, it falls short from giving any estimate of their masses. In fact, the charged states could be so heavy that they are not included in the low-energy theory. Therefore, to quantify the completeness of spectrum it would be nice to have an upper bound on the mass of the supposed charged particles. The Weak Gravity Conjecture (or WGC for short) tries to answer this question \cite{Arkani-Hamed:2006emk}. This Swampland conjecture has been extensively studied due to its strong theoretical consequences, however, it does not have as clean of a formulation as the previous Swampland conjectures. The general idea is that if you have a gauge symmetry such as electromagnetic $U(1)$, identical charged states experience repulsive forces in addition to the universal gravitational attraction. The Weak Gravity Conjecture proposes that there must always exist charged objects where the repulsion is stronger than (or equal to) the gravitational attraction. In our universe this is evident. The electrostatic repulsion between two electrons is much stronger than the gravitational attraction between them, 43 orders of magnitude stronger to be more precise! 

The comparison of these two forces between two identical charged particles comes down to the comparison between the charge and mass of the particle in Planck units. So we would like the final inequality to take the form $q/m$ in Planck units to be greater than some constant $c$ at least for some charged states. As we will see later, black hole offer a natural candidate for the constant $c$ which is the charge-to-mass ratio of large extremal black holes.

But before discussing the precise statement or consequences of such a conjecture, let us take a step back and ask why should there be a statement like this at all? Electric charge represents the coupling of the gauge field. Can we not just take the coupling constant to zero? As far as field theory is concerned, this is in fact a very desirable limit where everything interacts weakly and perturbative calculations are more convergent\footnote{Note that by charge we mean IR coupling. In other words, we compare the forces at long distances. There are scale dependent version of Weak Gravity Conjecture as well.}. Why should gravity say we cannot make a certain coupling too weak? 

One argument could be that as the gauge coupling $g$ goes to $0$, the theory gets closer and closer to having a global symmetry. At $g=0$, the gauge symmetry becomes a global symmetry. Therefore, this statement is trying to quantify the no-global symmetry conjecture, just like it quantifies the completeness of spectrum conjecture. 

Now let us look at this conjecture in String theory. String theory naturally incorporates supersymmetry, and supersymmetry implies the BPS bound which states that mass is bigger than or equal to the (central) charge. At first, supersymmetry and Weak Gravity Conjecture seem to be in contradiction. BPS bounds apply to all states while Weak Gravity Conjecture can be satisfied by only \textit{some} states. This leaves a way for the two statements to be compatible in a marginal way. This is usually achieved by the existence of BPS states which saturate the BPS bound. Therefore, in supersymmetric theories, we can think of Weak Gravity Conjecture as the statement that BPS states \textit{must} exist. 

The stability of BPS states is protected by supersymmtry. In fact, there is a natural connection between stability and the states that satisfy WGC. Consider an unstable state with a charge $q$ and mass $m$ that decays into some lighter particles with charges and masses of $(q_1,m_1),~(q_2,m_2),~...,(q_N,m_N)$. From conservation of charge and energy we find
\begin{align}
    &q_1+q_2+\hdots+q_N=q,\nonumber\\
    &m_1+m_2+\hdots+m_N\leq m,
\end{align}
where the last inequality is due to positivity of the kinetic energy of the product particles. From the two equations above we find
\begin{align}
    \frac{|q|}{m}\leq \frac{|\sum q_i|}{\sum_i m_i}\leq \frac{\sum_i|q_i|}{\sum_i m_i}\leq \max_i\frac{|q_i|}{m_i}.
\end{align}
Therefore, in the outcome of any decay, there are always particles that satisfy the WGC better than the initial particle. In other words, stabler particles tend to satisfy WGC better. Even BPS states which often marginaly satisfy WGC are also stable. In fact in many string theory examples any stable charged state seems to satisfy WGC. Let us study some examples in string theory.

\subsection{Evidence from string theory}

Let us consider the type IIA theory in $d=10$, there is a gauge field and a single $D0$ brane is charged under the gauge field. The charge-to-mass ratio of $D_0$ branes turns out to be equal to that of extremal black holes. Therefore, they saturate the WGC inequality. Even N $D_0$ branes can form a bound state which still saturate the WGC inequality.

As another example, consider the toroidal compactification of Heterotic theory on $T^d$ down to $\mathbb{R}^{10-d}$. There is a Narain lattice $\Gamma^{16+d,d}$. Before studying the string spectrum, note that there is a charge lattice of a group with rank $16+2d$ labled by $(P_L,P_R)$. The BPS formula tells us that $m(P_L,P_R)\geq |P_R|$. The mass formula from the Heterotic string takes the form $\frac{1}{2}m^2=\frac{1}{2}P_R^2+N_R=\frac{1}{2}P_L^2+N_L-1$. This shows that the BPS bound $m\geq P_R$ is automatically satisfied and the equality only holds for $N_R=0$. For the BPS states that saturate the bound we have 
\begin{align}
    \frac{1}{2}(P_R^2-P_L^2)=N_L-1,
\end{align}
which implies that $\frac{1}{2}(P_R^2-P_L^2)\geq -1$ in terms of charges. In fact, the Heterotic string occupies all of this lattice for arbitrary $P_R$ with BPS states. However, if $P_L$ is large compared to $P_R$, the inequality gets violated and the supersymmetry is broken. BPS bound gives us a WGC-like inequality for $P_R$, but what about other directions of the charge lattice? In particular, the large $P_L/P_R$ direction which also breaks the supersymmetry? In fact, Heterotic string theory mass formula gives us a WGC-like inequality in terms of $P_L$ that is $m^2\geq P_R^2-2$. As you can see, when this inequality is saturated, it satisfies the WGC even better than the BPS states. This is a universal observation that in directions of charge lattice where SUSY is broken (since $m<|P_R|$ BPS states are absent), the WGC is satisfied even stronger.

Let us study the above example from a dual perspective. Heterotic theory on $T^3$ is dual to M-theory on K3. In that case, the Narain lattice is the lattice of 2-cycles in K3. There is a polarization from the metric of K3 that splits them into 19 self-dual and 3 anti-self-dual cycles. The dulaity maps the wrapping of M2 branes around 2-cycles to $P_L$ and $P_R$ on the Heterotic side. The minimal mass (minimal area) configurations that saturates the BPS bound are M2 branes arapped around holomorphic cycles. The charge of a 2-cycle is calculated by
\begin{align}\label{ncm}
    k_{i\bar j}(\partial X^i\bar\partial X^{\bar j}-\bar \partial X^i\partial X^{\bar j}).
\end{align}
This is because M-theory three form could be written in terms of the Kähler form as 
\begin{align}
    C_{\mu i \bar j}=k_{i\bar j}A_\mu
\end{align}
However, \eqref{ncm} is not the area (mass) which is given by a similar expression except with a plus sign in parentheses. However, the two expressions are the same when $\bar\partial X^i=0$ which is realized for the holomorphic cycles. Thus, holomorphic cycles have masses proportional to their charges and saturate the BPS bound. 

We can even recover the $P_R^2-P_L^2\geq -2$ which is obvious from the mass formula on the heterotic side. To see that, one must look at the self intersection of the holomorphic curve which is $\Sigma\cdot\Sigma=P_R^2-P_L^2$. However, we can also calculate the self-intersection differently. For a Riemann surface in K3 which has a vanishing first Chern class (line bundle cotangent to the Riemann surface) we have $\Sigma\cdot \Sigma=2g-2$ where $g\geq 0$ is the genus of the Riemann surface. This is because $\Sigma\cdot\Sigma=\chi(\Sigma)$ as the local geometry of $\Sigma$ in K3 is $T^*\Sigma$. Thus, we get the same inequality $P_R^2-P_L^2=2g-2\geq-2$ as the one on the Heterotic side. 

All this evidence shows that BPS objects are just holomorphic curves with some choice of complex structure on K3. We found the mass relation for BPS objects and they satisfy the WGC just like their Heterotic counterparts. However, we do not know how to calculate the non-holomorphic mass relations because we do not know the metric of K3, and so we cannot check the non SUSY prediction of WGC. 

Let us consider one last example before formulating weak gravity conjecture. Consider the compactification of the type II on $M\times S^1/\mathbb{Z}_2$. The $\mathbb{Z}_2$ acts freely on $M$ and maps $\theta$ to its antipodal point $\theta+\pi$ on the $S^1$. If n is the winding number, in string units we get the mass formula $m=nR$ for winding states which saturates $M=|Q|$. You might expect after modding out we can wrap half of $S^1$ so we get half masses. However, since the $\mathbb{Z}_2$ acts freely on $M$, the two endpoints of the half-wrapped strings are separated on $M$. Because of this stretching, the masses of the half-integer winding strings goes beyond the $M=Q$ line in Planck units. Therefore, we get an infinite lattice of odd windings that does not satisfy the WGC. However, the ineuqality is still saturated by the infinite subblattice corresponding to the even windings around $S^1$. There are stronger versions of WGC that state there must be infinite number of bound states that satisfy the WGC \cite{Andriolo:2018lvp}.

Naively you would have guessed that we can push the minimal charge that strictly satisfies WGC even higher by modding out by $Z_n$ rather than $Z_2$. However, in all string theory construction the isometry groups of compact spaces are always bounded. For example, this leads to a precise (but non-trivial) mathematical proposal that the order of free symmetry group of Calabi--Yaus of a given dimension is bounded.

Now we are ready to formulate the basic version of the Weak Gravity Conjecture.

\subsection{Weak gravity conjecture: formulation}

\begin{statement5*}
Consider a $U(1)$ gauge field. The charged black holes have a mass charge formula that prevents naked singularities and is given by $Q\leq M_{ext}(Q)$. The conjecture states that there is always a "small" charged particle with charge $q$ and mass $m$ such that 
\begin{align}
    \frac{m}{q}\leq\lim_{Q\rightarrow\infty} \frac{M_{ext}(Q)}{Q}.
\end{align}
\end{statement5*}

Here are some natural questions that arise about the above conjecture that lead to some generalizations of the above formulation.

\begin{itemize}
    \item \textbf{How small is $q$?} Note as we will see in some string theory examples, $q$ does not have to be the fundamental charge. The word small makes the formulation a bit imprecise. Perhaps a more precise formulation would be that there is a universal dimension dependent constant $N_d$ such that there exists a charged particle with $q<N_d$ that satisfies the above conjecture. 
    
    \item {How about higher-form gauge symmteries?} Usually any higher-form gauge symmetry can lead to a 0-form gauge symmetry in a lower dimensional theory. Thus, it is natural that a similar statement must hold for higher-form symmetries as well. The higher dimensional generalization of the WGC replaces extremal black holes with extremal black branes and replaces mass with tension. 
    
    \item {What if there are no large black holes?} As we will see, there are cases that where a modulus couples to the fields in a way that there are no large extremal black hole solutions. For example, if we compactify IIB on a conifold singularity (zero-size 3-cycle). When the size of the 3-cycle is non-zero, the black hole solutions have non-zero mass that scales with the volume of the 3-cycle. However, in the conifold limit, the black hole mass goes to zero. But there is still a non-trivial statement here! Naively, WGC would tell us that there must be a stable charged particle with $q/m$ greater than or equal to that of extremal black holes which in this case is $\infty$. So, if we take the conjecture at its face value, it predicts that there must be a stable massless charged particle which turns out to be true. Let us take a closer look at this example.
    
    Consider type IIB  on $CY3$ with a confiold singularity (shrinking $S^3$). If you take the four form gauge field and write it as $D=\Omega\wedge A$ where $\Omega_{abc}$ is a three form on the Calabi--Yau and $A_\mu$ is a one-form in the non-compact spacetime. In the lower dimensional theory, $A_\mu$ is a gauge field and the object that is charged under it is a D3-brane wrapped around the 3-cycle. There is a BPS bound $m\geq \frac{|n\vol(S^3)|}{\lambda}$  where $n$ is the charge and $\lambda$ is the string coupling. However, it turns out the once wrapped brane ($n=1$), is the only bound state which also happens to saturate both the BPS bound and satisfies the WGC formula if the radius of $S^3$ is fixed. However, given that the radius of $S^3$ is set dynamically, and it flows to 0 for black hole solutions, both the once wrapped D3-brane as well as would-be extremal black holes are massless. In this example, the Weak Gravity Conjecture still has a non-trivial consequence even though there are no black holes. 

    Note that the fact that $n=1$ is the only bound state is not obvious. If you take two D3 branes, you expect to get an $SU(2)$ gauge group which lives on the $S^3$. If you look at thelow-energy configurations, they have transverse directions and it turns out there is no normalizable states due to this.

    \item {What about de Sitter?} In de Sitter, the universe has a finite size which makes it impossible to talk about infinitely large charged black holes. We will come back to de Sitter spaces later. 
\end{itemize}

So far we have argued why a statement like WGC is natural to have in string theory. However, WGC is a sharp statement in the sense that all the constants in the statement are determined by black hole physics. In the following, we review the motivation for this conjecture.

\subsection{Motivation}

Take a massive charged extremal black hole with mass $M$ and charge $Q$ that satisfies the extremality condition $M_{ext}(Q)=AQ$, where $A$ is some constant. Let us assume that the black hole is not BPS. We do not think such black holes are stable, because if they are, the black hole entropy formula counts an exponentially large number of microstates which are stable without any more fundamental reason to protect their stability. So let us assume that there are only finite number of stable non supersymmetric objects. In that case the black hole should decay into another black hole with mass $M-m$ and charge $Q-q$ by emitting a particle of energy $m$ and charge $q$. The new black hole must satisfy the extremality inequality $M-m\geq A(Q-q)$. Since the old black hole was extremal ($M=AQ$), we find $m\leq Aq$. This inequality means the emitted particle must satisfy the Weak Gravity Conjecture. Thus, the black hole extremality formula motivates the Weak Gravity Conjecture \cite{Arkani-Hamed:2006emk}. 

Black holes are the natural extension of particles for large masses where gravitational self energy is significant \cite{Bedroya:2022twb}. Therefore, it is reasonable to expect that the extremal bound $M>AQ$ has a a generalization that extends to particles as well. However, there are corrections to the action that modify the extremality curve for black holes. Weak Gravity Conjecture suggests that the curve must bend in the $M_{Ext}\leq Q$ direction to ensure that there can be particles that satisfy $M<AQ$. This is called the mild form of Weak Gravity Conjecture. The mild form of the WGC also implies that large extremal black holes themselves satisfy the WGC.

In four dimensions, some of the higher derivative terms that affect the extremality bound are $(a/M_{Pl}^4)(F^2)^2$ and $(b/M_{Pl}^2)F_{\mu\nu}F_{\alpha\beta}W^{\mu\nu\alpha\beta}$. In fact, all other such terms can be rewritten in terms of these terms using the equations of motion. The mild WGC then leads to $4a-b\geq0$ \cite{Hamada:2018dde,Kats:2006xp}.

The mild form of the WGC has been extensively tested in low-energy theories in string theory Landscape \cite{Kats:2006xp}. Moreover, there have been arguments that suggest the mild form of the WGC may be related to more fundamental principles such as unitarity and causality \cite{Bellazzini:2015cra,Adams:2006sv,Bellazzini:2019xts,Hamada:2018dde,Arkani-Hamed:2021ajd,Cheung:2018cwt}. 

However, the extremality bound

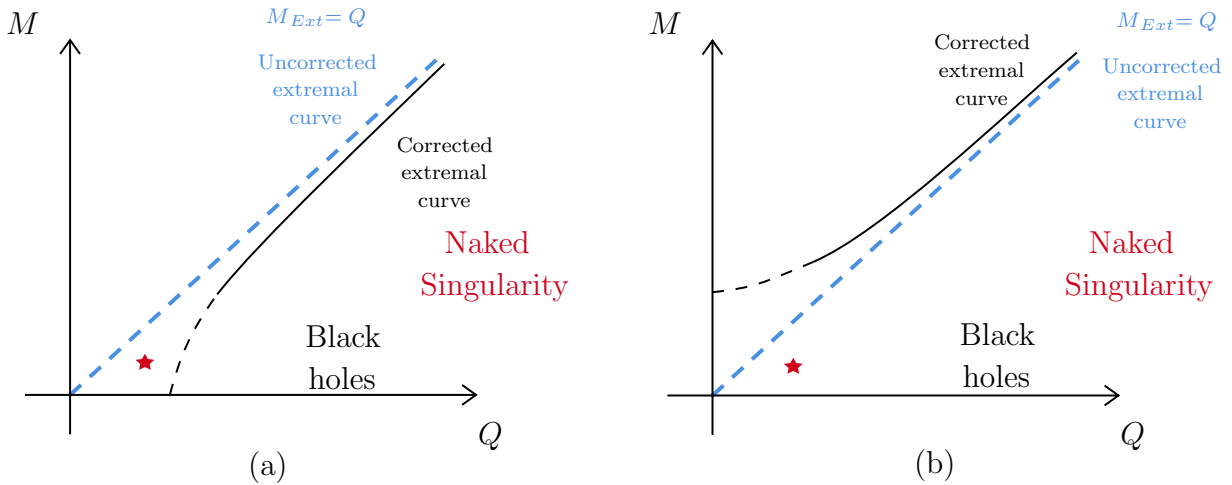
\begin{figure}[H]
    \centering

\tikzset{every picture/.style={line width=0.75pt}} 

\begin{tikzpicture}[x=0.75pt,y=0.75pt,yscale=-1,xscale=1]

\draw [color={rgb, 255:red, 74; green, 144; blue, 226 }  ,draw opacity=1 ][line width=1.5]  [dash pattern={on 5.63pt off 4.5pt}]  (45.75,243.1) -- (230.5,74) ;
\draw  (23,243.1) -- (250.5,243.1)(45.75,64) -- (45.75,263) (243.5,238.1) -- (250.5,243.1) -- (243.5,248.1) (40.75,71) -- (45.75,64) -- (50.75,71)  ;
\draw    (120,192.33) .. controls (140,166.33) and (201.5,108) .. (234.5,76) ;
\draw [color={rgb, 255:red, 74; green, 144; blue, 226 }  ,draw opacity=1 ][line width=1.5]  [dash pattern={on 5.63pt off 4.5pt}]  (369.75,243.39) -- (554.5,74.29) ;
\draw  (347,243.39) -- (574.5,243.39)(369.75,64.29) -- (369.75,263.29) (567.5,238.39) -- (574.5,243.39) -- (567.5,248.39) (364.75,71.29) -- (369.75,64.29) -- (374.75,71.29)  ;
\draw    (420.46,176.02) .. controls (455.46,161.02) and (516.5,102.29) .. (553.5,70.29) ;
\draw  [draw opacity=0][fill={rgb, 255:red, 208; green, 2; blue, 27 }  ,fill opacity=1 ] (410.5,224.29) -- (411.97,226.97) -- (415.26,227.4) -- (412.88,229.49) -- (413.44,232.43) -- (410.5,231.04) -- (407.56,232.43) -- (408.12,229.49) -- (405.74,227.4) -- (409.03,226.97) -- cycle ;
\draw  [dash pattern={on 4.5pt off 4.5pt}]  (96,243.33) .. controls (101,224.33) and (106,211.33) .. (120,192.33) ;
\draw  [dash pattern={on 4.5pt off 4.5pt}]  (369.5,191.29) .. controls (390.46,189.02) and (394.46,187.02) .. (420.46,176.02) ;
\draw  [draw opacity=0][fill={rgb, 255:red, 208; green, 2; blue, 27 }  ,fill opacity=1 ] (83.5,222.29) -- (84.97,224.97) -- (88.26,225.4) -- (85.88,227.49) -- (86.44,230.43) -- (83.5,229.04) -- (80.56,230.43) -- (81.12,227.49) -- (78.74,225.4) -- (82.03,224.97) -- cycle ;

\draw (12,48.4) node [anchor=north west][inner sep=0.75pt]    {$M$};
\draw (250,255.4) node [anchor=north west][inner sep=0.75pt]    {$Q$};
\draw (143,46.4) node [anchor=north west][inner sep=0.75pt]  [font=\scriptsize]  {$\textcolor[rgb]{0.29,0.56,0.89}{M}\textcolor[rgb]{0.29,0.56,0.89}{_{Ext}}\textcolor[rgb]{0.29,0.56,0.89}{=Q}$};
\draw (209,105) node [anchor=north west][inner sep=0.75pt]  [font=\scriptsize] [align=left] {\begin{minipage}[lt]{36.06pt}\setlength\topsep{0pt}
\begin{center}
Corrected \\extremal\\curve
\end{center}

\end{minipage}};
\draw (336,48.69) node [anchor=north west][inner sep=0.75pt]    {$M$};
\draw (574,255.69) node [anchor=north west][inner sep=0.75pt]    {$Q$};
\draw (480,53.29) node [anchor=north west][inner sep=0.75pt]  [font=\scriptsize] [align=left] {\begin{minipage}[lt]{36.06pt}\setlength\topsep{0pt}
\begin{center}
Corrected \\extremal\\curve
\end{center}

\end{minipage}};
\draw (221,159) node [anchor=north west][inner sep=0.75pt]  [color={rgb, 255:red, 208; green, 2; blue, 27 }  ,opacity=1 ] [align=left] {\begin{minipage}[lt]{50.35pt}\setlength\topsep{0pt}
\begin{center}
Naked \\Singularity
\end{center}

\end{minipage}};
\draw (546,159) node [anchor=north west][inner sep=0.75pt]  [color={rgb, 255:red, 208; green, 2; blue, 27 }  ,opacity=1 ] [align=left] {\begin{minipage}[lt]{50.35pt}\setlength\topsep{0pt}
\begin{center}
Naked \\Singularity
\end{center}

\end{minipage}};
\draw (163,196) node [anchor=north west][inner sep=0.75pt]   [align=left] {\begin{minipage}[lt]{27.66pt}\setlength\topsep{0pt}
\begin{center}
Black\\holes
\end{center}

\end{minipage}};
\draw (493,195) node [anchor=north west][inner sep=0.75pt]   [align=left] {\begin{minipage}[lt]{27.66pt}\setlength\topsep{0pt}
\begin{center}
Black\\holes
\end{center}

\end{minipage}};
\draw (139,63) node [anchor=north west][inner sep=0.75pt]  [font=\scriptsize,color={rgb, 255:red, 74; green, 144; blue, 226 }  ,opacity=1 ] [align=left] {\begin{minipage}[lt]{43.59pt}\setlength\topsep{0pt}
\begin{center}
Uncorrected \\extremal\\curve
\end{center}

\end{minipage}};
\draw (572,48.4) node [anchor=north west][inner sep=0.75pt]  [font=\scriptsize]  {$\textcolor[rgb]{0.29,0.56,0.89}{M}\textcolor[rgb]{0.29,0.56,0.89}{_{Ext}}\textcolor[rgb]{0.29,0.56,0.89}{=Q}$};
\draw (565,65) node [anchor=north west][inner sep=0.75pt]  [font=\scriptsize,color={rgb, 255:red, 74; green, 144; blue, 226 }  ,opacity=1 ] [align=left] {\begin{minipage}[lt]{43.59pt}\setlength\topsep{0pt}
\begin{center}
Uncorrected \\extremal\\curve
\end{center}

\end{minipage}};
\draw (134,270) node [anchor=north west][inner sep=0.75pt]   [align=left] {(a)};
\draw (470,269) node [anchor=north west][inner sep=0.75pt]   [align=left] {(b)};

\end{tikzpicture}

    \caption{The dashed blue line is the usual uncorrected mass-charge relation for extremal black holes. The solid black line is the corrected curve for black holes and the dashed black line is its extension for particles. All the particles and black holes are expected to be above the black curve. The red star represents a particle that satisfies the WGC ($m\leq q$). If the corrections tilts the extrmal curve upward (as in (b)), the WGC particle would be  in the prohibited region. Thus, assuming an extension of the extremality bound for particles, the WGC suggests that corrections to extremality bound will tilt it downward, as in (a).}
    \label{Extremal}
\end{figure}

\subsection{Festina lente}

In de Sitter space there is a natural IR length scale $\frac{1}{H}$. Suppose, the cosmological constant is so small that there is a very large scale separation between the UV and IR scales. In that case, we expect the flat space WGC to hold in de Sitter as well. However, there is some new interesting features due to the finite size of de Sitter. If we put a black hole, it cannot be bigger than the Hubble scale. This puts an upper bound on how massive a black hole can be. The black holes that saturate this bound are called the Nariai black holes. The region of the allowed charged black holes in de Sitter is shown in the figure below. 

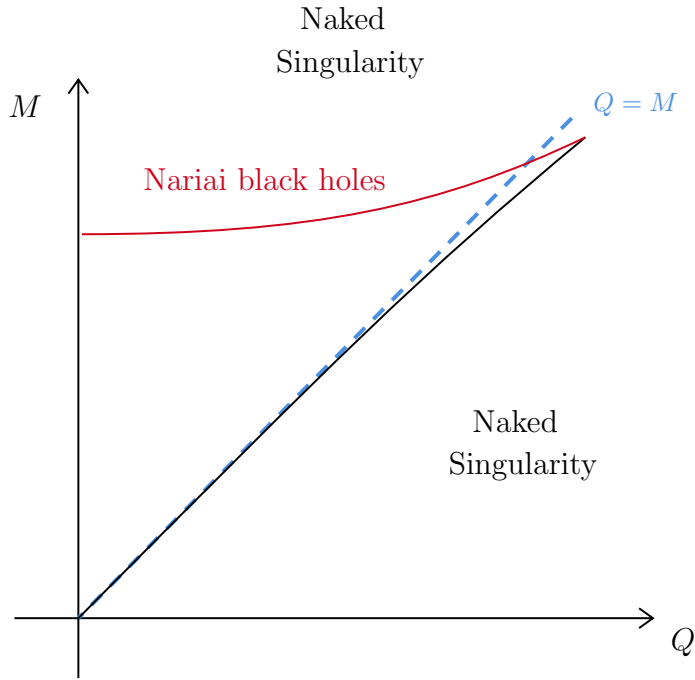
\begin{figure}[H]
    \centering

\tikzset{every picture/.style={line width=0.75pt}} 

\begin{tikzpicture}[x=0.75pt,y=0.75pt,yscale=-1,xscale=1]

\draw  (194.7,363) -- (194.7,61)(484.5,332.8) -- (162.5,332.8) (199.7,68) -- (194.7,61) -- (189.7,68) (477.5,337.8) -- (484.5,332.8) -- (477.5,327.8)  ;
\draw [color={rgb, 255:red, 74; green, 144; blue, 226 }  ,draw opacity=1 ][line width=1.5]  [dash pattern={on 5.63pt off 4.5pt}]  (443.54,80.39) -- (194.7,332.8) ;
\draw    (194.7,332.8) .. controls (322.93,203.52) and (366.17,159.89) .. (450.37,90.09) ;
\draw [color={rgb, 255:red, 208; green, 2; blue, 27 }  ,draw opacity=1 ]   (196.5,139) .. controls (313.5,139) and (377.5,125) .. (450.37,90.09) ;

\draw (158,67.4) node [anchor=north west][inner sep=0.75pt]    {$M$};
\draw (492,336.4) node [anchor=north west][inner sep=0.75pt]    {$Q$};
\draw (226,104) node [anchor=north west][inner sep=0.75pt]  [color={rgb, 255:red, 208; green, 2; blue, 27 }  ,opacity=1 ] [align=left] {Nariai black holes};
\draw (453,66.4) node [anchor=north west][inner sep=0.75pt]  [font=\footnotesize,color={rgb, 255:red, 74; green, 144; blue, 226 }  ,opacity=1 ]  {$Q=M$};
\draw (380,226) node [anchor=north west][inner sep=0.75pt]   [align=left] {\begin{minipage}[lt]{50.35pt}\setlength\topsep{0pt}
\begin{center}
Naked \\Singularity
\end{center}

\end{minipage}};
\draw (293,23) node [anchor=north west][inner sep=0.75pt]   [align=left] {\begin{minipage}[lt]{50.35pt}\setlength\topsep{0pt}
\begin{center}
Naked \\Singularity
\end{center}

\end{minipage}};

\end{tikzpicture}
    \caption{Extremal curve in de Sitter space. The extremal black holes on the upper edge are the Nariai black holes whose horizons are of the order of Hubble horizon.}
    \label{sharkfin}
\end{figure}

We know that the lower edge of the triangle is similar to the extremal curve in flat space which leads to the WGC. But how about the upper edge? This edge turns out to give us an opposite inequality! For Nariai black holes in 4d, we have $M\sim Q\sim R\sim \frac{1}{H}$. The electric field at the horizon goes like $E=Q^2/R^2\sim H$. Suppose there is a particle with  $qH\geq m^2$, we will have a large Schwinger pair production which will lead to a large flow of charge from inside the black hole to outside the horizon. However, such a process will force the black to exit the allowed region and create a naked singularity. So we find $m\geq \Lambda^\frac{1}{4} q^\frac{1}{2}$. This bound is called \textit{Festina Lente}\cite{Montero:2019ekk}.  Note that this bound is satisfied by electron in our universe.

Interestingly, in our universe, the $\Lambda^\frac{1}{4}$ scale is the mass scale of neutrino. Note that Festina Lente does not apply to broken symmetries. However, if there is an unbroken phase, you could use this in that phase. Therefore, an unbroken electroweak symmetry which requires masses to be zero would have been inconsistent with a positive cosmological constant.

\subsection{Applications}

In four dimensions, we can apply the WGC to magnetic charges. If the electric charge is $g$, the magnetic charge is a multiple of $1/g$. Moreover, the masses of 't Hooft–Polyakov monopoles go like $\Lambda/g^2$ where the $\Lambda$ is some UV energy scale where a symmetry is sponatiously broken and the monopoles are created. Therefore we find $\Lambda\leq gM_{P}$ which implies the UV cutoff is not necesarily all the way at Planck scale in weakly coupled theories. 

Now let us consider another application to axions. Axion couples to instantons through $\theta(x) F\wedge F$. If $\theta$ is not a field, we have free parameters which will violate the cobordism conjecture. If we promote $\theta$ to a field and include other parts of its action we find 
\begin{align}
    f^2(\partial \theta)^2+V(\theta)+i\theta F\wedge F.
\end{align} 
The potential $V(\theta)$ receives non-perturbative corrections by instantons in the form of a cosine term. The discrete shift symmetry of axion is a $-1$ form symmetry which is gauged due to existence of instantons. If we think of this system as a $-1$ form gauge symmetry with instantons being the charges objects, we can apply the WGC and find
\begin{align}
    S_{ins}\lesssim\frac{M_{P}}{f},
\end{align}
where $1/f$ plays the role of charge. For instanton actions bigger than 1, we have $f<M_P$.

WGC is naturally connected to an older conjecture called the cosmic censorship conjecture which is also based on avoiding naked singularities \cite{Crisford:2017gsb}. The cosmic censorship conjecture roughly states that naked singularities cannot arise due to natural physical dynamics. However, recently a counteraxample was found in pure Einstein-Maxwell setup in AdS \cite{Crisford:2017zpi}. The idea is to gradually turn on the electric field over time until a naked singularity appears. However, the Swampland conditions immediately tell us that we must have charged states with small masses which will screen large electric fields. Therefore, the WGC resolved the puzzle, including the precise numerical factors. 

There are more applications to cosmology and particle phenomenology that involve relationships between the mass of the neutrino and the cosmological constant. For example, if you take the strong version of WGC that says everything strictly satisfying WGC is unstable, the non-supersymmetric AdS must be unstable. This is because a non-supersymmetric brane whose near horizon geometry is that AdS (carries the same fluxes) would be unstable \cite{Ooguri:2016pdq}. Now we can use this for compactification of our universe on a circle which leads to an inequality for the mass of neutrino \cite{Gonzalo:2021zsp}. 

In all the well-known non-supersymmetric constructions of AdS, there have been found some instantons that create instabilities (e.g. \cite{Antonelli:2019nar}). 

\section{Swampland IV: Distance conjectures}

\subsection{Introduction}

In the previous sections we talked about how the low-energy field theories typically have finite field ranges. The idea was that by varying the scalar field, at some point the potential energy might increase so much that it surpasses the cut-off scale. However, if one increase the cut-off, previously distinct low-energy theories might unify in the sense that they are realized in different corners of the UV theory's field space (see Figure from section \ref{secintro}). 

On the other hand, increasing the field range, lowers the EFT cut-off by bringing in new light states which were not part of EFT. In this section we will try to quantify this observation from examples in string theory. But first, we need to clarify our terminologies. Let us start with the notion of moduli space. 

\subsection{Moduli space}

There are different notions of moduli space. Sometimes it is taken to be the space parametrized by scalar fields. However, for us it will represent the space of different vacua. For example, suppose we have a background with a set of scalar fields $\phi^i$ and an effective action $\Gamma$. The expectation value of the scalar fields extremizes the effective action,
\begin{align}
    \frac{\delta}{\delta\phi^i}\Gamma(\langle{\phi^j}\rangle)=0.
\end{align}
Once, we take a true quantum mechanically stable background, we can study perturbation around that background. Suppose the effective action at low energies can be approximated with the following local action 
\begin{align}
    \Gamma_{local}=\int d^Dx [\frac{1}{2}\eta^{\mu\nu}g_{ij}(\partial_\mu \phi^i)(\partial_\nu\phi^j)-V_{eff}(\phi^i)]+\hdots
\end{align}
where $\hdots$ represents the terms containing fields other than scalar fields. We are interested in values of scalar fields where the potential is minimum. This might not be a single point, but rather a manifold. Since the potential is constant in the directions of field space that potential stays minimum, these are called the flat directions of the field space. We can think of these values as boundary conditions for the scalar fields at infinity, each of which defines a distinct vacuum. For now, we will assume that the minimum value of the potential is zero which is to say the background is Minkowski. We will consider non-flat backgrounds later.

The $V_{eff}=0$ subspace of the field space is called the moduli space and it represents the space of different vacua. Suppose the moduli space is locally parametrized by scalar fields $\phi^I$. The restriction of the effective action to the moduli space takes the following form.
\begin{align}
    \int d^Dx \frac{1}{2}\eta^{\mu\nu} g_{IJ}(\Phi)(\partial_\mu \phi^I)(\partial_\nu\phi^J).
\end{align}
One can show that under a reparmetrization of the scalar fields $\phi^I$, $g_{IJ}$ transforms as a symmetric rank 2 tensor. In other words, $g_{IJ}$ can be viewed as a metric on the field space. The above action is called the non-linear sigma model and is simply the generalization of free scalar field theory to the case where the field target space is geometrically non-trivial. 

We will refer to the metric $g_{IJ}$ as the canonical metric on the moduli space. Using this metric, we can talk about geometric quantities associated with the moduli space, such as distance $\int dl\sqrt{g_{IJ} \frac{d}{dl}\phi^I\frac{d}{dl}\phi^J}$ or volume $\int d^N\phi \sqrt{g}$.

Note that the effective action generally has non-local terms, but at low energies (compared to some UV scale) we expect a local description to be valid. 

Now let us consider a special class of theories; supersymmetric theories. 

\subsection{Supergravities}

Given that global supersymmetry is only a symmetry at $V=0$, one might think that supersymmetry always protects $V=0$, however this is not true. For example, an $\mathcal{N}=1$ supergravity theory with some chiral fields has a scalar potential,
\begin{align}
    V=e^{-K}(|\mathcal{D}W|^2-3|W|^2),
\end{align}
where $W$ is the superpotential and $K$ is the Kähler potential. In string theory almost always $\mathcal{N}=1$ theories come with non-zero superpotentials unless they are "secretly" even more supersymmmetric. When we have 8 or more supercahrges, which corresponds to $\mathcal{N}\geq2$ in four dimensions, there is no scalar potential in an ungauged supergravity. However, through gauging we might get a scalar potential. 

As we explained in the previous section, the moduli space $\mathcal{M}$ comes with a canonical metric that can be read off from the kinetic term. But what do we know about the geometric properties of $\mathcal{M}$? Let us start with its dimension. It turns out $\mathcal{M}$ could be zero-dimensional. For example, one (and possibly only) example in Minkowski background is 11 dimensional supergravity. Other examples in AdS backgrounds are $AdS_7\times S^4$ or $AdS_4\times S^7$. However the $AdS_5\times S^5$ IIB background has the IIB coupling as a modulus. 
\vspace{5pt}

\noindent\textbf{Exercise 1: }Show there are scalars in M-theory that minimize their potential at the 11d supergravity corner. 
\vspace{10pt}

Now let us consider 10d supergravity theories which have more interesting moduli spaces. All of these theories have one modulus in common, the dilaton. However, while the type IIA, type I, and Heterotic theories all have a real dilaton, the type IIB theory has a complex coupling constant $\tau$. Note that in all of these examples the target space of dilaton is non-compact. We can also compute the distances and see that it goes off to infinity at the asymptotes. This is thanks to the special form of dilaton's kinetic term,
\begin{align}
    S_{10d}\propto\int (\frac{\partial\lambda}{\lambda})^2+\hdots
\end{align}
For type IIB theory where the coupling is complex, the metric takes a slightly different form $ds^2=d\tau d\bar\tau/(\tau_2^2)$. However, due to the $SL(2,\mathbb{Z})$ identification the topology of the moduli space is not like a plane. After the identifications, the moduli space has an infinite diestance limit and two cusp points. 
\begin{figure}[H]
    \centering
\tikzset{every picture/.style={line width=0.75pt}} 
\begin{tikzpicture}[x=0.75pt,y=0.75pt,yscale=-1,xscale=1]
\draw  [draw opacity=0][fill={rgb, 255:red, 155; green, 155; blue, 155 }  ,fill opacity=1 ] (182.5,11) .. controls (184.5,9) and (182.29,194) .. (182.39,193) .. controls (182.5,192) and (158.5,179) .. (127.5,178) .. controls (96.5,177) and (70.5,194) .. (72.5,193) .. controls (74.5,192) and (69.5,14) .. (72.5,11) .. controls (75.5,8) and (180.5,13) .. (182.5,11) -- cycle ;
\draw [fill={rgb, 255:red, 74; green, 80; blue, 226 }  ,fill opacity=1 ]   (72.5,11) -- (72.5,193) ;
\draw [fill={rgb, 255:red, 74; green, 80; blue, 226 }  ,fill opacity=1 ]   (182.5,11) -- (182.39,193) ;
\draw [draw opacity=0][fill={rgb, 255:red, 74; green, 80; blue, 226 }  ,fill opacity=1 ]   (72.5,11) -- (182.5,11) ;
\draw  [draw opacity=0] (72.5,193) .. controls (88.67,183.7) and (107.44,178.38) .. (127.45,178.38) .. controls (147.46,178.38) and (166.22,183.7) .. (182.39,193) -- (127.45,288.17) -- cycle ; \draw   (72.5,193) .. controls (88.67,183.7) and (107.44,178.38) .. (127.45,178.38) .. controls (147.46,178.38) and (166.22,183.7) .. (182.39,193) ;
\draw  (25.45,289) -- (239.45,289)(127.5,87.17) -- (127.5,299) (232.45,284) -- (239.45,289) -- (232.45,294) (122.5,94.17) -- (127.5,87.17) -- (132.5,94.17)  ;
\draw    (61.01,156.23) .. controls (5.88,176.74) and (40.28,218.96) .. (127.5,217) .. controls (214.72,215.04) and (238.07,175.62) .. (195.21,155.22) ;
\draw [shift={(192.5,154)}, rotate = 23.27] [fill={rgb, 255:red, 0; green, 0; blue, 0 }  ][line width=0.08]  [draw opacity=0] (10.72,-5.15) -- (0,0) -- (10.72,5.15) -- (7.12,0) -- cycle    ;
\draw [shift={(64.5,155)}, rotate = 161.57] [fill={rgb, 255:red, 0; green, 0; blue, 0 }  ][line width=0.08]  [draw opacity=0] (10.72,-5.15) -- (0,0) -- (10.72,5.15) -- (7.12,0) -- cycle    ;
\draw    (253,170) -- (398.5,169.01) ;
\draw [shift={(400.5,169)}, rotate = 179.61] [color={rgb, 255:red, 0; green, 0; blue, 0 }  ][line width=0.75]    (10.93,-3.29) .. controls (6.95,-1.4) and (3.31,-0.3) .. (0,0) .. controls (3.31,0.3) and (6.95,1.4) .. (10.93,3.29)   ;
\draw  [draw opacity=0] (509.45,216.42) .. controls (509.45,216.42) and (509.45,216.42) .. (509.45,216.42) .. controls (509.45,216.42) and (509.45,216.42) .. (509.45,216.42) .. controls (529.46,216.42) and (548.22,221.74) .. (564.39,231.04) -- (509.45,326.21) -- cycle ; \draw   (509.45,216.42) .. controls (509.45,216.42) and (509.45,216.42) .. (509.45,216.42) .. controls (509.45,216.42) and (509.45,216.42) .. (509.45,216.42) .. controls (529.46,216.42) and (548.22,221.74) .. (564.39,231.04) ;  
\draw    (509.5,22) -- (509.45,216.42) ;
\draw    (518.5,20) -- (564.39,231.04) ;
\draw [color={rgb, 255:red, 74; green, 144; blue, 226 }  ,draw opacity=1 ]   (509.5,108) .. controls (520.5,112) and (531.5,108) .. (537,100) ;
\draw [color={rgb, 255:red, 74; green, 144; blue, 226 }  ,draw opacity=1 ] [dash pattern={on 4.5pt off 4.5pt}]  (509.5,108) .. controls (517.5,101) and (524.5,97) .. (537,100) ;
\draw [color={rgb, 255:red, 74; green, 144; blue, 226 }  ,draw opacity=1 ]   (509.09,71.15) .. controls (518.13,74.77) and (525.89,71.15) .. (528.82,63.91) ;
\draw [color={rgb, 255:red, 74; green, 144; blue, 226 }  ,draw opacity=1 ] [dash pattern={on 4.5pt off 4.5pt}]  (509.09,71.15) .. controls (514.09,64.82) and (518.79,61.2) .. (528.82,63.91) ;
\draw  [fill={rgb, 255:red, 0; green, 0; blue, 0 }  ,fill opacity=1 ] (506.7,216.42) .. controls (506.7,214.9) and (507.93,213.67) .. (509.45,213.67) .. controls (510.97,213.67) and (512.2,214.9) .. (512.2,216.42) .. controls (512.2,217.94) and (510.97,219.17) .. (509.45,219.17) .. controls (507.93,219.17) and (506.7,217.94) .. (506.7,216.42) -- cycle ;
\draw  [fill={rgb, 255:red, 0; green, 0; blue, 0 }  ,fill opacity=1 ] (560.64,228.29) .. controls (560.64,226.77) and (561.87,225.54) .. (563.39,225.54) .. controls (564.91,225.54) and (566.14,226.77) .. (566.14,228.29) .. controls (566.14,229.81) and (564.91,231.04) .. (563.39,231.04) .. controls (561.87,231.04) and (560.64,229.81) .. (560.64,228.29) -- cycle ;
\draw    (528.5,285) -- (510.1,226.91) ;
\draw [shift={(509.5,225)}, rotate = 72.43] [color={rgb, 255:red, 0; green, 0; blue, 0 }  ][line width=0.75]    (10.93,-3.29) .. controls (6.95,-1.4) and (3.31,-0.3) .. (0,0) .. controls (3.31,0.3) and (6.95,1.4) .. (10.93,3.29)   ;
\draw    (537.5,285) -- (557.7,238.83) ;
\draw [shift={(558.5,237)}, rotate = 113.63] [color={rgb, 255:red, 0; green, 0; blue, 0 }  ][line width=0.75]    (10.93,-3.29) .. controls (6.95,-1.4) and (3.31,-0.3) .. (0,0) .. controls (3.31,0.3) and (6.95,1.4) .. (10.93,3.29)   ;
\draw (110,61.4) node [anchor=north west][inner sep=0.75pt]    {$Im\ \tau $};
\draw (216,299.4) node [anchor=north west][inner sep=0.75pt]    {$Re\ \tau $};
\draw (100,229) node [anchor=north west][inner sep=0.75pt]   [align=left] {Identify};
\draw (486.4,93) node [anchor=north west][inner sep=0.75pt]  [rotate=-270]  {$Im\ \tau \rightarrow \infty $};
\draw (509,288) node [anchor=north west][inner sep=0.75pt]   [align=left] {Cusps};
\draw (545.4,82) node [anchor=north west][inner sep=0.75pt]  [rotate=-270]  {$\textcolor[rgb]{0.29,0.56,0.89}{l\rightarrow 0}$};
\end{tikzpicture}
    \caption{The moduli space of type IIB theory is the $\mathbb{H}/SL(2,\mathbb{Z})$. The moduli space has an infintie distance limit and two cusp singularities.}
    \label{TFR}
\end{figure}
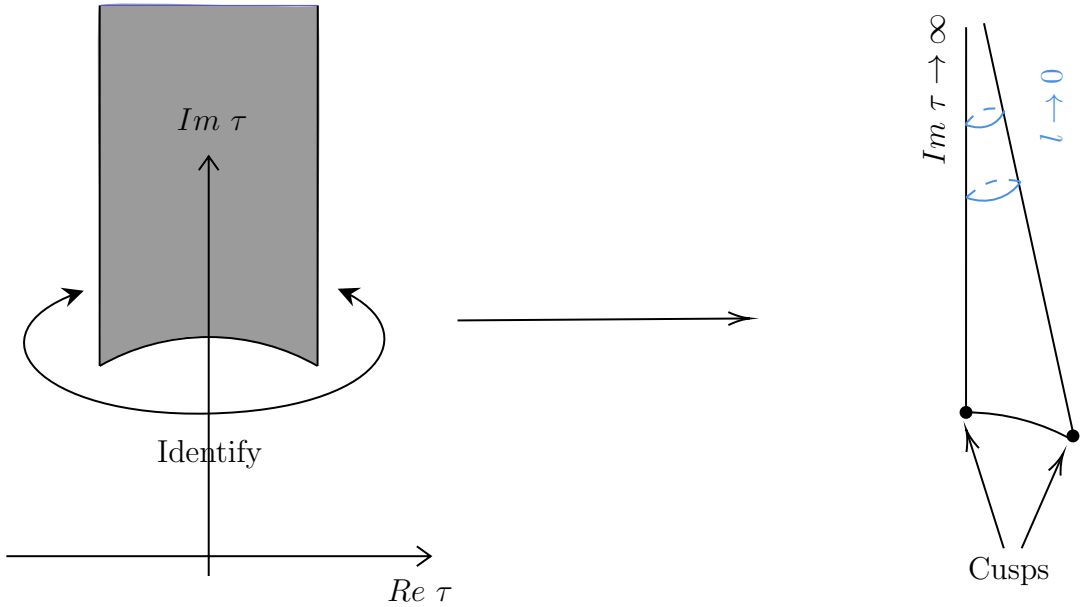
Note that as we go up in the infinite distance limit $\text{Im}(\tau)\rightarrow\infty$, the size of the cycle that wrap around the throat go to zero. Therefore, any geodesic can be shrunk to an arbitrarily small length with continuous deformation. In other words, if we include the asymptotic points, there is no non-trivial cycle. Also, the cusp points are usually when we get enhanced gauge symmetries $\mathbb{Z}_4$ and $\mathbb{Z}_6$. Note that the diameter of the moduli space is still infinite but this time the volume is finite as opposed to the IIA or Heterotic case.

\subsection{Dualities and infinite distances limits}

There is very close connection between infinite distance limits in the moduli space and dualities in quantum gravity. When we say two theories are dual, it means they have the same moduli space and there is a correspondence between the physical objects in the two theories. In perturbative dualities (usually T-dualities), the perturbative degrees of freedom are mapped to each other. However, in non-perturbative (strong-weak) dualities, we must include non-perturbative objects in each theory to complete the corresopndence. Some of our most powerful insights into the non-perturbative features of quantum gravity come from dualities. In that sense, understanding dualities is as fundamental as understanding quantum gravity. 

Moreover, usually, the dualities are between theories that have a well understood perturbative description in some corner of their moduli space. In all the known examples, these "corners" are always some infinite distance limits. In other words, the dualities are between perturbative descriptions of theories at two different infinite distance limits of the moduli space. Two most important examples of such infinite distance limits are (1) when string coupling goes to 0, (2) when the scalar field corresponding to the size of internal geometry goes to infinity. In the first example, higher string excitations must be included and the pertuabtion becomes more convergent. In the second example, higher KK modes must be included which can be achieved by using the higher dimensional theory. In both cases, there is a rich physical structure that enters and changes the IR description at the infinite distance limits. Moreover, the nature of what happens is closely connected with the type of dualities that connect that corner with other corners. Thus, the goal of understanding the infinite distance limits of the moduli space is as deep as understanding the nature of dualities in quantum gravity. In fact, we will see that just like dualities that fall into some universal classes, infinite distance limits in string theory also seems to fall into some universal classes. 

We only deeply understand infinite distance limits of the moduli space since we can use perturbation theory as the coupling gets weak in some duality frame. Usually interesting things happen at these infinite distance limits in the form of some states becoming light. The reason these infinite distance limits are important is because these tells us more about dualities which are some of the big mysteries about string theory. Note that from the field theory perspective, it is very unnatural to expect any rich structure to appear at the infinite distance limit in a flat direction.

With the above explanation in mind, let us now consider some more complicated examples of the connection between dualities and infinite distance limits in string theory. Consider type II theories on Calabi--Yau threefolds. The resulting theory is a four dimensional $\mathcal{N}=2$ theory. The moduli space of such theories is a direct product of the Coulomb branch (vector multiplet scalars) and the Higgs branch (hypermultiplet scalars). From the string theory perspective, the moduli of the Calabi--Yau become the moduli of the four dimensional theory. For example, if we compactify IIB, the complex structure moduli become Coulomb branch moduli and the Kähler moduli mix with R–R-fields to give the Higgs branch moduli.

On the other hand, if we compactify the heterotic theory on a $d$ dimensional torus we find that the moduli space is
\begin{align}
SO(d+16,d;\mathbb{Z})\text{\textbackslash} \frac{SO(d+16,d)}{SO(d+16)\times SO(d)}\times\mathbb{R}^+.
\end{align}
This space has many infinite distance limits. For example, you can take the limit where one compactification radius goes to infinity.
\begin{align}
SO(16+d,d;\mathbb{Z})\rightarrow SO(16+d-1,d-1;\mathbb{Z})\times SO(1,1).
\end{align}
The leftover moduli would be the moduli space of the heterotic theory in one higher dimension as expected.

Now let us consider another example. Consider supersymmetric theories in 6d with $(1,0)$ supersymmetry. There are various ways of getting such a theory. For example we can put Heterotic on K3, F-theory on elliptic CY threefold, or M theory on $K3\times I$. If you are given a point in this moduli space, if the duality picture is correct, you should be able to go to the inifinte distance limits of the moduli space to discover all these different corners. For example, from the M theory perspective, one can shrink the the instantons at the end of the interval and move them to the inside to get NS5 branes from the Heterotic perspective. 

Let us take a step back. The infinite distance limits are very unnatural to have any interesting physical characteristic from field theory perspective. However, in quantum gravity, infinite distance limits have rich physics. The question of understanding infinite distance limits in string landscape is as deep as the question of understanding dualities in string theory. 

Let us consider another example. Consider the type II theories (IIA and IIB) on CY threefold. Both theories will give us a 4d $\mathcal{N}=2$ theory. Duality between IIA and IIB suggests a pairing between Calabi--Yau threefolds such that IIA on one is the same as IIB on the other. This symmetry is called the mirror symmetry and the two manifolds are called the mirror pairs. The mirror symmetry has been very useful in studying superymmetric theories. For exmaple, if you compactify on a threefold with ADE singularities you get ADE gauge theories as we discussed before. If we compactify IIB on a threefold with an ADE singularity, the D1 branes wrapped around the 2-cycle give non-perturbative corrections. In fact, such non-pertubative corrections are so large that the formally infinite distance limit of a shrinking CY beomes a finite distance point in the moduli space. Instead of dealing with the instanton sum in IIB picture, it is easier to work in the S-dual IIB picture. In the dual picture, $D1$ branes get mapped to the worldsheet instantons and the perturbative calculations are trustable and automatically take the summed up worldsheet instantons into account, which leads to the above claim.

Now we try to see what are the intrinsic way of seeing these infinite distance limits in the lower dimensional theory without knowing the UV construction behind the theory.

\subsection{Universal properties of infinite distance limits}

The first observation is that at infinite distance limits we always get a tower of light states ($m\ll M_{pl}$). And the second observation is that the tower is always weakly coupled. These properties have been tested for many infinite distance limits in the known string constructions \cite{Grimm:2019ixq,Corvilain:2018lgw,Grimm:2018ohb}.

Let us start with string excitations in 10d. If we use $M_P^8=M_S^8/g_s^2$ to go to the Einstein frame, we see that mass of the string states in Planck units scales like $g_s^\frac{1}{4}$. Therefore, in the $g_s\rightarrow 0$ limit, the string states are light in Planck units, and by definition weakly coupled. Therefore, the string excitations become stable light particles at the infinite distance limit. 

Similarly, in the stroung coupling limit of type IIB when $\tau\rightarrow 0$, the D1 branes give rise to a light tower of weakly coupled states. However, in type IIA the situation is slightly different. In the strong coupling limit $g_s\rightarrow\infty$. The tower is the KK modes of M theory on circle corresponding to D0 branes, the infinite distance limit is not longer a 10d theory. One might wonder why IIA D0 branes are weakly coupled? This is because when we compactify M theory, the coefficient of $\mathcal{R}$ is $R_{11}$ and its inverse is some power of coupling $D_0$ brane coupling. 

Now let us move to the two heterotic strings. Interestingly, the strong coupling limits of the two theories are very different. In one ($SO(32)$) we get an exponential number of states (string tower) and in the other ($E_8\times E_8$) we get much less number of states (KK tower).

\begin{figure}[H]
    \centering

\tikzset{every picture/.style={line width=0.75pt}} 

\begin{tikzpicture}[x=0.75pt,y=0.75pt,yscale=-1,xscale=1]

\draw    (249.08,81.83) -- (318.75,202.5) ;
\draw    (318.75,202.5) -- (457.5,202.5) ;
\draw    (457.17,203.06) -- (388.16,82.02) ;
\draw    (388.16,82.02) -- (249.08,81.83) ;

\draw  (111.5,202) -- (564.75,202)(318.5,15.5) -- (318.5,316.5) (557.75,197) -- (564.75,202) -- (557.75,207) (313.5,22.5) -- (318.5,15.5) -- (323.5,22.5)  ;
\draw  [fill={rgb, 255:red, 0; green, 0; blue, 0 }  ,fill opacity=1 ] (380.83,163.96) .. controls (380.83,163.06) and (381.56,162.33) .. (382.46,162.33) .. controls (383.36,162.33) and (384.08,163.06) .. (384.08,163.96) .. controls (384.08,164.86) and (383.36,165.58) .. (382.46,165.58) .. controls (381.56,165.58) and (380.83,164.86) .. (380.83,163.96) -- cycle ;
\draw  [dash pattern={on 4.5pt off 4.5pt}]  (111.08,81.83) -- (180.75,202.5) ;
\draw    (180.75,202.5) -- (319.5,202.5) ;
\draw  [dash pattern={on 4.5pt off 4.5pt}]  (250.16,82.02) -- (111.08,81.83) ;
\draw  [dash pattern={on 4.5pt off 4.5pt}]  (318.08,201.83) -- (387.75,322.5) ;
\draw  [dash pattern={on 4.5pt off 4.5pt}]  (387.75,322.5) -- (526.5,322.5) ;
\draw  [dash pattern={on 4.5pt off 4.5pt}]  (526.17,323.06) -- (457.16,202.02) ;
\draw  [fill={rgb, 255:red, 0; green, 0; blue, 0 }  ,fill opacity=1 ] (245.5,138) .. controls (245.5,137.1) and (246.23,136.38) .. (247.13,136.38) .. controls (248.02,136.38) and (248.75,137.1) .. (248.75,138) .. controls (248.75,138.9) and (248.02,139.63) .. (247.13,139.63) .. controls (246.23,139.63) and (245.5,138.9) .. (245.5,138) -- cycle ;
\draw  [dash pattern={on 4.5pt off 4.5pt}]  (180.08,201.83) -- (249.75,322.5) ;
\draw  [dash pattern={on 4.5pt off 4.5pt}]  (387.75,322.5) -- (248.67,322.31) ;
\draw    (383.46,162.33) .. controls (374.54,99.32) and (284.41,77.22) .. (247.06,137.09) ;
\draw [shift={(246.5,138)}, rotate = 301.24] [color={rgb, 255:red, 0; green, 0; blue, 0 }  ][line width=0.75]    (10.93,-3.29) .. controls (6.95,-1.4) and (3.31,-0.3) .. (0,0) .. controls (3.31,0.3) and (6.95,1.4) .. (10.93,3.29)   ;

\draw (453,212.4) node [anchor=north west][inner sep=0.75pt]    {$1$};
\draw (224,57.4) node [anchor=north west][inner sep=0.75pt]    {$\tau =\omega $};
\draw (384.46,168.98) node [anchor=north west][inner sep=0.75pt]    {$p$};
\draw (228.46,139.98) node [anchor=north west][inner sep=0.75pt]    {$\omega p$};
\draw (337,93.4) node [anchor=north west][inner sep=0.75pt]    {$\times \omega $};

\end{tikzpicture}
    \caption{Each $T^2$ is constructed by identifying $z\in \mathds{C}$ with $z+n+m\omega$ where $n$ and $m$ are integers. The complex plane has a $\mathbb{Z}_3$ symmetry which acts as $z\rightarrow z\times\omega$. Since this symmetry maps the lattice $\{n+m\omega|n,m\in \mathbb{Z}\}$ to itself, it is also a symmetry of the torus.}
    \label{Z3T2}
\end{figure}
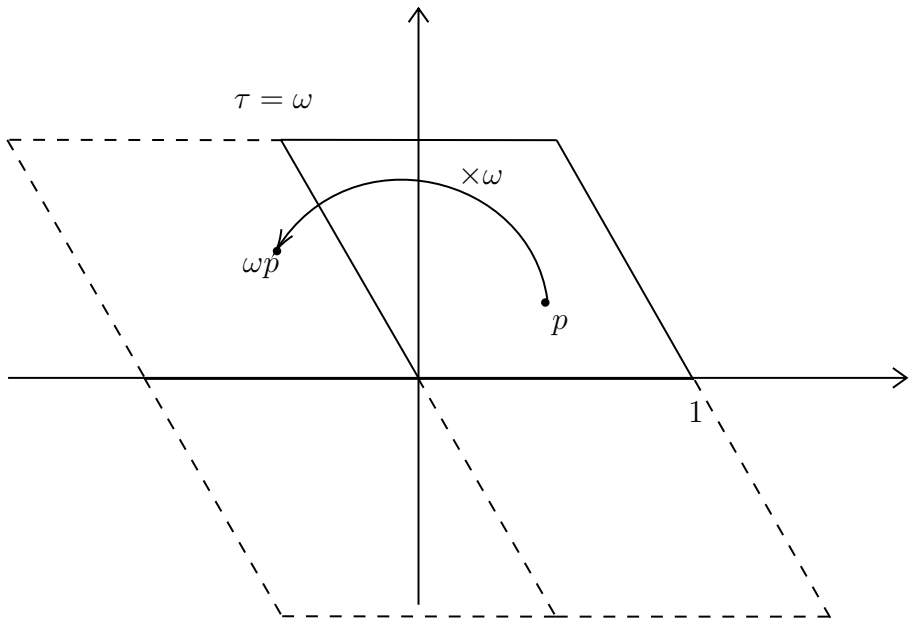

It seems infinite distance limits correspond to infinite number of light states. However, can we say the opposite? (\emph{i.e.} infinite number of states only appear at infinite distances?) The answer is no! Take M-theory on $T^6/\mathbb{Z}_3$. The $T^6$ is the product of three identical $T^2$ with Teichmüller variable of $\omega$ and the $\mathbb{Z}_3$ acts by $\times\omega_3$ on each $T^2$ where $\omega=\exp(2\pi i/3)$ is the third root of unity (Fig \ref{Z3T2}). 

This theory is at finite distance due to the finite distance resolution of the singularity. However, there are infinite number of light particles. For example, if you take the $P^2$ from the resolution of the singularity and wrap M2 branes around it, you get infinite number of light particles. A degree-d curve in $\mathds{P}^2$ is the Riemann surface. 

\begin{align}
    p(z_1^d+z_2^{d}+z_1^{d-1}z_2+...)=0
\end{align}
with a genus $g=\frac{(d-1)(d-2)}{2}$ where $\mathds{P}^2$ is the projective space $(z_1,z_2)\sim\lambda\times(z_1,z_2)$. In this limit you get a conformal field theory with tensionless strings coming from M5 branes wrapping $\mathds{P}^2$ and light particles corresponding to M2 branes wrapping surfaces that interact. In CFT with more than four dimensions we almost always get tensionless strings. However, this tower is not a weakly coupled string which distinguishes it from the tower of light states in the infinite distance limits. Given this amount of supersymmetry, this is the only kind of new phases (or critical phenomena) that you can have. In all these limits the gravity is decoupled meaning the volume of the manifold can be taken to infinity. 

Note that when we say a tower is weakly coupled, we measure the coupling in the original description. For example, the KK tower is weakly coupled in the lower dimensional theory but the theory could be strongly coupled in the higher dimensional description. 

There is even more quantifiable structure to these infinite distance towers. You can compute distance using the canonical metric. And you can ask how fast do the towers become light? It turns out that the mass scale of the tower always goes to zero exponentially as
\begin{align}
    m_{tower}\sim e^{-\alpha \text{dist.}}
\end{align}
where dist. is measured in Planck units and $\alpha$ is some $\mathcal{O}(1)$ constant. In all the known examples $\alpha\geq \frac{1}{\sqrt{d-2}}$ \cite{Etheredge:2022opl}.
\vspace{10pt} 

\noindent\textbf{Exercise 2: }Show that type IIA theory has a tower of states with exponentially decreasing masses $m\sim\exp(-\alpha\varphi)$ in the $g_s\rightarrow \infty$ limit with $\alpha=\frac{3}{2\sqrt{2}}$ where $\varphi$ is a scalar field with canonical kinetic term $-\frac{1}{2}(\partial_\mu\varphi)^2$ in the Einstein frame.
\vspace{10pt}

\noindent\textbf{Exercise 3: }Similar to the previous exercise, show that in type IIB string theory, in both limits $g_s\rightarrow \infty$ or $g_s\rightarrow 0$, there is a tower of states with exponentially decreasing masses with a decay rate of $\alpha=\frac{1}{2\sqrt{2}}$. 
\vspace{10pt}

The exponential rate is the same for the two infinite distance limits in IIB but not the same in IIA.

Another example is the $SO(16)\times SO(16)$ Heterotic theory which is non-supersymmetric but we know how to describe it at weak coupling. In that case too, in the infinite distance limit, we get an exponentially light tower of states. 

Now we are ready to give a more precise formulation of the distance conjecture. First, we give the basic version, then we state stronger versions with additional conditions. 

\begin{statement6*}
\textbf{Basic version:} At any infinite limit in the moduli space, a light tower of light states emerges with $m\sim e^{-\alpha \cdot dist.}$ \cite{Ooguri:2006in}. Additional versions include
\vspace{5pt}

\textbf{1)} The tower is weakly coupled. 
\vspace{5pt}

\textbf{2)} If the moduli space is not a point, it is non-compact. 
\vspace{5pt}

\textbf{3)} The first homology is always trivial\footnote{This statement sounds like a consequence of the no-global symmetry conjecture given that a non-trivial homotopy leads to a topological charge. However, the conservation could be violated by moving the closed curve in the direction of the massive scalars which are not part of the moduli space.}.  
\vspace{5pt}

\textbf{4)} There is always a dual description in the infinite distance limit. However, it is difficult to make this statement precise.
\end{statement6*}

There is a refinement of the first and last conditions which states that there are only two possibilities for the leading tower: either KK tower or a tensionless fundamental string. This refinement is known as the emergent string conjecture \cite{Lee:2019wij}. This conjecture is very non-trivial because we could have had tensionless membranes in the infinite distance limits, but such membranes always turn out to remain relatively heavy.  

The constant $\alpha$ is expected to be of order one, but the lack of a precise lower bound for it in the formulation of the conjecture makes the conjecture a bit less precise than some other conjectures, such as the WGC. However, we will see in the next section that de Sitter conjectures suggest a precise lower bound for $\alpha$ that seems to be true for all stringy examples. There are two natural ways to relate the mass of the tower to the scalar potential which are $m^2\sim V$ and $m^d\sim V$ \cite{Bedroya:2019snp,Andriot:2020lea,Bedroya:2020rmd}. We will discuss these two options in the next section and show that they lead to the following natural candidates for $\alpha$.
\begin{align}
    \alpha\geq \frac{1}{\sqrt{d-2}}\text{ or } \frac{2}{d\sqrt{d-2}}.
\end{align}

This is consistent with all known cases where $\alpha\geq \frac{1}{\sqrt{d-2}}$. The sharpened version of the distance conjecture \cite{Etheredge:2022opl} states that $1/\sqrt{d-2}$ is indeed the lower bound for $\alpha$. In \cite{Etheredge:2022opl}, it was shown that this proposal is invariant under dimensional reduction and is saturated in toroidal compactifications of supergravities. Moreover, the authors in \cite{Etheredge:2022opl} made the following interesting observation in string theory examples that $\alpha=1/\sqrt{d-2}$ if and only if one of the leading towers of states is a string tower and commented on its connection with emergent string conjecture. In the following, we explain why the two conjectures are indeed related. In particular, we show that the sharpened distance conjecture follows from the emergent string conjecture.

According to the emergent string conjecture, every infinite distance limit in the moduli space is either a fundamental string limit or a decompactification limit. Let us first assume that a limit is a perturbative string limit. A limit with a fundamental string is a limit where the scattering of all weakly coupled particles are given by a fixed worldsheet theory (in string units) with a coupling that goes to 0. Therefore, this infinite distance limit, by definition, corresponds to a limit where all scalars are kept fixed in the string frame with the exception of the string coupling $\exp(\phi)$, which is taken to 0.

By definition, $\phi$ couples to the worldsheet via $\int \mathcal{R}\phi$ in the string frame. Similarly, the metric couples as $\int g^{\mu\nu}\partial X^\mu\cdot \partial X^\nu$ in the string frame. Since we assume the string is fundamental, we can apply the machinery of the perturbative string theory to the above vertex operators to read off the tree-level amplitudes of graviton and $\phi$. This will give us the standard string theory result for the effective action in the string frame \cite{Polchinski:1998rq}. 
 
\begin{align}
S=\frac{M_s^{d-2}}{2}\int e^{-2\phi}(\mathcal{R}+4(\partial\phi)^2+\hdots).
\end{align}
After going to the Einstein frame, we find
\begin{align}
S=\int \frac{M_P^{d-2}}{2}\mathcal{R}+\frac{1}{2}(\hat\phi)^2+\hdots,
\end{align}
where $M_P^{d-2}=M_s^{d-2}\exp(-2\phi)$ and $\phi=\hat\phi\cdot\frac{\sqrt{d-2}}{2}$. If we combine the two, we find
\begin{align}
M_s=M_Pe^{-\frac{1}{\sqrt{d-2}}\kappa\hat\phi}.
\end{align}
Therefore, whenever the light states are described by a fundamental string, the coefficient in the distance conjecture is exactly $1/\sqrt{d-2}$.

Now suppose the leading tower of light states is described by a KK reduction of a higher dimensional field theory. If we take a $D$ dimensional theory and compactify it down to $d$ dimensions, the mass of the KK tower will go like $m\sim \exp(-\sqrt\frac{D-2}{(D-d)(d-2)}\hat\rho)$ where $\hat\rho$ is the canonically normalized volume modulus. Note that the coefficient $\sqrt{(D-2)/[(D-d)(d-2)]}$ is always greater than $1/\sqrt{d-2}$ and it saturates it at $D\rightarrow\infty$. Therefore, for any KK tower, if we move in the direction of the corresponding volume modulus, the tower satisfies $\lambda>1/\sqrt{d-2}$. In the following, we show that the emergent string conjecture implies that no mixing of the volume modulus with other moduli can bring this coefficient below $1/\sqrt{d-2}$. We provide an algorithmic procedure that changes the infinite distance limit in a way that strictly decreases the coefficient of the distance conjecture. Then we show that the endpoint of the algorithm is either a string limit or a KK limit where only the volume modulus is taken to infinity. Since the coefficient of the distance conjecture in both cases is $\geq 1/\sqrt{d-2}$, We find that the coefficient of any tower is $\geq1/\sqrt{d-2} $. The argument also trivially implies that the only towers that saturate this bound are string towers.
\vspace{5pt}

\noindent\textbf{The procedure}

The limit that takes the volume modulus to infinity while keeping other moduli fixed yields the largest coefficient of distance conjecture among theories which decompactify to a particular theory. This is because it avoids any unnecessary change of moduli to which the KK tower is insensitive. Let us try to lower the coefficient of distance conjecture from $\sqrt\frac{(D-2)}{(D-d)(d-2)}$ by gradually changing the direction of the limit in a region where the theory still decompactifies to the same theory. The smallest coefficient for the KK tower (the largest mixing of the volume modulus) must be achieved on the boundary of the region where the higher dimensional description holds. According to the emergent string conjecture, there are two possible scenarios for the new description at the boundary, we either decompactify to an even larger dimension, or we get a string tower. If we find a string limit, we know that the coefficient of the distance conjecture is $1/\sqrt{d-2}$. Therefore, we have found a strict lower bound for the coefficient of the KK towers inside the decompactification region. Otherwise, if we decompactify to an even larger internal geometry, we can repeat the algorithm to lower the coefficient of the distance conjecture. There are only two ways the process stops: we either find a string tower, which as we explained, provides a strict lower bound of $1/\sqrt{d-2}$ for the initial KK tower. Or we end up with a rigid decompactification limit, i.e. it only occurs in one limit. In that case, that limit must correspond to taking the volume modulus to infinity, and the corresponding coefficient must be $\sqrt\frac{(D-2)}{(D-d)(d-2)}$ for some $D$. Again, we find that the coefficient of the tower is strictly greater than $1/\sqrt{d-2}$.

The above argument shows that string towers satisfy $\alpha=1/\sqrt{d-2}$, while for KK towers $\alpha$ is strictly greater than $1/\sqrt{d-2}$. As we will explain in the next section, this sharpened bound is also motivated by holography \cite{Bedroya:2022tbh}.

Another way to interpret the distance conjecture is that for a given cut-off, the field range of any EFT is finite. Because if we move the vev of the moduli too much, there will be new light states below the cut-off that were previously integrated out. Since these new states are light enough to be excited, the EFT that assumes they are in their vacuum is no longer a good approximation. Thus, the effective field theory breaks down. 

Suppose the cutoff is $\Lambda$ and the scale of the tower is $m$. The distance conjecture tells us that if we traverse $\Delta\phi\sim|\ln(m/\Lambda)|$ in the moduli space in Planck units, the effective field theory will break down. 

The finiteness of the moduli space is not surprising from the EFT perspective, however the size of it is. In any effective field theory, we neglect some higher order irrelevant operators in the perturbative expansion of the action. For example, operators that are proportional to $(\phi/\Lambda)^n$. However, such operators will become significant at large values of $\phi$. But the difference between quantum gravity and non-gravitational field theory is that EFT puts a polynomial upper bound in $\Lambda$ on the field range rather than a logarithmic one. 

It is always nice to connect Swampland conditions to black holes since we know some of their universal features that must be true in any theory of quantum gravity. For distance conjecture, there is a heuristic connection to black holes \cite{Hamada:2021yxy}. Consider the field configuration that is $\phi=0$ at the origin and $\phi=\phi_0$ at some fixed radius. It turns out for large enough field ranges of $\mathcal{O}(1)$ in Planck units, this configuration collapses into a black hole. Therefore, the EFT breaks down where we want to probe some regions of field range with light modes.

Sometimes in the infinite distance limit it appears that we get a tower of light instantons instead of a tower of light states. In these cases it turns out the instantons correct the action in such a way that the formally infinite distance point becomes a finite distance point at the moduli space \cite{Marchesano:2019ifh}. In general, tower of light instantons drastically modify the geometry of the moduli space and their appearance signals that we are not working in the correct perturbative duality frame. 

For example, consider the type IIA on a Calabi--Yau threefold. If we take the radius modulus for the CY to zero, because of the $\propto(\frac{dr}{r})^2$ kinetic term,  we would naively think the field space distance goes to infinity. So both zero and infinite size are at infinite distance limit. However, at the zero size, the worldsheet instantons that wrap aroung CY give large corrections to the metric and make the infinite distance finite. This point is T-dual to conifold singularity in IIB which we discussed earlier. You can even continue past the 0 size and analytically continue to negative volumes until the space eventually comes to an end at the Landau-Ginzburg point with no massless modes and enhanced gauge symmetry. After analytic continuation, the moduli space complexifies to a complex plane with Landau-Ginzburg point at the 0 and the zero size CY at 1.

Another example of instantons making a formally infinite distance limit finite is M theory on quintic. The instantons corresponding to M5 branes wrapped around threefold make the formally infinite distance limit of zero size quintic finite.

\begin{statement7*}
Tower of light particles (0+1 d states) come with infinite distance and tower of light instantons erase infinite distance limits. 
\end{statement7*}

A very natural question to ask about Swampland condition is that how do they change under compactification? Suppose you have a tower of states at an infinite distance, would you still get the same tower at the corresponding limit in the lower dimensional theory? The answer is not necessarily! Because the $0+1$ dimensional objects of the higher dimensional tower can now wrap around the compact circle and become instantons that drastically correct the geometry of moduli space.

We saw how the tower of light instantons erase infinite distance limits. One might wonder, could it be that tower of weakly interacting light states create the infinte distance limits and that is why they are associated with each other? There have been several attempts to approach distance conjecture from this perspective. In particular, to assume that the moduli space is always morally compact and it can only become non-compact (have an infintie distance limit) due to corrections caused by light tower of states. In other words, the tower of states cause the infinite distance rather than the other way around. We will come back to this point and what morally compact means in the last section on finiteness conjecture. 

For example, in field theory if we integrate out a fermion, all the terms involving scalars that interact with that fermion receive corrections. Let us estimate this correction. Consider an interaction like $\bar\psi m(\phi)\psi$.

\begin{figure}[H]
    \centering

\tikzset{every picture/.style={line width=0.75pt}} 

\begin{tikzpicture}[x=0.75pt,y=0.75pt,yscale=-1,xscale=1]

\draw  [dash pattern={on 4.5pt off 4.5pt}]  (147,86) -- (277.5,87) ;
\draw   (277.5,87) .. controls (277.5,57.87) and (301.12,34.25) .. (330.25,34.25) .. controls (359.38,34.25) and (383,57.87) .. (383,87) .. controls (383,116.13) and (359.38,139.75) .. (330.25,139.75) .. controls (301.12,139.75) and (277.5,116.13) .. (277.5,87) -- cycle ;
\draw  [dash pattern={on 4.5pt off 4.5pt}]  (383,87) -- (513.5,88) ;
\draw    (286,101) .. controls (304.13,142.16) and (355.39,140.1) .. (374.38,102.36) ;
\draw [shift={(375.5,100)}, rotate = 114.23] [fill={rgb, 255:red, 0; green, 0; blue, 0 }  ][line width=0.08]  [draw opacity=0] (10.72,-5.15) -- (0,0) -- (10.72,5.15) -- (7.12,0) -- cycle    ;
\draw    (286.63,74.12) .. controls (305.76,28.14) and (360.76,31.88) .. (373.5,75) ;
\draw [shift={(285.5,77)}, rotate = 290.17] [fill={rgb, 255:red, 0; green, 0; blue, 0 }  ][line width=0.08]  [draw opacity=0] (10.72,-5.15) -- (0,0) -- (10.72,5.15) -- (7.12,0) -- cycle    ;
\draw  [fill={rgb, 255:red, 0; green, 0; blue, 0 }  ,fill opacity=1 ] (273.75,87) .. controls (273.75,84.93) and (275.43,83.25) .. (277.5,83.25) .. controls (279.57,83.25) and (281.25,84.93) .. (281.25,87) .. controls (281.25,89.07) and (279.57,90.75) .. (277.5,90.75) .. controls (275.43,90.75) and (273.75,89.07) .. (273.75,87) -- cycle ;
\draw  [fill={rgb, 255:red, 0; green, 0; blue, 0 }  ,fill opacity=1 ] (379.25,87.25) .. controls (379.25,85.18) and (380.93,83.5) .. (383,83.5) .. controls (385.07,83.5) and (386.75,85.18) .. (386.75,87.25) .. controls (386.75,89.32) and (385.07,91) .. (383,91) .. controls (380.93,91) and (379.25,89.32) .. (379.25,87.25) -- cycle ;

\draw (162,58.4) node [anchor=north west][inner sep=0.75pt]    {$\phi $};
\draw (493,60.4) node [anchor=north west][inner sep=0.75pt]    {$\phi $};
\draw (325,108.4) node [anchor=north west][inner sep=0.75pt]    {$\psi $};
\draw (324,45.4) node [anchor=north west][inner sep=0.75pt]    {$\psi $};
\draw (214.25,89.9) node [anchor=north west][inner sep=0.75pt]    {$q$};
\draw (450.25,90.9) node [anchor=north west][inner sep=0.75pt]    {$q$};
\draw (323.25,146.9) node [anchor=north west][inner sep=0.75pt]    {$p$};
\draw (310.25,11.4) node [anchor=north west][inner sep=0.75pt]    {$p+q$};
\draw (246,61.4) node [anchor=north west][inner sep=0.75pt]  [font=\footnotesize]  {$\partial _{\phi } m$};
\draw (390,64.4) node [anchor=north west][inner sep=0.75pt]  [font=\footnotesize]  {$\partial _{\phi } m$};

\end{tikzpicture}
    \caption{Correction to the propagator of $\phi$ from integrating out a missive spinor $\psi$ which interacts with $\phi$ via a term $\bar\psi m(\phi)\psi$.}
    \label{fig:my_label3}
\end{figure}
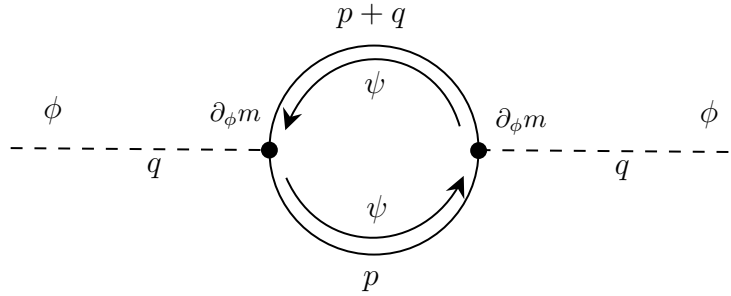

The correction to the fermion propagator comes from the following integral
\begin{align}
    \sim\int \frac{d^dp}{(\slashed p+m)^2}(\partial_\phi m)^2.
\end{align}
The correction to the metric takes the form of $\sum m_i^{d-4}(\partial_\phi m)^2$. Under certain assumptions, one can show that corrections of this type can create an infinite distance limit for the scalar field $\phi$. For example, if we assume the tower to be uniform with jumps of $\sim\Delta m(\phi)$ (e.g. KK tower), we end up with a metric of the form $ds^2\sim \frac{d(\Delta m)^2}{(\Delta m)^2}\sim (d\ln\Delta m)^2$ which indeed has an infinite distance limit. 

Another natural question to ask is how the two characteristics of the tower (coupling strength and mass) are related? Suppose the tower is a charged light tower. In that case, the WGC (if satisfied by the tower) tells us that $m\lesssim g$. Thus, it is not a coincidence that both the coupling and mass are going to zero; one explains the other, including the exponential dependence since $g\sim e^{-\beta\phi}$. However, the other piece of the distance conjecture still remains a mystery. Why weak coupling has something to do with infinite distance? Usually the coupling is exponential with some modulus. In other words, the gauge coupling, if promoted to a field, takes a kinetic form like $\sim (\partial g/g)^2$. A satisfying explanation for this behavior is still missing. Usually in the web of the Swampland conjectures everything fits nicely together, but there is always some missing piece that stops us from completely deriving one conjecture from another.

In the above argument about the connection between the distance conjecture and the WGC, we assumed that the tower is charged. However, not every tower is always charged. For example, the strong coupling limit of $E_8\times E_8$ is M theory on interval where the tower is a KK tower with no gauge charge. In this case there seems to be an approximate or Higgsed U(1).

The distance conjecture has an important cosmological implication. In general, the distance conjecture is in tension with conventional slow-roll inflation which you need a $5-10 M_p$ field range for the inflation. In that case, the distance conjecture tells us that that you get a tower of states that break down the EFT.

\subsection{AdS and CFT distance conjectures}

In the following, we use distance conjecture to motivate two other conjectures, one about AdS and the other about CFTs. In all the well-controlled AdS constructions in string theory, the spacetime takes the form $AdS_d\times S^p\times M^r$ where $M^r$ is some compact manifold. and the length scale of the extra pieces ($S^p$ and $M^r$) scale like the AdS scale. 
\begin{align}
    \Lambda_{AdS}\sim -\frac{1}{l_{S^p}^2}.
\end{align}

In holography there is a relation between the mass $m$ of the particle in the bulk and and dimension $\Delta$ of the corresponding operator on the boundary \begin{align}
    (ml_{AdS})\sim\Delta.
\end{align}

Suppose we could find a case where the radii of the extra dimension decouples from the AdS scale. In that case, we could find a huge gap in the dimesnions of the CFT operators. If there were to be no gaps, we must always have masses which go like $m\sim |\Lambda|^\frac{1}{2}\sim1/l_{Ads}$. This last equality, is observed in all the examples with well-controlled de Sitter constructions. But this sounds very much like the distance conjecture. In fact, in string theory, the cosmological constant is usually exponential in some modulus
\begin{align}\label{dsc1}
    \Lambda\sim e^{-c\phi},
\end{align}
and the distance conjecture predicts a tower of states with masses
\begin{align}\label{dsc2}
    m\sim e^{-c'\phi}.
\end{align}
More generally, we can define distance in the space of metric configurations $g_{\mu\nu}$ and one can see $\propto|\ln(\Lambda)|$ is a natural notion of of distance to be used instead of $\phi$. Another way of deriving the same result is to combine the two equations \eqref{dsc1} and \eqref{dsc2}. We find that in the limit $\Lambda\rightarrow 0$, there must be a tower of states with masses that are polynomial in $\Lambda$. This is called the \textit{AdS distance conjecture} \cite{Lust:2019zwm}. The stronger version of the conjecture (for SUSY case) fixes the exponent and claims that there is always a tower of states with masses that go like
\begin{align}
    m\sim |\Lambda|^\frac{1}{2}.
\end{align}

This conjecture in particular means that there is no pure AdS quantum gravity. For example, it rules out the holographic dual of pure $AdS_3$.

If we apply this to dS in our universe with small $\Lambda$, it tells us that in our universe we should expect a tower of states $m\sim |\Lambda|^\alpha$ with $\alpha\sim \mathcal{O}(1)$. In other words, we must have a hierarchy problem as a consequence of the cosmological constant problem. Moreover, they have to be weakly coupled which could be a candidate for the dark sector. This has recently lead to the dark dimension scenario with $\alpha=\frac{1}{4}$ \cite{Montero:2022prj,Gonzalo:2022jac}.
\vspace{10pt}

\noindent\textbf{Exercise 4: }For each massive particle in the standard model, assuming that it is in a tower of states satisfying $m\propto \Lambda^\alpha$, find the value of $\alpha$.
\vspace{10pt}

The AdS conjecture implies that we cannot have an arbitrarily large mass hierarchy in AdS space. If we have a large AdS, the first excited states will have a mass of order $\Lambda^{1/2}$ which in string theory compactifications correspond to the extra dimensions. In other words, the AdS space cannot be studied in isolation and it always acompanies the extra dimensions, because there cannot be a limit where the AdS scale goes down but the kk modes of the extra dimensions or any analagous excitations are gapped enough to not be considered. 

Let us try to come up with a counterexample for this statement. Consider an AdS construction with a sphere as a compact manifold extra dimension. Suppose we want to keep the AdS scale fixed while increase the masses of the KK modes. This would correspond to keeping the curvature of the sphere fixed while decreasing its diameter. This is because the mass of the KK modes correspond to the eigenvalues of the Laplacian which increase as the diameter of the space increases. One might think an easy way to do that would be to replace the sphere with some spherical orbifold where the sphere is moded out by a symmetry subgroup. The simplest example would be $\mathbb{Z}_n$ rotation group. However, if we mod out sphere by this group, the diameter does not change. 

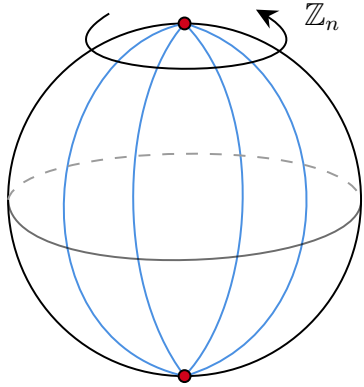
\begin{figure}[H]
    \centering

\tikzset{every picture/.style={line width=0.75pt}} 

\begin{tikzpicture}[x=0.75pt,y=0.75pt,yscale=-1,xscale=1]

\draw   (238,153) .. controls (238,103.85) and (277.85,64) .. (327,64) .. controls (376.15,64) and (416,103.85) .. (416,153) .. controls (416,202.15) and (376.15,242) .. (327,242) .. controls (277.85,242) and (238,202.15) .. (238,153) -- cycle ;
\draw [color={rgb, 255:red, 74; green, 144; blue, 226 }  ,draw opacity=1 ]   (327,242) .. controls (238,222) and (254,79) .. (327,64) ;
\draw [color={rgb, 255:red, 74; green, 144; blue, 226 }  ,draw opacity=1 ]   (327,242) .. controls (410,218) and (408,87) .. (327,64) ;
\draw [color={rgb, 255:red, 74; green, 144; blue, 226 }  ,draw opacity=1 ]   (327,242) .. controls (362,205) and (370,108) .. (327,64) ;
\draw [color={rgb, 255:red, 74; green, 144; blue, 226 }  ,draw opacity=1 ]   (327,242) .. controls (290,193) and (295,100) .. (327,64) ;
\draw [color={rgb, 255:red, 0; green, 0; blue, 0 }  ,draw opacity=0.58 ]   (238,153) .. controls (240,198) and (418,189) .. (416,153) ;
\draw [color={rgb, 255:red, 155; green, 155; blue, 155 }  ,draw opacity=1 ] [dash pattern={on 4.5pt off 4.5pt}]  (238,153) .. controls (245,124) and (408,120) .. (416,153) ;
\draw    (289,59) .. controls (227.62,96.62) and (430.87,96.02) .. (365.08,58.16) ;
\draw [shift={(363,57)}, rotate = 28.44] [fill={rgb, 255:red, 0; green, 0; blue, 0 }  ][line width=0.08]  [draw opacity=0] (10.72,-5.15) -- (0,0) -- (10.72,5.15) -- (7.12,0) -- cycle    ;
\draw  [fill={rgb, 255:red, 208; green, 2; blue, 27 }  ,fill opacity=1 ] (324,64) .. controls (324,62.34) and (325.34,61) .. (327,61) .. controls (328.66,61) and (330,62.34) .. (330,64) .. controls (330,65.66) and (328.66,67) .. (327,67) .. controls (325.34,67) and (324,65.66) .. (324,64) -- cycle ;
\draw  [fill={rgb, 255:red, 208; green, 2; blue, 27 }  ,fill opacity=1 ] (324,242) .. controls (324,240.34) and (325.34,239) .. (327,239) .. controls (328.66,239) and (330,240.34) .. (330,242) .. controls (330,243.66) and (328.66,245) .. (327,245) .. controls (325.34,245) and (324,243.66) .. (324,242) -- cycle ;

\draw (386,52.4) node [anchor=north west][inner sep=0.75pt]    {$\mathbb{Z}_{n}$};

\end{tikzpicture}
    \caption{Quotienting out a sphere by its $\mathbb{Z}_n$ subgroup will decrease its volume, but not its diameter. The distance between the north and south remains unchanged.}
    \label{fig:my_label4}
\end{figure}

But $\mathbb{Z}_n$ is not the only symmetry of hypersphere. So, the AdS conjecture is making a mathematical prediction: consider a sphere with radius $1$ and mod it out by an isometry subgroup $\Gamma$. There must be a minimum on the diamter of $S/\Gamma$. This is infact a true statement. For example, it was show that for the case of three sphere the minimum diameter is achieved by the icosahedral subgroup.

This has been shown to be true in an arbtirary Sasaki-Einstein manifolds coming from CY.

Now let us apply the AdS conjecture to holographically realized CFTs. Conformal field theories in dimensions less than five can have a moduli space which is called a conformal manifold. If the CFT realized holographically, their moduli space matches with the moduli space of massless scalars in the bulk.

Consider a symmetry generator in the CFT which has spin $J$ and dimension greater than $d-2+J$ (according to unitarity). Particularly, when $J=2$, you get the energy-momentum tensor which has dimension $d$. But how about higher spin generators? Can we have them and can they saturate the unitarity bound? These generators, if they exist, are called higher spin symmetries. If they exist in a CFT, they would include $J=4$ and infinitely many more. So far, these are all claims that can be shown from CFT data. 

Moreover, it can be shown that if such higher spin symmetries exist, there is a sector of the theory that is a free CFT. This points in the conformal manifold are higher symmetry (HS) points. We expect to find infinitely many weakly coupled currents as we get closer and closer to an HS point. But, this sounds very similar to the statement of distance conjecture! In fact, the only known examples of HS points are realized at the infinite distance in the conformal manifold. For example in $\mathcal{4}$ SYM, you take the $\tau\rightarrow 0$, the theory becomes free. 

\textit{The CFT distance conjecture} states that the higher symmetry points are always at inifnite distance limits in the conformal manifold and they are free \cite{Perlmutter:2020buo}. 

Now if we take the anomalous dimension $\gamma_4=-(J+d-2)$ how fast does it go to zero? It scales like 
\begin{align}
    [\text{diam}(\mathcal{M})]\sim\beta\ln_\epsilon \gamma_4
\end{align}

which is consistent with the distance conjecture. In the CFT context the conjecture says that theere is a tower of higher spin currents that comes down with at least an exponent of $1/4$.

Note that the CFT distance conjecture is more general than just an application of the AdS distance conjecture because some CFTs might not have a holographic dual.

\section{Swampland V: de Sitter conjectures}

So far, all the string theory examples we have studied were either Minkowski or Anti-de Sitter spacetimes. Similar to how Minkowski spacetime is the maximally symmetric spacetime with zero cosmological constant, de Sitter (dS) and Anti-de Sitter (AdS) spaces are respectively the maximally symmetric solutions with positive and negative cosmological constants. In this section, we move away from Minkowski and AdS, and study dS spacetimes. The symmetry algebra of the Minkowski, AdS, and dS are different which lead to very different properties among the three spaces. For example, the symmetry algebra of Minkowski and AdS can be extended to a supersymmetry algebra, however, as we will see, the same cannot be done for de Sitter. Therefore, any potential dS construction in string theory is non-supersymmetric. Given that there is no known stable non-supersymmetric examples in string theory, constructing de Sitter, if at all possible, is much more challenging than Minkowski or AdS in string theory. 

Studying de Sitter spaces is particularly important, because our universe seems to be approximately de Sitter at its current cosmological stage. This could be realized in different ways. The cosmological constant, which is the vacuum energy density, could be the value of the scalar field potential $V(\phi)$. The fact that this number seems to be almost constant means that the universe is:
\begin{itemize}
    \item stuck at a local minimum of $V(\phi)$, or
    \item $V(\phi)$ has a very small slope which makes it look almost constant, or
    \item significantly fine-tuned to be at top of the potential
\end{itemize}

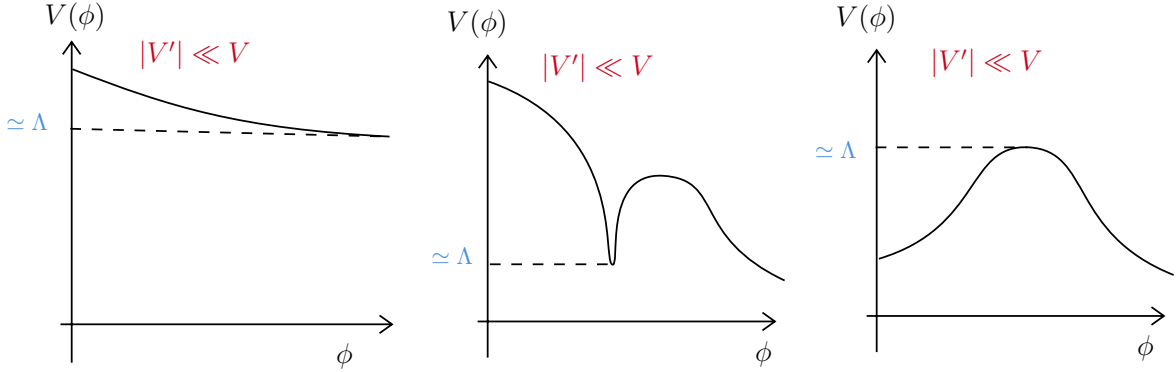
\begin{figure}[H]
    \centering
\tikzset{every picture/.style={line width=0.75pt}} 
\scalebox{.9}{
\begin{tikzpicture}[x=0.75pt,y=0.75pt,yscale=-1,xscale=1]
\draw    (38.93,56.17) .. controls (93.55,77.48) and (132.64,89.58) .. (216.62,93.97) ;
\draw  (32.22,199.24) -- (218,199.24)(38.49,40.86) -- (38.49,220.78) (211,194.24) -- (218,199.24) -- (211,204.24) (33.49,47.86) -- (38.49,40.86) -- (43.49,47.86)  ;
\draw  [dash pattern={on 4.5pt off 4.5pt}]  (37.71,89.58) -- (216.62,93.97) ;
\draw    (271.98,62.99) .. controls (349.07,90.3) and (335.76,161.45) .. (341.2,165.58) .. controls (346.65,169.72) and (334.87,113.9) .. (369.64,115.94) .. controls (404.4,117.98) and (385.37,149.86) .. (438,174.68) ;
\draw  (265.96,197.65) -- (432.58,197.65)(271.59,48.58) -- (271.59,217.92) (425.58,192.65) -- (432.58,197.65) -- (425.58,202.65) (266.59,55.58) -- (271.59,48.58) -- (276.59,55.58)  ;
\draw  [dash pattern={on 4.5pt off 4.5pt}]  (272.84,165.58) -- (341.2,165.58) ;
\draw    (490.84,162.49) .. controls (549,145) and (539.64,97.85) .. (575,100) .. controls (610.36,102.15) and (597,148) .. (656,171.59) ;
\draw  (483.96,194.56) -- (650.58,194.56)(489.59,45.49) -- (489.59,214.83) (643.58,189.56) -- (650.58,194.56) -- (643.58,199.56) (484.59,52.49) -- (489.59,45.49) -- (494.59,52.49)  ;
\draw  [dash pattern={on 4.5pt off 4.5pt}]  (489,100) -- (574,100) ;
\draw (74.58,38.73) node [anchor=north west][inner sep=0.75pt]  [color={rgb, 255:red, 208; green, 2; blue, 27 }  ,opacity=1 ]  {$| V'|\ll V$};
\draw (21.93,17.57) node [anchor=north west][inner sep=0.75pt]    {$V( \phi )$};
\draw (185.23,208.61) node [anchor=north west][inner sep=0.75pt]    {$\phi $};
\draw (0.38,79.52) node [anchor=north west][inner sep=0.75pt]  [font=\footnotesize,color={rgb, 255:red, 74; green, 144; blue, 226 }  ,opacity=1 ]  {$\simeq \Lambda $};
\draw (300.34,46.13) node [anchor=north west][inner sep=0.75pt]  [color={rgb, 255:red, 208; green, 2; blue, 27 }  ,opacity=1 ]  {$| V'|\ll V$};
\draw (247.28,21.49) node [anchor=north west][inner sep=0.75pt]    {$V( \phi )$};
\draw (419.97,206.96) node [anchor=north west][inner sep=0.75pt]    {$\phi $};
\draw (238.55,154.67) node [anchor=north west][inner sep=0.75pt]  [font=\footnotesize,color={rgb, 255:red, 74; green, 144; blue, 226 }  ,opacity=1 ]  {$\simeq \Lambda $};
\draw (518.34,43.04) node [anchor=north west][inner sep=0.75pt]  [color={rgb, 255:red, 208; green, 2; blue, 27 }  ,opacity=1 ]  {$| V'|\ll V$};
\draw (465.28,18.4) node [anchor=north west][inner sep=0.75pt]    {$V( \phi )$};
\draw (633.97,203.87) node [anchor=north west][inner sep=0.75pt]    {$\phi $};
\draw (453.55,95.58) node [anchor=north west][inner sep=0.75pt]  [font=\footnotesize,color={rgb, 255:red, 74; green, 144; blue, 226 }  ,opacity=1 ]  {$\simeq \Lambda $};
\end{tikzpicture}}
    \caption{Three possible explanation for the slow variation of the cosmological constant. }
\end{figure}

As we will discuss later, the scalar potential is believed to not have absolute positive minimum. Thus, if the first scenario happens, the potential will likely be lower somewhere else in the field space. When this happens, the scalar field can tunnel through the potential barrier to decrease the cosmological constant. Two examples of such processes are Coleman-Deluccia instantons and Hawking-Moss instantons \cite{Coleman:1980aw,Hawking:1981fz}. Hence, no matter which scenario happens, the value of cosmological constant is thought to eventually decrease. In fact, in all known examples in string theory, loss of supersymmetry comes with an instability even in Minkowski or AdS spaces. These observations suggest that the correct question to ask is: How stable or unstable de Sitter space can be? 

We can understand de Sitter space as a hypersphere embedded in a higher dimensional flat space. Consider the solution to the following equation:
\begin{align}
    -X_0^2\pm X_1^1+X_2^2+...+X_d^2=\pm R^2,
\end{align}
where the metric in the ambient space is $ds^2=-dX_0^2\pm dX_1^2+\sum_{i>1}dX_i^2$. The plus sign corresponds to de Sitter space while the minus sign corresponds to Anti-de Sitter space. The symmetries of the two spaces are respectively $SO(1,d)$ and $SO(2,d-1)$. One of the main differentiating features of dS is that it cannot support a global notion of positive conserved energy. This is also where the tension with supersymmetry comes in. Supersymmetry algebra tells us there must be a positive definite bosonic symmetry operator that can be written in terms of square of supersymmetry generators. To be more precise, suppose $Q$ is one of the supersymmetry generators, $\{Q,Q^\dagger\}$ is a positive definite bosonic symmetry generator. Moreover, $\{Q,Q^\dagger\}$ transforms as a tensor product of two spin $1/2$ representations under rotation, therefore, there must be a vector representation in the decompositition of $\{Q,Q^\dagger\}$.
\begin{align}
    \{Q,Q^\dagger\}=\sigma_\mu P^\mu+...,
\end{align}
where $P^\mu$ is the density flow of a positive-semidefinite conserved quantity $H$.
\begin{align}
    H=\int_\Sigma \star P,
\end{align}
where $\Sigma$ is a space-like Cauchy surface. Therefore, if de Sitter geometry cannot support a global positive deifinite energy, it cannot support supersymemtry. 

If such an $H$ exists, it would generate an isometry in de Sitter space. However, all the symmetries in de Sitter are generated by Killing vectors of the type $X^i\frac{\partial}{\partial X^0}+X^0\frac{\partial}{\partial X^i}$ or $X^i\frac{\partial}{\partial X^j}-X^j\frac{\partial}{\partial X^i}$, where $i,j\neq 0$. The Noether's conserved current has a term $\Delta\mathcal{L}$ which captures the deformation of Lagrangian under the symmetry transformation. Due to the $X^i$ prefactor in Killing vector fields, the term $\Delta\mathcal{L}$ will linearly depend on the coordinate. Note that this does not happen in flat space because the Killing vector $\partial_t$ does not have such a prefactor. Therefore, the corresponding conserved current is of the form $(\partial_t)^2/2+\vec\nabla^2/2+V$ with no $X^i$ prefactor which allows it to be manifestly positive semidefinite. However, in de Sitter, the sign of the current cannot be semidefinite and will depend on the sign of some combination of coordinates. For example, for the conserved current under $X^i\frac{\partial}{\partial X^0}+X^0\frac{\partial}{\partial X^i}$, the corresponding conserved current can have a different sign depending on the sign of $X^i$. \vspace{10pt}

\noindent \textbf{Exercise 1:} Show that AdS avoids this problem. 
\vspace{10pt}

In $\mathcal{N}=1$ supergravities, the scalar potential looks like \begin{align}
    V=e^K(|\mathcal{D}W|^2-3|W|^2).
\end{align}
Supersymmetry requires $\mathcal{D}W=0$ but it does not require $W=0$, which is why it can give us non-positive cosmological constants. 

\subsection{(Non)-supersymmetry and (in)stability}

The fact that we need to study de Sitter or quasi de Sitter backgrounds means we need to study non-supersymmetric backgrounds. Let us go back to the Minkowski case and search for non-supersymmetric Minkowski background. How can we break the supersymmetry? 
\vspace{10pt}

\noindent\textbf{Scherk-Schwarz mechanism}
\vspace{5pt}

One way to break the supersymmetry completely is to impose anti-periodic boundary condition on fermions in a circle. This boundary condition breaks all supersymmetry, however, at the same time, for small radius, it creates a Tachyon from winding string which is the familiar stringy Tachyon in the NS sector. This is the Scherk-Schwarz mechanism for breaking the supersymmetry \cite{Scherk:1978ta}.

One might ask what is the ultimate fate of this universe? First, we can calculate the effective potential for the scalar field corresponding to the radius. In supersymmetric theories, the one-loop amplitudes vanish. However, now that supersymmetry is broken, the one-loop contribution to the effective potential is no longer zero and therefore the potential gets corrected. After going to the Einstein frame, we get a potential like $V\sim -R^{-n_d}$ where $n_d$ is a positive constant that only depends on the spacetime dimension $d$ (e.g. $n_4=6$ \cite{vonGersdorff:2003rq}). Hence, the quantum corrections generate a potential that shrinks the size of the circle. But this is not all. The winding string has a scalar mode that at small enough radius $R$ becomes Tachyonic. Suppose we call this field $\phi_w$, we see that our perturbative potential does not have a minimum in the $R,\phi_w$ plane which means the theory is unstable.

Another way of seeing this instability is through non-pertubative processes. Witten showed that in a theory with antiperiodic frmions, there is a finite action bounce solution that creates a bubble of nothing which expands and eats the universe \cite{Witten:1981gj}. 
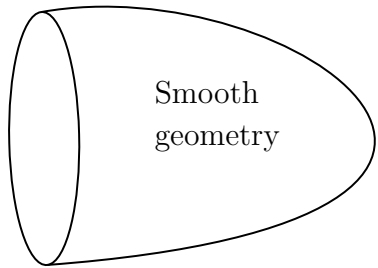
\begin{figure}[H]
    \centering

\tikzset{every picture/.style={line width=0.75pt}} 

\begin{tikzpicture}[x=0.75pt,y=0.75pt,yscale=-1,xscale=1]

\draw   (260.14,139.75) .. controls (259.25,104.51) and (266.42,74.93) .. (276.14,73.69) .. controls (285.86,72.44) and (294.47,100.01) .. (295.36,135.25) .. controls (296.25,170.49) and (289.08,200.07) .. (279.36,201.31) .. controls (269.64,202.56) and (261.03,174.99) .. (260.14,139.75) -- cycle ;
\draw    (276.14,73.69) .. controls (404.64,50.69) and (578.5,179) .. (279.36,201.31) ;


\draw (332,107) node [anchor=north west][inner sep=0.75pt]   [align=left] {Smooth\\geometry\\};

\end{tikzpicture}
    \caption{By capping the geometry in a smooth way, we allow the spacetime to end on a domain wall. This is Witten's bubble of nothing.}
\end{figure}

Note that if you put periodic boundary condition, the spin structure would be nontrivial and shrinking the circle and capping the geometry would be impossible.

Thanks to the earlier perturbative analysis we knew that the solution is unstable and thanks to the non-perturbative analysis, one can see what the fate of the instability is; the space disappears! 

Let us consider other examples of non-supersymmetric string theories.

\vspace{10pt}

\noindent\textbf{Type 0 theories}
\vspace{5pt}

In construction of type II string theories, we apply specific GSO projections to the spectrum of closed superstring to find a consistent theory that is modular invariant, satisfies level matching, and has a mutually local and closed OPE. It turns out that there are two more restrictions of the full spectrum other than type IIA and type IIB that satisfy all these consistency conditions. These theories are called 0A and 0B theories. The spectrum of these theories is made up of NS-NS and R-R sectors (no mixed sectors). Both theories have NS-NS states with the equal right moving and left moving worldsheet fermion numbers projection. However, type 0A has the R-R states with $(-)^F=-(-)^{\tilde F}$ while the type 0B theory has states with $(-)^F=(-)^{\tilde F}$. It is easy to see that type 0B theory is chiral while type 0A theory has a parity invariant spectrum. These theories have different spectrums than type II threories. Since there are no NS-R or R-NS sectors, this theory has no spacetime fermions. Thus, it does not have any supersymmetry. At the same time, it has the NS-NS tachyon. This is yet another example of how the lack of supersymmetry is accompanied with an instability. Our next examples are non-supersymmetric orbifolds. 

\vspace{10pt}

\noindent\textbf{Non-SUSY orbifolds}
\vspace{5pt}

We can consider orbifolds $T^d/G$ with tori with periodic boundary conditions for fermions (no Scherk-Schwarz mechanism) with a symmetry $G$ that does not preserve any supersymmetry. As an example, we will study compactifications on $T^2/\mathbb{Z}_3$. Consider a torus $\mathds{C}/\Gamma$ such that the lattice $\Gamma$ is generated by shifting $z$ by numbers $r$ and $\omega\cdot r$ where $r\in\mathbb{R}$ and $\omega$ is the third root of unity.

The lattice $\Gamma$ is invariant under multiplication by $\omega$ which makes this actions a $\mathbb{Z}_3$ symmetry of the torus. This action has three fixed points $\{0,\exp(\pi i/6)r/\sqrt{3},ir/\sqrt{3}\}$.

\begin{figure}[H]
    \centering

\tikzset{every picture/.style={line width=0.75pt}} 

\begin{tikzpicture}[x=0.75pt,y=0.75pt,yscale=-1,xscale=1]

\draw    (215.08,91.83) -- (284.75,212.5) ;
\draw    (284.75,212.5) -- (423.5,212.5) ;
\draw    (423.17,213.06) -- (354.16,92.02) ;
\draw    (354.16,92.02) -- (215.08,91.83) ;

\draw  (184.5,212) -- (530.75,212)(284.5,25.5) -- (284.5,237) (523.75,207) -- (530.75,212) -- (523.75,217) (279.5,32.5) -- (284.5,25.5) -- (289.5,32.5)  ;
\draw  [fill={rgb, 255:red, 208; green, 2; blue, 27 }  ,fill opacity=1 ] (358.5,171.79) .. controls (358.5,170.25) and (359.75,169) .. (361.29,169) .. controls (362.83,169) and (364.08,170.25) .. (364.08,171.79) .. controls (364.08,173.33) and (362.83,174.58) .. (361.29,174.58) .. controls (359.75,174.58) and (358.5,173.33) .. (358.5,171.79) -- cycle ;
\draw  [fill={rgb, 255:red, 208; green, 2; blue, 27 }  ,fill opacity=1 ] (281.5,123.79) .. controls (281.5,122.25) and (282.75,121) .. (284.29,121) .. controls (285.83,121) and (287.08,122.25) .. (287.08,123.79) .. controls (287.08,125.33) and (285.83,126.58) .. (284.29,126.58) .. controls (282.75,126.58) and (281.5,125.33) .. (281.5,123.79) -- cycle ;
\draw  [fill={rgb, 255:red, 208; green, 2; blue, 27 }  ,fill opacity=1 ] (281.71,211.71) .. controls (281.71,210.17) and (282.96,208.92) .. (284.5,208.92) .. controls (286.04,208.92) and (287.29,210.17) .. (287.29,211.71) .. controls (287.29,213.25) and (286.04,214.5) .. (284.5,214.5) .. controls (282.96,214.5) and (281.71,213.25) .. (281.71,211.71) -- cycle ;

\draw (419,263.4) node [anchor=north west][inner sep=0.75pt]    {$1$};
\draw (190,67.4) node [anchor=north west][inner sep=0.75pt]    {$\tau =\omega $};

\end{tikzpicture}
    \caption{The red points are the fixed point under the $\mathbb{Z}_3$ orbifold action}
\end{figure}
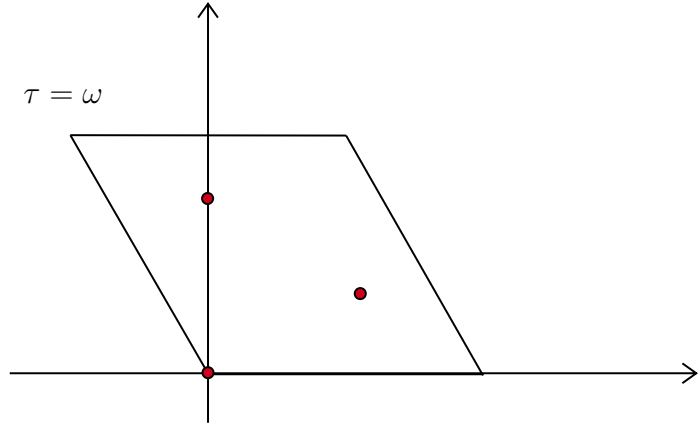

To study the spectrum of this theory, it is easiest to work in the Green-Schwarz formalism where gauge is fixed to lightcone and the spacetime supersymemtry is manifest. Let us quickly review the Green-Schwarz formalism. For bosnonic string, we can go to lightcone gauge $X^+=2\alpha'p^+\tau$ and write the action in terms of the transverse coordinates $X^i~(1\leq i\leq d-2)$ as
\begin{align}
    S=-\frac{1}{4\pi\alpha'}\int d\sigma d\tau \eta^{\alpha\beta}\sum_i\partial_\alpha X^i\partial_\beta X^i,
\end{align}
where $\eta$ is 2d Minkowski metric. The advantage of lightcone gauge is that all the remaining oscillators are physical. Note in the lightcone gauge, the worldsheet fields furnish representations of the little group of $d$ dimensional massless particles which contains an $SO(d-2)$ subgroup corresponding to the rotations in the transverse directions. For the bosonic worldsheet fields, the remaining coordinates $X^i$ furnish a vector representation of $SO(d-2)$. We can generalize the lightcone gauge to superstrings. In that case, we should incorporate some spacetime fermionic "coordinates" which will give rise to worldsheet oscilating modes. This spacetime fermions are nothing other than the supercharges. This is why in the Green-Schwarz formalism, the spacetime supersymmetry is more manifest. For type II theories where we have two spacetime supersymmetries, the superstring action in the lightcone gauge is \cite{Green:1981yb,Green:1983wt}
\begin{align}
        S=\int d\sigma d\tau \frac{-1}{4\pi\alpha'}\sum_i\partial_\alpha X^i\partial^\alpha X^i+\frac{i}{4\pi} \bar S\gamma^-\rho^\alpha\partial_\alpha S,
\end{align}
where $S^{A\alpha}$ is both a worldsheet fermion and a spacetime fermion. The index $A$ is a 2-valued worldsheet index while the index $\alpha$ is a 32-valued spacetime index. $\gamma^\mu$ are 32-dimensional spacetime Dirac matrices and $\rho^\alpha$ are two-dimensional worldsheet Dirac matrices (Pauli matrices). Lastly, $\bar S$ is defined as
\begin{align}
    \bar S_{B\beta}=S^\dagger_{A\alpha}(\gamma^0)^\alpha_\beta(\rho^0)^A_B.
\end{align}
Note that although $S^{1\alpha}$ and $S^{2\alpha}$ are 32-dimensional, they do not live in a 32-dimensional space. In fact, they only have 8 degrees of freedom. This is because each one is a 10d Majorana--Weyl spinor which is a 16-dimensional real representation. Moreover, after going to the lightcone gauge, they must satisfy the lightcone condition
\begin{align}
    (\gamma^+)S^A=0.
\end{align}
These reduce the number of degrees of freedom in $S^1$ and $S^2$ to eight. In fact, we can think of each one of them as the minimal irreducible spinor representation of the $SO(8)$ subgroup of the little group. The action can be then rewritten in terms of 8d spinors $S^{Aa}$ where $1\leq a\leq 8$ and $S^1$ and $S^2$ transform in the 8d spinor representation of $\text{Spin}(8)$. Each one of these components, just like the $X^i$, gives rise to raising and lowering operators $S^a_n$ that live in $\bf 8_s$ (or $\bf 8_c$ depending on parity) of $\text{Spin}(8)$ for any fixed $n$. Therefore, we have two sets of raising and lowering operators that build up the spectrum:
\begin{itemize}
    \item Bosonic modes $\alpha^\mu_{-n}$ and $\tilde\alpha^{\mu}_{-n}$ living in $\bf 8_v$ representation of $\text{Spin}(8)$.
    \item Fermoinic modes $S^{1\alpha}_{-n}$ and $S^{2\alpha}$ living in $\bf 8_s$ or $\bf 8_c$ representation of $\text{Spin}(8)$.
\end{itemize}
Note that in the Green-Schwarz formalism, as opposed to the RNS formalism, worldsheet fermions map to spacetime fermions. 

Now let us study the weight lattice of each one of the two representations. Cartan of $SO(8)$ (or $\text{Spin}(8)$) is four dimensional. We can take the four generators to be rotations in the four planes, $\{(X^1,X^2),(X^3,X^4),(X^5,X^6),(X^7,X^8)\}$. The corresponding eigenvectors of the representations under the elements of Cartan are:
\begin{align}
   &\textbf{8}_v:~(\pm 1,0,0,0),(0,\pm1,0,0,0),(0,0,\pm 1,0),(0,0,0,\pm1)\nonumber\\
    &\textbf{8}_c:~\{(s_1,s_2,s_3,s_4)|\forall i\in\{1,2,3,4\}: s_i=\pm \frac{1}{2}~\&~\Pi_{i}s_i=+1\}    \nonumber\\
        &\textbf{8}_s:~\{(s_1,s_2,s_3,s_4)|\forall i\in\{1,2,3,4\}: s_i=\pm \frac{1}{2}~\&~\Pi_{i}s_i=-1\}.   
\end{align}

Now let us go back to our $T^2/\mathbb{Z}_3$ orbifold. The action of $\mathbb{Z}_3$ is an order-3 rotation in the $(X^1,X^2)$ plane. Given that all the spinors have half-integer weight under this rotation, none of them will be mapped to themselves under a $\theta=4\pi/3$ rotation. Note that the spinors $S^a$ are the spacetime supersymmetry charges, therefore, there are no fermionic zero modes in the untwisted sector, and no supersymmetry will be preserved under the $\mathbb{Z}_3$ rotation. 

Now let us look at the twisted sector. The winding modes in the twisted sectors are typically massive, however, near the fixed points, they can shrink to zero size and give us a light spectrum. Therefore, we expect three copies of a light spectrum coming from the degrees of freedom of the twisted sectors
localizing around the fixed points. Let us start with the $(X^1,X^2)=(0,0)$ fixed point. We consider the perturbations around the string solution where $(X^1(\sigma,\tau),X^2(\sigma\tau))=(0,0)$ and for $i>2$, $X^i(\sigma,\tau)$ is constant. These perturbation are subject to the twisted boundary condition $\partial_\sigma[X^1+iX^2](\sigma,\tau)=\omega[X^1+iX^2]\partial_\sigma(\sigma+2\pi,\tau)$. These boundary conditions shift the frequencies $n$ from integers by $\pm1/3$. Similarly, the twisted boundary conditions affect the fermionic oscillators $S^a$ with suitable $(-)^F$ action on $S$ to match the orbifold action order 3. However, given that all of them are affected by the $\mathbb{Z}_3$. The frequencies of all of them are shifted by $\pm 1/3$. Shifting the frequencies has a two-fold impact: it changes the 2d casimir energy and changes the mass of the first excitations. Therefore, the previosuly massless excitations, get gapped. Usually, light states are expected to dominate the contribution to the casimir energy. In this case, there are six remaining bosonic light states while all the other 2d single particle states are gapped. Since bosonic degrees of freedom lower the casimir energy while fermionic degrees of freedom raise it, we expect the final result to be negative which means the theory has a spacetime tachyon. Now let us do a more precise calculation. The theory, before compactifying on an orbifold, was supersymmetric and tachyon-free. Therefore, the casimir energy was zero. Given that two of the bosonic modes ($X^1$ and $X^2$) have been gaped exactly like the fermionic modes (in terms of the shift in the frequency), their effect cancel that of two of the fermionic modes. Therefore, all that is left to calculate is the shift in the casimir energy from six fermionic modes. If the modes numbers are shifted by something which is equal to $0<\eta<1$ modulo 1, the casimir energy is shifted by each mode by $(-1)^F\frac{1}{4}\eta(1-\eta)$ where $F$ is the worldsheet fermion number. Therefore, the total shift is equal to 
\begin{align}
    6\cdot(-1)\cdot[\frac{1}{4}\frac{1}{3}(\frac{2}{3})]=-\frac{1}{3}<0.
\end{align}
The casimir energy of the 2d theory translates to the mass squared of the single particle state with lowest $m^2$ in spacetime. Therefore, this theory has tachyons in spacetime. 

\noindent \textbf{Exercise 2:} Consider orbifolds with a singularity of the type $\mathds{C}^2/\mathbb{Z}_n$ where the $\mathbb{Z}_n$ acts by multiplication by some powers of n-th root of unity on each $\mathds{C}$. Show that if the $\mathbb{Z}_n$ preserves supersymmetry, there is no Tachyon but if it does not, there is always a Tachyon. 

Now let us give some good news! There is a 10d theory \cite{Alvarez-Gaume:1986ghj,Dixon:1986iz} where there is no tachyon, no supersymmetry, and even though it is chiral, it does not have gauge, gravitational, or mixed anomalies thanks to the Green-Schwarz mechanism. 

\vspace{10pt}

\noindent\textbf{$O(16)\times O(16)$ Heterotic string}
\vspace{5pt}

Since this theory is a modification of the Heterotic construction, let us first review the $E_8\times E_8$ Heterotic string theory. The Heterotic string theory can be described in two ways. One is in the Green-Schwarz formalism \cite{Green:1981yb,Green:1983wt} where there is 16 lef-moving bosonic coordinates, 10 ordinary bosonic coordinates (left and right pair), and 16 fermionic coordinates $S^a$ furnishing the 10d right-handed Majorana--Weyl representation. The $S^a$ are nothing other than 10d supersymmetry generators. The 16 left-handed bosnonic coordinates have a 16d Euclidean Narain lattice $\Gamma^{16}$. There are only two possibilities; the root lattice of $SO(32)$ and the root lattice of $E_8\times E_8$. Now take the  $E_8\times E_8$ theory. The compact bosons have a fermionic description as well. To see that we should go to RNS formalism. Although the supersymmetric Heterotic strings are easier to construct in Green-Schwarz formalism, the non-supersymmetric versions are easier to work with in the RNS formalism. In the $E_8\times E_8$ theory in the light cone, we have three sets of 16 worldhseet Majorana--Weyl fermions. Two sets are left movers (which equivalently describe the compact bosons) and one set is the usual right-moving sector. Each of these fermions can have NS or R boundary condition. We allow all 8 posiibilities. However, we do a GSO projection to only keep the states that have even number of worlsheet fermion of each set. Moreover, we only keep the states that respect the level-matching condition. 
\begin{align}\label{E8E8S}
\{NS+,R+\}_L\times\{NS+,R+\}_L\times \{NS+,R+\}_R/\text{level-mathcing}.
\end{align}
There is a left-mover tachyon which is thrown out by the level matching condition. We can think of the above spectrum as a $\mathbb{Z}_2\times\mathbb{Z}_2$ orbifold of a theory with same boundary conditions for all fermions. Before orbifolding, this theory is the heterotic analogue of the type 0 theory, and similar to that is tachyonic. Let us call this theory the type 0 Heterotic theory. The spectrum of type 0 Heterotic theory is 
\begin{align}
    (NS~\mathcal{F}_{L1},NS~\mathcal{F}_{L2},NS~\mathcal{F}_{r})\cup (R~\mathcal{F}_{L1},R~\mathcal{F}_{L2},R~\mathcal{F}_{R}),
\end{align}
where $\{\mathcal{F}_{L1},\mathcal{F}_{L1},\mathcal{F}_{R}\}$ are the worldsheet fermion numbers and 
\begin{align}
    \mathcal{F}_{L1}+\mathcal{F}_{L2}+\mathcal{F}_{R}\equiv 2\mod 2.
\end{align} 
The $E_8\times E_8$ Heterotic theory can be viewed as a $\mathbb{Z}_2\times\mathbb{Z}_2$ orbifold of the type 0 theory. The first $\mathbb{Z}_2$ multiplies the first left-moving fermion by $(-1)^\mathcal{F}_{L1}$ and the second $\mathbb{Z}_2$ does the same for the second left-moving fermion. The untwisted sector of this orbifold is just $(NS+,NS+,NS+)\cup (R+,R+,R+)$. However, in the twisted sector, the boundary conditions can mix between NS and R. This leads to the spectrum \eqref{E8E8S} of the $E_8\times E_8$ theory. This theory is modular invariant and consistent. However, there is a $\mathbb{Z}_2$ ambiguity known as discrete torsion. This is to say, we can sum over the twisted sectors with specific weights that can be thought of as $\exp(i\int B)$. Suppose $\alpha$ is the generator of the first $\mathbb{Z}_2$ and $\beta$ is the generator of the second $\mathbb{Z}_2$.  We want to change the $\beta$ such that it acts non-trivially on $\alpha$, however we should make sure we can assign appropriate signs to the one-loops twisted sectors to maintain modular invariance. Suppose $\epsilon(p,q)$ where $p,q\in\mathbb{Z}_2\times\mathbb{Z}_2$ is the sign of the one-loop twisted sector where there is a $p$ twist in the $\sigma$ direction and a $q$ twist in the $\tau$ direction. In the $E_8\times E_8$ theory, we take $\epsilon=+1$ for all sectors. However, there is another modular invariant choice which has $\epsilon(\alpha,\beta)=-1$ while $\epsilon(1,1)=+1$. This choice gives a new non-Tachyonic orbifold theory \cite{Alvarez-Gaume:1986ghj,Dixon:1986iz}. If one works out the $\epsilon(p,q)$ and does the level-mathcing carefully, one sees that this theory has the following massless 10d spectrum.
\begin{align}
    &\text{Bosonic: }&&g_{\mu\nu}, B_{\mu\nu}, A_\mu, \phi\nonumber\\
    &\text{Fermionic: }&& \Psi_+:(\textbf{16},\textbf{16}),~\Psi_-:(\textbf{128},\textbf{1})\oplus(\textbf{1},\textbf{128}),
\end{align}
where the numbers show the representation of Majorana--Weyl fermions under a $O(16)\times O(16)$. The subscript represent the chirality of the fermions. The massless vector fields $A_\mu$ are in the adjoint of $O(16)\times O(16)$. Therefore, there is a $O(16)\times O(16)$ gauge symmetry in the theory. Note that this theory is chiral because the left and right handed fermions are in different representations of the gauge group. Chiral theories often have gravitational, gauge, or mixed anomalies. However, this theory, although chiral, has the same number of left and right handed 10d Majorana--Weyl spinors (256 of each). Therefore, there is no gravitational anomalies. As for gauge and mixed anomalies, they are canceled via the Green-Schwarz mechanism. This theory is a non-supersymmetric and non-tachyonic theory which is perfectly fine at tree-level. Whenever we give up supersymemtry, the magical cancelation of one-loop vacuum amplitude might not occur. This would coorect the effective potential. After computing the one-loop vacuum amplitude, it turns out the cosmological constant is positive! In fact, we could expect this from the fact that there are more bosons than fermions. 
\vspace{10pt}

\noindent \textbf{Exercise 3:} Count the number of massless degrees of freedom in Heterotic $O(16)\times O(16)$ and show there are more fermionic degrees of freedom than bosonic. 
\vspace{10pt} 

Even though the vacuum energy is positive, this solution is not the usual de Sitter. The reason for this discrepancy is that de Sitter is a solution to an effective with a positive constant when the action is written in the Einstein frame where the coefficient of $\mathcal{R}$ is constant. However, the string calculation give us the effective action in string frame where the Ricci scalar has a dilaton dependent prefactor. We must redefine the metric to go from one frame to another, and when we do that, we see that the previosly constant term in the action, becomes dilaton dependent. In other words, what we have found is not a cosmological constant, but a positive potential for dilaton. However, the potential turns out to be exponential in dilaton.
\vspace{10pt}

\noindent \textbf{Exercise 4:} Suppose the $O(16)\times O(16)$ theory has a cosmological constant in the string frame. Go to the Einstein frame and show that the potential has exponential dependence in dilaton $V=e^{-\alpha\phi}$. Compute $\alpha$.
\vspace{10pt}

Another way of seeing the exponential behavior of the potential is that the normal scale for the vacuum energy in string theory is $M_s^D$ which depends on the dilaton field in the Einstein frame via
\begin{align}
    M_s^{D}\sim M_P^Dg_s^\frac{2D}{D-2}\sim M_P^De^{-\frac{4D}{D-2}\Phi}.
\end{align}
However, note that $\Phi$ is not canonically normalized in the Einstein frame yet which is why the exponent is slightly different. The $g_s$ dependence signifies that this effective potential is generated by quantum corrections from from string excitations.

To summarize, the $O(16)\times O(16)$ theory does not live in de Sitter space since the effective potential has no minimum. Therefore, the effective value of cosmological constant will keep descreasing.

Note that when $V(\phi)$ goes like $e^{-\alpha\phi}$, $|V'|/V\simeq \alpha$ is an $\mathcal{O}(1)$ constant (in Planck units). 

If we put this theory on a circle it turns out to be connected to Heterotic on a circle with anti-periodic boundary condition with a suitable Wilson lines around the circle. As we know from the Scherk-Scwarz mechanism, that theory is tachyonic. Therefore, we cannot use the $O(16)\times O(16)$ theory to find similar lower dimensional theories.

Note that even though the $O(16)\times O(16)$ theory averted tachyonic instability, it has a runaway instability. Other proposals for non-supersymmetric tachyon-free perturbative string backgrounds also have runaway instabilities \cite{Sagnotti:1996qj,Sugimoto:1999tx}. The instability, in one form or another, seems to be a universal feature of non-supersymmetric theories. There is no existing proposal for a non-supersymmetric permanently stable vacuum. The question is, how stable can a non-supersymmetric vacuum be? For example, in the Tachyonic examples this is captured by the the imaginary mass of the tachyon.  In the examples we mentioned, the mass of the tachyon ($im_\phi$), which is captured by second derivative of scalar potential, is at least of the same order as the vacuum energy produced by quantum corrections. In other words, $|V''/V|$ is at least order one in Planck units. The reason is that in all the examples, the quantum corrected potential goes like $V\sim M_s^D$ where $D$ is the spacetime dimension and $|V''|= m_{Tachyon}^2/2\sim M_s^2 $. Therefore, we have
\begin{align}
    \frac{|V''|}{V}\sim M_s^{-(D-2)}\gtrsim M_P^{2-D},
\end{align}
where in the last line we used $D\geq 2$ and $M_s<M_P$.

To summarize, in all the examples we reviewed, a positive cosmological constant either comes with a runaway instability ($|V'|\gtrsim V$) or an unstable equilibrium ($V''\lesssim V$) where the inequalities are written in Planck units. Both of these inequalities suggest that de Sitter are not too stable. We will try to understand and explain these case-based observations in the following sections.

\subsection{de Sitter and tree-level string theory}

In this section we review a general argument from \cite{Obied:2018sgi} that shows the observation from the last section were not just coincedences and they are in fact true at tree-level weakly coupling limit of M-theory compactification. There are similar arguments that apply to the same or other corners of the string theory landscape \cite{Dine:1985he,Maldacena:2000mw,Hertzberg:2007wc,Steinhardt:2008nk,Wrase:2010ew,Andriot:2019wrs}. 

Consider M-theory and compactify it on an arbitrary manifold $\mathcal{M}$ with a non-vanishing $G$-flux. This setup is the most generic M-theory construction. The lower dimensional theory technically has a scalar potential with infinitely many scalars corresponding to infinitely many possible deformation of the internal manifold and its fluxes. One might be tempted to think that surely this infinite dimensional space has some critical point which is a local minimum of the tree-level potential. However, it turns out as long as we can trust the classical supergravity description (curvature and fluxes are sub-Planckian), there can be no critical points. In fact, there is a stronger constraint than $\nabla V\neq 0$. We show there is a universal order one lower bound on $|\nabla V|/V$ which prevents $V$ from becoming too flat. 

When we compactify the higher dimensional theory on $\mathbb{M}$, the integral of the higher dimensional action over $\mathcal{M}$ show up as an effective potential in the lower dimensional theory. In the reduced Planck units, the potential reads
\begin{align}\label{PMNNG}
    V\simeq-\int_{\mathcal{M}} \sqrt{g}(\mathcal{R}-\frac{1}{2}|G|^2).
\end{align}
Suppose we write the metric as a direct sum of a Minkowski metric with a warped internal geometry.
\begin{align}
ds^2=dx^\mu dx^\nu\eta_{\mu\nu}+e^{2\rho}ds_I^2,
\end{align}
where $\rho$ only depends on $x$ and is a scalar field in the lower dimensional theory which controls the overall size of the internal manifold. We will show that $\rho$ is an unstable mode in the sense that the potential monotonically decreases as $\rho$ increases. The monotonicity of the potential makes $\rho$ continually descrease without ever stopping at a stable value. 

Note that both terms in \eqref{PMNNG} depend exponentially on $\rho$. For non-constant $\rho(x)$, the 11-dimensional action $\int_{\mathcal{M}\times\mathds{R}^{d}}\sqrt{G} \mathcal{R}$ gives a kinetic term for $\rho$ which is not canonically normalized (it is not of the form $1/2(\partial_\mu\rho)^2$). To find the canonically normalized field, we should first rewrite the action in the Einstein frame in which the lower dimensional Ricci scalar has a constant coeffitient. The change of frame will change the coefficient of the kinetic term of $\rho$. Finally, we normalize the field $\rho$ such that the kinetic term is $1/2(\partial_\mu\hat\rho)^2$. After some careful calculation, one can see that the canonically normalized scalar is $\hat\rho=\rho\sqrt{9(11-d)/(d-2)}$. 

\noindent \textbf{Exercise 5:} Show that in M theory compactification to d dimensions, the potential is proportional to $V_R e^{-\lambda_1\hat\rho}+V_G e^{-\lambda_2\hat\rho}$ where 
\begin{align}
    \lambda_1=\frac{6}{\sqrt{(d-2)(11-d)}},\quad
    \lambda_2=\frac{2(d+1)}{\sqrt{(d-2)(11-d)}}
\end{align}
and $\hat \rho$ is the canonically normalized volume modulus. 
\vspace{10pt}

Note that $V_G$ is always positive since the contribution of the flux is always positive. However, the contribution of $V_R$ could be negative or positive. 

Now let us compute the slope in the $\hat\rho$ direction in the regions where $V$ is positive. If both contributions to the potential are positive.
\begin{align}
    \frac{|V'|}{V}=\frac{\lambda_1 e^{-\lambda_1\hat\rho}V_R+\lambda_2 e^{-\lambda_2\hat\rho}V_G}{ e^{-\lambda_1\hat\rho}V_R+ e^{-\lambda_2\hat\rho}V_G}\geq \frac{\lambda_1 e^{-\lambda_1\hat\rho}V_R+\lambda_1 e^{-\lambda_2\hat\rho}V_G}{ e^{-\lambda_1\hat\rho}V_R+ e^{-\lambda_2\hat\rho}V_G}\geq \lambda_1,
\end{align}
where we used $\lambda_2>\lambda_1$ for $d>2$. Now let us consider the case where $V_R$ is negative. Assuming $V$ is positive, we have
\begin{align}
    \frac{|V'|}{V}=\frac{-\lambda_1 e^{-\lambda_1\hat\rho}|V_R|+\lambda_2 e^{-\lambda_2\hat\rho}V_G}{ -e^{-\lambda_1\hat\rho}|V_R|+ e^{-\lambda_2\hat\rho}V_G}\geq \frac{-\lambda_2 e^{-\lambda_1\hat\rho}|V_R|+\lambda_2 e^{-\lambda_2\hat\rho}V_G}{ -e^{-\lambda_1\hat\rho}|V_R|+ e^{-\lambda_2\hat\rho}V_G}\geq \lambda_2.
\end{align}
Again, we used $\lambda_2>\lambda_1$. Note that in both cases, as long as $V$ is posotive, we find a lower bound on $|V'|/V$ which is either $\lambda_1$ or $\lambda_2$. Either of these numbers are universal $\mathcal{O}(1)$ constants that only depend on the dimension of spacetime. Is this a proof that $|V'|/V\gtrsim \mathcal{O}(1)$? Not quite. This proof only takes the tree-level action into account and neglects the quantum effects. However, given that the classical piece is unable to produce a de Sitter, it is safe to say our only hope to get a de Sitter space in string theory is via quantum effects.

Note that there might be directions in which the potential has a local minimum but this shows that there is always a direction in which the potential is monotonically decreasing. In other words, the direction of steepest descent is never too flat. 
\begin{align}
    \frac{|\nabla V|}{V}>\lambda_1.
\end{align}

\subsection{de Sitter conjectures}

In the previous sections, we considered non-supersymmetric examples in string theory and observed that they always come with an instability which is either in the form of a local maximum (tachyons) or a rolling potential. In both cases, the potential was never too flat by which we mean either $|V''|\gtrsim V$ or $|V'|\gtrsim V$. Then we reviewed a general argument that shows the inequality of the type $|V'|\gtrsim V$ holds in any well-controlled regime of M-theory moduli space where quantum corrections are small. In this section, we formulate these observations into concrete conjectures, and we study their consequences\footnote{For other refinements of the de sitter conjecture motivated by tachyonic de Sitter solutions \cite{Andriot:2021rdy}, see \cite{Andriot:2018wzk,Andriot:2018mav}.}. 
\begin{statement8*}
The effective scalar potential satisfies one of the following two inequalities at every point in the field space \cite{Obied:2018sgi,Ooguri:2018wrx,Garg:2018reu}:
\begin{align}
|\nabla V|\geq c_1{V}~~~~~\text{or}~~~~~\min_{i,j}(\nabla_i\nabla_j V)\leq -c_2V,
\end{align}
where $\min_{i,j}(\nabla_i\nabla_j V)$ is the minimum eigenvalue of the Hessian in an orthonormal basis and $c_1$ and $c_2$ are $\mathcal{O}(1)$ constants in Planck units. This orthonormal basis, as well as the size of the gradient vector $|\nabla V|$, are defined with respect to the canonical metric on field space ($g_{ij}$) which is determined by the kinetic terms of the scalar fields

\begin{align}
    \mathcal{L}_{Kinetic}=-\frac{1}{2}g_{ij}\partial_\mu\phi^i\partial^\mu\phi^j.
\end{align}
\end{statement8*}
\vspace{10pt}

\text{A few quick remarks:}
\begin{itemize}
    \item The conjecture is trivially satisfied for $V\leq 0$. 
    \item The de Sitter conjecture forbids a positive valued, local minimum for scalar potential. Because at such a point, we will have $V>0, |\nabla V|=0$, and $\min_{i,j}(\nabla_i\nabla_j V)>0$. These signs will violate both of the inequalities in the de Sitter conjecture.
    \item The two conditions are very similiar in spirit. They both say that any solution with a positive cosmological constant is sufficiently unstable. If the first condition is satisfied, the instability is given by a steep rolling direction in the field space. And if the second condition is satisfied, we have a a steep tachyonic direction. 
    \item Even though the motivations for the conjecture were completely unrelated to inflation, the statement of the conjecture is almost equivalent to saying that inflation cannot happen. For positive potentials, we can rewrite the two conditions in the following form.
    \begin{align}
        \epsilon=\frac{1}{2}(\frac{|\nabla V|^2}{V})>\mathcal{O}(1)~~~~~\text{or}~~~~~\eta=\frac{-\min_{i,j}(\nabla_i\nabla_j V)}{V}>\mathcal{O}(1).
    \end{align}
    The two parameters $\epsilon$ and $\eta$ are called slow-roll parameters in inflationary cosmology. In many conventional inflationary models their value is required to be small to avoid an initial condition fine-tuning problem. If the values of both parameters is small, there is an attractor solution called the slow-roll trajectory which removes the fine tuning problem. However, de Sitter conjecture seems to be directly against such inflationary models. However, the constant numbers $c_1$ and $c_2$ are not explicitly known. One could speculate that the universal values are $\sim 0.01$ due to some numerical factors. However, it is fair to say that conventional inflationary models are at least in tension with the de Sitter conjecture. 
    \item  The no-go theorems for M-theory \cite{Obied:2018sgi,Maldacena:2000mw} and type II theories \cite{Hertzberg:2007wc,Andriot:2019wrs,Andriot:2020lea} tell us that at least there is a very strong evidence for this conjecture in the classical regime. But the classical regimes are nothing more than the infinite distance limits in the moduli space. In fact, in those limits, the first inequality in the dS conjecture seems to suffice and $V''/V\lesssim -1$ is not needed. Although the de Sitter conjecture is believed to be true in the infinite distance limits of the moduli space, we do not have much evidence for its validity in the interior of the moduli space, due to strong coupling. 
\end{itemize}

The de Sitter conjecture is telling us that potential must decay exponentially in the asymptotic of the field space. It is reasonable to assume de Sitter conjecture is related to the distance conjecture which also concerns infinite distance limits. Consider an infinite distance limit. The distance conjecture tells us that there is a tower of light states with exponentially decaying masses $m\sim \exp(-\alpha \phi)$. Normally, we would expect the same physics that leads to the mass of the tower states to contribute to the vacuum energy which has a simension of $M^d$. Thus, it is natural to expect that $V\sim m^d\sim \exp(-\alpha d\phi)$.However, we know some examples in flux compactifications where the contribution to the potential comes from the flux terms $|F|^2\sim g^2$ where $g$ is the gauge coupling. We can use Weak Gravity Conjecture to estimate $g$ with $m$ and conclude that the tower particles may contribute to the potential as $m^2$ instead of $m^d$. These arguments connect the undetermined universal constants in de Sitter conjecture and distance conjecture to each other. If we had a sharper rationale for de Sitter conjecture, we could not only fix the constants in the de Sitter conjecture, but even fix the constant for the distance conjecture.

In the following we review another Swampland conjecture which can serve as a potential rationale underlying the de Sitter conjecture.

Consider a homogeneous and isotropic expanding d-dimensional universe. Such a spacetime has a $E(d-1)$ symmetry group. Such a spacetime is called an FRW solution and the metric can be written as $ds^2=-dt^2+a(t)^2dx^2$ where $H=\frac{\dot a}{a}$ is called the Hubble parameter and $a(t)$ is called the scale factor. This solution is of significant phenomenological interest given that our universe seems to be isotropic and homogeneous to a good degree. 

The dynamics of $a(t)$ is determined by the content of the theory and the initial conditions. For example, for a universe with a cosmological constant $\Lambda$, we have $\frac{(d-1)(d-2)}{2}H^2\propto \Lambda$. The constant $H\sim \sqrt\Lambda$ corresponds to de Sitter space. As one can see from the metric, positive value of $H$ corresponds to an expanding universe and if $H$ is constant, the expansion is exponential ($a(t)\propto \exp(Ht)$). This fast expansion gives multiple physical meanings to the natural length scale in this universe, which is $1/H$ and is called the Hubble horizon. If $H$ remains constant, any two points with spacelike seperation $|\Delta x|>\frac{1}{H}$ will be out of each other's light cone. Therefore, there is a horizon at $r=\frac{1}{H}$ from the perspective of the observer who is moving at $r=0$. But there is a second feature which in some ways is very unique to de Sitter space. All the fields in a de Sitter background have some quantum fluctuations, just as they would have in any background. However, due to the exponential expansion in de Sitter space, these quantum fluctuations exponentially expand almost at the same rate as the scale factor. At some point the fluctuations exist the Hubble horizon and the expansion overcomes becomes so fast that even the peaks and troughs of the perturbation exit each other's lightcone. Therefore, the mode does not have enough time to react to the expansion and it effectively "freezes". This means the amplitudes of the fluctuations change but their profile does not. Modes with frequencies $k\gg H$ oscillate but the ones with $k\ll H$ freeze. In addition to the freezing, these quantum flcuctuations classicalize upon their exit from the Hubble horizon. This means, the Wigner distribution of the physical observables becomes such that the quantum fluctuations can be well-estimated with a classical ensemble. Intrestingly, these fluctuations remain classical even if at some point the expansion stops and they re-enter the Hubble horizon. 

\begin{figure}[H]
    \centering

\tikzset{every picture/.style={line width=0.75pt}} 

\begin{tikzpicture}[x=0.75pt,y=0.75pt,yscale=-1,xscale=1]

\draw  [color={rgb, 255:red, 74; green, 144; blue, 226 }  ,draw opacity=1 ][line width=1.5]  (209.19,75.97) .. controls (215.11,87.22) and (220.77,97.94) .. (227.34,97.94) .. controls (233.91,97.94) and (239.57,87.22) .. (245.49,75.97) .. controls (251.41,64.71) and (257.07,54) .. (263.63,54) .. controls (270.2,54) and (275.86,64.71) .. (281.78,75.97) .. controls (287.7,87.22) and (293.36,97.94) .. (299.93,97.94) .. controls (306.5,97.94) and (312.16,87.22) .. (318.08,75.97) .. controls (324,64.71) and (329.66,54) .. (336.23,54) .. controls (342.8,54) and (348.46,64.71) .. (354.38,75.97) .. controls (360.3,87.22) and (365.96,97.94) .. (372.53,97.94) .. controls (379.09,97.94) and (384.76,87.22) .. (390.68,75.97) .. controls (396.59,64.71) and (402.26,54) .. (408.82,54) .. controls (415.39,54) and (421.05,64.71) .. (426.97,75.97) .. controls (432.89,87.22) and (438.55,97.94) .. (445.12,97.94) .. controls (451.69,97.94) and (457.35,87.22) .. (463.27,75.97) .. controls (464.24,74.12) and (465.21,72.28) .. (466.17,70.51) ;
\draw  [color={rgb, 255:red, 208; green, 2; blue, 27 }  ,draw opacity=1 ][line width=1.5]  (269.5,222.43) .. controls (271.87,235.56) and (274.13,248.06) .. (276.76,248.06) .. controls (279.39,248.06) and (281.65,235.56) .. (284.02,222.43) .. controls (286.39,209.3) and (288.65,196.8) .. (291.28,196.8) .. controls (293.91,196.8) and (296.17,209.3) .. (298.54,222.43) .. controls (300.91,235.56) and (303.17,248.06) .. (305.8,248.06) .. controls (308.42,248.06) and (310.69,235.56) .. (313.06,222.43) .. controls (315.42,209.3) and (317.69,196.8) .. (320.32,196.8) .. controls (322.94,196.8) and (325.21,209.3) .. (327.58,222.43) .. controls (329.94,235.56) and (332.21,248.06) .. (334.84,248.06) .. controls (337.46,248.06) and (339.73,235.56) .. (342.09,222.43) .. controls (344.46,209.3) and (346.73,196.8) .. (349.35,196.8) .. controls (351.98,196.8) and (354.25,209.3) .. (356.61,222.43) .. controls (358.98,235.56) and (361.25,248.06) .. (363.87,248.06) .. controls (366.5,248.06) and (368.76,235.56) .. (371.13,222.43) .. controls (373.5,209.3) and (375.76,196.8) .. (378.39,196.8) .. controls (381.02,196.8) and (383.28,209.3) .. (385.65,222.43) .. controls (386.97,229.74) and (388.26,236.86) .. (389.59,241.74) ;
\draw    (397.93,237.07) .. controls (400.84,162.01) and (428.42,121.74) .. (486.5,72.31) ;
\draw    (265.81,240.73) .. controls (268.72,165.67) and (241.13,121.74) .. (188.86,74.14) ;
\draw    (158.5,242) -- (156.52,14) ;
\draw [shift={(156.5,12)}, rotate = 89.5] [color={rgb, 255:red, 0; green, 0; blue, 0 }  ][line width=0.75]    (10.93,-3.29) .. controls (6.95,-1.4) and (3.31,-0.3) .. (0,0) .. controls (3.31,0.3) and (6.95,1.4) .. (10.93,3.29)   ;
\draw    (264,262.99) -- (401.5,262.01) ;
\draw [shift={(403.5,262)}, rotate = 179.6] [color={rgb, 255:red, 0; green, 0; blue, 0 }  ][line width=0.75]    (10.93,-3.29) .. controls (6.95,-1.4) and (3.31,-0.3) .. (0,0) .. controls (3.31,0.3) and (6.95,1.4) .. (10.93,3.29)   ;
\draw [shift={(262,263)}, rotate = 359.6] [color={rgb, 255:red, 0; green, 0; blue, 0 }  ][line width=0.75]    (10.93,-3.29) .. controls (6.95,-1.4) and (3.31,-0.3) .. (0,0) .. controls (3.31,0.3) and (6.95,1.4) .. (10.93,3.29)   ;

\draw (133,141) node [anchor=north west][inner sep=0.75pt]  [rotate=-270] [align=left] {time};
\draw (249,273) node [anchor=north west][inner sep=0.75pt]   [align=left] {Wavelength of fluctuations};

\end{tikzpicture}
    \caption{As trans-Planckian modes exit the Hubble horizon, they freeze out and classicalize.}
\end{figure}
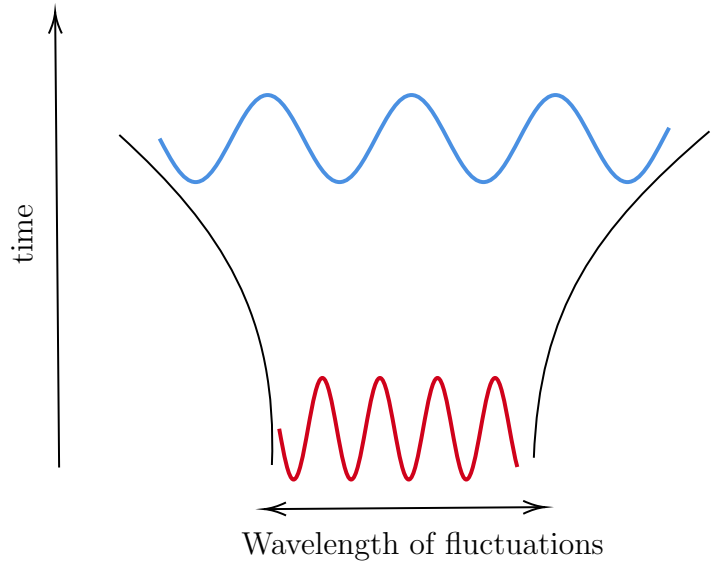

The freezing of the quantum fluctuation is a proposed mechanism to produce the fluctuations in CMB. However, there is something bizarre about classical imprint of trans-Planckian physics when they exit the Hubble horizon. If the expansion is sufficiently long, the quantum fluctuations that eventually exit the Hubble horizon and classicalize can be as small as Planck-sized. The Trans-Planckian Censorship Conjecture (TCC) states that this should not be possible. 

\begin{statement9*}
In a consistent theory of quantum gravity, a long-lasting expansionary solution in which Planck-sized fluctuations exist the Hubble horizon does not exist \cite{Bedroya:2019snp}. 
\begin{align}
    \frac{a_f}{a_i}l_p<\frac{1}{H_f}.
\end{align}
\end{statement9*}

Let us say a few quick remarks about TCC.

\begin{itemize}
    \item The immediate consequence of TCC is that the expansion at a constant Hubble parameter cannot last longer than $\tau_{TCC}\sim\frac{1}{H}\ln(\frac{M_P}{H})$. More concretely, TCC implies that regardless of the dynamics of $H$, a significant change to the rate expansion must happen before $\tau_{TCC}$ (See \cite{Bedroya:2020rmd} for an overview of possibilities).
    \item The age of our universe is only a few orders of magnitude smaller than $\tau_{TCC}$. Thus, our universe marginally passes the test all thanks to the $\log$ term in $\tau_{TCC}$. In a universe with a cosmological constant, you can only measure the Hubble parameter using experiments that have a Hubble size scale. For example, this could correspond to measurements of the light coming from a Hubble time in the past. But according to TCC, as soon as the universe is old enough to measure the Hubble parameter, it is approaching the end of an era.
    \item If we have a scalar field with a monotonically decreasing potential, it cannot be too flat over very long field ranges. The equation of motion for the scalar field gives
    \begin{align}
    \frac{(d-1)(d-2)}{d}H^2&=\frac{1}{2}\dot\phi^2+V(\phi)\nonumber\\
    \ddot\phi+(d-1)H\dot \phi+V'&=0.
\end{align}
From this we find
\begin{align}
    \frac{H}{\dot\phi}>\frac{1}{\sqrt{(d-1)(d-2)}}.
\end{align}
Now we can use TCC and change the integration variable to find 
\begin{align}
    \int \frac{H}{\dot\phi} d\phi<-\ln(H_f).
\end{align}
By combining the two equations, we find
\begin{align}
    H_f\lesssim e^{-\Delta\phi/\sqrt{(d-1)(d-2)}}
\end{align}
Since $V\lesssim H^2$, we find 
\begin{align}
    V\lesssim e^{-2\Delta\phi/\sqrt{(d-1)(d-2)}}
\end{align}
In the exponential case, we find 
\begin{align}\label{TCCdSc}
    |\frac{V'}{V}|\geq \frac{2}{\sqrt{(d-1)(d-2)}}.
\end{align}
The above inequality is for trajectories that are driven by an exponential potentials in addition to some extra positive contribution to the Hubble energy for example from a tower of states. If we ignore the extra contribution of the tower and consider the trajectories that are driven purely by the exponential potential, the TCC leads to 
\begin{align}\label{sTCCc}
        |\frac{V'}{V}|\geq \frac{2}{\sqrt{(d-2)}},
\end{align}
which is a stronger constraint\footnote{This condition was also suggested based on prohibiting eternal accelerated expansion. For cosmologies that are driven by exponential potentials, TCC is satisfied if and only if the expansion is decelerated \cite{Bedroya:2019snp}. This follows trivially from $\frac{a_f}{a_i}<\frac{M_P}{H_f}$ since the right side is linear in time.}. In \cite{Rudelius:2022gbz}, it was argued that the emergent string conjecture implies that the mass scale of any tower of light states is always higher than the Hubble scale. In that case, the states of the tower are too massive to get excited and contribute to the cosmological evolution, and the TCC is equivalent to the above condition on the asymptotic behavior of the potential \cite{Bedroya:2019snp}. This looks like the de Sitter conjecture, but this time with an explicit coefficient. Therefore, TCC gives a strong justification for de Sitter conjecture in the asymptotics of the moduli space. In fact, there is no known counterexample to this inequality in the known string theory constructions (see \cite{Bedroya:2019snp,Andriot:2020lea,Andriot:2022xjh} for tests of TCC in string theory and \cite{Mishra:2022fic} for Karch-Randall setup). This fixes the coefficients of both the de Sitter conjecture and distance conjecture through its relation to the de Sitter conjecture. Note that, TCC is formulated based on the expansionary trajectory rather than any instantanious configuration. Therefore, the consequences of TCC are usually constraints on the shape of the potential over long field ranges rather than pointwise implications. This seperates TCC from the de Sitter conjecture. For example, TCC does not imply $|V'|/V\gtrsim \mathcal{O}(1)$ at every point in the moduli space. This makes TCC a more relaxed constraint for late time cosmology. However, TCC is very restrictive for early universe cosmology \cite{Bedroya:2019tba}. The conventional inflationary models are either inconsistent with TCC or highly fine-tuned too explain all the observational data. 
\item It is also worth mentioning that as opposed to the de Sitter conjecture, TCC allows meta-stable de Sitter vacua. However, it requires their lifetime to be smaller than $\tau_{TCC}$. Consider an expansionary trajectory that is sourced by a cascade of tunnellings between metastable vacua. 

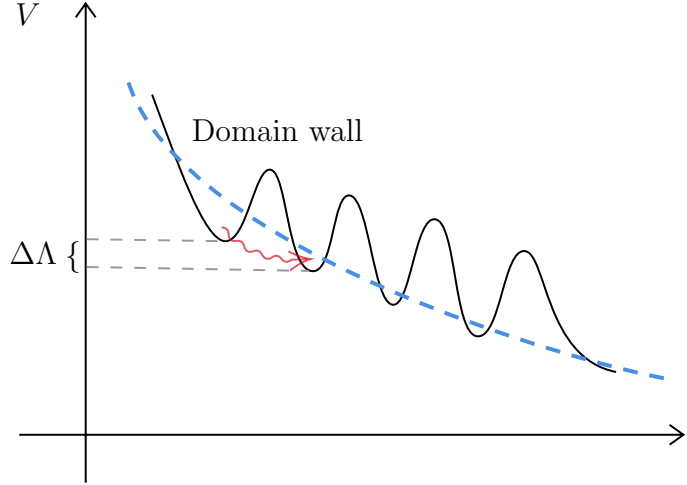
\begin{figure}[H]
    \centering

\tikzset{every picture/.style={line width=0.75pt}} 

\begin{tikzpicture}[x=0.75pt,y=0.75pt,yscale=-1,xscale=1]

\draw    (237.43,75.11) .. controls (249.43,107.11) and (263.43,149.11) .. (274.43,149.11) .. controls (285.43,149.11) and (289.39,110.65) .. (297.43,113.11) .. controls (305.47,115.58) and (305.43,161.11) .. (317.43,164.11) .. controls (329.43,167.11) and (328.08,123.68) .. (337.43,126.11) .. controls (346.77,128.55) and (348.92,178.87) .. (358.43,181.11) .. controls (367.94,183.36) and (369.84,135.95) .. (380.43,138.11) .. controls (391.02,140.28) and (389.43,196.11) .. (401.43,197.11) .. controls (413.43,198.11) and (414.4,152.84) .. (425.43,154.11) .. controls (436.46,155.38) and (435.43,208.11) .. (471.43,215.11) ;
\draw  (170.43,246.81) -- (505.43,246.81)(203.93,29.11) -- (203.93,271) (498.43,241.81) -- (505.43,246.81) -- (498.43,251.81) (198.93,36.11) -- (203.93,29.11) -- (208.93,36.11)  ;
\draw [color={rgb, 255:red, 208; green, 2; blue, 27 }  ,draw opacity=0.66 ]   (272.43,142.11) .. controls (274.81,142.21) and (276.01,143.44) .. (276.04,145.8) .. controls (276.21,148.2) and (277.44,149.26) .. (279.73,148.98) .. controls (282.04,148.62) and (283.43,149.59) .. (283.9,151.89) .. controls (284.61,154.22) and (286.08,155) .. (288.3,154.23) .. controls (290.41,153.32) and (291.96,153.92) .. (292.94,156.03) .. controls (294.19,158.13) and (295.81,158.56) .. (297.82,157.31) .. controls (299.74,155.96) and (301.36,156.22) .. (302.67,158.07) .. controls (304.19,159.88) and (305.86,160) .. (307.69,158.42) -- (307.98,158.43) -- (315.72,158.25) ;
\draw [shift={(317.43,158.11)}, rotate = 175.03] [color={rgb, 255:red, 208; green, 2; blue, 27 }  ,draw opacity=0.66 ][line width=0.75]    (10.93,-4.9) .. controls (6.95,-2.3) and (3.31,-0.67) .. (0,0) .. controls (3.31,0.67) and (6.95,2.3) .. (10.93,4.9)   ;
\draw [color={rgb, 255:red, 155; green, 155; blue, 155 }  ,draw opacity=1 ] [dash pattern={on 4.5pt off 4.5pt}]  (204.43,148.11) -- (274.43,149.11) ;
\draw [color={rgb, 255:red, 155; green, 155; blue, 155 }  ,draw opacity=1 ] [dash pattern={on 4.5pt off 4.5pt}]  (204.43,162.11) -- (316.43,164.11) ;
\draw [color={rgb, 255:red, 74; green, 144; blue, 226 }  ,draw opacity=1 ][line width=1.5]  [dash pattern={on 5.63pt off 4.5pt}]  (225.5,69) .. controls (253.5,154) and (451.5,211) .. (499.5,219) ;

\draw (167,27.4) node [anchor=north west][inner sep=0.75pt]    {$V$};
\draw (256,86) node [anchor=north west][inner sep=0.75pt]   [align=left] {Domain wall};
\draw (192,147.4) node [anchor=north west][inner sep=0.75pt]    {$\{$};
\draw (164,148.4) node [anchor=north west][inner sep=0.75pt]    {$\Delta \Lambda $};

\end{tikzpicture}
    \caption{The blue curve is the effective potential which effectively captures the dynamics of the universe as a result of a cascade of tunnellings between nearby vacua.}
\end{figure}

Suppose we can effectively describe the above expansion using an effective monotonic potential $V_{eff}$. It turns out that applying TCC for the individual tunnelings implies that $|V'_{eff}|>V^{3/2}$ in Planck units \cite{Bedroya:2020rac}. Interestingly, this is exactly the inequality that must be violated to get eternal inflation \cite{Rudelius:2019cfh}. Thus, TCC suggests that there cannot be an eternal inflation in the interior of the moduli space. Moreover, if the potential is generated by a top-form gauge potential, TCC implies the higher-dimensional generalization of Weak Gravity Conjecture for the domain wall between neighboring vacua \cite{Bedroya:2020rac}. 

\item Let us go back to the connection to the distance conjecture. As we explained before, we expect the potential to be either $\sim m^2$ or $\sim m^d$ in the asymptotic limits of field space where $m$ is the mass scale of the lightest tower \cite{Bedroya:2019snp,Andriot:2020lea,Bedroya:2020rmd}. If we want to be on the conservative side, we can apply $m^d\sim V$ to \eqref{TCCdSc} which leads to the following bound on the coefficient of the distance conjecture is
\begin{align}
    \alpha\geq \frac{2}{d\sqrt{d-2}}.
\end{align}
and on the stronger side, based on $V\sim m^2$ TCC suggests that 
\begin{align}\label{TCCSDC}
    \alpha\geq \frac{1}{\sqrt{d-2}}.
\end{align}
Remarkable, the above inequality coincides with the sharpened version of the distance conjecture \cite{Etheredge:2022opl}. We can also motivate this relation via the Higuchi bound \cite{Higuchi:1986py}, which states that particles with spin $s\geq 2$ are heavier than the Hubble scale. If we apply this bound to the particles of the lightest tower, we find $m\gtrsim H\sim\sqrt{V}$. Assuming the sharpened distance conjecture $m\lesssim\exp(-\kappa\Delta\phi/\sqrt{d-2})$ this leads to $V\lesssim\exp(-\kappa\cdot2\Delta\phi/\sqrt{d-2})$ which agrees with \eqref{sTCCc}. Note that this is only a heuristic derivation because we applied the Higuchi bound to rolling backgrounds which are not de Sitter spaces. 

\item There is an interesting connection between TCC and the holographic principle \cite{Bedroya:2022tbh}. To understand the holographic argument for TCC, let us review the holographic principle. In quantum gravity, the notion of spacetime is expected to be emergent. A nice example of this emergence is T-duality, where depending on the size of the compact dimension, the spacetime that provides the best semiclassical description can change\footnote{Note that, in T-dual descriptions, a local wavepacket in the compact manifold in one picture maps to a winding state in the other picture. Therefore, there is no direct mapping between the points in two spacetimes.}. If the notion of spacetime is emergent in quantum gravity, true physical observables cannot rely on a definition of spacetime. However, this raises the question that then what is the meaning of an effective field theory in quantum gravity, given that it is fundamentally a theory of local observables, e.g. fields. At its most basic form, the holographic principle is the statement that physical observables in quantum gravity are defined on the boundary of spacetime, and the right effective field theory is the one that best produces such boundary observables. For example, in Minkowski spacetime these boundary observables are scattering amplitudes, and in AdS space, they are boundary correlators. 

Now, one can apply the holographic principle to expanding  universes with polynomial expansion ($a(t)\sim t^p$). These backgrounds are ubiquitous in string theory given that exponential potentials lead to polynomial expansions. In \cite{Bedroya:2022tbh}, it was shown that an effective field theory in such backgrounds could produce non-trivial boundary observables if and only if $p\leq 1$, which is equivalent to TCC\footnote{A similar argument is presented in \cite{Rudelius:2021azq} which makes an extra assumption about the physical observables and draws stronger conclusions. The argument in  \cite{Rudelius:2021azq} assumes that the physical observables are accessible to a bulk observer. For example, an eternal de Sitter space can have dS/CFT boundary observables \cite{Strominger:2001pn}, but those observables will not be fully measurable by any bulk observer. Therefore, eternal de Sitter, although not ruled out by the argument we reviewed, does not meet the extra criterion assumed in \cite{Rudelius:2021azq}.}. In other words, the holographic principle can be satisfied if and only if TCC is satisfied. 

The holographic principle also has non-trivial consequences for the mass of the weakly coupled particles. If the mass of such a particle does not satisfy $m\lesssim t^{1-2p}$, its correlation functions will freeze out to a delta function at future infinity \cite{Bedroya:2022tbh}. Therefore, such a field does not yield any non-trivial boundary data and violates the holographic principle. The condition $m\lesssim t^{1-2p}$ can be expressed as
\begin{align}
m\lesssim e^{-\alpha(\lambda)\phi};~~~~~\alpha(\lambda)=\frac{4}{(d-2)\lambda}-\frac{\lambda}{2},
\end{align}
 where $V\sim\exp(-\lambda\phi)$ is driving the expansion. For $\lambda<\sqrt{8/(d-2)}$, $\alpha(\lambda)$ is positive and the above result implies the distance conjecture. Moreover, the strongest bound is realized for $\lambda=2/\sqrt{d-2}$ which is
 \begin{align}
 \alpha_{\max}=\frac{1}{\sqrt{d-2}}.
 \end{align}
The above coefficient matches with the heuristic bound \eqref{TCCSDC} which is also the coefficient of the sharpened distance conjecture \cite{Etheredge:2022opl}.
 
\end{itemize}

TCC provides a partial explanation for the de Sitter and distance conjectures by phrasing them in terms of a physical process. TCC is well-supported in the asymptotic region of the field space based on 1) non-trivial consistency in string theory, 2) connection to holography \cite{Bedroya:2022tbh}, and 3) consistency with Weak Gravity Conjecture among other Swampland conjectures \cite{Bedroya:2020rac}. However, there is less evidence for de Sitter conjecture in the interior of the moduli space which makes them less rigorous. Among the Swampland conjectures, the ones that are less rigorous are typically more phenomenologically powerful. But even though some conjectures are less supported than others, all the conjectures seem to fit together nicely. The main challenge with making de Sitter conjectures more rigorous is that there is no supersymmetry in de Sitter and we lose analytic control. Given that the potentials that lead to non-zero cosmological constants usually diverge ($+\infty$) at some limit and vanish at some other limit. Based on the minimum number of inflection points required, we can see that to get a de Sitter vacuum, we need an interplay of at least three terms, unlike the AdS vacua, which only need two terms\footnote{A refinement of TCC in terms of the potential leads to similar non-trivial results as TCC for negative potentials with AdS minima \cite{Andriot:2022brg}.}. 

\pagebreak

\section{Swampland VI: Finiteness and string lamppost}

The question that we study in this section is whether there could be infinitely many theories of quantum gravities? From the EFT perspective this question sounds unnatural because there is always a large number of inequivalent field theories. However, we will see that in string theory the number of possibilities seems to be finite which motivates this question in quantum gravity. 

\subsection{String lamppost and finiteness principles}

Before proceeding further, let us distinguish two related, but separated question. \begin{itemize}
    \item \textbf{Finiteness:} is the number of possible low energy EFTs in quantum gravity finite?
    \item \textbf{String Lamppost Principle: }Are all consistent quantm gravity theories low-energy limit of some string theory compacfitication? 
\end{itemize}

These are both very important questions. We draw a lot of our intuition about quantum gravity from string theory.  However, if there are other theories of quantum gravities, we might be getting the wrong kind of conclusions. This objection is often called the string lamppost effect. It refers to the possibility that string theory is like a lamppost than only lights up a small area, and searching within this bright spot might make us miss out on important physics outside of the lamppost's domain. However, if there is indeed only one theory of quantum gravity, then the we do not have to worry about this objection. The postulate that string theory is the only theory of quantum gravity is called the \textit{String Lamppost Principle (SLP)}.

The hypothesis that the answer to the first question is yes is called the \textit{Finiteness Principle} and the hypothesis that quantum gravity is unique is called \textit{String Lamppost Principle}. 

In the following we will try to gain some intuition about these questions and see how well-motivated finiteness principle and string lamppost principle are. 

let us start our search of theories with the highly supersymmetric theories in Minkowski space where we have more tools to limit the theory space. Let us start with theories with 32 supercharges which is the maximum number of supersymmetry\footnote{In theories with more supersymmetry, any massless multiplet would contain a massless particle with helicity $h>2$ which violates the Weinberg-Witten theorem \cite{Weinberg:1980kq}.}. 

\noindent \textbf{$N_{SUSY}=32$:}
\vspace{10pt}

The highest dimension for theories with 32 supercharges is $d=11$.

For $d=11$, supersymmetry completely determines the low-energy effective action. Although this proves the uniqueness of the low-energy theory, one could speculate that there could be infinitely many theories that are different at high energies but identical at low energies. This is a valid point and we cannot be sure about the uniquness of M theory, however, the fact that the webs of string dualities which usually capture non-perturbative aspects of the theory are pointing to the existence of a unique theory provides a strong reason to believe M theory is unique.

Note that the question of uniqueness of quantum gravity (String Lamppost Principle) cares about the UV physics, however, the question of finiteness is purely about the low-energy physics. Thus, as far as finiteness is concerned, we call two theories the same if they agree on the low-energy physics. In that sense, the 11d supergravity with $N=32$ is unique and the same is true in all the smaller dimensions because the supersymmetry fixes the content of the theory completely. 

In 10d, we can make two choices for the chirality of the supercharges. One corresponds to type IIA and the other to type IIB. The supersymmetry argument shows the finiteness for all theories with 32 supercharges, but how about the SLP? It turns out we get all these supergravities in string theory. They are just toroidal compactifications of M-theory. In fact, not only we see all the theories with 32 supercharges, but we see all of them in the same moduli space! For example, from the IIB/M-theory duality we know that even the type IIB supergravity corresponds to a particular limit in the moduli space of M-theory compactifications of M theory on $T^2$ where the area of the  $T^2$ goes to $0$. 

In string theory, theories of different dimensions can share the same moduli space. A $d+1$ dimensional theory can be thought of as a decompactification limit of a $d$ dimensional theory at which the $SO(1,d-1)$ Lorentz symmetry enhances to $SO(1,d)$ symmetry.

So we saw that for $N=32$, quantum gravity is finite and SLP is true. Now let us move on to a more non-trivial case.

\noindent \textbf{$N_{SUSY}=16$:}
\vspace{10pt}

Theories with 16 supercharges in a Minkowski background only exist in dimensions $d\leq10$. Let us start with $d=10$. The supercharges must form a Weyl representation of the Lorentz group. Thus, the supercharges and hence the spectrum of the theory is chiral. However, chiral theories typically have gauge, gravity, and mixed local anomalies and cancellation of all of these anomalies puts a strong constraint on the theories. 

There are only two types of multiplets in 10d supergravity; vector multiplet and the gravity multiplet. Therefore, the spectrum is uniquely determined by the gauge group. The anomaly cancellation puts a very strong bound on the structure of the gauge group. There are only 4 possibilities: $\{E_8\times E_8, SO(32), U(1)^{248}\times E_8, U(1)^{496}\}$. Only the first two are realized in string theory, so the SLP is making a prediction that the last two are inconsistent. We will come back to this prediction later. 

Let us go down another dimension to $d=9$. The two gauge theories that we know in 10d both have rank 16. If we compactify any of them on a circle, the matter (non-gravitational multiplets) of the 9d theory will have a gauge group of rank 17 (the extra rank comes from the $U(1)$ symmetry of the internal geometry). There are other ways to get a 9d supersymemtric theory as well. For example, we can compactify the $E_8\times E_8$ Heterotic theory on $S^1/\mathbb{Z}_2$, where the $\mathbb{Z}_2$ switches the two $E_8$s and acts on the circle by $x\sim x+\frac{1}{2}$. Note that the $\mathbb{Z}_2$ does not have a fixed point, therefore, this is not a compactification on an interval. Otherwise, we could not preserve the supersymmetry because the Heterotic theory is chiral. This theory will have rank $9$. We can also compactify M-theory on $S^1\times S^1$ moded out by $\mathbb{Z}_2$ which acts by a minus sign on the first $S^1$ and by a half circle shift on the second $S^1$. The internal manifold in this case is a Klein bottle and the resulting theory is a 9d N=16 theory with rank 1 \cite{Aharony:2007du}. We can compactify these 9d theories on  circle to find $8d$ supersymmetric theories with ranks $r\in \{2,10,18\}$. All of the 8d constructions have a natural F-theory embedding where F-theory is compactified on an elliptic K3. But the possible values for the ranks remains the same. So we observe that:

\begin{itemize}
    \item Supersymmetric 9d theories only have ranks 1,9, and 17.
    \item Supersymmetric 8d theories only have ranks 2,10, and 18.
\end{itemize}
But what about all of the missing ranks? Is the string lamppost missing them?

In 8 and 9 dimensions there is a nice argument based on global anomalies that fixes the rank of gauge group modulo 2. Note that here by global anomaly, we do not mean an anomaly of a global symmetry, but anomaly of a gauge transformation that cannot be continuously deformed to the identity transformation. It turns out in dimensions 8 and 9, there are diffeomorphisms that do not change the boundary, however, they could affect the sign of the fermion measure. Let us see how this might happen. Imagine having a Dirac spinor. If we integrate out a fermionic mode we find
\begin{align}
    \mathcal{Z}=\int \mathcal{D}\phi\mathcal{D}\mathcal{A}\hdots \int \mathcal{D}\psi\mathcal{D}\bar\psi e^{iS}=\int \mathcal{D}\phi\mathcal{D}\mathcal{A}\hdots e^{iS_\psi}\det[i\slashed{D}],
\end{align}
where $S_\psi$ does not depend on $\psi$ anymore. The $\det[i\slashed{D}]$ is the regularized product of all the eigenvalues of $i\slashed{D}$ on the background. If we replace the Dirac spinor with a Mjorana spinor, the numnber of degrees of freedom will be cut in half. Thus we find 
\begin{align}
    \mathcal{Z}=\int \mathcal{D}\phi\mathcal{D}\mathcal{A}\hdots e^{iS_\psi}\sqrt{\det[i\slashed{D}]}.
\end{align}
Suppose the fermionic measure $\det[i\slashed{D}]$ picks up a phase under a diffeomorphism which is not continually connected to identity. Take the two configurations that are mapped to each other under this diffeomorphism. They must be both included in the path integral. This is because although they cannot go to each other continuously in the gauge orbit, they can still continuously deform to each other outside their gauge orbit. Therefore, there is no sharp way to impose a selection rule that would only keep one representative out of every equivalency class.  But this could potentially create a problem. Suppose the $\sqrt{\det[i\slashed{D}]}$ factor picks up a $-1$ under the symmetry transformation. In that case the two gauge equivalent configurations that are both included will cancel each other out and the path integral will vanish! This is called a global anomaly and for a theory to be consistent, the fermionic part of the path integral measure must be invariant under any gauge symmetry transformation that is not homotopic to the identity transformation. 

In the case of $d=8$ and $d=9$, the undesired scenario that we described happens for some large diffeomorphism that is homotopic to the identity. However, the phase change is only non-zero for odd number of Majorana fermions. Thus, the total number of Majorana fermions in these dimensions must be even. In $d=8,9$ with 16 supercharges, the matter content is very constrained. The only multiplets are gravity multiplet and vector multiplet, therefore, we must have exactly $r$ vector multiplets where $r$ is the rank of the gauge group. This allows us to calculate number of the Majorana fermions in terms of $r$. Then we can use the global anomaly cancelation condition and we find $r\equiv 0~\mod2$ in $d=8$ and $r\equiv 1~\mod 2$ in $d=9$ \cite{Alvarez-Gaume:1983ihn}. 

This argument partially explains the limited set of ranks that we observe in the string theory constructions. But what about all the other missing ranks? 

Let us think about finiteness now. We observe that the list of gauge group ranks always ends at some number. Is that because we are missing infinitely many good theories, or is it that there is a fundamental reason why the list must be finite. 

Let us see if we can find a rule of thumb for the maximum allowed rank of the gauge group. From the 9d and 8d case it seems that a toroidal compactification is the most optimal option to obtain the maximum rank. If that is true, we would expect the rank of the gauge group of any gravitational theories in Minkowski space with 16 supercharges to be bounded by $r\leq 26-d$ where $d$ is the dimension of spacetime. 

This upper bound is satisfied in all known string theory constructions. But could there be a more basic explanation for it?  These are all important questions and we will come back to answer them later.

How about 6d theories? There are two kinds of theories with 16 supercharges, $\mathcal{N}=(2,0)$ or $(1,1)$. The first case is chiral and could have gauge and gravitational anomalies. In fact the anomalies are so constraining for the theory that they uniquely determine the low-energy theory. The next question is, does it belong to the Landscape? The answer turns out to be yes. This is type IIB on K3. You can also get it from M theory by putting M theory on $T^5/\mathbb{Z}_2$ where the $\mathbb{Z}_2$ flips all the circles \cite{Witten:1995em}. However, in that case you would need to assign half-unit of flux to each one of the fixed points of $T^5/\mathbb{Z}_2$ and have 16 $M_5$ branes in the $T^5$ to cancel the fluxes. The moduli of the theory in the M-theory picture is determined by the position of the M5 branes. 
\vspace{10pt}

\noindent \textbf{Exercise 1:} Using the dualities that we have discussed in the class, show that type IIB on K3 and M-theory on $T^5/\mathbb{Z}_2$ where the $\mathbb{Z}_2$ flips the signs of all the five circles are dual to each other. 
\vspace{10pt}

\noindent \textbf{$N_{SUSY}=8$:}
\vspace{10pt}

Now let us reduce the supersymemtry even further. Consider the minimally supersymmetric theory in 6d which has 8 supercharges. We can construct these theories in string theory by putting F-theory on elliptic CY threefolds. We can also construct these theories by putting Heterotic on K3, or type IIB on K3 orientifold. However, using the F-theory/Heterotic duality described before, it is easy to see that all of these constructions have F-theory embedding. 

The finiteness priniple, if true, would imply that there must be finite number of elliptic CY threefolds. In fact, this is a true but very non-trivial mathematical statement. So finiteness leads us on the right track in this case.

How about 5d theories? Eight supercharges is the minimal number of supercharges in 5d. Thus such a theory would be an $\mathcal{N}=1$ theory in 5d. To construct 5d $\mathcal{N}=1$ theories, we can put M theory on a Calabi--Yau threefold. Similar to the 6d case, the finiteness principle would tell us that the number of Calabi--Yau threefolds must be finite. This is a very non-trivial and famous conjecture by Yau. As of now, there is no known infinite family of Calabi--Yau threefolds. 

If we look at the moduli space of the CY's, there could be singularities with some loci (e.g. conifold). The Higgs mechanism allows us to go from one CY moduli space to another by traversing finite distance in the field space and crossing the loci of singular manifolds. Therefore, one can think of the moduli space of the CY compactification as a collection of many individual moduli spaces that are glued together on lower dimensional loci corresponding to singular CYs.

The uniqueness of quantum gravity (\emph{i.e.} string lamppost principle) would imply that the moduli space is unique and connected and thus all of the Calabi--Yau manifolds are connected via geometric transitions through singular manifolds. This is another non-trivial math conjecture known as Reid's fantasy.

Now let us think about the low energy EFTs. Usually the field theories around the singular loci have a limited field range. Therefore, if we impose a cutoff, the big moduli of string theory gets chopped off into many smaller pieces, each of which is well-approximated by a low-energy effective field theory. 

Note that these EFTs from the low-energy perspective look completely different and disconnected. However, a single theory of quantum gravity seem to be requiring many such EFTs to be patched together.

So far we only talked about the Minkowski space, now let us discuss the finitness principle and the string lamppost principle in AdS.

\subsection{Finiteness principle in AdS} 

There is an obvious potential counterexample to the naive version of the finiteness principle in AdS. There are infinitely many CFTs and infinitely many of them are expected to have holographic dual, so we get infinitely many quantum gravities. For example different $\mathcal{N}=4$ SYM theories with different $SU(N)$ gauge groups all are holographic dual to $AdS_5\times S^5$ quantum gravities. However, the cosmological constant of these theories is different since $l_{AdS}\sim N^2$. Another example is $AdS_7\times S^4$ which is holographic dual of $N$ parallel $M5$ branes. But one can show that in that case too the cosmological constants are different for different values of $N$.

\vspace{10pt}

\noindent \textbf{Exercise 2:} In the M-theory construction of $AdS_7\times S^4$, estimate $l_{AdS}$ in terms of $N$ (the number of $M5$ branes).
\vspace{10pt}

Consider $AdS_7\times S^4/\mathbb{Z}_n$ where the $\mathbb{Z}_n$ acts on the sphere by rotation. The $\mathbb{Z}_n$ action has two fixed points at the poles. So we have two $A_{n-1}$ singularities on the sphere. Putting M-theory on $S^4/\mathbb{Z}_n$ gives us an $SU(n)\times SU(n)$ gauge theory where each $SU(n)$ corresponds to one of the $A_{n-1}$ singularities. Since the curvature of the sphere is the same for all these theories, they have the same AdS scale. Therefore, we have an infintie familiy of solutions that have different matter content but same cosmological constant. One might think this is surely a counter example to the finiteness principle in AdS! 

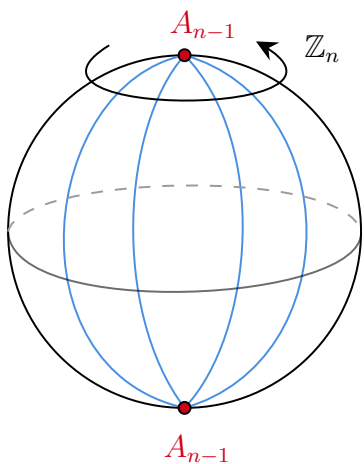
\begin{figure}[H]
    \centering

\tikzset{every picture/.style={line width=0.75pt}} 

\begin{tikzpicture}[x=0.75pt,y=0.75pt,yscale=-1,xscale=1]

\draw   (238,153) .. controls (238,103.85) and (277.85,64) .. (327,64) .. controls (376.15,64) and (416,103.85) .. (416,153) .. controls (416,202.15) and (376.15,242) .. (327,242) .. controls (277.85,242) and (238,202.15) .. (238,153) -- cycle ;
\draw [color={rgb, 255:red, 74; green, 144; blue, 226 }  ,draw opacity=1 ]   (327,242) .. controls (238,222) and (254,79) .. (327,64) ;
\draw [color={rgb, 255:red, 74; green, 144; blue, 226 }  ,draw opacity=1 ]   (327,242) .. controls (410,218) and (408,87) .. (327,64) ;
\draw [color={rgb, 255:red, 74; green, 144; blue, 226 }  ,draw opacity=1 ]   (327,242) .. controls (362,205) and (370,108) .. (327,64) ;
\draw [color={rgb, 255:red, 74; green, 144; blue, 226 }  ,draw opacity=1 ]   (327,242) .. controls (290,193) and (295,100) .. (327,64) ;
\draw [color={rgb, 255:red, 0; green, 0; blue, 0 }  ,draw opacity=0.58 ]   (238,153) .. controls (240,198) and (418,189) .. (416,153) ;
\draw [color={rgb, 255:red, 155; green, 155; blue, 155 }  ,draw opacity=1 ] [dash pattern={on 4.5pt off 4.5pt}]  (238,153) .. controls (245,124) and (408,120) .. (416,153) ;
\draw    (289,59) .. controls (227.62,96.62) and (430.87,96.02) .. (365.08,58.16) ;
\draw [shift={(363,57)}, rotate = 28.44] [fill={rgb, 255:red, 0; green, 0; blue, 0 }  ][line width=0.08]  [draw opacity=0] (10.72,-5.15) -- (0,0) -- (10.72,5.15) -- (7.12,0) -- cycle    ;
\draw  [fill={rgb, 255:red, 208; green, 2; blue, 27 }  ,fill opacity=1 ] (324,64) .. controls (324,62.34) and (325.34,61) .. (327,61) .. controls (328.66,61) and (330,62.34) .. (330,64) .. controls (330,65.66) and (328.66,67) .. (327,67) .. controls (325.34,67) and (324,65.66) .. (324,64) -- cycle ;
\draw  [fill={rgb, 255:red, 208; green, 2; blue, 27 }  ,fill opacity=1 ] (324,242) .. controls (324,240.34) and (325.34,239) .. (327,239) .. controls (328.66,239) and (330,240.34) .. (330,242) .. controls (330,243.66) and (328.66,245) .. (327,245) .. controls (325.34,245) and (324,243.66) .. (324,242) -- cycle ;

\draw (386,52.4) node [anchor=north west][inner sep=0.75pt]    {$\mathbb{Z}_{n}$};
\draw (318,40.4) node [anchor=north west][inner sep=0.75pt]  [color={rgb, 255:red, 208; green, 2; blue, 27 }  ,opacity=1 ]  {$A_{n-1}$};
\draw (315,254.4) node [anchor=north west][inner sep=0.75pt]  [color={rgb, 255:red, 208; green, 2; blue, 27 }  ,opacity=1 ]  {$A_{n-1}$};

\end{tikzpicture}
    \caption{If we mod out the $S^4$ by $\mathbb{Z}_n$, we get two $A_n$ singularities at the poles. Each of these singularities will contribute an $SU(n)$ gauge theory.}
\end{figure}

Let us take a moment to study the holographic dual of the above theory. Consider M-theory on $\mathds{C}^2/\mathbb{Z}_n$. The locus of the singularity is seven dimensional. Suppose we want to probe the singularity with a stack of $N$ parallel M5 branes. The M5 brane's wolrdvolume is 6 dimensional, therefore, it has one normal dimension in the locus of the singuliary. We can show the normal direction to the M5 brane in the singularity as a line and the stack of M5 branes on the singulairy as a point on that line.

Since the locus is the fixed point of the $\mathbb{Z}_n$ action, the theory on the singular locus has a global $\mathbb{Z}_n$ symmetry. After placing the stack of $N$ M5 branes, we will have two $SU(n)$ actions, one one the left half-line and the other on the right half-line. These actions turn out to induce an $SU(n)\times SU(n)$ global symmetry on the worldvolume theory of the stack of $N$ M5 branes. This SCFT is the holographic dual of M-theory on $AdS_7\times S^4/\mathbb{Z}_n$. 

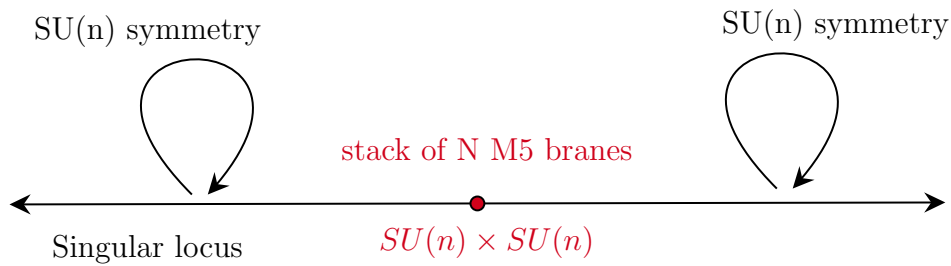
\begin{figure}[H]
    \centering

\tikzset{every picture/.style={line width=0.75pt}} 

\begin{tikzpicture}[x=0.75pt,y=0.75pt,yscale=-1,xscale=1]

\draw    (108,156.99) -- (574,156.01) ;
\draw [shift={(577,156)}, rotate = 179.88] [fill={rgb, 255:red, 0; green, 0; blue, 0 }  ][line width=0.08]  [draw opacity=0] (10.72,-5.15) -- (0,0) -- (10.72,5.15) -- (7.12,0) -- cycle    ;
\draw [shift={(105,157)}, rotate = 359.88] [fill={rgb, 255:red, 0; green, 0; blue, 0 }  ][line width=0.08]  [draw opacity=0] (10.72,-5.15) -- (0,0) -- (10.72,5.15) -- (7.12,0) -- cycle    ;
\draw    (197,152) .. controls (107.45,62.45) and (287.19,60.02) .. (206.24,150.63) ;
\draw [shift={(205,152)}, rotate = 312.4] [fill={rgb, 255:red, 0; green, 0; blue, 0 }  ][line width=0.08]  [draw opacity=0] (10.72,-5.15) -- (0,0) -- (10.72,5.15) -- (7.12,0) -- cycle    ;
\draw    (492,149) .. controls (402.45,59.45) and (582.19,57.02) .. (501.24,147.63) ;
\draw [shift={(500,149)}, rotate = 312.4] [fill={rgb, 255:red, 0; green, 0; blue, 0 }  ][line width=0.08]  [draw opacity=0] (10.72,-5.15) -- (0,0) -- (10.72,5.15) -- (7.12,0) -- cycle    ;
\draw  [fill={rgb, 255:red, 208; green, 2; blue, 27 }  ,fill opacity=1 ] (337.5,156.5) .. controls (337.5,154.57) and (339.07,153) .. (341,153) .. controls (342.93,153) and (344.5,154.57) .. (344.5,156.5) .. controls (344.5,158.43) and (342.93,160) .. (341,160) .. controls (339.07,160) and (337.5,158.43) .. (337.5,156.5) -- cycle ;

\draw (125,171) node [anchor=north west][inner sep=0.75pt]   [align=left] {Singular locus};
\draw (116,60) node [anchor=north west][inner sep=0.75pt]   [align=left] {SU(n) symmetry};
\draw (463,57) node [anchor=north west][inner sep=0.75pt]   [align=left] {SU(n) symmetry};
\draw (271,122) node [anchor=north west][inner sep=0.75pt]  [color={rgb, 255:red, 208; green, 2; blue, 27 }  ,opacity=1 ] [align=left] {stack of N M5 branes};
\draw (290,166.4) node [anchor=north west][inner sep=0.75pt]    {$\textcolor[rgb]{0.82,0.01,0.11}{SU( n) \times SU( n)}$};

\end{tikzpicture}
    \caption{We consider a stack of N M5 branes that probe an $A_n{n-1}$ singularity which is 7 dimensional. The line represents the normal direction to the M5 branes in the locus of singularity.}
\end{figure}
One resolution to the apparent violation of the finiteness principle is that all of these AdS constructions, there is an extra space which is as big as the AdS itself. Since the scales of the AdS is always correlated with the extra space, we should always view them together and cannot think of it as a lower dimensional theory. In that sense, the finiteness principle, should be understood as the finiteness of the higher dimensional low-energy EFT which is correct in all these examples. From this perspective, infinite families of lower dimensional theories correspond to infinite families of defects   (singularities and branes) in the same theory, which is fine. 

Note that this resolution only applies if the scale of AdS and some length scale of the internal geometry are correlated. If one could take the internal space to be arbitrarily small while the AdS scale is kept fixed, the resulting theory would truly be lower dimensional. Such an AdS is called a scale-separated AdS which is in tension with the AdS distance conjecture. 

There is another resolution which leads to a sharper formulation of finiteness principle in AdS. Up to now we did not consider the cut-off of the low-energy EFT. We can think of the finiteness principle as the following statement. For a fixed EFT cut-off $\Lambda_{\text{cut-off}}$, there are finitely many AdS with cut-off $\Lambda_{\text{cut-off}}$. 

We can use the AdS distance conjecture and holography to state the finiteness principle in the CFT language. The AdS distance conjecture tells us that there is a tower of state with masses $m\sim \Lambda^\alpha$ in Planck units. For the EFT description to work, we need $\Lambda_{\text{cut-off}}\lesssim m \sim \Lambda^\alpha$. On the other hand, we can express the cosmological constant in terms of the central charge of the dual CFT as $\Lambda\sim c^{-\frac{2}{d-2}}$. Combining these equations leads to 
\begin{align}\label{csb}
c<\Lambda_{\text{cut-off}}^{-\frac{d-2}{2\alpha}}.
\end{align}

The EFT cut-off also bounds the CFT central charge. This clarifies how imposing a cut-off can make the number of theories finite. Even though the inequality \eqref{csb} is only for CFTs with gravitational holographic dual, it is plausible that it is correct for all CFTs. If so, the finiteness principle is making a prediction, that the number of CFTs with a central charge lower than a cut-off is finite. 

In retrospect, the introduction of cut-off was necessary. To see why, consider a large number of non-supersymmetric compactifications such that their internal geometries are different by small perturbations. For small-wavelength perturbations to the internal geometry, the lower-dimensional EFTs will only change in the UV. Therefore, imposing a cut-off would make the number of low-energy theories finite. 

We provided strong evidence for finiteness principle in Minkowski and AdS spaces. In the following subsection, we go back to the SLP. In the first subsection, there were a few examples of supergravity theories that were non-anomalous but did not have any string theory realization. In the following, we show why Swampland conditions rule out those theories. 

\subsection{String lamppost principle from brane probes}

Let us start with the 10d supergravities. Anomaly allows a theory with $U(1)^{496}$ gauge symmetry. What could be the problem with it? We show that we can a lot of constraints from the consitency of the branes in quantum gravity. But first, how do we know we have branes? 

The branes are usually required by the Swampland completeness princinple. We saw many examples of this requirement in the section on Cobordism conjecture and also the completeness hypothesis. However, it is not so easy to just add a brane. When we add a brane, we want the  QFTs on the worldvolume of the brane be unitary and consistent. It turns out that the unitarity and other consistency conditions on the brane rule out many possibilities. In some sense, the brane probes bootstrap the Swampland program. 

In this section we use a stronger version of the completeness principle for supersymmetric theories. We assume that if a brane is required by completeness, and it can be BPS, then such a BPS brane has to exist. This is the generalization of completeness of spectrum of gauge theories to gauged supersymmetry (supergravity). There is no known counterexample to this statement. For example, in Narain compactification of the Heterotic string theory, in every direction BPS of charge lattice, there is always a BPS particle. 

In dimensions higher than 6, the supergravity string (that coupled to $B_{\mu\nu}$ in the gravity multiplet can be BPS. We will consider the theory living on the worldsheet of this BPS 1-brane. Let us review a few key properties of 2d CFT. Consider a 2d CFT with left and right central charges and some current algebras corresponding to global symmetry $G$. The current algebra gives $c=k\dim(G)/(k+\hat h_G)$. The current algebra could be either on the left or right moving part. Suppose it is on the left, we get $c_G\leq c_L$. This is because any extra piece that satisfies unitarity has positive contribution to $c$. So, if we have an argument that the gauge group appears as a current algebra on the worldsheet theory and that there is a bound for the Virasoro central charge, we find a bound on the rank of the gauge group.

Also, if we have a representation, the dimension of the representation is $c_2(R)/(k+\hat h_G)$. So, if we have a representation that is Higgsed, it should appear as a relevant operator on the worldsheet theory with dimension less than 1 to be a relevant deformation. 

Let us also review anomalies in 2d theories. We consider two classes of anomalies, gravitational anomalies and global symmetry anomalies. Let us start with the anomaly of global symmetry. The global symmetry we consider is the gauge symmetry of bulk which is realized as a global symmetry on the brane. The anomaly diagram is a two point function. Therefore, the change in the effective action of the brane under a symmetry transformation with gauge parameter $\epsilon$ is
\begin{align}
    S_{2d}\rightarrow S_{2d}-K_{gauge}\int \tr[\epsilon {F}],
\end{align}
where $K_{gauge}$ is some number and ${F}$ is the gauge field tensor on the worldsheet. This anomaly is due to the non-invariance of the path integral measure of the 2d theory. On the other hand, we know that the 2d theory is coupled to $B_{\mu\nu}$. Thus the worldsheet action has an external coupling $\int B$. For the worldsheet theory to be non-anomalous, we need the change of $B$ under the gauge transformation be $K_{gauge}\tr[\epsilon \mathcal{F}]$. If we take the exterior derivative of both expression we find
\begin{align}
    \delta{H}=K_{gauge}\delta{\tr[A\wedge {F}]},
\end{align}
where $\delta$ denotes the change under a gauge transformation in the bulk. Taking another exterior derivative from both sides gives
\begin{align}
        \delta{dH}=K_{gauge}\delta(\tr[{F}\wedge{F}]).
\end{align}
Similarly, for the gravitational anomaly, we find 
\begin{align}
    \delta(dH)=K_{gravity}\delta(\tr[\mathcal{R}\wedge\mathcal{R}]).
\end{align}
In fact the equations of motion in the bulk tell us that
\begin{align}
dH=\frac{1}{2}(\tr\mathcal{R}\wedge\mathcal{R}-\tr F\wedge F).
\end{align}
The right hand side are contributions of the topological terms in the supergravity action to the equations of motion. The presence of these terms are required by anomaly cancellation. By mathcing the coefficients of $\mathcal{R}\wedge\mathcal{R}$ and $F\wedge F$ in the equations, we find
\begin{align}
    K_{gauge}=-\frac{1}{2},\quad
    K_{gravity}=\frac{1}{2}.
\end{align}
On the other hand, $K_{gauge}$ and $K_{gravity}$ can be calculated in terms of the worldsheet theory. Their values are
\begin{align}
    K_{gauge}=\frac{k_R-k_L}{2},\quad
    K_{gravity}=\frac{c_L-c_R}{24}.
\end{align}
So, we find
\begin{align}\label{amcp}
    k_L-k_R=1,\quad
    c_L-c_R=12.
\end{align}

The study of brane probes via cancellation of the anomalies of the bulk symmetries is called the anomaly inflow which is a very powerful method.

\noindent \textbf{Exercise 3:} Consider minimal supergravity in 10d. Show that if the supergravity string (\emph{i.e.} the string that couples to two-form $B_{\mu\nu}$ in the gravity multiplet) is supersymmetric, it will have $(0,8)$ supersymmetry. 
\vspace{10pt}

\noindent \textbf{Exercise 4:} Suppose you have a p-brane where p is odd. Suppose the p-brane is charged under a p-form gauge potential $A_{p+1}$ and $dF_{p+2}$ is not identically zero (is some function of curvature, etc.). In that case $F_{p+2}-dA_{p+1}$ could be a non-zero topological term. Moreover, this term might not be gauge invariant in the sense that when integrated on a $p+2$ dimensional surface $\mathcal{M}$ with a $p+1$ dimensional boundary $\Sigma$, the result changes under gauge transformation as 
\begin{align}
    \delta_\epsilon\int_\mathcal{M}F_{p+2}-dA_{p+1}=\int_\Sigma \epsilon X_{p+1}
\end{align} 
Show that $X_{p+1}$ must match the anomaly of worldvolume theory on the brane under the action of the spacetime gauge group which realizes as a global symmetry on the brane. 
\vspace{10pt}
 
Now let us go back to the 10d $\mathcal{N}=(1,0)$ with a gauge group $G$. The anomaly inflow tells us $k_l-k_r=1$ and $c_L-c_R=12$. Note that we have $(0,8)$ supersymmetry which has an SO(8) R-symmetry on the righmoving side. Moreover, the level of the R-symmetry is proportional to the central charge. 
\begin{align}
    c_R=12 \kappa,
\end{align}
where $\kappa$ is the level of the R-symmetry. For the supergravity string, the R-symmetry has a very physical meaning. It corresponds to the SO(8) rotation in the transverse direction to the string. In fact, we can find $\kappa$ by looking at the induced action from the $\mathcal{R}\wedge\mathcal{R}$ term. The coefficient of this term turns out to be the level $\kappa$. Thus, we find $\kappa=1$ and $c_R=12$. Note that $c_R=12$ is exactly the contribution of the 8 bosonic transverse modes plus their fermionic counterparts on the worldsheet. Plugging this into \eqref{amcp} gives
\begin{align}
    c_L=24~~~\&~~~c_R=12.
\end{align}
Suppose we subtract the contribution of the transverse oscilations from the central charges to define $\hat c_L=c_L-8$ and $\hat c_R=c_R-12$. The difference is due to the fact that there is no supersymmetry on the leftmoving sector. The reduced central charges are
\begin{align}
        \hat c_L=16~~~\&~~~\hat c_R=0.
\end{align}
From $c_R=0$ we find that the theory on the rightmoving sector is trivial. Therefore, $k_R=0$. Plugging this into \eqref{amcp} leads to
\begin{align}
        \hat k_L=1~~~\&~~~\hat k_R=0.
\end{align}
Therefore, the worldsheet theory is completely leftmoving. Moreover, from $c=k\dim(G)/(k+\hat h_G)$ we know $\text{rank}(G)\leq c$ which implies $\text{rank}(G)\leq 16$. Therefore, the $G=U(1)^{496}$ theory or the $G=E_8\times U(1)^{248}$ are inconsistent and belong to the Swampland. 

We can try to do the same analysis in lower dimensions. The supergravity equations of motion tell us $dH=\frac{\kappa}{2} \tr \mathcal{R}\wedge \mathcal{R}-\frac{1}{2}\tr F\wedge F$. Similar arguments as before tells us that $c_R=12\kappa$ and $c_L=24\kappa$. The theory on supergravity string is still $(0,8)$ and we have at least a $U(1)$ R-symmetry which corresponds to rotation in the transverse direction. The highest R-charge $S$ that appears on the string is related to the level by $S=2\kappa$. Since the R-symmetry is the spacetime rotation, the R-charge is the spin. Now we argue that $\kappa\leq1$. Let us compactify the theory on a circle. We can wind the string around the circle to get a BPS particle. In the limit where the circle shrinks to zero size, the mass of the string excitations goes to zero. Suppose this limit exists (postulated by a strong version of distance conjecture), the limiting theory has a massless particle with spin $2\kappa$. According to Weinberg-Witten theorem, the spin of this massless particles must be less than or equal to $2$. Thus, $\kappa\leq 1$.

There are two possibilities, $\kappa=0$ or $\kappa=1$. If $\kappa=1$, we have $c_L=24$ and $c_R=12$. In d spacetime dimensions, we have $d-2$ transverse dimensions. Therefore, after subtracting the contribution from the center of mass, we find $c_L=26-d$. Since the central charge is less than $\text{rank}(G)$, we find $\text{rank}\leq 26-d$. The highest rank is realized by Narain compactification of Heterotic theory.

If $\kappa=0$, we have a contradiction, because the central charges are $0$. The resolution is that all of our calculations was based on the assumption that the supergravity string has $(0,8)$ supersymmetry. However, it could be that the worldsheet theory has enhanced supersymmetry in the infrared. For $\kappa=0$, the IR theory on the string must be a $(8,8)$ theory. In this case, we have two R-symmetries. Using the spin argument above, we can say the level of the R-symmetry on each side $\kappa_{R,L}$ is at most 1 and the central charge of each side is $c_{L,R}=12\kappa_{L,R}$. Due to the center of mass modes on each side, the central charge cannot be $0$. So we find $\kappa_{L,R}=1$ and $c_{L,R}=12$. If we subtract the contribution of the center of mass modes, we find $\hat c_{L,R}=10-d$ which must be greater than the rank of the gauge group. So we find 
\begin{align}
    \kappa=0&:~~\text{rank}(G)\leq10-1\nonumber\\
    \kappa=1&:~~\text{rank}(G)\leq26-d.    
\end{align}

Both of these bounds can be saturated. For example, 9d string theories with $\text{rank}=1$ have $\kappa=0$ (M-theory on Klein bottle).

Note that, a stack of $N$ D3 branes with arbitrarily high $N$ is not a counter example to this result. Because, in that case, the gravity is not confined to the brane. The inequality $\text{rank}\leq 26-d$ applies to theories where the gravity is $d$ dimensional.

The anomaly inflow argument gave us an upper bound on the rank, but as we discussed before, the list of the available ranks in string theory is much more restricted. In 9d, the available ranks are $\{1,9,17\}$ and in 8d the available ranks are $\{2,10,18\}$. How can we explain the absence of the other ranks? 

An explanation for this list was given in \cite{Montero:2020icj} based on the cobordism conjecture. The argument has many details but we summarize the main idea here. In \cite{Montero:2020icj}, it was argued that supergravities with $d>6$ must have parity symmetries (sometimes more than 1). For this parity symmetry to be broken, we must be able to compactify the theory on non-orientable manifolds with appropriate $\text{Pin}$ structure. We focus on compactifications on non-orientable 2d manifolds. There are 8 different cobordism classes for these 2d manifolds that are generated by  $\mathbb{R}\mathds{P}^2$. Cobordism conjecture tells us that each cobordism class must be trivializable. This means, we must be able to have an end of the universe wall for the $\mathbb{R}\mathds{P}^2$ compactification. Such a wall, would be a 7d defect. Moreover, since, we can detect the presence of this defect from large distances (due to the $\mathbb{R}\mathds{P}^2$ boundary in the transverse directions) it must carry a $\mathbb{Z}_8$ gauge charge. Therefore, if we put 8 of these defects together to cancel the gauge charge, we should be able to find a compact singular 3-manifold with 8 defects which is an allowed internal geometry. This is nothing other than $T^3/\mathbb{Z}_2$. Each one of the 8 fixed points correspond to one of the defects. Therefore, the Cobordism conjecture implies that supergravity theories in dimensions greater than 6 must have a consistent compactification on $T^3/\mathbb{Z}_2$. The condition that the resulting lower dimensional theory be anomaly free imposes a strong constraint on the matter content. 
\begin{align}
    d=9:&~~\text{rank}\equiv1~\mod 8\nonumber\\
    d=8:&~~\text{rank}\equiv2~\mod 8\nonumber\\
    d=7:&~~\text{rank}\equiv1~\mod 2.
\end{align}

This is a remarkable consistency check for SLP. However, the EFTs that we get in string theory have more structure to them than the rank of the gauge group. For example, consider the 8d theories. All of the known 8d theories in string theory have an F-theory construction. That means every 8d theory comes with an elliptic K3. But where is the K3 in the low-energy theory? 

The theory in 8d has a gauge group which has $3+1$ dimensional instantons. We can study the classical moduli space of the gauge instantons. There are two branches in the moduli space, Higgs branch and the Coulomb branch. The Coulomb branch corresponds to zero size instantons (also called small instantons). The theory living on the small instantons is an $\mathcal{N}=2$ 4d theory. We assume that the rank of the gauge theory which is the dimension of the couomb branch is 1. Thus, there is a $U(1)$ living on the brane with a coupling $\tau$. 

From supersymmetry, we know that the geometry is hyperKähler and has one complex dimension. Moreover, the geometry of the coulomb branch must be compact. Suppose it is not compact. If we compactify the theory on $T^3$, the eigenvalues of the Laplacian on the moduli space correspond to the spectrum of particles in lower dimension. If the moduli space is non-compact, the spectrum of the laplacian will be continious which is in contradiction with the finiteness of black hole entropy. Therefore, the moduli space is a compact hyperKähler and $\tau$ gives you an elliptic fibration of K3 over the Coulomb branch. We have thus found the internal geometry of string theory in the EFT using Swampland conditions.

\subsection{Finiteness of light species}

The black hole entropy formula suggsests there is only a finite number of states at any given energy. If we have infinite or arbitrarily large number of species, the entropy formula would look problematic. So the fact that we do not have arbitrarily large number of species also seem to be related to black holes. However, there is a loophole in this argument as we now discuss.

Suppose the highest energy scale where the EFT is valid is $\Lambda$. The smallest black hole we can describe using EFT must have a curvature $\mathcal{R}\lesssim \Lambda^{2}$ where $\mathcal{R}$ here represents the order of magnitude of the Riemann curvature tensor. Suppose the entropy of this black hole is $S$, we find
\begin{align}
    \Lambda<\frac{1}{S^\frac{1}{d-2}}
\end{align} 
If we have $N_{species}$ species of particles with masses below $\Lambda$ that are sufficiently gapped (like a KK tower), the number of high energy states is grows exponentially with $\gtrsim N$. Therefore, we find $S\geq N$ and 
\begin{align}\label{SB}
    \Lambda<\frac{1}{N_{species}^\frac{1}{d-2}}.
\end{align} 
This therefore suggests that the black hole entropy is not a good argument for bounding $N_{species}$. 

The number of species below energy $E$ is defined such that the number of high-energy states grows like $\gtrsim \exp(c N_{species}(E))$. If the tower of state is very dense (like a the tower of string excitations) the number of species is much smaller than the number of particles. For string tower, it is known that the number of states with energy $E$ grows like $\exp(cE/M_s)$. Therefore, $N_{species}\propto E/M_s$. This is while the number of one-particle string excitations is exponential. What happens for string tower is that the numebr of string excitations grows so rapidly that the majority of the states at any given energy scale $\gg M_s$ are dominated by one-particle states. This is different from the KK tower where the number of KK particles grows polynomially with energy. However, in KK tower too we have $N_{species}\propto E/M_{KK}$. The inequality \eqref{SB} is called the species bound. It suggests that we can have arbitrarily large number of species as long as the cut-off is small enough. We can use the inequality in the opposite direction, which is to bound $\Lambda$ from the number of species. The highest EFT cutoff $\Lambda$ is often called the species scale. 

If we compactify a $D$ dimensional manifold down to $d$ dimensions, if the manifold is big we get a KK tower of light species. $M_{P,D}^{D-2}\vol(M)=M_{P,d}^{d-2}$. The higher dimensional $M_P$ is below the lower dimensional $M_P$. So, there is a scale lower than the $M_P$ which prevents us from going all the way to the higher dimensional $M_P$. This could serve as the species scale. The number of KK light states is given by $N\sim(M_{P,D}/M_{KK})^{D-d}$. Then we find $NM_{P,D}^{D-2}=M_{P,d}^{d-2}$ which exactly saturates the species bound. Therefore, the species bound predicts the correct cut-off for KK theories.

Another example is 10d string theory. We have $M_P^8=M_S^8/g_S^2$. We expect the species scale to be string scale. The species scale is the radius where particles become black holes and that is exactly the string scale. Following species bound, we find $N(E=\Lambda)=\frac{1}{g_s^2}$. But why should that be true in string theory. 

The Hagedorn entropy tells us that $S(E)\sim E/M_s$. At the correspondence point $E\sim M_s/g_s^2$ so we get the $S\sim 1/g_s^2$, and since this is the scale where the black hole description should take over, everything hangs together. 

Therefore, the species inequality is saturated by KK reduction and weakly coupled string theory. Suppose we try to push the species scale up. The species bound tells us that in the asymptotic of the field space where the number of species is large, the species scale would be small. Therefore, to push up the cut-off as high as possible, we should explore the interior of the moduli space. The biggest gap we can hope for is Planck mass which is realized for 11d supergravity. In M-theory, we have a desert in term of the mass spectrum of the particles. Species bound tells us that we should aim for a very small number of species. Let us motivate this observation independently.

Consider F theory on elliptic threefolds. We can use $D3$ brane and wrap it around two dimensional Riemann surfaces to get low dimensional BPS particles. The tension of the branes will go like the area which is some function of Kähler moduli. In Planck units we have $C_{ij}t_i t_j\sim 1$ where $C_{ij}$ is the metric on the on the Kähler moduli space. Suppose we have $N$ Kähler classes and a diagonal $C_{ij}$,  we find $Nt^2\sim 1$ where $t$ is the average volume of a 2-cycle in Planck units. So the masses of the excitations of the BPS string go like $m\sim T^{-1}\sim N^{-\frac{1}{4}}$ where $T$ is the string tension. This inequality saturates the species bound \eqref{SB} in $d=6$. 

\vspace{10pt}

\noindent \textbf{Exercise 5:} Take M-theory on a Calabi--Yau threefold. The M2 branes can wrap around Riemann surfaces and M5 branes can wrap around divisors. Give a heuristic argument for why you might expect the masses of the corresponding BPS particles to go like $m\sim N^{-\frac{1}{3}}$ where $N$ is the number of Kähler classes. 
\vspace{10pt}

\noindent\textbf{Exercise 6:} consider D3 branes around cycles of CY threefods in 4d. The complex structures do not receive quantum corrections. Show that the correponding particles saturate the species bound too. 

\vspace{10pt}

In the argument we assumed that $C_{ij}$ is diagonal which is not necessarily true. You can check explicitly for the known examples, that the mass of the BPS states vs the number of species follows a sharp line in the log-log plot but with a different slope than that of the species bound \cite{Long:2021jlv}.

\begin{figure}[H]
    \centering
    \includegraphics{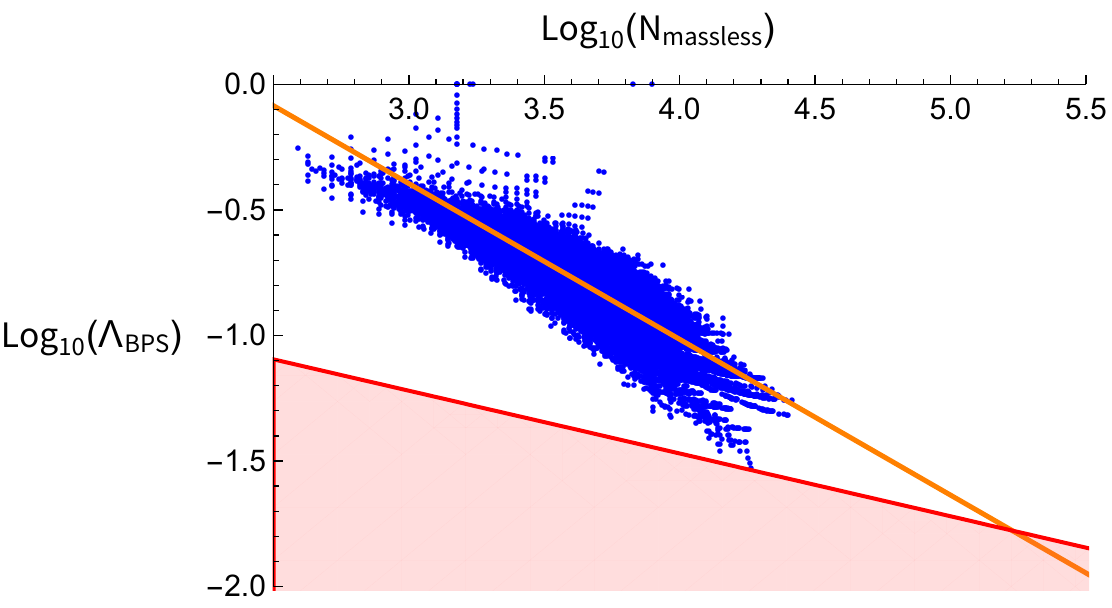}
    \caption{Plot taken from \cite{Long:2021jlv} shows the number of light species vs the mass scale of the BPS states for a large family of Calabi--Yau compactifications. The red region is excluded by the species bound.}
    \label{fig:my_label5}
\end{figure}

If the curve of the BPS states continues passed its intersection with the species curve, we get a contradiction with the species bound. Therefore, the species bound, together with observed CY examples, seems to suggest that the curve must stop and the number of possibilities are finite. Thus, the finiteness of vacua and massless species may indeed be related to the finiteness of the black hole entropy.

It might be tempting to postulate that the finiteness of the quantum gravity path integral is related to the finiteness principle. We claim that is in fact correct. Consider a cutoff $\Lambda$. Compactify your theory all the way to 1 dimension of time. Given that the number of non-compact dimensions is small, the moduli no longer freeze and there is no superselection. Therefore, we have to integrate over everything. If the number of Calabi--Yau spaces is infinite, the zero modes would likely give rise to a divergent path integral. Therefore, the finiteness of theories of quantum gravity is motivated by the finiteness of the quantum gravity path-integral. 

\section*{Acknowledgement}

The research of A.~B.~and C.~V.~is supported by a grant from the Simons Foundation (602883, CV) and by the NSF grant PHY-2013858. M.~J.~K.~is supported by a Sherman Fairchild Postdoctoral Fellowship and the U.S.~Department of Energy, Office of Science, Office of High Energy Physics, under Award Number DE-SC0011632.

\bibliographystyle{unsrt}
\bibliography{References}
\end{document}